\documentclass{ifgw}

\usepackage{lmodern}

\usepackage{amsmath}
\usepackage{cite}
\usepackage{lipsum}
\usepackage{pdfpages}

\usepackage{float}
\usepackage{amssymb}

\usepackage{tikz}								

\usepackage{tabularx}
\usepackage{booktabs,multirow}

\def\lsim{\raise0.3ex\hbox{$<$\kern-0.75em\raise-1.1ex\hbox{$\sim$}}}

\def\gsim{\raise0.3ex\hbox{$>$\kern-0.75em\raise-1.1ex\hbox{$\sim$}}}

\newcommand{\slashed}[1]{\!\not{\!#1}}

\author{Reinaldo Francener}
\city{Campinas}
\cosupervisor[Coorientador]{Donato Giorgio Torrieri}
\logo[align=left]{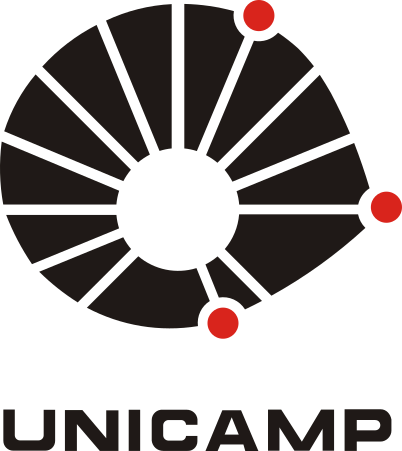}
\supervisor[Orientador]{Diego Rossi Gratieri}

\title{
    english={Lepton interactions from GeV to EeV},
    portuguese={Interações de léptons de GeV à EeV},
}

\subtitle{
    english={},
    portuguese={},
}

\institute{
    english={Gleb Wataghin Institute of Physics (IFGW)},
    portuguese={Instituto de Física Gleb Wataghin (IFGW)},
}

\program{
    english={Physics Graduate Program},
    portuguese={Programa de Pós-Graduação em Física},
}

\university{
    english={University of Campinas (UNICAMP)},
    portuguese={Universidade Estadual de Campinas (UNICAMP)},
}

\backcover{
    portuguese declaration={
        Tese apresentada ao \theinstitute~da \theuniversity~como
        parte dos requisitos exigidos para a obtenção do título de
        \uppercase{Doutor em Ciências}, na Área de \uppercase{Física}.
    },
    english declaration={
        Thesis presented to the \theinstitute[e]~of the \theuniversity[e]~in
        partial fulfillment of the requirements for the de\-gree of
        \uppercase{Doctor of Science}, in the area of \uppercase{Physics}.
    },
    version={
        Es\-te tra\-ba\-lho cor\-res\-pon\-de à ver\-são fi\-nal da
        te\-se de\-fen\-di\-da.
    },
    language=english,
}

\begin{document}

\maketitle
\makebackcover

\includepdf[pages=-]{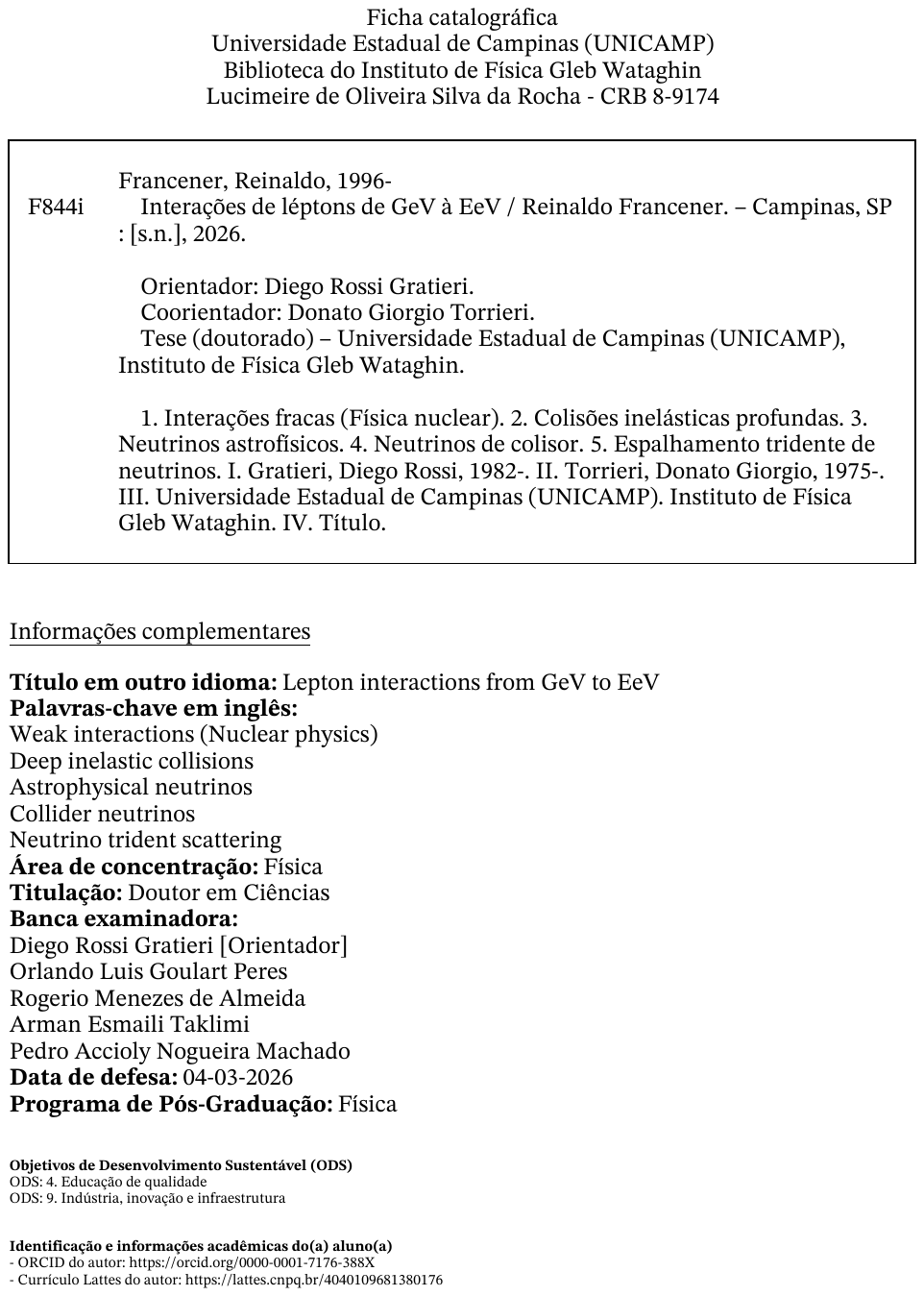}
\includepdf[pages=-]{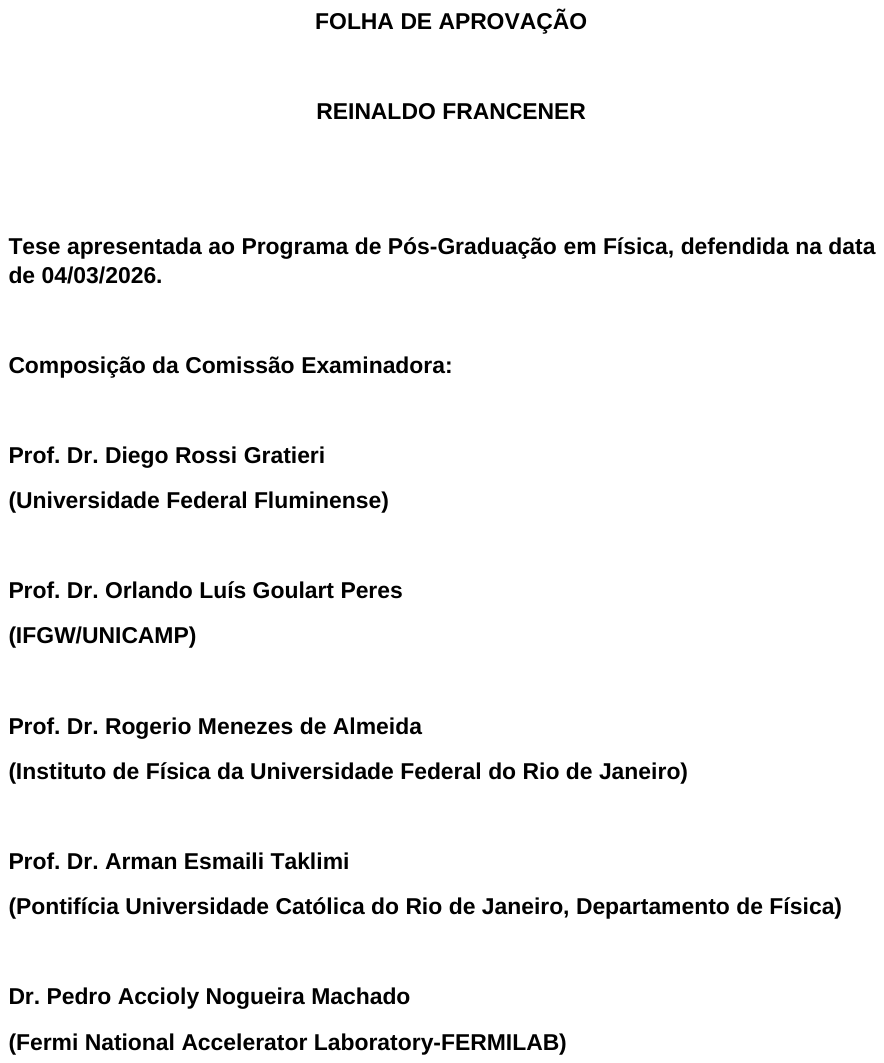}

\textbf{Note}: This manuscript is an English translation of the PhD thesis defended and approved at the University of Campinas (UNICAMP). The official version of the thesis is the Portuguese document submitted to the university. This translation was prepared by the author for broader dissemination of the work.


\begin{abstract}[name=Acknowledgments]

First and foremost, I would like to thank my advisors, Diego Rossi Gratieri and Donato Giorgio Torrieri, for agreeing to supervise me, for entrusting me with this project, for their guidance and mentorship, for their partnership throughout the four years of my doctoral studies, even during difficult periods, and for their unwavering efforts to help me.

I thank Professor Victor Paulo Barros Gonçalves for his friendship, for all the shared teachings and collaboration that extends back to my master's degree, for always being available for meetings and discussions, for entrusting me with challenging projects, and for being an exemplary researcher and a reference in my academic journey. 

I would also like to thank Professor Juan Rojo, my supervisor during my time at Nikhef in Amsterdam. I am immensely grateful for entrusting me with one of his projects, and for all the effort he made so that I could return to Brazil with the project developed. I also thank all my colleagues at Nikhef for the incredible experience I had during the three months I spent in the Netherlands, especially Peter Krack, for all his help and patience in the collaboration we developed during this period. I also thank Felix Kling for all the discussions about the projects and for his collaboration.

I would like to extend a special thank you to the friends and colleagues with whom I shared my time at UNICAMP. Especially to Gabriel Rabelo Soares and Paulo Henrique de Moura. I am extremely grateful for the knowledge, experiences, and coffees we shared.

I am immensely grateful to my girlfriend and partner, Jennyfer Castro da Silva, for all the affection, love, and companionship during this period. I thank her for always supporting me and encouraging me to make the most difficult decisions. I also thank my family, especially my mother Grasiela and my father Edilson, for all their support.

I also want to thank everyone who contributed to my education, starting with my undergraduate studies at the Federal Institute of Santa Catarina. In particular, I would like to thank professors Antônio João Fidélis, João Carlos Xavier, and Bruno Duarte da Silva Moreira for encouraging me to pursue an academic career and for all the knowledge they shared.

Finally, I would like to thank the funding agencies for their financial support. This study was financed in part by the Coordenação de Aperfeiçoamento de Pessoal de Nível Superior - Brasil (CAPES) - Finance Code 001, process number 88887.668071/2022-00. This study was partially financed by the Ministry of Science, Technology and Innovation and the National Council for Scientific and Technological Development – CNPq, process number 161770/2022-3. I would also like to thank the DERI/Santander Program for International Mobility of postgraduate students for the support it provided during my sandwich period in the Netherlands.

\end{abstract}

\begin{abstract}

Neutrinos are the unique fermions in the Standard Model whose only known interaction is the weak force. This makes neutrinos the most difficult fermions to detect, therefore having the least known and tested properties. In this work, we investigate the phenomenological consequences of neutrino and muon interactions with matter. In our studies, we focused in phenomenological predictions for two experiments: FASER and IceCube. FASER is a detector located at the LHC that measures neutrinos produced in proton-proton collisions. LHC neutrinos have an observed flux with energies on the order of TeV, and non-negligible components for the three flavors of the Standard Model. A new version of FASER, FASER2, has been proposed to operate in the Forward Physics Facility during the high-luminosity regime of the LHC, and the prediction is for event rates leveraged by a factor of approximately 200 compared to FASER. The intense flux of tau neutrinos expected at FASER2 motivated us to study the polarization effects of the tau produced in charged current interaction. Our results show that the produced taus will not be completely polarized, and that the precise measurement of the momentum of the pion produced in the tau decay will be sensitive to its polarization state. We also show that the structure function $F_5$ plays an important role in the total and differential cross sections of tau neutrinos at the LHC. Among the Standard Model particles, only neutrinos and muons produced in proton-proton collisions at the LHC can reach FASER. The muon flux reaching this detector is treated as an inconvenient background for the search for neutrino-initiated events. In our study, we show that muon-initiated events can reveal interesting nucleon properties, such as nuclear effects and the existence of an intrinsic charm. The high number of events induced by neutrinos at FASER motivated us to study rare processes in neutrino interaction, such as the neutrino trident, a process of producing pairs of charged leptons in the fusion of gauge bosons in neutrino-nucleus interactions. Our results indicate that the neutrino trident process can be observed at FASER2 for the first time with more than 5$\sigma$ of statistical significance, and will be sensitive to effects beyond the Standard Model in this process at a FASER2 operating in the Future Circular Collider. We have also studied muon trident at the LHC, and we showed that tau pair production can be observed for the first time in this reaction. In contrast to neutrinos detected at the LHC, the neutrinos observed at IceCube come from natural sources, being mainly atmospheric and astrophysical neutrinos. IceCube is capable of observing neutrinos across a wide energy spectrum, ranging from a few GeV to beyond PeV. Our focus in this work is on events classified as High Energy Start Events, which are events initiated within the observatory and that meet selection criteria with the intention of obtaining a set of astrophysical neutrino events with greater purity. We show that the study of these events can contribute to our understanding of the structure of target hadrons, as well as the search for physics effects beyond the Standard Model in the propagation of these neutrinos in the universe until they reach the Earth.

\end{abstract}

\begin{abstract}[name=Resumo]

Neutrinos são os únicos férmions do Modelo Padrão cuja única interação conhecida é a força fraca. Isso faz dos neutrinos os férmions mais difíceis de serem detectados, portanto tendo suas propriedades menos conhecidas e testadas. Neste trabalho, investigamos as consequências fenomenológicas de interações de neutrinos e de múons com a matéria. Em nossos estudos, damos ênfase à predições fenomenológicas para dois experimentos: o FASER e o IceCube. O FASER é um detector localizado no LHC que está medindo neutrinos produzidos em colisões próton-próton. Neutrinos do LHC possuem fluxo observado com energia da ordem de TeV, e componentes não desprezíveis para os três sabores do Modelo Padrão. Uma nova versão do FASER, o FASER2, foi proposto para operar no Forward Physics Facility durante o regime de alta luminosidade do LHC, e a previsão é de taxas de eventos alavancadas por um fator de aproximadamente 200 comparado com o FASER. O intenso fluxo de neutrinos tauônicos esperado no FASER2 nos motivou a estudar efeitos da polarização do tau produzido em interação de corrente carregada. Nossos resultados mostram que taus produzidos não serão completamente polarizados, e que a medida precisa no momento do píon produzido no decaimento do tau será sensível ao seu estado de polarização. Mostramos também que a função de estrutura $F_5$ desempenha papel importante nas seções de choque totais e diferenciais de neutrinos tauônicos no LHC. Dentre as partículas do Modelo Padrão, apenas neutrinos e múons produzidos em colisões próton-próton no LHC podem chegar ao FASER. O fluxo de múons que atinge esse detector é tratado como um inconveniente background para a busca de eventos iniciados por neutrinos. Em nosso estudo, mostramos que eventos iniciados por múons podem revelar interessantes propriedades do núcleon, como efeitos nucleares e a existência de um charm intrínseco. O alto número de eventos iniciados por neutrinos no FASER também nos motivou a estudar processos raros na interação de neutrinos, como o neutrino tridente, processo de produção de pares de léptons carregados na fusão de bósons de gauge em interações neutrino-núcleo. Nossos resultados indicam que o processo neutrino tridente poderá ser observado no FASER2 pela primeira vez com mais de 5$\sigma$ de significância estatística, bem como será sensível a efeitos além do Modelo Padrão neste processo em um FASER2 operando no Future Circular Collider. Também estudamos o processo tridente induzido por múons do LHC, e mostramos que a produção de pares de taus pode ser observada pela primeira vez nessa reação. Em contraste aos neutrinos detectados no LHC, os neutrinos observados no IceCube provém de fontes naturais, sendo principalmente neutrinos atmosféricos e neutrinos astrofísicos. O IceCube é capaz de observar neutrinos em um extenso espectro de energia, variando de poucos GeV até além de PeV. Nosso foco neste trabalho está nos eventos classificados no High Energy Start Events, que são eventos iniciados dentro do observatório e que atendem à critérios de seleção com a intenção de obter um conjunto de eventos de neutrinos astrofísicos com maior pureza. Mostramos que o estudo desses eventos pode contribuir para o nosso entendimento da estrutura dos hádrons alvos, bem como pela busca de efeitos de física além do Modelo Padrão na propagação destes neutrinos no universo até chegarem na Terra.

\end{abstract}


\renewcommand{\contentsname}{
\begin{center}
        \MakeUppercase{Contents}
\end{center}
}
\tableofcontents

\plainchapter{Introduction}
\addcontentsline{toc}{chapter}{Introduction}

It is known that the quest to understand the composition of matter has sparked human interest since at least the 7th century BC in ancient Greece. It was there that the philosophers Leucippus, Democritus, and Epicurus introduced the concept of the atom, understood as the fundamental and indivisible unit constituent of matter \cite{AsimovAtom}. This idea remained largely unexplored for centuries, and was only taken up again in the 19th century by Dalton, who formulated the first modern atomic theory in 1808 \cite{Dalton1808}. Another important advance occurred in 1897 \cite{Thomson1897} with the discovery of the electron and the description of what we know as Thomson's atomic model, which began to incorporate subatomic constituents \cite{Thomson1904}. The quest to understand the particles that make up the atom led Geiger and Marsden, under the guidance of Ernest Rutherford, to conduct experiments scattering alpha particles in a thin gold foil \cite{GeigerMarsden1909}. The result revealed the existence of the compact and positively charged atomic nucleus, establishing the basic structure of the modern atom \cite{Rutherford:1911zz}. 

In this context, the study of radioactive decay processes, particularly beta decay, became a central problem in nuclear physics. To avoid violating the conservation of energy in this process, Wolfgang Pauli proposed the existence of neutrinos \cite{Pauli:1930pc}, whose theoretical description was later formalized by Enrico Fermi \cite{Fermi:1934hr}. Experimental confirmation of neutrinos came about two decades later, with the experiment by Cowan and Reines, who detected electron antineutrinos from nuclear reactors for the first time, through inverse beta decay \cite{Cowan:1956rrn}.

The study of neutrino interaction with matter plays a fundamental role in understanding particle physics, astrophysics, and cosmology. In recent years, with the beginning of the era of precision neutrino measurements, there has been unprecedented progress in the experimental validation of models and in the observation of rare processes predicted by the Standard Model of particle physics. Neutrinos are the second most abundant particle in the universe, but their properties of low mass (no mass predicted by the Standard Model) and interaction only via the weak force make their observation rare and require precise instruments and techniques for their identification.

Current detectors and observatories measure a wide range of neutrino energies, from MeV solar neutrinos to astrophysical neutrinos of hundreds of PeV. Within this vast energy spectrum, neutrinos originating from different sources are observed. In addition to the aforementioned solar and astrophysical sources, there is great interest in experiments measuring neutrinos from accelerators, colliders, supernovae, reactors, atmosphere, etc. Recently, the KM3NeT Observatory measured the most energetic neutrino ever detected by humans, with an estimated energy of over 200 PeV \cite{KM3NeT:2025npi}. The origin of this and other observed high-energy neutrinos remains a mystery, but measurements from the IceCube and Baikal-GVD Observatories indicate with more than 5$\sigma$ of statistical significance that it is a neutrino flux of extragalactic origin \cite{IceCube:2013low,IceCube:2020wum,GVD:2025lya,Baikal-GVD:2025kbe}, called astrophysical neutrinos. Of the hundreds of neutrino events classified as being of astrophysical origin at IceCube, only two sources have been identified in joint measurements with other multi-messenger astrophysics experiments: Blazar TXS 0506+056 \cite{IceCube:2018dnn}, and the Active Galactic Nucleus NGC 1068 \cite{IceCube:2022der}. In addition to observatories dedicated to observing neutrinos, there are large-area cosmic ray observatories, such as the Pierre Auger Observatory \cite{PierreAuger:2015eyc}, which are also sensitive to the astrophysical neutrino flux at ultra-high energies, through the detection of extensive atmospheric showers produced by neutrino interactions in the atmosphere or the Earth's crust, having set upper limits for the existence of this flux \cite{PierreAuger:2015ihf}.

Neutrinos with slightly lower energies than those astrophysical observed, with energies on the order of TeV, have been measured at the Large Hadron Collider (LHC) \cite{SNDLHC:2023pun,FASER:2023zcr,FASER:2024hoe,FASER:2024ref}. Since the 1980s, it has been known that hadronic collisions are expected to produce an intense and collimated neutrino flux in the frontal region of these collisions \cite{DeRujula:1984ns}. To observe this neutrino flux, and also to search for processes beyond the Standard Model, the experiments SND@LHC \cite{SNDLHC:2022ihg} and FASER \cite{Feng:2017uoz,FASER:2018ceo,FASER:2018eoc,FASER:2018bac,FASER:2019aik,FASER:2019dxq,FASER:2020gpr} were built, located approximately 480~m from the interaction point in the ATLAS detector. The result of the operation of these new detectors at the LHC is the first observation of neutrinos from colliders and consequently the detection of the most energetic neutrinos produced by humans. With these neutrinos, it has been possible not only to extract neutrino-nucleus cross sections on the TeV scale that had never been measured before, but also to extract the neutrino fluxes \cite{FASER:2024ref,John:2025qlm} produced in the frontal region of proton-proton collisions, which allows the study of hadron production in this region, since the observed neutrinos come from the decay of these particles.

Despite recent advances in neutrino detection across a wide energy range, several fundamental aspects associated with their interaction with matter remain poorly understood. In particular, neutrino-nucleus cross sections at energies on the order of TeV and above still exhibit significant uncertainties, stemming from both a scarcity of experimental data and limited knowledge of parton distributions in poorly explored kinematic regimes. These uncertainties directly impact the interpretation of atmospheric and astrophysical neutrino measurements made by large observatories, hindering the distinction between Standard Model effects and potential signals of new physics.

The study of neutrinos produced at the LHC is of great interest for improving our understanding of astrophysical neutrinos, as these are the human-produced neutrinos with energies closest to those of extraterrestrial origin. Therefore, they can contribute to our knowledge of cross sections in the TeV range and beyond, which can help to reduce current uncertainties in the cross sections and fluxes of extragalactic and atmospheric neutrinos. There is also the possibility of studying the same processes in both, but at different energies, such as the production of heavy mesons and muon propagation, and extracting complementary information from the data.

In this context, this thesis investigates processes involving neutrino interactions in experiments located at the LHC, with a particular focus on the FASER and SND@LHC detectors. Processes sensitive to both neutrino-nucleus cross sections and neutrino fluxes produced in the frontal region of proton-proton collisions are studied. Through realistic simulations and the incorporation of experimental efficiencies, this thesis provides quantitative estimates of the expected number of events, assesses the impact of theoretical and experimental uncertainties, and explores the potential of these experiments to constrain relevant parameters of the Standard Model and scenarios beyond it.

The results presented in this thesis also explore direct implications for neutrino physics in large neutrino observatories such as IceCube. The interpretation of astrophysical neutrino measurements in this experiment crucially depends on knowledge of neutrino-nucleus cross sections at energies on the order of TeV and above, the determination of the topology associated with individual events, as well as the modeling of atmospheric and astrophysical neutrino fluxes. In particular, the studies developed throughout this thesis contribute to a more detailed understanding of the interaction channels relevant to event topologies at IceCube, aiding in the identification and classification of events and in distinguishing between different components of the observed neutrino flux.

\section*{Thesis structure}

This thesis will be organized as follows. In Chapter \ref{cap:MP} we will briefly describe the Standard Model of particle physics, describing some fundamental properties of the electroweak theory and Quantum Chromodynamics (QCD). In this section we will also address neutrinos and their characteristic sources in a general way. Chapter \ref{cap:cs} will be the last one of theoretical foundation. In it we will describe the main interaction processes that we will be interested in the following chapters, namely Deep Inelastic Scattering (DIS), Glashow resonance and lepton trident scattering. In this chapter we will also present the coupled differential equations that describe the propagation of neutrinos in the material medium, necessary for calculations of the attenuation of neutrino flux by the Earth, and which, in the limit of few interactions, recover the equations for the number of events in detectors and observatories. Finally, still in Chapter \ref{cap:cs}, we will describe the IceCube Observatory and the FASER$\nu$ detector located at the LHC, which are the experiments whose data we will be analyzing and for which we will be constructing phenomenological predictions starting from Chapter \ref{cap:Ice}.

In Chapter \ref{cap:Ice} we will present the results published in \cite{Francener:2024bfm,Francener:2024ljx,Francener:2025apz}, in which we discuss processes of the Standard Model and beyond and their implications for the IceCube neutrino observatory. In particular, in reference \cite{Francener:2024bfm} we investigate the phenomenological consequences of adopting simplified descriptions of the Earth's structure to describe neutrino propagation. We compare predictions constructed using simplified versions with three, two, and one layers of the Earth, with more realistic predictions using the PREM (Preliminary Reference Earth Model) \cite{Dziewonski:1981xy}. Also in Chapter \ref{cap:Ice}, we estimate the impacts of subdominant processes on the description of events characterized as tracks at IceCube. These processes, such as the production of heavy flavors mesons and the $W^{\pm}$ boson, are usually disregarded in analyses of this event topology. Our results show that they account for more than 20\% of events with the track topology \cite{Francener:2024ljx}. Finally, also in Chapter \ref{cap:Ice}, we will study the impact of the existence of a new gauge boson, $Z'$, predicted by the $L_\mu - L_\tau$ theory \cite{He:1990pn,He:1991qd}, on the astrophysical neutrino flux and consequently on the events observed at IceCube. Our results indicate that the events collected over 12 years by IceCube slightly favor this theory in some scenarios compared to the Standard Model, and that IceCube-Gen2 will be able to probe the existence of this new gauge boson in regions of mass and coupling never before tested for the $Z'$ \cite{Francener:2025apz}.

Starting in Chapter \ref{cap:pol}, we present our results exploring the potential of muons and neutrinos produced at ATLAS for different observable processes, particularly at the FASER$\nu$ detector and its proposed upgrade, FASER$\nu$2. In Chapter \ref{cap:pol}, we will explore some effects of the neutrino deep inelastic interaction that are most important in the case of tau neutrinos \cite{Francener:2024ney}. This neutrino flavor is the least studied and measured fermion of the Standard Model, having its properties less tested and potentially being sensitive to several models that include physics beyond the Standard Model. In particular, we will study how the polarization states of the tau affect observables such as total and differential cross sections for TeV neutrinos, which is the neutrino focus of the LHC. In addition to polarization effects, we will study the impact of the structure function $F_5$ \cite{Francener:2024euo}, which has never been measured to date. Furthermore, in the description of the neutrino-nucleon interaction amplitude, this structure function appears multiplied by terms dependent on the mass of the charged lepton produced, which makes it practically impossible to extract in other processes with neutrinos from lighter leptons.

In Chapter \ref{cap:MuonDIS} we will present our studies for muon and neutrino DIS at the FASER$\nu$ as an important tool in describing hadronic structure. As main applications, we will show that muon-initiated events at FASER$\nu$, even after realistic identification cuts in the detector, can prove the existence of an intrinsic charm quark component in the nucleon wavefunction \cite{Francener:2025pnr}. The intrinsic charm was proposed as early as the 1980s \cite{Brodsky:1980pb} as an alternative to explain the excess of large Bjorken-$x$ events observed by the EMC collaboration \cite{EuropeanMuon:1982fow}. These measurements reported by the EMC are still the subject of intense debate today, but there is other experimental evidence in favor of the existence of an intrinsic charm component in the wavefunction of the nucleon \cite{Ball:2022qks}. In addition to proving its existence, FASER$\nu$, in conjunction with the FASER electronic detector, can search for asymmetries in the intrinsic charm in Bjorken-$x$, which would be unequivocal evidence of non-perturbative effects on the parton distribution function (PDF) of the charm quark. Lepton-induced events at these LHC frontal detectors can also contribute to the understanding and improvement of the description of nuclear PDFs, which currently have much greater uncertainties than those of free protons. We will show that in some scenarios the expected statistical uncertainties for the data are smaller than the current uncertainties of the PDFs \cite{Francener:2025tyh}.

In Chapter \ref{cap:tridente} we discussed the process called neutrino trident. This scattering is characterized by three leptons in the final state, two of which are electrically charged, meaning that they can be measured. This process has not yet been observed with 5$\sigma$ statistical significance, and the candidate events all come from measurements from the last century \cite{CCFR:1991lpl,CHARM-II:1990dvf,NuTeV:1999wlw}. Some recent works have shown the possibility of using this scattering to search for phenomena beyond the Standard Model, as well as for a better determination of the couplings of the $Z^0$ boson. Our study will focus on the trident process with LHC neutrinos at the FASER$\nu$2. We will obtain the neutrino-nucleus cross sections in this energy regime, as well as the number of events expected for this process with different combinations of charged lepton flavors in the final state. Our results indicate that neutrino trident scattering can, in principle, be observed at the LHC during its high-luminosity regime \cite{Francener:2024wul}. In addition to the trident process in the Standard Model, we will study how a specific model of new physics, the $L_\mu - L_\tau$ theory \cite{He:1990pn,He:1991qd}, will affect trident data that can be collected both at the LHC and at the FCC in proton-proton collisions \cite{Francener:2024jra}.

Chapter \ref{cap:MuonTridente} presents our results and discussions regarding the trident process, found in reference \cite{Francener:2025wzh}, now initiated by LHC muons reaching frontal detectors. The difference compared to the neutrino trident is the presence of a muon in the initial state and three charged leptons in the final state; therefore, we are only interested in neutral current interactions. Although the exchange of a $Z^{0}$ boson is possible for both the muon and the target nucleus, we will focus only on the case of photons being exchanged by both particles involved in the interaction, given that the exchange of the $Z^0$ boson is suppressed by its high mass. This process has already been measured with muons for electron pairs \cite{Roe:1959zz} and for muon pairs produced in the final state \cite{Russell:1971mf,Maciuc:2006xb,SNDLHC:2024bzp}. Our results indicate that the observation of tau pairs can be made for the first time with projectile muons at FASER$\nu$ during the Run 3 of the LHC. In addition to the open lepton pair produced, we will estimate the cross sections and event rates for lepton bound states, focusing on true muonium, a bound state between muon and antimuon, which has not yet been observed. Its event rates will only be significant at FASER$\nu$2, and we will discuss the associated backgrounds that make its observation so hard.

Finally, in Chapter \ref{cap:conclusions} we will present our main conclusions, and discuss some avenues we could pursue in future research.

\section*{Publications list}

This section presents a list of publications in which the student participated during their doctoral studies. These include:

\vspace{0.5 cm}
\noindent 1. FRANCENER, Reinaldo; GONCALVES, Victor P.; GRATIERI, Diego R. Sensitivity of the neutrino transmission coefficient at high energies to the Earth's density profile. \textbf{Journal of Physics G: Nuclear and Particle Physics}, 2024.

\vspace{0.5 cm}
\noindent 2. FRANCENER, Reinaldo; GONÇALVES, Victor P.; GRATIERI, Diego R. Track signals at IceCube from subleading channels. \textbf{Physical Review D}, v. 110, n. 5, p. 053011, 2024.

\vspace{0.5 cm}
\noindent 3. FRANCENER, Reinaldo; GONÇALVES, Victor P.; GRATIERI, Diego R. Probing a low-mass $Z'$ gauge boson at IceCube and prospects for IceCube-Gen2. \textbf{Physical Review D}, v. 111, n. 9, p. 095005, 2025.

\vspace{0.5 cm}
\noindent 4. FRANCENER, Reinaldo; GONÇALVES, Victor P.; GRATIERI, Diego R. Tau polarization in neutrino-nucleus interactions at the LHC energy range. \textbf{Physical Review D}, v. 109, n. 11, p. 113005, 2024.

\vspace{0.5 cm}
\noindent 5. FRANCENER, Reinaldo; GONÇALVES, Victor P.; GRATIERI, Diego R. Neutrino trident scattering at the LHC energy regime. \textbf{The European Physical Journal C}, v. 84, n. 9, p. 923, 2024.

\vspace{0.5 cm}
\noindent 6. FRANCENER, Reinaldo; GONÇALVES, Victor P.; GRATIERI, Diego R. Tau polarization effects in $\nu_\tau/\bar{\nu}_\tau$-tungsten interactions at the LHC energies. \textbf{Physical Review D}, v. 110, n. 7, p. 073006, 2024.

\vspace{0.5 cm}
\noindent 7. FRANCENER, Reinaldo; GONÇALVES, Victor P.; GRATIERI, Diego R. Probing a $Z'$ gauge boson via neutrino trident scattering in the Forward Physics Facility at the LHC and FCC. \textbf{The European Physical Journal C}, v. 85, n. 5, p. 601, 2025.

\vspace{0.5 cm}
\noindent 8.  FRANCENER, Reinaldo et al. Deep-inelastic scattering at TeV energies with LHC muons. \textbf{The European Physical Journal C}, v. 85, n. 10, p. 1098, 2025.

\vspace{0.5 cm}
\noindent 9. FRANCENER, Reinaldo; GONCALVES, Victor P.; GRATIERI, Diego R. Investigating nuclear effects in lepton-ion DIS at the LHC. \textbf{Journal of High Energy Physics}, v. 1, n. 149, 2026.

\vspace{0.5 cm}
\noindent 10. FRANCENER, Reinaldo; GONCALVES, Victor P.; RABELO-SOARES, Gabriel. Muon trident process at far-forward LHC detectors. \textbf{Nuclear Physics B}, v. 1025, n. 117396, 2026.

\vspace{0.5 cm}
\noindent 11.  ANCHORDOQUI, Luis A. et al. Letter of Intent: The Forward Physics Facility. arXiv preprint \textbf{arXiv:2510.26260}, 2025.

\vspace{0.5 cm}
\noindent 12. FRANCENER, Reinaldo et al. Photoproduction of QED bound states in future electron-ion colliders. \textbf{Physics Letters B}, v. 854, p. 138753, 2024.

\vspace{0.5 cm}
\noindent 13. BERTULANI, Carlos A. et al. Particle production by $\gamma - \gamma$ interactions in future electron-ion colliders. \textbf{Physical Review C}, v. 111, n. 2, p. 025201, 2025.

\vspace{0.5 cm}
\noindent 14. FRANCENER, Reinaldo; GONCALVES, Victor P.; MARTINS, Daniel E. Investigating the exclusive toponium production at the LHC and FCC. \textbf{Physical Review D}, v. 112, n. 11, p. 094050, 2025.

\vspace{0.5 cm}
\noindent 15. RABELO-SOARES, Gabriel; FRANCENER, Reinaldo; RAMOS, Gabriel S.; TORRIERI, Giorgio. QCD Wehrl and entanglement entropies in a gluon spectator model at small-$x$. arXiv preprint \textbf{arXiv:2512.24855}, 2025.

\vspace{0.5 cm}

Among the publications presented above, the results of works 1 to 10 will be discussed in this thesis: \cite{Francener:2024bfm,Francener:2024ljx,Francener:2025apz,Francener:2024wul,Francener:2024jra,Francener:2025pnr,Francener:2025tyh,Francener:2024ney,Francener:2024euo,Francener:2025wzh}. The first three relate to the phenomenology of high and ultra-high energies neutrinos at IceCube, and the others relate to the interaction of neutrinos and muons in frontal experiments at the LHC, such as FASER$\nu$. All ten references mentioned have already undergone peer review in journals and are published. As a result of the work carried out, I was invited to join as a co-author in the reference 11, which is the Letter of Intent for the Forward Physics Facility \cite{FPF:2025bor}. Works 12-14 refer to the research topic I studied in my master's degree, and which I continued in collaborations during my doctorate \cite{Francener:2024eep,Bertulani:2024vpt,Francener:2025tor}. Finally, work 15 relates to a new topic recently developed with collaborators \cite{Rabelo-Soares:2025ams}.

\chapter{The Standard Model of particle physics}
\label{cap:MP}

The Standard Model of particle physics is the theory responsible for grouping elementary particles and describing their fundamental interactions. This theory was initially proposed in 1961 by Sheldon Glashow, in a first attempt to unify the weak and electromagnetic interactions \cite{Glashow:1961tr}. Also in the 1960s, Steven Weinberg \cite{Weinberg:1967tq} and Abdus Salam \cite{Salam:1968rm}, in independent works, applied the Higgs mechanism \cite{Higgs:1964pj,Englert:1964et,Guralnik:1964eu} to generate the masses of weak bosons and formulated what we know today as the Electroweak Theory. In the 1970s, Quantum Chromodynamics was developed \cite{Han:1965pf,Fritzsch:1973pi}. The Standard Model is based on the formalism developed in the 1950s by Yang and Mills \cite{Yang:1954ek}, which considers non-abelian gauge theories of the SU(N) group. The fundamental particles, whose properties are shown in Figure \ref{cap1:1}, are divided into two main sectors: fermions, which have half-integer spin; and bosons, which have integer spin.

Fermions are divided into leptons and quarks. Leptons do not interact via the strong force and can be electrically charged (electron, muon, and tau) or uncharged (corresponding charged leptons' neutrinos). Charged leptons have mass, while neutrinos in the Standard Model are considered massless. Currently, it is known that neutrinos also have mass, but their origin is still unknown. Unlike leptons, quarks interact via the strong force. Quarks are not seen free in nature due to the confinement property of this interaction, and their bound states are called hadrons. States with two valence quarks are called mesons, while states with three valence quarks form baryons. Recently, bound states of the strong interaction with four \cite{LHCb:2011zzp} and five \cite{LHCb:2015yax} quarks were discovered.

The boson sector of the Standard Model contains spin-1 bosons: the photon, gluon, $Z^0$, and $W^{\pm}$, and one spin-0 boson, the Higgs boson \cite{Higgs:1964pj,Englert:1964et,Guralnik:1964eu}. Spin-1 bosons are called gauge bosons and are responsible for carrying information about the forces between elementary particles. Photons carry information about the electromagnetic interaction, which occurs between particles carrying electric charge. The theory that describes this interaction at a fundamental level is Quantum Electrodynamics (QED). Gluons are the mediators of the strong interaction, Quantum Chromodynamics, which occurs between particles carrying color charge. The $Z^0$ and $W^\pm$ bosons are the only massive gauge bosons and are responsible for the weak interaction. Throughout this chapter, we will describe in more detail the interactions between the particles of the Standard Model. Finally, the scalar boson of the Standard Model, the Higgs, was the last of the fundamental particles to be observed experimentally, its discovery being confirmed at the Large Hadron Collider (LHC) by the ATLAS \cite{ATLAS:2012yve} and CMS \cite{CMS:2012qbp} experiments. The Higgs boson plays a crucial role in the Standard Model, being responsible for generating the mass of the other elementary particles.

\begin{figure}[t]
\centering

\begin{tikzpicture}[thick, scale=1]

\draw[fill=pink] (0,0) rectangle (2.5,2.5);
\draw[fill=pink] (2.8,0) rectangle (5.3,2.5);
\draw[fill=pink] (5.6,0) rectangle (8.1,2.5);
\draw[fill=red] (8.4,0) rectangle (10.9,2.5);

\draw[fill=pink] (0,2.8) rectangle (2.5,5.3);
\draw[fill=pink] (2.8,2.8) rectangle (5.3,5.3);
\draw[fill=pink] (5.6,2.8) rectangle (8.1,5.3);
\draw[fill=red] (8.4,2.8) rectangle (10.9,5.3);
\draw[fill=blue!50] (11.2,2.8) rectangle (13.7,5.3);

\draw[fill=green] (0,-2.8) rectangle (2.5,-0.3);
\draw[fill=green] (2.8,-2.8) rectangle (5.3,-0.3);
\draw[fill=green] (5.6,-2.8) rectangle (8.1,-0.3);
\draw[fill=red] (8.4,-2.8) rectangle (10.9,-0.3);

\draw[fill=green] (0,-5.6) rectangle (2.5,-3.1);
\draw[fill=green] (2.8,-5.6) rectangle (5.3,-3.1);
\draw[fill=green] (5.6,-5.6) rectangle (8.1,-3.1);
\draw[fill=red] (8.4,-5.6) rectangle (10.9,-3.1);

\draw[fill=white] (0,6) rectangle (8.1,6.7);
\draw[fill=white] (8.4,6) rectangle (13.7,6.7);

\draw[fill=white]  (1.5,1.5) node {\Huge {d}} (1.5,1.5)circle (0.6); 
\draw[fill=white]  (4.3,1.5) node {\Huge {s}} (4.3,1.5)circle (0.6); 
\draw[fill=white]  (7.1,1.5) node {\Huge {b}} (7.1,1.5)circle (0.6);
\draw[fill=white]  (9.9,1.5) node {\Huge {$\gamma$}} (9.9,1.5)circle (0.6);

\draw[fill=white]  (1.5,4.3) node {\Huge {u}} (1.5,4.3)circle (0.6);
\draw[fill=white]  (4.3,4.3) node {\Huge {c}} (4.3,4.3)circle (0.6);
\draw[fill=white]  (7.1,4.3) node {\Huge {t}} (7.1,4.3)circle (0.6);
\draw[fill=white]  (9.9,4.3) node {\Huge {g}} (9.9,4.3)circle (0.6);
\draw[fill=white]  (12.7,4.3) node {\Huge {H}} (12.7,4.3)circle (0.6);

\draw[fill=white]  (1.5,-1.3) node {\Huge {e}} (1.5,-1.3)circle (0.6);
\draw[fill=white]  (4.3,-1.3) node {\Huge {$\mu$}} (4.3,-1.3)circle (0.6);
\draw[fill=white]  (7.1,-1.3) node {\Huge {$\tau$}} (7.1,-1.3)circle (0.6);
\draw[fill=white]  (9.9,-1.3) node {\Huge {Z}} (9.9,-1.3)circle (0.6);

\draw[fill=white]  (1.5,-4.1) node {\Huge {$\nu_{e}$}} (1.5,-4.1)circle (0.6);
\draw[fill=white]  (4.3,-4.1) node {\Huge {$\nu_{\mu}$}} (4.3,-4.1)circle (0.6);
\draw[fill=white]  (7.1,-4.1) node {\Huge {$\nu_{\tau}$}} (7.1,-4.1)circle (0.6);
\draw[fill=white]  (9.9,-4.1) node {\Huge {W}} (9.9,-4.1)circle (0.6);

\draw [thick, black] (0.2,0.2) node[right] {\resizebox{!}{0.3cm}{Down}};
\draw [thick, black] (3,0.2) node[right] {\resizebox{!}{0.3cm}{Strange}};
\draw [thick, black] (5.8,0.2) node[right] {\resizebox{!}{0.3cm}{Bottom}};
\draw [thick, black] (8.6,0.2) node[right] {\resizebox{!}{0.3cm}{Photon}};

\draw [thick, black] (0.2,3) node[right] {\resizebox{!}{0.3cm}{Up}};
\draw [thick, black] (3,3) node[right] {\resizebox{!}{0.3cm}{Charm}};
\draw [thick, black] (5.8,3) node[right] {\resizebox{!}{0.3cm}{Top}};
\draw [thick, black] (8.6,3) node[right] {\resizebox{!}{0.3cm}{Gluon}};
\draw [thick, black] (11.4,3) node[right] {\resizebox{!}{0.3cm}{Higgs}};

\draw [thick, black] (0.2,-2.6) node[right] {\resizebox{!}{0.3cm}{Electron}};
\draw [thick, black] (3,-2.6) node[right] {\resizebox{!}{0.3cm}{Muon}};
\draw [thick, black] (5.8,-2.6) node[right] {\resizebox{!}{0.3cm}{Tau}};
\draw [thick, black] (8.6,-2.6) node[right] {\resizebox{!}{0.3cm}{$Z^)$ boson}};

\draw [thick, black] (0.2,-5.4) node[right] {\resizebox{!}{0.3cm}{neutrino}};
\draw [thick, black] (3,-5.4) node[right] {\resizebox{!}{0.3cm}{neutrino}};
\draw [thick, black] (5.8,-5.4) node[right] {\resizebox{!}{0.3cm}{neutrino}};
\draw [thick, black] (8.6,-5.4) node[right] {\resizebox{!}{0.3cm}{$W^{\pm}$ boson}};
\draw [thick, black] (0.2,-5.1) node[right] {\resizebox{!}{0.3cm}{Electronic}};
\draw [thick, black] (3,-5.1) node[right] {\resizebox{!}{0.3cm}{Muonic}};
\draw [thick, black] (5.8,-5.1) node[right] {\resizebox{!}{0.3cm}{Tauonic}};

\draw [thick, pink] (-0.4,0) node[right,rotate=90] {\resizebox{!}{0.5cm}{QUARKS}};
\draw [thick, green] (-0.4,-5.6) node[right,rotate=90] {\resizebox{!}{0.5cm}{LEPTONS}};
\draw [thick, red] (11.3,-5.6) node[right,rotate=90] {\resizebox{!}{0.5cm}{GAUGE BOSONS}};
\draw [thick, blue!50] (13.3,-3.1) node[right,rotate=90] {\resizebox{!}{0.5cm}{SCALAR BOSONS}};

\draw [thick, black] (0,6.3) node[right] {\resizebox{!}{0.34cm}{Three fermionic generations}};
\draw [thick, black] (8.4,6.3) node[right] {\resizebox{!}{0.34cm}{Bosons}};

\draw [thick, black] (1,5.65) node[right] {\resizebox{!}{0.36cm}{1ª}};
\draw [thick, black] (3.8,5.65) node[right] {\resizebox{!}{0.36cm}{2ª}};
\draw [thick, black] (6.6,5.65) node[right] {\resizebox{!}{0.36cm}{3ª}};

\draw [thick, black] (-1.1,5.1) node[right] {\resizebox{!}{0.2cm}{Mass}};
\draw [thick, black] (-1.1,4.8) node[right] {\resizebox{!}{0.2cm}{Charge}};
\draw [thick, black] (-1.1,4.5) node[right] {\resizebox{!}{0.2cm}{Spin}};
\draw [thick, black] (-1.1,4.2) node[right] {\resizebox{!}{0.2cm}{Color}};
\draw [thick, black] (-1.1,3.9) node[right] {\resizebox{!}{0.2cm}{I, I$_{3}$}};

\draw [thick, black] (0,2.3) node[right] {\resizebox{!}{0.2cm}{4.7\;MeV}};
\draw [thick, black] (2.8,2.3) node[right] {\resizebox{!}{0.2cm}{96\;MeV}};
\draw [thick, black] (5.6,2.3) node[right] {\resizebox{!}{0.2cm}{4.66\;GeV}};
\draw [thick, black] (8.4,2.3) node[right] {\resizebox{!}{0.2cm}{0}};

\draw [thick, black] (0,5.1) node[right] {\resizebox{!}{0.2cm}{2.2\;MeV}};
\draw [thick, black] (2.8,5.1) node[right] {\resizebox{!}{0.2cm}{1.27\;GeV}};
\draw [thick, black] (5.6,5.1) node[right] {\resizebox{!}{0.2cm}{173.2\;GeV}};
\draw [thick, black] (8.4,5.1) node[right] {\resizebox{!}{0.2cm}{0}};
\draw [thick, black] (11.2,5.1) node[right] {\resizebox{!}{0.2cm}{125\;GeV}};

\draw [thick, black] (0,-0.5) node[right] {\resizebox{!}{0.2cm}{0.511\;MeV}};
\draw [thick, black] (2.8,-0.5) node[right] {\resizebox{!}{0.2cm}{105.7\;MeV}};
\draw [thick, black] (5.6,-0.5) node[right] {\resizebox{!}{0.2cm}{1.777\;GeV}};
\draw [thick, black] (8.4,-0.5) node[right] {\resizebox{!}{0.2cm}{91.19\;GeV}};

\draw [thick, black] (0,-3.3) node[right] {\resizebox{!}{0.2cm}{<2.0\;eV}};
\draw [thick, black] (2.8,-3.3) node[right] {\resizebox{!}{0.2cm}{<2.0\;eV}};
\draw [thick, black] (5.6,-3.3) node[right] {\resizebox{!}{0.2cm}{<2.0\;eV}};
\draw [thick, black] (8.4,-3.3) node[right] {\resizebox{!}{0.2cm}{80.38\;GeV}};

\draw [thick, black] (0,2) node[right] {\resizebox{!}{0.2cm}{-1/3}};
\draw [thick, black] (2.8,2) node[right] {\resizebox{!}{0.2cm}{-1/3}};
\draw [thick, black] (5.6,2) node[right] {\resizebox{!}{0.2cm}{-1/3}};
\draw [thick, black] (8.4,2) node[right] {\resizebox{!}{0.2cm}{0}};

\draw [thick, black] (0,4.8) node[right] {\resizebox{!}{0.2cm}{2/3}};
\draw [thick, black] (2.8,4.8) node[right] {\resizebox{!}{0.2cm}{2/3}};
\draw [thick, black] (5.6,4.8) node[right] {\resizebox{!}{0.2cm}{2/3}};
\draw [thick, black] (8.4,4.8) node[right] {\resizebox{!}{0.2cm}{0}};
\draw [thick, black] (11.2,4.8) node[right] {\resizebox{!}{0.2cm}{0}};

\draw [thick, black] (0,-0.8) node[right] {\resizebox{!}{0.2cm}{-1}};
\draw [thick, black] (2.8,-0.8) node[right] {\resizebox{!}{0.2cm}{-1}};
\draw [thick, black] (5.6,-0.8) node[right] {\resizebox{!}{0.2cm}{-1}};
\draw [thick, black] (8.4,-0.8) node[right] {\resizebox{!}{0.2cm}{0}};

\draw [thick, black] (0,-3.6) node[right] {\resizebox{!}{0.2cm}{0}};
\draw [thick, black] (2.8,-3.6) node[right] {\resizebox{!}{0.2cm}{0}};
\draw [thick, black] (5.6,-3.6) node[right] {\resizebox{!}{0.2cm}{0}};
\draw [thick, black] (8.4,-3.6) node[right] {\resizebox{!}{0.2cm}{$\pm 1$}};

\draw [thick, black] (0,1.7) node[right] {\resizebox{!}{0.2cm}{1/2}};
\draw [thick, black] (2.8,1.7) node[right] {\resizebox{!}{0.2cm}{1/2}};
\draw [thick, black] (5.6,1.7) node[right] {\resizebox{!}{0.2cm}{1/2}};
\draw [thick, black] (8.4,1.7) node[right] {\resizebox{!}{0.2cm}{1}};

\draw [thick, black] (0,4.5) node[right] {\resizebox{!}{0.2cm}{1/2}};
\draw [thick, black] (2.8,4.5) node[right] {\resizebox{!}{0.2cm}{1/2}};
\draw [thick, black] (5.6,4.5) node[right] {\resizebox{!}{0.2cm}{1/2}};
\draw [thick, black] (8.4,4.5) node[right] {\resizebox{!}{0.2cm}{1}};
\draw [thick, black] (11.2,4.5) node[right] {\resizebox{!}{0.2cm}{0}};

\draw [thick, black] (0,-1.1) node[right] {\resizebox{!}{0.2cm}{1/2}};
\draw [thick, black] (2.8,-1.1) node[right] {\resizebox{!}{0.2cm}{1/2}};
\draw [thick, black] (5.6,-1.1) node[right] {\resizebox{!}{0.2cm}{1/2}};
\draw [thick, black] (8.4,-1.1) node[right] {\resizebox{!}{0.2cm}{1}};

\draw [thick, black] (0,-3.9) node[right] {\resizebox{!}{0.2cm}{1/2}};
\draw [thick, black] (2.8,-3.9) node[right] {\resizebox{!}{0.2cm}{1/2}};
\draw [thick, black] (5.6,-3.9) node[right] {\resizebox{!}{0.2cm}{1/2}};
\draw [thick, black] (8.4,-3.9) node[right] {\resizebox{!}{0.2cm}{1}};

\draw [thick, black] (-0.1,1.4) node[right] {\resizebox{!}{0.2cm}{R,G,B}};
\draw [thick, black] (2.7,1.4) node[right] {\resizebox{!}{0.2cm}{R,G,B}};
\draw [thick, black] (5.5,1.4) node[right] {\resizebox{!}{0.2cm}{R,G,B}};
\draw [thick, black] (8.4,1.4) node[right] {\resizebox{!}{0.2cm}{0}};

\draw [thick, black] (-0.1,4.2) node[right] {\resizebox{!}{0.2cm}{R,G,B}};
\draw [thick, black] (2.7,4.2) node[right] {\resizebox{!}{0.2cm}{R,G,B}};
\draw [thick, black] (5.5,4.2) node[right] {\resizebox{!}{0.2cm}{R,G,B}};
\draw [thick, black] (8.4,4.2) node[right] {\resizebox{!}{0.2cm}{C$\overline{\mathrm{C}}$}};
\draw [thick, black] (11.2,4.2) node[right] {\resizebox{!}{0.2cm}{0}};

\draw [thick, black] (0,-1.4) node[right] {\resizebox{!}{0.2cm}{0}};
\draw [thick, black] (2.8,-1.4) node[right] {\resizebox{!}{0.2cm}{0}};
\draw [thick, black] (5.6,-1.4) node[right] {\resizebox{!}{0.2cm}{0}};
\draw [thick, black] (8.4,-1.4) node[right] {\resizebox{!}{0.2cm}{0}};

\draw [thick, black] (0,-4.2) node[right] {\resizebox{!}{0.2cm}{0}};
\draw [thick, black] (2.8,-4.2) node[right] {\resizebox{!}{0.2cm}{0}};
\draw [thick, black] (5.6,-4.2) node[right] {\resizebox{!}{0.2cm}{0}};
\draw [thick, black] (8.4,-4.2) node[right] {\resizebox{!}{0.2cm}{0}};

\draw [thick, black] (-0.1,1.1) node[right] {\resizebox{!}{0.18cm}{1/2,-1/2}};
\draw [thick, black] (2.7,1.1) node[right] {\resizebox{!}{0.18cm}{1/2,-1/2}};
\draw [thick, black] (5.5,1.1) node[right] {\resizebox{!}{0.18cm}{1/2,-1/2}};
\draw [thick, black] (8.4,1.1) node[right] {\resizebox{!}{0.18cm}{0}};

\draw [thick, black] (0,3.9) node[right] {\resizebox{!}{0.18cm}{1/2,1/2}};
\draw [thick, black] (2.8,3.9) node[right] {\resizebox{!}{0.18cm}{1/2,1/2}};
\draw [thick, black] (5.6,3.9) node[right] {\resizebox{!}{0.18cm}{1/2,1/2}};
\draw [thick, black] (8.4,3.9) node[right] {\resizebox{!}{0.18cm}{0}};
\draw [thick, black] (11.2,3.9) node[right] {\resizebox{!}{0.18cm}{0}};

\draw [thick, black] (-0.1,-1.7) node[right] {\resizebox{!}{0.18cm}{1/2,-1/2}};
\draw [thick, black] (2.7,-1.7) node[right] {\resizebox{!}{0.18cm}{1/2,-1/2}};
\draw [thick, black] (5.5,-1.7) node[right] {\resizebox{!}{0.18cm}{1/2,-1/2}};
\draw [thick, black] (8.3,-1.7) node[right] {\resizebox{!}{0.18cm}{1, 0}};

\draw [thick, black] (0,-4.5) node[right] {\resizebox{!}{0.18cm}{1/2,1/2}};
\draw [thick, black] (2.8,-4.5) node[right] {\resizebox{!}{0.18cm}{1/2,1/2}};
\draw [thick, black] (5.6,-4.5) node[right] {\resizebox{!}{0.18cm}{1/2,1/2}};
\draw [thick, black] (8.4,-4.5) node[right] {\resizebox{!}{0.18cm}{1, $\pm 1$}};

\end{tikzpicture}


\caption{Standard Model of particle physics. Figure from \cite{Francener:2022sfw}.}

\label{cap1:1}
\end{figure}

\section{Neutrinos and the electroweak theory}
\label{sec_MP:neutrinos}

The existence of neutrinos was postulated by Wolfgang Pauli in 1930 \cite{Pauli:1930pc} and formally theorized by Enrico Fermi in 1934 to explain the fraction of energy lost in beta decays \cite{Fermi:1934hr}. Due to their rare interaction with matter, neutrinos were first detected only in 1956 by the Reines-Cowan experiment \cite{Cowan:1956rrn}. About 70 years after their discovery, neutrinos remain the least known fermions of the Standard Model, and some of their properties, such as the origin of their mass and whether they are Dirac or Majorana fermions, remain open questions and play an important role in the search for physics beyond the Standard Model.

\subsection{Lagrangian of the electroweak theory}
\label{subsec_MP:LagrangianEW}

In the Standard Model, neutrinos interact only via the weak force, which is described by the electroweak unification (unification between the electromagnetic and weak interactions) \cite{Glashow:1961tr,Weinberg:1967tq,Salam:1968rm}. The electroweak theory is a local gauge theory with symmetry group SU(2)$_I \times$U(1)$_Y$, where $I$ is the weak isospin and $Y$ is the hypercharge. The electroweak theory also introduces the Higgs mechanism \cite{Higgs:1964pj,Englert:1964et,Guralnik:1964eu}, which is responsible for generating the masses of particles that couple to this field through the mechanism called spontaneous symmetry breaking.

We will describe here the formalism for weak interactions for a generic charged lepton $l$ and its corresponding neutrino $\nu_l$. In weak interactions, these particles appear in left-handed isospin doublets: $(\nu_L, l_L)$. The right-handed components of the Standard Model are not involved in weak interactions. In the Standard Model there are no right-handed neutrinos, and right-handed charged leptons are represented as isospin singlets. The weak isospin doublet and singlet components are represented by
\begin{eqnarray}
    \psi_L := 
    \begin{pmatrix}
    \nu_L \\
    l_L
    \end{pmatrix} 
\end{eqnarray}
e
\begin{eqnarray}
    \psi_R := l_R \, ,
\end{eqnarray}
respectively. The hypercharge $Y$ is defined by 
\begin{eqnarray}
    Y = 2(Q - I_3) \, ,
\end{eqnarray}
where $Q$ is the electric charge of the particle and $I_3$ is its third weak isospin component. The values of $Q$ and $I_3$ that define the hypercharge in the equation above are shown in Figure \ref{cap1:1} for each of the elementary particles. The quark sector of the Standard Model also interacts via the weak force. Analogously to leptons, left-handed quarks form doublets within each generation, while right-handed quarks are singlets.

The kinetic part of the electroweak Lagrangian before spontaneous symmetry breaking is

\begin{eqnarray}
    \mathcal{L} = 
    i \bar{\psi}_L \slashed{\partial} \psi_L + i \bar{l}_R \slashed{\partial}l_R 
    - \frac{1}{4} B_{\mu \nu} B^{\mu \nu} - \frac{1}{4} W_{\mu \nu} W^{\mu \nu} \, ,
    \label{eq_1:LagrangianaEW}
\end{eqnarray}
where 
\begin{eqnarray}
    B_{\mu\nu} = \partial_\mu B_{\nu} - \partial_\nu B_{\mu} \, ,
    \label{eq_1:tensorEM}
\end{eqnarray}
and
\begin{eqnarray}
    W_{\mu\nu}^{i} = \partial_\mu W_{\nu}^{i} - \partial_\nu W_{\mu}^{i} + g\epsilon_{ijk} W^{j}_{\mu} W^{k}_{\nu} \, ,
    \label{eq_1:tensorG}
\end{eqnarray}
are the tensors that describe the gauge field associated with the abelian group U(1$)_Y$ ($B_{\mu\nu}$), and the massless vector fields $W^{i}$ of the group SU(2)$_I$ ($W^{i}_{\mu\nu}$). Note that the Lagrangian of Equation (\ref{eq_1:LagrangianaEW}) does not completely describe the electroweak theory, since both bosons and fermions are considered massless, i.e., it is the Lagrangian before the spontaneous symmetry breaking and introduction of the Higgs mechanism.

The Lagrangian of the Equation (\ref{eq_1:LagrangianaEW}) is invariant under local transformations
\begin{eqnarray}
    U_{1} = \mathrm{e}^{i g T_{j}\cdot \mathbf{\alpha}_{j}(x)}
    \label{eq_1:U1}
\end{eqnarray}
and
\begin{eqnarray}
    U_{2} = \mathrm{e}^{i\frac{g'}{2}Y\cdot \alpha(x)} \, ,
    \label{eq_1:U2}
\end{eqnarray}
being $Y$ a Hermitian generator and $\alpha (x)$ being local rotation parameters. $T_j$ are the generators of the SU(2) group, which are defined in terms of the Pauli matrices ($T_j = \sigma_j / 2$) and satisfy the commutation relation
\begin{eqnarray}
    [T_{a}, T_{b}] = i \epsilon _{abc} T_c \, .
    \label{eq_1:PauliComutator}
\end{eqnarray}

Although the mechanism of spontaneous symmetry breaking can be carried out in several ways with different scalar fields, the standard solution consists of introducing a Higgs doublet composed of charged and neutral scalar fields, which couple to gauge bosons through the covariant derivative, and which also interact with fermionic fields through Yukawa couplings \cite{Yukawa:1935xg}.
The Higgs sector is introduced into the Standard Model by means of a self-interaction term (potential), which is essential because it leads to spontaneous symmetry breaking: the vacuum state does not preserve the SU(2)$_I$ $\times$ U(1)$_Y$ symmetry. After the symmetry violation, the vacuum expected value of the Higgs field gives rise to the masses of the other massive fields in the theory.
Furthermore, the gauge bosons $W^{3}_\mu$ and $B_{\mu}$ must be rotated to obtain the neutral bosons, while combinations of $W^{1}_\mu$ and $W^{2}_\mu$ give rise to the physical states $W^{\pm}_\mu$ as follows:
\begin{eqnarray}
    A_{\mu} = \mathrm{sen}\,\theta_W W^{3}_\mu + \mathrm{cos}\,\theta_W B_{\mu} \, ,
    \label{eq_1:Amu}
\end{eqnarray}
\begin{eqnarray}
    Z^{0}_{\mu} = \mathrm{cos}\,\theta_W W^{3}_\mu - \mathrm{sen}\,\theta_W B_{\mu} \, ,
    \label{eq_1:Zmu}
\end{eqnarray}
and
\begin{eqnarray}
    W^{\pm}_{\mu} = \frac{1}{\sqrt{2}}(W^{1}_\mu \mp i W^{2}_{\mu}) \, ,
    \label{eq_1:Wmu}
\end{eqnarray}
where $\theta_W$ is the Weinberg angle that can be obtained through the couplings $g$, $g'$ and $e$ (electric charge):
\begin{eqnarray}
    e = g' \mathrm{cos}\, \theta_W = g\, \mathrm{sen}\,\theta_W \, .
    \label{eq_1:thetaW}
\end{eqnarray}

In terms of the physical gauge boson fields of electroweak theory, the covariant derivative is written as
\begin{equation}
D_\mu = \partial_\mu 
  + i e Q A_\mu 
  + i \frac{g}{\sqrt{2}} \left( T^+ W_\mu^+ + T^- W_\mu^- \right)
  + i \frac{g}{\cos \theta_W} \left( T^3 - Q \sin^2 \theta_W \right) Z_\mu^0,
\label{eq_1:covariant_derivative}
\end{equation}
where $T^\pm = T_1 \pm i T_2$.

Using the expression above, the Lagrangian of the electroweak theory can be rewritten as the sum of the parts of the electromagnetic (EM), charged current (CC), and neutral current (NC) interactions in the form
\begin{equation}
\mathcal{L}_{\text{EW}} = \mathcal{L}_{\text{EM}} + \mathcal{L}_{\text{CC}} + \mathcal{L}_{\text{NC}},
\end{equation}
where
\begin{equation}
\mathcal{L}_{\text{EW}} = 
  - e A_\mu J^\mu_{\text{EM}}
  - \frac{g}{\sqrt{2}} \left( W_\mu^+ J^{\mu}_{\text{CC}} + W_\mu^- J^{\mu\,\dagger}_{\text{CC}} \right)
  - \frac{g}{\cos \theta_W} Z_\mu J^\mu_{\text{NC}} \, .
\label{eq_1:Lew}
\end{equation}
Leptonic currents are given by
\begin{align}
J^\mu_{\text{EM}} &= - \bar{l} \gamma^\mu l \nonumber \\
J^\mu_{\text{CC}} &= \bar{\nu}_l \gamma^\mu (1 - \gamma^5) l \\
J^\mu_{\text{NC}} &= 
  \frac{1}{2} \bar{\nu}_l \gamma^\mu (1 - \gamma^5) \nu_l
  - \frac{1}{2} \bar{l} \gamma^\mu (1 - \gamma^5) l
  + \sin^2 \theta_W \, \bar{l} \gamma^\mu l \nonumber \, .
\end{align}

These currents from weak and electromagnetic interactions are essential in the study of fundamental interactions. They act on particle states and generate their scatterings. The currents are used in the derivation of Feynman's rules, which allow us to treat interactions in a perturbative way and obtain scattering matrices, which are indispensable in cross section calculations.

\subsection{Neutrino sources}
\label{subsec_MP:neutrinoSource}

Neutrinos are the second most abundant Standard Model particle in the universe, after photons. Since neutrinos have no electric charge, it is not possible to control neutrino beams with the same precision in energy and position as in charged particle colliders. However, it is possible to produce fluxes of unstable particles in the laboratory that decay to produce neutrinos, or to use neutrino fluxes from natural sources, both those originating on Earth, such as atmospheric neutrinos and those from radioactive particle decay, and those of extraterrestrial origin, such as solar neutrinos and those from supernova explosions.

\begin{figure}
	\centering
				\begin{tabular}{ccc}
	\includegraphics[width=0.7\textwidth]{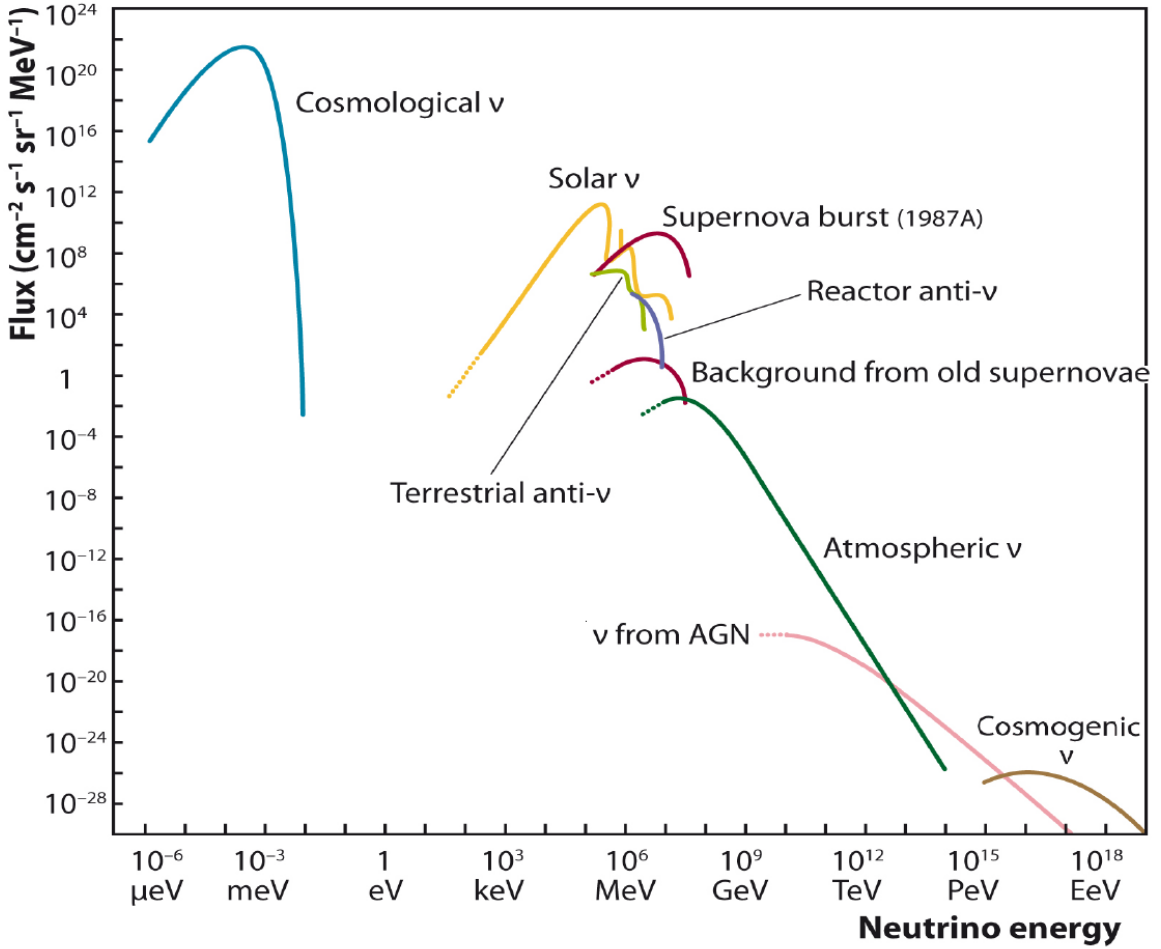}
				\end{tabular}
	\caption{ Neutrino spectrum and its main sources. Figure taken from the reference \cite{FernandezMenendez:2017ccn}. }
	\label{fig_cap1:neutrinoSources}
\end{figure}

\begin{figure}
	\centering
				\begin{tabular}{ccc}
	\includegraphics[width=0.8\textwidth]{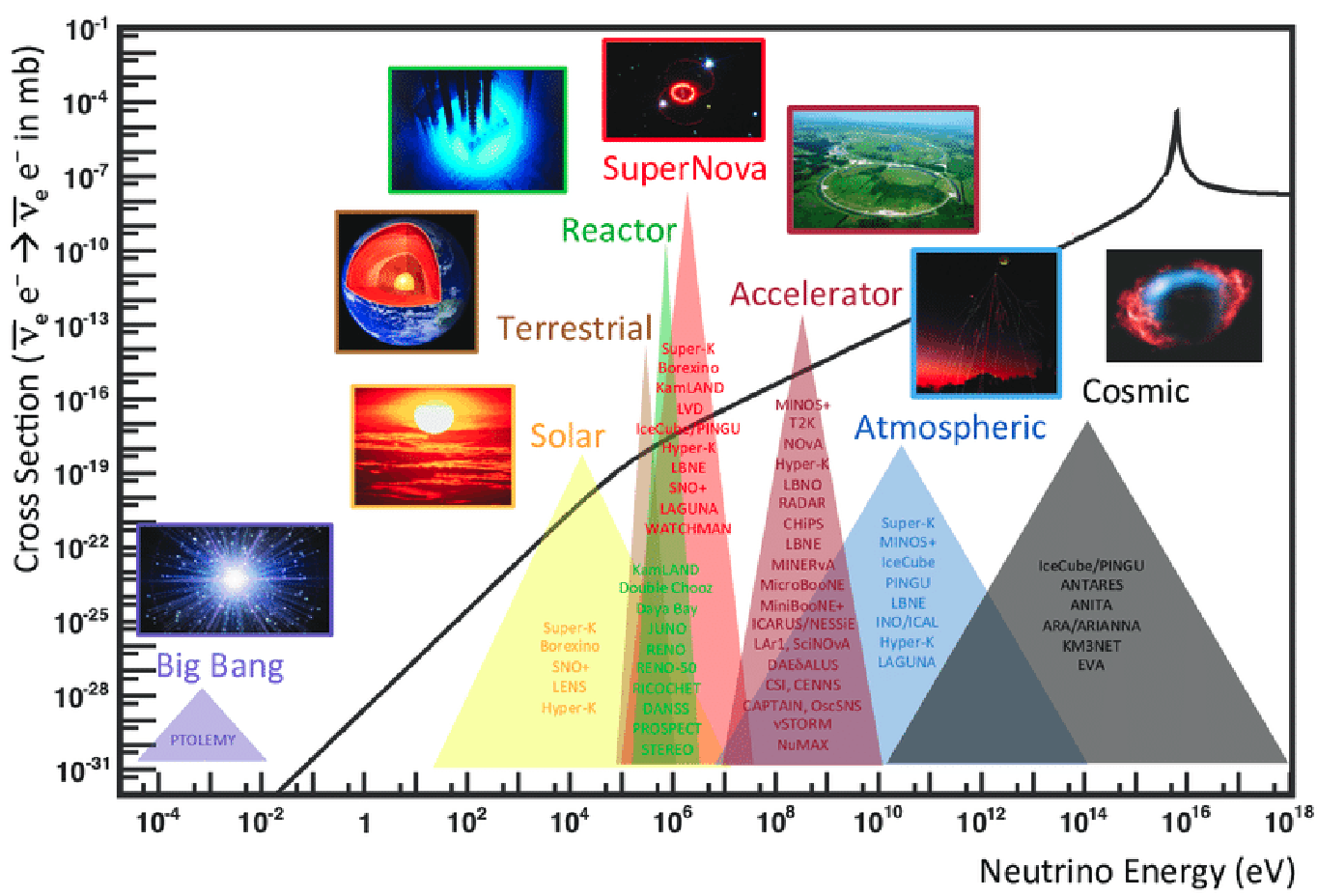}
				\end{tabular}
	\caption{ Antineutrino-electron cross section as a function of the incident antineutrino energy. In each different energy regime there are different characteristic sources for neutrinos. We also show some of the main existing and planned experiments for each energy range. Figure taken from the reference \cite{Hewett:2014qja}. }
	\label{fig_cap1:neutrinoInteractions}
\end{figure}

In Figure \ref{fig_cap1:neutrinoSources} we show the main neutrino sources and their fluxes as a function of energy. It is important to note that neutrino fluxes become smaller as neutrino energy increases, which reflects the difficulty of producing increasingly energetic processes. Complementing Figure \ref{fig_cap1:neutrinoSources}, we present Figure \ref{fig_cap1:neutrinoInteractions}, which shows the antineutrino-electron cross section for elastic scattering as a function of the incident antineutrino energy. The cross section increases with energy throughout the energy regime shown, except after the peak of the Glashow resonance \cite{Glashow:1960zz}. Along with the cross section, Figure \ref{fig_cap1:neutrinoInteractions} shows the typical neutrino sources in each interval, as well as the main detectors and observatories specialized in observing these neutrinos.

Broadly speaking, we can separate neutrino sources into natural and artificial sources. The most common natural sources are:

\begin{itemize}
\item \textbf{Cosmological neutrinos:} these are neutrinos produced shortly after the Big Bang, in the early stages of the universe \cite{Giunti:2007ry}. As the universe cooled, these neutrinos decoupled as the rate of expansion of the universe became greater than their rate of interaction. When produced, these neutrinos were very energetic, but with the expansion of the universe they became the least energetic neutrinos existing. The temperature of the cosmic neutrino background is about 1.95~K, which corresponds to neutrinos with energies on the order of $10^{-4}$ eV \cite{Kumar:2022htk}. Given this low energy, they are the only non-relativistic neutrinos that can be measured, but this has not yet been done. There is a proposal to measure them in the future with the PTOLEMY experiment \cite{Betts:2013uya}.

\item \textbf{Geoneutrinos:} they are terrestrial neutrinos resulting from the decay of radioactive isotopes present on Earth \cite{Bellini:2021sow}. These neutrinos usually have energies on the order of a few MeV and have already been observed by two experiments: KamLAND \cite{Araki:2005qa} and Borexino \cite{Borexino:2015ucj}. The study of geoneutrinos can be a powerful tool for measuring the amount of radioactive isotopes on our planet.

\item \textbf{Solar neutrinos:} solar neutrinos are produced inside the Sun by different nuclear reactions. About 98\% of these neutrinos come from proton-proton fusion processes, all being electronic neutrinos and antineutrinos. These neutrinos have a spectrum that starts at less than 1 MeV and goes up to about 15 MeV. Their discovery occurred in 1968 by the Homestake experiment located in a gold mine in South Dakota, United States \cite{Davis:1968cp}. We can also highlight the SNO experiment, which measured solar neutrinos with neutral current interactions \cite{SNO:2002tuh}, showing that when summing all neutrino flavors, the measured flux agrees with theoretical predictions.

\item \textbf{Supernova neutrinos:} certain stars heavier than the Sun, when they reach a more advanced stage of fuel burning through nuclear fusion, have their cores collapse due to gravitational attraction, triggering a supernova explosion \cite{Giunti:2007ry}. During the explosion, protons and electrons fuse, giving rise to neutrons and neutrinos ($\sim 10^{58}$) with energies on the order of 10 MeV. In 1987, the Kamiokande II \cite{Kamiokande-II:1987idp}, IMB \cite{Bionta:1987qt}, and BAKSAN \cite{Alekseev:1988gp} detectors were operating when they observed, within a 12-second interval, approximately 24 neutrino events originating from the supernova SN1987A in the Large Magellanic Cloud. About three hours after the neutrinos were observed, the light from the supernova was spotted by the Las Campanas Observatory \cite{Kunkel:1987zz} in Chile.

\item \textbf{Atmospheric neutrinos:} Earth is constantly bombarded with primary cosmic rays from different regions of the universe, in an energy range extending from a few MeV to about $10^{20}$ eV \cite{Fermi:1949ee,Bhattacharjee:1999mup}. These cosmic rays are usually protons and helium nuclei, and when they interact with the upper atmosphere they can produce extensive atmospheric showers. In hadronic showers, large quantities of pions and kaons are produced, which decay into neutrinos and muons, the latter subsequently decaying to produce more neutrinos. Atmospheric neutrinos were first observed in 1965 by the Indian Kolar experiment \cite{Achar:1965cha,Achar:1965ova}, concurrently with observations in a mine in South Africa \cite{Reines:1965qk}. Currently, there are several experiments focused on observing these neutrinos.

\item \textbf{Astrophysical neutrinos:} astrophysical neutrinos were first observed in 2013 by the IceCube observatory \cite{IceCube:2013low} located in Antarctic ice, and are the most energetic neutrinos ever observed by humans. Recently, the Baikal-GVD Observatory also reported the existence of a neutrino flux of extragalactic origin \cite{GVD:2025lya}, in agreement with IceCube's measurements. Although these neutrinos were discovered more than a decade ago, little is known about their origin, with only two sources identified so far: the Blazar TXS 0506+056 \cite{IceCube:2018dnn}, and the Active Galactic Nucleus NGC 1068 \cite{IceCube:2022der}. The KM3NeT Observatory recently began partial operations and measured the most energetic neutrino ever observed by humans, of extraterrestrial origin, with an estimated energy of over 200~PeV \cite{KM3NeT:2025npi}.
\end{itemize}

Artificial neutrinos are those produced directly or indirectly by humans. Their most common sources are:

\begin{itemize}
\item \textbf{Reactor neutrinos:} electronic antineutrinos are produced in the beta decay reactions of radioactive isotopes in nuclear power plants. A reactor with an operating power of approximately 1~GW produces about $2\times 10 ^{20}$ electronic antineutrinos per second. They were the first neutrinos observed by humans, by Reines and Cowan in 1956, through inverse beta decay \cite{Cowan:1956rrn}. These neutrinos typically have energies in the MeV range.

\item \textbf{Accelerator neutrinos:} particle accelerators are used to produce neutrino fluxes, usually muonic neutrinos. Beams of accelerated particles are collided with material targets to produce unstable particles that decay into neutrinos. The study of accelerator neutrinos is an important tool for understanding phenomena such as neutrino oscillation, since they are the neutrino fluxes most easily controlled from an experimental point of view. As an example of experiments that measure this class of neutrinos, we highlight MINOS \cite{MINOS:1995txm,MINOS:2005iiy,MINOS:2006foh}, T2K \cite{T2K:2011qtm,T2K:2011ypd}, MINERvA \cite{MINERvA:2013zvz,MINERvA:2013kdn}, NOvA \cite{NOvA:2004blv,NOvA:2019cyt} and MicroBooNE \cite{MicroBooNE:2016pwy,MicroBooNE:2015bmn}.

\item \textbf{Collider neutrinos:} recently, the FASER and SND@LHC Collaborations reported the first observations of collider neutrinos \cite{SNDLHC:2023pun,FASER:2023zcr,FASER:2024hoe,FASER:2024ref}, produced in proton-proton collisions at the LHC. These neutrinos have energies on the order of TeV, and are the most energetic artificial neutrinos observed by humans. These neutrinos, like accelerator and atmospheric neutrinos, result from the decay of unstable particles produced in hadronic interactions.
\end{itemize}

The observation of solar and atmospheric neutrinos led to the discovery of the neutrino oscillation phenomenon \cite{Super-Kamiokande:1998kpq,SNO:2002tuh,SNO:2011hxd}. The most widely accepted explanation for this phenomenon is the existence of massive neutrinos, with flavor eigenstates being linear combinations of mass eigenstates. Since neutrinos originally had no mass in the Standard Model, their oscillation constitutes the first unequivocal sign of physics beyond the Standard Model.

In the context of this thesis, we are particularly interested in astrophysical neutrinos, which are observed at IceCube and other high-energy neutrino observatories. In addition to these neutrinos from natural sources, we will explore the physics of high energy neutrinos from colliders, particularly those produced at the LHC.

\section{Quantum chromodynamics}
\label{sec_MP:QCD}

Quantum Chromodynamics is the quantum field theory responsible for describing the strong interaction. The strong interaction occurs between quarks and gluons, and the bound states of this interaction are hadrons. The charge of the interaction is the color charge, which has multiplicity three: red, blue, and green, in addition to the corresponding anticolors. Gluons are the messenger bosons of the strong interaction, but they are also color charge carriers, meaning that gluons can interact directly with each other. This property is said to make QCD a non-abelian gauge theory, and mathematically it means that the generators of the group do not commute.

There are two basic experimentally observed properties that are the foundations that QCD needs to describe: asymptotic freedom and confinement  \cite{Gross:1973id,Politzer:1973fx}. At large distances, or at low energies, confinement manifests itself, causing the interactions to be very intense. As a result, quarks and gluons are not observed free in nature, but always in bound states that are color-neutral. This characteristic requires that the QCD coupling constant ($\alpha_s$) be large at low energies and large distances. Conversely, the coupling constant becomes small in the high-energy, short-distance regime, leading to asymptotic freedom, which allows us to consider hadrons as a set of asymptotically free quarks and gluons in this regime.

In this section we will introduce QCD and its main characteristics, which will be important in the development of deeply inelastic scattering that will be discussed in the following chapters.

\subsection{QCD Lagrangian}
\label{subsec_MP:LagrangianQCD}

The classic Lagrangian that describes QCD is written as \cite{Fritzsch:1973pi,Gross:1973id,Gross:1973zrg,Weinberg:1973un,Kovchegov:2012mbw,Peskin:1995ev}
\begin{eqnarray}
    \mathcal{L}_{\mathrm{classic}} = 
   - \frac{1}{4} F^{a\mu \nu} F_{\mu \nu}^{a} + \bar{\psi} (i \slashed{D} - m) \psi \, ,
    \label{eq_1:LQCDClassica}
\end{eqnarray}
being $\psi$ the quark fields, which are given by
\begin{eqnarray}
    \psi =  
    \begin{pmatrix}
    \psi_r \\
    \psi_b \\
    \psi_g
    \end{pmatrix} \, , 
\end{eqnarray} 
where the indices $r,b,g$ denote the three color charges present in QCD. We also have that the covariant derivative of QCD is given by $D_\mu = \partial_\mu - i g A_{\mu}$ with $A_{\mu} = A^{a}_{\mu} t^{a}$, and $A_\mu$ being the gluon fields and $t^{a}$ the generators of the SU(3) group. The index $a$ takes values from 1 to 8, which implies eight gluons present in QCD. Finally, $g$ is the coupling constant and $F^{\mu\nu} = F^{a \mu\nu} t^{a}$ is defined by $-i g F^{\mu\nu}\psi = [D^{\mu}, D^{\nu}] \psi$.

QCD is a gauge transformation invariant theory, therefore it is necessary that
\begin{eqnarray}
    \mathcal{L}(\psi ', \bar{\psi}', A_{\mu}') =  \mathcal{L}(\psi, \bar{\psi}, A_{\mu}) \, .
\end{eqnarray} 
The gauge transformations that preserve the invariance of the Lagrangian above are
\begin{eqnarray}
    \psi(x) \rightarrow \psi '(x) = U(x) \psi(x) \, ,
\end{eqnarray}
\begin{eqnarray}
    \bar{\psi}(x) \rightarrow \bar{\psi} '(x) = \bar{\psi}(x) U^{-1}(x) \, ,
\end{eqnarray} 
and
\begin{eqnarray}
    A_{\mu}(x) \rightarrow A_{\mu} '(x) = U(x) A_{\mu}(x) U^{-1}(x) - \frac{i}{g}(\partial_{\mu} U(x)) U^{-1}(x) \, ,
\end{eqnarray} 
where the operator $U(x) = \mathrm{e}^{i t^{a}\theta^{a}(x)}$ and $\theta^{a}(x)$ are real functions.

The generators of the SU(3) group must satisfy the relations 
\begin{eqnarray}
    \mathrm{Tr}(t^{a} t^{b}) = \frac{1}{2} \delta^{ab}
\end{eqnarray}
and
\begin{eqnarray}
    t^{a} t^{b} = \frac{\delta^{ab}}{2N_c}+\frac{1}{2}(d^{abc} + i f^{abc})t^{c} \, ,
\end{eqnarray}
which comes from the structure of the Lie algebra of the SU(3) group \cite{lie1893theorie,lie1891vorlesungen,lie1893vorlesungen}.

From the definition of $F_{\nu \mu}$, we have that
\begin{eqnarray}
    F_{\nu\mu} = 
    \partial_{\mu} A_{\nu} - \partial_{\nu} A_{\mu} -ig [A_{\mu}, A_{\nu}] \, .
    \label{eq_1:fmunu}
\end{eqnarray}
Since this is a non-abelian group theory, the commutator above does not vanish. Starting from the definition we gave to $A_{\mu}$, we obtain
\begin{eqnarray}
    [A_{\mu}, A_{\nu}] = A_{\mu}^{a} A_{\nu}^{b} i f^{abc} t^{c} \, ,
    \label{eq_1:comutadorA}
\end{eqnarray}
where $f^{abc}$ are structure functions of the SU(3) group. Combining the Equations (\ref{eq_1:fmunu}) (\ref{eq_1:comutadorA}), we obtain
\begin{eqnarray}
    F^{a}_{\mu\nu}  = \partial_{\mu} A_{\nu}^{a} - \partial_{\nu} A_{\mu}^{a} + g f^{abc} A_{\mu}^{b} A_{\nu}^{c} \, .
    \label{eq_1:FamunuFinal}
\end{eqnarray}
Unlike the electromagnetic field tensor in QED, the last term in the equation above highlights the non-abelian nature of the theory. This term accounts for the interaction between gluons at three- and four-gluon vertices, which arises because gluons carry color charge.

Equation (\ref{eq_1:LQCDClassica}), together with Equation (\ref{eq_1:FamunuFinal}), describes QCD classically. This means that the Lagrangian cannot yet be quantized due to redundant degrees of freedom in $A^{a}_{\mu}$ to describe the same physical state. This problem can be corrected by introducing a gauge condition and adding the term to the Lagrangian
\begin{eqnarray}
    \mathcal{L}_{\mathrm{fix}} = - \frac{1}{2\chi} (\partial^{\mu} A_{\mu}^{a})^{2} \, .
    \label{eq_1:Lfix}
\end{eqnarray}
Inserting Equation (\ref{eq_1:Lfix}) into the Lagrangian of the QCD gives rise to non-physical degrees of freedom that need to be removed by adding the Fadeev-Popov ghost fields \cite{Faddeev:1967fc}, written as
\begin{eqnarray}
    \mathcal{L}_{\mathrm{ghost}} = (\partial_{\mu} c^{a *}) (\delta^{ac} \partial^{\mu} + g f^{abc} A_{\mu}^{b}) c^{c} \, ,
\end{eqnarray}
where $c^{c} (x)$ are the ghost fields.

Therefore, the Lagrangian of the complete QCD is written as
\begin{eqnarray}
    \mathcal{L}_{\mathrm{QCD}} = \mathcal{L}_{\mathrm{classic}} + \mathcal{L}_{\mathrm{fix}} + \mathcal{L}_{\mathrm{ghost}} \, .
    \label{eq_1:LQCD}
\end{eqnarray}

Feynman rules for QCD can be derived using the Lagrangian of Equation (\ref{eq_1:LQCD}) assuming a specific gauge. The choice of gauge does not alter the observables, but it does alter the form of the resulting Feynman rules.

\subsection{Asymptotic freedom}
\label{subsec_MP:assFred}

The QCD, like other quantum field theories, exhibits divergences in higher-order calculations in perturbative expansions of the scattering matrix. To remove these divergences, which are non-physical results, renormalization methods are used. A quantum field theory is said to be renormalizable when we can absorb the divergences into a finite number of redefined physical parameters, such as masses, charges, and coupling constants.

When we use renormalization methods, a scale naturally emerges in the theory, called the renormalization scale, $\mu^{2}$, causing renormalized quantities to explicitly depend on this scale. However, physical observables such as cross sections must be independent of the chosen $\mu^{2}$.

To understand the dependence of the QCD coupling constant $\alpha_s$ on the chosen scale $\mu^{2}$, we will consider a dimensionless observable $R$, involving only the scale $Q^{2}$. The observable $R$ will be a function of $Q^{2}/\mu^{2}$ and $\alpha_s(\mu^{2})$, where $\alpha_{s}(\mu^{2}) = g^{2}/(4\pi)$ is the renormalized coupling constant on the scale $\mu^{2}$. Given that $R$ cannot depend on $\mu^{2}$, its total derivative with respect to this scale must be zero:
\begin{eqnarray}
    \mu^{2}\frac{\mathrm{d}R(Q^{2}/\mu^{2}, \alpha_s (\mu^{2}))}{\mathrm{d}\mu^{2}} = 0 \, , 
    \label{eq_1:R}
\end{eqnarray}
which leads to the renormalization group equation: 
\begin{eqnarray}
    \mu^{2} \left(
    \frac{\partial}{\partial \mu^{2}} + \frac{\partial \alpha_s}{\partial \mu^{2}} \frac{\partial}{\partial \alpha_s}
    \right) R = 0 \, .
    \label{eq_1:renorma}
\end{eqnarray}
If we define the beta function of QCD as
\begin{eqnarray}
    \beta (\alpha_s) = \mu^{2} \frac{\partial \alpha_s}{ \partial \mu^{2}}
    \label{eq_1:beta}
\end{eqnarray}
we can write the Equation (\ref{eq_1:renorma}) in the form
\begin{eqnarray}
    \left(
    \mu^{2} \frac{\partial}{ \partial \mu^{2}} + \beta(\alpha_s) \frac{\partial}{\partial \alpha_s}
    \right) R = 0 \, .
    \label{eq_1:renorma2}
\end{eqnarray}
The equation above shows us that any change in the renormalization scale implies a change in the coupling constant, leaving the observable $R$ invariant, therefore
\begin{eqnarray}
    R(Q^{2} / \mu^{2} , \alpha_s (\mu^{2})) = R(Q^{2} / \mu'^{2} , \alpha_s (\mu'^{2})) \, .
    \label{eq_1:R2}
\end{eqnarray}
Introducing the variable $t$ as $\mathrm{ln} (\mu'^{2} / \mu^{2})$, the beta function can be rewritten as
\begin{eqnarray}
    \beta (\alpha_s(\mu'^{2})) = \frac{\partial \alpha_s(\mu'^{2})}{\partial t} \, ,
    \label{eq_1:beta2}
\end{eqnarray}
which has a solution given by
\begin{eqnarray}
    t = \int^{\alpha(\mu'^{2})}_{\alpha_s(\mu^{2})} \frac{\mathrm{d} x}{\beta(x)} \, .
    \label{eq_1:t}
\end{eqnarray}

The beta function can be calculated perturbatively in QCD, and at the lowest order in $\alpha_s$ it is written as \cite{Prosperi:2006hx}
\begin{eqnarray}
    \beta(x) = -\frac{33 - 2n_f}{12 \pi} x^{2} \, ,
    \label{eq_1:beta3}
\end{eqnarray}
where $n_f$ is the number of active quark flavors. With Equations (\ref{eq_1:t}) and (\ref{eq_1:beta3}) it is possible to show that
\begin{eqnarray}
    \alpha_s (\mu'^{2}) = \frac{\alpha_s(\mu^{2})}{1 + \frac{33-2n_f}{12\pi} \alpha_s(\mu^{2}) \mathrm{ln}\, (\mu'^{2}/\mu^{2})} \, .
    \label{eq_1:alphas}
\end{eqnarray}
It is common to rewrite the equation above in terms of the QCD confinement scale, $\Lambda_{\mathrm{QCD}}$ ($\approx 200-300$~MeV), defined by
\begin{eqnarray}
    \mathrm{ln}\, \Lambda_{\mathrm{QCD}}^{2} = \mathrm{ln}\, \mu^{2} - \frac{1}{\frac{33 - 2n_f}{12\pi} \alpha_s(\mu^{2})}  \, .
    \label{eq_1:lambdaQCD}
\end{eqnarray}
Setting $Q^{2} = \mu^{2}$, the equation (\ref{eq_1:alphas}) becomes
\begin{eqnarray}
    \alpha_s (Q^{2}) = \frac{4 \pi}{1 + \frac{33-2n_f}{12\pi} \mathrm{ln} (Q^{2}/\Lambda^{2}_{QCD})} \, .
    \label{eq_1:alphas2}
\end{eqnarray}

\begin{figure}
	\centering
				\begin{tabular}{ccc}
	\includegraphics[width=0.6\textwidth]{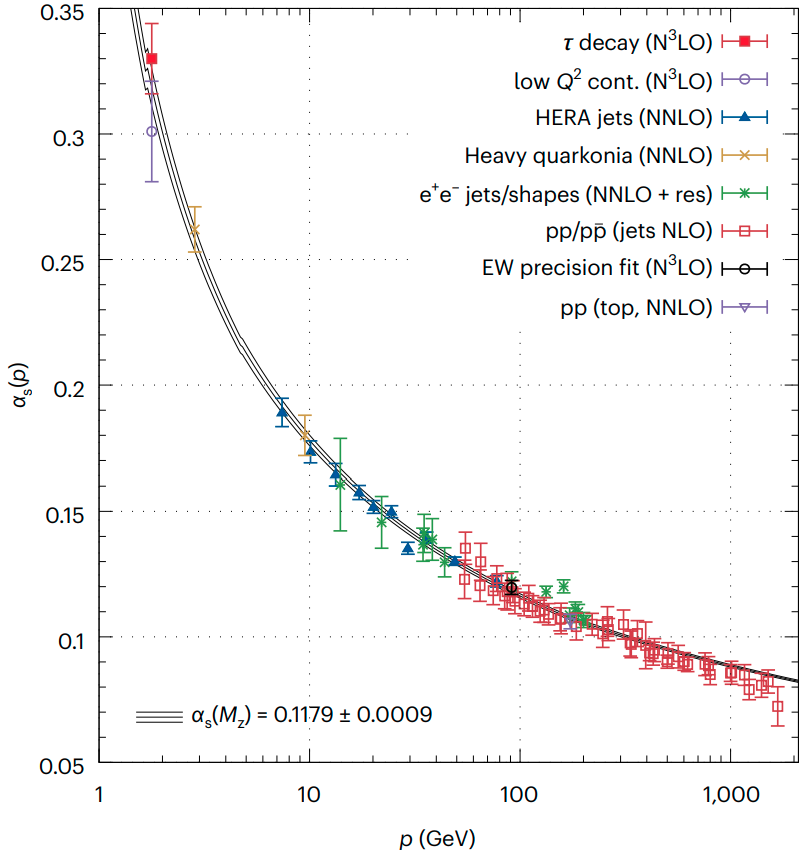}
				\end{tabular}
	\caption{ QCD coupling constant as a function of the magnitude of the transferred four-momentum $Q=p$ for various distinct processes. Figure taken from the reference \cite{Boito:2023lzf}. }
	\label{fig_cap1:QCD-coupling}
\end{figure}

The equation above shows that the coupling constant is small for $Q^{2} \gg \Lambda^{2}_{QCD}$ ($\Lambda^{2}_{QCD} \sim$ 200-300 MeV \cite{ParticleDataGroup:2024cfk}), which allows the use of perturbation theory to describe the interactions. When $Q^{2} \rightarrow \infty$ the coupling constant tends to zero, evidencing the asymptotic freedom \cite{Gross:1973id,Politzer:1973fx}. Conversely, for $Q^{2} \rightarrow \Lambda^{2}_{QCD}$, the coupling constant of Equation~\ref{eq_1:alphas2} diverges and the interactions between quarks and gluons become very intense, a phenomenon known as confinement. This behavior of the QCD coupling constant as a function of $Q$ is shown in Figure \ref{fig_cap1:QCD-coupling} with several experimental results. In the small $Q^{2}$ regime, perturbative calculations are not suitable, and it is necessary to use non-perturbative techniques, such as those used in lattice QCD\footnote{Lattice QCD is a non-perturbative formulation of quantum chromodynamics in which spacetime is discretized into a lattice, allowing the numerical calculation of observables through computer simulations.}.

\section{Conclusions}
\label{sec_MP:conclusion}

In this chapter, we present some basic theoretical concepts regarding the Standard Model of particle physics, mainly focused on the electroweak theory and quantum chromodynamics. We present the Lagrangian of both theories, which are fundamental elements in the derivation of Feynman rules to describe the interactions between elementary particles. Focusing on the electroweak sector, we describe neutrinos beyond their properties, highlighting their main sources and the energy regimes in which they leave signals in detectors. For QCD, we went beyond its Lagrangian and presented one of its most important characteristics in understanding deeply inelastic scattering: the asymptotic freedom. These concepts will be indispensable in understanding and constructing the following chapters, where we will address some phenomenological aspects of the Standard Model and beyond.

\chapter{Lepton interactions and propagation}
\label{cap:cs}

One of the main goals of this thesis is the study of lepton interactions and propagation. In this chapter, we will describe some interaction processes between leptons and targets that will be studied in later chapters. We will also describe neutrino propagation through interacting media, both for neutrinos reaching Earth, traversing Earth, and traversing detectors. We will connect neutrino propagation with the number of events observed in detectors and observatories.

\section{Deep inelastic scattering}
\label{sec_cs:DIS}

\begin{figure}
	\centering
				\begin{tabular}{ccc}
	\includegraphics[width=0.5\textwidth]{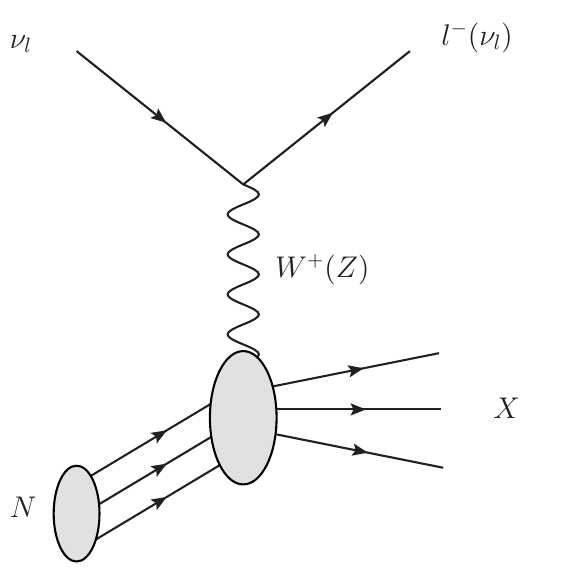}
				\end{tabular}
	\caption{ Feynman diagram for the neutrino-nucleon interaction of DIS by exchanging a boson $W^\pm (Z)$. Figure created by the author. We emphasize here that all subsequent figures throughout the thesis without reference were created by the thesis author. }
	\label{fig_cs:diagramaDIS}
\end{figure}

In this section, we will describe the deep inelastic scattering initiated by a lepton in a target nucleon. We will formally present the neutrino-initiated charged current DIS, but we will also discuss the cases of neutral current interaction of neutrino and charged leptons incidents. The metric used throughout this thesis follows the signature $(1, -1, -1, -1)$. The process of our interest is the interaction of a neutrino with four-momentum $k = (E_\nu , 0, 0, E_\nu)$ with a nucleon of four-momentum $p = (m_N , 0, 0, 0)$, producing a charged lepton of four-momentum $k' = (E_{l}', p_{l,x}, p_{l,y}, p_{l,z})$ and an unknown final hadronic state, $X$, of four-momentum $p'$:
\begin{eqnarray}
\nu_l(k) + N(p) \rightarrow l^{\pm}(k') + X(p') \, .
\label{eq_cs:processo}
\end{eqnarray}
The Feynman diagram corresponding to this process is shown in Figure \ref{fig_cs:diagramaDIS}.

DIS is the process in which a gauge boson is exchanged between the lepton and the target, this boson being energetic enough to fragment the hadronic target by interacting with its constituents. DIS experiments at SLAC with electron beams striking proton targets showed that the proton has point-like particles inside it \cite{Bloom:1969kc,Breidenbach:1969kd}. These point-like particles were initially called partons by Feynman \cite{Feynman:1969ej}. Feynman's parton model assumes the limit where the negative of the squared four-momentum of the exchanged gauge boson ($Q^2 = -q^2 = -(k - k')^{2}$) tends to infinity, and in this limit the DIS data point to asymptotically free partons within the nucleon, as discussed in the previous chapter in Subsection \ref{subsec_MP:assFred}, where we addressed asymptotic freedom. Partons have never been observed free in nature, only in bound states called hadrons. This phenomenon is called confinement. The combination of these two properties, asymptotic freedom and confinement, is a fundamental aspect in the formulation of Quantum Chromodynamics.

The double differential cross section of the DIS can be written as \cite{Paschos:2001np,Kretzer:2002fr,Haider:2016zrk}
\begin{eqnarray}
\frac{\mathrm{d}\sigma^{\nu(\bar{\nu})}}{\mathrm{d}\Omega_l' \mathrm{d}E_l'}=
\left(
\frac{G_F}{2 \pi}
\right)^{2}
\frac{|\vec{k}'|}{|\vec{k}|} L^{\mu \nu}W_{\mu \nu} \, ,
\label{eq_cs:DIS1}
\end{eqnarray}
where $G_F$ is the Fermi constant ($G_F = 1.166 \times 10^{-5}$~GeV$^{-2}$), $\Omega_l'$ is the solid angle of the scattered lepton with energy $E_l'$, and $L^{\mu \nu}$ and $W_{\mu \nu}$ are the leptonic and hadronic tensors, respectively. In addition to $Q^2$, it is common to describe the cross section of the DIS in terms of the Lorentz invariant quantities $y$ and $x$, called inelasticity and Bjorken-$x$, respectively. The Bjorken-$x$ variable is defined as $x = Q^2 / 2 p\cdot q$, and is interpreted as the fraction of momentum of the nucleon carried by the parton in the parton model. Inelasticity is defined as $y = p\cdot q / p\cdot k$, and in the rest frame of the target nucleon it becomes the fraction of the incident lepton's energy that is transferred to the nucleon. Another observable of interest from DIS is the invariant mass of the hadronic final state, $W^2 = (p + q)^2$, which in terms of the scalars $x$ and $Q^2$ becomes $W^{2} = m_N^{2} + Q^2(1/x - 1)$.

The leptonic tensor can be obtained directly from Feynman rules for the electroweak theory, and for unpolarized leptons in the final state it is given by \cite{Paschos:2001np,Ansari:2020xne}
\begin{eqnarray}
L^{\mu \nu } = k^{\mu}k'^{\nu} + k'^{\mu} k^{\nu} - g^{\mu \nu} k\cdot k' \pm i\epsilon ^{\mu \nu \alpha \beta} k'_{\alpha }k_{\beta} \, ,
\label{eq_cs:tensorL}
\end{eqnarray}
with $\epsilon$ being the Levi-Civita symbol. The sign of the antisymmetric term is positive for neutrinos and negative for antineutrinos. In the neutral current interaction by the exchange of a photon, this term disappears, as it is related to parity violation in the weak interaction.

The hadronic tensor describes the process that occurs in the nucleon, and can be written as \cite{Paschos:2001np,Kretzer:2002fr,Haider:2016zrk,Ansari:2020xne}
\begin{eqnarray}
W_{\mu \nu } = 
\frac{1}{2m_N} \sum_{X} \langle N(p) | j_\mu (0) | X(p') \rangle \langle X(p') | j_\nu (0) | N(p) \rangle
 \, ,
\label{eq_cs:tensorW}
\end{eqnarray}
where $m_N$ is the mass of the target nucleon and $j(0)$ is the hadronic current. The hadronic tensor cannot be easily obtained analytically like the leptonic tensor of Equation (\ref{eq_cs:tensorL}), because the vertex term has extra complications in the case of interaction of the gauge boson with a particle that has internal structure. To construct the hadronic tensor, we will start from its Lorentz invariance: the hadronic tensor must be written in terms of the metric $g_{\mu \nu}$ and the four-vectors $p$ and $q$. The most general possible form for the hadronic tensor is
\begin{eqnarray}
\begin{aligned}
W_{\mu \nu } = 
-g_{\mu \nu} W_1(x, Q^2) + \frac{p_{\mu}p_{\nu}}{m_N^2} W_2(x, Q^2) - \frac{i}{m_N^2} \epsilon_{\mu \nu \alpha \beta} p^{\alpha} q^{\beta} W_3(x, Q^2) + \frac{q_\mu q_\nu}{m_N^{2}} W_4(x,Q^2) + \\
+\frac{(p_\mu q_\nu + q_\mu p_\nu)}{m_N^{2}} W_5(x,Q^2) + \frac{i(p_\mu q_\nu - q_\mu p_\nu)}{m_N^{2}} W_6(x,Q^2)
 \, .
\end{aligned}
\label{eq_cs:tensorW2}
\end{eqnarray}
$W_i(x,Q^2)$ are structure functions that need to be determined. The term with $W_6$ disappears when we contract the hadronic tensor with the leptonic tensor. It is common to write the structure functions $W_i(x,Q^2)$ in terms of the dimensionless structure functions $F_i(x,Q^2)$ as \cite{Paschos:2001np,Kretzer:2002fr}
\begin{eqnarray}
\begin{aligned}
F_1(x,Q^2) = & W_1(x,Q^2); \\
F_2(x,Q^2) = & \frac{Q^2}{2xm_N^2}W_2(x,Q^2); \\
F_3(x,Q^2) = & \frac{Q^2}{xm_N^2}W_3(x,Q^2); \\
F_4(x,Q^2) = & \frac{Q^2}{2m_N^2}W_4(x,Q^2); \\
F_5(x,Q^2) = & \frac{Q^2}{2xm_N^2}W_5(x,Q^2)
 \, .
\end{aligned}
\label{eq_cs:Fi}
\end{eqnarray}
The structure function $F_1$ is related to the contribution of exchanged bosons with transverse polarization, while $F_2$ takes into account both transverse and longitudinal polarizations of the exchanged boson \cite{Devenish:2004pb,Badelek:2022cgr}. When the contribution of longitudinal polarization is neglected, the Callan-Gross relation ($F_2 = 2x F_1$) arises \cite{Callan:1969uq}. The structure function $F_3$ is related to parity violation in the interaction, arising from interference between the vector and axial parts of the weak current. The structure functions $F_4$ and $F_5$ appear due to longitudinal components of the weak current, but only for the axial part of the current. In particular, $F_4$ is sensitive to the partial non-conservation of the weak axial current \cite{LlewellynSmith:1971uhs}.

The double differential cross section for the process of Equation (\ref{eq_cs:DIS1}) represented in Figure \ref{fig_cs:diagramaDIS} is commonly written in differential form in the variables $x$ and $y$ with the transformation $\mathrm{d}x\,\mathrm{d}y = [E_l'/(m_N \nu)] \mathrm{d}\Omega ' \mathrm{d}E_l'$, with $\nu = E_\nu - E_l'$. With such a transformation, and contracting the leptonic tensor of Equation (\ref{eq_cs:tensorL}) with the hadronic tensor of Equation (\ref{eq_cs:tensorW}) written in terms of the structure functions $F_i$ of Equation (\ref{eq_cs:Fi}), we obtain the double differential cross section given by \cite{Paschos:2001np,Kretzer:2002fr}
\begin{eqnarray}
\begin{aligned}
\frac{\mathrm{d}\sigma^{\nu(\bar{\nu})}}{\mathrm{d}x\mathrm{d}y}=
\frac{G_F^2E_\nu m_N}{\pi}
\left( \frac{M_W^2}{Q^2+M_W^2} \right)^2
\left\{
\left(
y^2x+\frac{m_l^2y}{2E_\nu m_N}
\right)F_1(x,Q^2)+\right. \\
\left(
1-y-\frac{m_l^2}{4E_\nu^2}-\frac{m_Nxy}{2E_\nu}
\right)F_2(x,Q^2)+ \\
 +(-)\left(
xy-\frac{xy^2}{2}-\frac{m_l^2y}{4E_\nu m_N}
\right)F_3(x,Q^2) +\\
\left. +\frac{m_l^2 (m_l^2+Q^2)}{E_\nu^2 m_N^2 x}F_4(x,Q^2)-\frac{m_l^2}{E_\nu m_N}F_5(x,Q^2)
\right\} 
  \, ,
\label{eq_cs:sigmaDISCC}
\end{aligned}
\end{eqnarray}
where $M_W$ is the mass of the $W^\pm$ boson and $m_l$ is the mass of the charged lepton produced in the final state. The total cross section can be obtained from Equation (\ref{eq_cs:sigmaDISCC}) integrating in $x$ and $y$ within their respective kinematically allowed ranges. Neglecting the masses of the target nucleon and the charged lepton produced, the integration intervals are from 0 to 1 in both $x$ and $y$. Taking into account the masses in the process, the limits become \cite{Paschos:2001np,Kretzer:2002fr}
\begin{eqnarray}
\begin{aligned}
\frac{m_l^2}{2m_N(E_\nu - m_l)} \leq x \leq 1, \\
a-b \leq y \leq a+b \, ,
\label{eq_cs:limites}
\end{aligned}
\end{eqnarray}
where
\begin{eqnarray}
\begin{aligned}
a = \frac{1-m_l^2\left( \frac{1}{2m_N E_\nu x} + \frac{1}{2 E_\nu^2} \right)}{2 \left( 1+\frac{m_N x }{2 E_\nu} \right) } \, , \\
b = \frac{\sqrt{\left( 1 - \frac{m_l^{2}}{2 m_N E_\nu x} \right)^{2}  - \frac{m_l^{2}}{E_\nu^{2}}}}{2 \left( 1+ \frac{m_Nx}{2 E_\nu} \right)}\, .
\label{eq_cs:limites2}
\end{aligned}
\end{eqnarray}

The structure functions of the Equation (\ref{eq_cs:Fi}) are defined by merging two formalisms: the parametrization of the hadronic tensor in terms of the structure functions described above, and the parton model plus collinear factorization in which the neutrino-parton cross section is calculated and summed over all partons of the nucleon. Under this approach, the structure functions are generically written as \cite{Kretzer:2002fr}
\begin{eqnarray}
F_i(x, Q^2) = 
\sum_{a = q, \bar{q}, g} \int_{x}^{1} \frac{\mathrm{d}z}{z} C_{i,a}\left( z, \alpha_s(\mu^{2}), \frac{Q^{2}}{\mu^{2}} \right) f_{a} \left( \frac{x}{z}, \mu^{2} \right) \, ,
\label{eq_cs:Fi2}
\end{eqnarray}
where $f_a(y, \mu^{2})$ is the parton distribution function of the $a$ parton, $\alpha_s$ is the QCD coupling constant, $\mu^{2}$ is the factorization scale, $z$ is a convolution variable, and $C_{i, a}$ are the QCD perturbative coefficients calculated in orders of $\alpha_s$. At leading order (LO), which corresponds to the tree-level process (gauge boson exchanged between the neutrino and the struck quark and does not take into account higher-order diagrams such as diagrams with gluon radiation by quarks), the PDFs $f_a(x)$ can be interpreted as the probability of finding a $a$ type quark in the nucleon carrying fraction $x$ of its longitudinal momentum. Also at LO, the coefficients $C_{i,a}$ are proportional to $\delta (1-z)$, and the structure functions become linear combinations of PDFs. The LO structure functions for a proton interacting with a(n) (anti)neutrino are
\begin{eqnarray}
\begin{aligned}
F_2^{\nu p}        = & 2x[d(x, Q^2)+s(x, Q^2)+b(x, Q^2)+\bar{u}(x, Q^2)+\bar{c}(x, Q^2)+\bar{t}(x, Q^2)] \\
xF_3^{\nu p}       = & 2x[d(x, Q^2)+s(x, Q^2)+b(x, Q^2)-\bar{u}(x, Q^2)-\bar{c}(x, Q^2)+\bar{t}(x, Q^2)] \\
F_2^{\bar{\nu} p}  = & 2x[u(x, Q^2)+c(x, Q^2)+t(x, Q^2)+\bar{d}(x, Q^2)+\bar{s}(x, Q^2)+\bar{b}(x, Q^2)] \\
xF_3^{\bar{\nu} p} = & 2x[u(x, Q^2)+c(x, Q^2)+t(x, Q^2)-\bar{d}(x, Q^2)-\bar{s}(x, Q^2)+\bar{b}(x, Q^2)] \, .
\end{aligned}
\label{eq_cs:Fi_CC}
\end{eqnarray}
Note that in the equations above we only defined the structure functions $F_2$ and $F_3$. At LO, the structure function $F_1$ can be written in terms of $F_2$ using the Callan-Gross relation ($F_2 = 2 x F_1$)  \cite{Callan:1969uq}. Also at LO, in the limit where quarks and target nucleon are massless, we can use the Albright-Jarlskog relations \cite{Albright:1974ts}, which make $F_4 = 0$ and $F_2 = 2 x F_5$. The term that multiplies the structure function $F_5$ is proportional to the square of the mass of the charged lepton produced, so it will only play an important role in tau neutrino interactions. The structure functions for a target neutron are obtained using isospin symmetry: $u \leftrightarrow d$.

We can construct a double differential cross section for the neutrino neutral current interaction using the same procedure for obtaining the Equation (\ref{eq_cs:sigmaDISCC}). Following this methodology, the double differential cross section is given by \cite{Paschos:2001np}
\begin{eqnarray}
\begin{aligned}
\frac{\mathrm{d}^{2}\sigma^{\nu(\bar \nu)}}{\mathrm{d} x\mathrm{d} y} = \frac{G^{2}_{F}m_{N}E_{\nu}}{\pi}\left(\frac{M^{2}_{W}}{Q^{2}+m^{2}_{W}} \right)^{2}\left\{    
xy^{2}F_{1}^{NC}(x,Q^{2})\right. \\
+\left.  \left[ 1   - \left( 1+\frac{m_{N}x}{2E_{\nu}}\right)y  \right] F_{2}^{NC}(x,Q^{2}) 
\right.   \\
\pm \left.  \left[ xy\left( 1 - \frac{y}{2} \right)   \right] F_{3}^{NC} (x,Q^{2}) \right\} \, .
\end{aligned}
\label{eq_cs:sigmaDISNC}
\end{eqnarray}
Compared to the charged current interaction, its main differences are due to the mass of the interaction-mediating boson, the negligible mass of the lepton in the final state (neutrino mass), and the structure functions, which in this LO process are given by \cite{Paschos:2001np}
\begin{eqnarray}
\begin{aligned}
F_2^{\nu p,\bar{\nu} p}(x,Q^2) = 2x(g_L^2 + g_R^2)[u(x,Q^2)+c(x,Q^2)+\bar{u}(x,Q^2)+\bar{c}(x,Q^2)] + \\ 2x(g_L^{'2} + g_{R}^{'2})[d(x,Q^2)+s(x,Q^2)+\bar{d}(x,Q^2)+\bar{s}(x,Q^2)] \\
xF_3^{\nu p, \bar{\nu} p}(x,Q^2) = 2x(g_L^2 - g_R^2)[u(x,Q^2)+c(x,Q^2)-\bar{u}(x,Q^2)-\bar{c}(x,Q^2)] + \\
2x(g_{L}^{'2} - g_{R}^{'2})[d(x,Q^2)+s(x,Q^2)-\bar{d}(x,Q^2)-\bar{s}(x,Q^2)] \, ,
\end{aligned}
\label{eq_cs:f_NC}
\end{eqnarray}
where $g_L$ and $g_R$ are the couplings of the $Z^{0}$ boson. The Callan-Gross relation remains valid at leading order calculation of the neutral current interaction cross section.

For completeness, the double differential cross section of the neutral current DIS with the incident lepton being electrically charged (exchange of a photon) is given by 
\begin{eqnarray}
\begin{aligned}
\frac{\mathrm{d}^{2}\sigma^{l^{\pm}}}{\mathrm{d} x\mathrm{d} y} =  \frac{8\pi \alpha^{2}E_l m_N}{Q^{4}} \left[ x y^{2} F_{1}^{\gamma}(x,Q^{2}) + (1-y) F_{2}^{\gamma}(x,Q^{2}) \right] \, .
\end{aligned}
\label{eq_cs:sigmaDISNCfoton}
\end{eqnarray}
where $\alpha$ is the fine-structure constant of electromagnetic coupling and $E_l$ is the energy of the incident lepton. The Callan-Gross relation is also valid at LO and the structure function $F_2^{\gamma}$ is the same as that presented in Equation (\ref{eq_cs:f_NC}) for the neutrino neutral current interaction except for a change in the coupling term to $q_i^{2}$, where $q_i$ is the charge of the charged quark of type $i$. The structure function $F_3$ does not exist in the electromagnetic interaction, since this term is related to the parity violation at the interaction vertex, which in the Standard Model is a property exclusive to weak interactions.

\subsection{PDFs and the DGLAP evolution equations}
\label{subsec_cs:PDFs}

The partonic distributions used above in the definition of the structure functions provide the distribution of quarks and gluons, such that $\int q(x, Q^2)\mathrm{d}x$ gives us the number of quarks $q$ in the interval of $x$ integrated for a given $Q^2$. It is possible to obtain the partonic distributions from first principles of QCD, as in QCD calculations on the lattice \cite{Cichy:2018mum}. In our work we use the description of partonic distributions from parameterizations of experimental data, especially from deeply inelastic electron-proton scattering, or even in hadronic collisions such as those performed at the LHC.

The experimental data currently available for DIS cover a limited region in $x$ and $Q^2$, insufficient for cross section calculations in certain energy regimes \cite{H1:2015ubc}. To estimate partonic distributions in unknown kinematic regions, there are solutions obtained from perturbative QCD that evolve the distributions to a different regime, starting from a distribution on a factorization scale that is experimentally parameterized. The equations that perform this evolution of the partonic distribution in $Q^2$ are the Dokshitzer-Gribov-Lipatov-Altarelli-Parisi (DGLAP) equations \cite{Gribov:1972ri,Altarelli:1977zs,Dokshitzer:1977sg}.

\begin{figure}
	\centering
				\begin{tabular}{ccc}
	\includegraphics[width=0.98\textwidth]{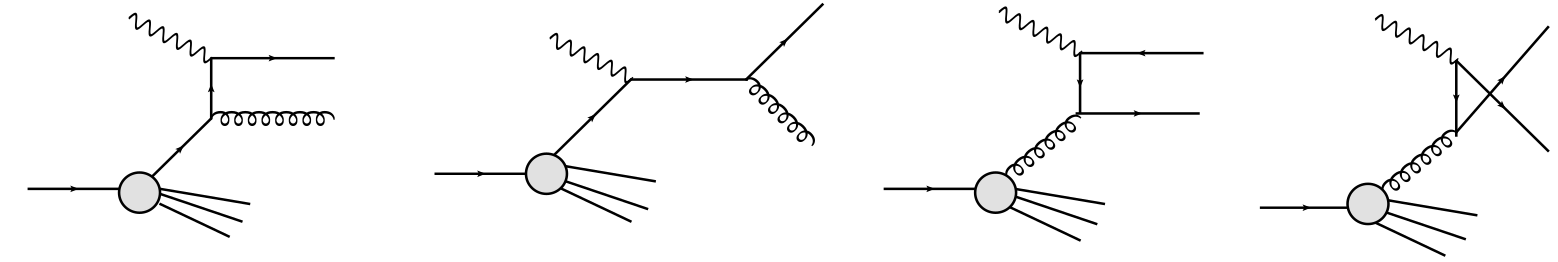}
				\end{tabular}
	\caption{ Feynman diagrams that contribute to the NLO for the DIS. Figure taken from the reference \cite{DuartedaSilvaMoreira:2017wvr}. }
	\label{fig_cs:NLO_DIS}
\end{figure}

To better understand the dynamics of the DGLAP equations, we will start by rewriting the Equation (\ref{eq_cs:Fi2}) for the structure function $F_2$ for a quark $q$:
\begin{eqnarray}
F_2(x, Q^2) = 
\sum_{a = q, \bar{q}} \int_{x}^{1} \frac{\mathrm{d}z}{z} C_{2,q}\left( z, \alpha_s(\mu^{2}), \frac{Q^{2}}{\mu^{2}} \right) f_{a} \left( \frac{x}{z}, \mu^{2} \right) \, .
\label{eq_cs:F2}
\end{eqnarray}
There are four diagrams at next to leading order (NLO) of QCD that contribute to DIS, they are shown in Figure \ref{fig_cs:NLO_DIS}. Of the four diagrams, two are for real gluon emission and two for fusion with virtual gluons. In the calculation of the matrix elements of these diagrams, singularities arise that need to be handled carefully. We will call collinear singularities those that arise in the case of real gluon emission diagrams. This singularity arises when the emitted gluon is collinear with the quark that emitted it. There are also singularities for soft gluon emission, but these cancel out when we sum the contributions of the real and virtual gluons from the diagrams in Figure \ref{fig_cs:NLO_DIS} \cite{Barone:2002cv}.

To regularize the collinear singularity, we will introduce a cut $k_0^{2}$ in the transverse momentum $k_{T}^{2}$ of the interacting quark. Then we can write the coefficient functions, which are calculated perturbatively up to the order $\mathcal{O}(\alpha_s)$, as 
\begin{eqnarray}
    C_{2,q} = e_q^{2}\delta(1-z) + \frac{\alpha_s}{2\pi} e_{q}^{2} z \left[ P(z) \mathrm{ln}\,\frac{Q^{2}}{k_0^{2}} + h(z) \right] \, ,
    \label{eq_cs:C2}
\end{eqnarray}
with $P(z)$ and $h(z)$ being finite functions. We can use regularization for the collinear singularity, introducing the factorization scale $\mu^{2}$
\begin{eqnarray}
    \mathrm{ln}\, \left( \frac{Q^{2}}{k_0^{2}} \right) = \mathrm{ln}\, \frac{Q^{2}}{\mu^{2}} + \mathrm{ln} \, \frac{\mu^{2}}{k_0^{2}} \, .
    \label{eq_cs:regularization}
\end{eqnarray}
We can also separate the function $h(z) = \bar{h}(z) + h'(z)$. Absorbing the singularity $\mathrm{ln} \, (\mu^{2} / k_0^{2})$ and the term $h'(z)$ into a redefinition of the parton distribution functions, we obtain
\begin{eqnarray}
    f_{a}(x, \mu^{2}) = f_a^{0} + \frac{\alpha}{2\pi} \int^{1}_x \frac{\mathrm{d}\xi}{\xi} f_{q}^{0}(\xi) \left[ P\left( \frac{x}{\xi} \right) \mathrm{ln}\,\frac{\mu^{2}}{k_0^{2}} + h' \left( \frac{x}{\xi}\right) \right] \, .
    \label{eq_cs: PDF_redefinidas}
\end{eqnarray}

Given that $F_2$ is a physical observable, it cannot depend on the factorization scale $\mu^{2}$ introduced. Therefore, we can impose the boundary condition.
\begin{eqnarray}
    \frac{\partial F_2}{\partial (\mathrm{ln}\, \mu^{2})} = 0 \, ,
    \label{eq_cs:contornoF2}
\end{eqnarray}
and when applying it to Equation~(\ref{eq_cs:F2}), after some algebra manipulations, it becomes
\begin{eqnarray}
    \frac{\partial f_a(x,\mu^{2})}{\partial (\mathrm{ln}\, \mu^{2})} = \frac{\alpha_s}{2 \pi} \int^{1}_{x} P\left( \frac{x}{\xi} \right)f_a(\xi , \mu^{2}) \, .
    \label{eq_cs:contornoPDF}
\end{eqnarray}
The equation above is known as the DGLAP equation \cite{Gribov:1972ri,Altarelli:1977zs,Dokshitzer:1977sg}. The function $P$ is known as the splitting function, and represents the probability of a quark emitting another quark with momentum fraction $x$. Although we have given an illustrative demonstration of the DGLAP equations for quarks, we can analogously generalize our result to the distribution of gluons in the nucleon. To show the more general DGLAP equations of partonic evolution, we first define the non-singlet distribution of quarks as
\begin{equation}
q_{\text{NS}}(x, Q^2) = q(x, Q^2) - \bar{q}(x, Q^2) \, ,
\label{eq_cs:qns_def}
\end{equation}
and the singlet distribution as
\begin{equation}
\Sigma(x, Q^2) = \sum_i \left[ q_i(x, Q^2) + \bar{q}_i(x, Q^2) \right] \, , 
\label{eq_cs:sigma_def}
\end{equation}
Therefore, we can show that the DGLAP evolution equations (setting $t = \ln(Q^2 / \mu^2)$) can be written in the form
\begin{equation}
\frac{\partial q_{\text{NS}}(x, t)}{\partial t}
  = \frac{\alpha_s(t)}{2\pi} 
    \int_x^1 \frac{d\xi}{\xi} \,
    P_{qq}\!\left(\frac{x}{\xi}\right)
    q_{\text{NS}}(\xi, t),
\label{eq_cs:dglap_ns}
\end{equation}
and
\begin{equation}
\frac{\partial}{\partial t}
\begin{pmatrix}
\Sigma(x, t) \\[3pt]
g(x, t)
\end{pmatrix}
=
\frac{\alpha_s(t)}{2\pi}
\int_x^1 \frac{d\xi}{\xi}
\begin{pmatrix}
P_{qq}\!\left(\frac{x}{\xi}\right) & 2n_f\, P_{qg}\!\left(\frac{x}{\xi}\right) \\[6pt]
P_{gq}\!\left(\frac{x}{\xi}\right) & P_{gg}\!\left(\frac{x}{\xi}\right)
\end{pmatrix}
\begin{pmatrix}
\Sigma(\xi, t) \\[3pt]
g(\xi, t)
\end{pmatrix} \, ,
\label{eq_cs:dglap_singlet}
\end{equation}
with $g(x,t)$ being the gluon distribution.

At leading order, it is possible to show that the splitting functions are\cite{Barone:2002cv}:
\begin{equation}
\begin{aligned}
P_{qq}(z) &= \frac{4}{3} \left[ \frac{1 + z^2}{(1 - z)_+} + \frac{3}{2}\, \delta(1 - z) \right] \, , 
P_{qg}(z) &= \frac{1}{2} \left[ z^2 + (1 - z)^2 \right] \, , 
P_{gq}(z) &= \frac{4}{3} \left[ \frac{1 + (1 - z)^2}{z} \right] \, , 
P_{gg}(z) &= 6 \left[ \frac{1-z}{(z)_+} + \frac{z}{(1-z)_+} 
            +  \frac{33 - 2 n_f}{6} \delta(1 - z) \right]  \, ,
\label{eq_cs:Pqq}\\[4pt]
\end{aligned}
\end{equation}
with the definition
\begin{equation}
\int_0^1 f(z)\,[g(z)]_+\, dz 
\;=\;
\int_0^1 \frac{f(z) - f(1)}{1 - z}\, dz \, .
\label{eq_cs:plus_distribution}
\end{equation}

DGLAP equations do not predict parton distribution functions for a given nucleon, they perform their evolution on the factorization scale $\mu^{2}$, starting from a known PDF. PDFs can be obtained in different ways. One of them, quite common in phenomenological studies of high energies, consists of assuming a certain parameterizing function for each parton and fixing the parameters according to experimental data. There are several collaborations that produce PDF fits using global experimental data to obtain PDFs of protons and nuclei, using different physical assumptions in their constructions, such as logarithmic resummation, mass effects, and corrections at different perturbative orders that currently take into account diagrams up to NNNLO (next to next to next to leading order).
Among the collaborations that provide these parton distributions from the solutions of the DGLAP equations, we highlight here CTEQ \cite{Botts:1993shg,CTEQ:1993hwr} and NNPDF \cite{DelDebbio:2004xtd,DelDebbio:2007ee}.

\begin{figure}
	\centering
			\begin{tabular}{ccc}
			\includegraphics[width=0.31\textwidth]{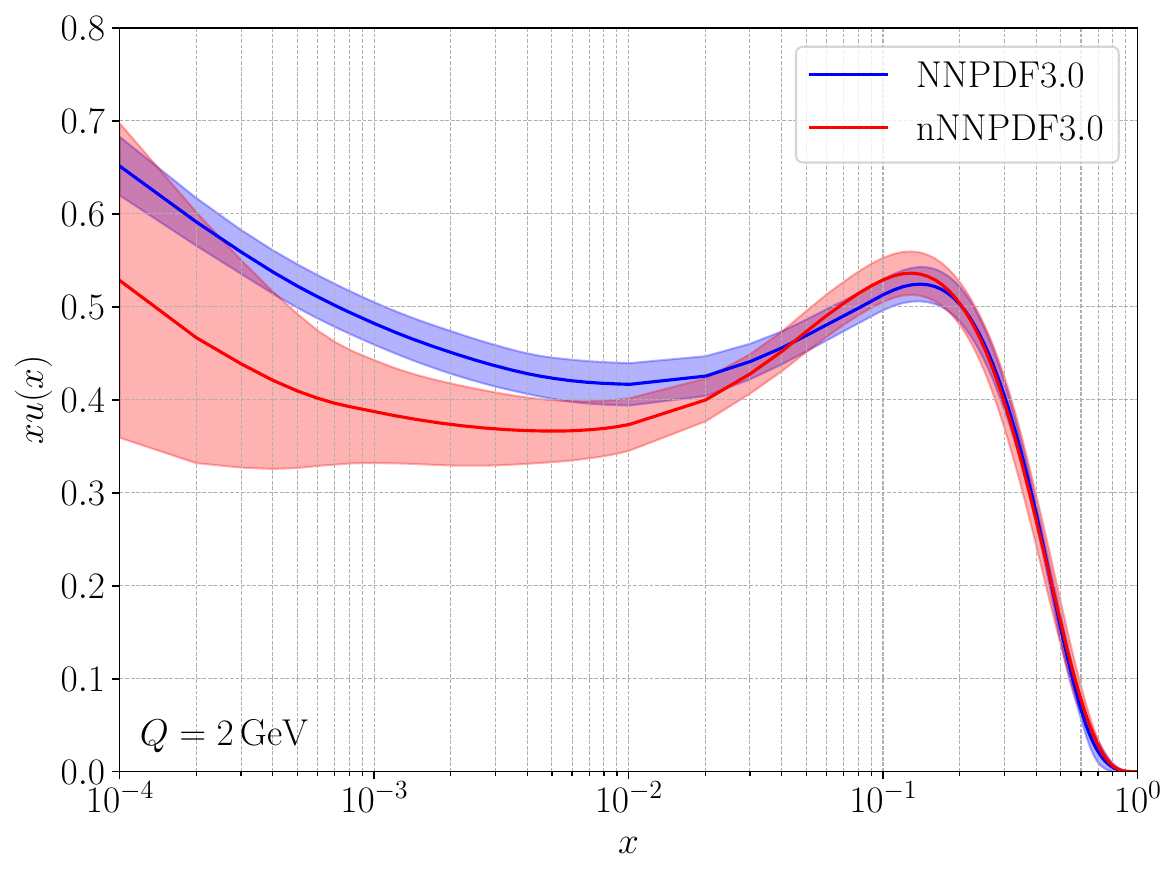} &
			\includegraphics[width=0.31\textwidth]{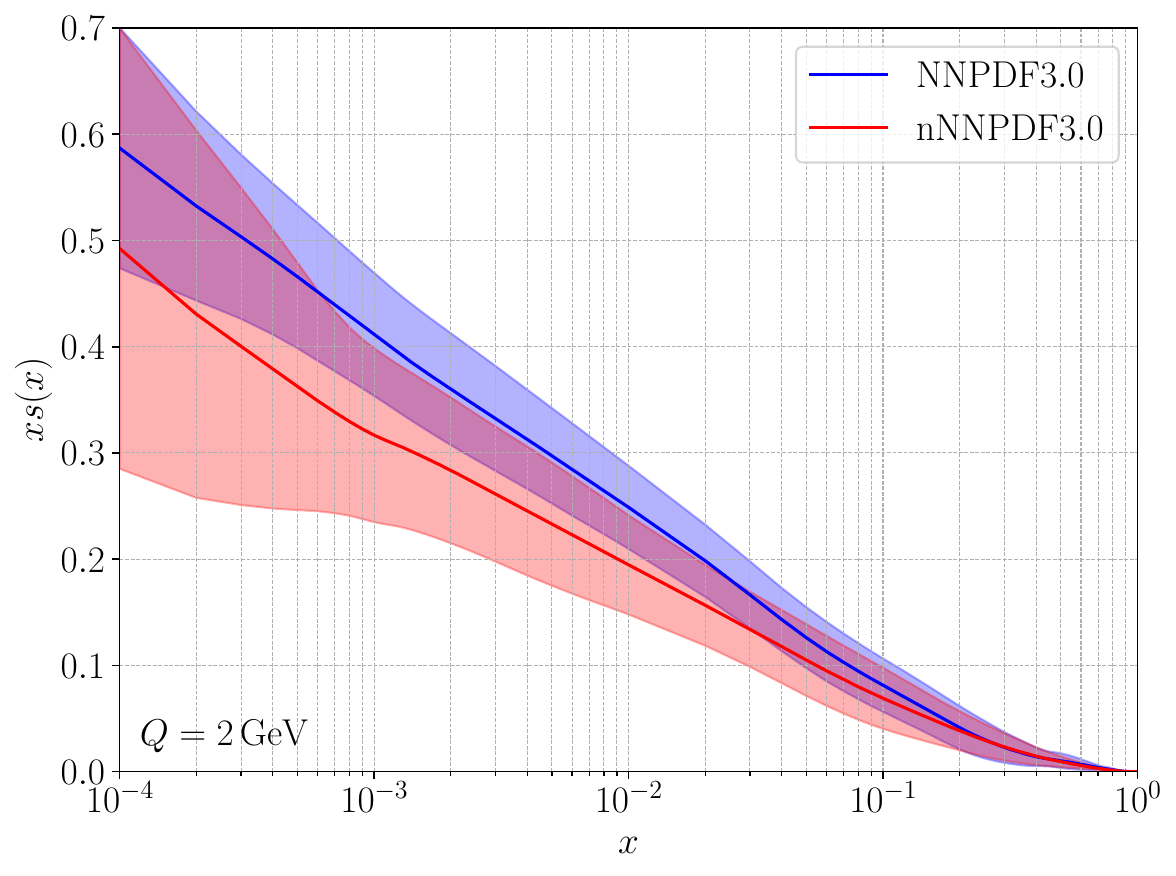} &
			\includegraphics[width=0.31\textwidth]{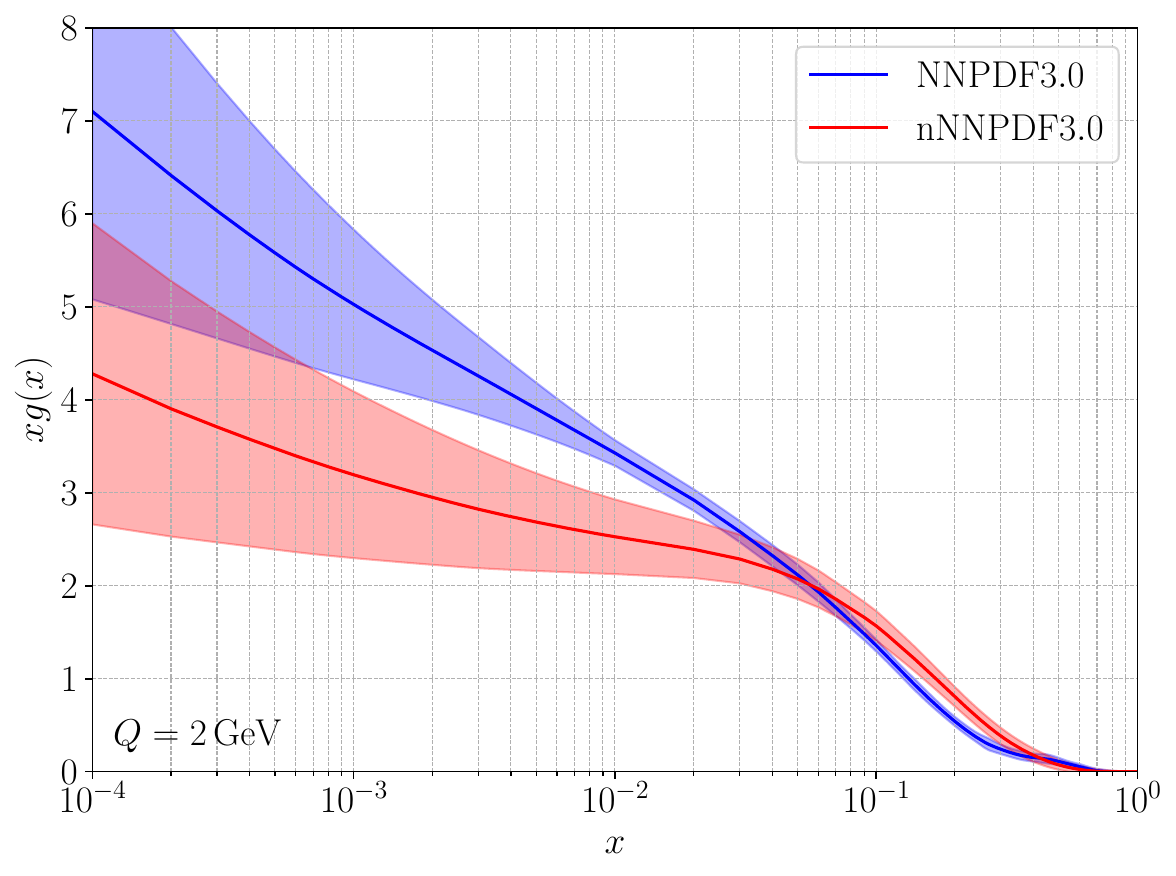}\\
			\includegraphics[width=0.31\textwidth]{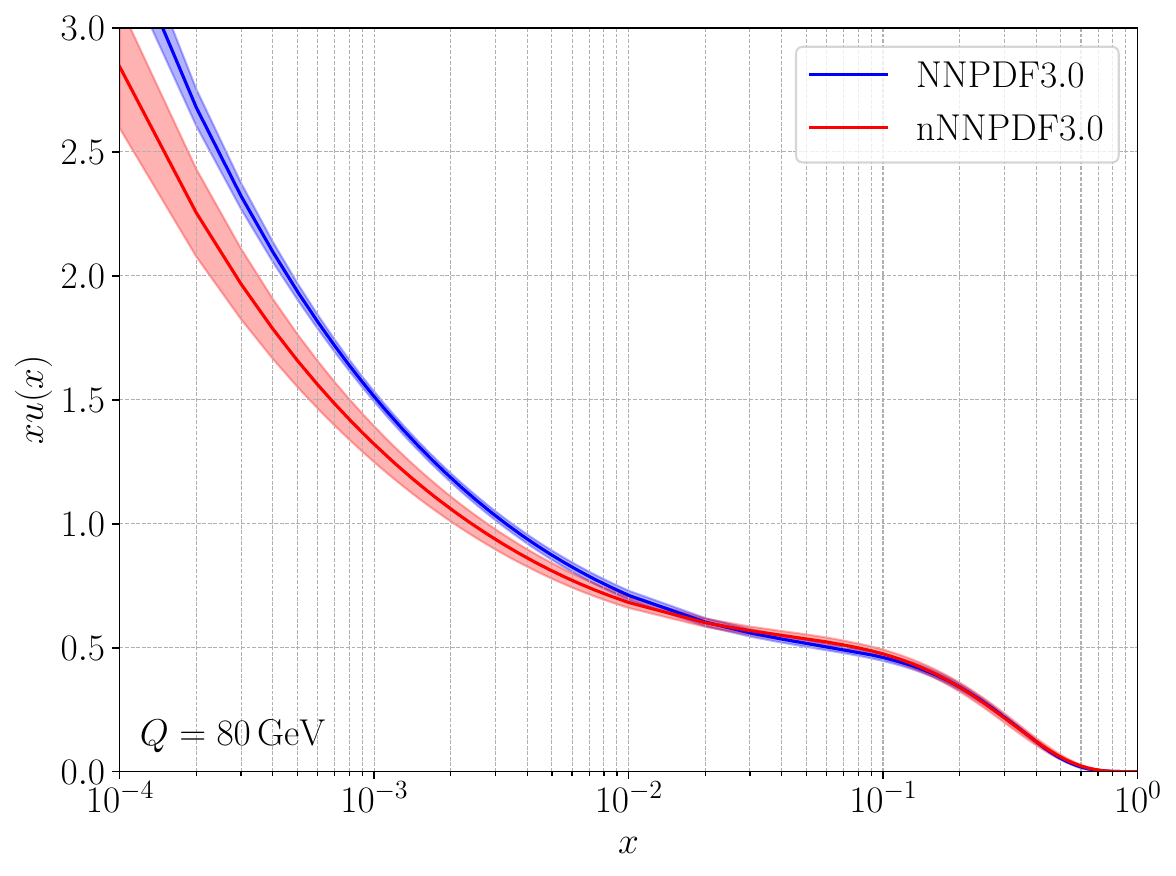} &
			\includegraphics[width=0.31\textwidth]{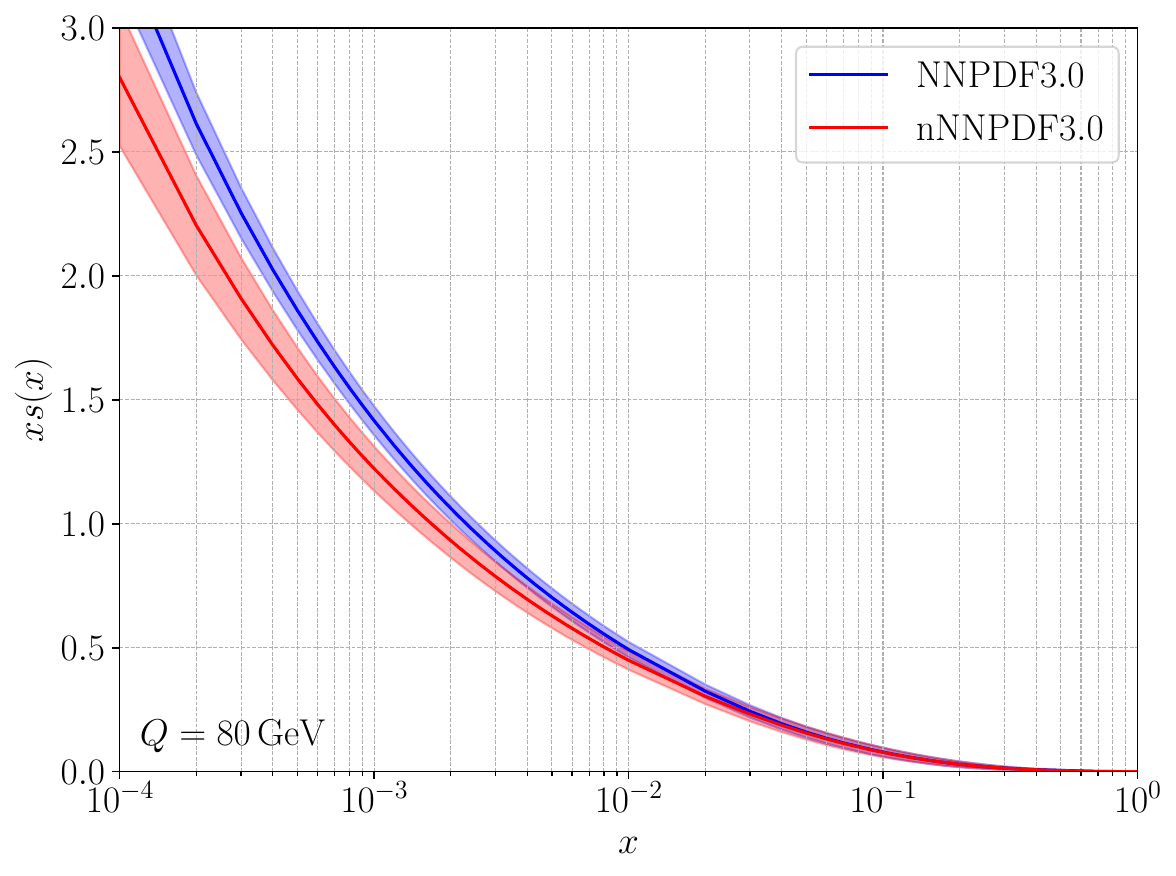} &
			\includegraphics[width=0.31\textwidth]{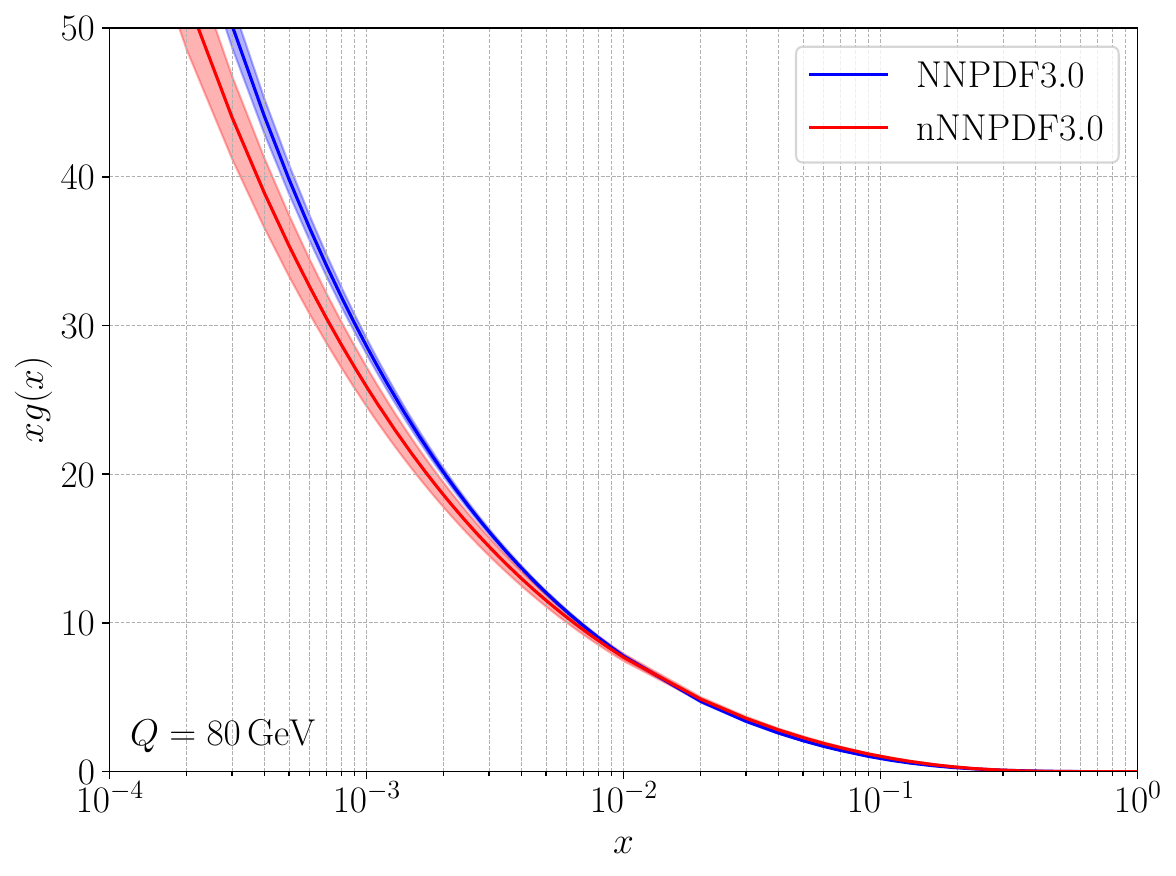}\\
			\includegraphics[width=0.31\textwidth]{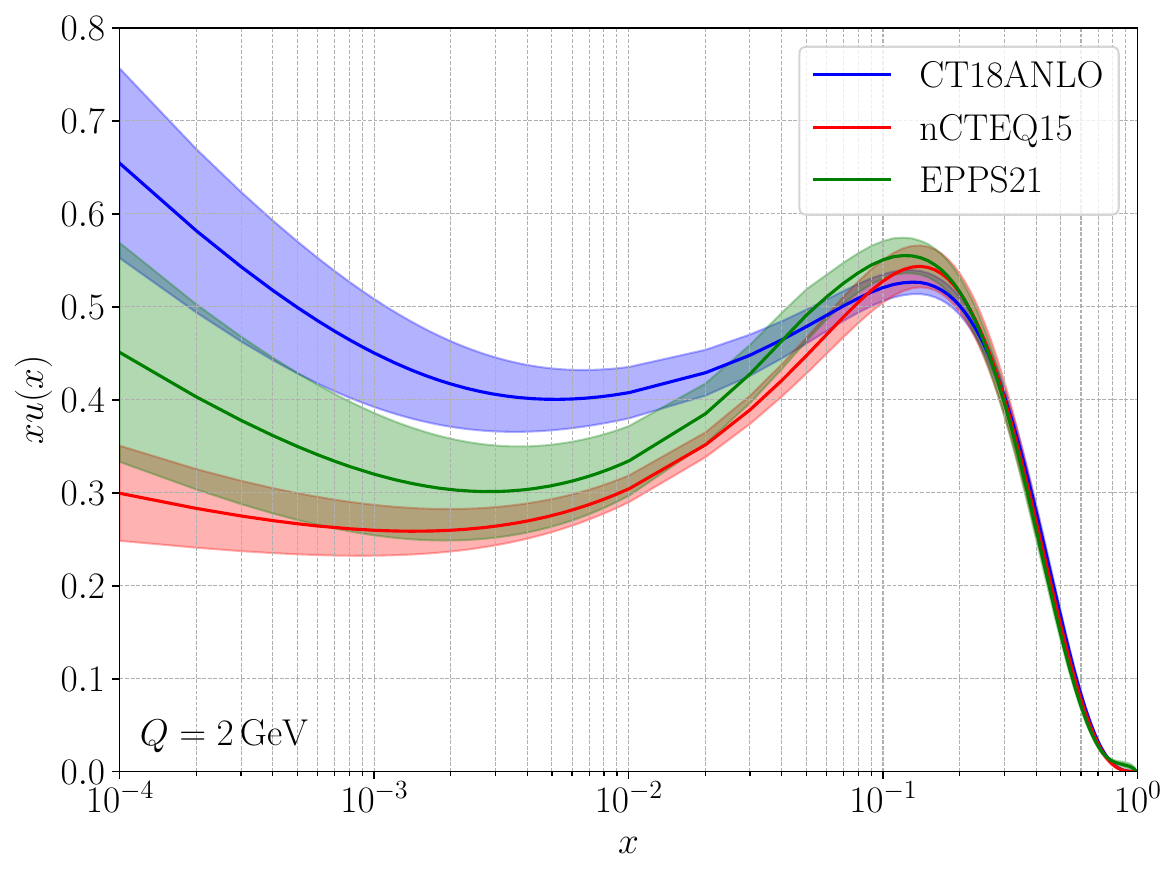} &
			\includegraphics[width=0.31\textwidth]{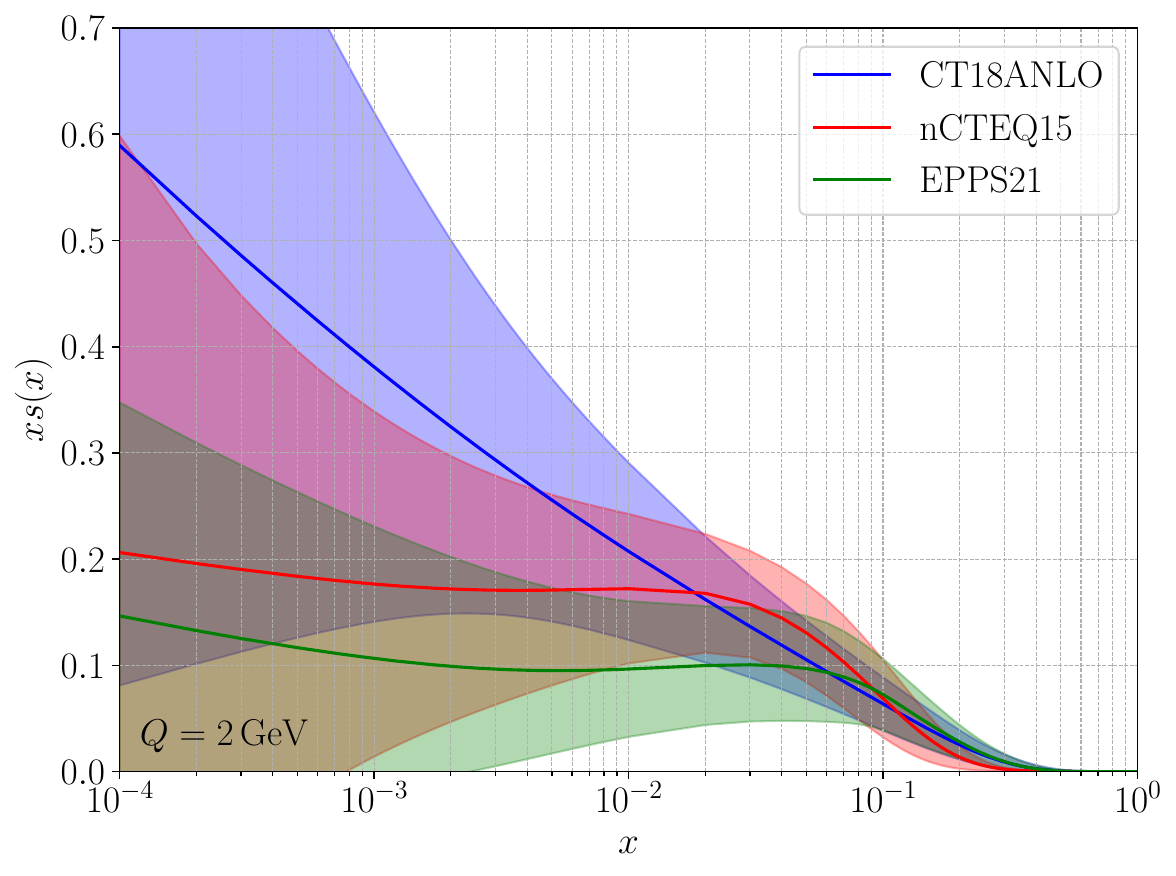} &
			\includegraphics[width=0.31\textwidth]{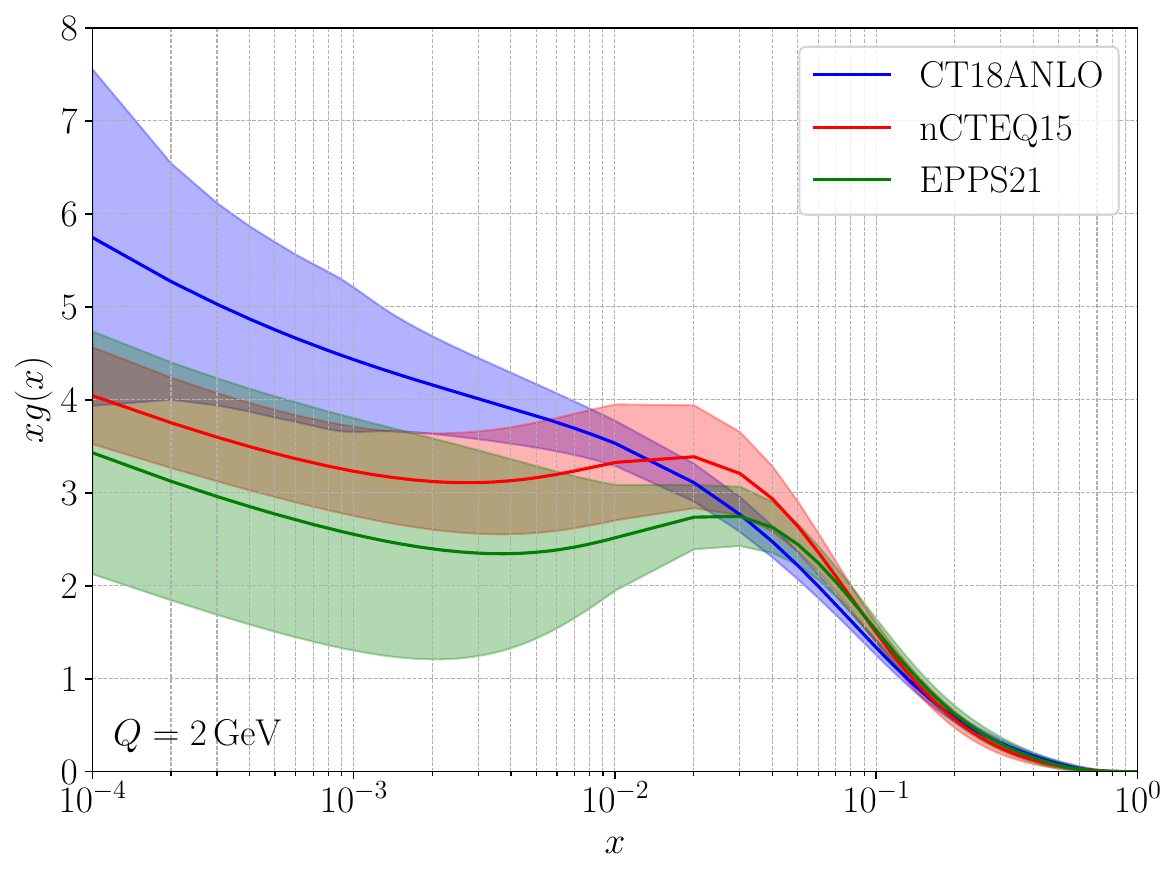}\\
			\includegraphics[width=0.31\textwidth]{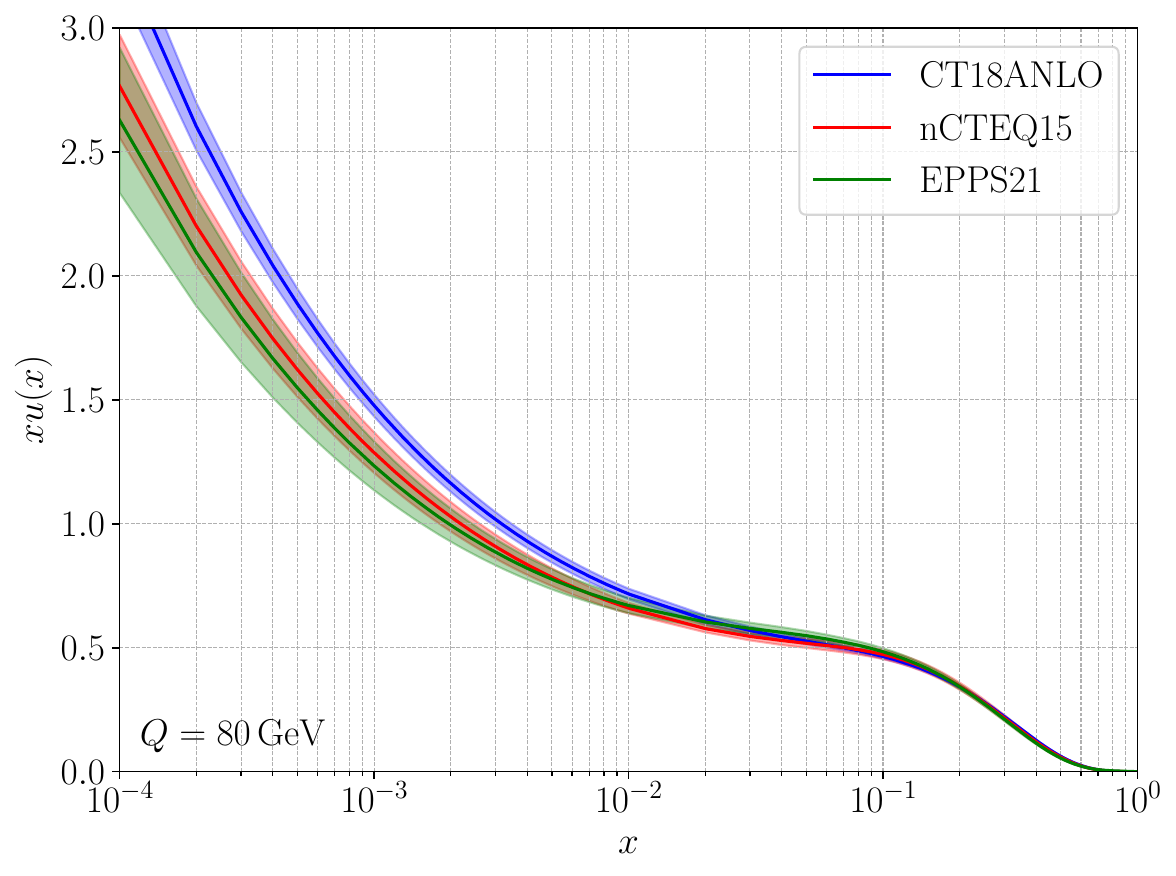} &
			\includegraphics[width=0.31\textwidth]{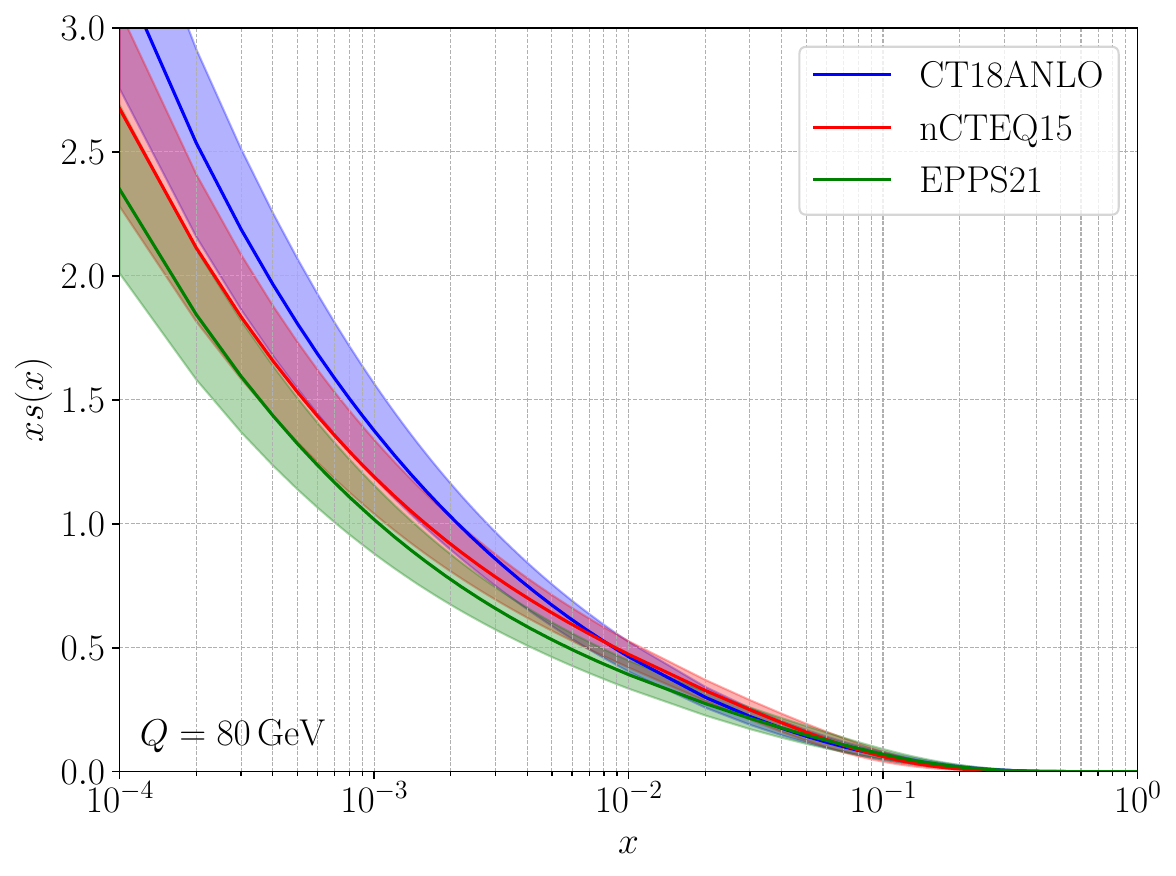} &
			\includegraphics[width=0.31\textwidth]{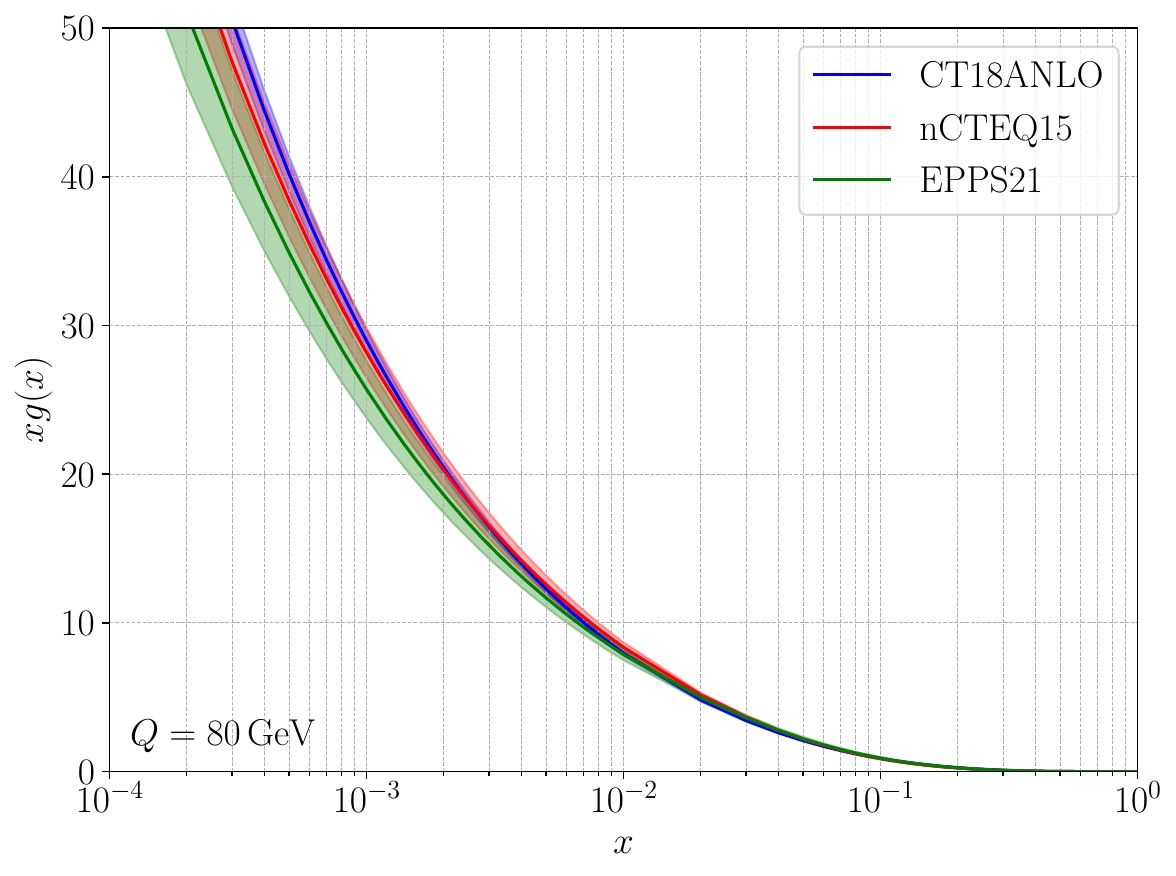}
		\end{tabular}
	\caption{Distributions of up quarks (left), strange quarks (center), and gluons (right) multiplied by $x$ in the average tungsten nucleon assuming $Q = 2\;\mathrm{GeV}$ ($Q = 80\;\mathrm{GeV}$) in the first and third (second and fourth) lines. To obtain these results, we used the parton distributions provided by the NNPDF, CT18ANLO, nCTEQ15HQ, and EPPS21 collaborations. }
	\label{fig_cs:pdf}
\end{figure}

Depending on the kinematic regime of interest, the parton distributions in a free nucleon compared to those in a bound nucleon in a nucleus can be similar. However, there are regimes in which nuclear effects become significantly important. For PDFs of bound nucleons in nuclei, there are also collaborations interested in parameterizing them, such as the CTEQ mentioned earlier, which provides the most recent version nCTEQ15HQ \cite{Kovarik:2015cma,Duwentaster:2022kpv,Muzakka:2022wey,nCTEQ:2023cpo}, and NNPDF, with its most current nuclear version called nNNPDF3.0 \cite{AbdulKhalek:2022fyi}. Another parameterization frequently used in similar works in the literature is EPPS \cite{Eskola:2007my,Eskola:2016oht,Eskola:2021nhw}. Unlike CTEQ and NNPDF, EPPS21 provides only the multiplicative factors that modify a given nucleon-free PDF to a bound nucleon PDF. In Figure \ref{fig_cs:pdf} we present the behavior for the PDFs of some parton types for $Q=2\,\mathrm{GeV}$ (first and third lines) and for $Q = 80\, \mathrm{GeV}$ (second and fourth lines). In the two upper lines of Figure \ref{fig_cs:pdf} we show the parameterizations of the NNPDF collaboration, while the two lower lines show the same PDFs for the CTEQ and EPPS parameterizations. We present a comparison of PDFs for a free nucleon using the NNPDF3.0p and CT18ANLO \cite{Hou:2019efy} parameterizations with the parameterizations for a nucleon bound in a tungsten nucleus according to the nNNPDF3.0, nCTEQ15, and EPPS21 parameterizations (the later used in conjunction with CT18ANLO). The uncertainty bands are in one sigma of statistical significance. We see that the nuclear PDFs have larger uncertainties, reflecting the smaller amount of data associated with scattering with nuclei.

There are four nuclear effects that are usually discussed in the literature in global data analyses and obtaining nuclear PDFs. These are the effects that modify the PDFs in Figure~\ref{fig_cs:pdf}, which indicates $q_A \lesssim q_N$ for momentum fractions $x \lesssim 0.1$ (shadowing region) and $0.3 \lesssim x \lesssim 0.7$ (EMC effect region), and that $q_A \gtrsim q_N$ for $0.1 \lesssim x \lesssim 0.3$ (anti-shadowing region) and $x \gtrsim 0.7$ (Fermi motion), where the indices $A$ and $N$ denote a bound nucleon in the atomic nucleus and a free nucleon, respectively. These nuclear effects are not fully understood from a theoretical point of view, with the exception of Fermi motion, which arises from a reference frame transformation of the nucleon when it is inside the nucleus. The other nuclear effects are generally interpreted as consequences of different physical mechanisms acting in distinct regions of $x$. At low values of $x$, the shadowing effect is attributed to the coherent interaction of hadronic fluctuations of the virtual photon with several nucleons of the nucleus \cite{Boroun:2023vuu}, which can be described in the Glauber–Gribov formalism and is closely related to diffractive processes \cite{Gribov:1968gs,Nikolaev:1990ja,Kopeliovich:2000ra}. In the intermediate region, anti-shadowing is often interpreted as a partial compensation for suppression at low $x$, associated with momentum conservation and constructive interference in multiple scattering processes \cite{Frankfurt:1988nt}. The EMC effect, on the other hand, is generally associated with genuine modifications of the nucleon's internal structure in the nuclear medium, such as virtuality effects of the bound nucleon, short-range correlations between nucleons, and possible multinuclear components \cite{Geesaman:1995yd,Weinstein:2010rt,Hen:2016kwk}.

The DGLAP equations perform the partonic evolution in $Q^{2}$, starting from an initial scale, $Q^{2}_{0}$. In the small $x$ regime, terms of the type $\alpha_s\mathrm{ln}\,1/x$ arise in the evolution of the gluon distribution functions, and initially these can generate convergence problems. Therefore, we need a mechanism that sums these terms, which leads to the BFKL evolution equations \cite{Kuraev:1977fs,Balitsky:1978ic}, obtained by Balitsky, Fadin, Kuraev and Lipatov, GLR \cite{Gribov:1983ivg}, obtained by Gribov, Levin and Ryskin, and BK \cite{Balitsky:1995ub,Kovchegov:1999yj}, obtained by Balitsky and Kovchegov. These equations perform partonic evolution in $x$ with $Q^{2}$ fixed, and perform the resummation of the aforementioned logarithms. The GLR and BK equations, in addition to the resummation of logarithms, take into account gluon recombination effects in high partonic density regimes, and are therefore called nonlinear evolution equations. Collaborations such as CTEQ and NNPDF, which perform parameterized partonic evolution in global analyses, use the DGLAP equations as standard. However, some specific results take into account aspects of the resummation of small $x$ terms and nonlinear effects \cite{Hou:2019efy,Ball:2017otu}.

As mentioned above, nuclear effects, as well as effects of high partonic densities, are still not well understood from a theoretical point of view. Regarding the experimental aspect, several experiments have measured nuclear effects, however, the parameterizations of different collaborations still differ significantly from each other, both in terms of magnitude and the position in $x$ of the nuclear effects, as seen in Figure \ref{fig_cs:pdf}, in addition to large uncertainty bands. The study of nuclear effects in deeply inelastic scattering with heavy nuclei remains an essential tool for quantifying PDFs and understanding their modifications in nuclei.

\section{Annihilation of electronic antineutrino with electron}
\label{sec_cs:glashow}

\begin{figure}
	\centering
				\begin{tabular}{ccc}
	\includegraphics[width=0.5\textwidth]{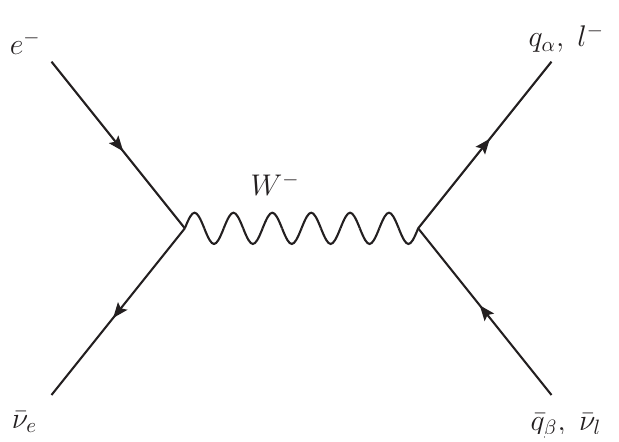}
				\end{tabular}
	\caption{ Feynman diagram for Glashow resonance. }
	\label{fig_cs:glashow}
\end{figure}

In addition to neutral and charged current neutrino-nucleon cross sections, to describe neutrino scattering with matter, we need to understand neutrino interactions with atomic electrons. In our work, we will focus on only one neutrino-electron process: the Glashow resonance \cite{Glashow:1960zz}, which consists of the annihilation of an electronic antineutrino with an electron, giving rise to a $W^{-}$ boson. The Feynman diagram for this process is shown in Figure \ref{fig_cs:glashow}. This process becomes important in the regime where the center-of-mass energy of the neutrino with the electron is close to the mass of the boson produced. In the electron's rest frame, this is equivalent to an incident neutrino with approximately 6.3 PeV. Glashow resonance is an important process in describing the propagation of electron antineutrinos on Earth and in describing ultra-high energy events at the IceCube Observatory. The cross section of this process is given by
\begin{eqnarray}
\sigma = 24\pi\Gamma^{2}_{W}\mathrm{Br}(W^-\rightarrow \bar{\nu}_e + e^-)
\frac{2E_\nu m_e/M_W^2}{(2E_\nu m_e - M_W^2)^2 + (M_W \Gamma_W)^2} \, ,
\label{eq_cs:glashow}
\end{eqnarray}
where $\Gamma_W$ is the decay width of the $W^{-}$ boson and $\mathrm{Br}(W^-\rightarrow \bar{\nu}_e + e^-)$ is its electron plus electronic antineutrino decay fraction. The Glashow resonance was observed at IceCube with a statistical significance of 2.3$\sigma$ in an event with a reconstructed deposited energy at the observatory of 6.05$\pm$0.72 PeV \cite{IceCube:2021rpz}. The observation of this process is of great interest to astrophysical neutrino physics, being sensitive to the antineutrino component of this flux \cite{Bhattacharya:2011qu,Xing:2011zm,Barger:2012mz,Bhattacharya:2012fh,Barger:2014iua,Biehl:2016psj,Huang:2023yqz,Liu:2023lxz,Liu:2024wmk}. Other processes of elastic neutrino scattering with electrons will not be considered here because they have subdominant cross sections in the neutrino-matter interaction \cite{Palomares-Ruiz:2015mka}.

\section{Neutrino trident scattering}
\label{sec_cs:tridente}

\begin{figure}
	\centering
				\begin{tabular}{ccc}
	\includegraphics[width=0.5\textwidth]{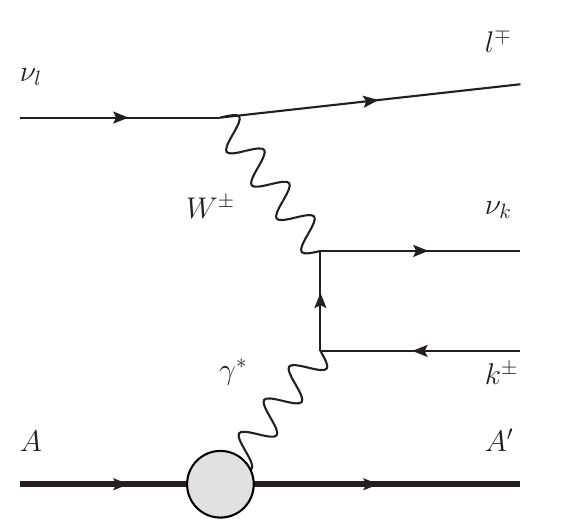} & 
	\includegraphics[width=0.5\textwidth]{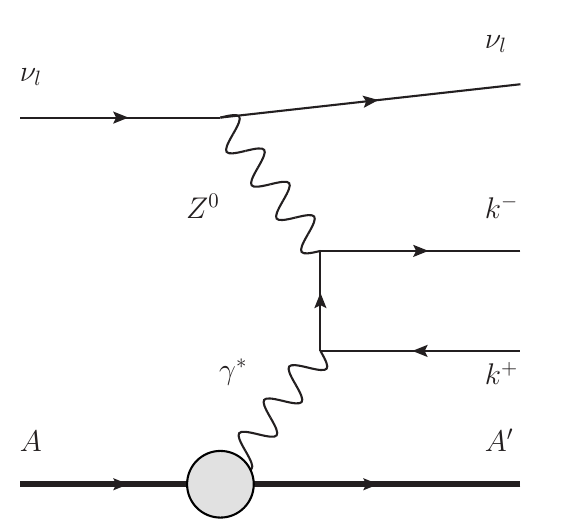}
				\end{tabular}
	\caption{Feynman diagrams for the neutrino trident scattering via the exchange of a $W^{\pm}$ (left) or $Z^0$ (right) bosons. }
	\label{fig_cs:diagramaTrident}
\end{figure}

The deep inelastic neutrino scattering in nuclear targets is the most probable process when we are interested in neutrinos with more than 10 GeV \cite{Formaggio:2012cpf}. With the advent of more sophisticated detectors capable of detecting more events, it becomes possible to study rarer events, involving more complex diagrams and unique final states. One of these processes is the neutrino trident scattering in nuclear targets. This process is characterized by three leptons in the final state, two of which are electrically charged. This process can occur through the exchange of either the $W^{\pm}$ boson or the $Z^0$ boson. Feynman diagrams for this process are shown in Figure \ref{fig_cs:diagramaTrident}.

In the interaction involving the exchange of the $Z^0$ boson, the two charged leptons are necessarily of the same flavor and opposite charges. While in the interaction mediated by the $W^\pm$ boson, the leptons do not need to be of the same flavor, but still respect the conservation of the lepton number within each family. There are currently three experiments that have candidates for muonic neutrino trident scattering with a muon pair final state: CCFR \cite{CCFR:1991lpl}, CHARM-II \cite{CHARM-II:1990dvf}, and NuTeV \cite{NuTeV:1999wlw}. CCFR observed $37\pm 12$ events, CHARM-II observed $55\pm 16$ events, while NuTeV observed only $7^{+17}_{-7}$ events. The large uncertainty in the number of events stemming from the background processes of heavy quark production and semileptonic decay does not allow us to say that the trident process was discovered in these experiments.

The cross section of the neutrino trident scattering has already been calculated by several authors and at different levels of approximation \cite{Ballett:2018uuc,Altmannshofer:2019zhy,Zhou:2019vxt,Zhou:2019frk}. Given that it is a process with four interaction vertices and four particles in the final state, its calculation is excessively long and we will not reproduce it here. 

\section{Neutrino transmission}
\label{sec_cs:transmission}

The (anti)neutrino-target cross section, described in previous sections, is one of the main ingredients for calculating the number of events and their distributions in neutrino detectors and observatories. Another key ingredient is the spectrum of neutrinos that reach the target. Given that neutrinos interact only via the weak force, neutrino fluxes are difficult to construct artificially like charged particle fluxes, and measuring the flux is an arduous task that needs to be done concurrently with cross section measurements. In certain contexts, such as the IceCube neutrino observatory, high-energy neutrinos propagate long distances, and depending on their direction, traverse a large amount of matter inside the Earth before reaching the observatory. Despite the low interaction cross sections, when traversing long columns of matter, the attenuation effects of the neutrino flux due to charged current interactions with the medium become significant, and to estimate the neutrino flux in the detector we need to know precisely its transmission coefficient from its emission at the source until it reaches the target and be detected.

The neutrino transport of flavor $l$ and energy $E_{\nu_l}$ traversing a column of matter $X$ on Earth is obtained with \cite{Nicolaidis:1996qu,Naumov:1998sf,Kwiecinski:1998yf,Iyer:1999wu,
Giesel:2003hj,Reya:2005vh,Rakshit:2006yi}
\begin{eqnarray}
\begin{aligned}
\dfrac{\partial \Phi (E_{\nu_l} , X)}{\partial X} = 
& -N_A [\sigma^{CC} (E_{\nu_l}) + \sigma^{NC}(E_{\nu_l})] \Phi (E_{\nu_l} , X) + \\
& + N_A \int_{0}^{1}\dfrac{\mathrm{d}y}{1+y}
\dfrac{\mathrm{d} \sigma^{NC}(E_{\nu_l} / (1-y), y)}{\mathrm{d}y} 
\Phi (E_{\nu_l} /(1-y) , X)+ \;\;\; \\
& + \frac{1}{(E_{\nu_l}/m_{\tau})\tau \rho (X)}\int^{1}_{0}\mathrm{d}y\frac{\mathrm{d}n_{\tau \rightarrow \nu_l}(1-y)}{\mathrm{d}y}\Phi_\tau(E_{\nu_l}/(1-y),X) \, ,
\end{aligned}
\label{eq:fluxonu}
\end{eqnarray}
where $N_{A}$ is Avogadro's number, $\tau$ is the tau lifetime, $\rho (X)$ is the Earth's profile density, $\Phi (E_{\nu_l} , X)$ and $\Phi_{\tau} (E_{\tau} , X)$ are the differential energy spectra of neutrinos and taus, respectively, in a column of matter $X$ defined by
\begin{eqnarray}
    X(\theta_z) = \int_{0}^{2R_{\mathrm{Earth}}\mathrm{cos}\,\theta_z}\rho(x)\mathrm{d}x\, ,
    \label{eq:column}
\end{eqnarray}
where $\theta_z$ is the angle between the direction of the incident neutrino and the Earth's axis of rotation (see Figure \ref{fig_1:attenuation}). In Equation (\ref{eq:fluxonu}) we also have the term $\mathrm{d}n_{\tau \rightarrow \nu_l}(z)/\mathrm{d}z$, which describes the energy distribution of $l$ flavor neutrinos resulting from the decay of tau with the initial tau energy fraction $z = E_{\nu_{l}}/E_{\tau}$. This quantity has already been calculated for different tau decay channels \cite{Kuhn:1990ad,Pasquali:1998xf,Bhattacharya:2016jce,Garg:2022ugd}.

Since taus are heavy and have short lifetimes, a very common approximation in the IceCube observation energy range is to consider the immediate decay of the tau, that is, the energy lost by the tau before decaying is negligible. In this context, the differential energy spectrum of the tau is obtained with
\begin{eqnarray}
\begin{aligned}
\dfrac{\partial \Phi_\tau (E_{\tau} , X)}{\partial X} = 
& - \frac{1}{(E_{\tau}/m_{\tau})\tau \rho (X)} \Phi_\tau(E_{\tau}, X) + \\
& + N_A \int^{1}_{0}\frac{\mathrm{d}y}{1-y}\dfrac{\mathrm{d} \sigma^{CC}(E_{\tau} / (1-y), y)}{\mathrm{d}y}\Phi_\tau(E_{\tau}/(1-y),X) \, .
\end{aligned}
\label{eq:fluxotau}
\end{eqnarray}

Equations (\ref{eq:fluxonu}) and (\ref{eq:fluxotau}) play an important role in neutrino physics in two contexts: in the propagation of neutrinos in columns of matter on the order of the neutrino interaction length, and in obtaining the number of events expected in detectors and observatories interested in measuring neutrino scattering, which will be described in the following sections.

\begin{figure}
	\centering
	\begin{tabular}{ccc}
	\includegraphics[width=0.5\textwidth]{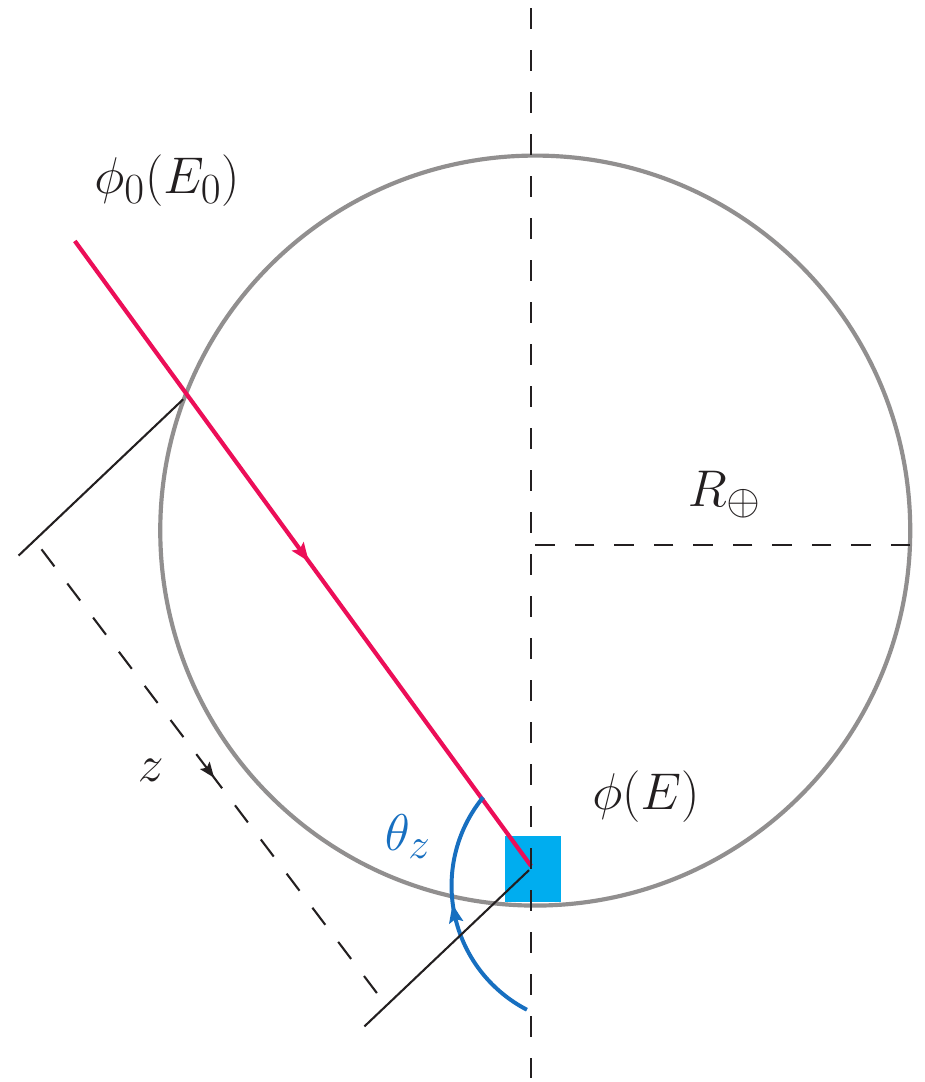}
		\end{tabular}
	\caption{Attenuation of neutrino flux by absorption by the Earth. }
	\label{fig_1:attenuation}
\end{figure}

\section{The IceCube neutrino observatory}
\label{sec_cs:IceCube}

The IceCube neutrino observatory was proposed in the early 2000s \cite{Goldschmidt:2002sv,IceCube:2002eys,IceCube:2003llu}, with the mission of searching for extragalactic neutrinos originating mainly from active galactic nuclei and gamma-ray bursts, as well as for physics beyond the Standard Model in the decay of weakly interacting massive particles (WIMPs) into neutrinos. The observatory consists of approximately 1 km$^3$ of ice volume at the south pole instrumented with digital optical modules (DOMs) equipped with photomultipliers capable of identifying relativistic charged particles through the Cherenkov radiation emitted by them in the blue region of the visible spectrum.

\begin{figure}
	\centering
	\begin{tabular}{ccc}
	\includegraphics[width=0.8\textwidth]{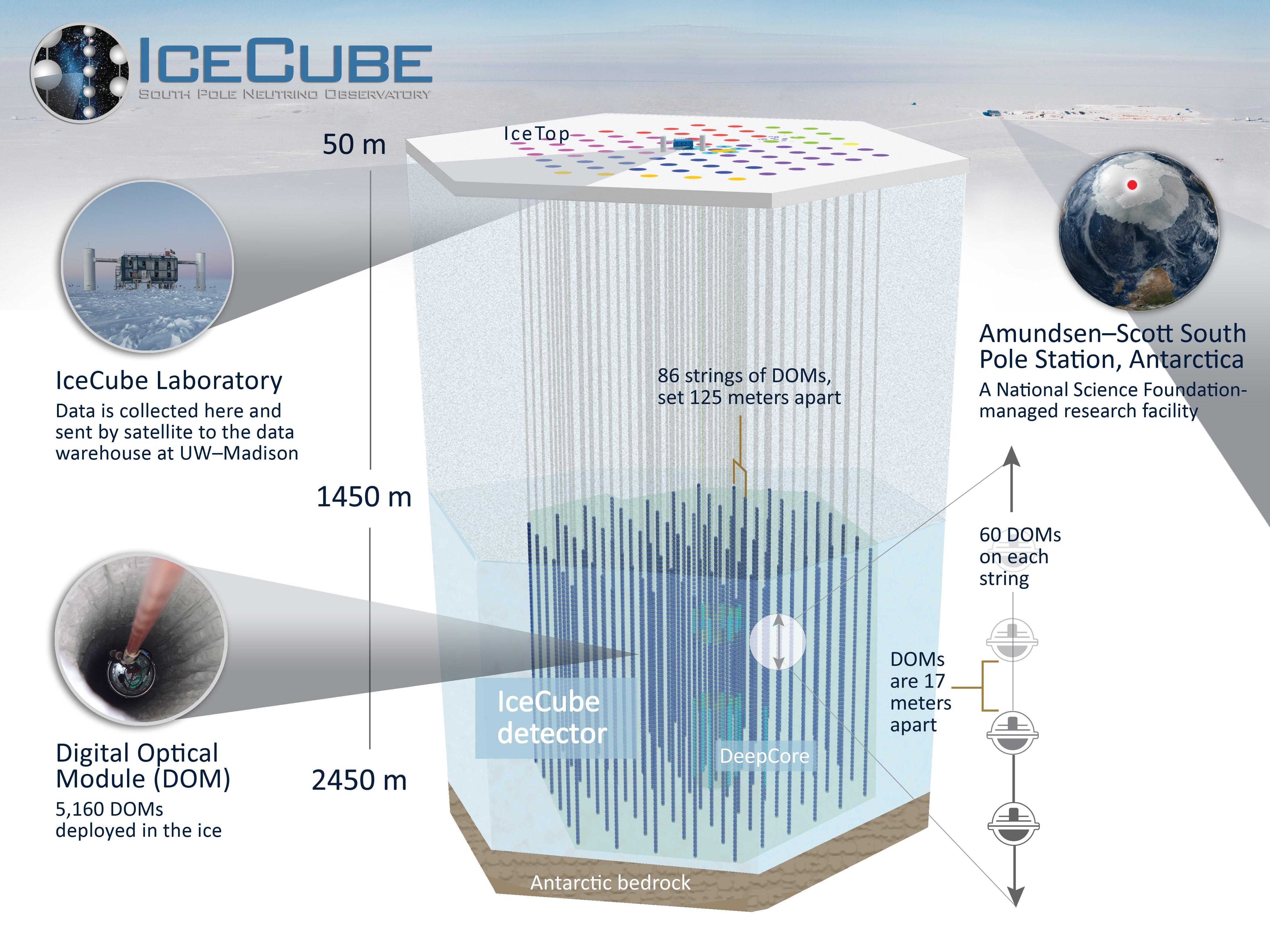}
		\end{tabular}
	\caption{Representation of the IceCube neutrino observatory. Figure taken from \cite{IceCubeWebsite}.}
	\label{fig_2:IceCube}
\end{figure}

Figure \ref{fig_2:IceCube} shows the IceCube Observatory \cite{IceCubeWebsite}. Currently, the observatory has 86 cables supporting a total of 5160 DOMs located between 1450 m and 2450 m above the Antarctic ice surface. The separation between the vertical cables is 125 m, while the DOMs are separated by 17 m on the cables. In addition to all the instrumented volume mentioned above, there are DOMs installed in the upper region of the observatory, between the surface and IceCube. This instrumented region is IceTop, and its main function is to help reject events initiated by muons and atmospheric neutrinos \cite{IceCube:2020wum}.

There is also a central sub-region of IceCube, called DeepCore, where the cable population has a horizontal separation of 70 m and a vertical separation between the DOMs of 7 m. DeepCore is located at approximately 2100 m, and its higher density of DOMs allows for much more precise measurements of the energy and direction of incidence of neutrinos. It also makes it possible to detect events with lower deposited energies.

The observatory began its operation in 2010, and with just under two years of data collection, it measured for the first time with more than 5 sigma of statistical significance an extragalactic neutrino flux \cite{IceCube:2013low}. Despite the great efforts of the IceCube collaboration, and the joint measurements with other multi-messenger astronomy experiments, only two sources of these extragalactic neutrinos have been identified so far: the Blazar TXS 0506+056 \cite{IceCube:2018dnn}, and the Active Galactic Nucleus NGC 1068 \cite{IceCube:2022der}. Among other remarkable results presented by IceCube, we highlight here the observation with 2.3$\sigma$ of the Glashow resonance \cite{Glashow:1960zz}, the annihilation process of an electronic antineutrino with an electron in a $W^{-}$ boson, and the construction of an image of the Milky Way with neutrinos \cite{IceCube:2023ame}.

\subsection{Event topologies at IceCube}
\label{subsec_cs:IceCubeTopologias}

\begin{figure}
	\centering
	\begin{tabular}{ccc}
	\includegraphics[width=0.31\textwidth]{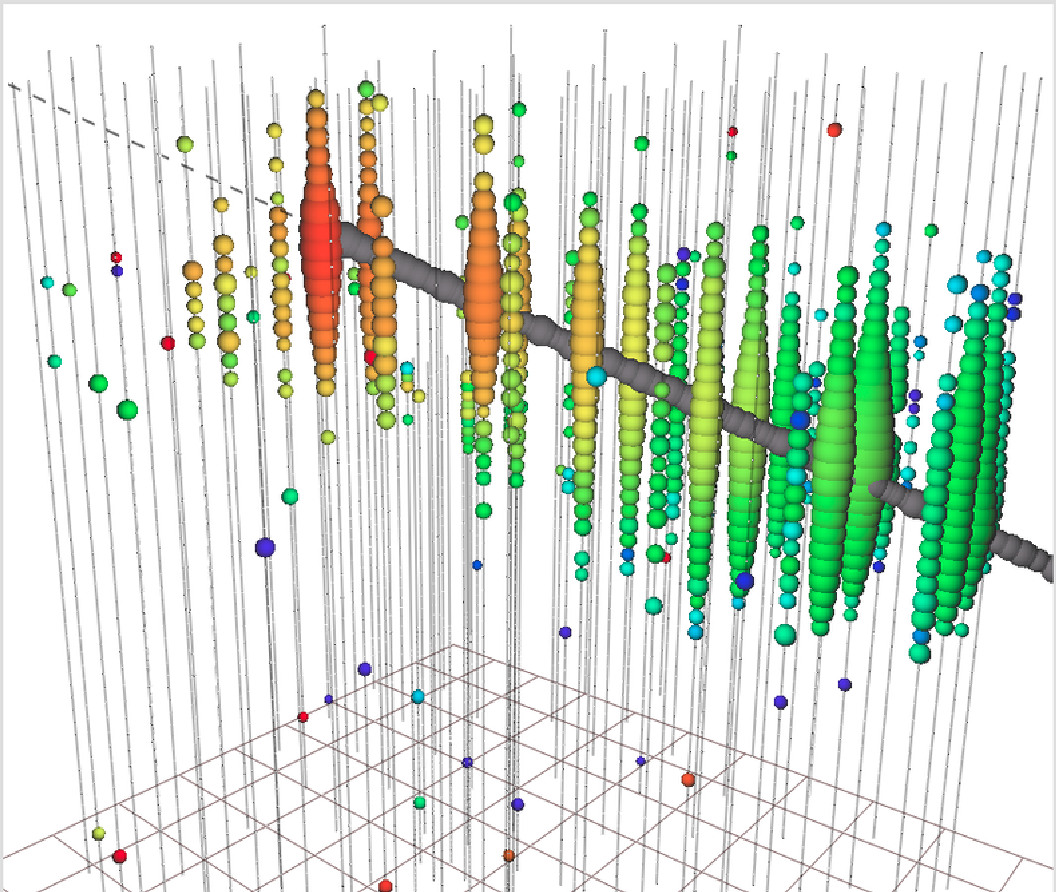} &
	\includegraphics[width=0.31\textwidth]{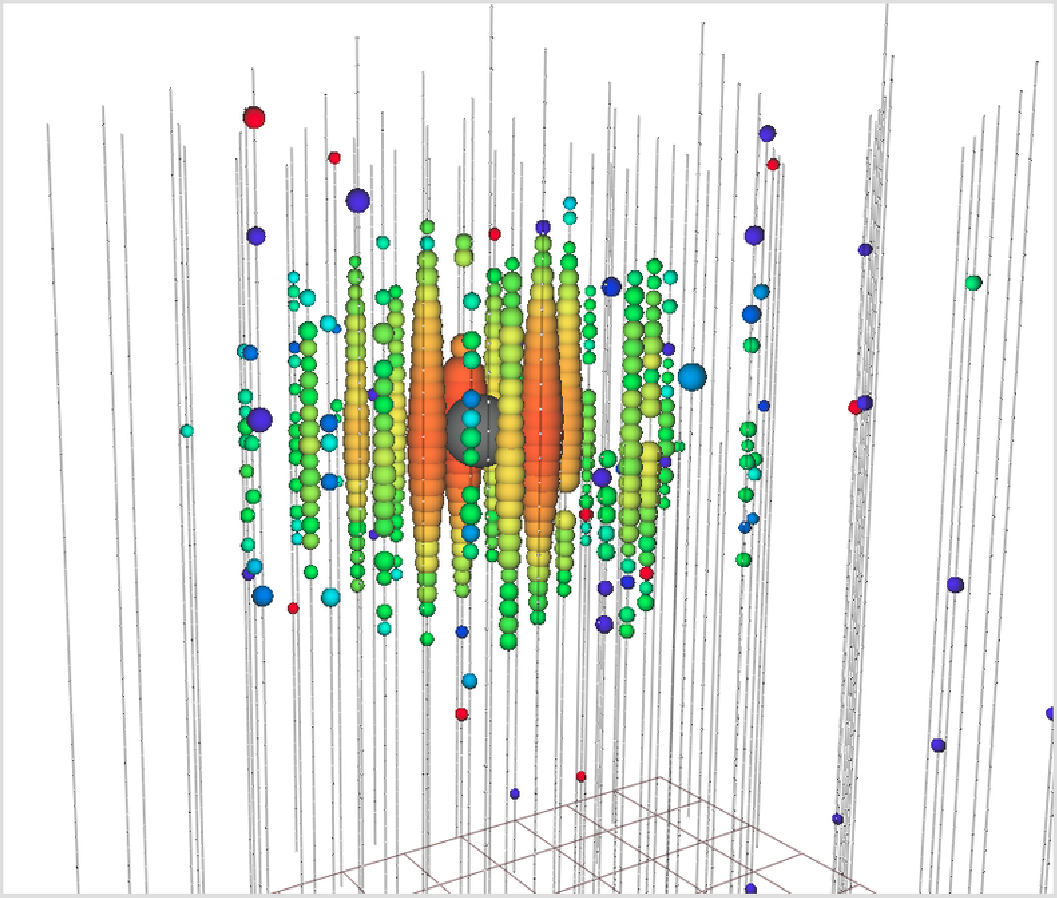} &
	\includegraphics[width=0.31\textwidth]{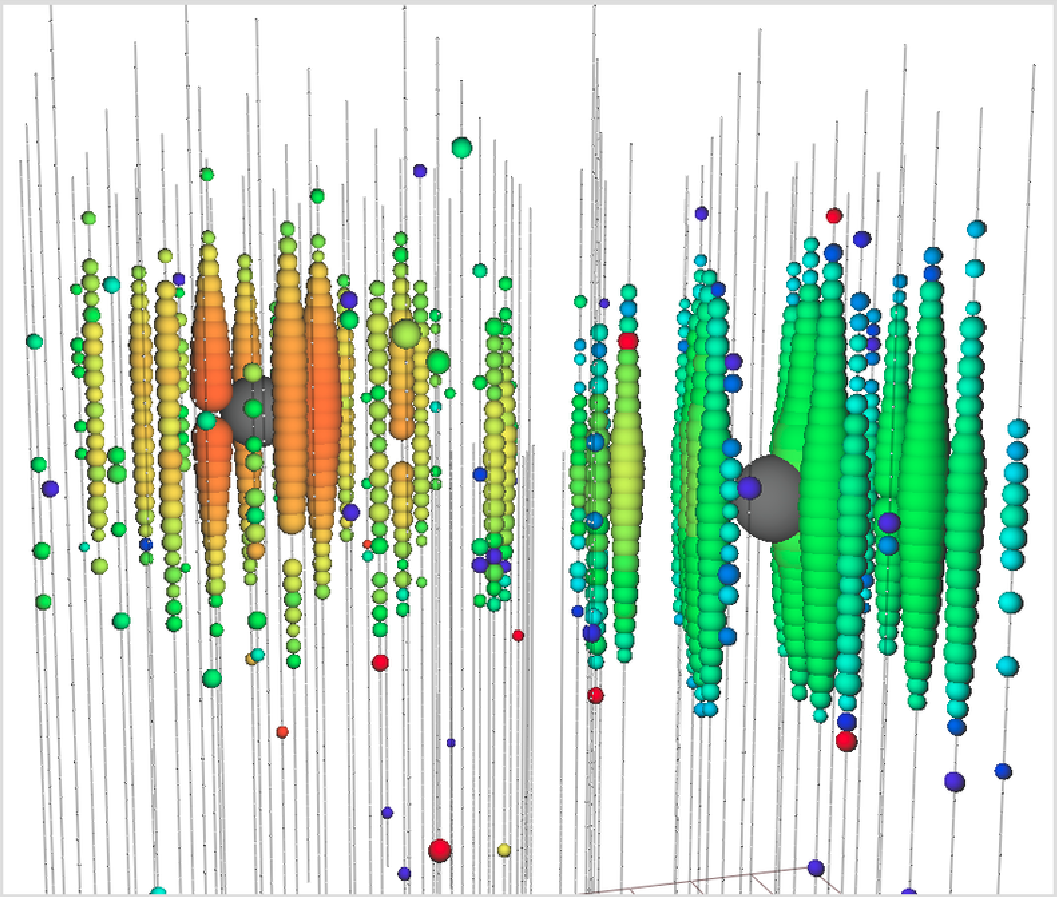} \\
	(a) &
	(b) &
	(c)
		\end{tabular}
	\caption{ Topologies of events observed in IceCube: (a) tracks, (b) cascades, and (c) double cascades. Figure taken from \cite{Usner:2018cel}.}
	\label{fig_2:Topologias}
\end{figure}

The IceCube observatory is capable of measuring neutrinos across a vast energy range, from atmospheric neutrino events that deposit a few GeV in Antarctic ice to events of several PeV, such as the Glashow resonance event. In addition to detecting neutrinos of diverse energies, IceCube can measure neutrinos and antineutrinos of the three different flavors existing in the Standard Model. However, separating neutrino-initiated events of different flavors is a difficult task, and in many cases can only be done statistically, given the observatory's limitations in identifying the charged lepton present in the final state. The events observed at IceCube can be separated into three topologies: cascades, tracks, and double cascades, as shown in Figure \ref{fig_2:Topologias} \cite{Usner:2018cel}. The color pattern in the figure represents the temporal resolution.

Events characterized as tracks are events that have a highly energetic muon in the final state, with energy greater than 70~GeV (see Figure \ref{fig_2:Topologias}a) \cite{Barger:2016deu}. These events can have different origins: they can be muonic (anti)neutrinos that produce an (anti)muon in the final state; tauonic (anti)neutrinos that produce (anti)taus that subsequently decay into muons; atmospheric muons resulting from the interaction of high-energy cosmic rays with the atmosphere; and even other subdominant processes of heavy quark flavor production that semileptonically decay into muons. Muons have a mass about 200 times that of an electron, and since they have a lifetime on the order of $\mu$s, these high-energy muons propagate for several kilometers until they lose all their energy and decay. This leaves a clear signal of coherent Cherenkov radiation emission in the detector, so muons can be identified when produced under these circumstances. Events characterized as tracks can also be \textit{Starting} or \textit{Through Going}. \textit{Starting}, as the name suggests, are events initiated within the instrumented volume. \textit{Through Going} are events initiated outside the instrumented region where the produced high-energy muon passes through the observatory. Given that track events have a long, straight trail left by the muon, these events have excellent angular resolution ($\sim 1^{\mathrm{o}}$). However, given that the muon does not deposit all of its energy within the observatory, reconstructing the energy of the incident neutrino becomes a more challenging task, and the uncertainties in the energy deposited in the instrumented region vary around 15\% \cite{IceCube:2013dkx}.

Cascades are energy deposits in a practically spherical shape on IceCube, as represented in Figure \ref{fig_2:Topologias}b. These events can be initiated by charged current interactions of electronic (anti)neutrinos, charged current interactions of tauonic (anti)neutrinos and subsequent decay of the tau into electrons or hadrons sufficiently close to the initial interaction vertex of the primary neutrino; or even neutral current interactions of all flavors. These events, by depositing all their energy in a short space, have better energy resolution than tracks, about 10\%. However, the angular resolution is less precise, with uncertainties of about 10$^{\mathrm{o}}$ \cite{IceCube:2017der}.

We still have events characterized as double cascades, shown in Figure \ref{fig_2:Topologias}c. This topology is a clear sign of charged current interactions of tau neutrinos, given that they are two approximately spherical depositions of energy, one originating from the hadronic shower at the vertex of the primary neutrino and the other from the shower initiated in the decay of the tau into electron or hadrons, separated by the distance propagated by the tau before decaying. These events have angular and energetic resolutions similar to cascades. Currently, there are 7 candidates of events classified as double cascades by the IceCube collaboration originating from astrophysical neutrinos \cite{IceCube:2024nhk}. One of the difficulties is that taus need to be highly energetic for separation between the two cascades to be possible, given that the mean free path of decay of a tau with energy $E_\tau$ and lifetime $\tau_0$ is $c\tau_0 E_\tau/m_\tau$ ($\sim$5 m for a tau of 100 TeV).

\subsection{Number of events at IceCube}
\label{subsec_cs:IceCubeEventos}

The IceCube observatory has reported atmospheric and astrophysical neutrino events in an energy regime never before observed in other experiments. With these measurements, the observatory has described a neutrino flux of extragalactic origin that has been parameterized as a power law of the type  \cite{IceCube:2013low,IceCube:2020wum}
\begin{eqnarray}
\Phi (E_\nu) = \Phi_0 \left( \dfrac{E_\nu}{ 100\;\mathrm{TeV}}\right) ^{-\gamma} \; 10^{-18} \mathrm{GeV}^{-1} \mathrm{s}^{-1} \mathrm{sr}^{-1}\mathrm{cm}^{-2}\, ,
\label{eq_cs:fluxo}
\end{eqnarray}
where $\Phi_0$ is the normalization of the flux and $\gamma$ is the spectral index. Both parameters are adjusted according to the observation of events. Recent results from the IceCube collaboration show that the flux above no longer describes the data satisfactorily. Several parameterization models for the astrophysical neutrino flux have been proposed and tested, and currently the one that best describes the data is the broken power-law parameterization \cite{IceCube:2025tgp,IceCube:2025ewu}. This parameterization considers that the flux has a power law with spectral index $\gamma_1$ up to a certain cutoff energy $E_{\mathrm{cut}}$ and from that energy onwards has spectral index $\gamma_2$. For the purposes of our work throughout this thesis, we will consider the astrophysical neutrino flux as a power law of the Equation (\ref{eq_cs:fluxo}), given that our interest lies in events initiated by neutrinos with more than 60~TeV, while current data point to a breakdown in the power law at $E_{\mathrm{cut}} \approx 25$~TeV \cite{IceCube:2025tgp,IceCube:2025ewu}.

A common approach in the literature is to consider that this flux is equal for all neutrino flavors and also equal between neutrinos and antineutrinos. This assumption is based on a neutrino flux dominated by the decay process of charged pions produced in proton-proton interactions in astrophysical sources \cite{Nunokawa:2016pop}. In this way, a muonic (anti)neutrino is produced in the decay of the positive (negative) pion, and another muonic (neutrino) antineutrino and an electronic (anti)neutrino in the decay of the antimuon (muon) resulting from the pion decay. Therefore, if we have symmetrical quantities of positive and negative pions being produced, we will have the same quantity of neutrinos and antineutrinos in the astrophysical flux. Having an initial flux with two muonic neutrinos and one electronic neutrino, after propagation over astrophysical distances, the decoherence\footnote{Neutrinos are produced and detected as flavor eigenstates, but propagate as mass eigenstates. Each mass eigenstate propagates at a different speed, and after propagation over long distances the wave packets cease to overlap. Without overlap, neutrino oscillation ceases, and the system becomes a classical mixture of mass eigenstates.} results in approximately similar quantities of neutrinos of each flavor reaching Earth. Recent work has explored the possibility of neutrino flux production through photon-proton interaction in the astrophysical source, which would lead to an asymmetry between neutrino and antineutrino fluxes \cite{Nunokawa:2016pop}. There are also studies exploring the initial flux in the source with different ratios between the flavors, induced by muons slowed by propagation in the astrophysical source, components resulting from neutron decay, and even effects of physics beyond the Standard Model in neutrino propagation in the extragalactic medium \cite{IceCube:2025cjy,IceCube:2025uyt}.

IceCube observatory measurements are usually reported in terms of energy distributions of observed events and also angular distributions. The events of interest in this study are those initiated within the detector volume, known as High Energy Start Events (HESE). Because they have an observed interaction vertex, these events exhibit better resolutions in reconstructed primary neutrino energy. Events to be classified as HESE must meet a series of criteria, including at least 5000 detected photoelectrons; not being accompanied by a high-energy atmospheric muon; the observatory's veto region having detected at most 1 photoelectron within 50 ns before the first photon associated with the event be detected; and being contained within a 200 ns time. For a complete and detailed review of the HESE event classification system, we recommend Section II of the reference \cite{IceCube:2020wum}.

To calculate the number of observed events, we define \cite{Chen:2013dza,Goncalves:2014woa,Goncalves:2021gcu,Goncalves:2022uwy}
\begin{eqnarray}
\frac{\mathrm{d}N_{\mathrm{events}}}{\mathrm{d}(E_{\mathrm{vis}})\mathrm{d}\Omega} = 
T \sum_{\alpha} N_{\mathrm{effe},\alpha}(E_{\nu})\times \Phi _{\nu _\alpha}(E_{\nu})\times \sigma _{\nu _\alpha N}(E_\nu)
\times T^{\nu}(E_\nu , \theta_{z}) \, ,
\label{eq_cs:eventos}
\end{eqnarray}
where $T$ is the exposure time during which the data were collected, $N_{\mathrm{effe},\alpha}$ is the effective number of $\alpha$ type targets in the detector (provided by the IceCube collaboration \cite{IceCube:2013low}), $\Omega$ is the solid angle of the incident neutrino direction, and $T^{\nu _{l}}$ is the transmission coefficient of the neutrino flux of flavor $l$ as it passes through the Earth as a function of neutrino energy and zenith angle ($\theta_z$).

It is important to emphasize that the distribution presented above refers to the visible energy of the events, not the energy of the incident neutrino. This is due to our lack of knowledge about neutrino energy, as we cannot distinguish all types of visible interactions, nor measure the energy of long-lived neutral particles produced in the interaction. For example, it is not possible to separate charged current interactions of electron neutrinos from neutral current interactions. However, charged current interactions of electron neutrinos deposit almost all the energy of the incident neutrino, while neutral current interactions deposit only the portion that is transferred to the target hadron. Therefore, it is necessary to understand how each type of process deposits visible energy in the detectors. For a complete review of the methodology we are adopting for energy deposition in each process, we recommend the reference \cite{Palomares-Ruiz:2015mka}.

\subsection{IceCube-Gen2 and future neutrino observatories}
\label{subsec_cs:IceCubeGen2}

IceCube-Gen2 is the next generation of the IceCube observatory. Its construction is planned to begin in 2027 and last approximately 10 years. IceCube-Gen2 will have an instrumented volume of 8~km$^3$ of Antarctic ice, with a horizontal separation between cables of 240 m and the same vertical separation between the IceCube DOMs, totaling 9600 DOMs \cite{IceCube-Gen2:2020qha}. In addition to all the instrumented ice volume, IceCube-Gen2 will have an area of 8~km$^{2}$ above the main observatory with the aim of identifying atmospheric showers produced by cosmic rays, and then separating atmospheric neutrino events associated with cosmic rays.

With an instrumented volume approximately 8 times that of the current IceCube, IceCube-Gen2 will be able to measure neutrinos with higher energies than currently measured and more frequently. There is also a desire for more precise measurements of the arrival direction of astrophysical neutrinos, which is expected to be sufficient for the identification of more sources of these extragalactic neutrinos.

In addition to IceCube and its planned upgrade in the coming years, there are other neutrino observatories operating, under construction, and planned for the coming years. The Baikal-GVD (Gigaton Volume Detector) neutrino telescope has been under construction since 2016 on Lake Baikal in Russia \cite{Baikal:1990lsj,BAIKAL:1997iok,Baikal-GVD:2018isr}. Like IceCube, it measures Cherenkov radiation from particles produced by neutrino interactions in water using its current 3465 optical modules. Current results from this neutrino telescope agree with IceCube on the existence of a diffuse astrophysical neutrino flux \cite{Baikal-GVD:2022fis,GVD:2025lya}, and also indicate a galactic neutrino flux above 200 TeV \cite{Baikal-GVD:2024kfx}.

Another neutrino observatory that uses water as a medium for neutrino interaction and propagation is KM3NeT, which is being installed but is already partially operating in the Mediterranean Sea \cite{Kappes:2007ci,KM3Net:2016zxf}. This experiment consists of two detectors, ARCA and ORCA, the first designed for the study of astrophysical neutrinos and the second with the main objective of studying neutrino oscillation. With its partial operation, KM3NeT observed the most energetic neutrino ever seen by humans, with an associated muon with a reconstructed energy of $120^{+110}_{-60}$~PeV and an estimated neutrino energy of 220~PeV \cite{KM3NeT:2025npi}. This event challenges our current understanding of astrophysical neutrino flux and may be a new window in the search for neutrino sources at this energy scale \cite{KM3NeT:2025vut,KM3NeT:2025bxl,KM3NeT:2025aps,KM3NeT:2025ccp,Dzhatdoev:2025sdi,DeLaTorreLuque:2025zsv} and processes beyond the Standard Model \cite{Satunin:2025uui,KM3NeT:2025mfl,Cattaneo:2025uxk,Yang:2025kfr,Borah:2025igh,Kohri:2025bsn}.

In addition to the neutrino telescopes mentioned that are in operation and/or currently under construction, there are others being planned/proposed for the next generation of observatories: P-ONE (Pacific Ocean Neutrino Experiment) \cite{P-ONE:2020ljt}, TRIDENT \cite{TRIDENT:2022hql}, NEON \cite{NEONTelescope:2023mak} and GRAND \cite{GRAND:2018iaj}.

\section{The FASER detector}
\label{sec_cs:FASER}

\begin{figure}
	\centering
	\begin{tabular}{ccc}
	\includegraphics[width=1\textwidth]{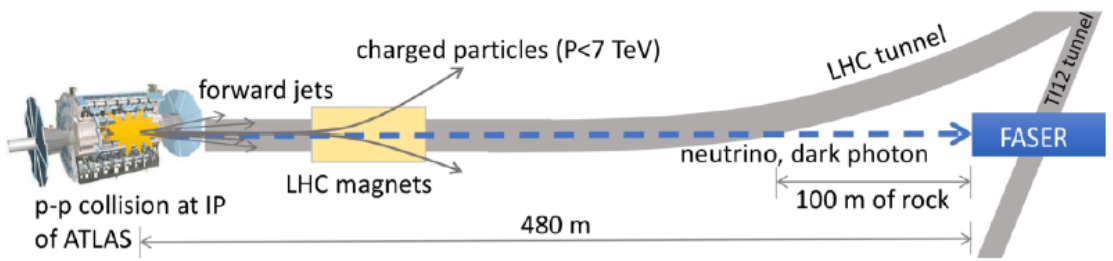}
		\end{tabular}
	\caption{Partial representation of the LHC with the ATLAS and FASER experiments. Figure taken from \cite{FASER:2019dxq}.}
	\label{fig_2:FASER_motivation}
\end{figure}

Since the 1980s, it has been known that high-energy hadron colliders are expected to produce dense neutrino fluxes resulting from the decay of hadrons produced mainly in the region parallel to the direction of the incident hadrons \cite{DeRujula:1984ns}. While the colliding hadrons follow their circular path due to deflections in magnetic fields, the neutrinos produced follow their straight path. The FASER experiment (ForwArd Search ExpeRiment), proposed in 2018 \cite{Feng:2017uoz}, was the first built with the intention of detecting these collider neutrinos. Also in 2018, a FASER pilot project, with only 11 kg of tungsten with emulsion films, was placed in the TI18 tunnel of the LHC for just four weeks. The result, published in 2021, indicated the first collider neutrino candidates detected by humans \cite{FASER:2021mtu}. The statistical significance of this measurement was 2.7$\sigma$, insufficient to claim the discovery of collider neutrinos. Figure \ref{fig_2:FASER_motivation} represents the LHC with the ATLAS and FASER experiments, highlighting FASER's ability to observe highly collimated particles with the colliding proton beam that cannot be measured by central detectors like ATLAS.

Between 2019 and 2022, two experiments were installed at the LHC approximately 480 m from the hadron interaction point in ATLAS with the aim of observing particles produced in the frontal collision region: the FASER \cite{Feng:2017uoz,FASER:2018ceo,FASER:2018eoc,FASER:2018bac,FASER:2019aik,FASER:2019dxq,FASER:2020gpr} and the SND@LHC (Scattering and Neutrino Detector at the LHC) \cite{SNDLHC:2022ihg}. These experiments have been collecting data since 2022 and have already reported the discovery of colliding neutrinos \cite{SNDLHC:2023pun,FASER:2023zcr,FASER:2024hoe,FASER:2024ref}. These experiments were initially proposed to operate until the end of Run 3 of the LHC, but both have already been approved to continue operating during Run 4 \cite{SNDLHC:2026why}.

\begin{figure}
	\centering
	\begin{tabular}{ccc}
	\includegraphics[width=0.8\textwidth]{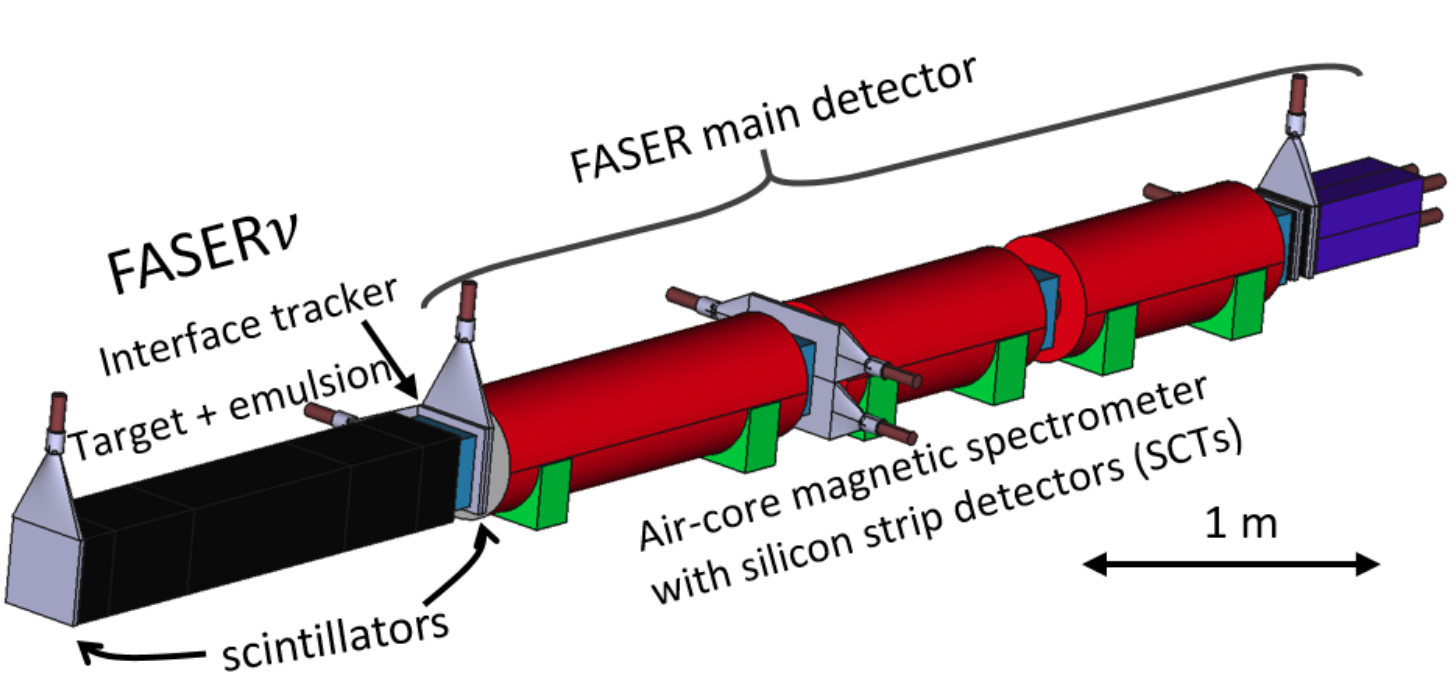}
		\end{tabular}
	\caption{Representation of the FASER experiment. Figure taken from \cite{FASER:2019dxq}.}
	\label{fig_2:FASER}
\end{figure}

The FASER detector is composed of several instruments necessary for measuring and observing different processes and excluding associated backgrounds. Figure \ref{fig_2:FASER} shows its structure. Our main interest is in the FASER$\nu$ neutrino detector, which consists of 730 tungsten plates of 1.1 mm $\times$ 25 cm $\times$ 30 cm interleaved with emulsion films \cite{FASER:2020gpr}. In total, FASER$\nu$ has 1.1 metric tons of tungsten. The detector is located at coordinates ($x$ = 1 cm, $y$ = -3.3 cm), and is capable of measuring particles produced in ATLAS with pseudorapidity $\eta \geq 9$. After the FASER$\nu$ detector, the apparatus has an electronic detector equipped with a spectrometer and magnetic fields that make it possible to identify particles and their electric charges. The SND@LHC detector, unlike FASER, is positioned outside the ATLAS collision axis, covering a pseudorapidity range of 7.1 to 8.4. Because it is outside the collision axis, SND@LHC measures a less dense neutrino flux than FASER, thus having lower total event rates.

\begin{figure}
	\centering
	\begin{tabular}{ccc}
	\includegraphics[width=0.8\textwidth]{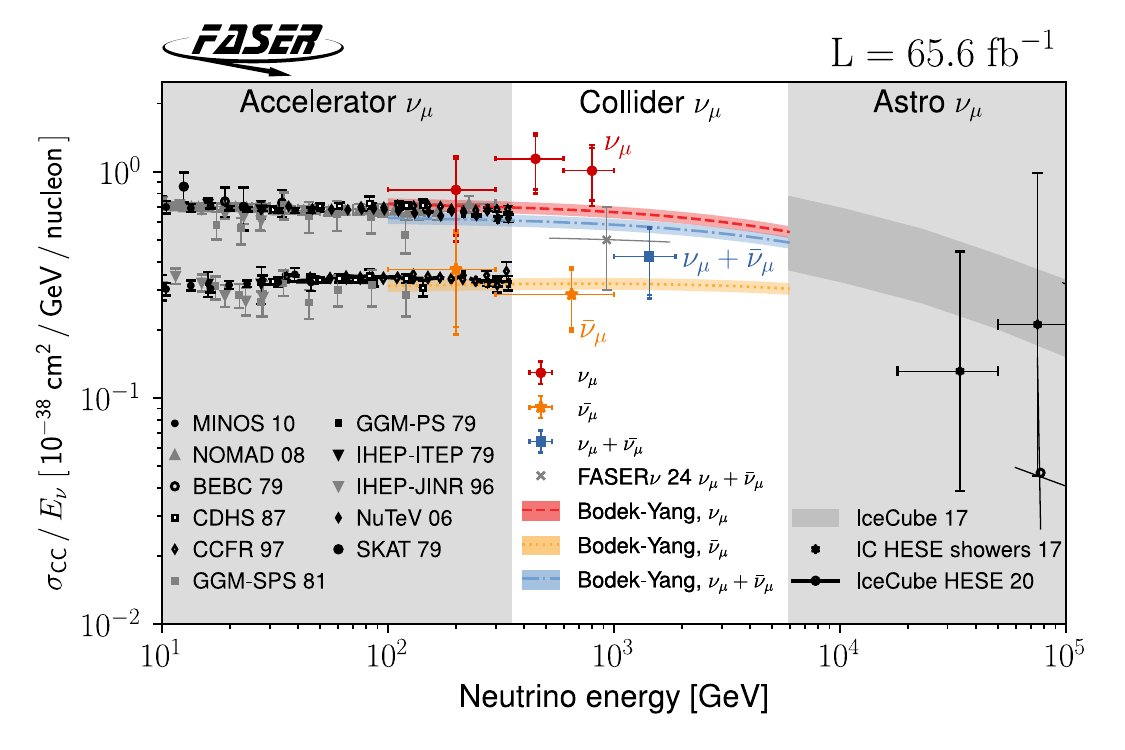}
		\end{tabular}
	\caption{Current measurements of energy-normalized neutrino cross sections for muonic neutrinos with target nucleon as a function of incident neutrino energy. The figure shows current accelerator neutrino measurements for $E_\nu \le 350$ GeV, current IceCube results for $E_\nu \ge 10$ TeV, and current FASER results for intermediate energies ($\sim 1$ TeV). Figure taken from \cite{FASER:2024ref}.}
	\label{fig_2:FASER_cs}
\end{figure}

The FASER and SND@LHC experiments measured the first collider neutrinos, which are the most energetic neutrinos from artificial sources detected by humans ($\approx$ TeV). Measuring these neutrinos has made it possible to extract neutrino-nucleon cross sections in an energy range not yet observed, filling the gap between accelerator neutrino data and astrophysical neutrinos measured by IceCube \cite{FASER:2024ref}. Figure \ref{fig_2:FASER_cs} shows the current status of FASER's contribution to measurements of muonic neutrino cross sections with nuclear targets. In addition to total cross sections, the FASER collaboration has reported measurements of cross sections and differential distributions in energy and pseudorapidity of neutrinos. These differential measurements are important for extracting the neutrino flux produced in the forward region of ATLAS, which in turn is sensitive to different approaches that describe the forward production of hadronic particles \cite{John:2025qlm}.

The FASER detector, in addition to being able to measure and identify neutrinos of all three flavors, is capable of searching for processes beyond the Standard Model, both those produced in hadronic collisions and those within the detector itself \cite{FASER:2023tle,FASER:2024bbl}. Among Standard Model particles produced in collisions at ATLAS, only neutrinos and muons reach FASER. Muons are usually treated as an inconvenient background for searching for neutrino-initiated events or processes predicted by new physics models. Recent studies have shown that these muons can be very useful both in describing hadronic structure and in searching for processes beyond the Standard Model \cite{Ariga:2023fjg,Batell:2024cdl,MammenAbraham:2025gai}.

The number of events in detectors like the FASER can be estimated using the equation
\begin{eqnarray}
N_{\mathrm{eventos}} = \sigma_{\nu A}\times  \phi(E_\nu) \times \rho L/m_{A} \, ,
\label{eq_cs:eventosFASER}
\end{eqnarray}
where $\sigma_{\nu A}$ is the neutrino-nucleon cross section for the process of interest, $\rho$ is the density of the target material in the detector ($19.3\,\mathrm{g/cm}^{3}$ for tungsten), $L$ is the length of the detector (0.80 m), $m_A$ is the mass of the target nucleus, and $\phi(E_\nu)$ is the time-integrated neutrino flux that will cross the detector's front surface.

\subsection{The Forward Physics Facility}
\label{subsec_cs:FPF}

The success of FASER and SND@LHC in measuring frontal particles from colliders motivated the proposal for a new cavern at the LHC to house a set of experiments searching for frontal particles from colliders, the Forward Physics Facility (FPF) \cite{Feng:2022inv,FPFWorkingGroups:2025rsc,FPF:2025bor}. The construction of a new cavern is justified by the inability of existing ones to accommodate larger detectors. The FPF is being proposed to house four new experiments: FLARE, FASER$\nu$2, FASER2, and FORMOSA. We highlight here FASER$\nu$2, which would be an upgrade of FASER$\nu$, with 3300 tungsten plates of 2~mm$\times$25~cm$\times$64~cm and a total mass of 20 metric tons. The expectation is that the FPF will operate during the high-luminosity regime of the LHC, which will have a cumulative luminosity\footnote{The luminosity mentioned refers to proton collisions at the LHC, and describes the number of particles per area available to interact in a given time interval. } of $\mathcal{L_{\mathrm{pp}}}=$3~ab$^{-1}$.

\section{Conclusions}
\label{sec_cs:conclusions}

In this chapter, we describe the main lepton-hadron interaction processes that we will be interested in our analyses in the following chapters. In particular, we address the deeply inelastic scattering of leptons in charged and neutral current interactions; the Glashow resonance scattering, which consists of the annihilation of electron antineutrinos with an electron in a $W^{-}$ boson; and the trident scattering of neutrinos in nuclei, which produces a lepton pair by the fusion of two gauge bosons. These scattering processes play an important role in describing current detector and observatory data, as well as in the search for new physics and rare processes predicted by the Standard Model. We also describe the general aspects of the experiments in which we will study the interactions: the IceCube neutrino observatory and the FASER$\nu$ detector. We show how the detector characteristics and cross sections connect to describe the main observables, which are the associated event numbers. In the next chapter, we will apply the formalism described here to the study of high-energy neutrinos at the IceCube observatory.


\chapter{Implications of Standard Model and Beyond processes at IceCube}
\label{cap:Ice}

In this chapter, we will present our results for scattering and propagation of high-energy neutrinos (> TeV) with phenomenological applications to the observables of the IceCube neutrino observatory and its planned upgrade to operate in the next decade, IceCube-Gen2. In particular, we will describe the results published by the thesis author in the references \cite{Francener:2024bfm,Francener:2024ljx,Francener:2025apz}. Neutrinos of this energy regime that interact with IceCube are usually atmospheric and astrophysical neutrinos, and their most common interaction is the deep inelastic scattering of charged and neutral currents, in addition to the Glashow resonance for electronic antineutrinos with energies close to 6.3~PeV.

Given that the IceCube is located at the geographic south pole of our planet, neutrinos coming from directions with $\theta_z$ greater than $90^\mathrm{o}$ pass through part of the Earth before reaching the observatory (see Figure~\ref{fig_1:attenuation}b). It is known that for high-energy neutrinos, the planet Earth is no longer transparent, as their interaction length is on the same order of magnitude as the Earth's radius. This allows us to use the Earth to attenuate the neutrino flux at this energy, and consequently make indirect measurements such as cross sections by measuring the remaining neutrino flux that reaches the observatory. Another possibility, which we explore here, is to verify the impacts of simplified models for the distribution of matter on Earth and how it affects the description of neutrino propagation and, consequently, the events measured at IceCube. This approach of using neutrinos to obtain information about the Earth's internal structure has been explored before in several works, both by neutrino oscillations and attenuation \cite{Volkova:1974xa,DeRujula:1983ya,Wilson:1983an,Jain:1999kp,Reynoso:2004dt,Gonzalez-Garcia:2007wfs,Borriello:2009ad,Denton:2020jft,Kumar:2021faw,Hajjar:2023knk,Donini:2018tsg,Denton:2021rgt,Krishnamoorthi:2025wie}. We are quantifying the attenuation of high-energy neutrinos with simplified models of the Earth's structure for the first time \cite{Francener:2024bfm}.

After neutrinos pass through the Earth towards IceCube and interact with the target ice, three event topologies are observed for events initiated within the observatory, as described in the previous chapter: tracks, cascades, and double cascades. Tracks are usually expected to be events initiated by muonic neutrinos in charged current interaction processes, and indeed this process is expected to be dominant. However, with the increase in the observatory's exposure time for data collection, other channels may be important in describing the events. In this chapter, we will explore the subdominant channels of taus decay, $W^{\pm}$ boson decay, and heavy-flavor quark decay for events characterized as tracks \cite{Francener:2024ljx}.

The high-energy neutrinos observed in the HESE dataset, which are events initiated within the observatory and that meet certain selection criteria aimed at excluding atmospheric neutrinos and muons, are mostly neutrinos of extragalactic origin. Commonly called astrophysical neutrinos or diffuse neutrino flux, their origin is still not well understood, with several theoretical-phenomenological works in the literature estimating their production from different types of sources \cite{Murase:2019vdl,Fiorillo:2025ehn}. However, to date, only two possible sources of these neutrinos have been identified. Given that these neutrinos travel astrophysical distances to be observed at IceCube, they may also be a tool for searching for phenomena beyond the Standard Model in their propagation through the universe. In this work, we estimate the effects on the alteration of the neutrino flux shape due to the existence of physics beyond the Standard Model that allows interaction between astrophysical neutrinos and cosmic neutrino background. We show that current data may provide evidence of new physics in neutrino propagation, and that IceCube-Gen2 will have high sensitivity to a new gauge boson with a mass of a few MeV \cite{Francener:2025apz}.

\section{Neutrino transmission on Earth with different density approximations}
\label{sec_Ice:trans}

The (anti) neutrino-nucleon cross section, described in Chapter \ref{cap:cs} (Equations (\ref{eq_cs:sigmaDISCC}) and (\ref{eq_cs:sigmaDISNC})), is one of the main ingredients for calculating the transmission of neutrino flux while it passes through the Earth, along with the density profile of our planet. The transmission of neutrino flux is fundamental to knowing the surviving flux that passes through the Earth until it reaches IceCube, as represented in Figure \ref{fig_1:attenuation}b.
The neutrino transport of $l$ flavor that pass through the Earth is obtained with \cite{Nicolaidis:1996qu,Naumov:1998sf,Kwiecinski:1998yf,Iyer:1999wu,Giesel:2003hj,Reya:2005vh,Rakshit:2006yi}
\begin{eqnarray}
\begin{aligned}
\dfrac{\partial \Phi (E_{\nu_l} , X)}{\partial X} = 
& - N_A [\sigma^{CC} (E_{\nu_l}) + \sigma^{NC}(E_{\nu_l})] \Phi (E_{\nu_l} , X) + \\
& + N_A \int_{0}^{1}\dfrac{\mathrm{d}y}{1+y}
\dfrac{\mathrm{d} \sigma^{NC}(E_{\nu_l} / (1-y), y)}{\mathrm{d}y} 
\Phi (E_{\nu_l} /(1-y) , X)+ \;\;\; \\
& + \frac{1}{(E_{\nu_l}/m_{\tau})\tau \rho (X)}\int^{1}_{0}\mathrm{d}y\frac{\mathrm{d}n_{\tau \rightarrow \nu_l}(1-y)}{\mathrm{d}y}\Phi_\tau(E_{\nu_l}/(1-y),X) \, ,
\end{aligned}
\label{eq_Ice:fluxonu}
\end{eqnarray}
where $N_{A}$ is Avogadro's number, $\tau$ is the tau lifetime, $\rho (X)$ is the density profile of the Earth, $\Phi (E_{\nu_l} , X)$ and $\Phi_{\tau} (E_{\tau} , X)$ are the differential energy spectra of neutrinos and taus, respectively, in a column of matter $X$ defined by
\begin{eqnarray}
    X(\theta_z) = \int_{0}^{2R_{\mathrm{Earth}}\mathrm{cos}\,\theta_z}\rho(x)\mathrm{d}x\, ,
    \label{eq_Ice:column}
\end{eqnarray}
where $\theta_z$ is the angle between the direction of the incident neutrino and the Earth's axis of rotation (See Figure \ref{fig_1:attenuation}b). In Equation (\ref{eq_Ice:fluxonu}) we also have the term $\mathrm{d}n_{\tau \rightarrow \nu_l}(z)/\mathrm{d}z$, which describes the energy distribution of $l$ flavor neutrinos resulting from the decay of the tau with the initial tau energy fraction $z = E_{\nu_{l}}/E_{\tau}$. For this quantity we are using the parameterizations provided in \cite{Garg:2022ugd}.

For the energy regime of interest in this work, we consider that the typical interaction distance of the tau is much greater than its average propagated distance until its decay. Therefore, the differential energy spectrum of the tau is obtained with
\begin{eqnarray}
\begin{aligned}
\dfrac{\partial \Phi_\tau (E_{\tau} , X)}{\partial X} = 
& - \frac{1}{(E_{\tau}/m_{\tau})\tau \rho (X)} \Phi_\tau(E_{\tau}, X) + \\
& + N_A \int^{1}_{0}\frac{\mathrm{d}y}{1-y}\dfrac{\mathrm{d} \sigma^{CC}(E_{\tau} / (1-y), y)}{\mathrm{d}y}\Phi_\tau(E_{\tau}/(1-y),X) \, .
\end{aligned}
\label{eq_Ice:fluxotau}
\end{eqnarray}

\begin{figure}
	\centering
	\begin{tabular}{ccc}
	\includegraphics[width=0.45\textwidth]{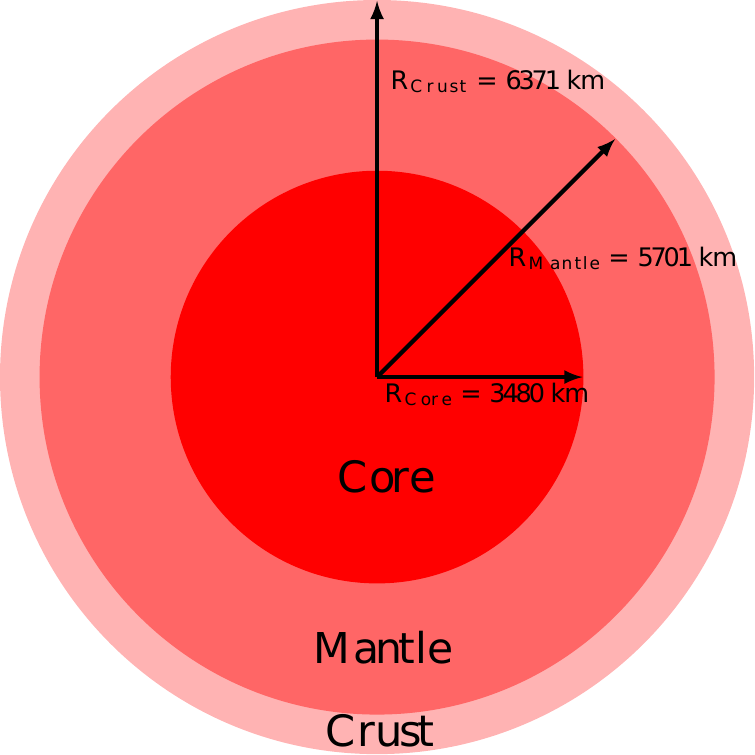} &
	\includegraphics[width=0.43\textwidth]{Images/Prop/prop2.pdf} \\
			(a) & (b)
		\end{tabular}
	\caption{(a) Three-layer Earth model, and (b) attenuation of neutrino flux by absorption by the Earth.}
	\label{fig_1:attenuation}
\end{figure}

\begin{figure}
	\centering
		\begin{tabular}{ccc}
			\includegraphics[width=0.48\textwidth]{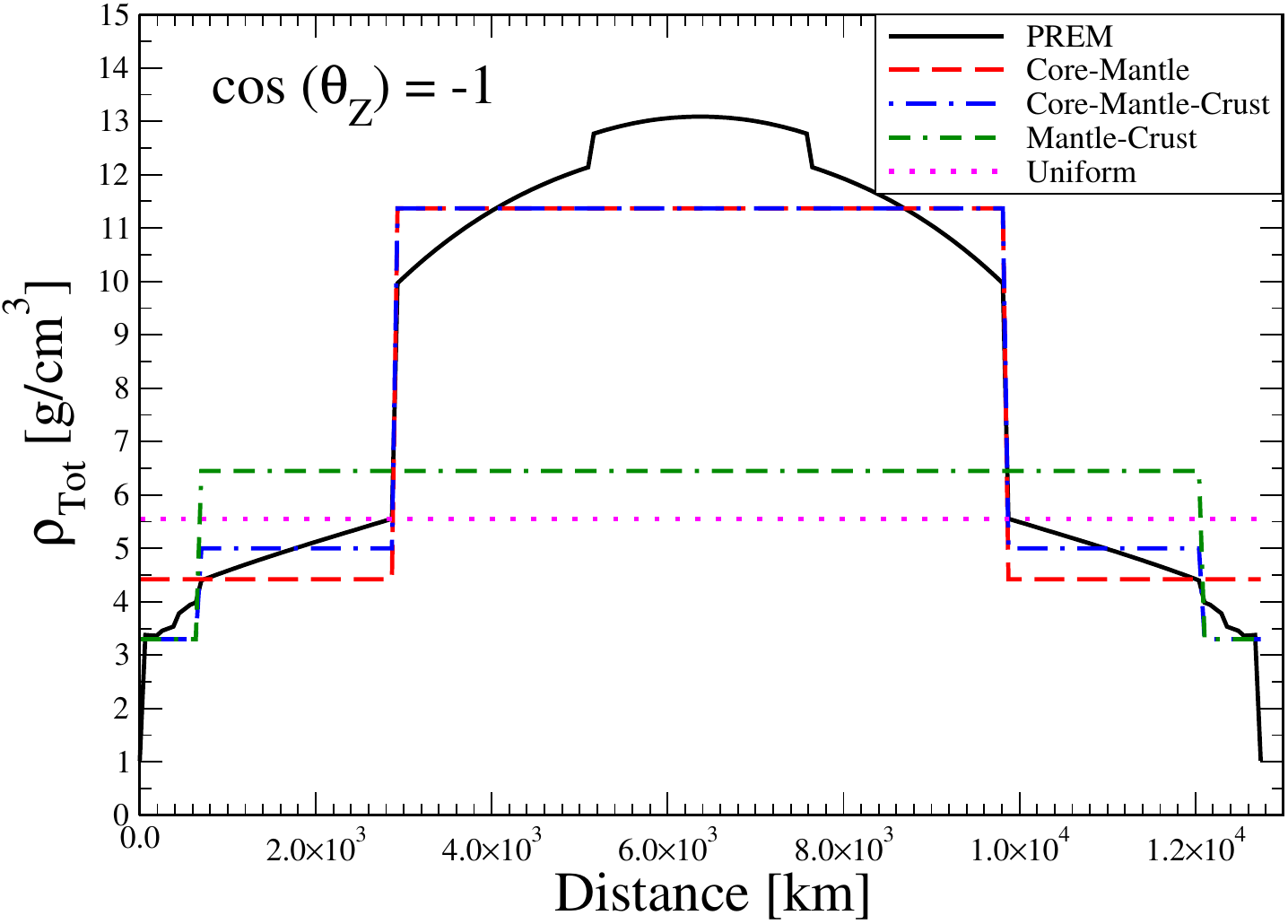} &
			\includegraphics[width=0.48\textwidth]{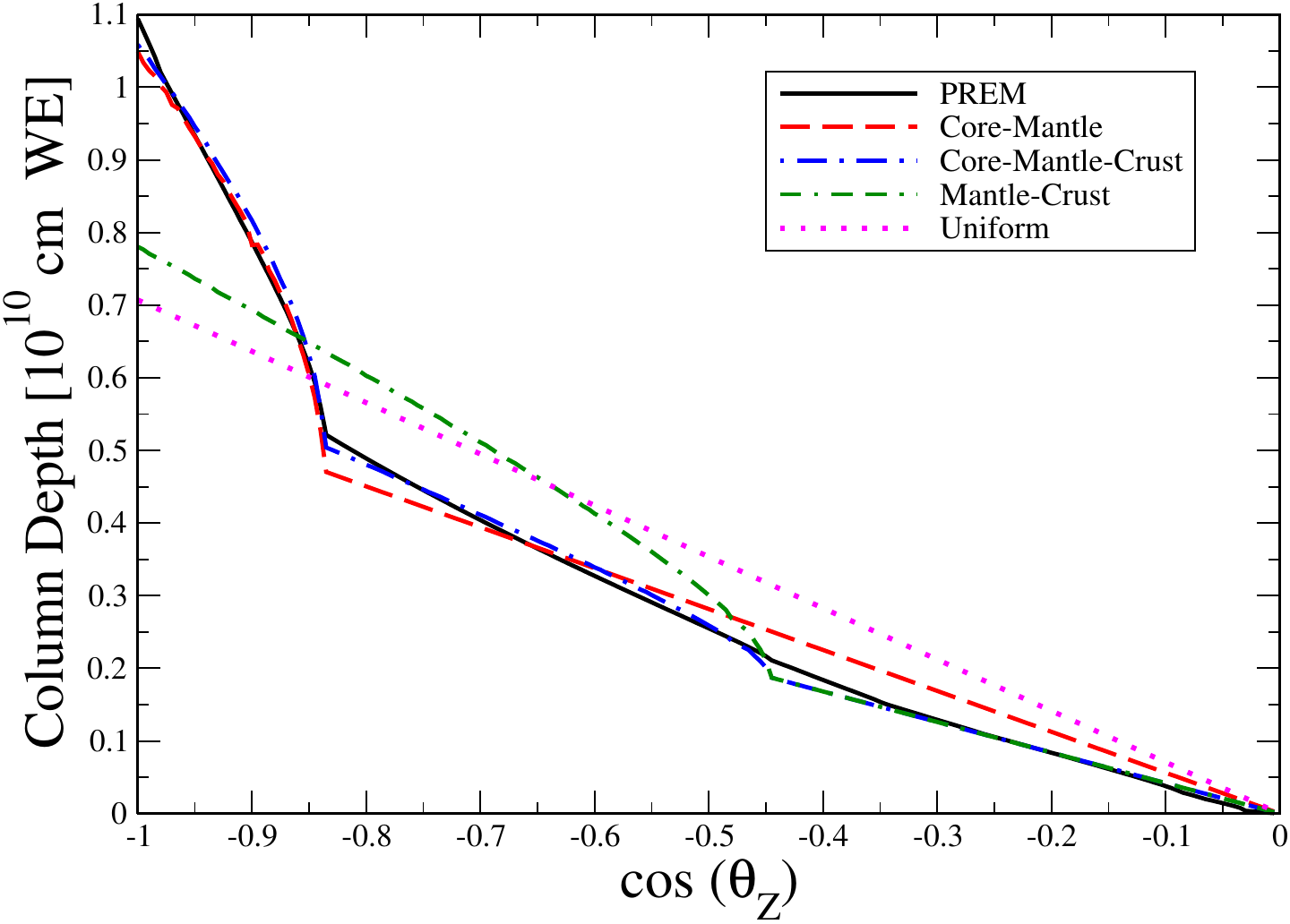} \\
			(a) & (b)
		\end{tabular}
	\caption{(a) Earth's density as a function of the distance traversed within it for different density profile models. (b) Column of matter traversed in centimeters of water as a function of the cosine of the angle of incidence for different density profile models.}
	\label{fig_1:terra}
\end{figure}

It is common for works similar to this one in the literature to use the Preliminary Reference Earth Model (PREM \cite{Dziewonski:1981xy}) for the Earth's density profile, a model based on seismic studies, where the propagation of seismic waves reveals the properties of matter distribution. One of our main results is the comparison of high-energy neutrino transmissions using PREM and three-, two-, and one-layer models for the Earth, similar to what was developed in \cite{Kumar:2021faw} for neutrino oscillation. These simplified Earth models consider only the core, mantle, and crust, in addition to constant densities in each of these layers (we will refer to these models as Core-, Mantle-, Crust-, and Uniform). Table~\ref{tab_Ice:profiles} shows the radius and density information associated with the different Earth structure models we are considering in our work. In Figure \ref{fig_1:attenuation}a we represent this three-layered Earth model with the radius information for each layer.

\begin{table}[t]
    \centering
    \begin{tabular}{|c|c|c|} \hline 
             Model       & Radius of the layers (km) & Layer density (g/cm$^3$) \\ \hline \hline
     PREM   & Multi  layers         & Multi densities                \\ \hline 
         Core-Mantle-Crust          & (0, 3480, 5701, 6371)             & (11.37, 5, 3.3)                     \\ \hline 
         Mantle-Crust          & (0, 5701, 6371)             & (6.45, 3.3)                     \\ \hline 
         Core-Mantle           & (0, 3480, 6371)             & (11.37, 4.42)                     \\ \hline 
         Uniform          & (0, 6371)             & (5.55)                     \\ \hline \hline
         
    \end{tabular}
    \caption{ Radii and density of the layers associated with the different models of Earth's structure considered in our analysis.}
    \label{tab_Ice:profiles}
    
\end{table}

To better understand the differences between Earth density profile models, we present Figure \ref{fig_1:terra}a. In it, we see the Earth's density as a function of the distance traveled by the neutrino, considering the case where the neutrino passes exactly through the Earth's center. In this figure, we compare the models mentioned earlier, and we see that, depending on the distance traveled, the densities present appreciably different values (PREM describes the density as being approximately twice that of the Mantle-Crust and Uniform models in the region of the Earth's core).

In neutrino transmission, it is important to describe the amount of matter that the neutrino flux traverses; that is, we need to know what is usually reported as surface mass density, or simply the column of matter traversed, as we are defining it here. This quantity was defined in Equation (\ref{eq_Ice:column}), and Figure \ref{fig_1:terra}b presents it for the different Earth models. Again, we see that there are angles of incidence of the neutrino flux where several models provide good estimates, but there are also regions where the simplified models differ by a factor of almost two compared to the PREM.

\subsection{Results for neutrino transmission}
\label{subsec_Ice:transResu}

\begin{figure}
	\centering
    \includegraphics[width=0.6\textwidth]{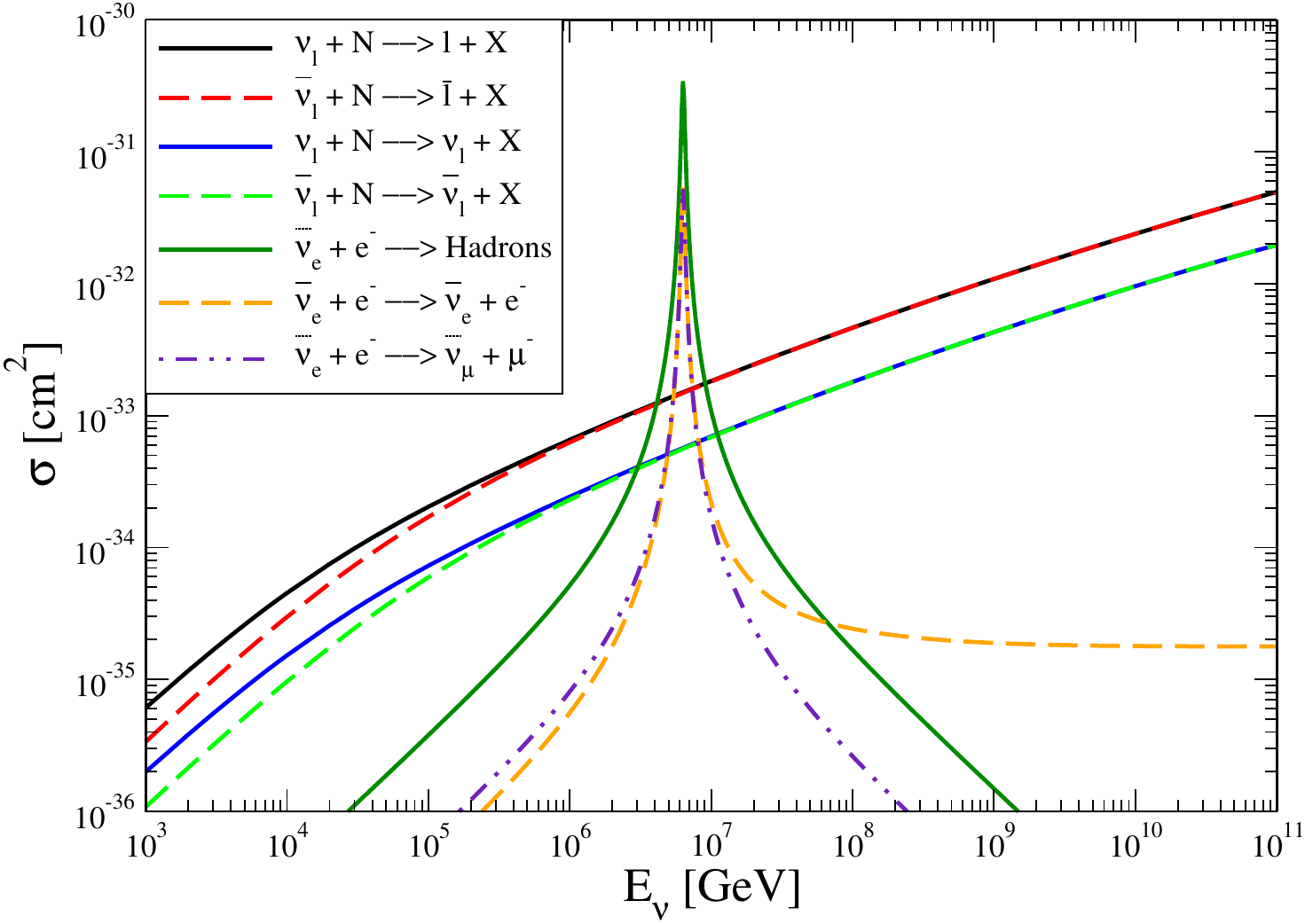} 
\caption{Neutrino cross section with different targets and in different types of interactions. }
\label{fig_1:sigma}
\end{figure}

Using Equations (\ref{eq_cs:sigmaDISCC}) and (\ref{eq_cs:sigmaDISNC}), we can estimate the total (anti) neutrino-nucleon cross sections for charged and neutral current interactions; the interaction of electronic antineutrinos with electrons can be calculated using Equation (\ref{eq_cs:glashow}). As discussed in the previous chapter, to calculate the neutrino-nucleon cross sections we need to define how we will describe the necessary structure functions. In Figure \ref{fig_1:sigma} we present the neutrino cross sections with different targets as a function of the incident neutrino energy. For interactions with hadrons, we consider an isoscalar target, that is, the average between free proton and neutron. In this result, we use the predictions of DGLAP \cite{Gribov:1972ri,Altarelli:1977zs,Dokshitzer:1977sg} for the partonic evolution in $Q^2$, parameterized by the CT14 collaboration at LO \cite{Dulat:2015mca}. We see that the charged current interaction is always dominant over the neutral current interaction, reflecting the different couplings of leptons with the $W^\pm$ and $Z^0$ bosons. For incident neutrino energies close to 6.3 PeV, the contribution of the Glashow resonance becomes significant and even dominant in a small energy range, showing that this process is important for describing high-energy neutrino data.

\begin{figure}
	\centering
	\begin{tabular}{ccc}
    \includegraphics[width=0.33\textwidth]{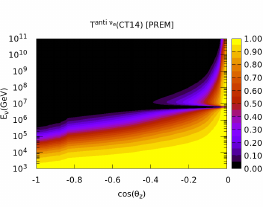} &\includegraphics[width=0.33\textwidth]{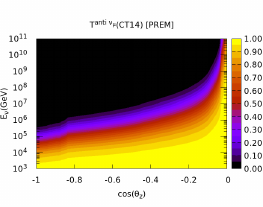} & \includegraphics[width=0.33\textwidth]{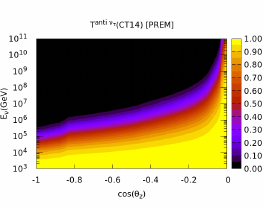} \\
    \includegraphics[width=0.33\textwidth]{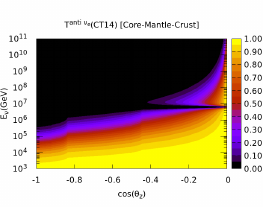} &\includegraphics[width=0.33\textwidth]{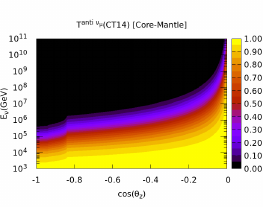} & \includegraphics[width=0.33\textwidth]{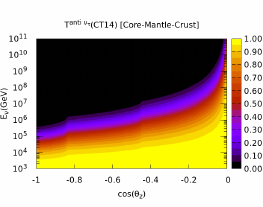} \\
    \includegraphics[width=0.33\textwidth]{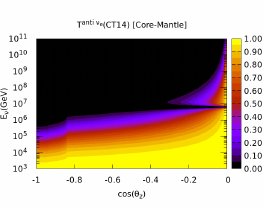} & \includegraphics[width=0.33\textwidth]{Images/Prop/Transm_an_m_CM.pdf} & \includegraphics[width=0.33\textwidth]{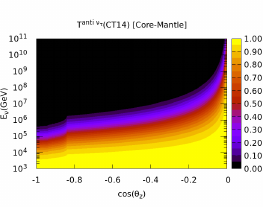} \\
    \includegraphics[width=0.33\textwidth]{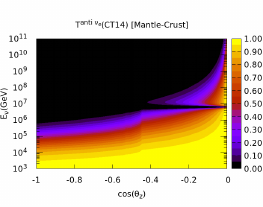} & \includegraphics[width=0.33\textwidth]{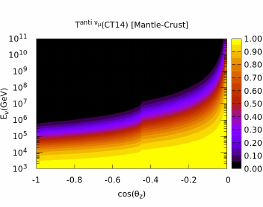} & \includegraphics[width=0.33\textwidth]{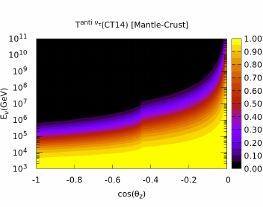} \\
    \includegraphics[width=0.33\textwidth]{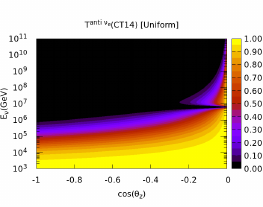} & \includegraphics[width=0.33\textwidth]{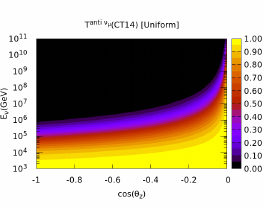} & \includegraphics[width=0.33\textwidth]{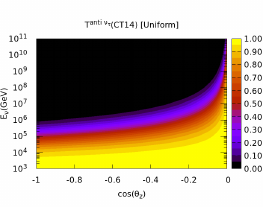} 
			\end{tabular}
\caption{ Transmission coefficient of electronic (left), muonic (middle) and tauonic (right) astrophysical antineutrino flux using different models for the Earth's density profile and a spectral index equal to 3.0. }
\label{fig_1:fluxo}
\end{figure}

Neutrino cross sections with targets on Earth are one of the necessary ingredients for estimating neutrino flux transmission along our planet. As described earlier (Equations (\ref{eq_Ice:fluxonu}) and (\ref{eq_Ice:fluxotau})), we also need to know the Earth's density profile. Solving the coupled differential Equations (\ref{eq_Ice:fluxonu}) and (\ref{eq_Ice:fluxotau}) for neutrino flux transmission along the Earth, we obtain Figure \ref{fig_1:fluxo}. In it, we show the flux transmission coefficient (ratio between the surviving flux and the initial flux) as a function of the neutrino's angle of incidence and its energy. Since we are using a model that takes into account the attenuation of the flux energy by neutral current interactions, and flux regeneration by the decay of taus produced by charged current interactions, we have to assume an initial neutrino flux. We are assuming a power-law flux, as Equation (\ref{eq_cs:fluxo}), which is usually assumed for astrophysical neutrino flux. In the figures presented here, we assume that the spectral index of the neutrino flux is $\gamma = $3.0, given that the latest IceCube update for HESE found $\gamma = 2.87^{+0.20}_{-0.19}$ for the astrophysical neutrino flux \cite{IceCube:2024fxo}. When we consider the neutral current interaction and the regeneration of the tau neutrino flux, we allow neutrinos of energy $E_\nu$ to become neutrinos of energy $(1-y)E_\nu$ and $(1-y)z\, E_\nu$, respectively. $y$, as we saw in the cross sections, is the inelasticity of the collision, so $(1-y)$ provides the fraction of the initial energy remaining with the lepton in the final state; and $z$ is the fraction of the tau energy that remains with the neutrino after tau decay. Therefore, when interacting via NC, or a tau decays, the higher-energy neutrino flux becomes a lower-energy flux, and the impact of this description is directly influenced by the spectral index of the flux.

In Figure \ref{fig_1:fluxo} we present the transmission coefficient of the electron (left), muon (middle), and tau (right) antineutrino flux. For comparison, we show these coefficients considering different Earth density profiles, with the models (from top to bottom) being PREM, Core-Mantle-Crust, Core-Mantle, Mantle-Crust, and Uniform. It is easy to see that these different models lead to distinct predictions depending on the kinematic region tested. We can also identify in the figures the appearance and disappearance of discontinuities in regions of cos $\theta_z = -$0.85 and $-$0.46. This reflects the discontinuities existing in the Earth density profile models in the regions between crust and mantle and between mantle and core.

\begin{figure}
	\centering
	\begin{tabular}{ccc}
	\includegraphics[width=0.4\textwidth]{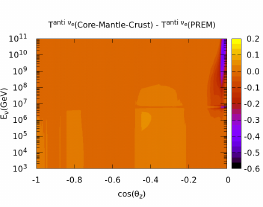} & \,\,\,\,\, & \includegraphics[width=0.4\textwidth]{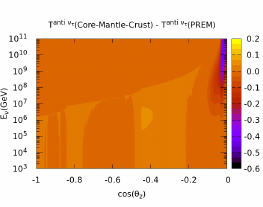} \\
	\includegraphics[width=0.4\textwidth]{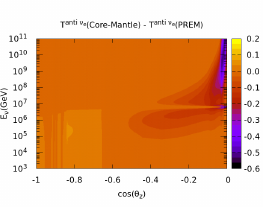} & \,\,\,\,\, & \includegraphics[width=0.4\textwidth]{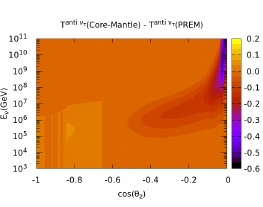} \\
	\includegraphics[width=0.4\textwidth]{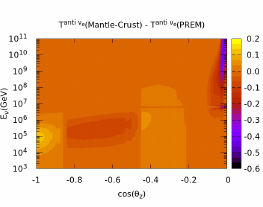} & \,\,\,\,\, & \includegraphics[width=0.4\textwidth]{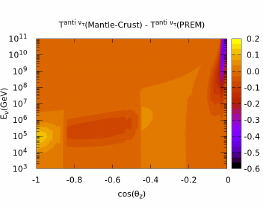} \\
	\includegraphics[width=0.4\textwidth]{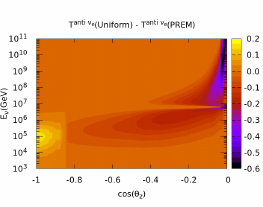} & \,\,\,\,\, & \includegraphics[width=0.4\textwidth]{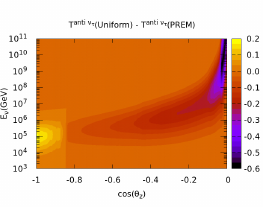} 
			\end{tabular}
\caption{ Difference between antineutrino transmission coefficients using simplified Earth density profile models and using PREM. We present the results for electronic antineutrinos (left) and tauonic antineutrinos (right). }
\label{fig_1:diff_fluxo}
\end{figure}

To better observe the impact of the effects of different Earth models, we present Figure \ref{fig_1:diff_fluxo}. In it, we show the differences between transmissions using simplified models and transmissions using PREM. On the left, we present this difference for electronic antineutrinos and on the right, for tauonic antineutrinos. The observed differences are of the same order of magnitude as the transmission coefficient itself. Their magnitude depends on the kinematic region observed and also on the models being compared. The differences can have positive and negative values. The largest differences occur when comparing the Mantle-Crust, Core-Mantle, and especially the Uniform models with PREM. The Core-Mantle-Crust model proved to be a good estimate when compared to the more precise model, differing only in the region of $\theta_z$ tending to zero, that is, when the neutrinos pass through a small layer of the Earth.

\begin{figure}
	\centering
	\begin{tabular}{ccc}
	\includegraphics[width=0.48\textwidth]{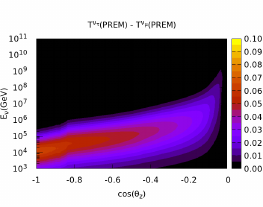} &  \includegraphics[width=0.48\textwidth]{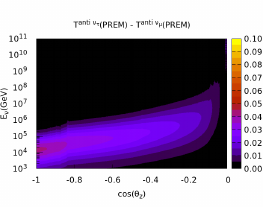} \\
	(a)  &  (b) 
			\end{tabular}
\caption{Difference between the transmission coefficient of tauonic and muonic (a) neutrinos and (b) antineutrinos. }
\label{fig_1:diff_taumuon}
\end{figure}

When a tau is produced, it can decay into hadrons plus tau neutrino; or into lighter leptons, the corresponding lighter lepton antineutrino, and a tau neutrino. In all channels, we have tau neutrinos in the final state, but in only about $34\%$ of cases do we have electron or muon neutrinos. This makes the regeneration of tau neutrinos much more significant than the regeneration of other neutrino flavors by the decay of tau produced in charged current neutrino interactions. In Figure \ref{fig_1:diff_taumuon} we show the difference between the transmission coefficients of tau neutrinos (left) and antineutrinos (right) and of muon neutrinos and antineutrinos. The differences are significant and more important for neutrinos and antineutrinos that traverse a greater amount of terrestrial matter. The difference between neutrino and antineutrino comparisons is also noteworthy, being more pronounced for neutrinos. The parameterizations we use for the fraction of the initial energy that carries the neutrino to the final state take into account fully polarized taus. Thus, when a tau decays, the neutrino is emitted in the direction of propagation of the initial tau, while in the decay of an antitau, the antineutrino is emitted in the opposite direction to the initial tau. Therefore, neutrinos carry a greater fraction of the initial tau's energy compared to antineutrinos, producing the asymmetry seen in Figure \ref{fig_1:diff_taumuon}.

\begin{figure}
	\centering
	\begin{tabular}{ccc}
	\includegraphics[width=0.48\textwidth]{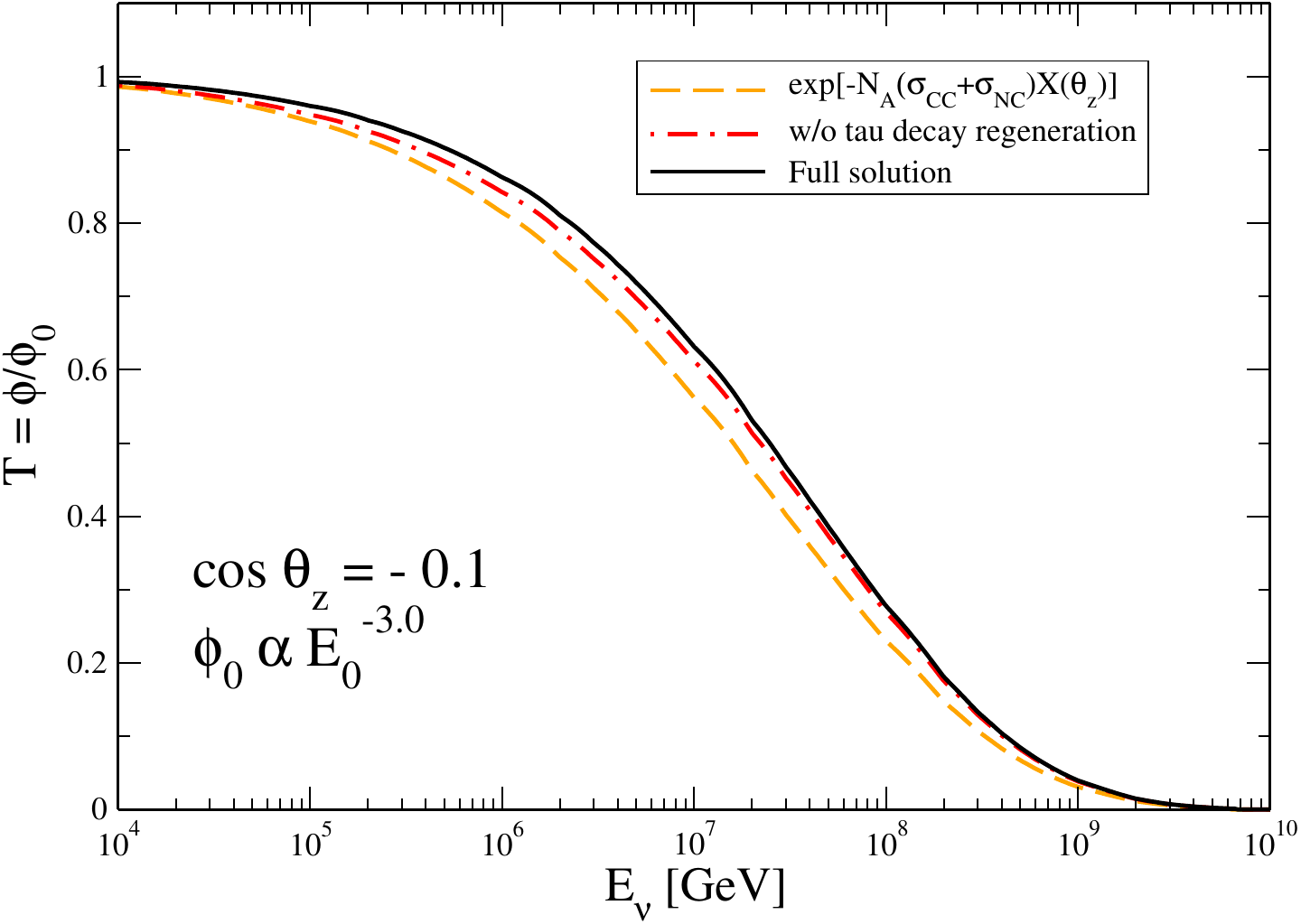} &  \includegraphics[width=0.48\textwidth]{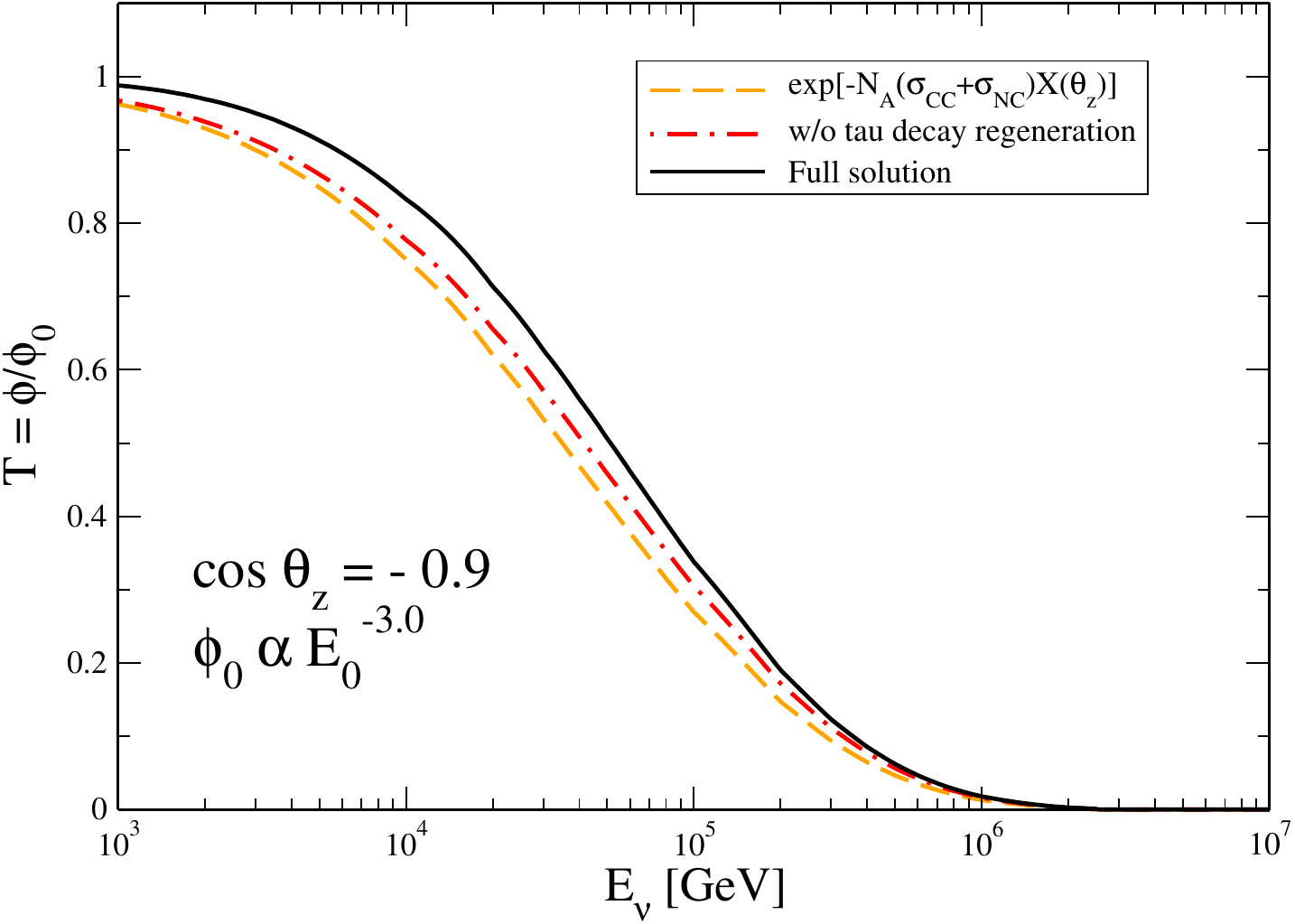} \\
	(a)  &  (b) 
			\end{tabular}
\caption{ Transmission coefficient of tauonic neutrinos as a function of energy considering different treatments for regeneration due to neutral current interactions and tau decay for (a) cos $\theta_z$ = -0.1 and (b) cos $\theta_z$ = -0.9. }
\label{fig_1:diff_taumuon_unidim}
\end{figure}

In Figure~\ref{fig_1:diff_taumuon_unidim}, we present a comparison between the predictions for the tauonic neutrino coefficient as a function of energy for (a) cos $\theta_z$ = -0.1 and cos $\theta_z$ = -0.9 (b). Let us consider different levels of approximation for flux regeneration: without taking into account any regeneration effect (dashed orange curve); considering only the regeneration from neutral current interaction (dashed-dotted red curve); and, finally, considering the complete solution of Equations (\ref{eq_Ice:fluxonu}) and (\ref{eq_Ice:fluxotau}), which takes into account both regeneration by neutral current interactions and by the decay of taus produced in charged current interactions. Our results indicate the importance of more accurate treatment with regard to regeneration effects, especially for neutrinos that pass through a greater amount of terrestrial matter before reaching observatories.

\begin{figure}
	\centering
	\begin{tabular}{ccc}
	\includegraphics[width=0.48\textwidth]{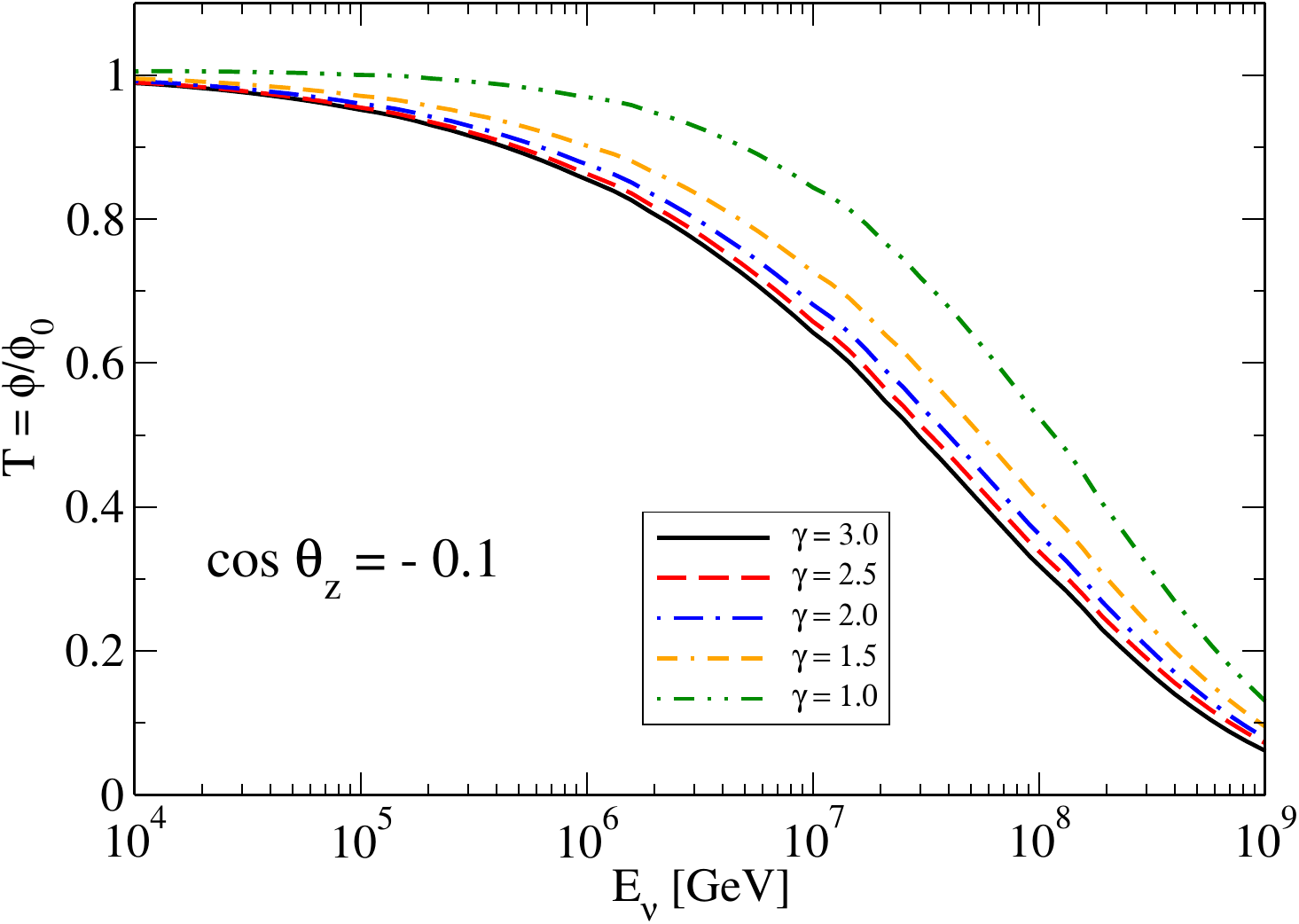} & \includegraphics[width=0.48\textwidth]{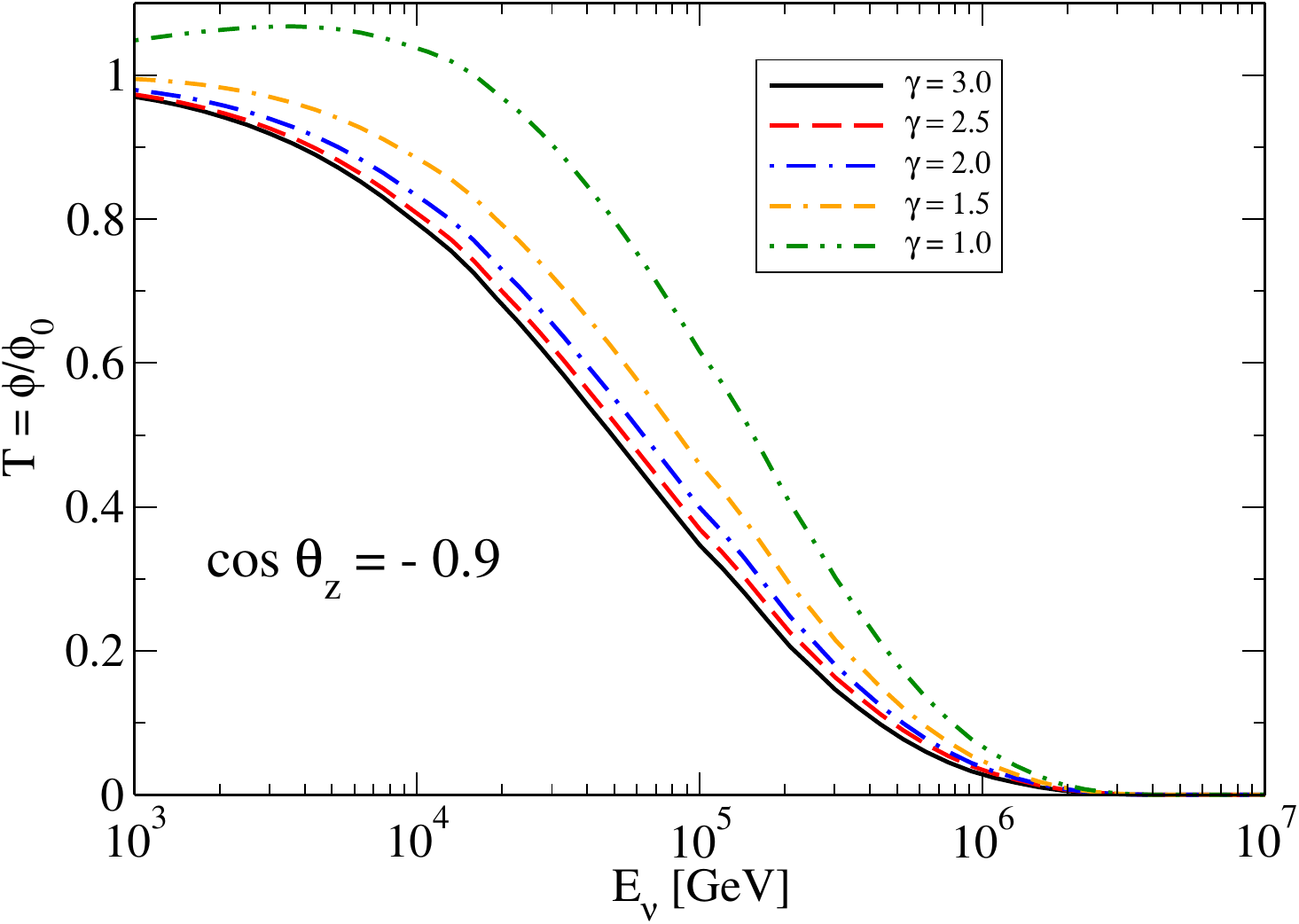} \\
	(a)  &  (b) 
			\end{tabular}
\caption{ Transmission coefficients of tauonic neutrinos as a function of energy considering different spectral indices for the incident neutrino flux for neutrinos incident with (a) $\cos \theta_z = -0.1$ and (b) $\cos \theta_z = -0.9$. }
\label{fig_1:diff_gammas}
\end{figure}

The results above were obtained considering a spectral index of 3.0, which is close to that obtained in analyses from HESE \cite{IceCube:2020wum}. However, analyses with other IceCube datasets, such as cascade-only data \cite{IceCube:2020acn} and track-only data \cite{IceCube:2024fxo}, yield slightly lower values for this parameter. In particular, the most recent IceCube results show $\gamma = 2.53$ and $\gamma = 2.58$ for cascade and track analyses, respectively. Motivated by these differences, we present Figure~\ref{fig_1:diff_gammas}, where we show the tau neutrino transmission coefficient considering different values for the spectral index between 1.0 and 3.0. Our results show that for index values of 2.5 and 3.0, the predictions for the flux transmission coefficient are similar, but these differences become significantly larger for smaller values of $\gamma$.

\subsection{Results for the number of events at IceCube}
\label{subsec_Ice:transEvents}

The transmission coefficients of astrophysical neutrino flux are essential for accurately understanding the surviving incident fluxes on neutrino observatories like IceCube. However, to describe the events observed at the IceCube observatory, we need to know the other quantities present in the Equation (\ref{eq_cs:eventos}), which we reproduce below
\begin{eqnarray}
\frac{\mathrm{d}N_{\mathrm{events}}}{\mathrm{d}(E_{\mathrm{vis}})\mathrm{d}\Omega} = 
T \sum_{\alpha} N_{\mathrm{effe},\alpha}(E_{\nu})\times \Phi _{\nu _\alpha}(E_{\nu})\times \sigma _{\nu _\alpha N}(E_\nu) 
\times T^{\nu}(E_\nu , \theta_{z}) \, .
\label{eq_Ice:eventos}
\end{eqnarray}
Given that one of our goals in this study is to build predictions for IceCube, we need to compare our results with those of the aforementioned collaboration to check if our methods are leading to similar results.

\begin{figure}
	\centering
	\begin{tabular}{ccc}
	\includegraphics[width=0.48\textwidth]{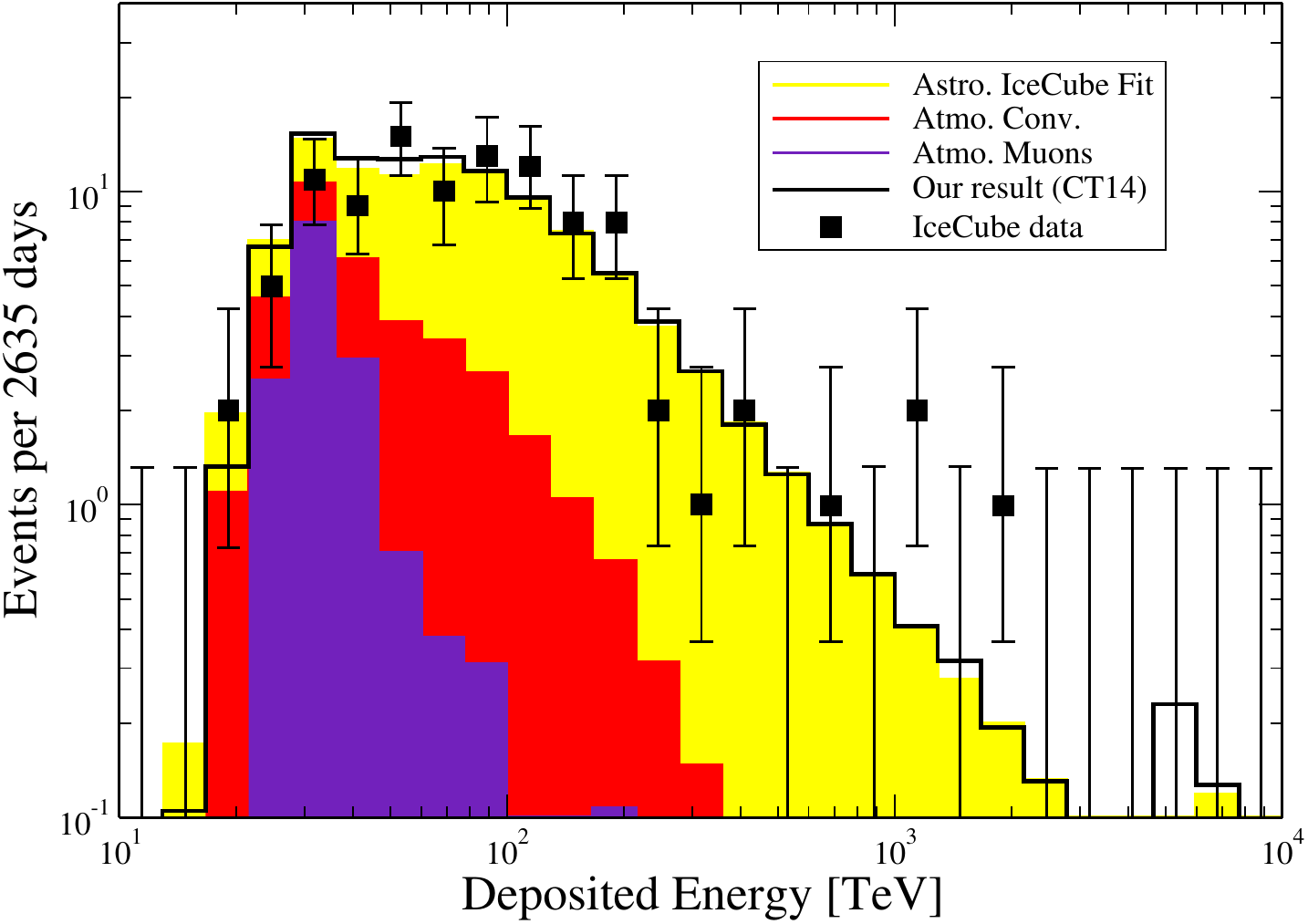} &  \includegraphics[width=0.48\textwidth]{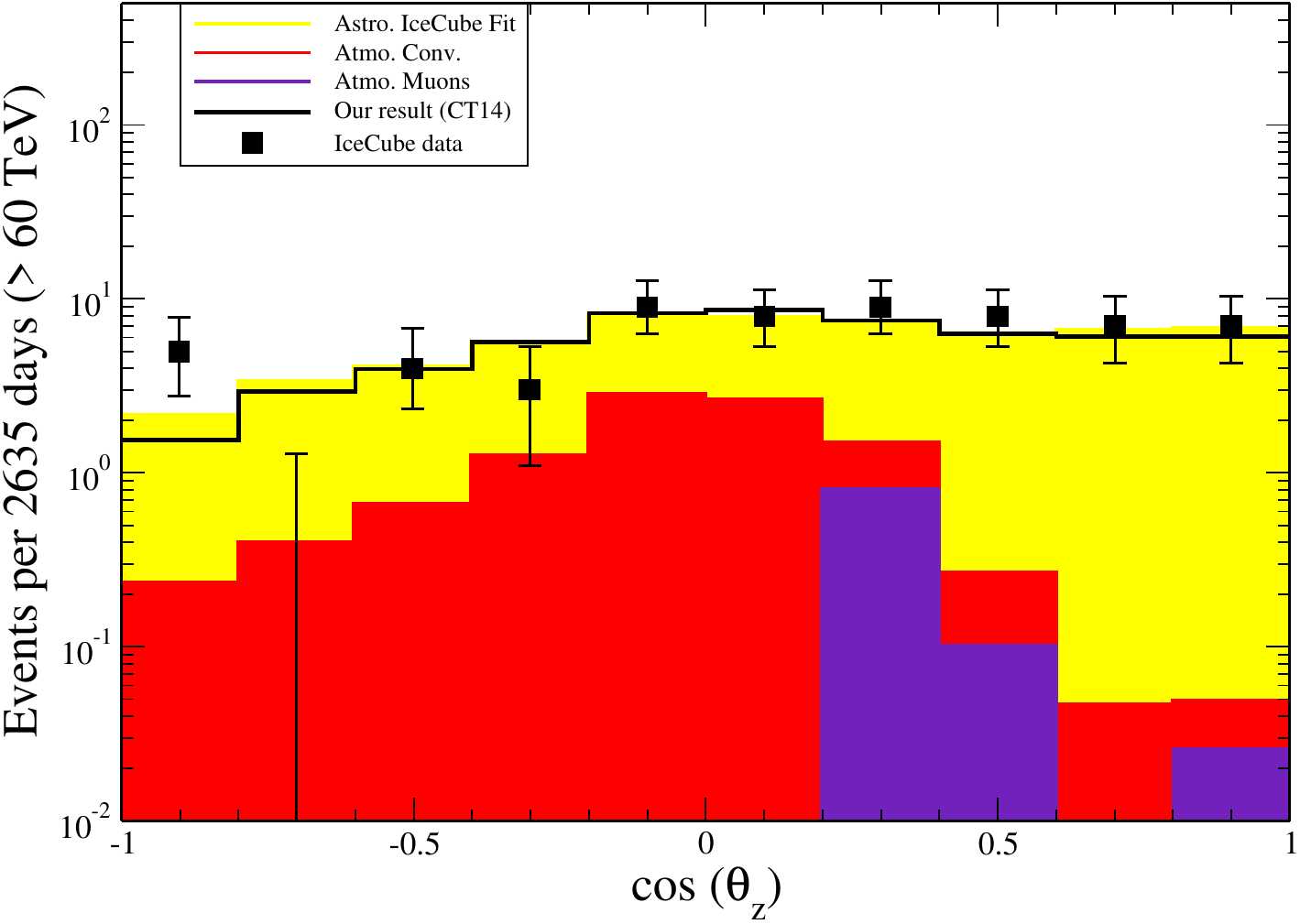} \\
	(a)  &  (b) 
			\end{tabular}
\caption{Energy (a) and angular (b) distributions of observed events for HESE at IceCube over 7.5 years. }
\label{fig_1:eventos}
\end{figure}

Using the parameterized cross sections with the results from the CT14 collaboration and presented in Figure \ref{fig_1:sigma}, the transmission coefficients obtained with PREM, and the effective number of IceCube targets provided by the collaboration itself, we obtained our results for the number of events presented in Figure \ref{fig_1:eventos}. This figure shows the energy (left) and angular (right) distributions of the observed events for those initiated within the observatory and meeting the HESE selection criteria over 7.5 years of data collection. In addition to the experimental data and our predictions, the figure presents IceCube's prediction (in yellow) for the events, and the atmospheric neutrino and atmospheric muon backgrounds in red and purple, respectively. Our predictions show good agreement with the predictions presented by the IceCube collaboration.

In addition to the nearly overlapping curves for the number of observed events, our predictions also show flux normalizations and spectral indices very close to those presented by the collaboration. While IceCube collaboration obtained the best fit with $\gamma = 2.87$ and $\Phi_0 = 1.06$, our results for these parameters were $\gamma = 2.89$ and $\Phi_0 = 1.25$. These good agreements in our results compared to those obtained by the IceCube collaboration indicate that our methodology is adequate and we can use it to test the models of interest to us for neutrino physics.

To obtain these neutrino flux parameter values that best describe the HESE events in IceCube, we used Poisson statistics \cite{Poisson:1837}: we assumed a Poisson distribution for each free flux parameter, which implies a likelihood function given by
\begin{eqnarray}
L(\vec{\theta}) = \prod_{i}^{N}\frac{\mu_{i}^{n_{i}}\mathrm{e}^{-\mu_{i}}}{n_{i}!}
    \label{eq_Ice:likelihood}
\end{eqnarray}
where $\vec{\theta}$ is the set of parameters we want to obtain that maximizes the likelihood function, $\mu_i$ is the number of expected events, and $n_i$ is the number of observed events in bin $i$. Using the Neyman-Pearson lemma for the statistical test \cite{Neyman:1933wgr}, $\lambda = -2\,\mathrm{ln}[L(\vec{\theta}_1)/L(\vec{\theta}_2)]$, the parameters $\vec{\theta}$ of the best fit curve will be those that minimize $\lambda$, which is given by
\begin{eqnarray}
\lambda = 2\sum_{i}^{N}\left[ \mu_{i} - n_{i} + n_{i}\,\mathrm{ln} \left( \frac{n_{i}}{\mu_{i}} \right) \right] + \sum_{j}^{m} \left( \frac{c_{j} - c_{j}^{*}}{\sigma_{j}} \right)^{2}\, .
    \label{eq_Ice:chi2}
\end{eqnarray}
The last term in the equation above refers to the Pull method \cite{Fogli:2002pt}, used here to adjust the parameters $c_i$ with expected values $c_{i}^{*}$, which include the renormalization of the atmospheric neutrino and muon backgrounds and the uncertainty in the energy deposited in each event. In our results, we are assuming the same allowed intervals for $c_i$ and $\sigma_j$ used in \cite{IceCube:2015gsk}. Finally, one could assume the validity of Wilks' theorem \cite{Wilks:1938dza}, which implies that $\lambda$ follows a $\chi^{2}$ distribution, if we wanted to obtain the allowed confidence intervals for the neutrino flux parameters, but we will not do so at this time. It is important to emphasize that, like the IceCube collaboration, we only use bins from Figure~\ref{fig_1:eventos} with energies deposited on the detector above 60 TeV, since it is in this region that expected astrophysical neutrino events dominate compared to atmospheric neutrino and muon events. The events are separated into bins of energy deposited in the event, which is not necessarily equal to the energy of the incident neutrino. In the next section we will discuss in more detail how each type of event deposits visible energy to the photomultiplier tubes.

\begin{figure}
	\centering
	\includegraphics[width=0.5\textwidth]{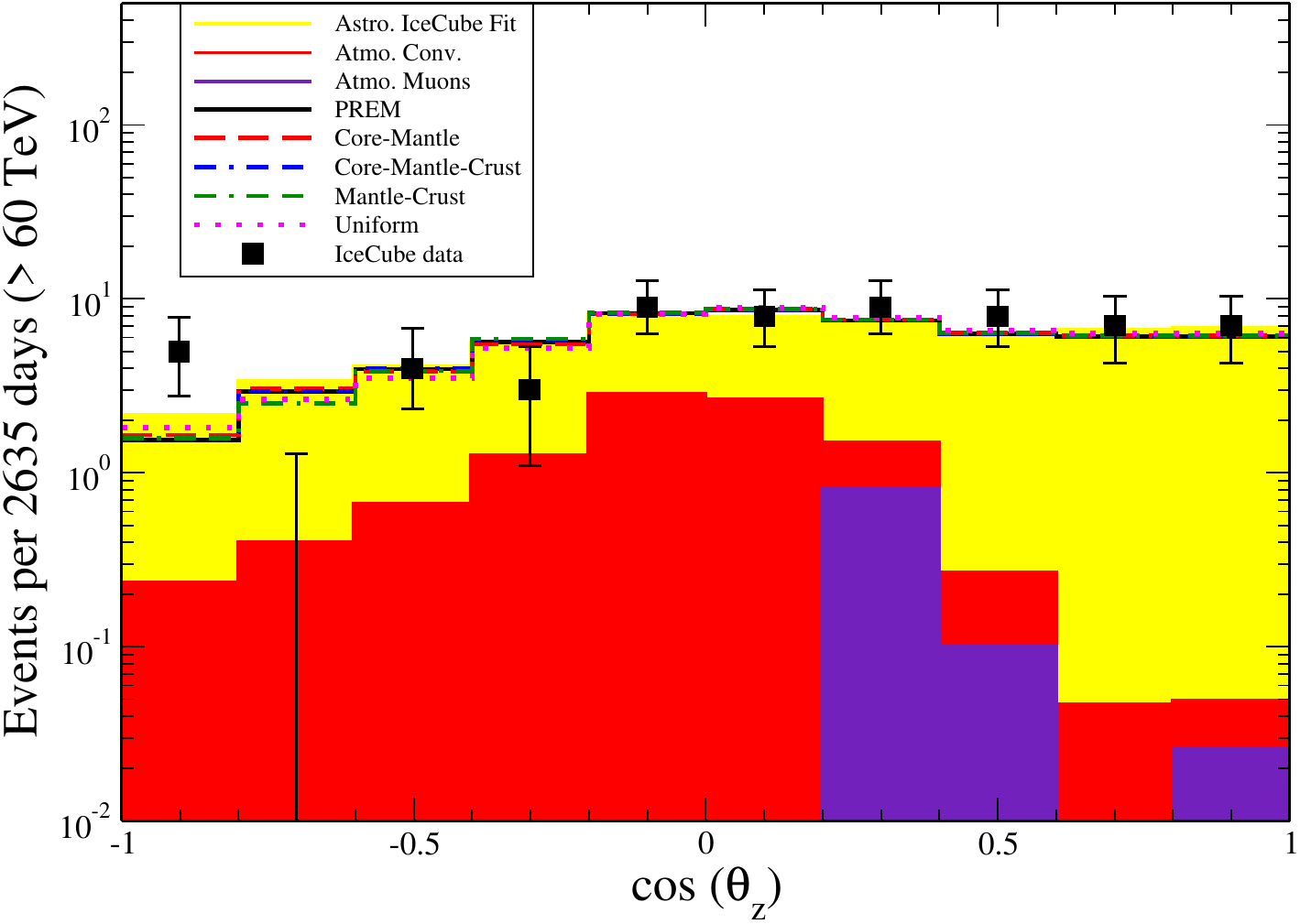} 
\caption{Angular distribution of observed events for HESE at IceCube over 7.5 years, considering different Earth density profile models. }
\label{fig_1:angularterra}
\end{figure}

In Subsection \ref{subsec_Ice:transResu}, where we describe the impacts of different models on the Earth's density profile, we saw that the transmission of neutrino flux is significantly affected by the use of different models. In Figure \ref{fig_1:angularterra}, we show how the angular distribution of events in IceCube is affected by the aforementioned different models. For the construction of this figure, we also used the cross sections from Figure~\ref{fig_1:sigma}, with partonic distributions parameterized by CT14.

Our results show that the number of events is little affected by the use of simplified Earth density profile models. The Uniform model differs most from PREM in almost all bins, but the separation between them is still much smaller than the error bars of the observed events. This indicates that for the current IceCube data (7.5 years of observation used here), simplified Earth density profile models are still good approximations to use. Furthermore, with the expected exposure for IceCube-Gen2 (8 times the size of the current observatory), the simplified models may cease to be good estimates, making the use of PREM necessary for more accurate predictions.

\section{Track signals from subleading processes at IceCube}
\label{sec_Ice:tracks}

IceCube can currently distinguish three different event topologies described in the previous chapter: cascades, tracks, and double cascades. Cascades are events where there is only one block of energy deposited and no visible muon track. Tracks are events where there is at least one very energetic muon in the final state, leaving the characteristic muon track on the detector. And double cascades originate from the decay of very energetic taus, which decay reasonably far (greater than a few meters) from the first energy deposition site of the interaction, leaving visible two sites with large energy deposits on the detector.

\begin{figure}[!t]
	\centering
	\begin{tabular}{ccccc}
	   \includegraphics[width=0.3\textwidth]{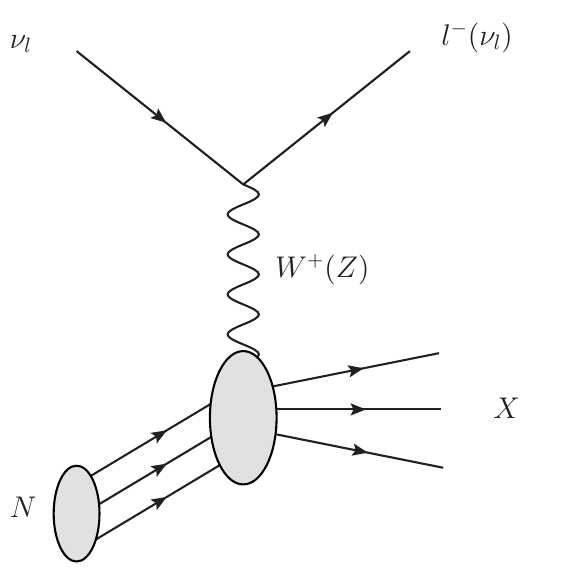} & 
          \includegraphics[width=0.3\textwidth]{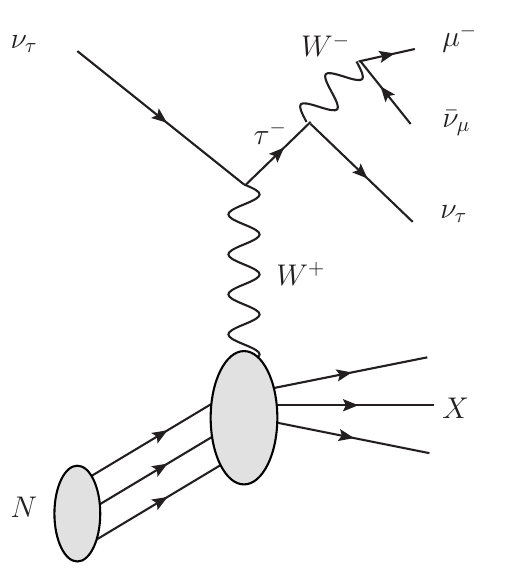} & 
          \includegraphics[width=0.3\textwidth]{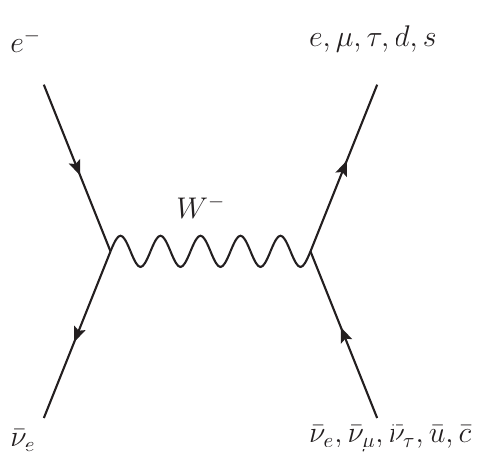} \\
          (a) & (b) & (c) \\
          \includegraphics[width=0.33\textwidth]{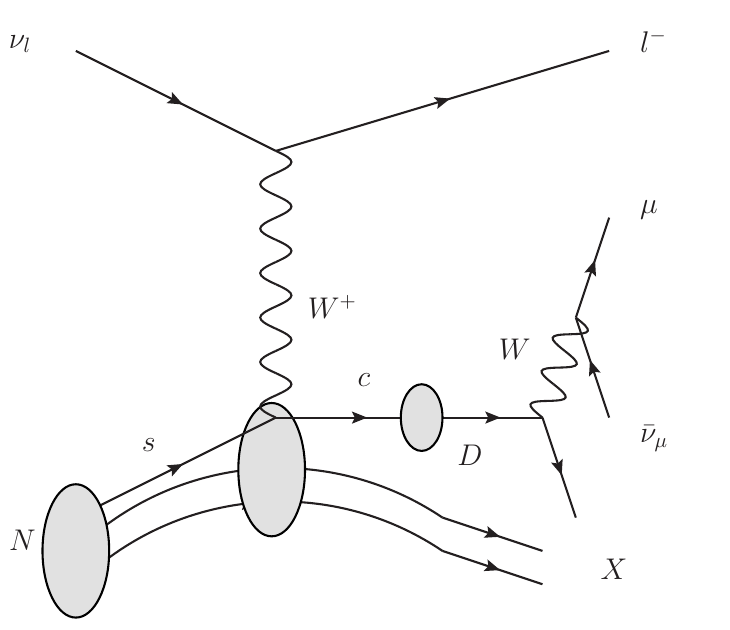} &
          \includegraphics[width=0.33\textwidth]{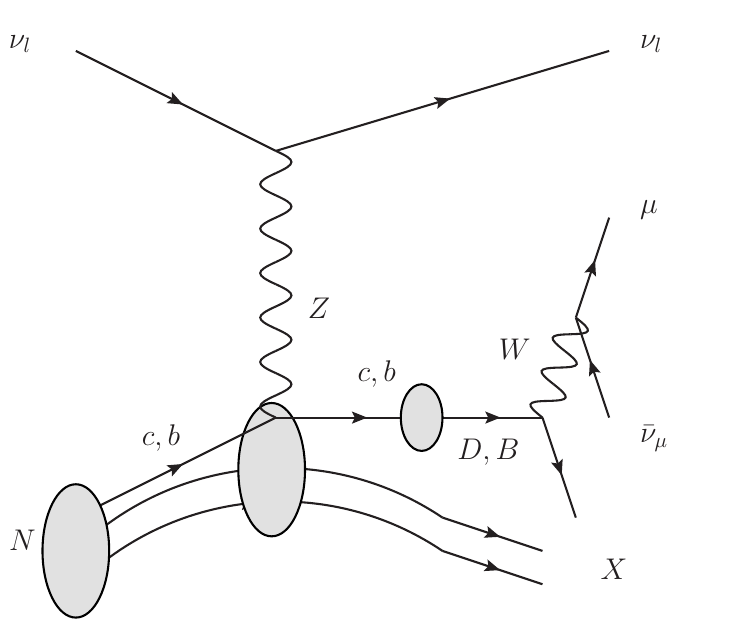} & 
          \includegraphics[width=0.33\textwidth]{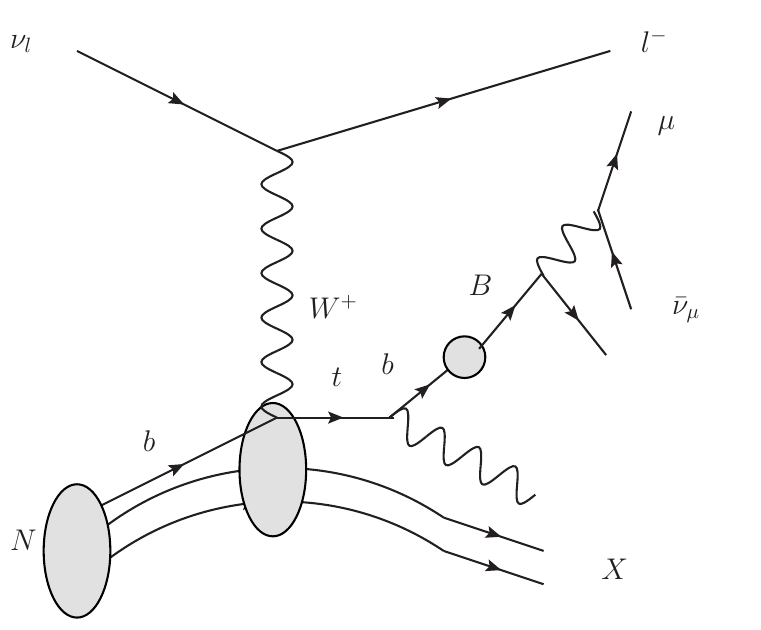} \\
          (d) & (e) & (f) \\
          \includegraphics[width=0.33\textwidth]{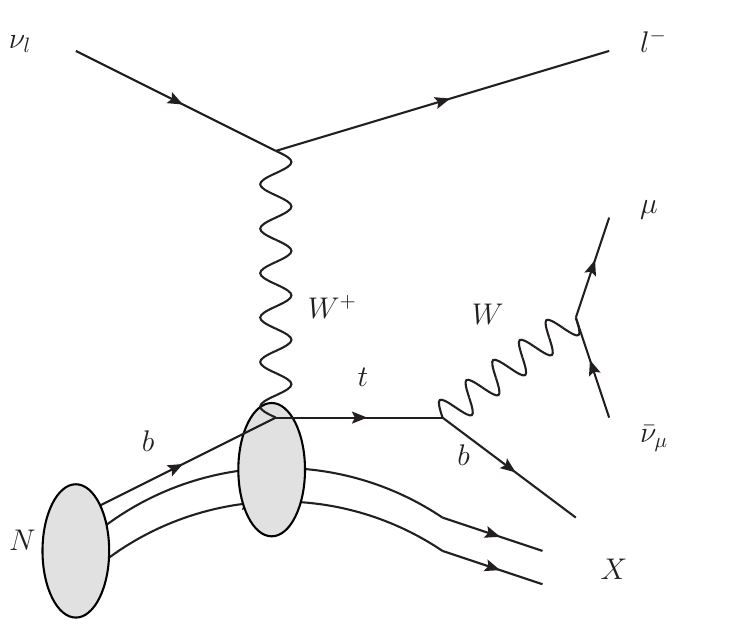} &
          \includegraphics[width=0.33\textwidth]{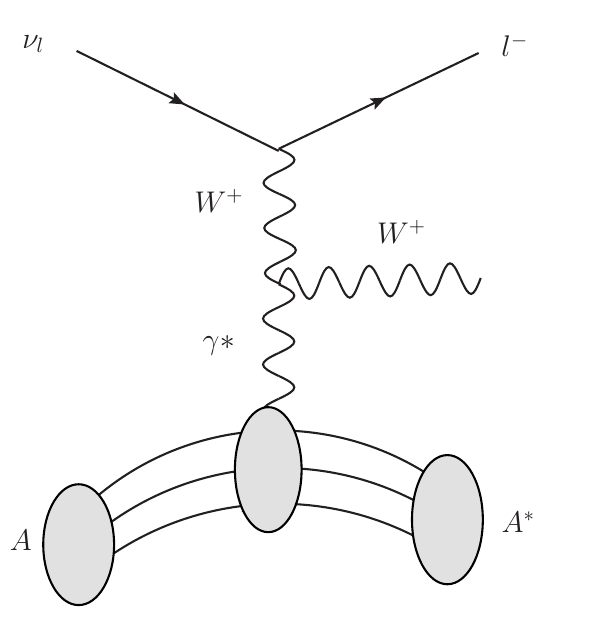} & 
          \includegraphics[width=0.33\textwidth]{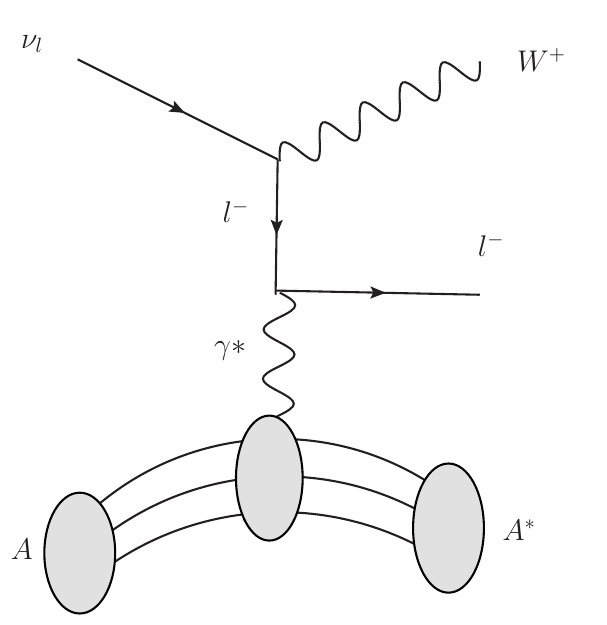} \\
          (g) & (h) & (i)
	\end{tabular}
\caption{ Typical neutrino interaction diagrams with the IceCube Observatory that give rise to energetic muons in the final state, which can be identified as tracks. }
\label{fig_2:diagramasTracks}
\end{figure}

HESE events are all those initiated within the detector, whether cascades, tracks, or double cascades, that meet the selection criteria of a minimum number of observed photons, as well as the position of the initial vertex inside the observatory. A common approach, widely used in the literature \cite{Chen:2013dza,Goncalves:2021gcu,Goncalves:2022uwy}, is to consider that tracks arise only from charged current interactions of muon neutrinos and from charged current interactions of tau neutrinos and subsequent decay of taus into muons, from diagrams (a) and (b) of Figure \ref{fig_2:diagramasTracks}. This approach is a good initial estimate, but it ignores tracks produced from the decay of heavy mesons $D$ and $B$ and top quarks (diagrams (d), (e), (f) and (g) of Figure \ref{fig_2:diagramasTracks}), in addition to the decays of bosons $W^\pm$ produced (diagrams (c), (h) and (i) of Figure \ref{fig_2:diagramasTracks}).

As the IceCube data statistics increase, a more complete description of the events that leave track signals in the detector becomes necessary. The main source of tracks due to meson decay are $D$ mesons, or charm mesons. These consist of a charm quark and some other quark, and about 10\% of the time they decay with a muon in the final state. The production of charm mesons becomes significant in charged current interactions, as they occur mainly through the transition from strange quark to charm quark, therefore being sensitive to the amount of strange quarks in the nucleon. In the results presented in this section, we are using Pythia6 \cite{Sjostrand:2006za} to simulate hadronization and the decay of heavy mesons, as well as the energy fractions carried by leptons and hadrons resulting from the decays.

Another process not usually taken into account for track events at IceCube is the production of $W^\pm$ bosons \cite{Zhou:2019vxt,Zhou:2019frk}. This process can occur through the emission of the $W^{\pm}$ boson by the neutrino, and then the boson or lepton interacts electromagnetically with the target nucleus (diagrams (h) and (i) of Figure \ref{fig_2:diagramasTracks}); or through the annihilation of an electron antineutrino with an electron, producing a $W^-$ boson (a process known as Glashow resonance, represented in diagram (c) of Figure \ref{fig_2:diagramasTracks}). The process of producing $W^{\pm}$ in the electromagnetic field of a nucleus was recently calculated for the IceCube energy regime \cite{Zhou:2019vxt,Zhou:2019frk} and is quite important for track signals, given that there are different decay channels of aforementioned boson into muons: direct decay into a muon, decay into a tau that decays into a muon, and decay into a charm-strange quark pair that hadronizes into a charm meson and subsequently decays into a muon.

It is important to point out here that many other processes produce many muons in IceCube, such as pion and kaon decays. But given that these lighter mesons have long lifetimes compared to heavier mesons, we are considering that light mesons lose a lot of its energies in scatterings before decaying, while in the case of heavy mesons we are considering that they do not lose energy before their decays. This approximation is frequently found in other works in the literature \cite{Barger:2016deu}.

\subsection{Cross sections of subleading channels}
\label{subsec_Ice:tracksCS}

\begin{figure}
	\centering
	\begin{tabular}{ccc}
	   \includegraphics[width=0.48\textwidth]{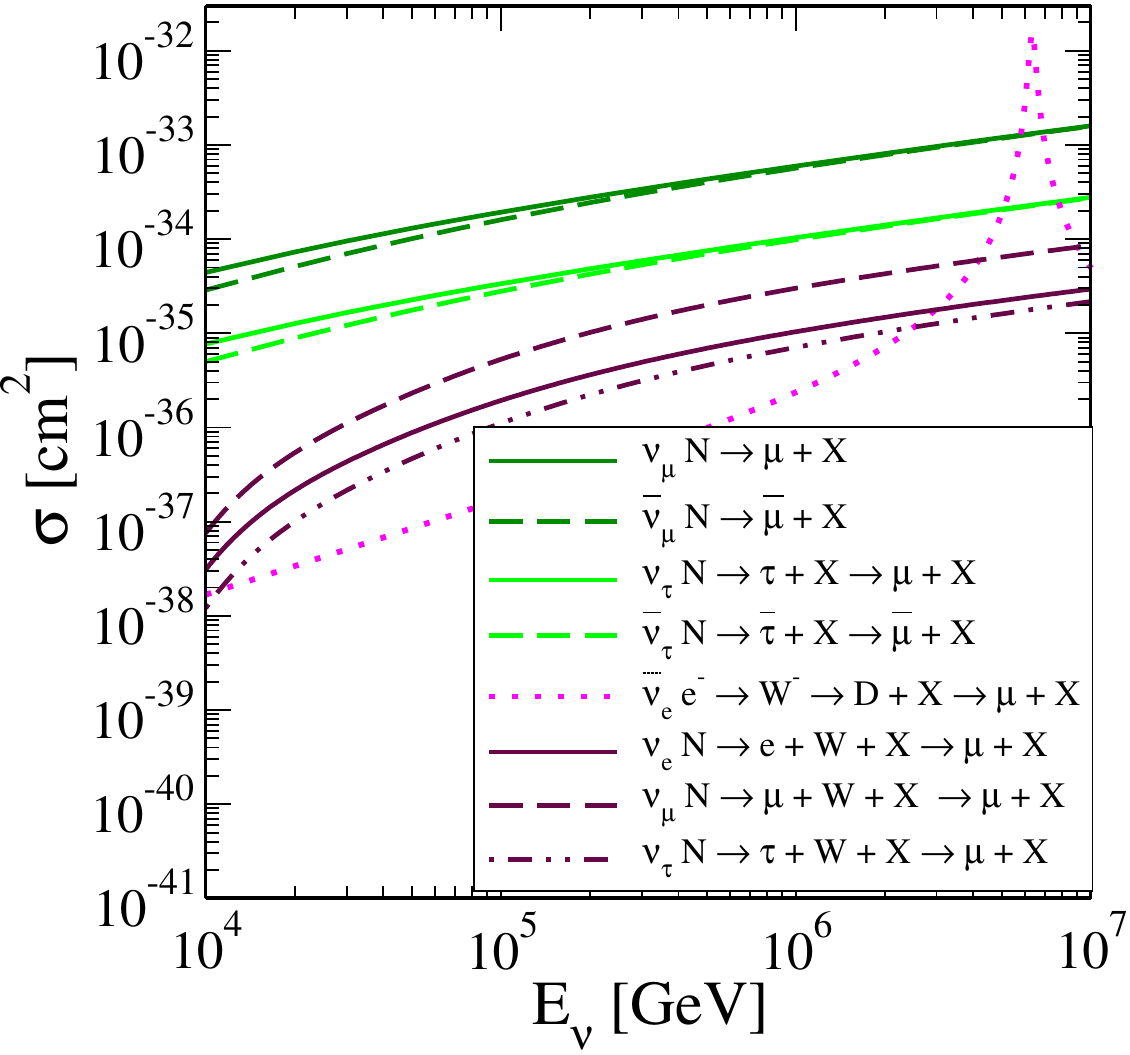} &
	   \includegraphics[width=0.48\textwidth]{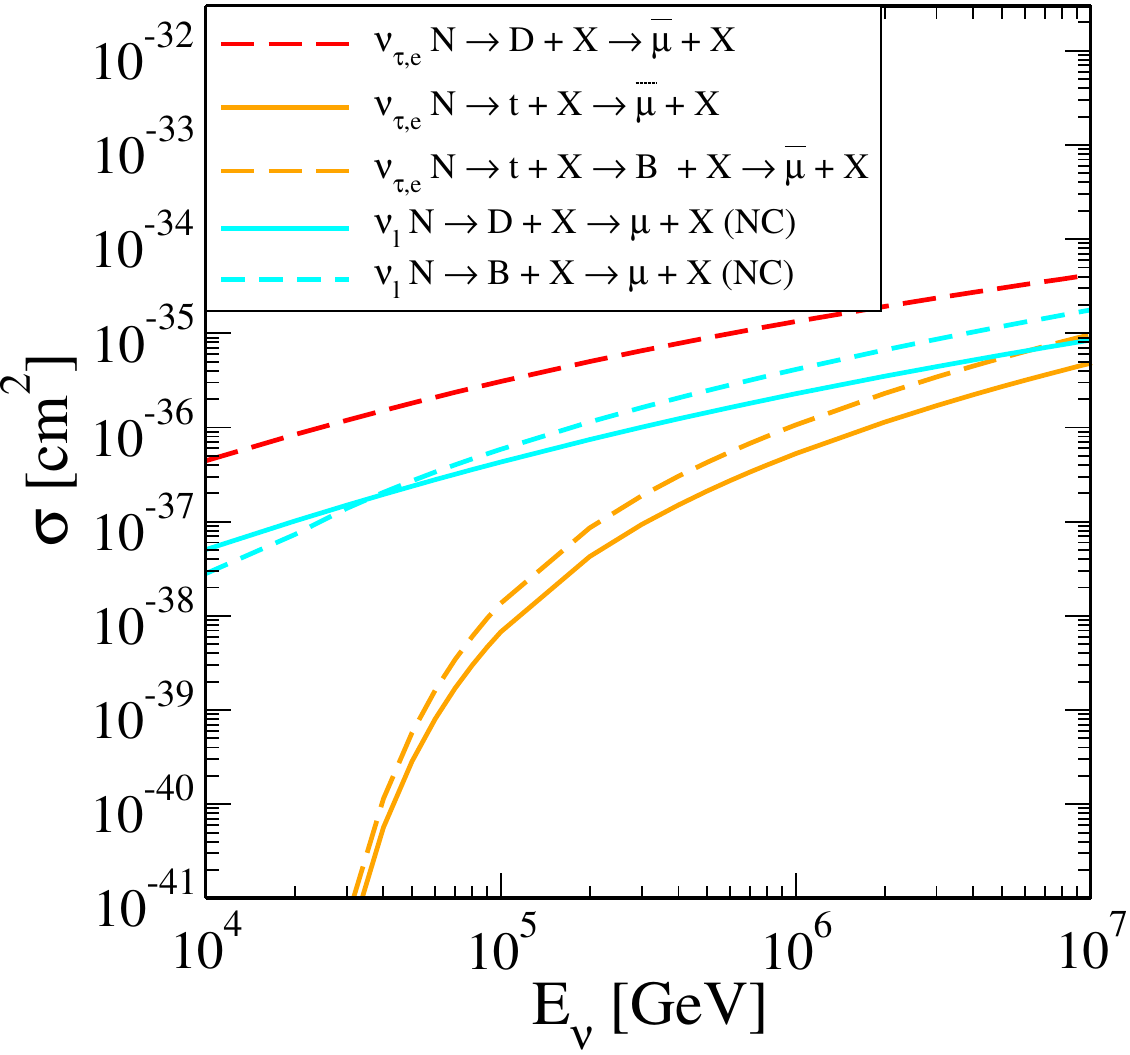}
	\end{tabular}
\caption{ Cross section per nucleon (per electron for the Glashow resonance) for the main channels that produce energetic muons that can be identified as tracks at IceCube. On the left, we have the muon processes arising from the muonic neutrino, tau  decay, and $W^{\pm}$ boson decay. On the right, we see the processes where muons arise from the decay of mesons and heavy quarks. }
\label{fig_2:sigma}
\end{figure}

As initial motivation for investigating the contribution of subleading channels to track events at IceCube, we present Figure \ref{fig_2:sigma}. In this figure, we show the cross section as a function of the incident neutrino energy for the main processes that give rise to tracks in the energy regime probed by IceCube. Our results show that subdominant channels have a cross section at least one order of magnitude below the main process, which is the charged current interaction of muonic neutrinos. With the increase in IceCube data statistics, as well as the advent of its successor, IceCube-Gen2, subdominant processes with cross sections only one order of magnitude below the dominant one become relevant. We highlight from Figure \ref{fig_2:sigma} the important contributions of $W^{\pm}$ boson production, and the decay of $D$ mesons produced in charged current interactions of electronic and tauonic neutrinos. Neutrino-nucleon cross sections were calculated using the DIS equations presented in Chapter \ref{cap:cs}, using the parton distribution functions parameterized by the CT14 collaboration at LO \cite{Dulat:2015mca}, while hadronization and decays were simulated with Pythia6. For the production channels of the $W^{\pm}$ boson, we used the cross sections provided in the references \cite{Zhou:2019vxt,Zhou:2019frk}.

The cross section is the main object of interest in our investigation, given that the number of events measured at IceCube and other observatories and detectors is directly proportional to it. In IceCube track events, the total cross section is not the only quantity we can extract. With the development of techniques for reconstructing the total energy of the muon present in the final state through energy loss during propagation (only a fraction of the muon's energy is deposited at the observatory, given that the muon of visible energies at IceCube travels several kilometers before decaying), and with the separation between the deposition of the muon's energy and that of other charged particles produced in the interaction, it is possible to reconstruct the inelasticity of each event \cite{IceCube:2018pgc}. Recently, the average inelasticity in astrophysical neutrino events has gained interest in the literature because, among other things, it is sensitive to the neutrino-antineutrino asymmetry of the incident flux \cite{Skrzypek:2025tmg}. As we saw in the previous chapter, the inelasticity $y$, in the rest frame of the target nucleon, represents the fraction of energy of the incident neutrino that is transferred to the nucleon. Therefore, $1-y$ is the fraction of the remaining energy of the lepton involved in the interaction. The IceCube collaboration recently reported measurements of reconstructed average inelasticity in processes characterized as tracks \cite{IceCube:2018pgc,IceCubeCollaborationSS:2025zgz}.

When it is possible to estimate the energies of the final hadronic and final muon states, we can extract the inelasticity of a charged current interaction event with a muonic neutrino. However, there are subdominant channels that produce the same muon plus hadrons final state, and in these cases the reconstructed inelasticity does not give us information about the fraction of the incident neutrino energy transmitted to the hadronic target. Let's discuss a specific case for illustration: an electron neutrino interacts via charged current, producing an electron plus a hadronic final state. The electron ended up with an energy fraction of $1-y$, hence the hadronic state with $y$. The interaction within the nucleon was with a strange quark, which became a charm quark that hadronized into a $D$ meson. If the $D$ meson decays semileptonically producing a final state muon, and this muon retains a fraction $z$ of the initial charm energy, the muon retains a fraction $yz$ of the initial neutrino energy. On the other hand, the associated interaction cascade will have a fraction of the electron energy $1-y$, plus $(y-2z)f_h$, where $f_h$ is used to account for the efficiency of hadron energy detection, which is lower than the efficiency in measuring electron energy, due to the possibility of producing neutral hadrons \cite{Palomares-Ruiz:2015mka}. The term $2z$ is necessary to subtract the energy of the muon and the muonic neutrino produced in the semileptonic decay of the $D$ meson.

\begin{figure}
	\centering
	\begin{tabular}{ccc}
	\includegraphics[width=0.36\textwidth]{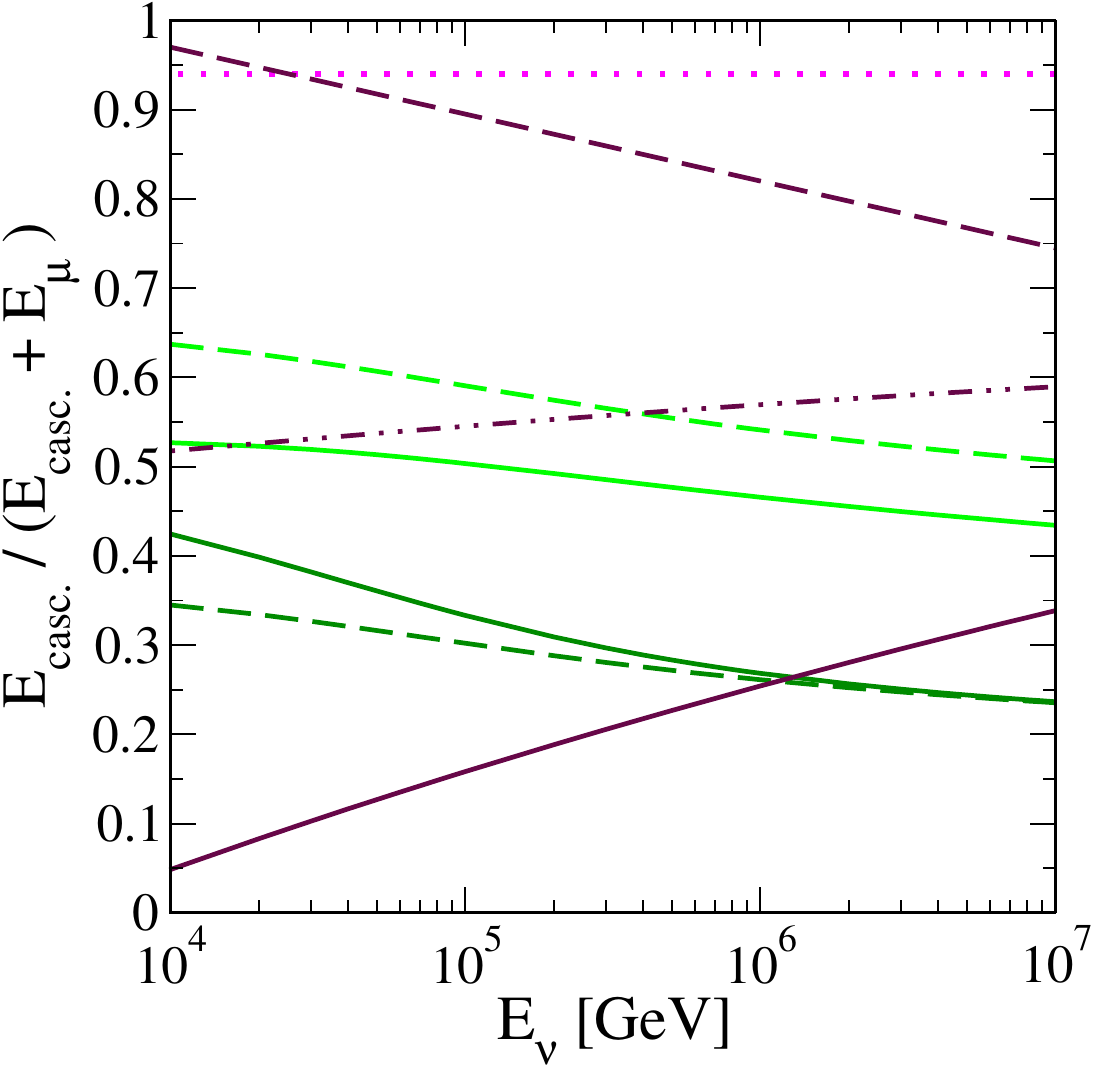}
	\includegraphics[width=0.618\textwidth]{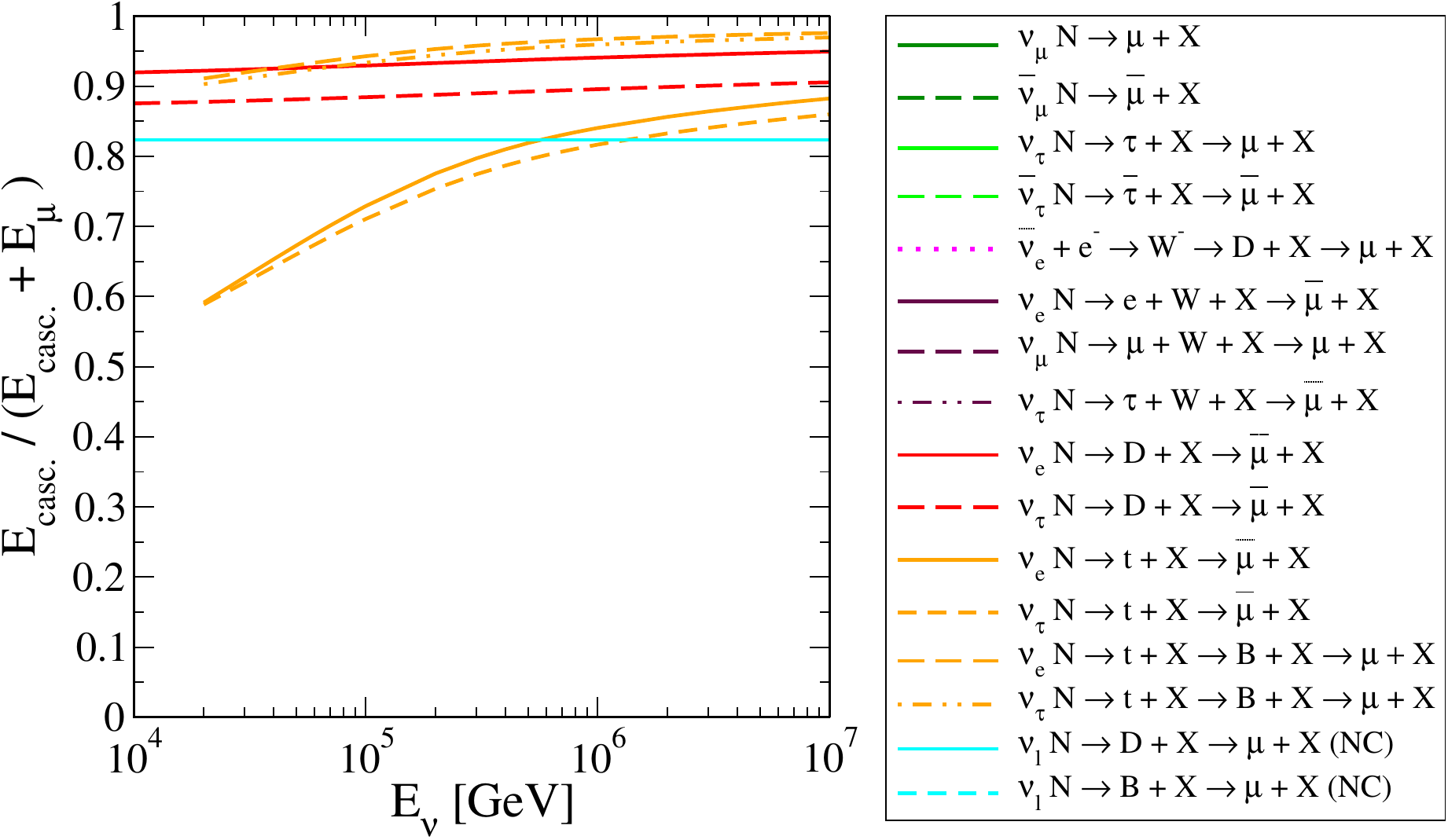} 
			\end{tabular}
\caption{ Average inelasticity reconstructed as a function of incident neutrino energy for different channels of final state energetic muon production. }
\label{fig_2:Ymedio}
\end{figure}

In Figure \ref{fig_2:Ymedio} we present the reconstructed average inelasticity for all energetic muon channels in the final state we are studying. We see that the subdominant channels, especially muons resulting from the decay of hadrons and quarks, have a much higher reconstructed average inelasticity than the $\nu_\mu N$ process, which may help in separating these processes in future observations. In particular, the $W^{\pm}$ boson production channels in the nuclear Coulomb field show quite distinct reconstructed average inelasticity from each other. Electronic neutrino induced $W^\pm$ production has an average inelasticity close to zero and increases with energy. This happens because at low electronic neutrino energies, only a small fraction of the energy remains with the electron, and a large part with the $W^{\pm}$ boson that decays into a muon plus its corresponding neutrino. The opposite effect is seen for interactions induced by muon neutrinos: the reconstructed average inelasticity is close to one for lower energies, and decreases with increasing energy of the incident neutrino. In this case, the muon originates from the vertex of the initial neutrino, which has little energy for less energetic neutrinos, while the cascade originates from the decay of the $W^{\pm}$ boson. In the case of events initiated by tau neutrinos, the reconstructed average inelasticity is almost constant, remaining close to 0.5. This is explained by the fact that muons originate from both the decay of the tau boson and the $W^{\pm}$ boson, causing this observable to behave as an average between the electronic and muonic neutrino channels discussed above.

\begin{figure}
	\centering
	\begin{tabular}{ccc}
    \includegraphics[width=0.36\textwidth]{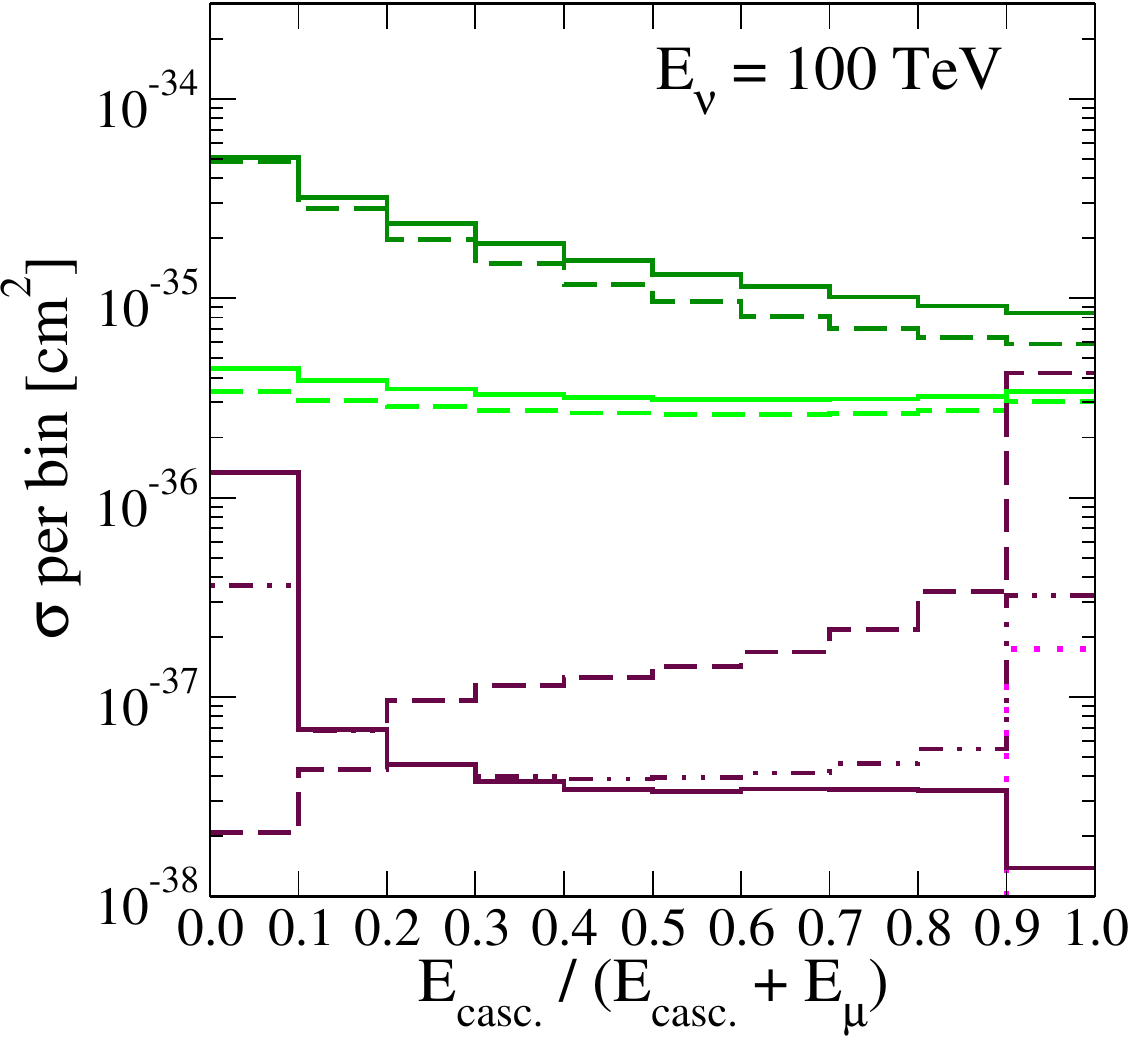} &
    \includegraphics[width=0.605\textwidth]{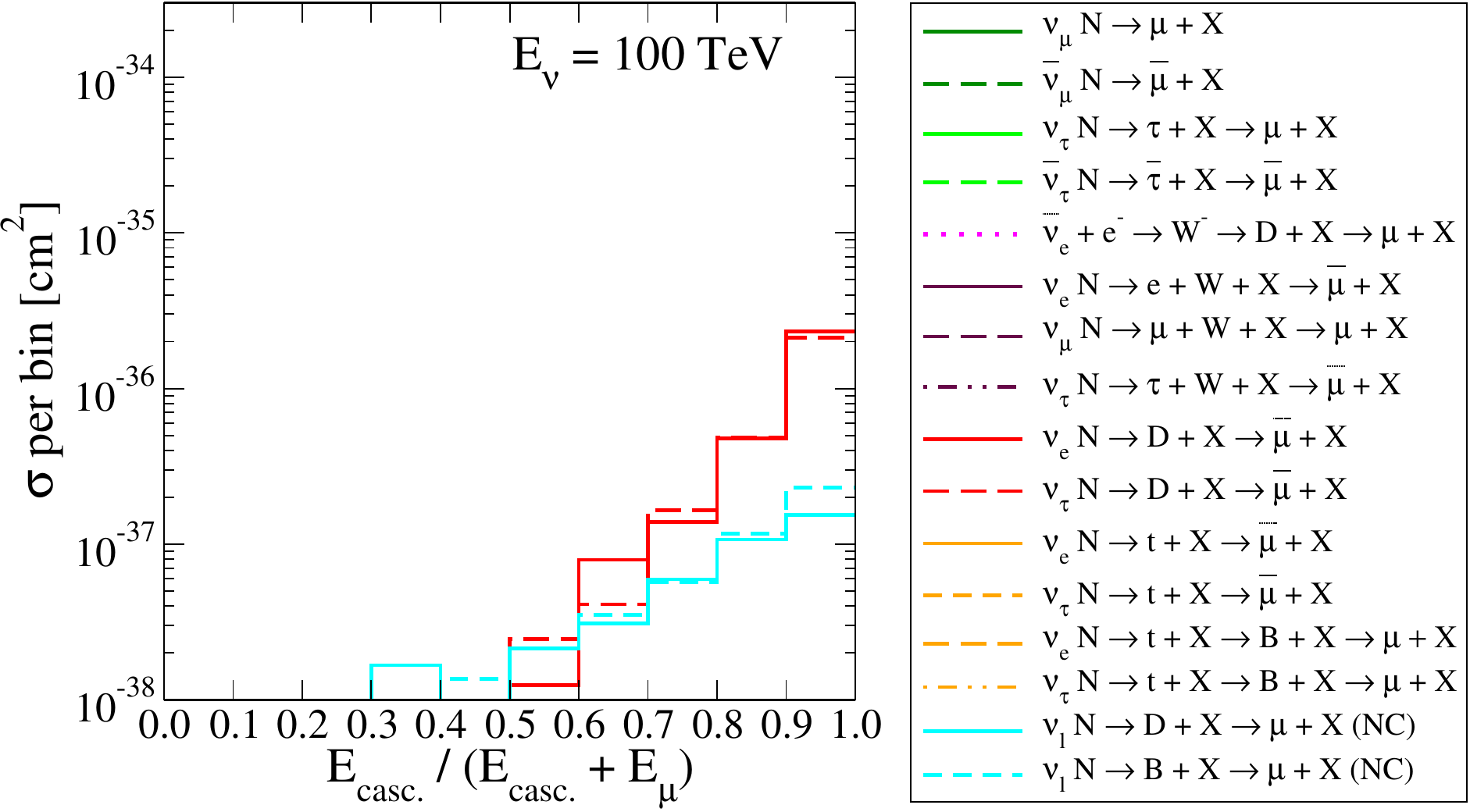} \\
    \,\,\, & \,\,\, \\
    \includegraphics[width=0.36\textwidth]{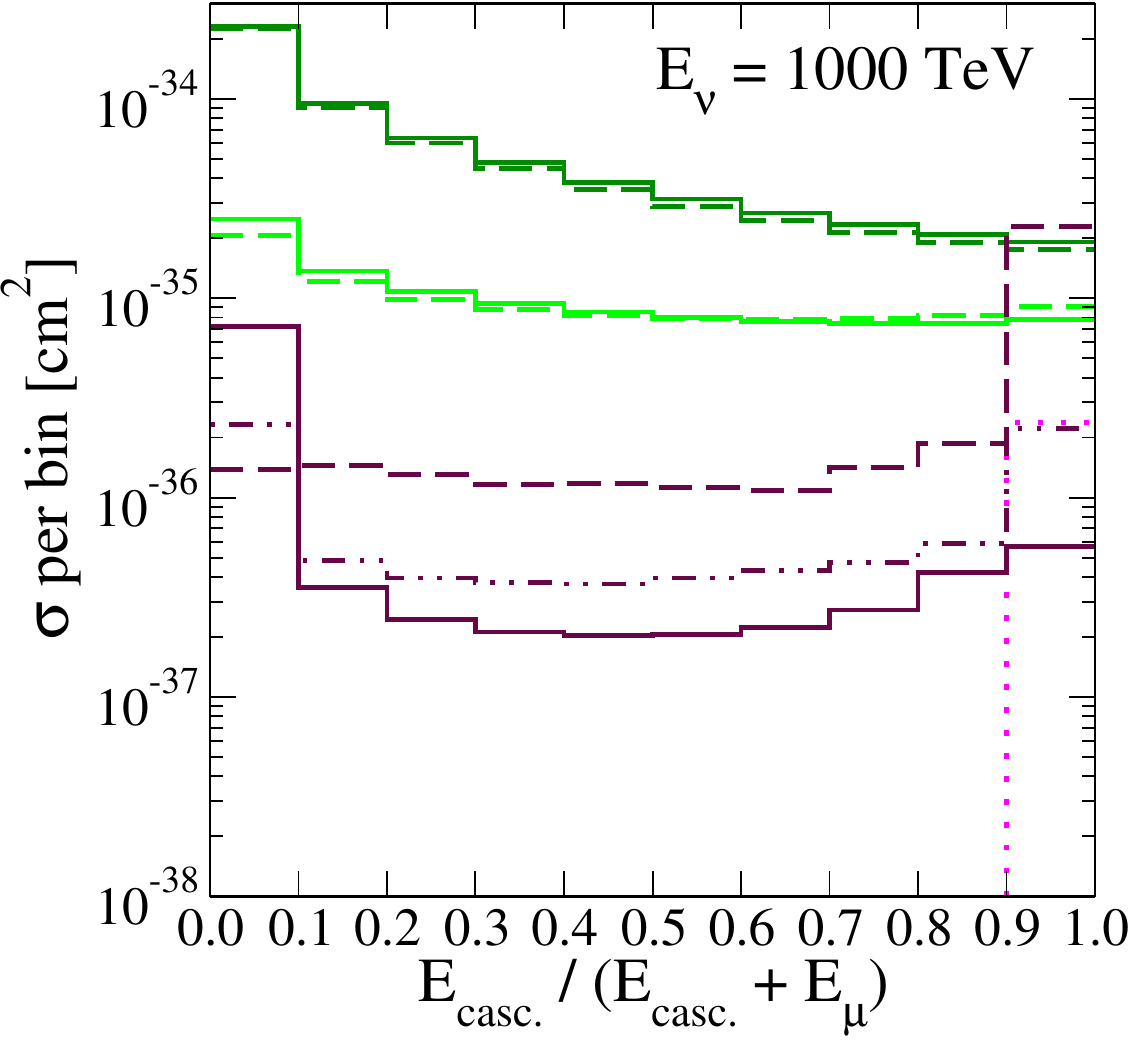} &
    \includegraphics[width=0.605\textwidth]{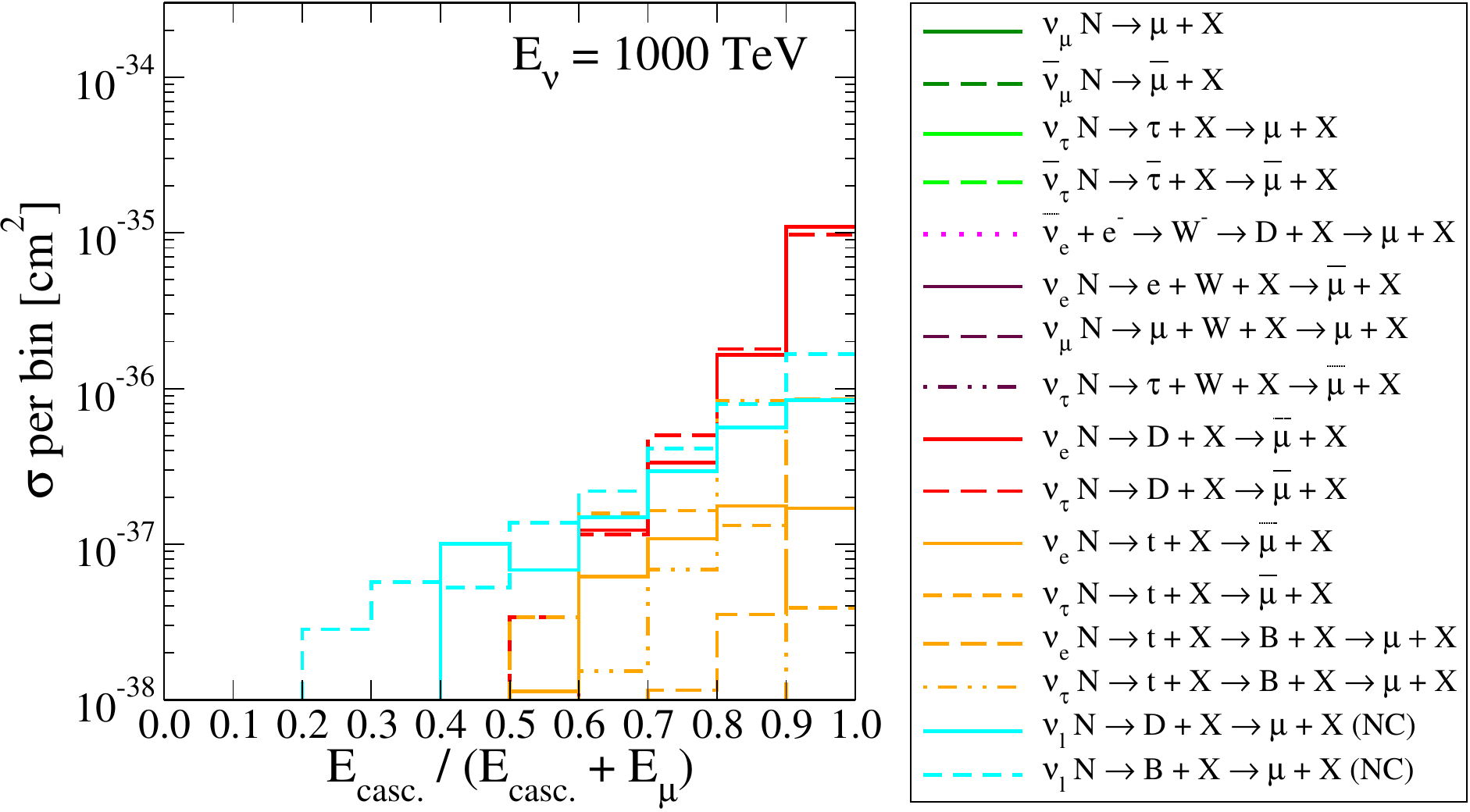}
			\end{tabular}
\caption{ Cross section as a function of average inelasticity reconstructed for several channels with muons in the final state. We consider incident neutrinos of 100 TeV (upper panels) and 1000 TeV (lower panels). }
\label{fig_2:histograma}
\end{figure}

To complete our analysis, in Figure \ref{fig_2:histograma} we present the cross section per bin for the different muon production channels in the final state as a function of the reconstructed average inelasticity. Our results are for incident neutrinos of 100 TeV and 1000 TeV, in the upper and lower panels, respectively. Our results show a greater importance of the subdominant channels discussed above at higher incident neutrino energies. We also see that the contribution of the dominant channel ($\nu_\mu N$ scattering) is more important for lower reconstructed average inelasticities, while hadron decay channels contribute significantly in the inelasticity region close to one. For the $W^{\pm}$ boson production channels, the contribution is strongly dependent on the flavor of the incident neutrino, as discussed in the previous paragraph.

\subsection{Number of track events originating from subdominant channels at IceCube}
\label{subsec_Ice:tracksEvents}

\begin{figure}
	\centering
	\begin{tabular}{ccc}
	\includegraphics[width=0.32\textwidth]{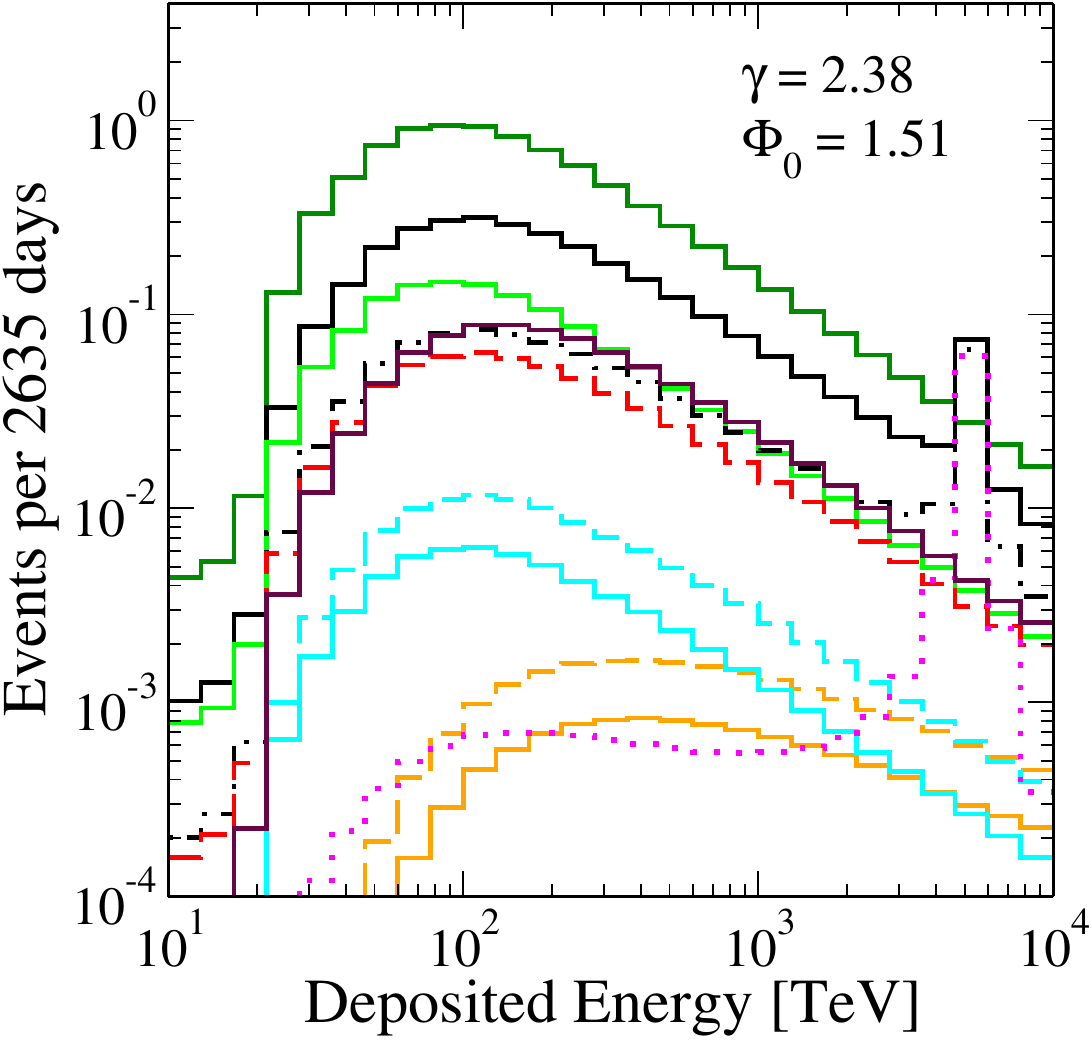} &     \includegraphics[width=0.65\textwidth]{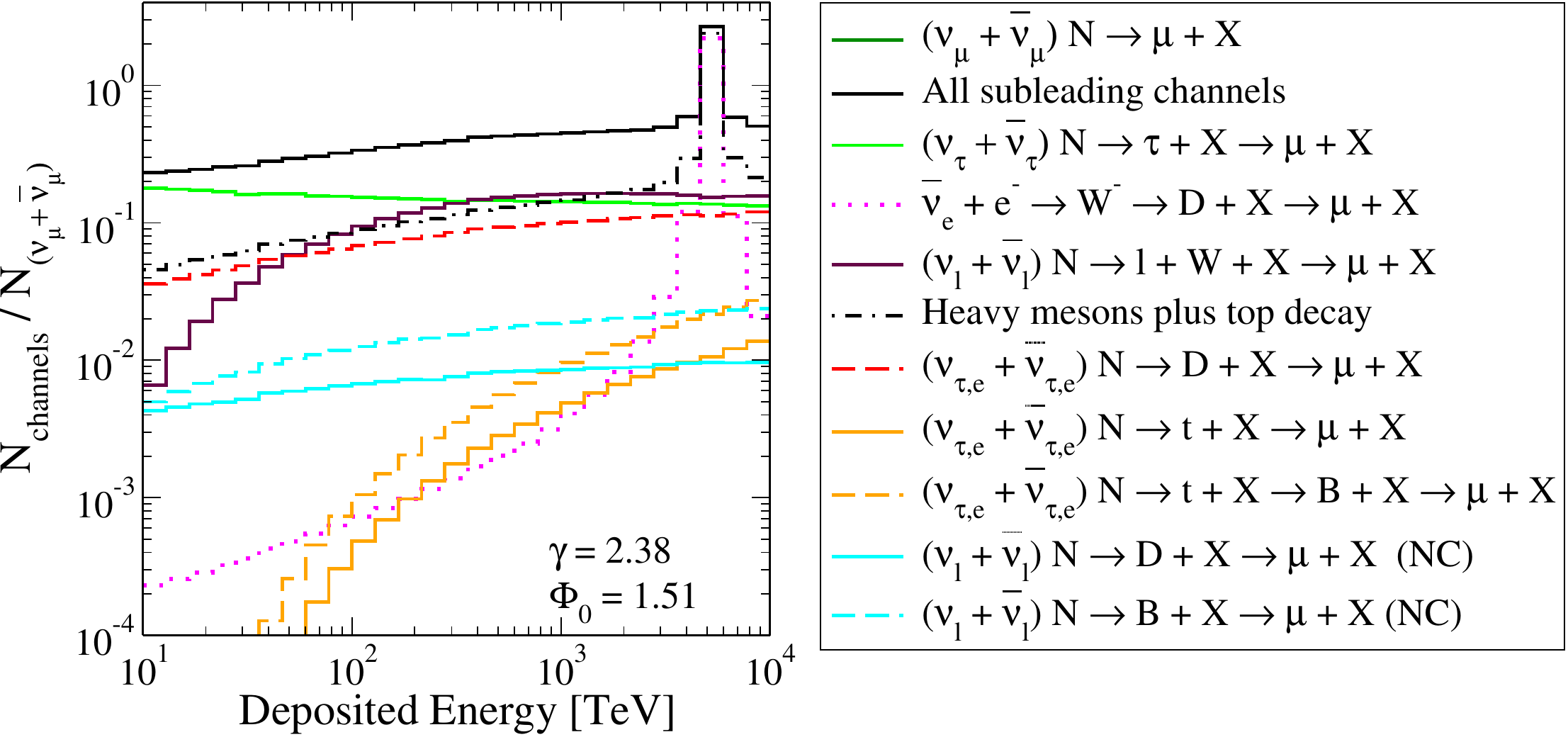}
			\end{tabular}
\caption{ {\bf Left:} Track events at IceCube as a function of deposited energy for different channels. We consider 7.5 years of observatory exposure for data collection and the astrophysical flux parameters given by $\gamma = $2.38 and $\Phi_0 = $1.51. {\bf Right:} Ratio between contributions to track events associated with subdominant channels and to $\nu_\mu$ CC interactions. }
\label{fig_2:events_channels}
\end{figure}

Our results so far show how important subdominant channels are for the cross section and average inelasticity reconstructed in the energy regime covered by IceCube. On the left side of Figure \ref{fig_2:events_channels}, we present the contribution of each of these channels to the events observed at IceCube as a function of the reconstructed event energy. The events are for a period of 7.5 years of astrophysical neutrino event data collection, with flux parameters used being $(\Phi_0 = 1.51, \gamma = 2.38)$, which are the parameters of the best fit curve obtained by the IceCube collaboration from the last published analysis for track events \cite{IceCube:2024fxo}. The reconstructed energy is the sum of the equivalent electromagnetic energy of cascades and tracks deposited by the particles in the instrumented volume of the observatory. On the right side of Figure \ref{fig_2:events_channels} we show the ratio of the number of events from the subdominant channels to the tracks with the dominant channel. Our results show that the sum of the subdominant channels is responsible for at least 20\% of the track events at IceCube in each bin, and in the bin with the highest contribution from the Glashow resonance, it surpasses the contribution of the dominant channel.

\begin{figure}
	\centering
	\begin{tabular}{ccc}
	\includegraphics[width=0.6\textwidth]{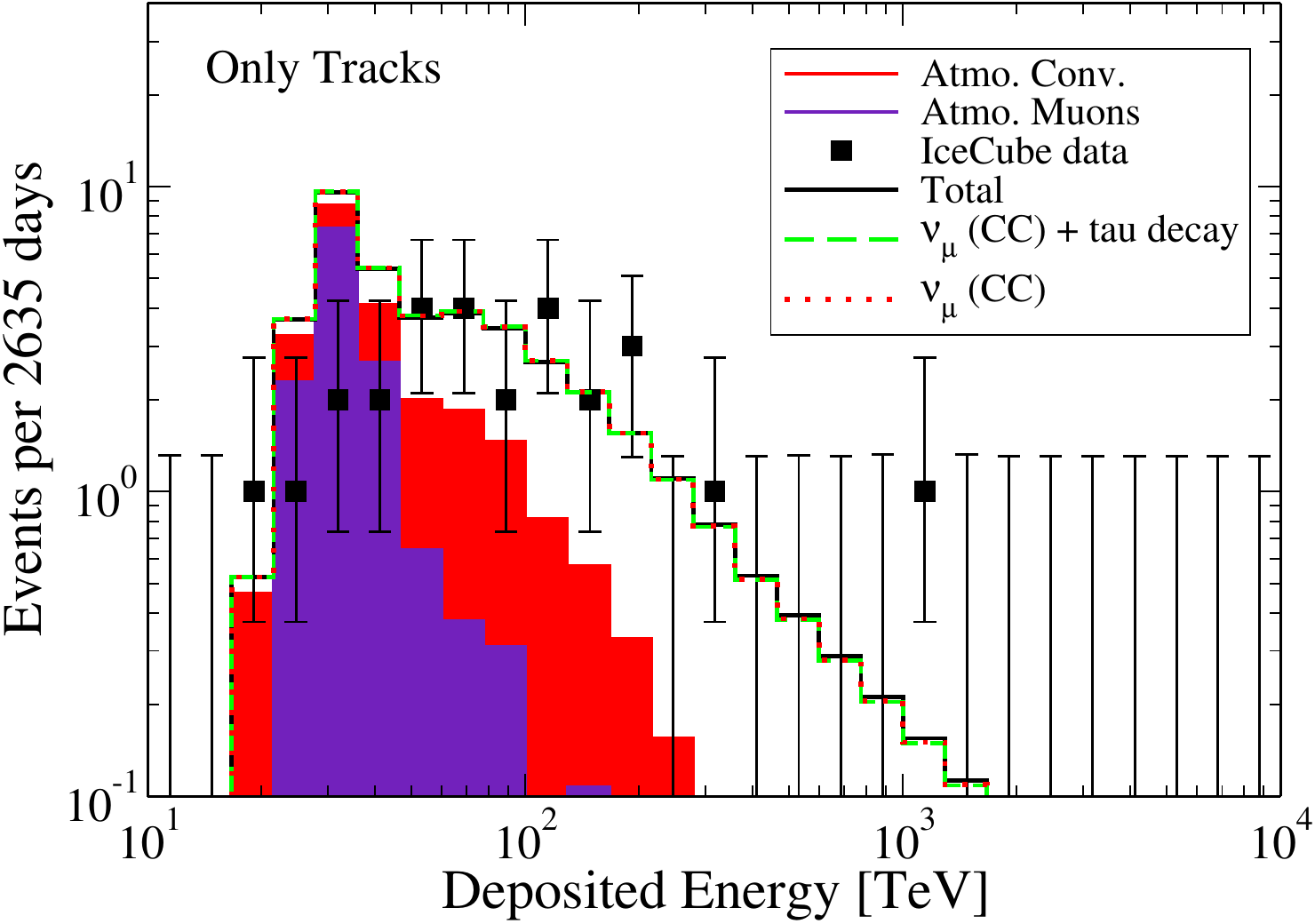}
	\end{tabular}
\caption{ Track events at IceCube for 7.5 years of HESE data for tracks \cite{IceCube:2020wum}. We obtain the histograms considering different sets of processes for the events characterized as tracks. }
\label{fig_2:events}
\end{figure}

Finally, we have investigated the impact of inserting subleading channels on the determination of astrophysical neutrino flux parameters. In Figure \ref{fig_2:events} we present our best fit curve for 7.5 years of IceCube data for HESE track events \cite{IceCube:2020wum}. Atmospheric neutrino and muon backgrounds were obtained from \cite{Schneider:2020tyd}. We constructed the best fit curves considering three distinct cases: Track events arising only from muon neutrino interactions; arising from muon neutrino interactions plus taus decays produced in charged current interactions; and considering all the channels presented earlier in the text. The best curve fits were obtained by maximizing likelihood with Poisson statistics, given the low count of events available so far, following the same procedure described in the previous section for fits with different models for the Earth's structure.

In Table \ref{table_2:parameters} we show the obtained values of $\chi^{2}_{\mathrm{min}}$ and the parameters of the best fit for the cases considered. We see that the spectral index of the flux is not greatly altered when we insert the subdominant channels, given that the cross section behavior of these additional channels is quite similar to the behavior for the dominant channel as a function of neutrino energy. However, the flux normalization factor drops significantly when we insert all the channels studied here.

\begin{center}
	\begin{table}
	\centering
		\begin{tabular}{|c|c|c|c|c|c|c|}
			\hline
			\hline 
		          	&  $\gamma$    &  $\Phi_0$  &   $\chi^{2}_{\mathrm{min}}$   \\
			\hline 	
			\hline 
Total                   &  2.56        &  1.16      &  9.49                \tabularnewline
			\hline 
$\nu_\mu$ + tau decay   &  2.55        &  1.26      &  9.49                \tabularnewline
			\hline 
$\nu_\mu$               &  2.55        &  1.45      &  9.49                \tabularnewline
			\hline 
			\hline 			
		\end{tabular}
		\caption{ Best fit parameters for the astrophysical neutrino flux obtained by considering different sets of processes for events characterized as tracks. }
		\label{table_2:parameters}
	\end{table}
\end{center}

\section{Impacts of the $L_{\mu} - L_{\tau}$ theory on current IceCube data and perspectives for IceCube-Gen2}
\label{sec_Ice:Zprime}

One of the main goals of the IceCube neutrino observatory is the discovery and characterization of the astrophysical neutrino flux. This goal was achieved in less than three years of data collection, with a result that described an approximately isotropic high-energy neutrino flux that can be described by a power law of the type $\phi_\nu \propto \Phi_0\, E_{\nu}^{-\gamma}$. With the increase in the observatory's exposure time, other important advances were made, such as the measurement of the neutrino-target cross section \cite{IceCube:2017roe,Bustamante:2017xuy,IceCube:2020rnc}, the identification of the first sources of these neutrinos \cite{IceCube:2018dnn,IceCube:2022der,IceCube:2023ame}, and, more recently, the description of the astrophysical neutrino flux with a broken power law \cite{IceCube:2025tgp,IceCube:2025ewu}.

Beyond the discoveries and measurements of Standard Model processes, IceCube's experimental results have motivated the search for various processes beyond the Standard Model in the production \cite{IceCube:2018tkk,Berlin:2024lwe,IceCube:2025fcu}, propagation \cite{Shoemaker:2015qul,Bustamante:2020mep,Agarwalla:2023sng,KA:2023dyz,Wang:2025qap,Esteban:2021tub} and interaction of these neutrinos on Earth \cite{IceCube:2024yaw}. One of the major interests has been the search for resonances that can be produced by astrophysical neutrinos with cosmic neutrino background, which would result in valleys in the neutrino flux in certain regions of its spectrum.

One of the simplest scenarios beyond the Standard Model that would have a major impact on the astrophysical neutrino flux is the $L_\mu - L_\tau$ theory \cite{He:1990pn,He:1991qd}. This model is constructed by assuming the difference between the leptonic number associated with the muon family, $L_\mu$, and the tau family, $L_\tau$, as a local symmetry. From this new symmetry arises a new massive neutral gauge boson, which we will call $Z'$. This new boson couples only with second- and third-generation leptons, and its associated interaction Lagrangian is written as
\begin{eqnarray}
    \mathcal{L} = g' Z_{\alpha}' [
    (\bar{\mu} \gamma^{\alpha} \mu) - 
    (\bar{\tau} \gamma^{\alpha} \tau) + 
    (\bar{\nu}_{\mu} \gamma^{\alpha} P_{L} \nu_{\mu}) - 
    (\bar{\nu}_{\tau} \gamma^{\alpha} P_{L} \nu_{\tau})
    ] \, ,
    \label{eq_Ice:lagrangianLmuLtau}
\end{eqnarray}
where $g'$ is the coupling of the new boson. The $L_\mu - L_\tau$ theory has been studied in various contexts and experiments in particle physics, and, in particular, has emerged as a possible explanation for the discrepancy in measurements of the anomalous magnetic moment of the muon $(g-2)_\mu$ existing at the time \cite{Keshavarzi:2018mgv}.

Several studies have highlighted the possibility of using IceCube data to probe the $L_\mu - L_\tau$ model \cite{Kamada:2015era,Araki:2015mya,DiFranzo:2015qea,Carpio:2021jhu,Hooper:2023fqn,delaVega:2024pbk}. In particular, these studies used 988 days of data collection from all topologies \cite{Kamada:2015era,Araki:2015mya,DiFranzo:2015qea} and six years of data collection from cascades \cite{Carpio:2021jhu,Hooper:2023fqn,delaVega:2024pbk}. In our study, we will extend these previous analyses to 12 years of observational data from HESE \cite{IceCube:2023sov}. Furthermore, we will extend the study to the sensitivity of IceCube-Gen2 to this model, also considering 12 years of data collection.

\subsection{Astrophysical neutrino flux in the presence of the $Z'$ boson}
\label{subsec_Ice:ZprimeFlux}

The propagation of astrophysical neutrinos through the intergalactic medium can be obtained from the Equation~(\ref{eq:fluxonu}), modifying the differential neutrino flux with respect to the matter column $X$ for the redshift $z$, then we obtain the differential flux from the set of differential equations \cite{Bhattacharjee:1999mup,Hooper:2023fqn}
\begin{eqnarray}
\begin{aligned}
& -(1+z)\frac{H(z)}{c}\frac{\mathrm{d}\phi_i}{\mathrm{d}z} = \\
& J_i(E_0,z)
- \phi_i\sum_{j} \langle n_{\nu_i}(z)\sigma_{ij}(E_0,z) \rangle
+ P_i\int_{E_0}^{\infty} \mathrm{d}E'\sum_{j,k} \phi_k 
\left\langle
n_{\nu_j}(z)\frac{\mathrm{d}\sigma_{kj}}{\mathrm{d}E_0}(E',z)
\right\rangle \, ,
    \label{eq_Ice:spectrum}
\end{aligned}
\end{eqnarray} 
where $P_i = \sum_{l} \mathrm{Br}(Z'\rightarrow \nu_l \nu_i)$, $n_{\nu_j}$ is the cosmic neutrino background density of mass eigenstate $j$, and $E_0$ is the neutrino energy measured on the Earth. The neutrino energy at the source is corrected for cosmological redshift and can be written in terms of $E_0$ as $E_\nu = E_0 (1+z)$. In our study, we are considering the cosmic neutrino background density $n_{\nu} = 336$ per cm$^{3}$, equally distributed between neutrinos and antineutrinos, as well as equally distributed among the three neutrino flavors. $H(z)$ is the Hubble rate, and in the $z$ range of our interest here it can be estimated by considering a universe dominated by dark matter and dark energy, which implies that
\begin{eqnarray}
    H(z) \simeq H_0 
    \sqrt{\Omega_\Lambda + \Omega_M(1+z)^{3}}\, ,
    \label{eq_Ice:hubble}
\end{eqnarray}
where $H_0$ is the current value of the Hubble rate, and we are assuming the approximation where $\Omega_\Lambda \approx 1-\Omega_M$. For the quantities present in Equation~(\ref{eq_Ice:hubble}), we will assume the best fit parameters obtained by the Planck collaboration \cite{Planck:2018vyg} ($H_0 = 67.4$ km s$^{-1}$ Mpc$^{-1}$ and $\Omega_M = 0.315$).

\begin{figure}[t]
	\centering
	\begin{tabular}{ccc}
	\includegraphics[width=0.7\textwidth]{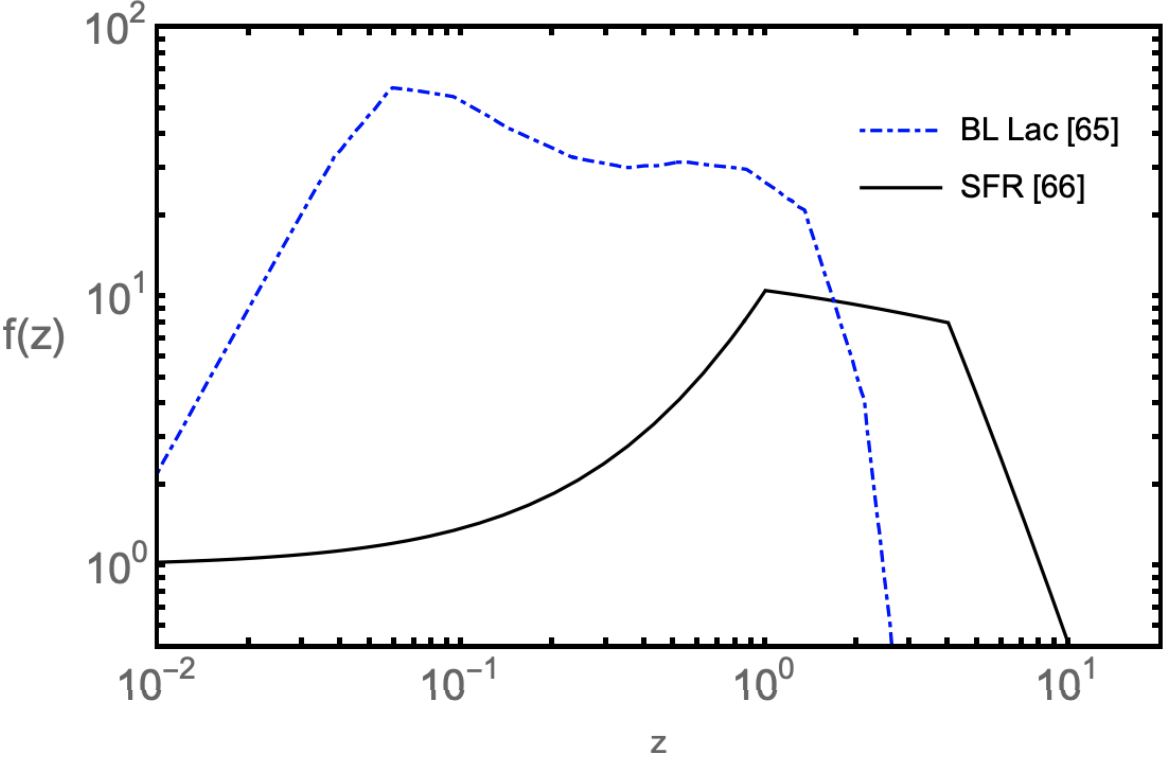}
			\end{tabular}
\caption{ Redshift distribution of astrophysical sources of high-energy neutrinos from star formation rate models (solid black line) and BL Lacertae objects (dashed-dashed blue line). Figure taken from reference \cite{Hooper:2023fqn}. }
\label{fig_Ice:fontes_astro}
\end{figure}

The term $J_i(E_0,z)$ in Equation (\ref{eq_Ice:spectrum}) is the astrophysical neutrino production term. Usually, $J_i$ is parameterized as the product of the power law of the astrophysical neutrino spectrum with the redshift distribution of the neutrino sources, as follows:
\begin{eqnarray}
    J_i(E_0,z) \propto \Phi_0 
    E_0^{-\gamma} f(z)\, .
    \label{eq:J}
\end{eqnarray}
In our study, we will consider two distinct cases for the redshift distribution of astrophysical neutrino sources: the distribution of the star formation rate (SFR) model \cite{Yuksel:2008cu}, and the distribution of BL Lacertae (BLL) objects \cite{Ajello:2013lka}, shown in Figure \ref{fig_Ice:fontes_astro}. The SFR model predicts a redshift distribution with a maximum at approximately $z\approx 1$, while the BLL distribution, which consists of a distribution of a subpopulation of blazars, with relatively closer sources, has a maximum in the distribution at $z\approx 0.06$.

The last two terms on the right side of Equation~(\ref{eq_Ice:spectrum}) describe the attenuation (penultimate term) and regeneration (last term) of the astrophysical neutrino flux when it interacts (penultimate term) in a medium where interactions play a significant role. In this study, we are interested in the scenario where astrophysical neutrinos can annihilate with a neutrino of cosmic background in the $Z'$ boson of the $L_\mu - L_\tau$ theory, and subsequently this boson decays into two neutrinos. The neutrino-antineutrino annihilation cross section, present in Equation~(\ref{eq_Ice:spectrum}), in a $Z'$ boson for neutrinos of mass eigenstates $i$ and $j$, is given by \cite{Hooper:2023fqn}
\begin{eqnarray}
    \sigma_{ij} = 
    \sigma(\nu_i\bar{\nu}_j \rightarrow \nu\bar{\nu}) = 
    \frac{2g'^{4}s(U_{\mu i}^{\dagger}U_{\mu j} - U_{\tau i}^{\dagger}U_{\tau j})^{2}}{3\pi [(s-m_{Z'}^{2})^{2}+m_{Z'}^{2}\Gamma_{Z'}^{2}]} \, ,
    \label{eq_Ice:csZprime}
\end{eqnarray}
where $U_{\alpha i}$ is the Pontecorvo-Maki-Nakagawa-Sakata (PMNS) matrix, and $s = 2E_\nu m_{j}$, with $m_j$ being the mass of the target neutrino. For the PMNS matrix elements, we are using the NuFIT-6.0 parameters \cite{Esteban:2024eli}. Previous studies \cite{Kamada:2015era,Araki:2015mya,DiFranzo:2015qea,Hooper:2023fqn,delaVega:2024pbk} demonstrated that IceCube can be sensitive to a low-mass $Z'$ boson ($\approx 1 - 10$ MeV). In this case, the boson produced does not have enough mass to decay into a pair of charged leptons of the second or third generation of fermions; therefore, its only permitted decay would be into neutrino-antineutrino pairs, with a decay width given by $\Gamma_{Z'} = g'^{2 }m_{Z'}/12\pi$. Some recent works also consider the $Z'$ boson of the $L_\mu - L_\tau$ theory as a possible mediating particle between Standard Model fermions and dark matter particles. There are phenomenological consequences associated with the neutrino regeneration term of Equation~(\ref{eq_Ice:spectrum}), but we will not explore these aspects in our study.

The average temperature present in the Equation (\ref{eq_Ice:spectrum}) represents
\begin{eqnarray}
    \langle n_{\nu_i}(z)\sigma_{ij}(E_0,z) \rangle = 
    \int \frac{\mathrm{d}^3\vec{p}}{(2\pi)^{3}}
    \frac{\sigma(E_0, z, \vec{p})}{\mathrm{e}^{|\vec{p}|/T_0(1+z)}+1} \; ,
    \label{eq:termalAverage}
\end{eqnarray}
with $T_0 \approx 1.95$~K being the temperature of the cosmic neutrino background when $z=0$. The cross section $\sigma(E_0, z, \vec{p})$ is obtained using Equation (\ref{eq_Ice:csZprime}), but considering the neutrino with momentum $\vec{p}$. In the case where target neutrinos have masses much larger than their momenta, we can assume that $\langle n_{\nu_i}(z)\sigma_{ij}(E_0,z) \rangle = n_{\nu_j} \sigma_{ij}$. In our study, we will assume this approximation, except when the lightest neutrino is considered massless. Currently, it is possible to obtain the difference between the masses of the different mass eigenstates of neutrinos using data from experiments focused on the neutrino oscillation mechanism. We will assume the values from reference \cite{Esteban:2024eli} for these quantities. Regarding the neutrino mass eigenvalues, we will consider two cases in particular: (a) that the neutrino of the lightest mass eigenstate is massless. This case will be labeled as inferior; and (b) that the sum of the masses associated with the three mass eigenstates is 0.12 eV, which is the maximum value allowed by observations from the Planck collaboration with 95\% confidence level \cite{Planck:2018vyg}. This case will be denoted hereafter as superior.

Once the $Z'$ boson is produced with the mass in the regime of our interest, it decays into neutrinos. The differential cross section $\frac{\mathrm{d}\sigma_{kj}}{\mathrm{d}E_0}(E',z)$ of Equation~(\ref{eq_Ice:spectrum}) gives us the distribution of neutrinos produced at an energy $E_0$ from the decay of the $Z'$ boson produced by an astrophysical neutrino of energy $E'$. This cross section can be obtained by modifying Equation~(\ref{eq_Ice:csZprime}) as \cite{Hooper:2023fqn}
\begin{eqnarray}
    \frac{\mathrm{d}\sigma_{kj}}{\mathrm{d}E_0}(E',z) = 
    \sigma_{kj} (E',z) f(E', E_0) \; ,
    \label{eq:distCS}
\end{eqnarray}
where
\begin{eqnarray}
    f(E',E_0) = \frac{3}{E'}
    \left[
    \left( \frac{E_0}{E'} \right)^{2}
    + \left( 1-\frac{E_0}{E'} \right)^{2}
    \right] \Theta(E' - E_0) \; .
    \label{eq_Ice:fZprime}
\end{eqnarray}
We will use this approximation in our calculations.

\begin{figure}[t]
	\centering
	\begin{tabular}{ccc}
	\includegraphics[width=0.325\textwidth]{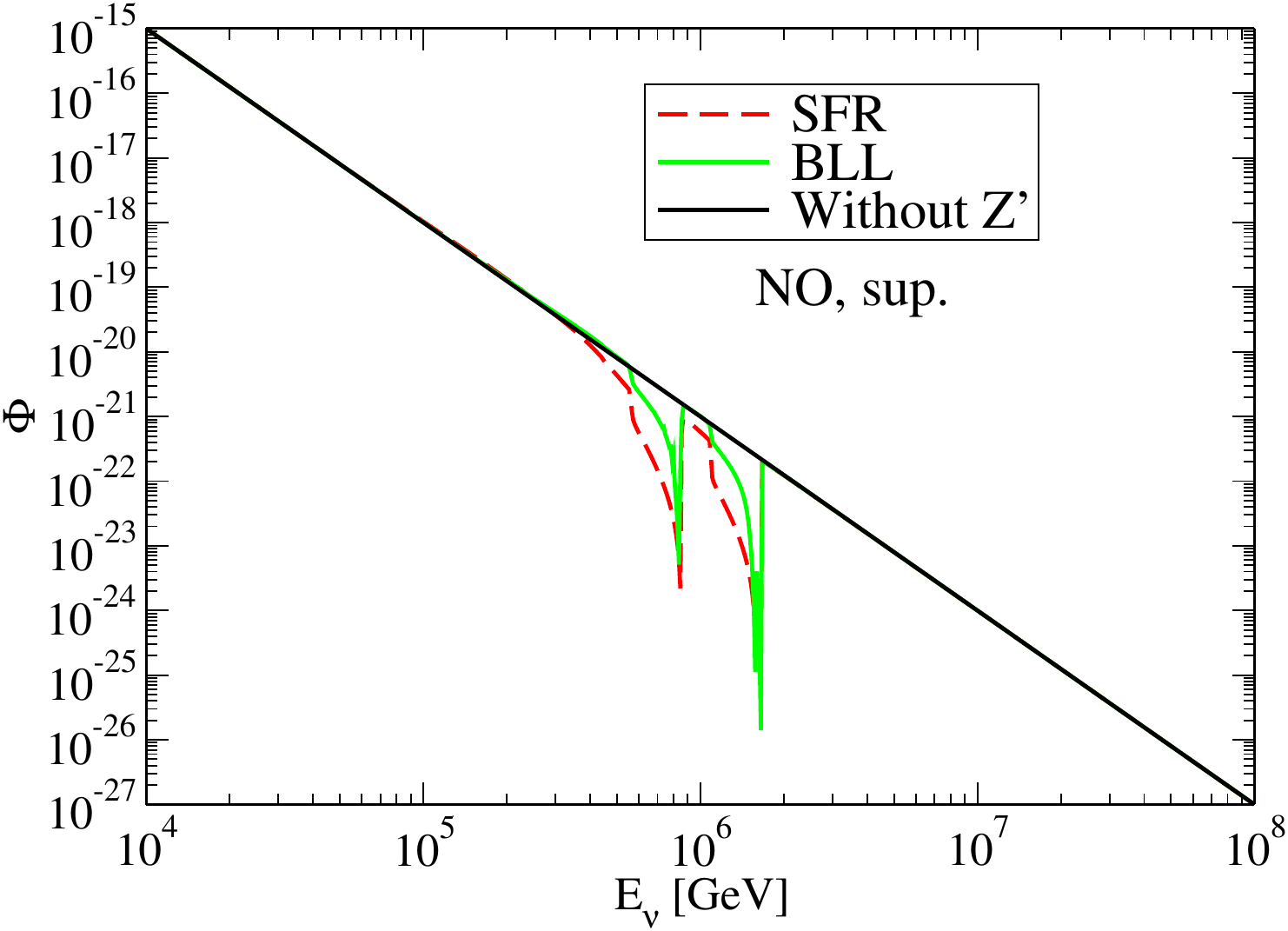}
	\includegraphics[width=0.325\textwidth]{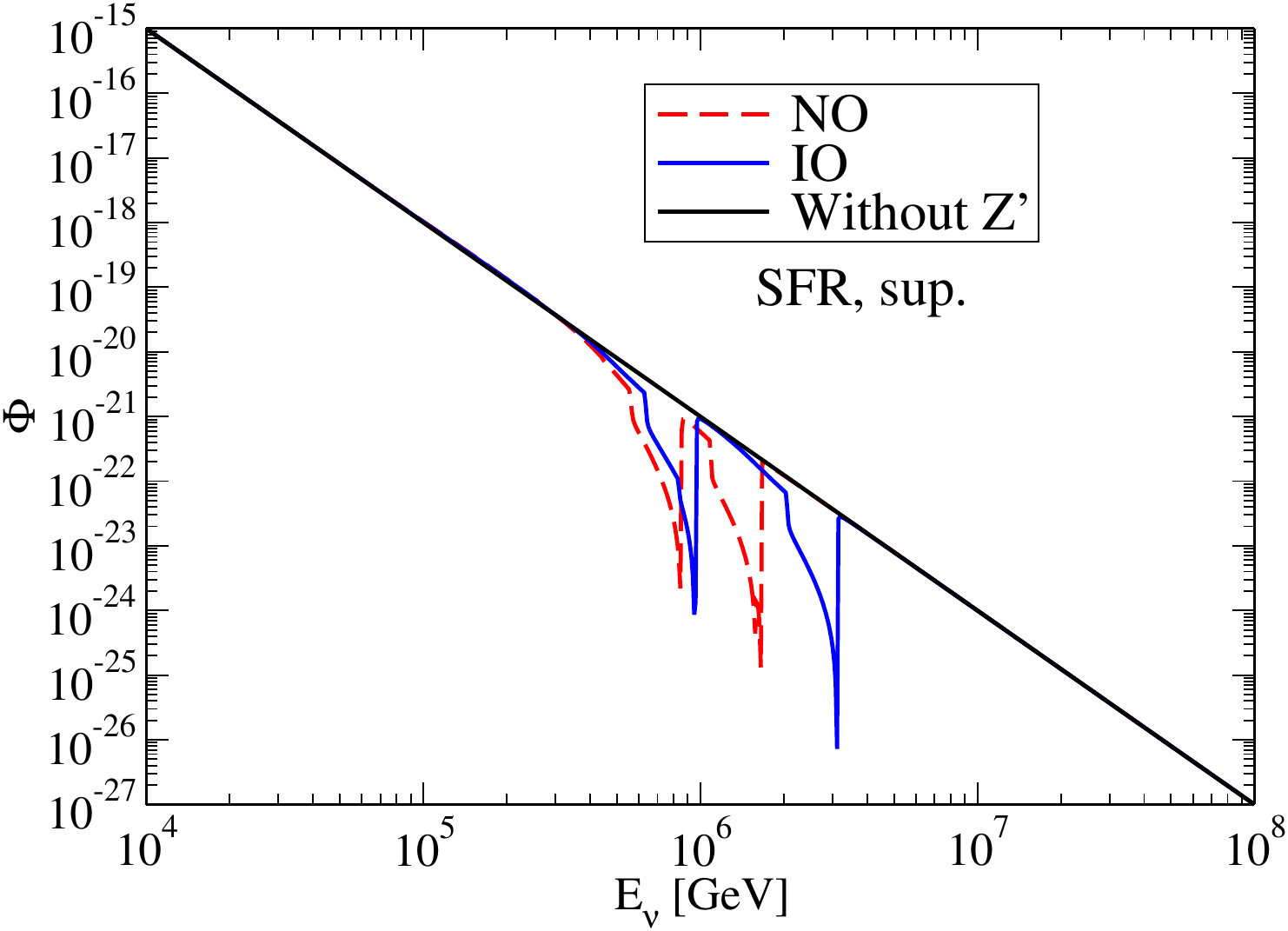} 
	\includegraphics[width=0.325\textwidth]{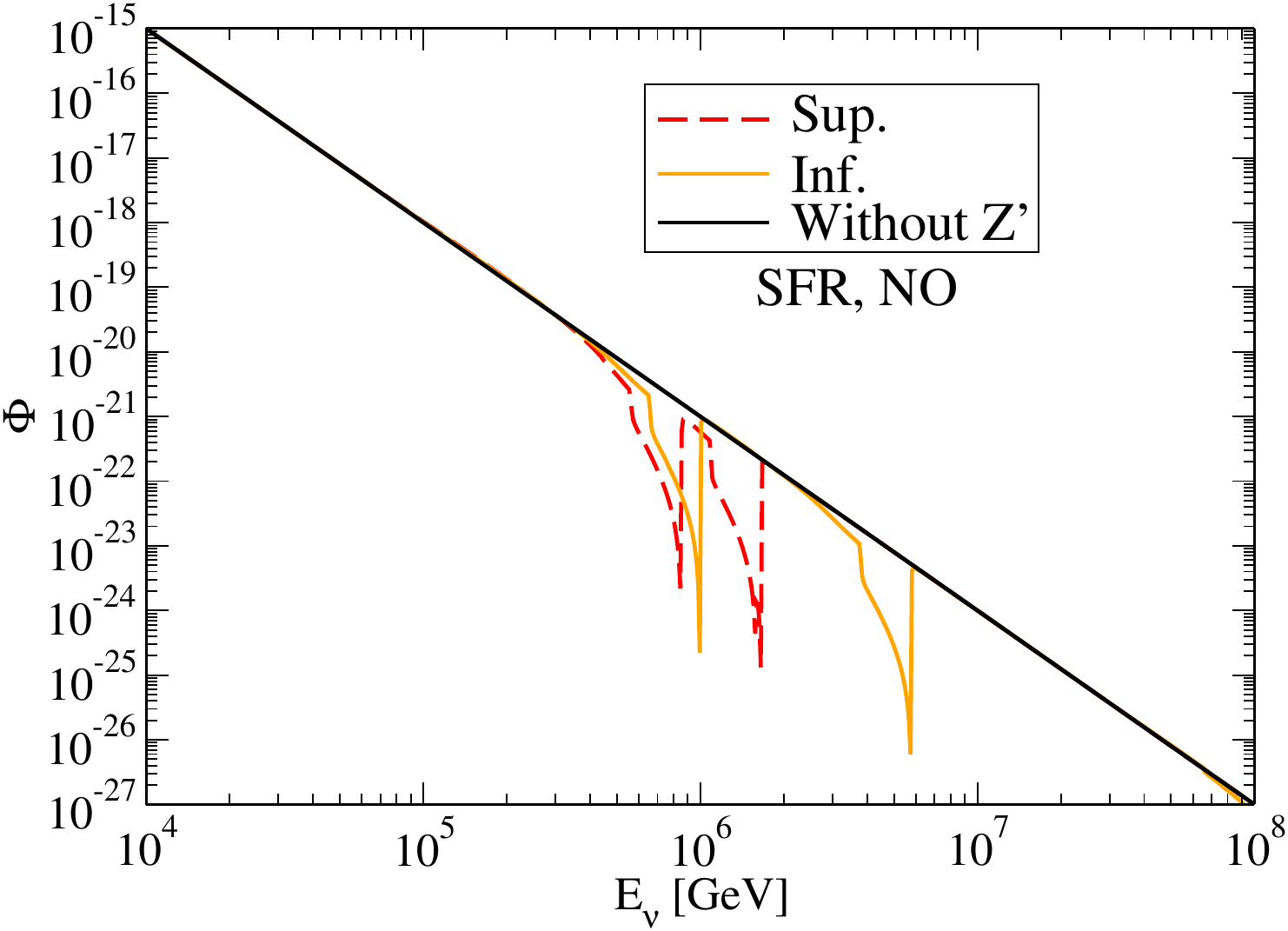} 
			\end{tabular}
\caption{ Neutrino flux as a function of energy observed on Earth, considering a boson $Z'$ with $m_{Z'} = 10$ MeV and spectral index $\gamma = 3.0$. {\bf Left:} Comparison between SFR and BLL redshift distributions for neutrino sources, assuming normal ordering and a superior limit for the sum of masses. {\bf Center}: Comparison between normal ordering (NO) and inverted ordering (IO), assuming the SFR redshift distribution and a superior limit for the sum of masses. {\bf Right:} Comparison between superior and inferior limits for the sum of neutrino masses, assuming the SFR redshift distribution for neutrino sources and normal ordering of masses. }
\label{fig_Ice:fluxZprime}
\end{figure}

As our first result in this section and initial motivation, we present the Figure~\ref{fig_Ice:fluxZprime}. In it, we present the astrophysical neutrino flux obtained from the solution of Equation~(\ref{eq_Ice:spectrum}) following a power law with spectral index $\gamma = $3.0, assuming a boson $Z'$ with a mass of 10~MeV and different configurations for the astrophysical neutrino sources, for the sum of the masses of the eigenstates, and for the ordering of the neutrino masses. On the left, we compare the neutrino flux assuming two distinct distributions for the astrophysical neutrino sources: SFR and BLL, in addition to assuming normal ordering and an superior limit for the sum of the masses. Given that the sources in the SFR model are more distant, the neutrinos propagate greater distances before reaching Earth. Consequently, the cosmological redshift of their neutrinos is greater, and the valleys in the spectrum are wider than in the case of the BLL source. In the central figure, we compare the attenuation in the flux considering normal and inverted ordering for the neutrino mass eigenstates, in addition to assuming SFR model as a source and superior limit for masses. In the case of normal ordering, the two smallest masses are close together, and in the case of inverted ordering, the two largest masses are close together. This effectively results in two valleys in the neutrino flux for both scenarios, as the close masses produce close valleys that merge. Finally, in the figure on the right, we compare the neutrino flux assuming superior and inferior limits for the sum of the masses, as well as SFR for neutrino sources and normal mass ordering. When we take the inferior limit of the masses, the lightest neutrino is massless, and its impact on the flux occurs at energies beyond those shown in the figure, in addition to making the relative distance between the masses of the two massive eigenstates larger, which further separates the valleys produced in the flux.

Figure~\ref{fig_Ice:fluxZprime} shows that the existence of a gauge boson $Z'$ from the $L_\mu - L_\tau$ theory with a mass of 10 MeV would impact the description of the high-energy neutrino flux in the range that is currently observed by IceCube and will be explored by IceCube-Gen2, as well as other future high and ultra-high-energy neutrino observatories.

\subsection{Impact of the $Z'$ boson on events observed at IceCube}
\label{subsec_Ice:ZprimeEvents}

\begin{figure}[t]
	\centering
	\begin{tabular}{ccc}
	\includegraphics[width=0.49\textwidth]{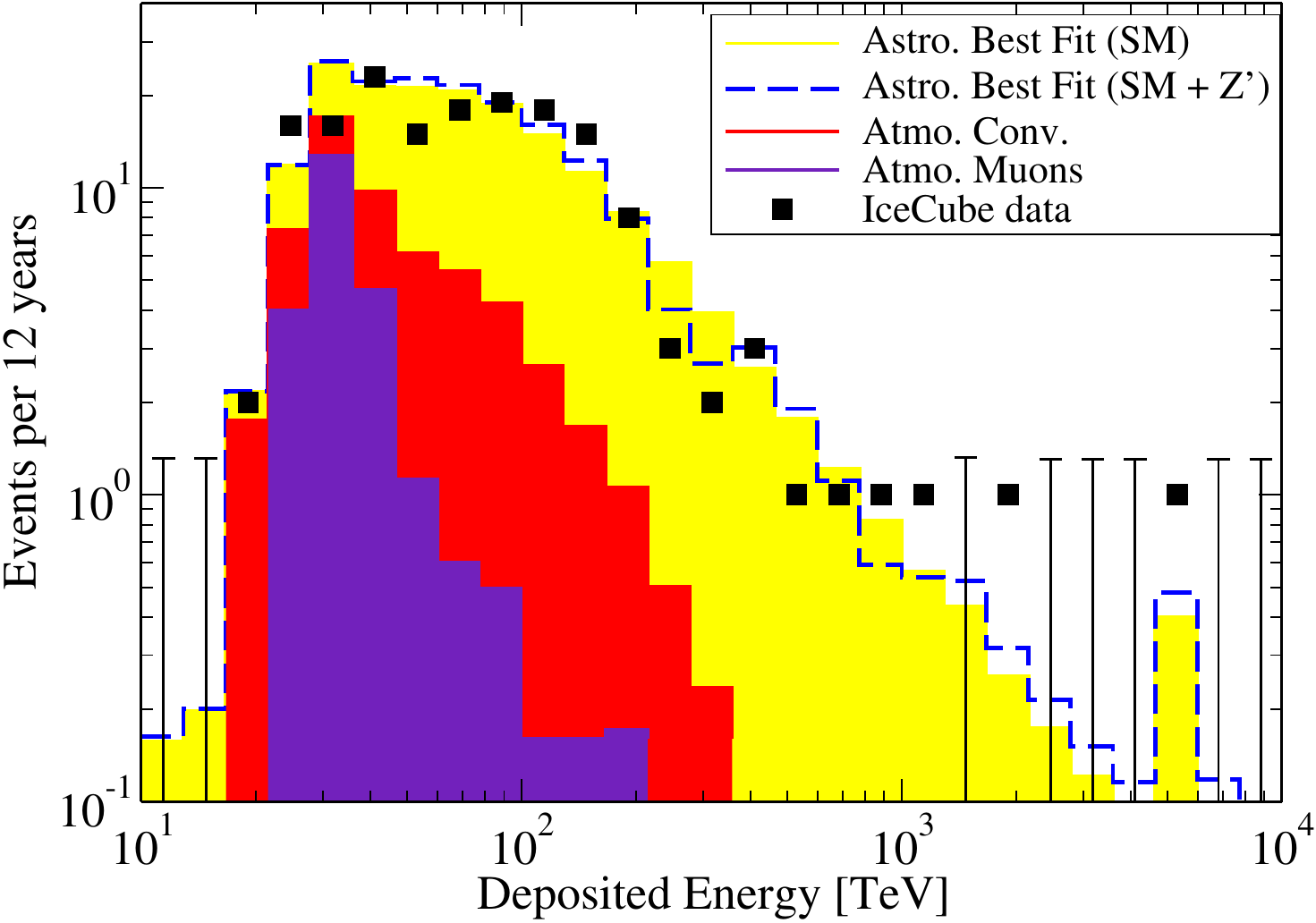}
	\includegraphics[width=0.49\textwidth]{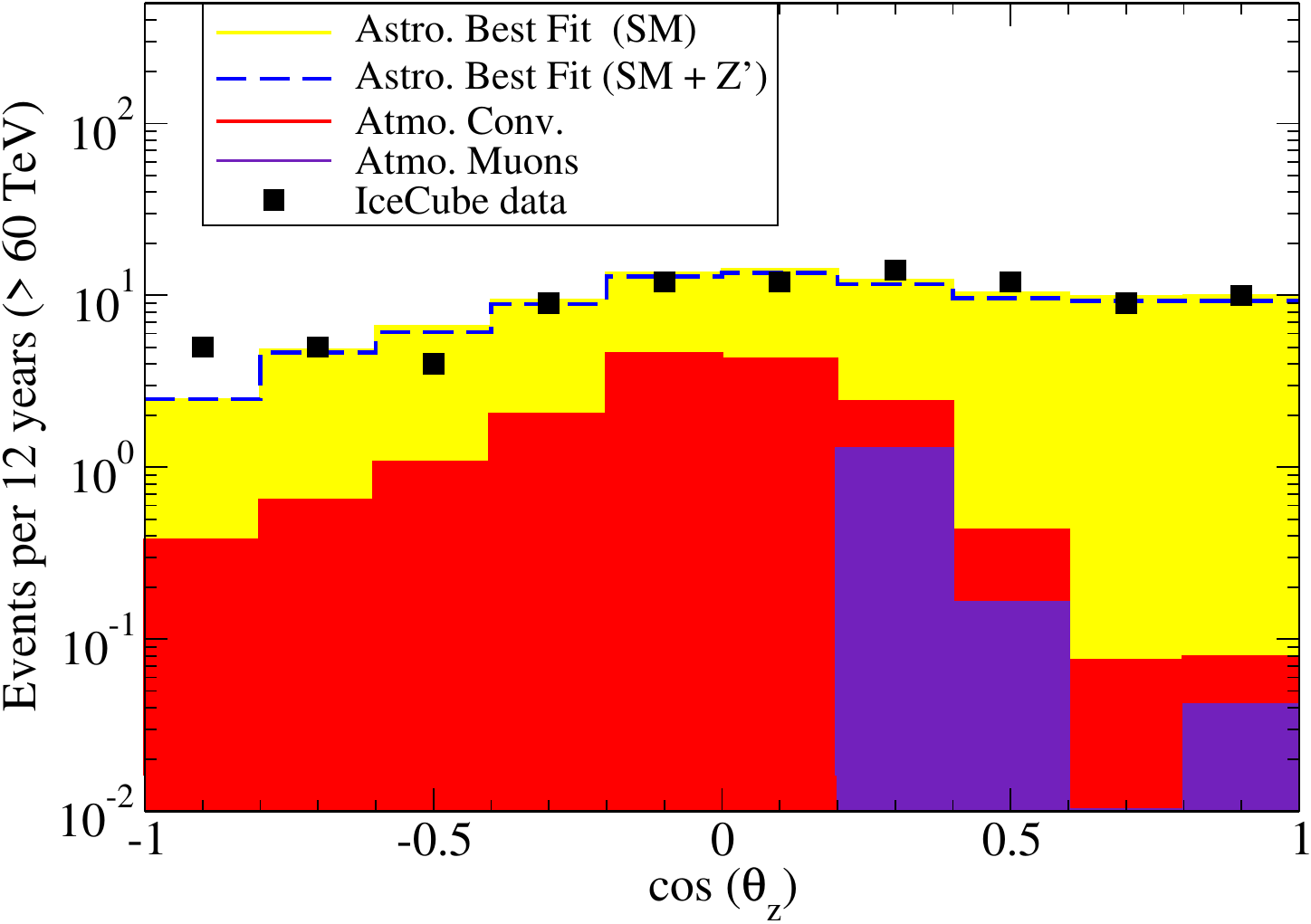} 
			\end{tabular}
\caption{ Number of HESE events at the IceCube observatory as a function of deposited energy (left) and azimuthal angle (right) for 12 years of data collection. Each color in the histogram specifies different contributions to the events, according to the best fit parameters: astrophysical neutrinos (yellow), atmospheric neutrinos (red), and atmospheric muons (purple). In the dashed blue line, we present our results for the best fit obtained considering a $Z'$ boson with a mass of 6 MeV, inverted ordering, BLL for the redshift distribution of neutrino sources, and a superior limit for the sum of masses. }
\label{fig_Ice:eventsZprime}
\end{figure}

One of the main goals of our study is to estimate the impact of the $Z'$ gauge boson on the neutrino flux and consequently on the astrophysical neutrino events observed at IceCube. In this study, we use new HESE data, now for 12 years of observation, which, in addition to the 4.5 additional years of data collection, recalibrated the reconstructed energies and directions with new a software \cite{IceCube:2023sov}. Initially, we estimated the parameters of the astrophysical neutrino flux assuming the power law for its spectrum, without inserting effects of physics beyond the Standard Model. We used the same method described in Section \ref{sec_Ice:trans}, when investigating the effects of the Earth's structure, assuming the same backgrounds of atmospheric neutrinos and muons. We again assume the cross section calculated at leading order with the PDFs parameterized by CT14 \cite{Dulat:2015mca}. For the best fit of event histogram, using only events with deposited energy greater than 60~TeV, we obtain $\Phi_0 = 1.33$ and $\gamma = 2.93$. The $\chi^{2}$ obtained was 9.54 via Wilks' theorem \cite{Wilks:1938dza}. In Figure~\ref{fig_Ice:eventsZprime} we show the number of HESE events as a function of the reconstructed energy deposited in the instrumented volume of the observatory (left) and the reconstructed direction of the incident neutrino (right). Along with IceCube data, we show the atmospheric neutrino and muon backgrounds, as well as our best fit curve for astrophysical neutrino events, in the yellow histogram.

Let's now consider the cases with the presence of the $Z'$ boson, taking into account the different scenarios discussed for neutrino sources, ordering, and mass sum. For each combination of scenarios, there is at least one local minimum of $\chi^{2}$ with a minimum smaller than that obtained assuming the Standard Model for neutrino propagation until they reach Earth. In Table \ref{table_Ice:parametersZprime}, we show the flux parameters for the best fit and the corresponding $\chi_{\mathrm{min}}^{2}$ for the Standard Model and for the combinations of different scenarios with $Z'$. Among all the scenarios tested, the one with the smallest $\chi^{2}$ is obtained using inverted ordering with an upper mass limit and a neutrino source following the BLL redshift distribution, in addition to a $Z'$ boson with a mass of 6~MeV and a coupling constant greater than $10^{-4}$. We have included the curve for the number of events in this scenario in Figure~\ref{fig_Ice:eventsZprime} in dashed blue line. In particular, this scenario better describes the data because it is sensitive to the decrease in observed events with deposited energies close to 300 TeV.

\begin{table}
\begin{center}
\begin{tabular}{|c|c|c|c|c|c|c|c|}
\hline
Setup & $m_{Z'}$ (MeV) &$ {\cal{X}}^{2}_{\mathrm{min}}$ & $\gamma$ & $\Phi_{0}$ & $k$ & $AIC$ &  $AIC_{c}$   \\
\hline
\hline
Standard Model & -- &  9.54 & 2.93 & 1.33 & 2 & 13.54 & 13.99 \\
\hline
\hline
 NO, inf, BLL & 6 & 7.51 & 2.90 & 1.44 & 4 & 15.51& 17.03  \\
\hline
NO, inf, SFR  & 6 & 8.06 & 2.90 & 1.66  & 4 & 16.06 & 17.66  \\
\hline
NO, sup, BLL  & 7 & 7.43 & 2.80 & 1.40 & 4 & 15.43 &  17.11 \\
\hline
NO, sup, SFR  & 7 & 8.54 & 2.80 & 1.61 & 4 & 16.54 &  18.14  \\
\hline
IO, inf, BLL  & 6 & 7.75 & 2.90 & 1.50 & 4 & 15.75 &  17.35 \\
\hline
IO, inf, SFR  & 9  & 8.57 & 2.80 & 1.28 & 4 & 16.57 &  18.17 \\
\hline
IO, sup, BLL  & 6 & 6.76 & 2.90 & 1.43 &  4 & 14.76 &   16.36 \\
\hline
IO, sup, SFR  & 6 & 7.96 & 2.80 & 1.60 &  4 & 15.96 & 17.56  \\
\hline
\end{tabular}
\caption{ Best fit parameters obtained for IceCube data, considering the Standard Model and assuming a $Z'$ boson. We considered different combinations of redshift distribution, ordering, and mass limit. We applied the AIC and AICc criteria to each model, considering the histograms presented in Figure \ref{fig_Ice:eventsZprime}. }
\label{table_Ice:parametersZprime}
\end{center}
\end{table}

Our results presented above agree with those derived in references \cite{Carpio:2021jhu,Hooper:2023fqn,delaVega:2024pbk} using 6 years of cascade collection data at IceCube, which also indicated that $\chi^{2}$ decreases when a $Z'$ is inserted with masses between approximately 7 and 16 MeV \cite{Carpio:2021jhu}, 5 and 15 MeV \cite{Hooper:2023fqn}, and 6 and 8 MeV \cite{delaVega:2024pbk} for $g' > 10^{-4}$. Our results indicated that the region between 6 and 9 MeV remains favored with 12 years of HESE data at IceCube. However, it is important to emphasize that this decrease in $\chi^2$ does not necessarily indicate an improvement in the quality of the fit, since we are adding two new parameters to the analysis. To perform a quantitative comparison of the models, we will apply the Akaike Information Criterion (AIC), which rewards the quality of the fit (evaluated by the likelihood function), but also includes a penalty associated with the increase in the number of estimated parameters. The AIC criterion is defined as \cite{Akaike:1974vps}
\begin{equation}
\text{AIC} = 2k - 2\ln({\cal L}_{\mathrm{max}}) = 2k + {\cal X}^{2}_{\mathrm{min}},
\label{eq_Ice:AIC}
\end{equation}
where $k$ is the number of parameters in the model, and ${\cal L}_{\mathrm{max}}$ is the maximum likelihood (the ${\cal{X}}^{2}_{\mathrm{min}}$). Furthermore, given the low number of events in each bin in the histograms of Figure \ref{fig_Ice:eventsZprime}, the AIC criterion must take into account the low event statistics and must be corrected as
\begin{equation}
AIC_{c} = AIC + \frac{2k(k+1)}{n-k-1} ,
\label{eq_Ice:AICc}
\end{equation}
where $n$ is the number of independent data points used to calculate the likelihood (i.e., the sum of the number of bins in each histogram). Then, it is possible to construct the estimator.
\begin{equation}
\Delta AIC_{c} = AIC^{Z'}_{c} - AIC^{SM}_{c} \,\,.
\label{eq_Ice:Delta_AICc}
\end{equation}
Following the reference \cite{burnham2002model}, if $0 \le \Delta AIC_{c} \le 2$ the model is strongly supported, while for $2 < \Delta AIC_{c} \le 4$ the model is still plausible, but with less support. Finally, if $\Delta AIC_{c} > 4$ the considered model will probably not be the best. In Table \ref{table_Ice:parametersZprime} we also present our results for the quantities $AIC$ and $AIC_{c}$ for the Standard Model case, as well as for the other models considered. The case ``IO, sup, BLL'' results in a minimum value of ${\cal X}^{2}_{\mathrm{min}}$ among all models (including the Standard Model prediction) and the minimum value of $AIC^{Z'}_{c}$ in relation to the hypothesis $Z'$. However, compared to the Standard Model prediction, we obtain $\Delta AIC_{c} = 2.37$, which means that this case may be almost as good as the Standard Model, but not a better option. As a consequence, the improvement of ${\cal X}^2_{\mathrm{min}}$ is not sufficient to compensate for the two extra parameters, and a stronger conclusion about the presence of a gauge boson $Z'$ is not yet possible.

\subsection{Regions of the $Z'$ parameter space probed by IceCube}
\label{subsec_Ice:ZprimeMapa}

\begin{figure}
	\centering
	\begin{tabular}{ccc}
	\includegraphics[width=0.49\textwidth]{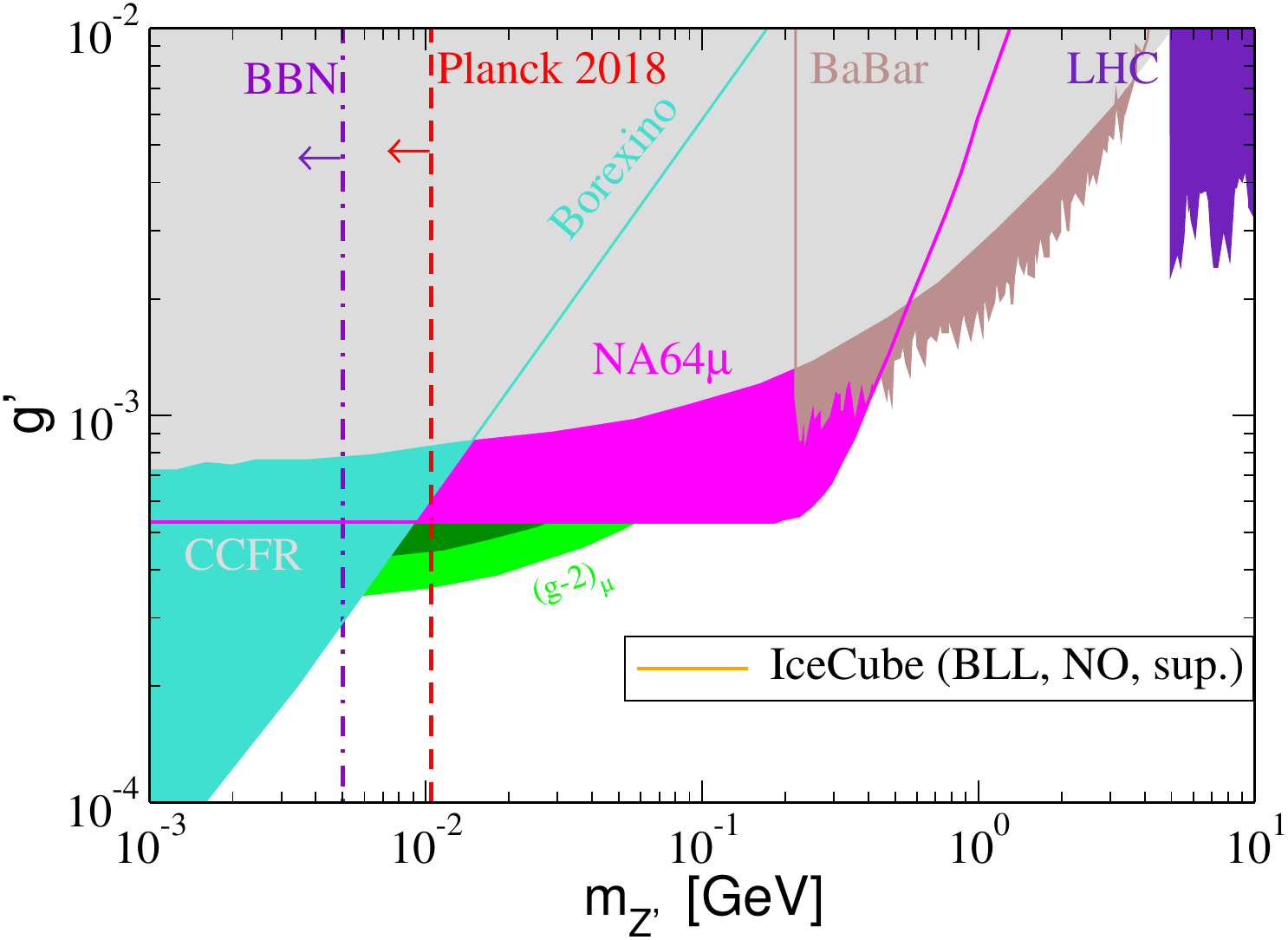}
	\includegraphics[width=0.49\textwidth]{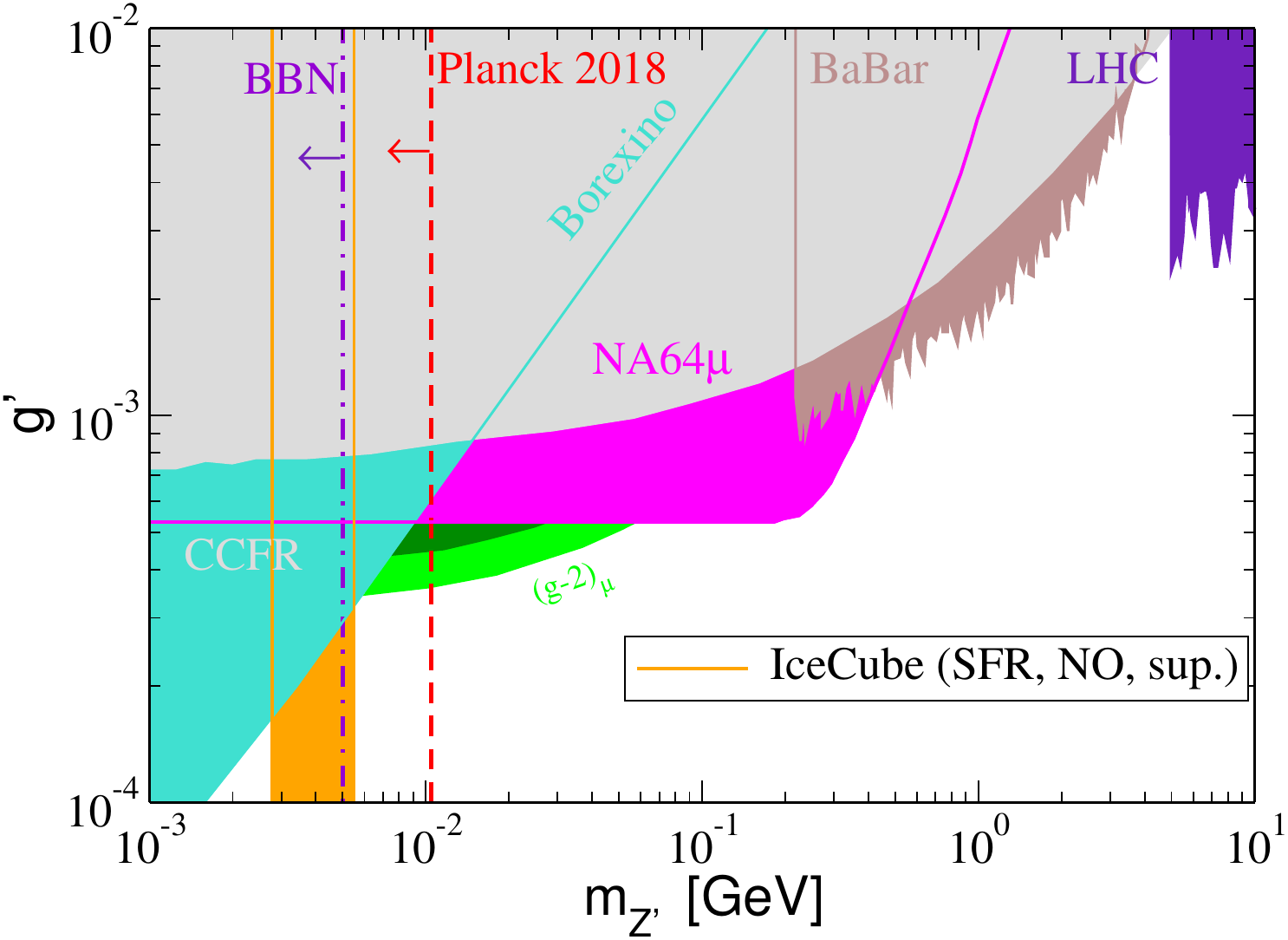} \\
    	\includegraphics[width=0.49\textwidth]{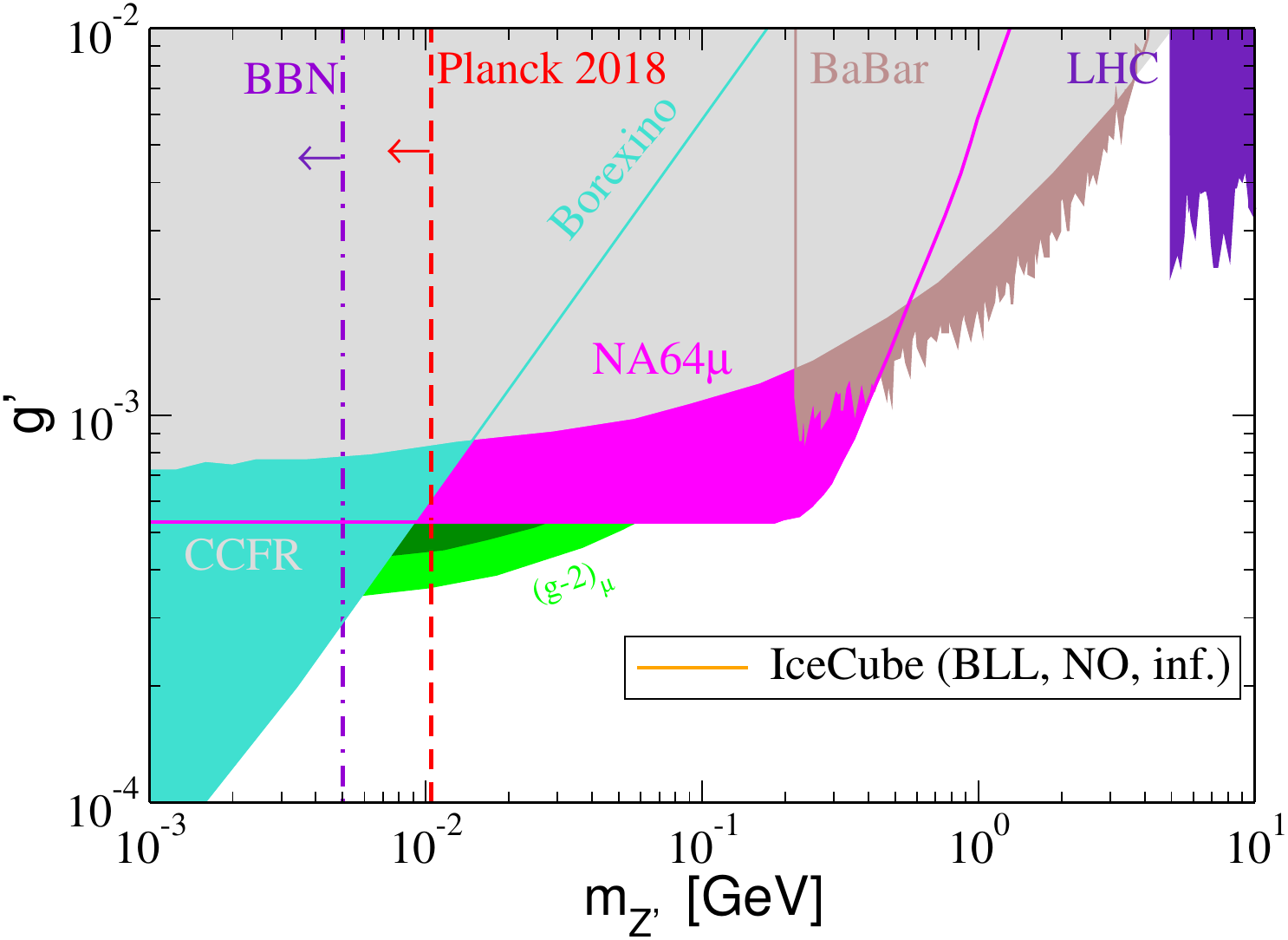}
	\includegraphics[width=0.49\textwidth]{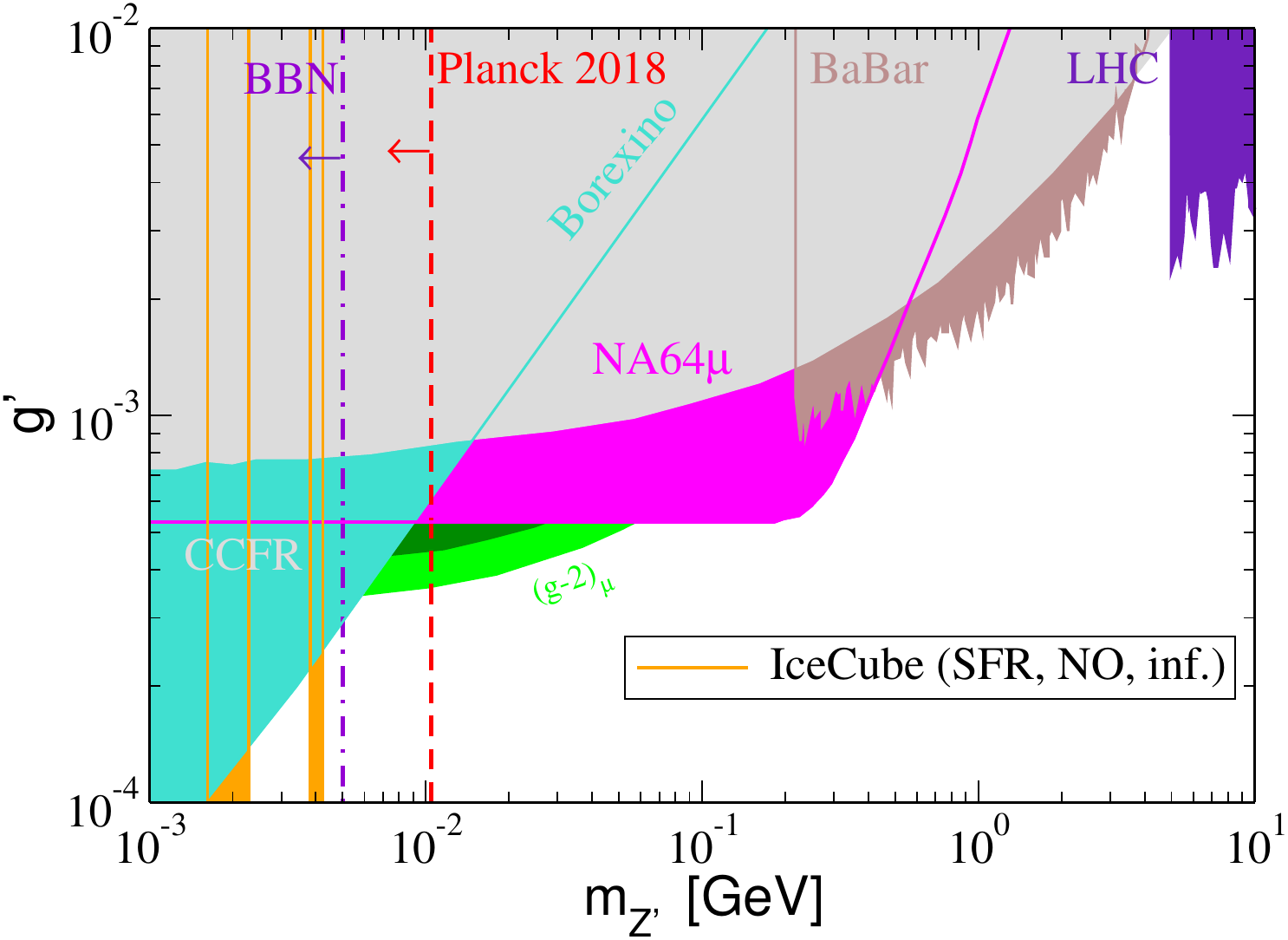}
    
			\end{tabular}
\caption{ Results for the sensitivity of HESE data at the IceCube Observatory to the astrophysical neutrino flux in the presence of the $Z'$ gauge boson predicted by the $L_\mu - L_\tau$ model, derived considering the normal ordering of neutrino masses. Predictions, at $2\sigma$ of statistical significance, were obtained considering the redshift distribution of BLL (left panels) and SFR (right panels), and the superior (upper panels) and inferior (lower panels) limits for the sum of neutrino masses. For comparison, existing constraints from other processes and experiments are also presented. The green bands represent the parameter space in which the presence of a $Z'$ resolves the muon magnetic moment anomaly with $1\sigma$ and $2\sigma$. }
\label{fig_Ice:spaceNO}
\end{figure}

\begin{figure}
	\centering
	\begin{tabular}{ccc}
	\includegraphics[width=0.49\textwidth]{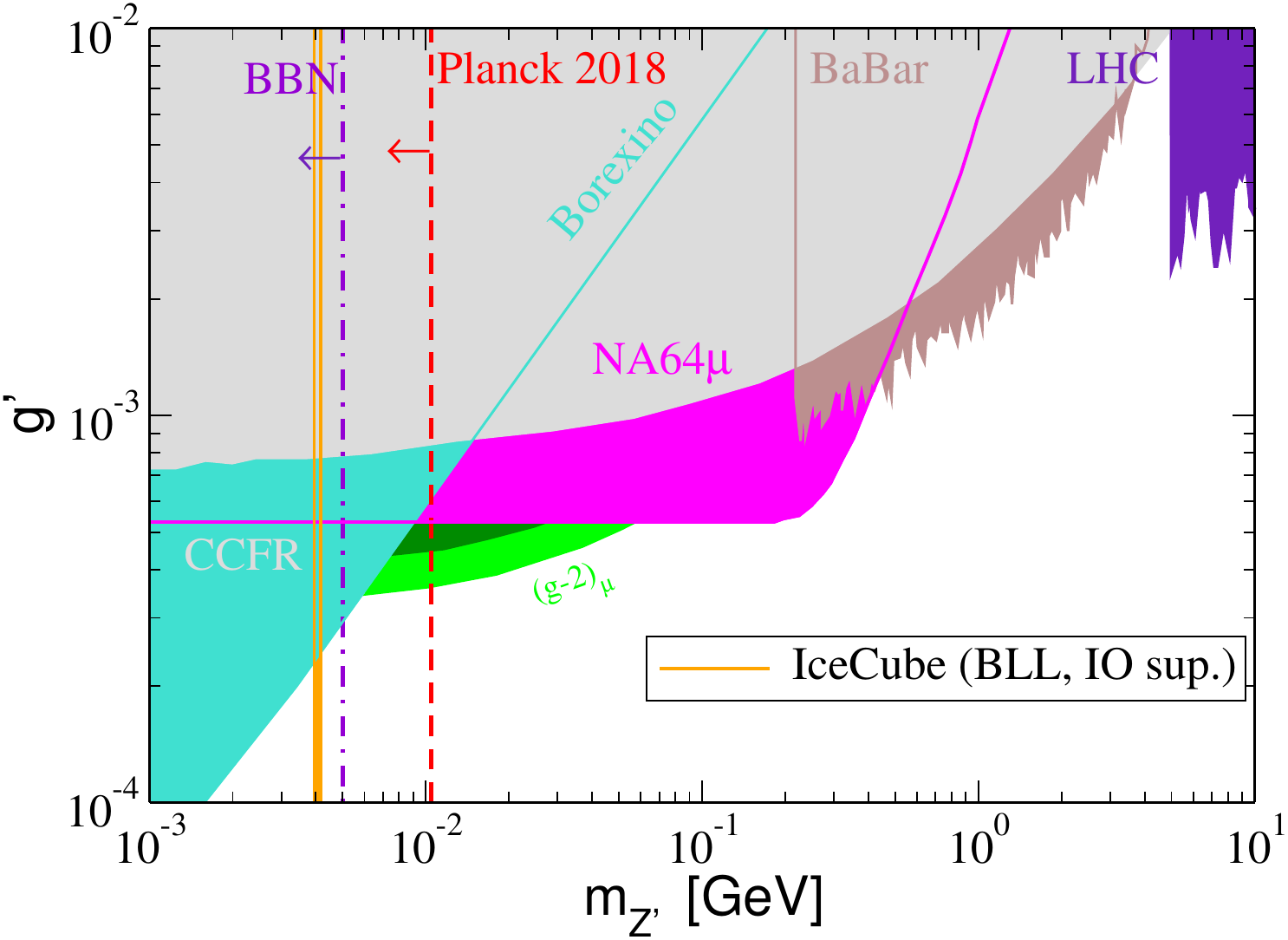}
	\includegraphics[width=0.49\textwidth]{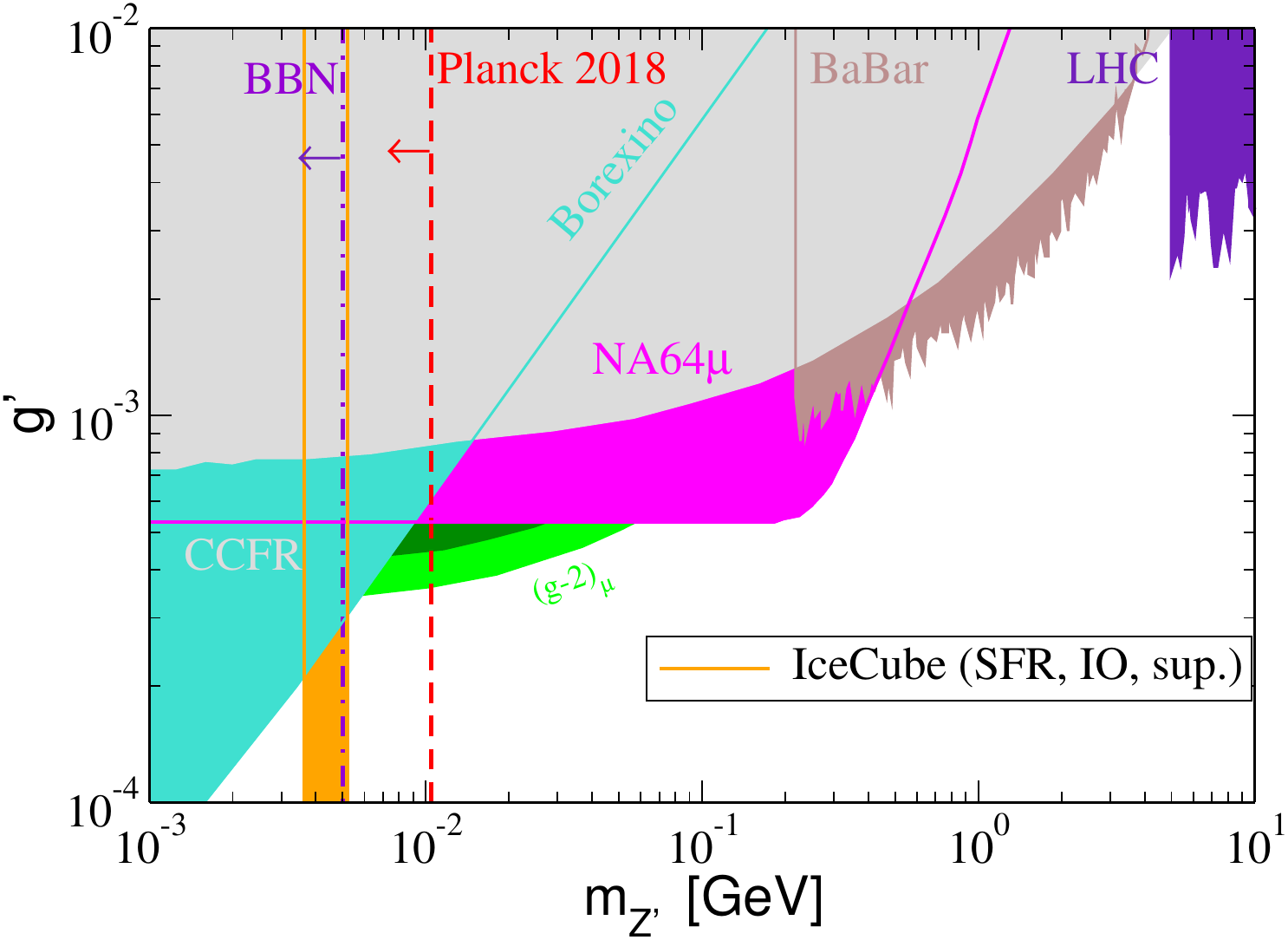} \\
   	\includegraphics[width=0.49\textwidth]{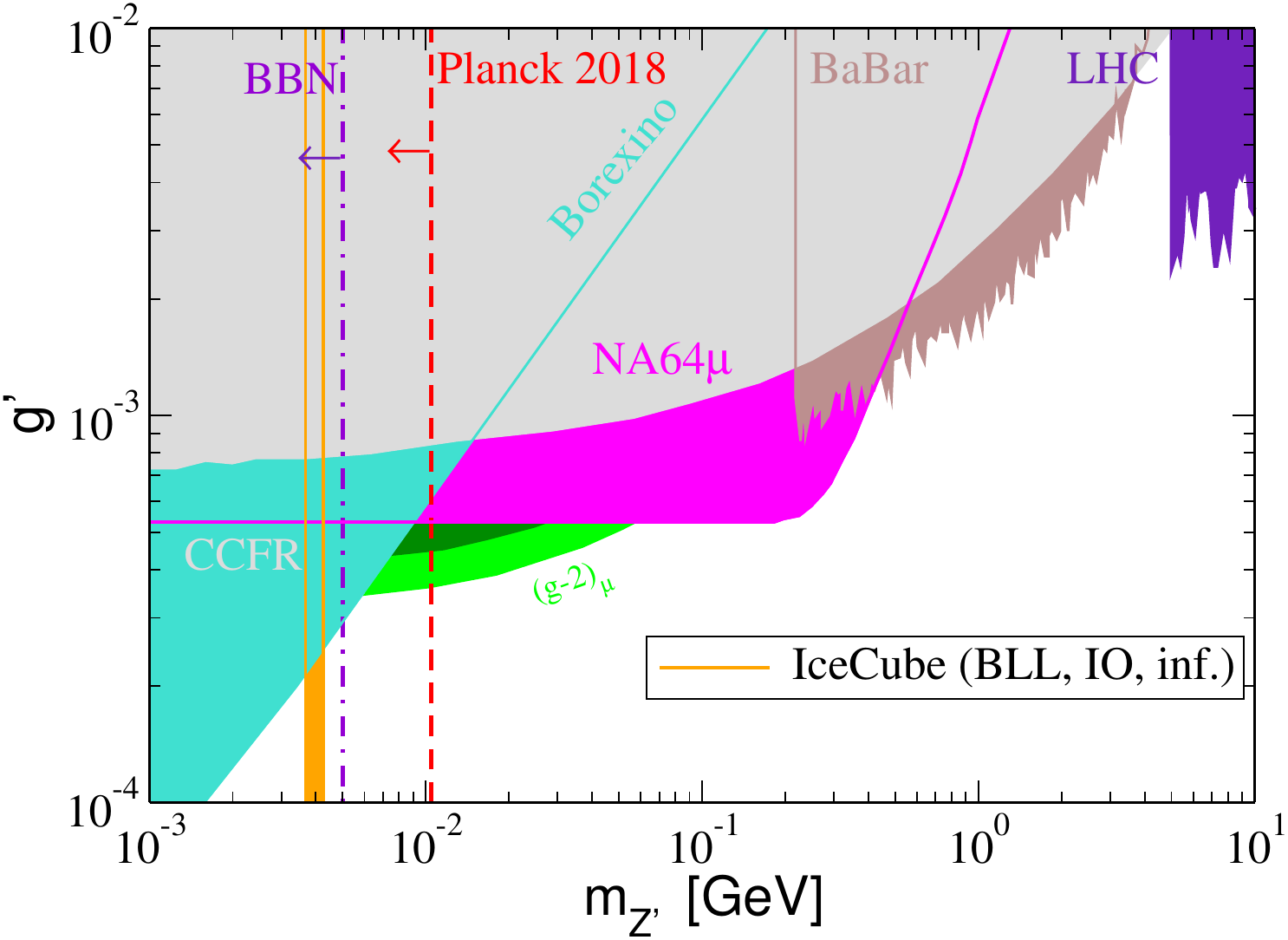}
	\includegraphics[width=0.49\textwidth]{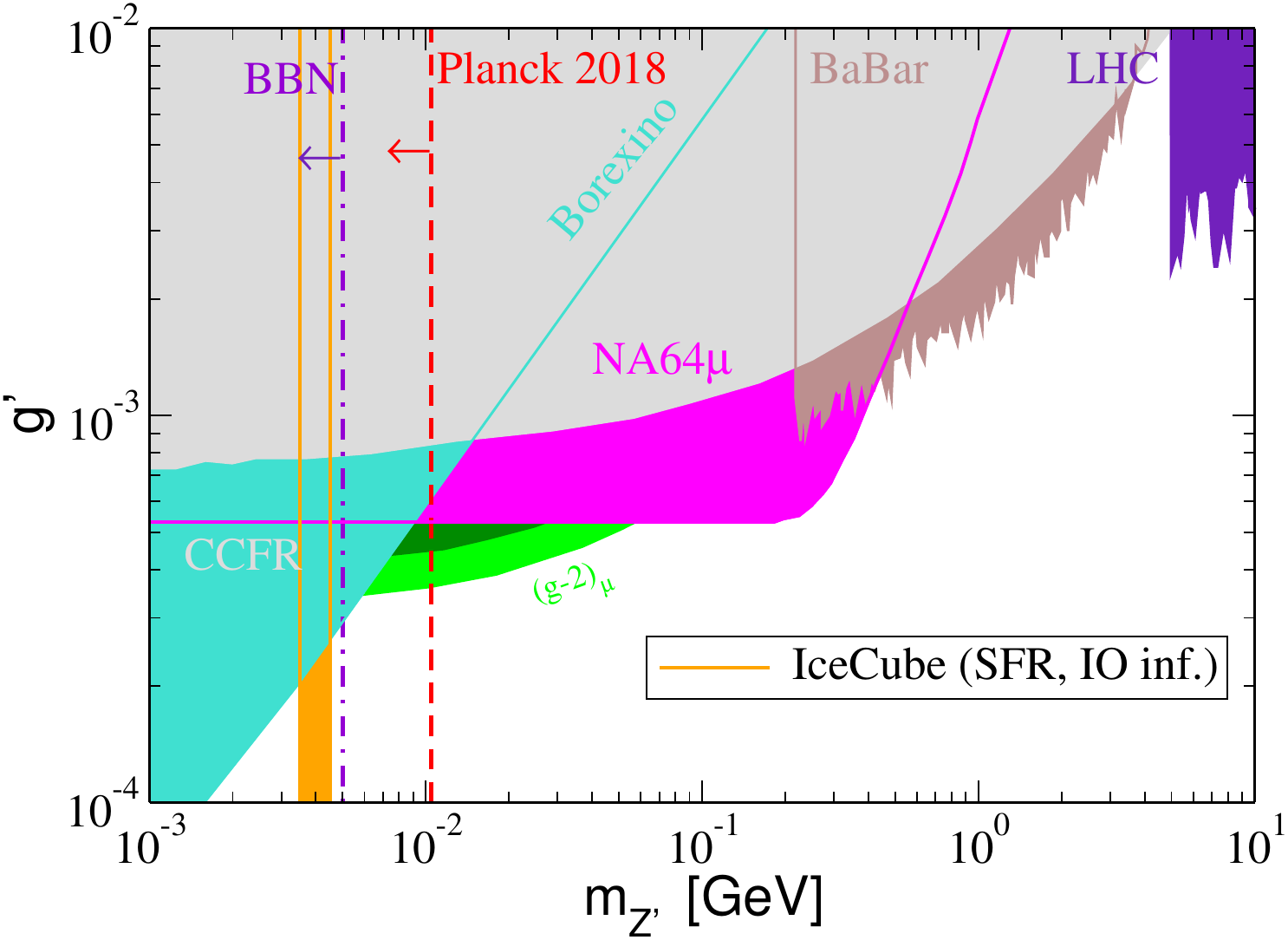}
    
			\end{tabular}
\caption{ Results for the sensitivity of HESE data at the IceCube Observatory to the astrophysical neutrino flux in the presence of the gauge $Z'$ boson predicted by the $L_\mu - L_\tau$ model, derived considering the inverted ordering of neutrino masses. Predictions, at $2\sigma$ of statistical significance, were obtained considering the redshift distribution of BLL (left panels) and SFR (right panels), and the superior (upper panels) and inferior (lower panels) limits for the sum of neutrino masses. For comparison, existing constraints from other processes and experiments are also presented. The green bands represent the parameter space in which the presence of a $Z'$ resolves the muon magnetic moment anomaly with $1\sigma$ and $2\sigma$. }
\label{fig_Ice:spaceIO}
\end{figure}

In Figure \ref{fig_Ice:spaceNO}, we present our results for the sensitivity of the IceCube observatory, derived considering the 12 years of HESE data and the normal ordering for the neutrino masses. Our goal is to derive the corresponding excluded regions in the mass and coupling parameter space of the $Z'$ boson with a 95\% confidence level. The results obtained assuming the BLL (SFR) distribution are presented in the left (right) panels. Finally, the upper (lower) panels represent the superior (inferior) limit of the sum of the masses. For comparison, existing limits derived using other experiments \cite{CMS:2018yxg,ATLAS:2023vxg,ATLAS:2024uvu,BaBar:2016sci,Bellini:2011rx,Harnik:2012ni,NA64:2024klw,Escudero:2019gzq,Ghosh:2024cxi,CCFR:1991lpl,Altmannshofer:2014pba} are also presented, as well as the still unexcluded region that explains $(g-2)_\mu$ (green bands) \cite{Keshavarzi:2018mgv}. It turns out that, for a BLL redshift distribution, IceCube is not able to exclude any region from the parameter space. In contrast, for an SFR distribution, a small region of coupling and masses, not covered by the Borexino experiment, begins to be excluded by IceCube. The covered region depends on the assumed ordering, as shown in Figure \ref{fig_Ice:spaceIO}, where we present the results obtained assuming the inverted ordering. In this case, IceCube is also able to be sensitive to a gauge boson $Z^{\prime}$ for a BLL distribution. It is important to emphasize that the regions excluded by IceCube are, in principle, excluded by cosmological constraints, but it is important to note that these constraints come from indirect measurements and depend on the assumptions present in the standard cosmology $\Lambda$CDM. As a consequence, the possibility of probing this region of the parameter space using a different process is opportune. Another important aspect to emphasize is that, in the analyzed region of the parameter space, the results are almost independent of the coupling constant $g'$. This result is implied by the maximum of the cross section $\nu_i \bar{\nu}_j$ as a function of the incident neutrino energy (see Equation \ref{eq_Ice:csZprime}) being independent of $g'$. However, since the width of the cross section distribution is proportional to $g'^{2}$, the distribution becomes narrower as $g'$ decreases and the impact of a $Z^{\prime}$ disappears when $g' \rightarrow 0$. Our results were obtained for $g' > 10^{-4}$, but recent analyses have shown that in the region of $g' < 10^{-4}$ the cross section is so narrow that the impact on neutrino flux attenuation becomes negligible (see references \cite{Carpio:2021jhu,Hooper:2023fqn,delaVega:2024pbk}).

\begin{figure}
	\centering
	\begin{tabular}{ccc}
	\includegraphics[width=0.49\textwidth]{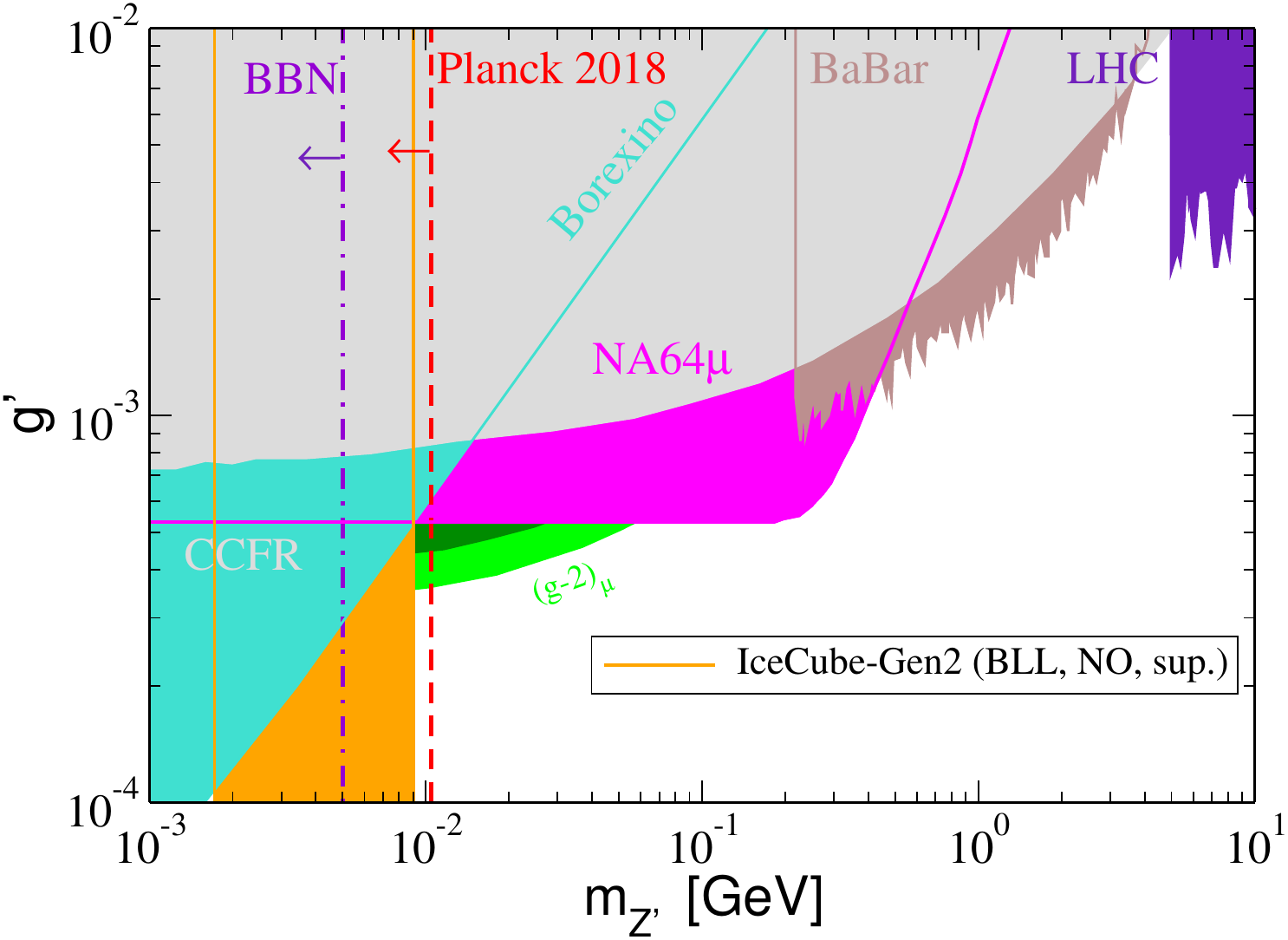}
	\includegraphics[width=0.49\textwidth]{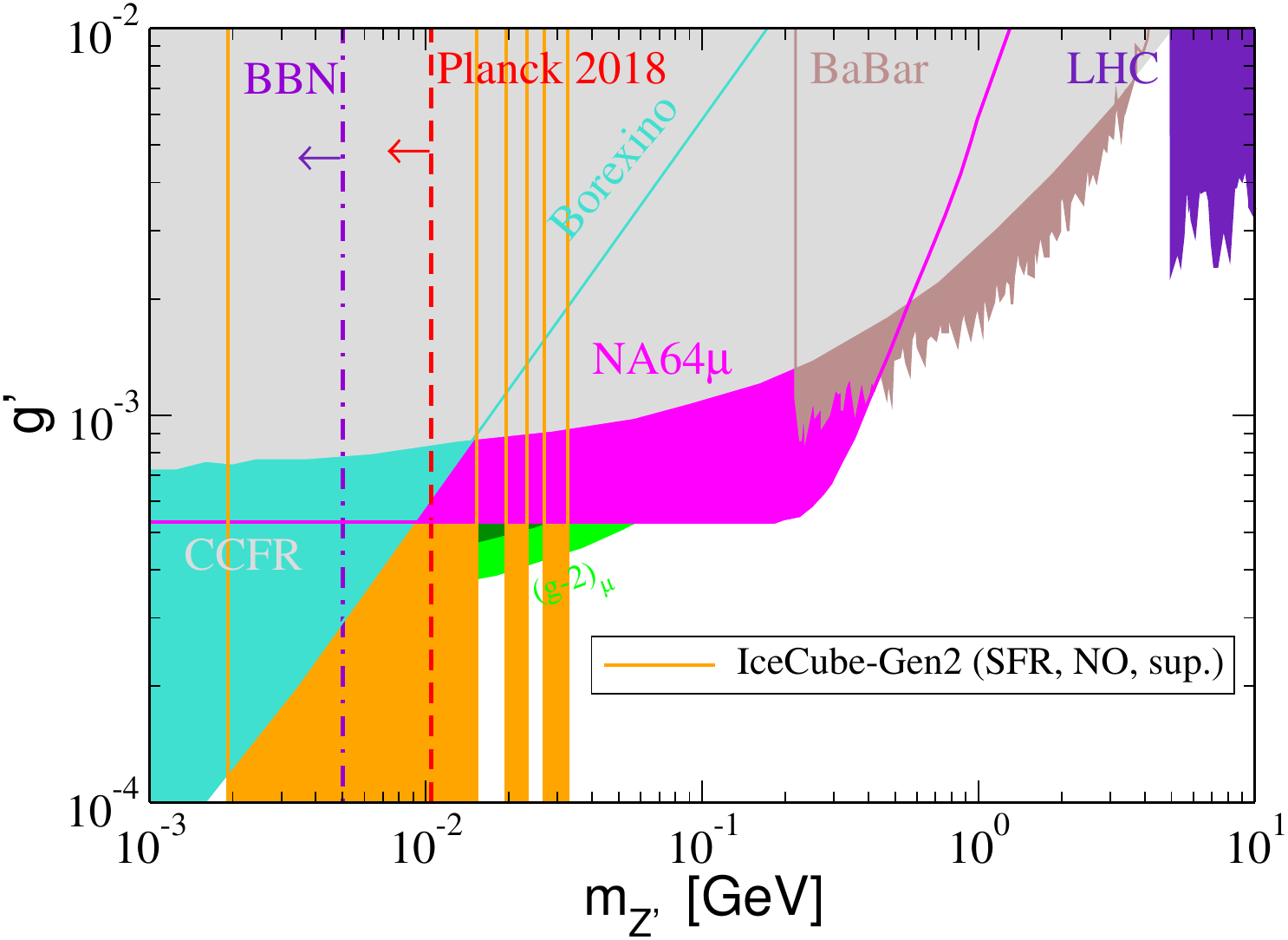} \\
   	\includegraphics[width=0.49\textwidth]{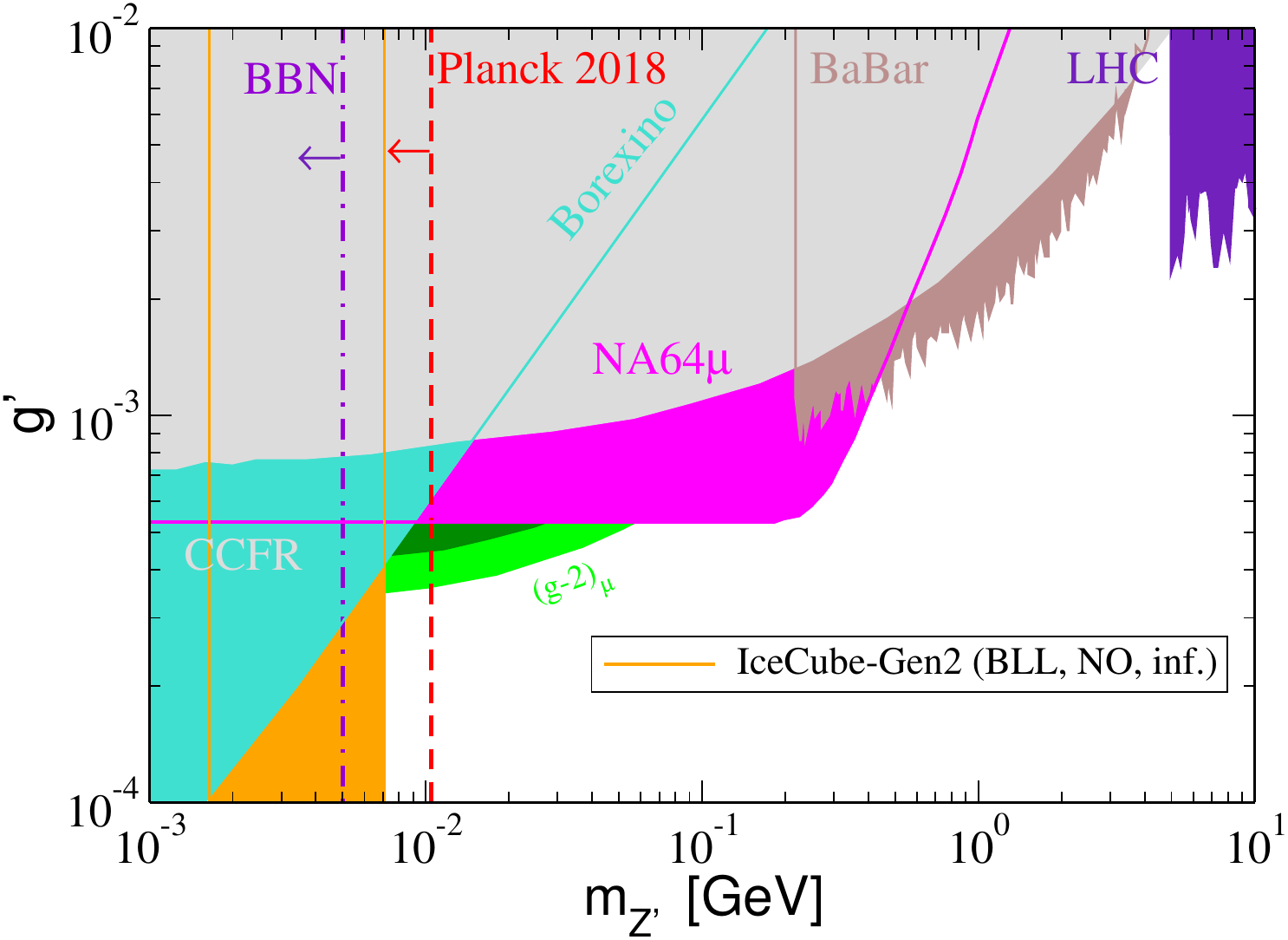}
	\includegraphics[width=0.49\textwidth]{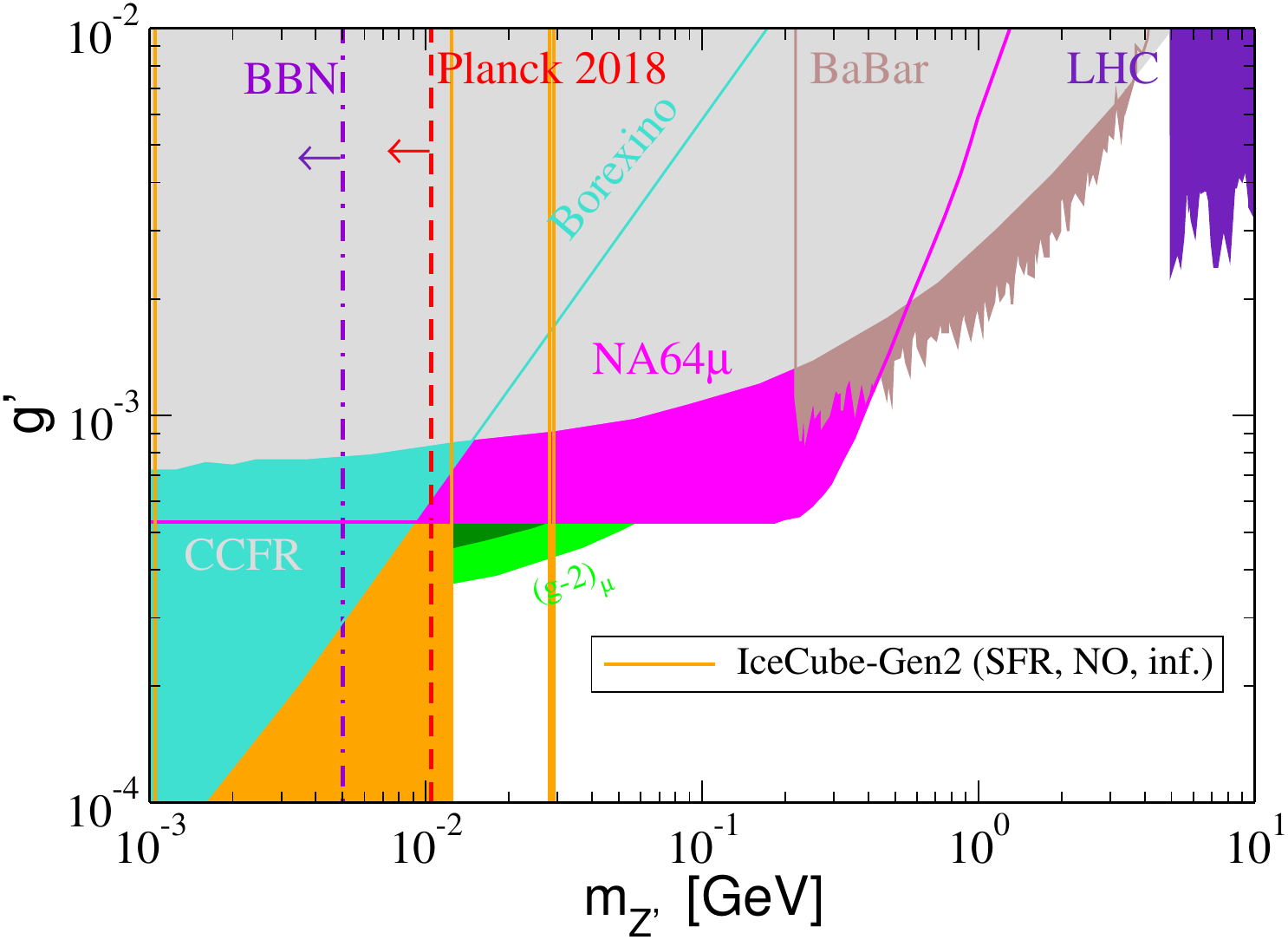}
    
			\end{tabular}
\caption{ Results for the sensitivity of HESE data expected at the IceCube-Gen2 Observatory for the astrophysical neutrino flux in the presence of the $Z'$ gauge boson predicted by the $L_\mu - L_\tau$ model, derived considering the normal ordering of neutrino masses. Predictions, at $2\sigma$ of statistical significance, obtained considering the redshift distribution of BLL (left panels) and SFR (right panels), and the superior (upper panels) and inferior (lower panels) limits for the sum of neutrino masses. For comparison, existing constraints from other processes and experiments are also presented. The green bands represent the parameter space in which the presence of a $Z'$ resolves the muon magnetic moment anomaly with $1\sigma$ and $2\sigma$. }
\label{fig_Ice:spaceNOgen2}
\end{figure}

\begin{figure}
	\centering
	\begin{tabular}{ccc}
	\includegraphics[width=0.49\textwidth]{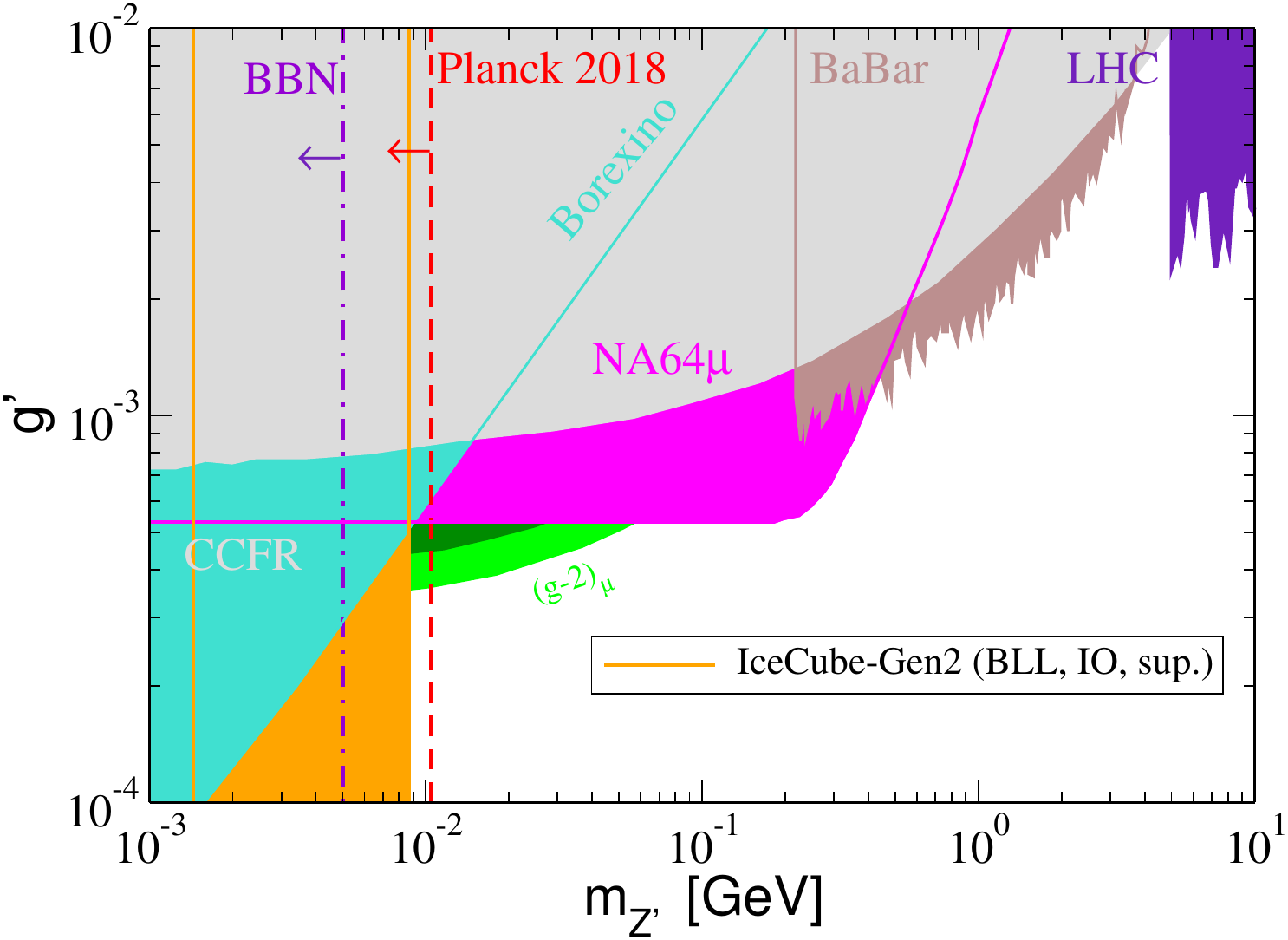}
	\includegraphics[width=0.49\textwidth]{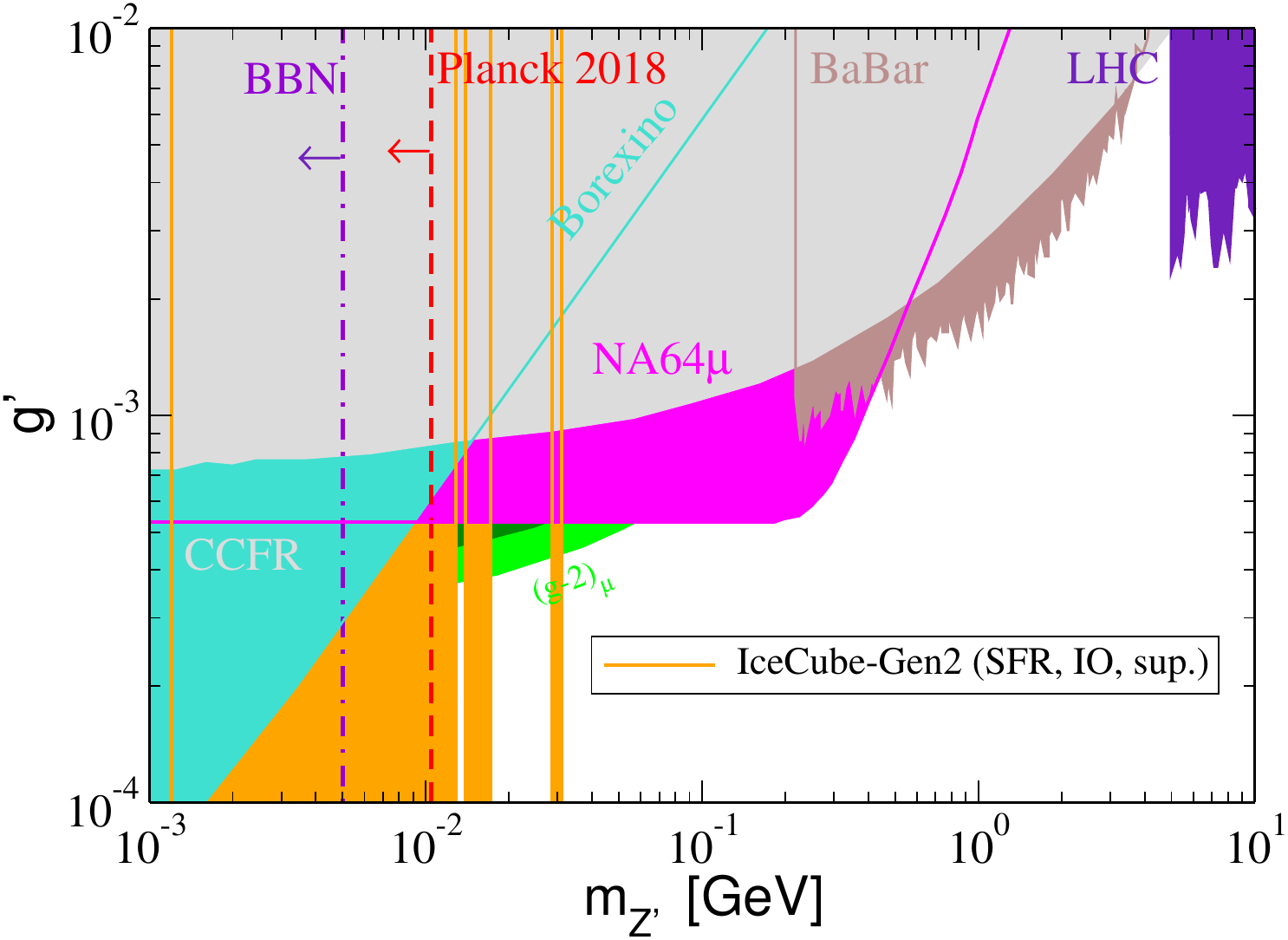} \\
    	\includegraphics[width=0.49\textwidth]{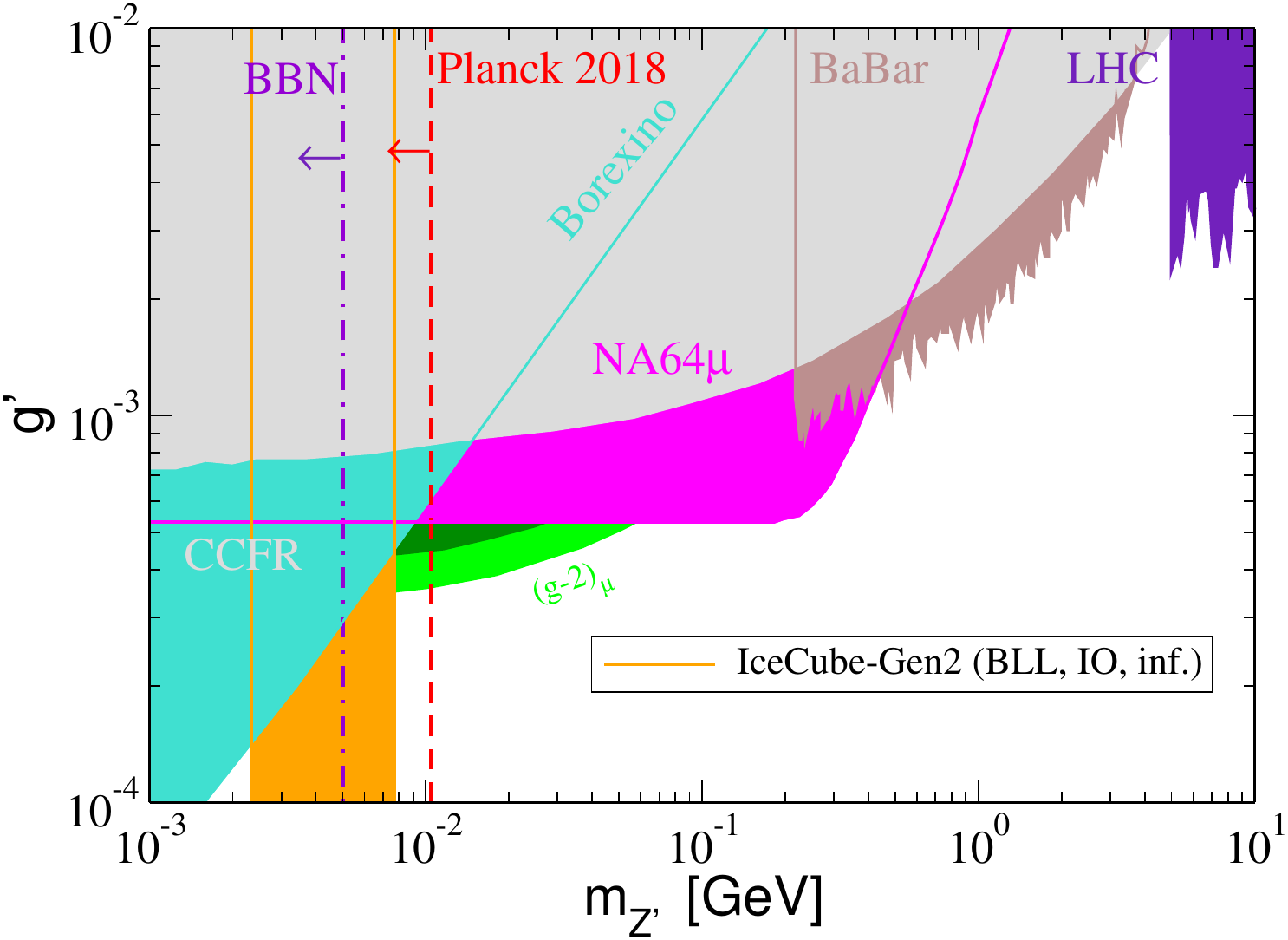}
	\includegraphics[width=0.49\textwidth]{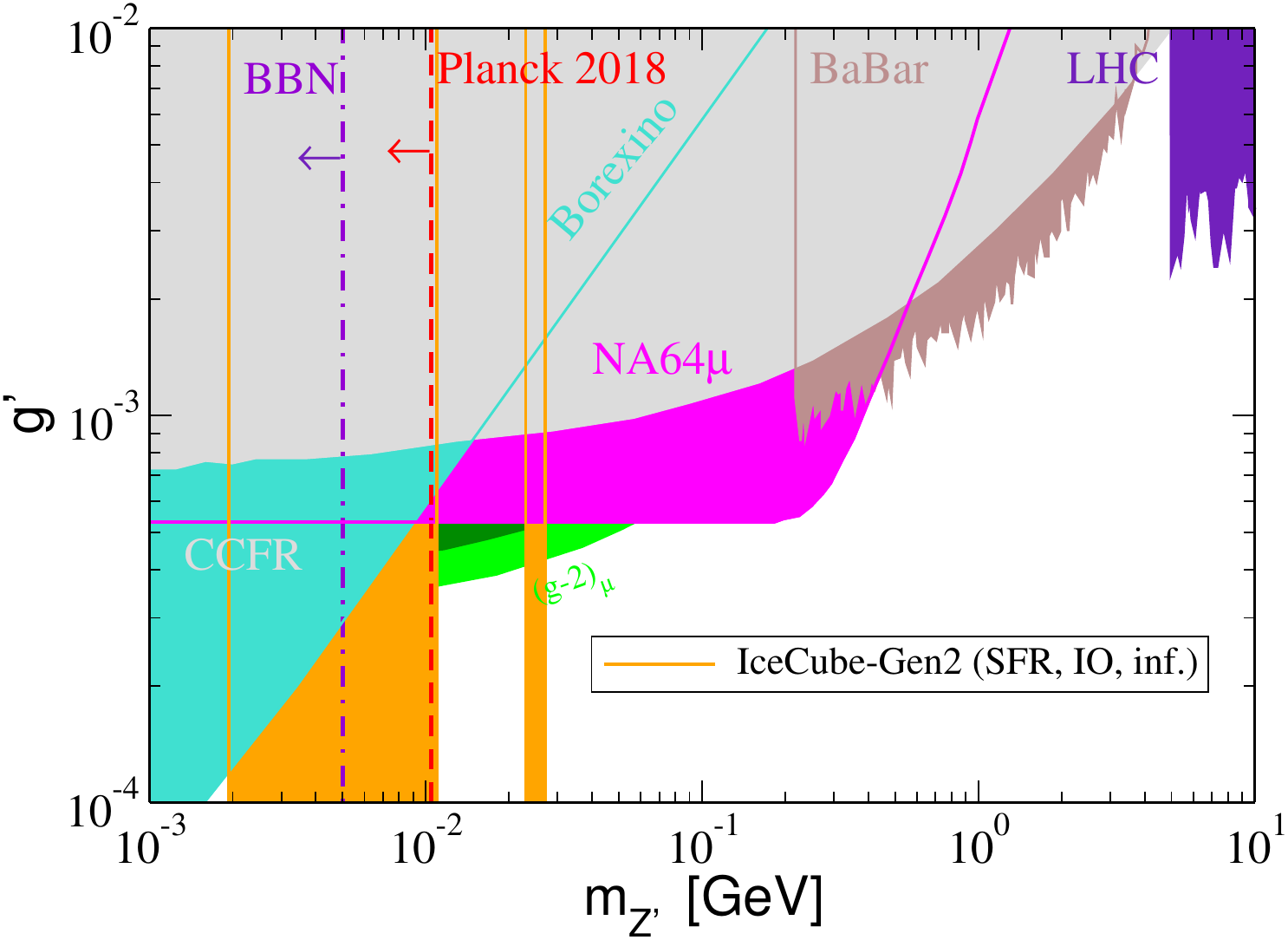}
    
			\end{tabular}
\caption{ Results for the sensitivity of HESE data expected at the IceCube-Gen2 Observatory for the astrophysical neutrino flux in the presence of the $Z'$ gauge boson predicted by the $L_\mu - L_\tau$ model, derived considering the inverted ordering of neutrino masses. Predictions, at $2\sigma$ of statistical significance, obtained considering the redshift distribution of BLL (left panels) and SFR (right panels), and the superior (upper panels) and inferior (lower panels) limits for the sum of neutrino masses. For comparison, existing constraints from other processes and experiments are also presented. The green bands represent the parameter space in which the presence of a $Z'$ resolves the muon magnetic moment anomaly with $1\sigma$ and $2\sigma$. }
\label{fig_Ice:spaceIOgen2}
\end{figure}

Finally, we will extend our analysis to IceCube-Gen2, which is the planned upgrade of the IceCube observatory that will expand the number of observed neutrino events. In our analysis, we will consider the same current exposure time currently at IceCube (12 years) and an increase in its volume by a factor of 8. Furthermore, we will consider the same parameters for systematic uncertainties in the Pull method used in IceCube. The corresponding results for the sensitivity of the IceCube-Gen2 Observatory are presented in Figures \ref{fig_Ice:spaceNOgen2} and \ref{fig_Ice:spaceIOgen2}, for normal and inverted ordering, respectively. Our results indicate that IceCube-Gen2 will cover a region in the parameter space not excluded by the Borexino and NA64$\mu$ experiments, regardless of the ordering considered. Furthermore, it will cover part of the region between 10 MeV and approximately 30 MeV for the mass of $Z'$, including part of the region that can explain $(g- 2)_\mu$, which currently remains unexcluded. In the case of the results derived considering the SFR distribution, the region covered by IceCube-Gen2 is not continuous. This is because the sensitivity to Glashow resonance covers an energy range that would not be excluded if only charged current interactions were taken into account. Our results further indicate that the excluded regions at IceCube-Gen2 will depend weakly on the ordering of the masses, but will be larger for the superior limit case.

\section{Conclusions}
\label{sec_Ice:conclusion}

In this chapter, we studied different processes related to the IceCube neutrino observatory, focusing on astrophysical neutrinos, which are the dominant events for energies greater than 60 TeV in the HESE dataset. We explore the possibility of probing the Earth's internal structure through the attenuation of high-energy neutrino flux, and show that simplified models of matter distribution on Earth, such as the three-layer model, can provide good estimates of neutrino flux attenuation. In the context of the Standard Model, we show that subdominant channels, usually neglected, can be important in describing processes characterized as tracks in IceCube, mainly those arising from the decay of $W^{\pm}$ bosons and $D$ mesons. In addition to contributing to the number of events characterized as tracks, we show that these subdominant channels have a reconstructed average inelasticity different from the dominant channel (muon production by interactions of muonic neutrinos and antineutrinos), indicating that these subdominant channels may be identified in the future through this observable. Finally, we studied the impacts of the $L_\mu - L_\tau$ theory and its associated $Z'$ gauge boson on the propagation of astrophysical neutrinos observed at experiments such as IceCube. Our results showed that IceCube-Gen2 will be able to probe regions of the parameter space of this new boson that have never been tested. In the following chapters, we will study slightly less energetic neutrinos, those from colliders, which are currently being measured at CERN.

\chapter{Tau polarization effects on tauonic neutrino processes at the LHC}	
\label{cap:pol}

Tauonic neutrinos remain the least studied fermion of the Standard Model to these days. They were predicted soon after the discovery of the tau lepton in the 1970s \cite{Perl:1975bf,Perl:1977se}. Only four experiments to date have evidence of their observation: DONuT \cite{DONUT:2000fbd,DONuT:2007bsg} and OPERA \cite{OPERA:2018nar} in a regime of a few GeVs, observing tau neutrinos in charged current interactions from accelerators by the decay of $D_s$ mesons and by the neutrino oscillation mechanism, respectively; SuperK observed events of this neutrino flavor originating from the atmospheric neutrino flux, also resulting from the flavor oscillation mechanism \cite{Super-Kamiokande:2017edb}; and IceCube has seven candidates for tau neutrino events of astrophysical origin \cite{IceCube:2019dqi}. Given that it is the least measured fermion in the Standard Model, tauonic neutrinos do not place significant restrictions on scenarios beyond the Standard Model, and the possibility of measuring these neutrinos in the near future has motivated several phenomenological studies in new physics scenarios \cite{DUNE:2024wvj,DeGouvea:2019kea,Dev:2023rqb,Bakhti:2023mvo,Yu:2025cyx,Ghoshal:2019pab,Machado:2020yxl,Cherchiglia:2025dnd}.

In this chapter, we will present and discuss our results for the polarization effects of the tau lepton produced in charged current interactions of tauonic neutrinos. Our results will focus on neutrinos with energies on the order of TeV, which are those currently measured at the LHC \cite{FASER:2023zcr,FASER:2024hoe,FASER:2024ref,SNDLHC:2023pun}. Several recent works have shown that tau leptons produced in neutrino interactions at energies in the range of a few GeV are not completely polarized \cite{Hagiwara:2003di,Hernandez:2022nmp,Zaidi:2023hdd,Isaacson:2023gwp}, although the incident tauonic neutrino and the final state tau are chiral eigenstates. These results are of interest in the search for the effects of tau polarization in experiments such as DUNE \cite{DUNE:2015lol}, which will measure a flux of tauonic neutrinos with energies of up to a few tens of GeV. One of the main motivations for these studies is that current Monte Carlo generators that simulate the production and decay of tau consider it completely polarized, as is the case with GENIE \cite{GENIE:2021zuu,Andreopoulos:2015wxa}, GiBUU \cite{Leitner:2006ww,Buss:2011mx}, NuWro \cite{Golan:2012rfa}, NEUT \cite{Hayato:2021heg} and TAUSIC \cite{Kudryavtsev:2008qh}. After events generated with fully polarized taus in the generators, these are usually coupled to TAUOLA \cite{Davidson:2010rw,Chrzaszcz:2016fte}, which simulates tau decay, and the polarization state affects the kinematics of the final state. Motivated by the large number of tau neutrino events expected in FASER$\nu$2 (between 2000 and 20000) during the high-luminosity regime of the LHC \cite{Anchordoqui:2021ghd,Feng:2022inv,FPFWorkingGroups:2025rsc,FPF:2025bor}, we extended previous studies to the polarization of the tau produced in the LHC energy regime ($10^2 - 10^3$ GeV) to understand if, in this regime where the tau mass is about two to three orders of magnitude below the incident neutrino energy, there are still effects of tau polarization on the observables in the detector.

In recent years, neutrino polarization effects have been explored in several works, both in the process mentioned above looking at the lepton produced \cite{Hagiwara:2003di,Hernandez:2022nmp,Zaidi:2023hdd,Isaacson:2023gwp}, and for polarized nuclear targets \cite{Forte:2001ph,Francener:2023wpr}. Recent results from the COMPASS \cite{COMPASS:2006mhr,COMPASS:2015mhb,COMPASS:2016jwv} and CLAS \cite{CLAS:2017qga} collaborations use polarized charged lepton beams to probe the structure functions of polarized targets, which have been the subject of intense debate since the results of the EMC collaboration known as the proton spin crisis \cite{EuropeanMuon:1987isl,EuropeanMuon:1989yki}. Since neutrinos are considered polarized due to their low mass and chiral eigenstates, there is a proposal to use neutrino beams produced at the LHC in experiments like COMPASS to extract polarized structure functions \cite{MammenAbraham:2024gun}.

An additional motivation for the study presented in this chapter, is the fact that tauonic neutrinos are expected to have the greatest sensitivity to the nuclear structure function $F_5$, given that this manifests itself in the cross section as a product with the mass of the lepton produced in the final state. This structure function has never been measured before, but there are experiments planned for the future interested in extracting data from it, such as the SHiP experiment at CERN \cite{DiCrescenzo:2015iza,SHiP:2015vad}. We will study here the impacts of this structure function on the cross sections of processes involving tauonic neutrinos from the LHC at energies of around 1 TeV.

\section{Tauonic neutrino interaction with nucleon}

Although we introduced the formalism and discussed deep inelastic scattering in Chapter \ref{cap:cs} in some detail, we will review its concepts again in this chapter, so that the aspects that differentiate charged current scattering of tauonic neutrinos from the interaction of other neutrino flavors become clear and didactic.

\begin{figure}
\begin{center}
\includegraphics[scale=0.45]{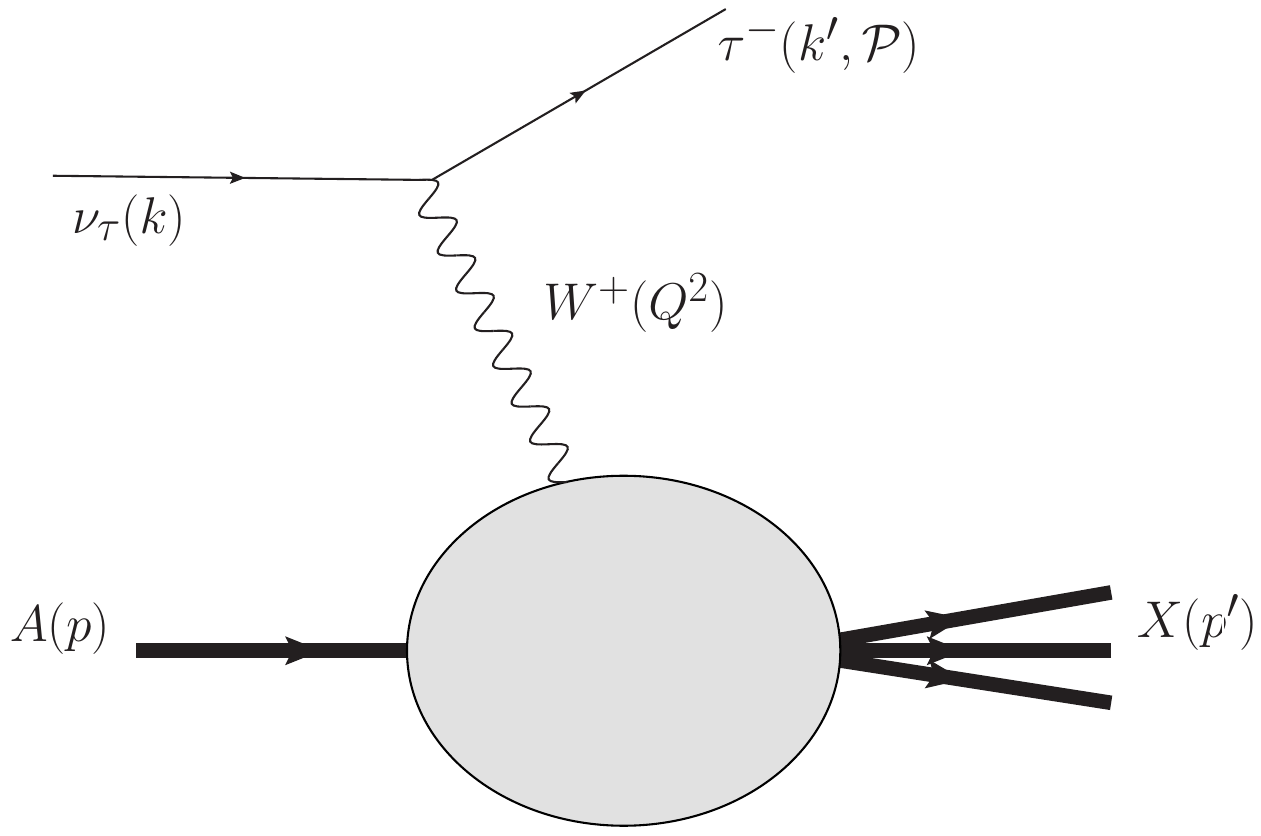} 
\caption{ Production of a tau lepton with four-momentum $k^{\prime}$ and polarization ${\cal{P}}$ in a charged current deep inelastic $\nu_{\tau} A$ scattering. }
\label{fig_4:diagram}
\end{center}
\end{figure}

In the energy regime of neutrinos produced by the LHC, the most probable interaction of these neutrinos with the detector is through the deep inelastic scattering. This scattering, represented in the diagram of Figure \ref{fig_4:diagram}, is characterized by the incidence of a high-energy neutrino with four-momentum $k$, exchanging a $W^{\pm}$ boson of virtuality $Q^{2}$ and becoming a tau lepton with four-momentum $k'$ in the final state, with polarization $\cal{P}$. The target nucleus $A$ goes to an unknown final state, characterized by $X$ and with four-momentum $p'$. For the calculation of the cross section of this process, we resort to collinear factorization, which allows us to separate the cross section into two parts: a perturbative part, from the interaction of the exchanged boson with an asymptotically free parton; and a non-perturbative part, which describes the parton density within the hadronic target.

As is standard practice in the literature, we write the double differential neutrino-nucleon cross section of the charged current interaction in the nucleon's rest frame, with respect to the energy of the tau produced, $E_\tau$, and its scattering angle, $\theta$, in terms of the contraction of the hadronic and leptonic tensors, as follows \cite{Zaidi:2023hdd}

\begin{eqnarray}
    \frac{\mathrm{d}^2\sigma_A}{\mathrm{d}E_\tau\mathrm{d\,cos}\,\theta} = 
    \frac{G_{F}^{2}|\vec{k}'|}{2\pi E_{\nu}\left( 1+\frac{Q^2}{M_{W}^{2}} \right)^{2}}
    L_{\mu\nu}W^{\mu\nu}_{A} \, ,
    \label{eq_4:sigma}
\end{eqnarray}
where $G F$ is the Fermi constant and $M_W$ is the mass of the boson $W^{\pm}$. The leptonic ($L_{\mu\nu}$) and hadronic ($W^{\mu\nu}_{A}$) tensors are given, respectively, by
\begin{eqnarray}
    L_{\mu\nu} = 8(k_\mu k'_\nu + k_\nu k'_\mu - k\cdot k' g_{\mu \nu} \pm i\epsilon_{\mu\nu\rho\sigma} k^\rho k'^\sigma )
    \label{eq_4:tensorL}
\end{eqnarray}
and
\begin{eqnarray}
\begin{aligned}
    W^{\mu\nu}_{A} = -g^{\mu \nu }F^A_{1}(x, Q^2) + 
    \frac{2x}{Q^2} p^{\mu} p^{\nu} F^A_{2}(x, Q^2) -
    \frac{i x}{Q^2}\epsilon^{\mu \nu\rho\sigma} p_{\rho} q_{\sigma} F^A_{3}(x, Q^2) + \\
    + \frac{2}{Q^2} q^{\mu} q^{\nu} F^A_{4}(x,Q^2) + 
    \frac{2x}{Q^2} (p^{\mu}q^{\nu} + q^{\mu}p^{\nu})F^A_{5}(x,Q^2) \, ,
    \label{eq_4:tensorW}
\end{aligned}
\end{eqnarray}
where $F^{A}_{i}$ are the structure functions of the hadronic target, which in turn are determined by the partonic distribution functions, as discussed in Chapter \ref{cap:cs}. In this chapter, we will use PDFs from different collaborations: CT14 \cite{Dulat:2015mca} for PDFs without nuclear effects of tungsten nucleons, and to insert nuclear effects we will use the nCTEQ15 \cite{Kovarik:2015cma} and EPPS21 \cite{Eskola:2021nhw} parameterizations. Contracting the hadronic and leptonic tensors, Equation (\ref{eq_4:sigma}) becomes \cite{Zaidi:2023hdd}
\begin{eqnarray}
    \begin{aligned}
    \frac{\mathrm{d}^2\sigma_A}{\mathrm{d}E_\tau\mathrm{d\,cos}\,\theta} = 
    \frac{G_F^2|\vec{k}'|}{2\pi E_\nu (1+Q^2/M_W^2)^2} \left\{ 2F^A_{1}(x,Q^2)(E_\tau -|\vec{k}'|\mathrm{cos}\,\theta) + \right.\\ 
    + F^A_{2}(x,Q^2)\frac{M_A}{\nu}(E_\tau+|\vec{k}'|\mathrm{cos}\,\theta)  + \\
    \pm F^A_{3}(x,Q^2)\frac{1}{\nu}[| \vec{k}' |^2 + E_\nu E_\tau - (E_\nu + E_\tau )|\vec{k}'|\mathrm{cos}\,\theta] + \\
    + F^A_{4}(x,Q^2)\frac{m_\tau^{2}}{\nu M_A x} (E_\tau - |\vec{k}'|\mathrm{cos}\,\theta) + \\
    \left. - F^A_{5}(x,Q^2)\frac{2m_\tau^2}{\nu} \right\} \, .
    \label{eq_4:sigma2}
    \end{aligned}
\end{eqnarray}
In order to obtain the total cross section of the process, we simply integrate with respect to $E_\tau$ and $\mathrm{cos}\,\theta$ in their respective kinematically allowed intervals.

\begin{figure}
	\centering
	\begin{tabular}{ccc}
	\includegraphics[width=0.48\textwidth]{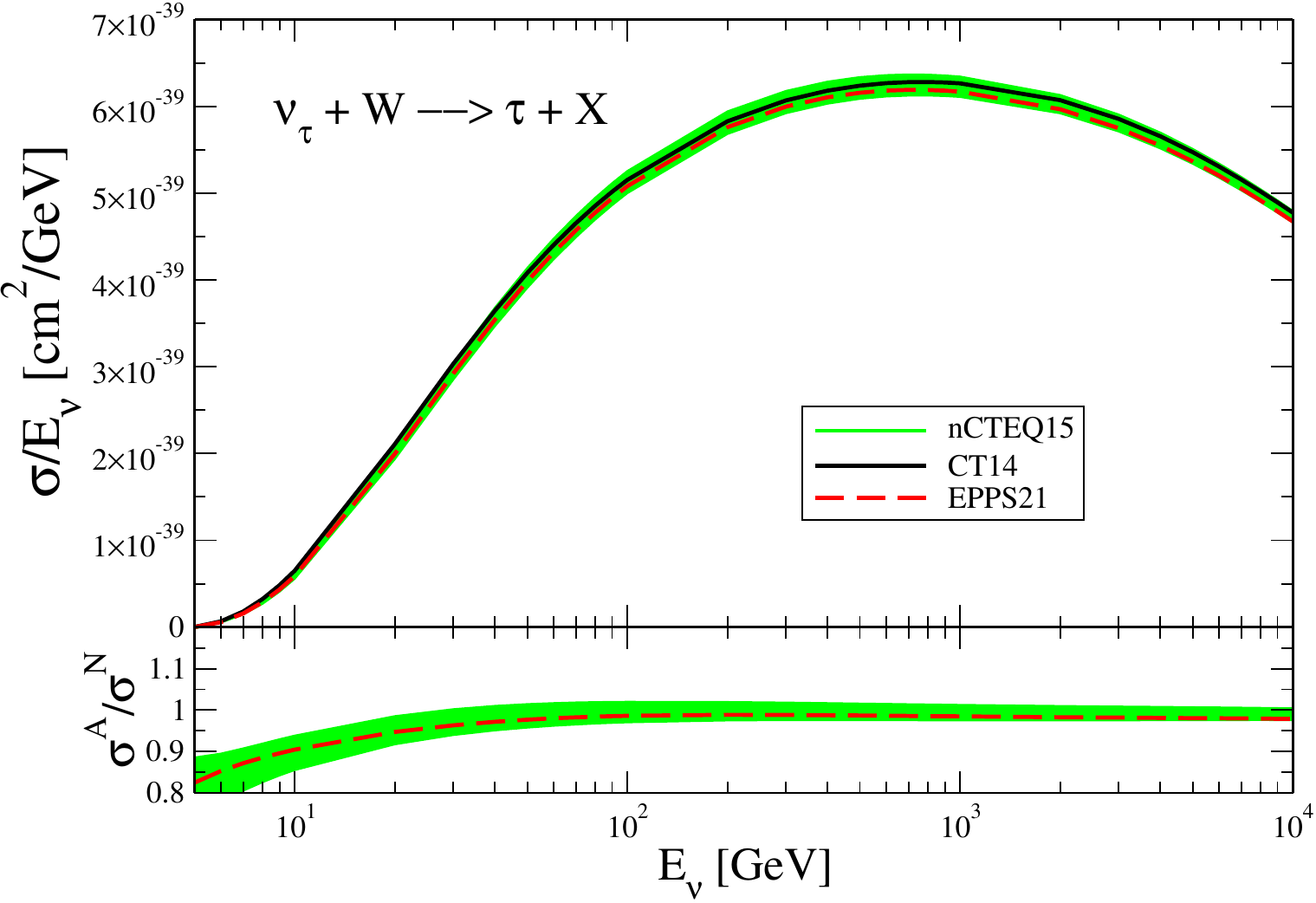} &  \includegraphics[width=0.48\textwidth]{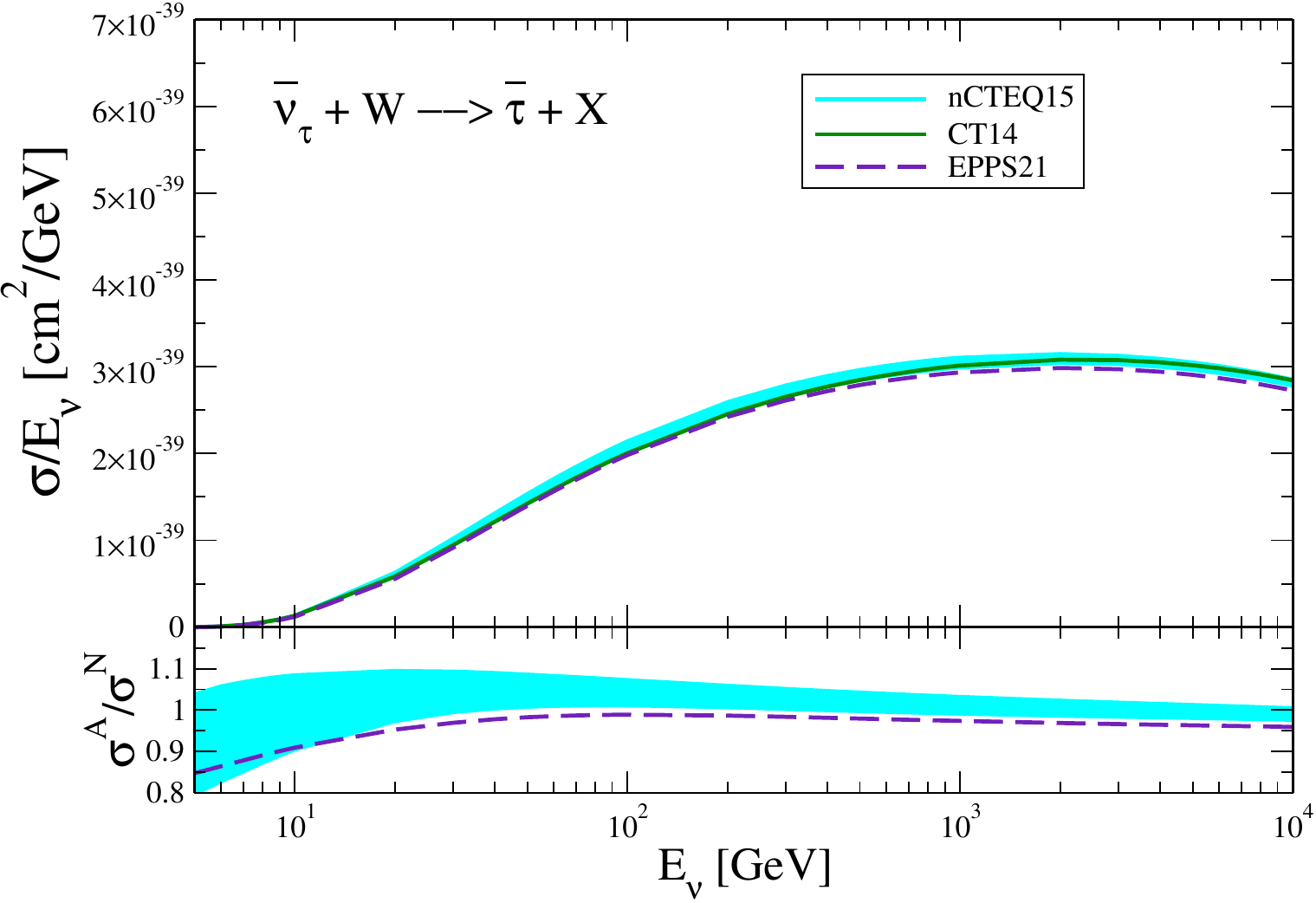} 
			\end{tabular}
\caption{ Predictions for the dependence of the ${\nu}_{\tau}W$ (left panel) and $\bar{\nu}_{\tau}W$ (right panel) cross sections on the energy of the incident tau neutrino. Results derived assuming different parameterizations for the nPDFs. Predictions for the ratio between the nuclear and nucleonic cross sections are presented in the lower panels. }
\label{fig_4:sigma}
\end{figure}

Our first result discussed here will be the total cross section, obtained from Equation (\ref{eq_4:sigma2}), presented in Figure \ref{fig_4:sigma}. We show the cross section per nucleon, divided by the energy of the incident neutrino, in the energy regime that will be covered by FASER$\nu$2 for interactions of tauonic neutrinos with the nucleon of the tungsten (W) nucleus. On the left, we present the results for neutrino interactions, and on the right, for antineutrino interactions. We compare the results using the three different parameterizations for the PDFs mentioned earlier. For the results using the nCTEQ15 parameterization, we constructed an uncertainty band with a 90\% confidence level to estimate whether the effects analyzed in the cross section are greater than the uncertainties present in the PDFs. In the lower panels, we have the ratio between the cross section with nuclear effect and the cross section without nuclear effect, from the CT14 parameterization. Our results indicate that nuclear effects are more important for incident neutrinos with energies below 10 GeV, and in general, the inclusion of nuclear effects decreases the total cross section. We can observe in the figure on the right that the prediction of the cross section using the EPPS21 parameterization is outside the uncertainty band of the prediction with the nCTEQ15 parameterization for energies greater than 20 GeV of the incident antineutrino.

\begin{figure}
	\centering
	\begin{tabular}{ccccccc}
     \includegraphics[width=0.23\textwidth]{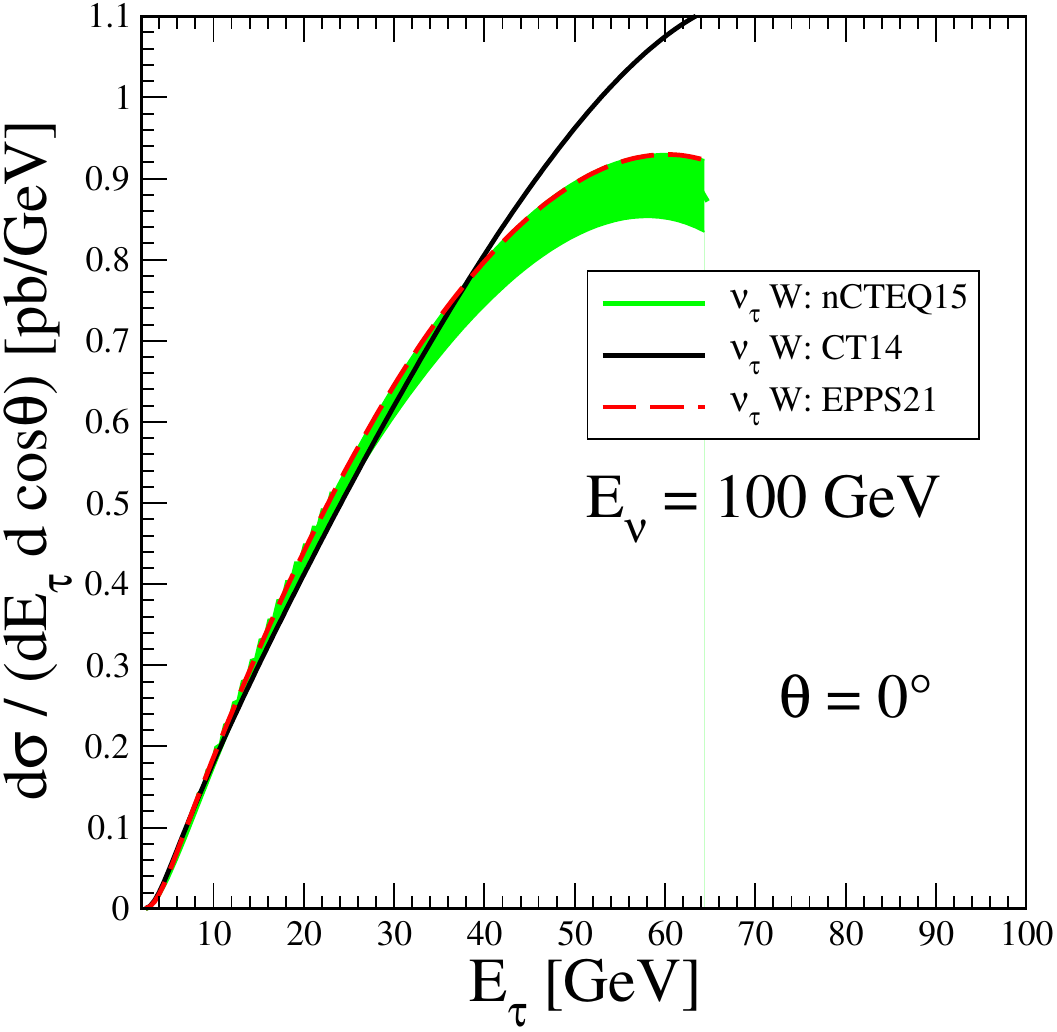} & \includegraphics[width=0.23\textwidth]{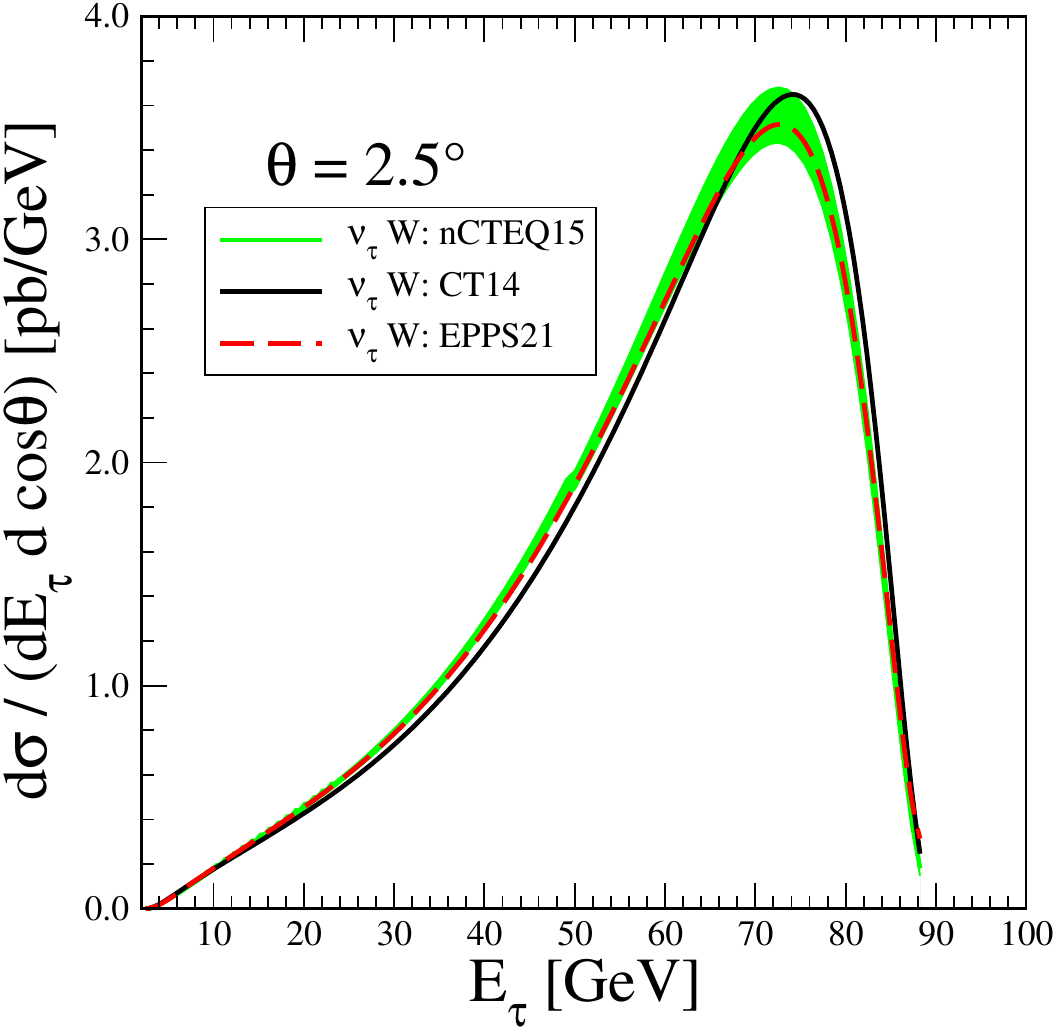} & 
     \includegraphics[width=0.23\textwidth]{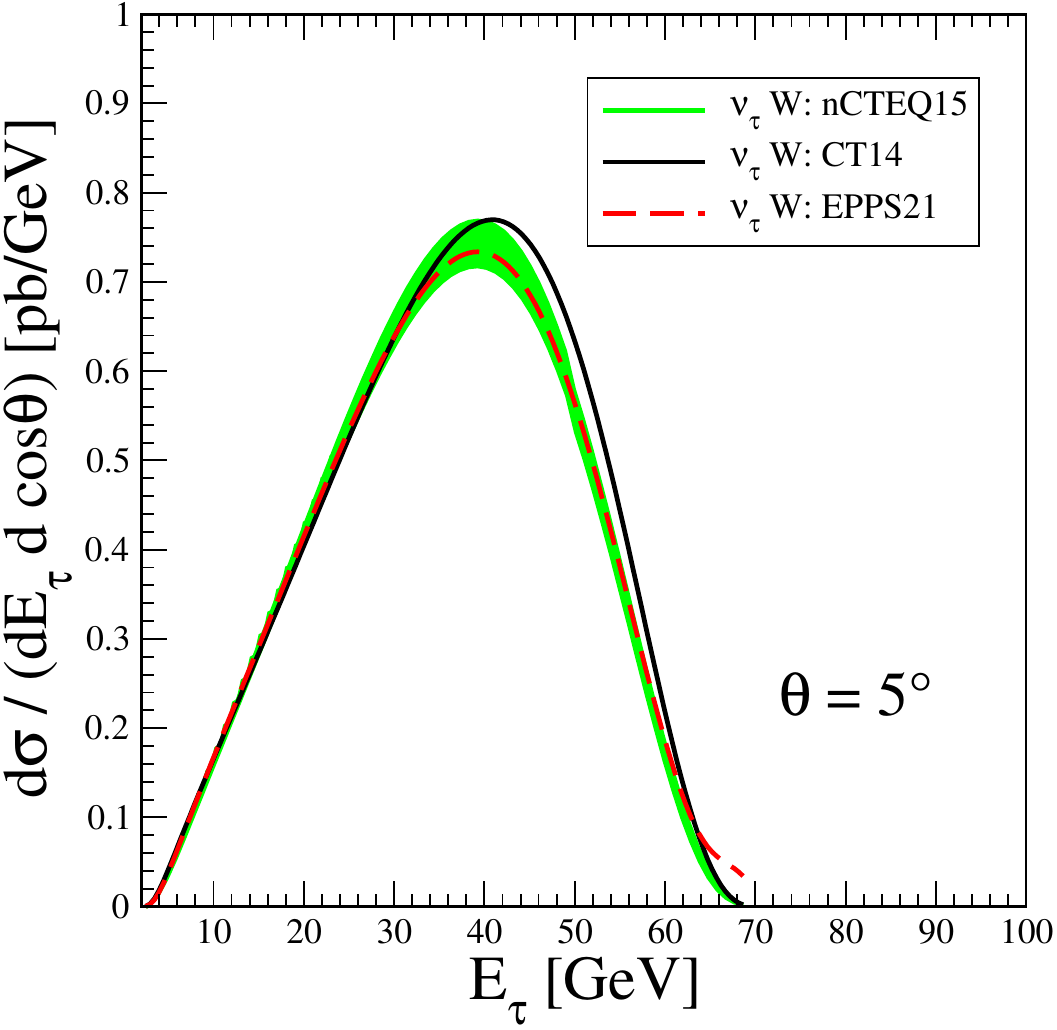} & \includegraphics[width=0.23\textwidth]{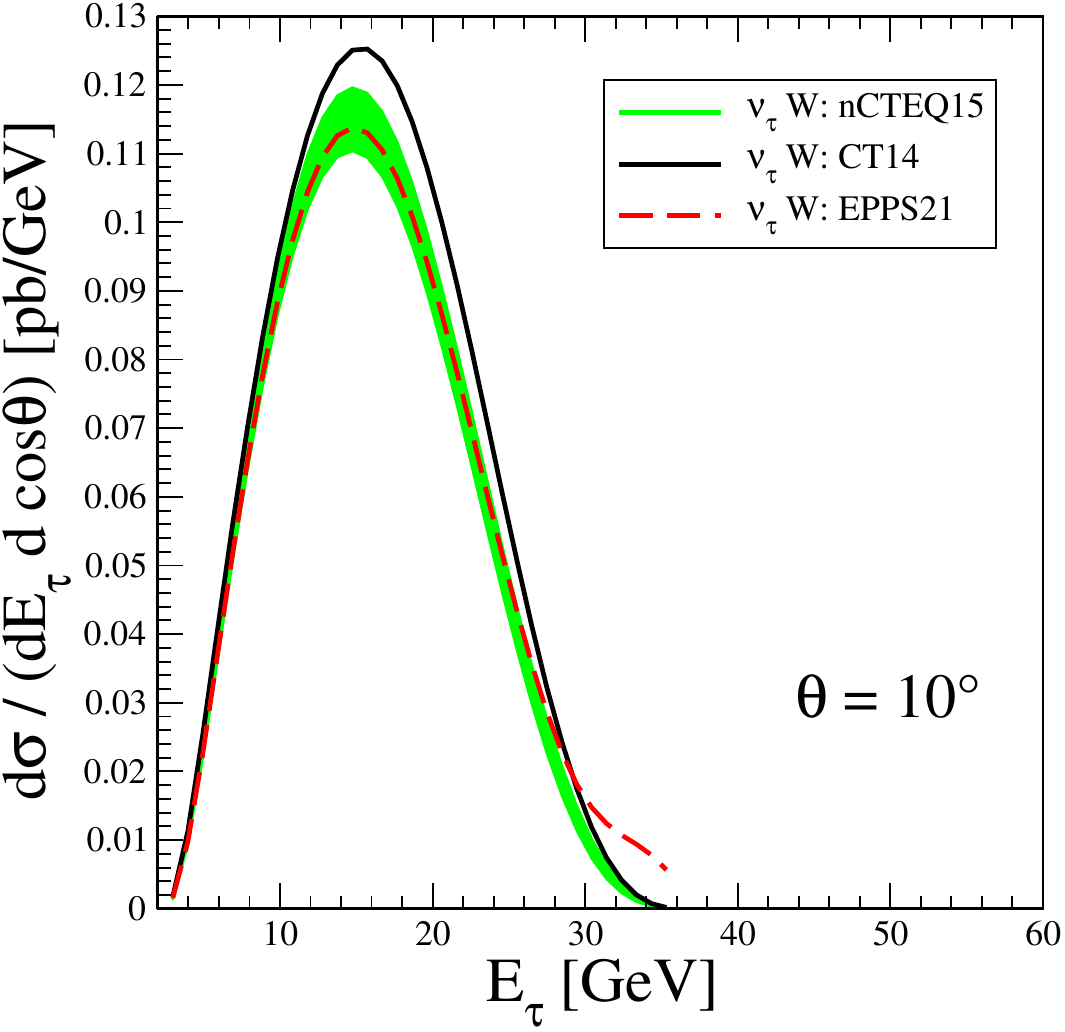}          \\
     \includegraphics[width=0.23\textwidth]{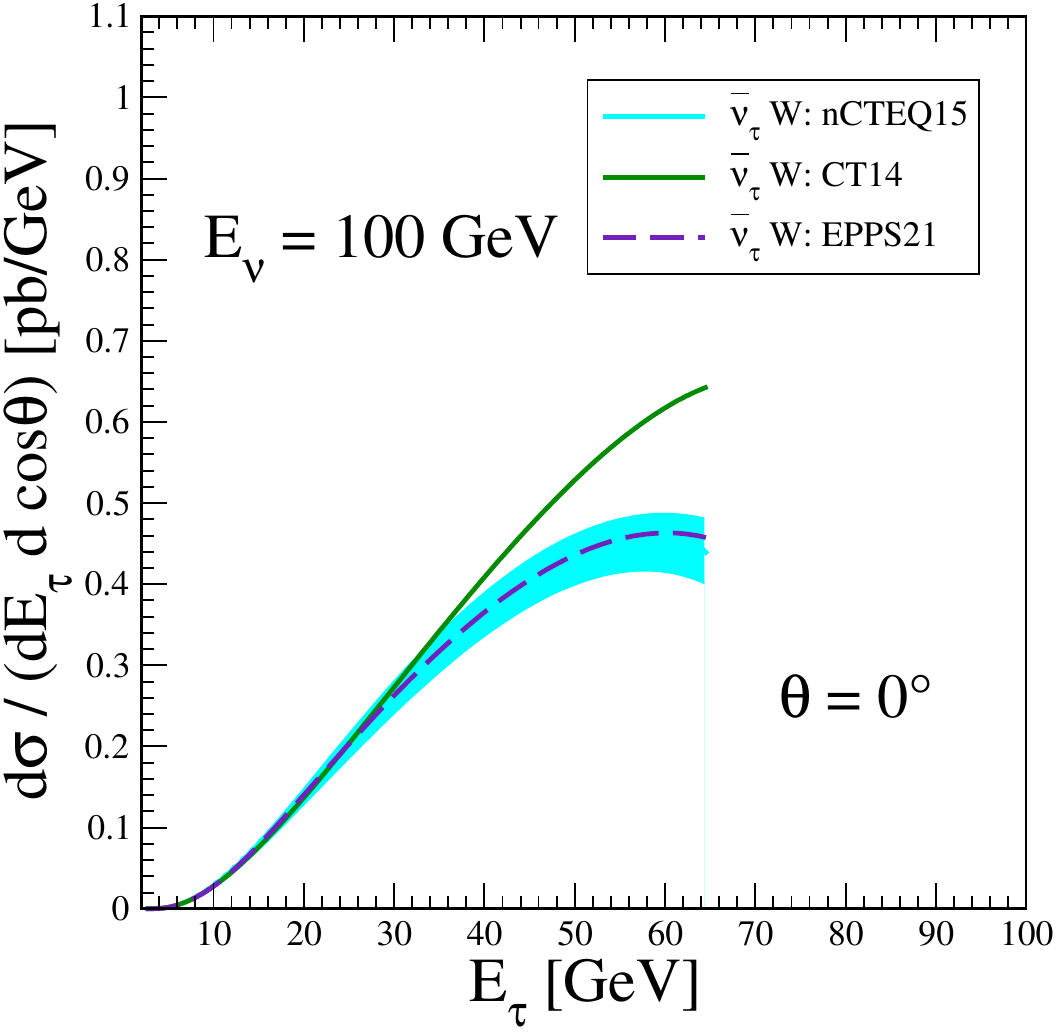} & \includegraphics[width=0.23\textwidth]{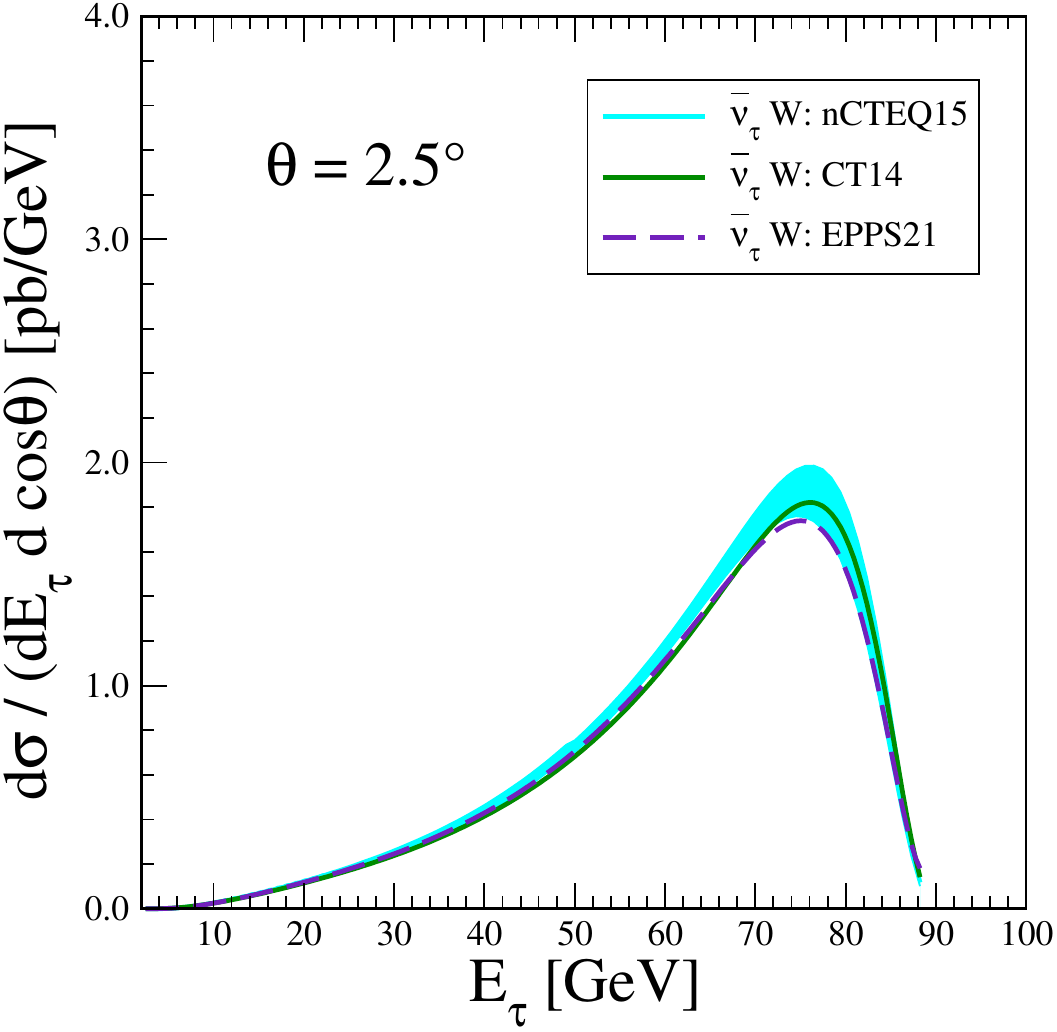} &  
     \includegraphics[width=0.23\textwidth]{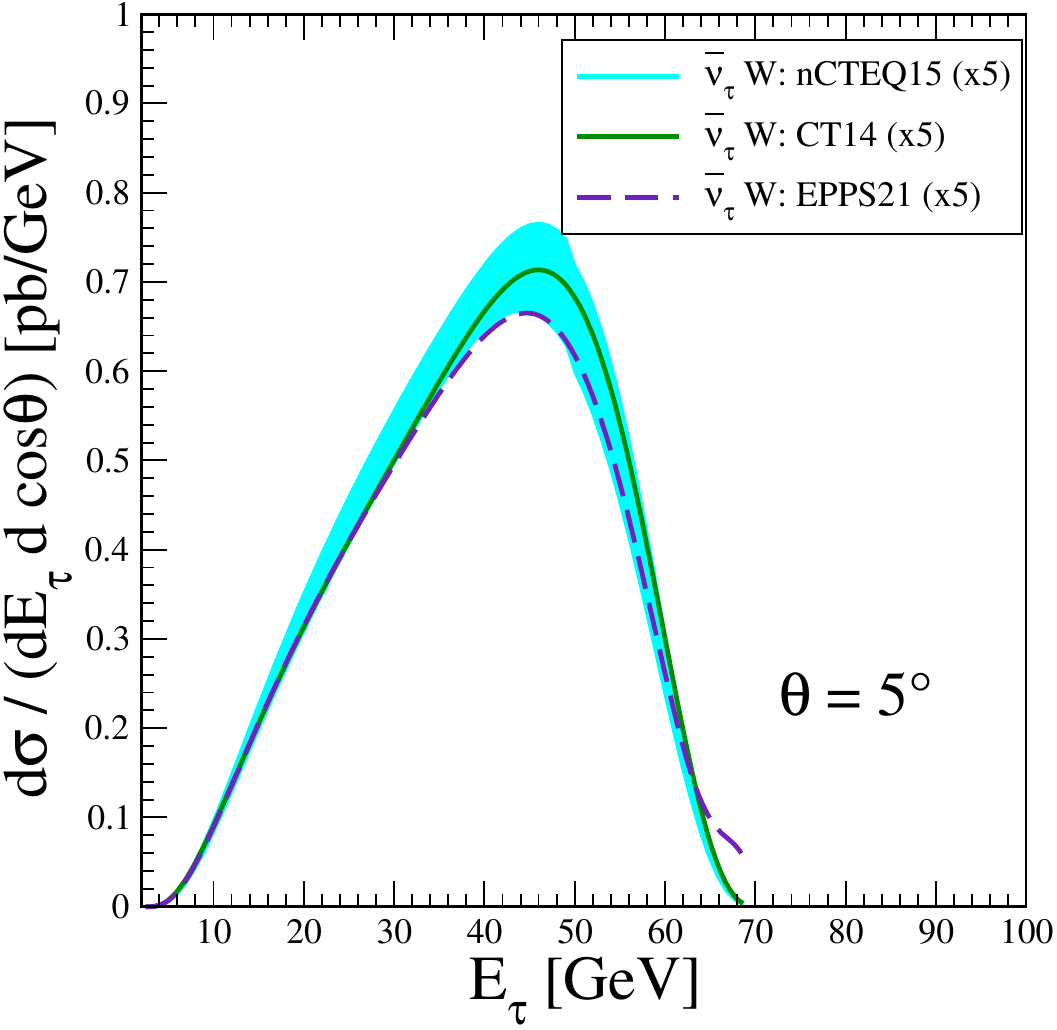} & \includegraphics[width=0.23\textwidth]{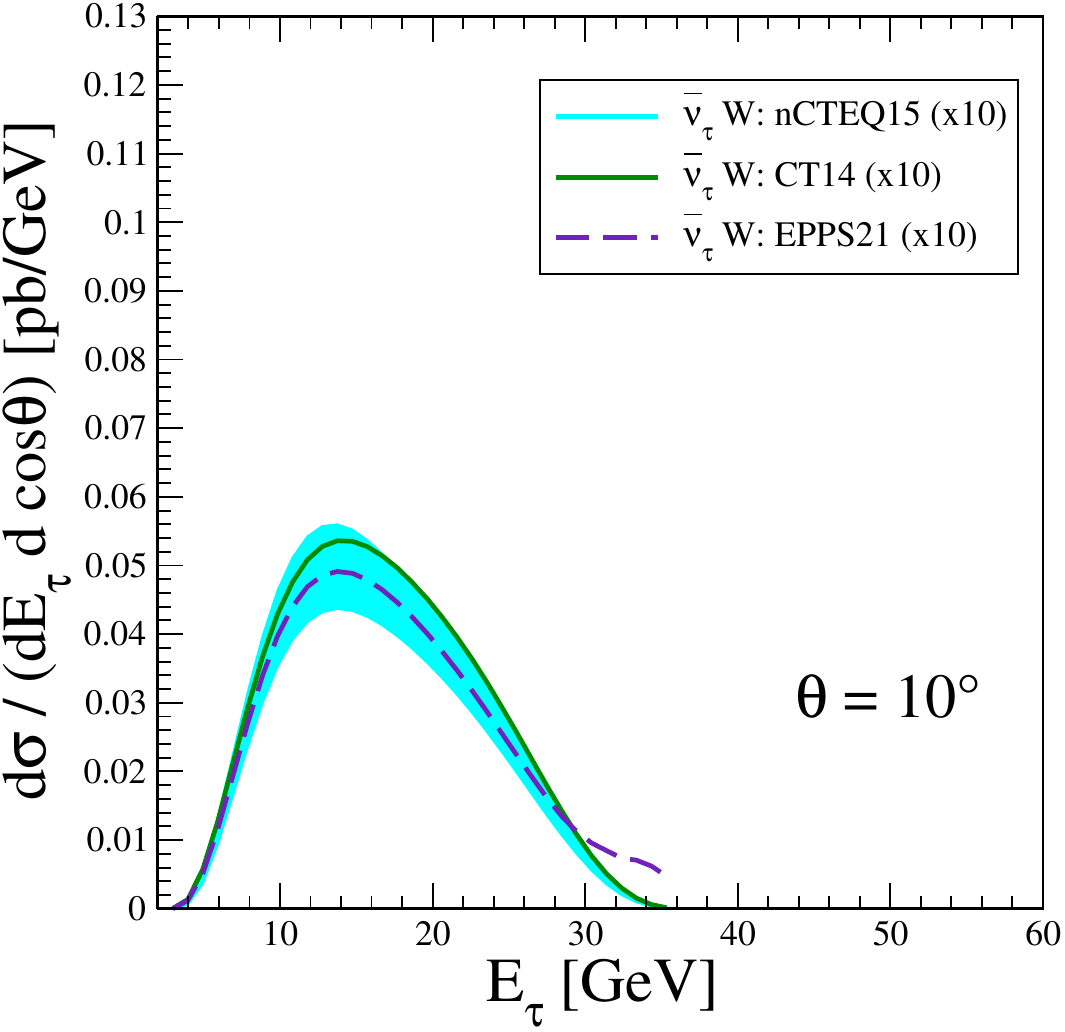}         
			\end{tabular}
\caption{ Double differential cross sections ${\nu}_{\tau}W$ (upper panels) and $\bar{\nu}_{\tau}W$ (lower panels) as a function of tau lepton energy for different values of the angle $\theta$. Results derived assuming different nPDFs and considering that the incident tau neutrino energy is equal to 100 GeV. Note the different scales of the ordinate axis in the distinct graphs. }
\label{fig_4:dsdE_nu100}
\end{figure}

\begin{figure}
	\centering
	\begin{tabular}{ccccccc}
    \includegraphics[width=0.23\textwidth]{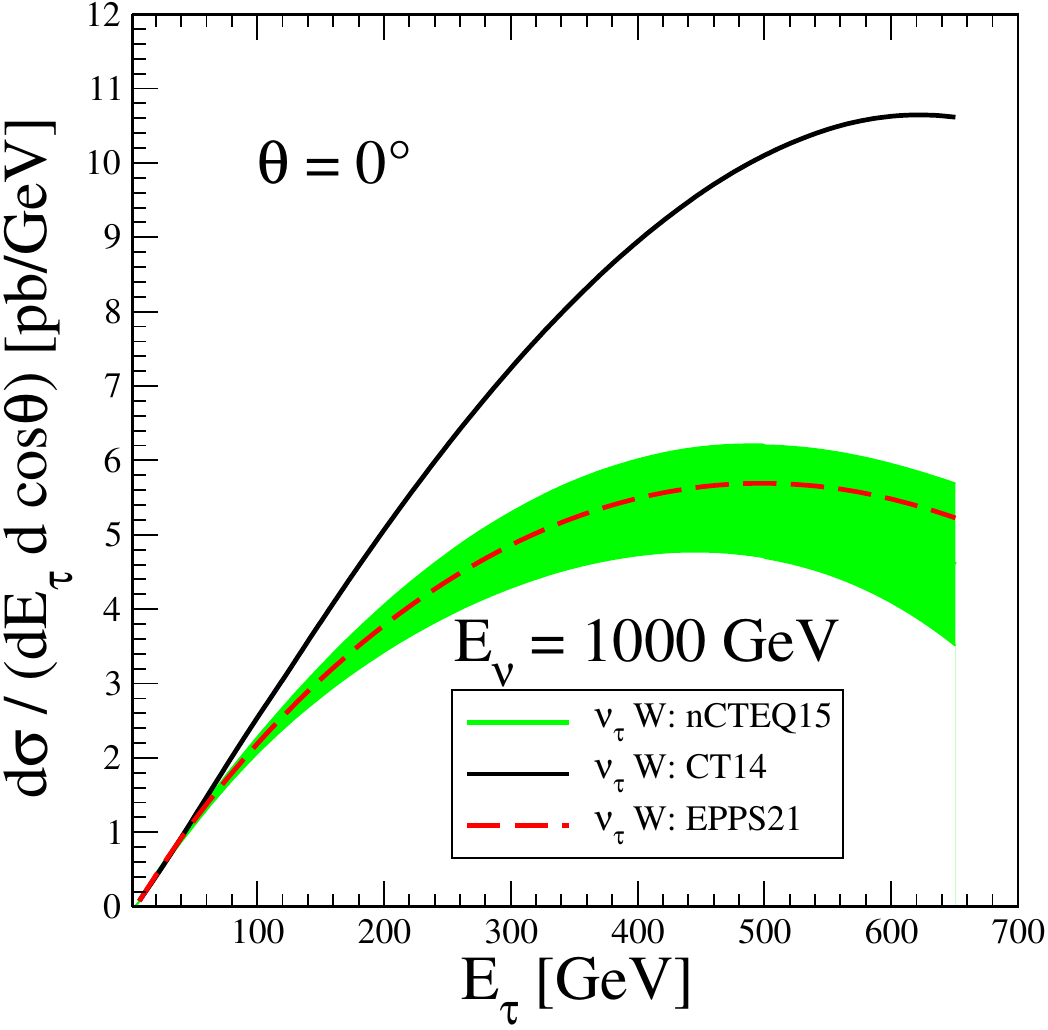} & \includegraphics[width=0.23\textwidth]{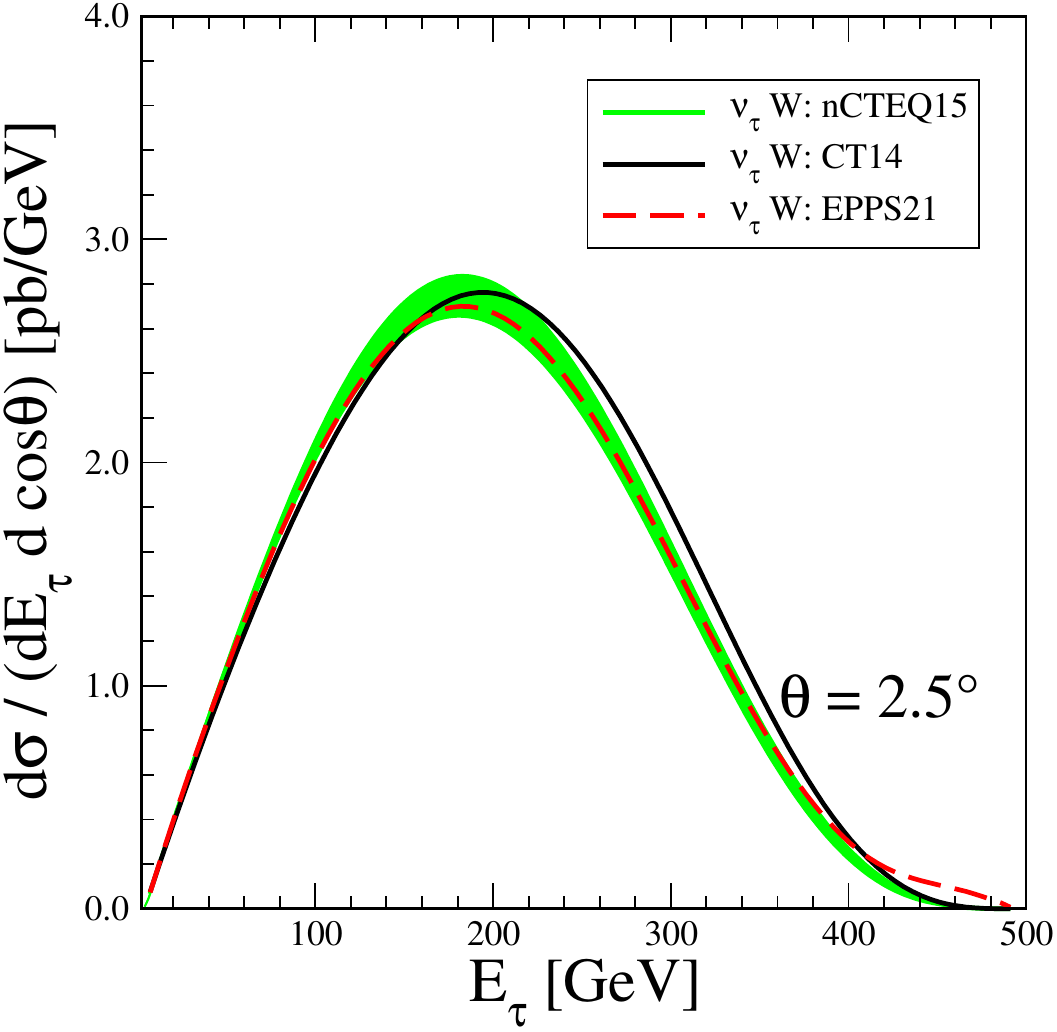} &
    \includegraphics[width=0.23\textwidth]{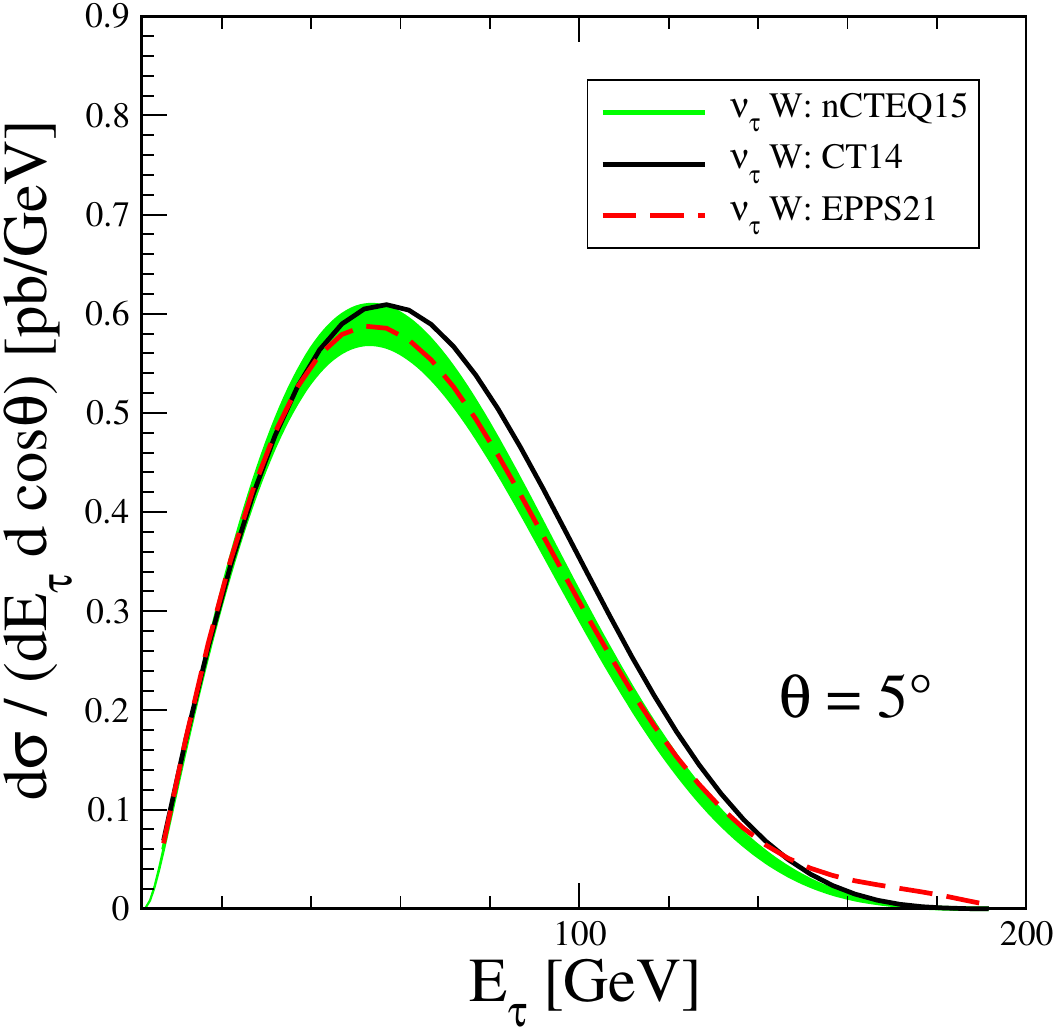} & \includegraphics[width=0.23\textwidth]{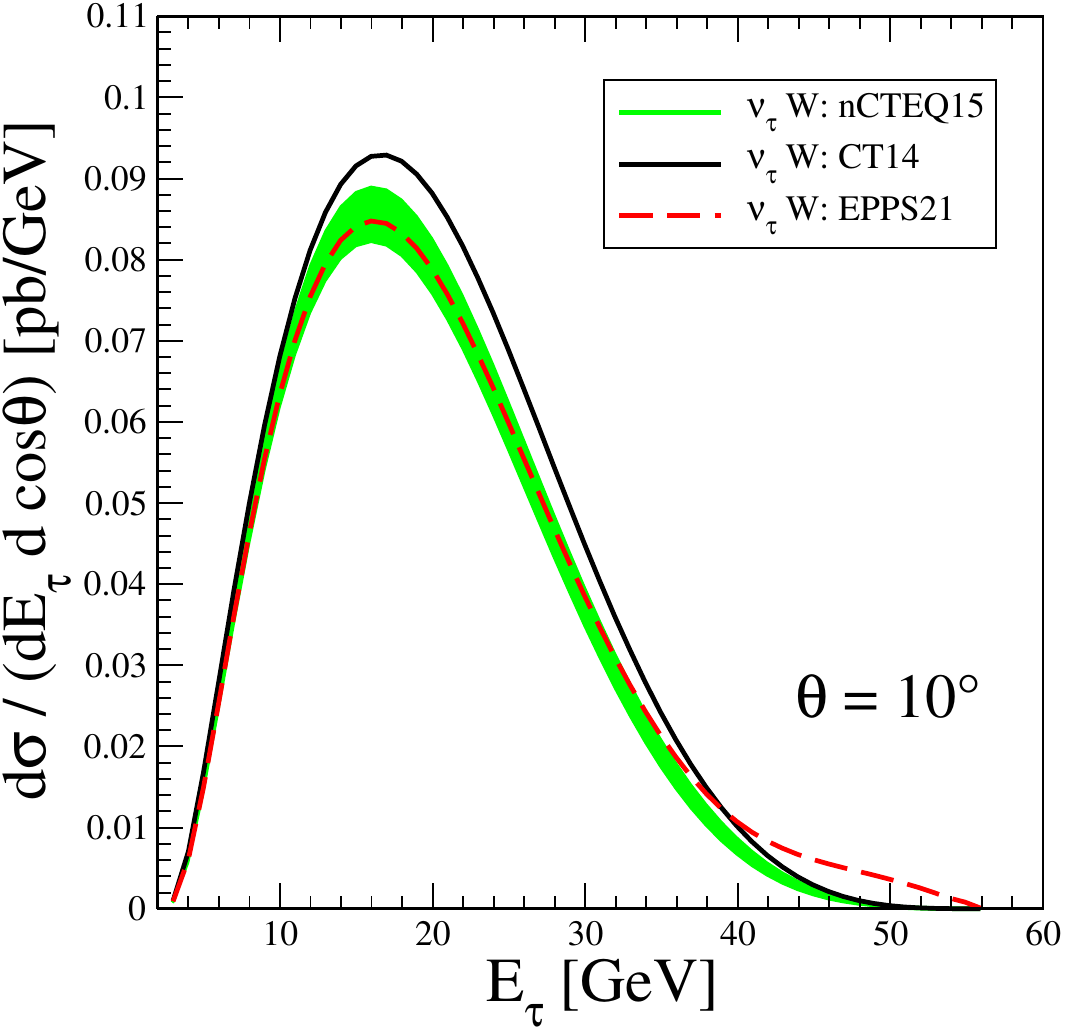}  \\
    \includegraphics[width=0.23\textwidth]{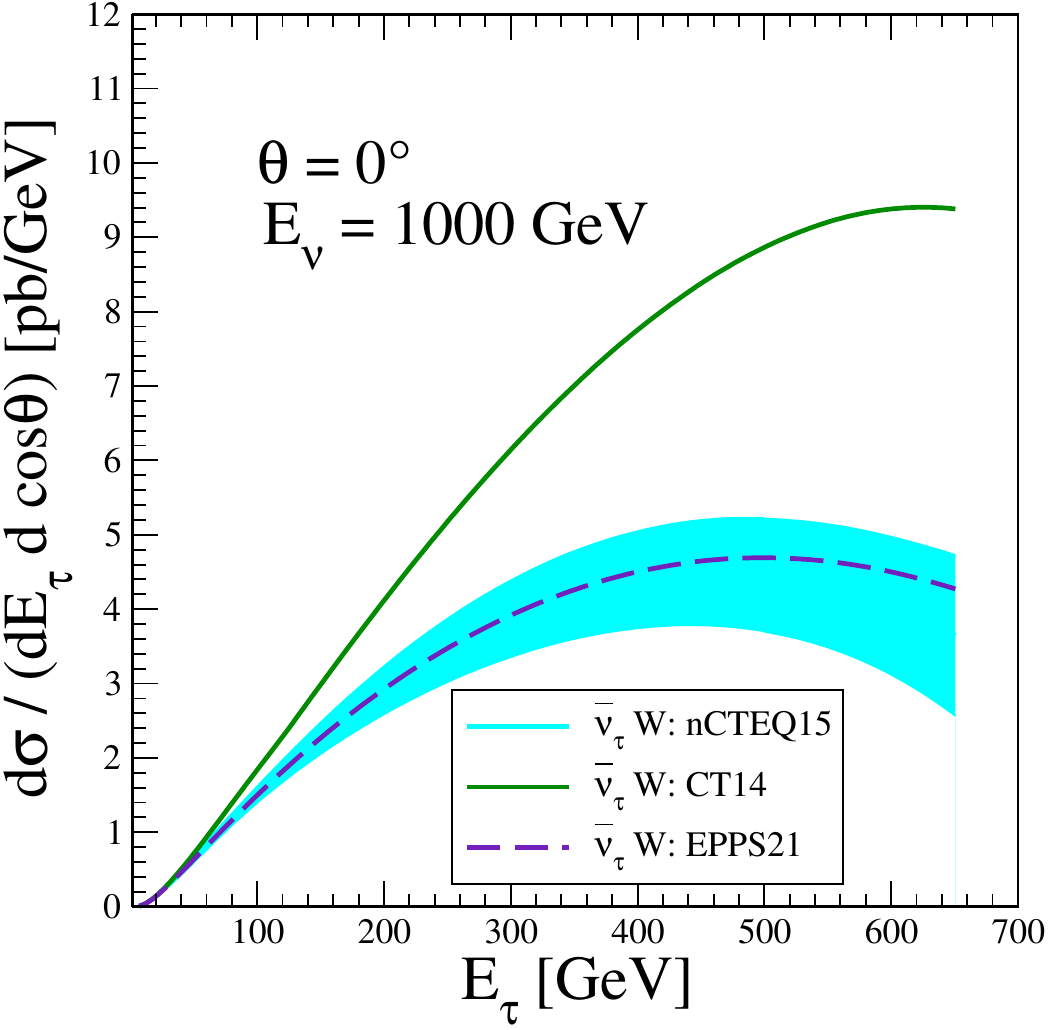} & \includegraphics[width=0.23\textwidth]{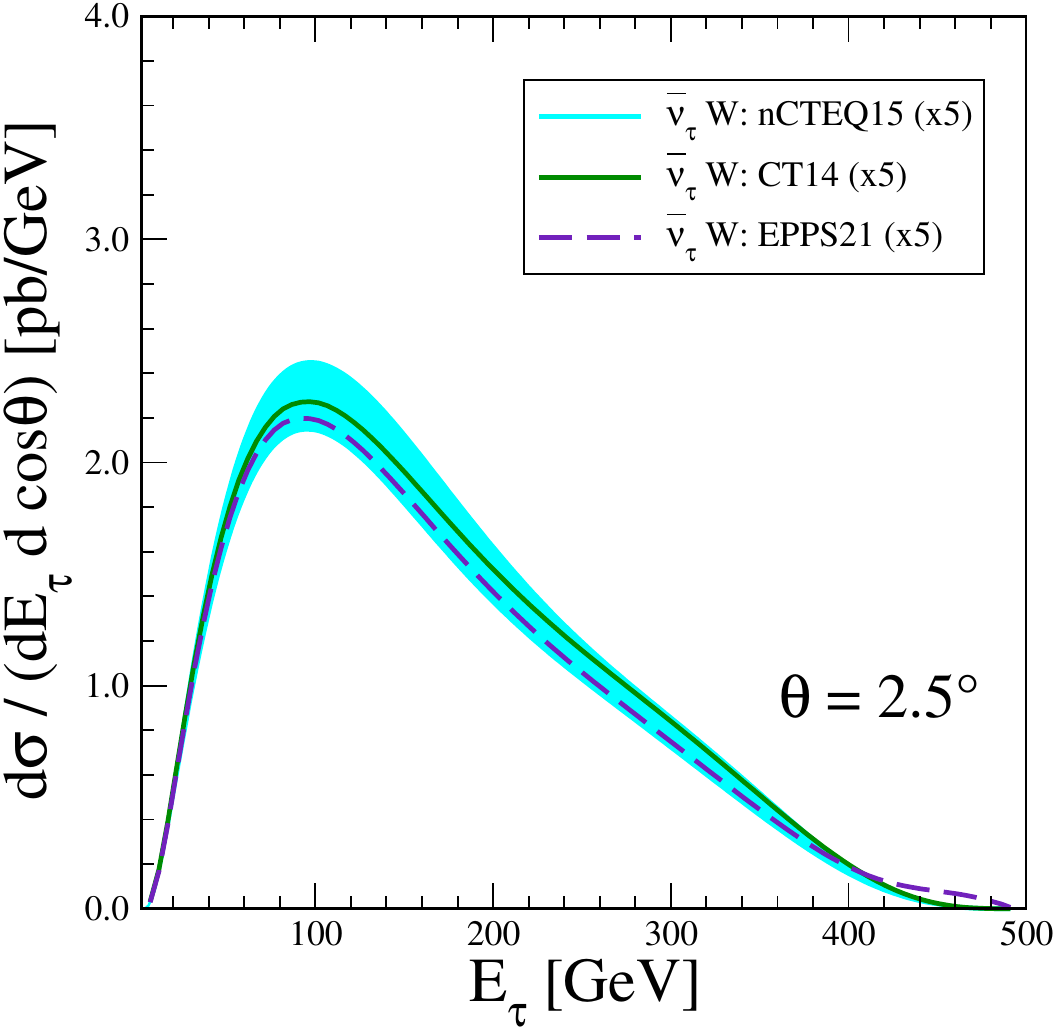} & 
    \includegraphics[width=0.23\textwidth]{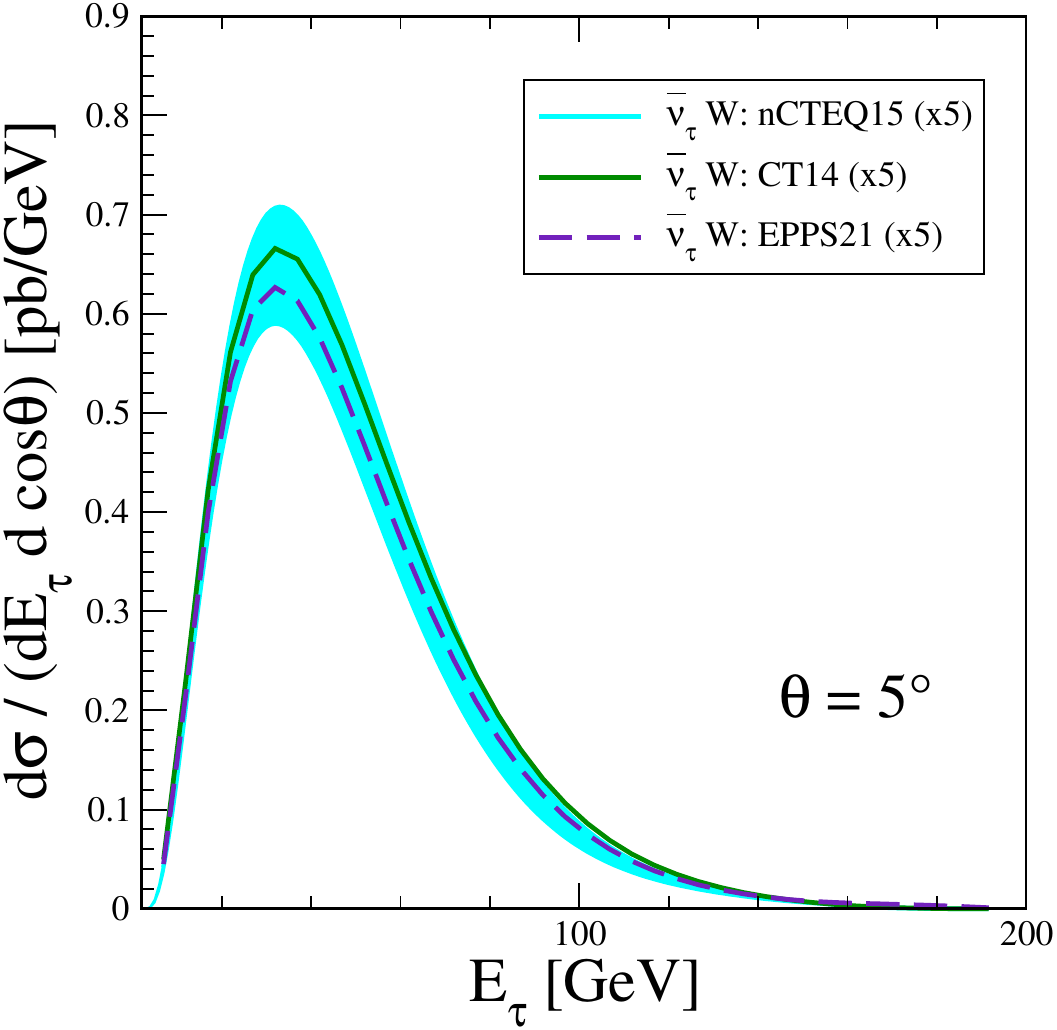} & \includegraphics[width=0.23\textwidth]{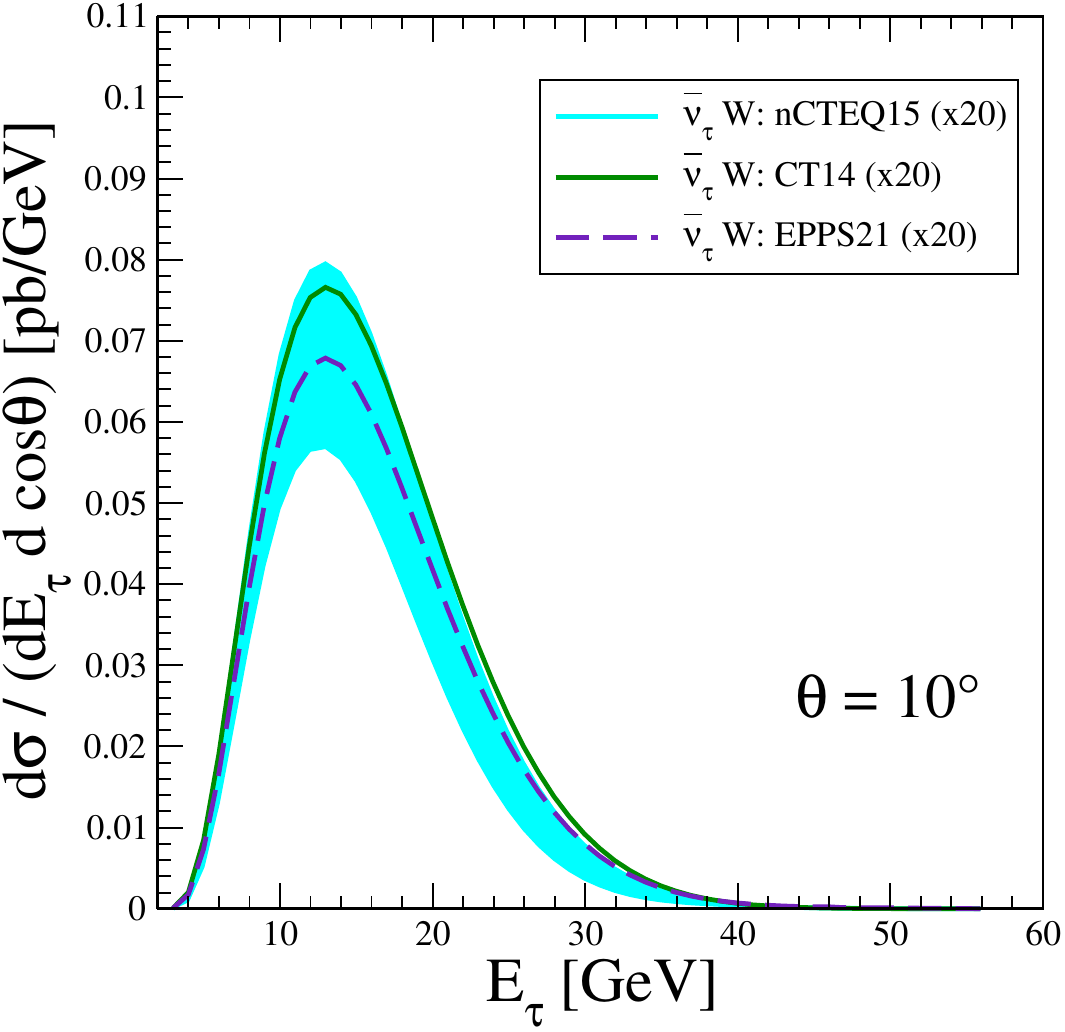}    
			\end{tabular}
\caption{ Double differential cross sections ${\nu}_{\tau}W$ (upper panels) and $\bar{\nu}_{\tau}W$ (lower panels) as a function of tau lepton energy for different values of the angle $\theta$. Results derived assuming different nPDFs and considering that the incident tau neutrino energy is equal to 1000 GeV. Note the different scales of the $y$ axis in the distinct graphs. }
\label{fig_4:dsdE_nu1000}
\end{figure}

The total cross section of tau neutrinos with nucleons did not show significant dependence on nuclear effects for either of the parameterizations we tested. This motivates us to look directly at Equation (\ref{eq_4:sigma2}), and calculate the differential cross section with respect to the energy of the tau produced in the interaction. In Figures \ref{fig_4:dsdE_nu100} and \ref{fig_4:dsdE_nu1000}, we present these observables for tau neutrinos of 100 GeV and 1000 GeV, respectively. The upper rows show the results for neutrinos, and the lower rows for antineutrinos. We also have, from left to right, the differential cross sections with respect to tau energy for different tau scattering angles: $0^\mathrm{o}$, $2.5^\mathrm{o}$, $5^\mathrm{o}$, and $10^\mathrm{o}$. Our results show that the differential cross section becomes sensitive to the nuclear effects predicted for tungsten at low tau scattering angles relative to the neutrino's direction of incidence: the effects are quite significant for $\theta = 0^\mathrm{o}$. In this kinematic region of small scattering angles, the cross sections with nuclear effects decrease drastically, becoming less than 50\% of the prediction without nuclear effects for 1000 GeV neutrinos and taus with energies greater than 500 GeV.

\section{ Tau polarization in neutrino interactions at the LHC}

One of the main objectives of our study is the investigation of the degree of polarization of the tau produced in charged current interactions of tauonic neutrinos. We will initially study the implications of assuming a polarized tau in the cross section expressions. The polarized tau in the final state alters the leptonic tensor of the scattering matrix element to \cite{Zaidi:2023hdd}
\begin{eqnarray}
    L^{pol}_{\mu\nu}(s,h) = \frac{1}{2} 
    [L^{}_{\mu\nu}
    \mp h m_\tau s^{\alpha}
    (
    k_\mu g_{\nu\alpha} + k_{\nu}g_{\mu\alpha}
    -k_{\alpha} g_{\mu\nu} 
    \pm i \epsilon_{\mu\nu\alpha\beta} k^{\beta}
    )
    ] \, ,
    \label{eq_4:tensorLpol}
\end{eqnarray}
where $s^{\alpha}$ is the spin four-vector of the tau and $h = \pm 1$ is its helicity. Contracting this new leptonic tensor with the hadronic tensor of Equation (\ref{eq_4:tensorW}), it is possible to show that the differential cross section in terms of the differential cross section for the unpolarized tau is given by \cite{Zaidi:2023hdd}
\begin{eqnarray}
    \frac{\mathrm{d}^2\sigma_A^{pol}}{\mathrm{d}E_\tau\mathrm{d\,cos}\,\theta} = 
    \frac{1}{2}(1+s_\mu P^{\mu}) 
    \frac{\mathrm{d}^2\sigma_A}{\mathrm{d}E_\tau\mathrm{d\,cos}\,\theta} \, ,
    \label{eq_4:sigmaPol}
\end{eqnarray}
where $P^{\mu}$ is the polarization four-vector of the tau. $P^{\mu}$ can be decomposed into longitudinal, transverse, and perpendicular components. The longitudinal direction, $P_L$, is in the direction of $k'$, the transverse direction, $P_T$, is orthogonal to the direction of the tau and contained in the neutrino-tau plane, and finally the perpendicular direction, $P_P$, is defined as being orthogonal to both the neutrino and tau directions. It can be shown that $P_{P}$ has all zero components. The non-zero components of the four-vector polarization of the tau in the rest frame of the hadronic target are \cite{Zaidi:2023hdd}
\begin{eqnarray}
    \begin{aligned}
    P_L = 
    \mp \frac{E_\nu}{L^{unpol}_{\mu\nu} W_{A}^{\mu\nu}} \left\{ \left[ 2F^A_{1}(x,Q^2) - F^A_{4}(x,Q^2)\frac{m_\tau^{2}}{\nu M_A x} \right] (|\vec{k}'| - E_\tau\mathrm{cos}\,\theta) +  \right.\\
    + F^A_{2}(x,Q^2)\frac{M_A}{\nu}(|\vec{k}'| - E_\tau\mathrm{cos}\,\theta) +  \\
    \pm \frac{F^A_{3}(x,Q^2)}{\nu}\frac{1}{\nu}[ (E_\nu + E_\tau )|\vec{k}'| - (| \vec{k}' |^2 + E_\nu E_\tau)\mathrm{cos}\,\theta] + \\ 
    \left. - F^A_{5}(x,Q^2)\frac{2m_\tau^2}{\nu} \mathrm{cos}\,\theta \right\}
    \label{eq_4:PL}
    \end{aligned}
\end{eqnarray}
and
\begin{eqnarray}
\begin{aligned}
    P_T = 
    \mp \frac{m_\tau \mathrm{sin}\,\theta E_\nu}{L^{unpol}_{\mu\nu} W_{A}^{\mu\nu}} \left[ 2F^A_{1}(x,Q^2) -  F^A_{2}(x,Q^2)\frac{M_A}{\nu} 
    \pm F^A_{3}(x,Q^2) \frac{E_\nu}{\nu} - F^A_{4}(x,Q^2)\frac{m_\tau^{2}}{\nu M_A x} + \right. \\
    \left. + F^A_{5}(x,Q^2)\frac{2E_\tau}{\nu} \right] \, .
     \label{eq_4:PT}
\end{aligned}
\end{eqnarray}	
We can see that $P_T$ is proportional to $m_\tau$ and $\mathrm{sen}\,\theta$, therefore it becomes zero for $\theta = 0^\mathrm{o}$ and negligible for tau energy much greater than its mass.

\begin{figure}
	\centering
	\begin{tabular}{ccccc}       
 \raisebox{0.2\height}{\includegraphics[width=0.2\textwidth]{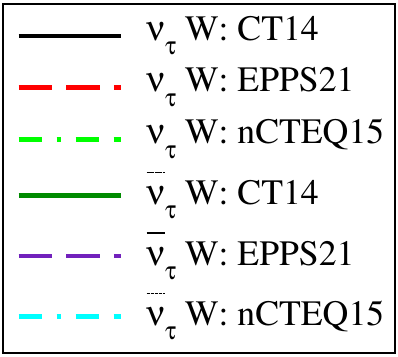}} &
 \includegraphics[width=0.23\textwidth]{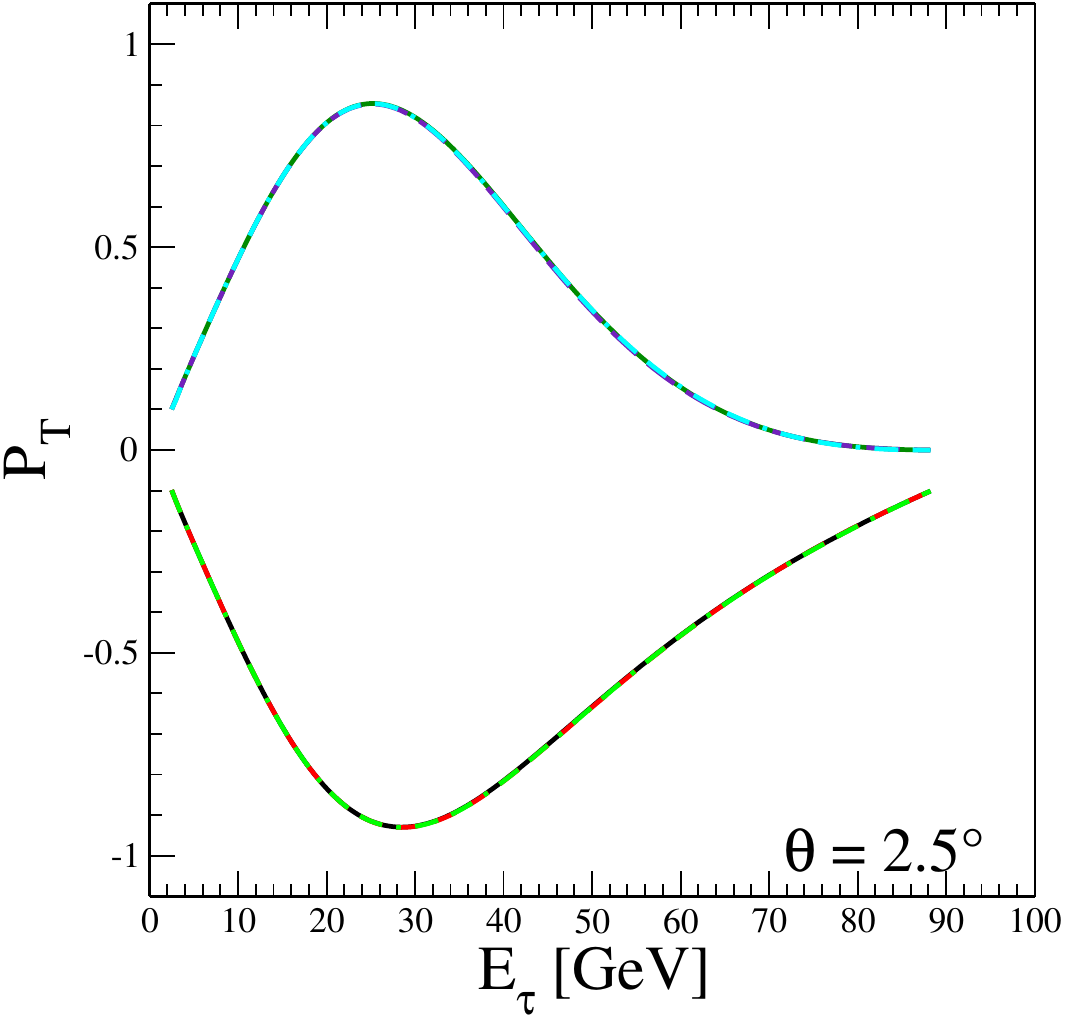} & 
 \includegraphics[width=0.23\textwidth]{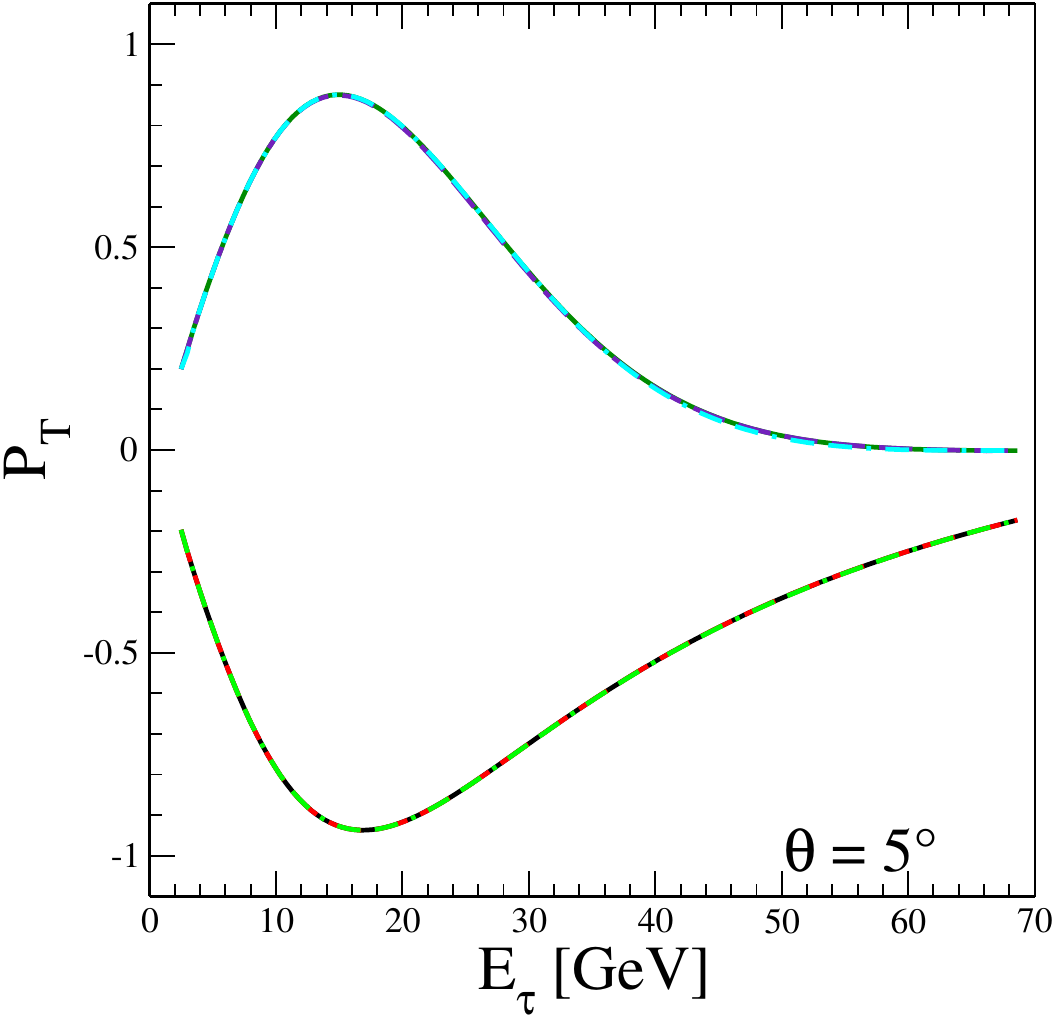} &  \includegraphics[width=0.23\textwidth]{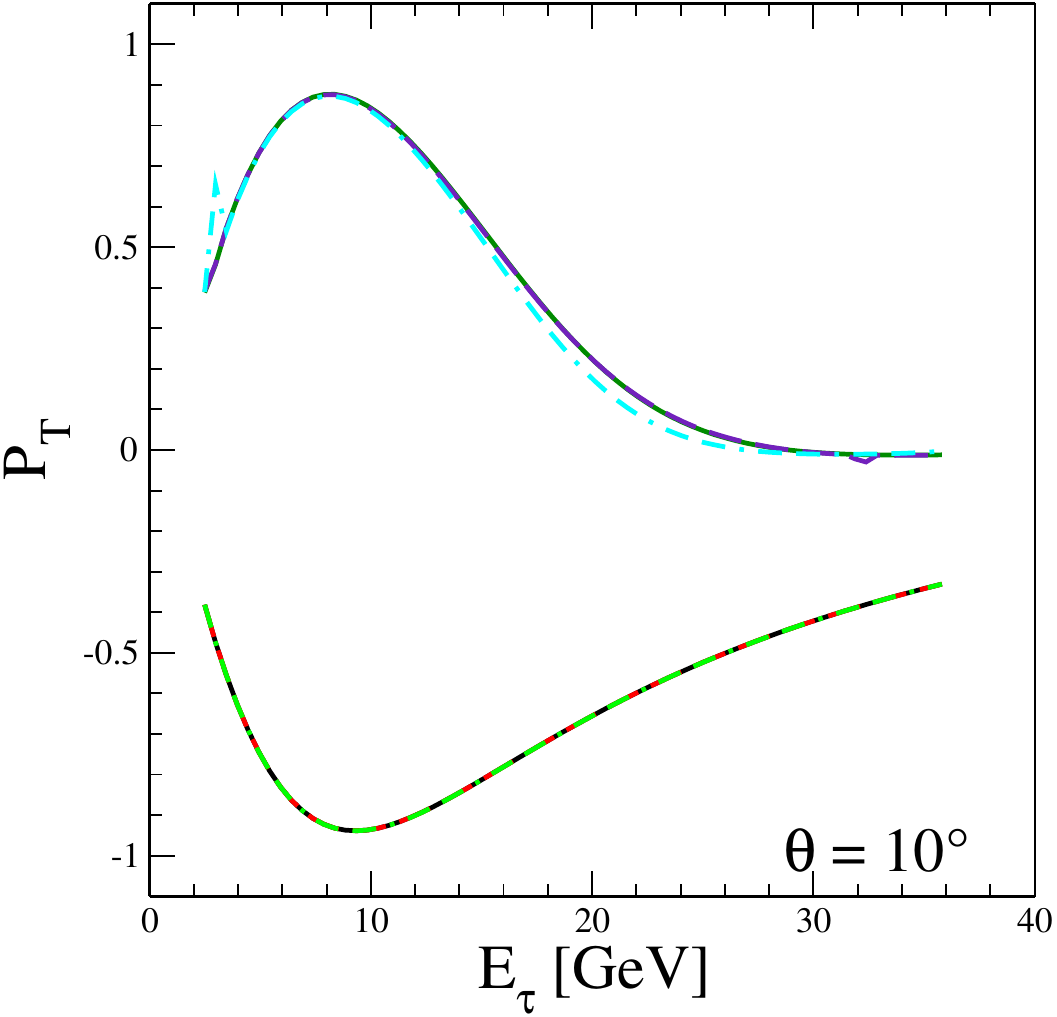} \\
 \includegraphics[width=0.23\textwidth]{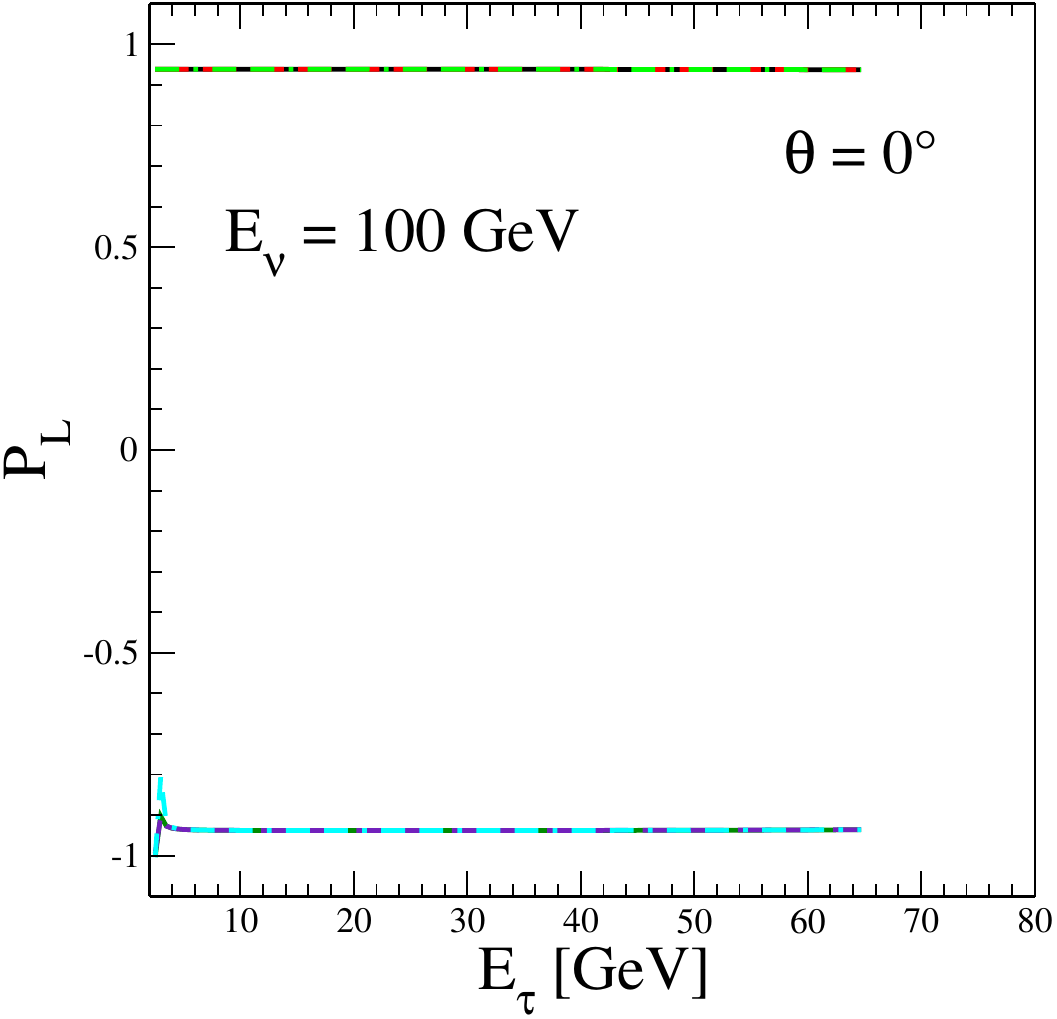} & \includegraphics[width=0.23\textwidth]{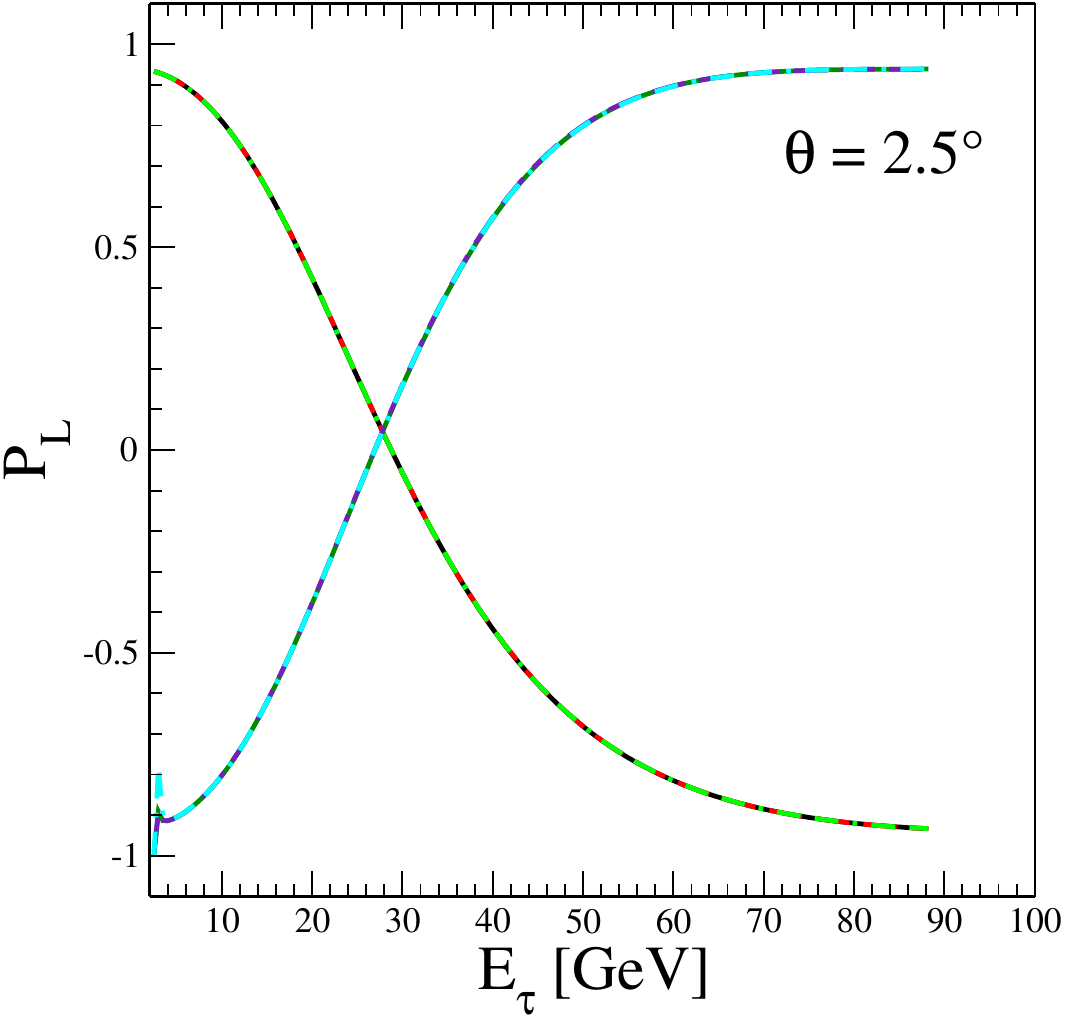} & 
 \includegraphics[width=0.23\textwidth]{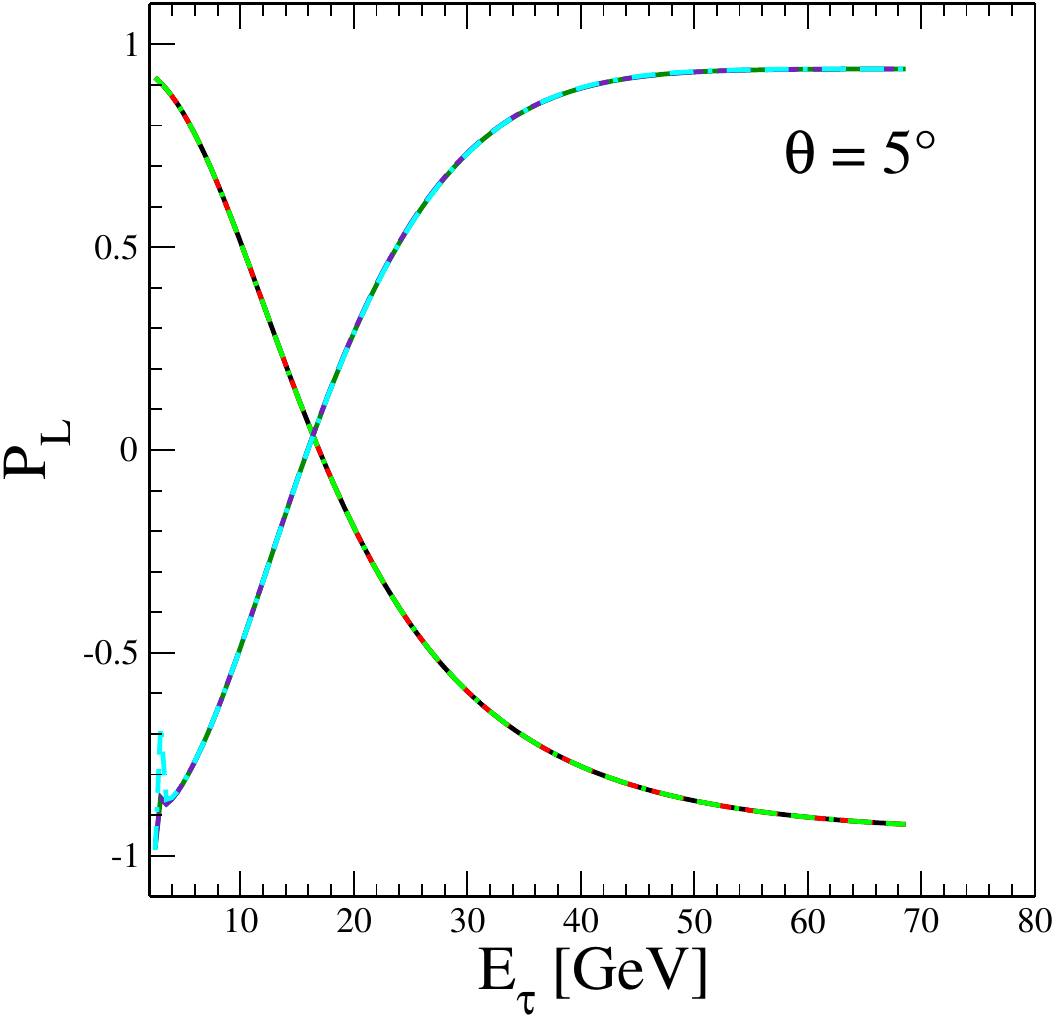} &  \includegraphics[width=0.23\textwidth]{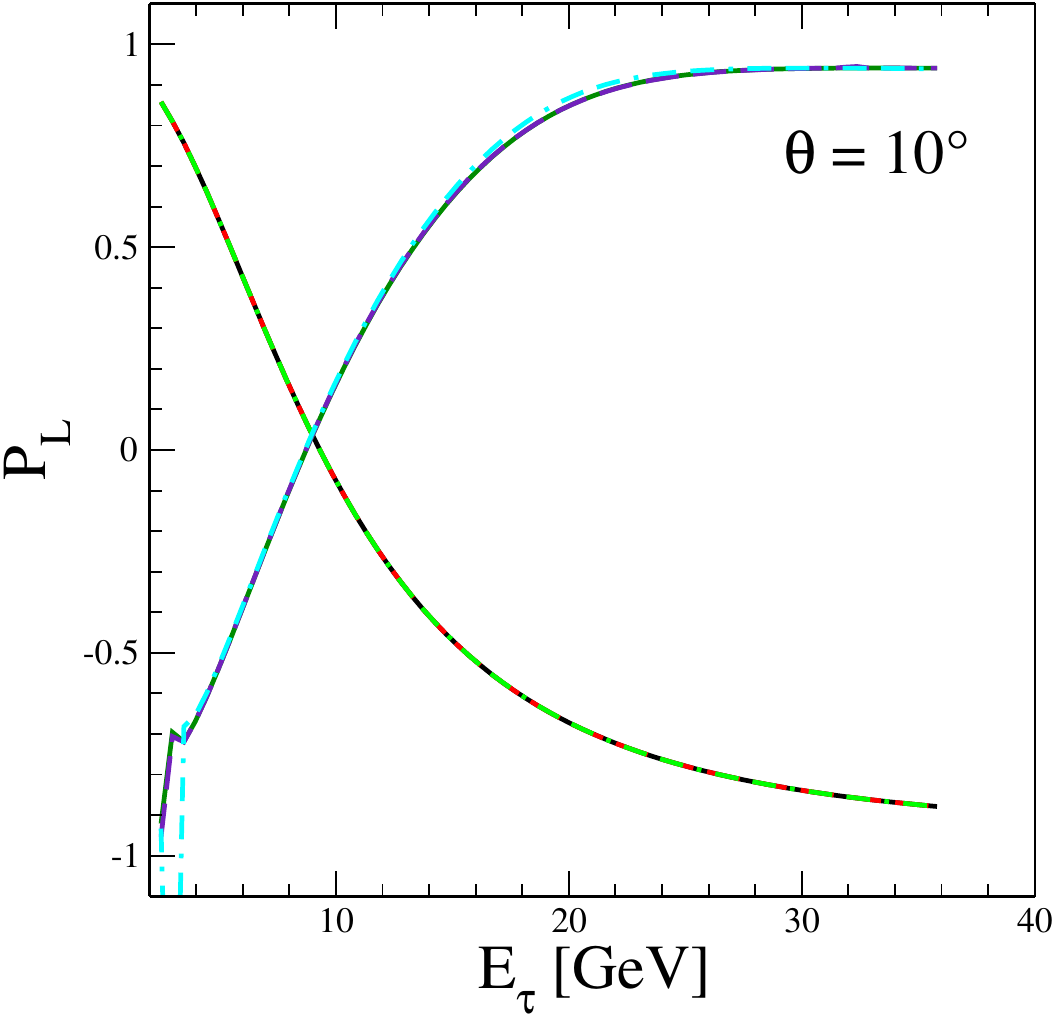}
			\end{tabular}
\caption{ Transverse (upper panels) and longitudinal (lower panels) components of the polarization vector as a function of tau lepton energy for an incident (anti)neutrino with an energy of 100 GeV. Results for different values of the angle $\theta$, derived assuming different nPDFs. }
\label{fig_4:PL100}
\end{figure}

\begin{figure}
	\centering
	\begin{tabular}{ccccc} 
    \raisebox{0.2\height}{\includegraphics[width=0.2\textwidth]{Images/Polarization/legenda.pdf}} &
    \includegraphics[width=0.23\textwidth]{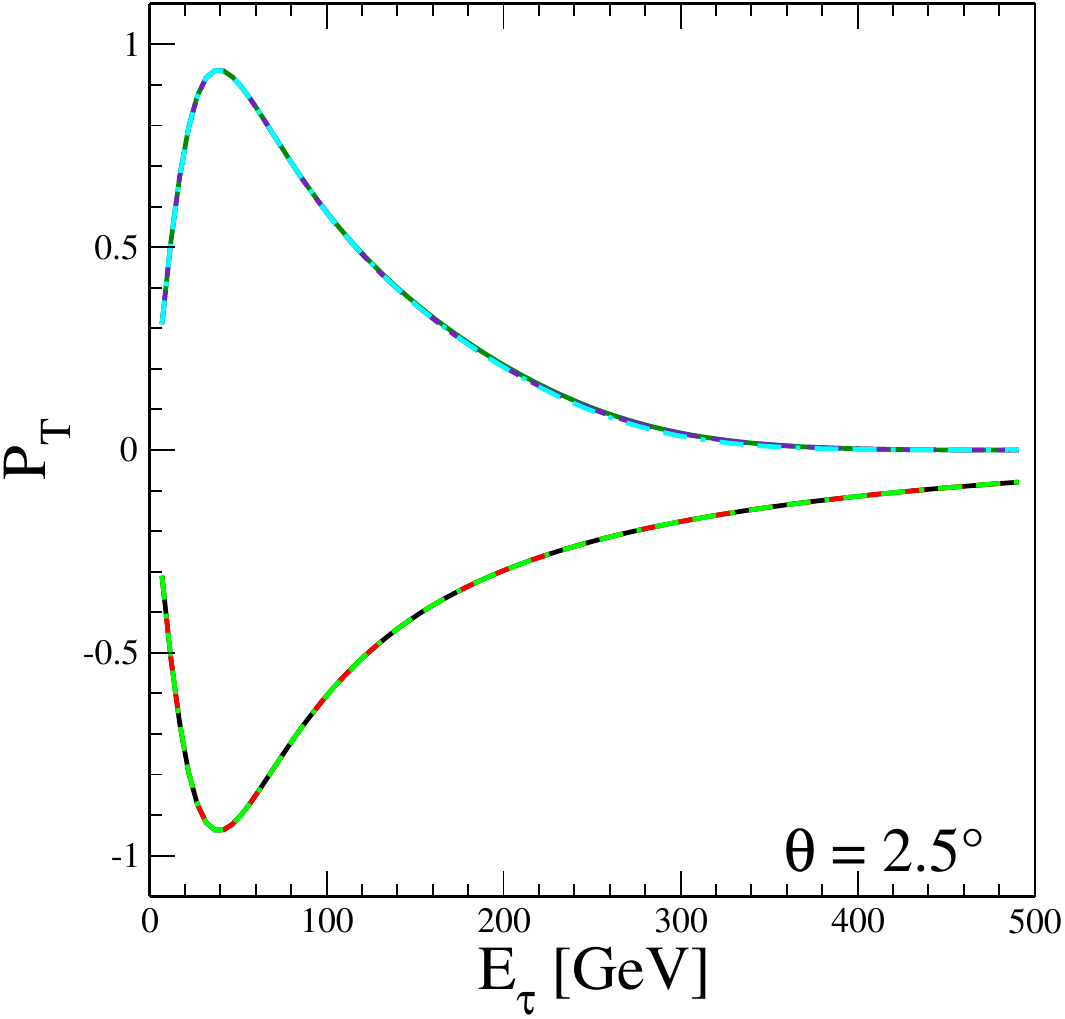} &  
    \includegraphics[width=0.23\textwidth]{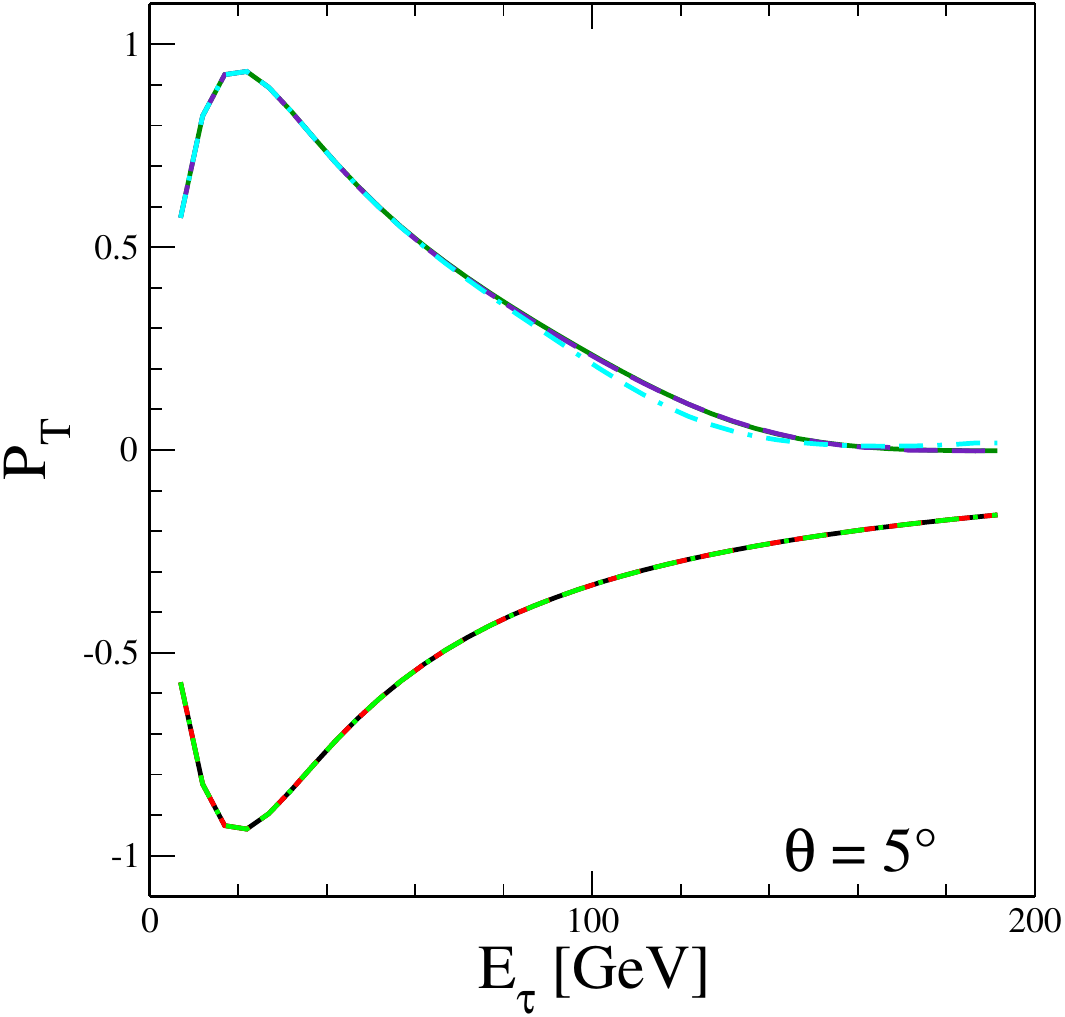} &  \includegraphics[width=0.23\textwidth]{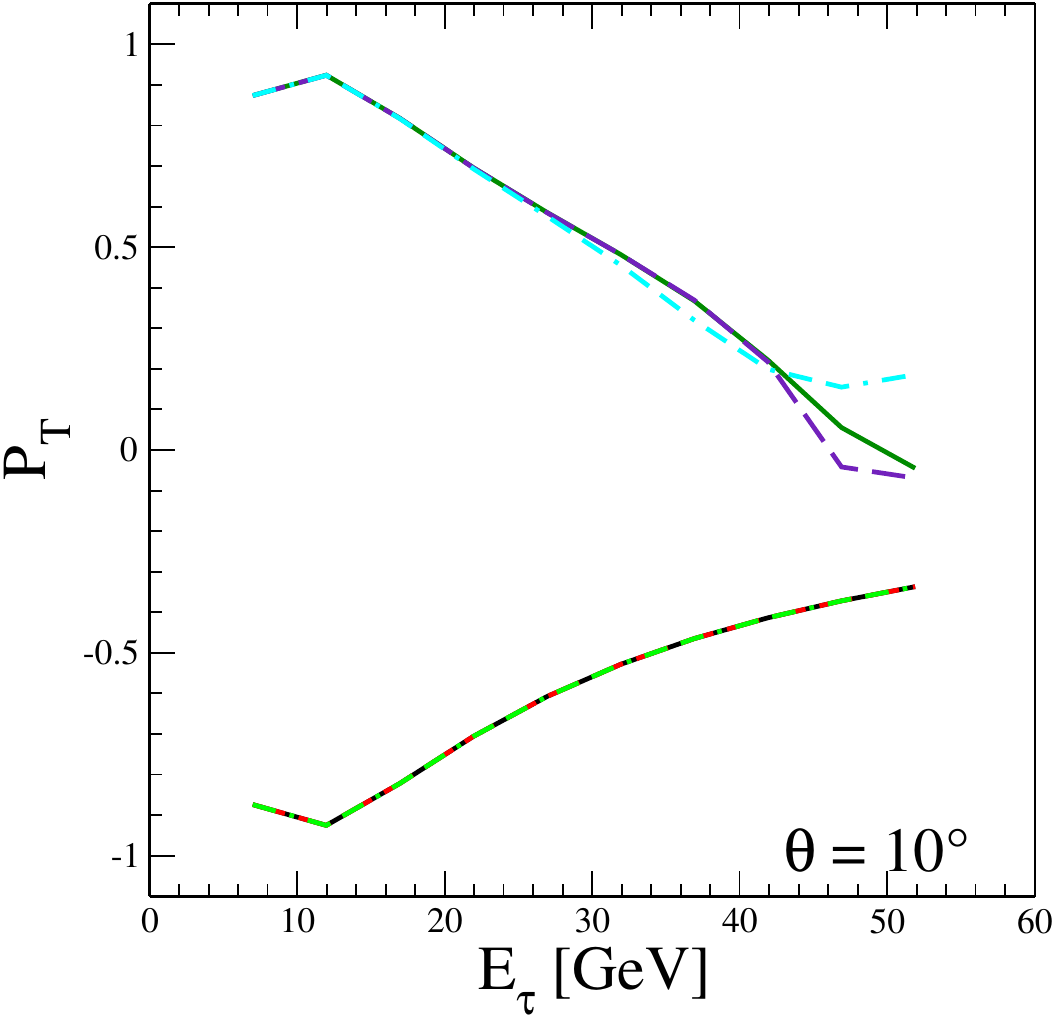} \\
    \includegraphics[width=0.23\textwidth]{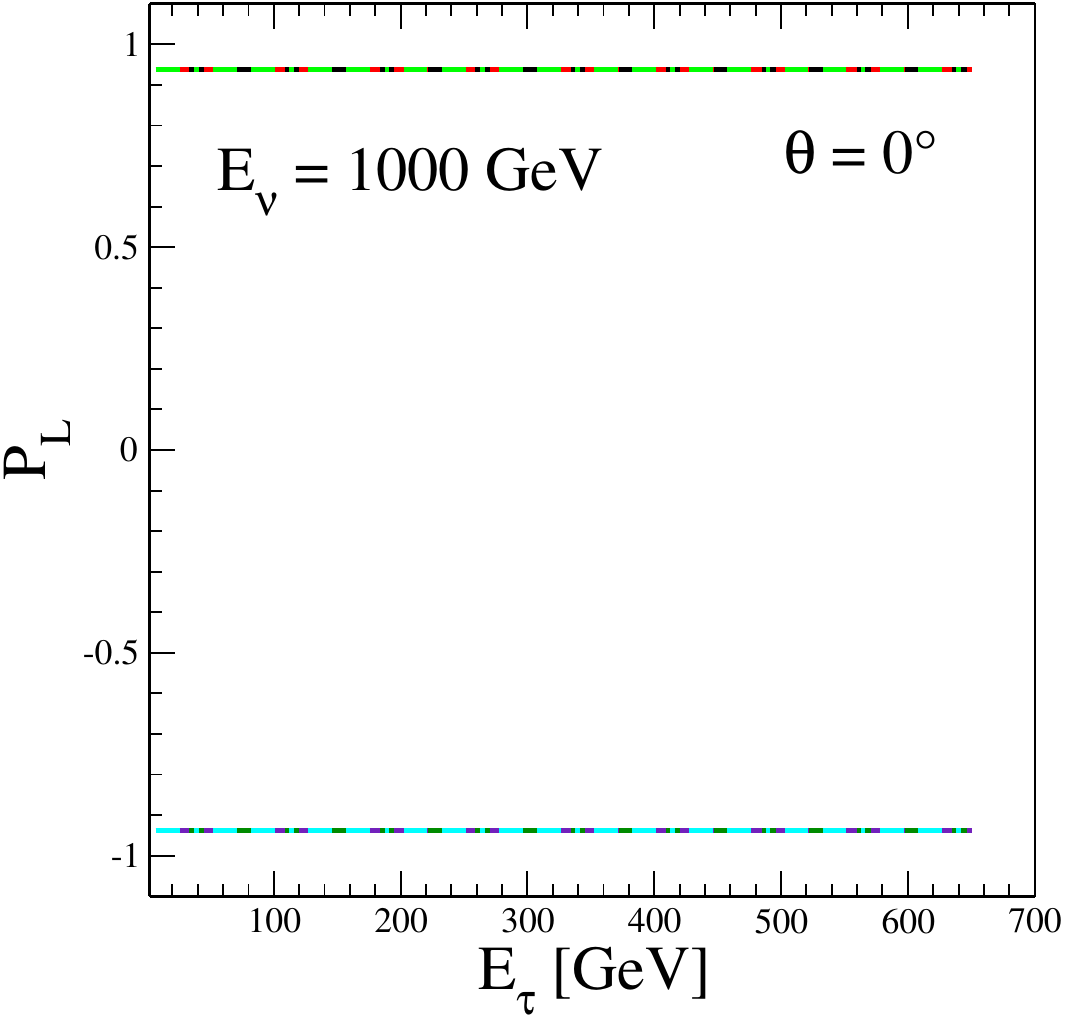} &
    \includegraphics[width=0.23\textwidth]{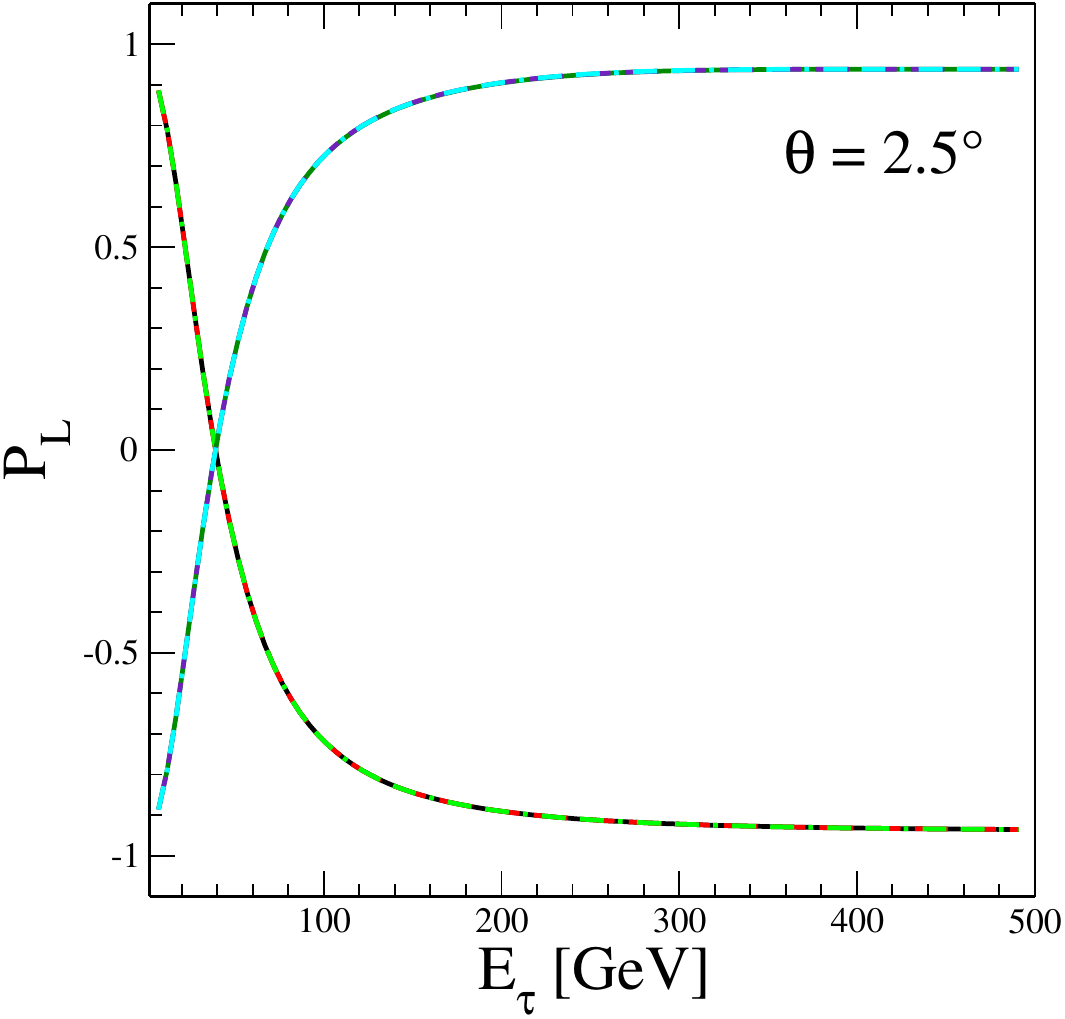} &  
    \includegraphics[width=0.23\textwidth]{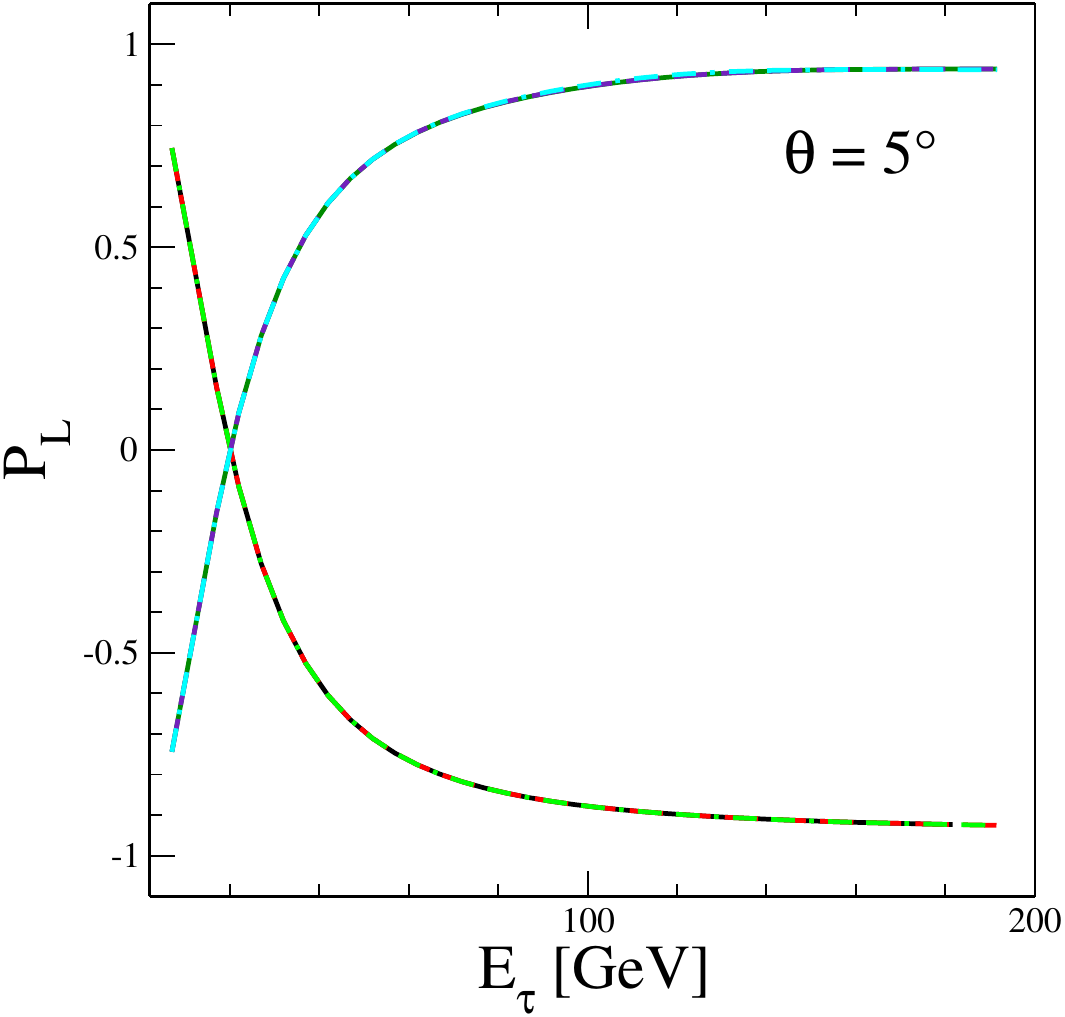} &  \includegraphics[width=0.23\textwidth]{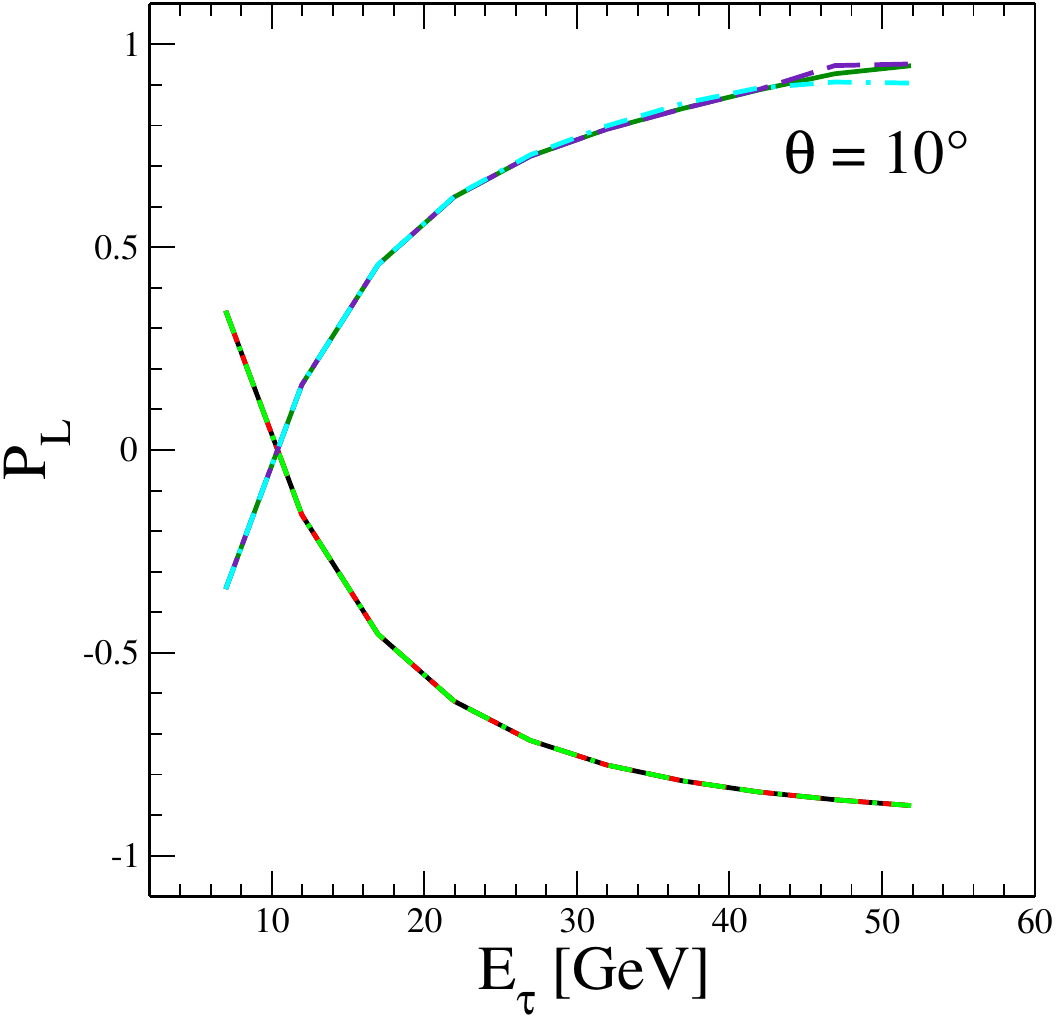}
			\end{tabular}
\caption{ Transverse (upper panels) and longitudinal (lower panels) components of the polarization vector as a function of tau lepton energy for an incident (anti)neutrino with an energy of 1000 GeV. Results for different values of the angle $\theta$, derived assuming different nPDFs. }
\label{fig_4:PL1000}
\end{figure}

Using Equations (\ref{eq_4:PL}) and (\ref{eq_4:PT}), we can evaluate the polarization components of the tau as a function of the tau energy for different incident neutrino energy values and different tau scattering angles. We present this analysis in Figures \ref{fig_4:PL100} and \ref{fig_4:PL1000}, for incident neutrinos with energies of 100 GeV and 1000 GeV, respectively. Again, we present our results for different scattering angles of the produced tau, which are, from left to right in the figure panel: $0^{\mathrm{o}}$, $2.5^{\mathrm{o}}$, $5^{\mathrm{o}}$, and $10^{\mathrm{o}}$. The transverse component of the tau polarization four-vector is always negative (positive) for neutrinos (antineutrinos), while the longitudinal component is negative (positive) for low neutrino (antineutrino) energies and reverses its sign as the tau energy increases. This reversal becomes faster as the incident neutrino energy increases. The figures also show that the polarization components are weakly dependent on the parameterizations of the nPDFs used. We also see that the uncertainty in these polarization components is small, so it is not visible in the lines presented for the nCTEQ15 parameterization.

\begin{figure}
	\centering
	\begin{tabular}{ccccccc}
    \includegraphics[width=0.23\textwidth]{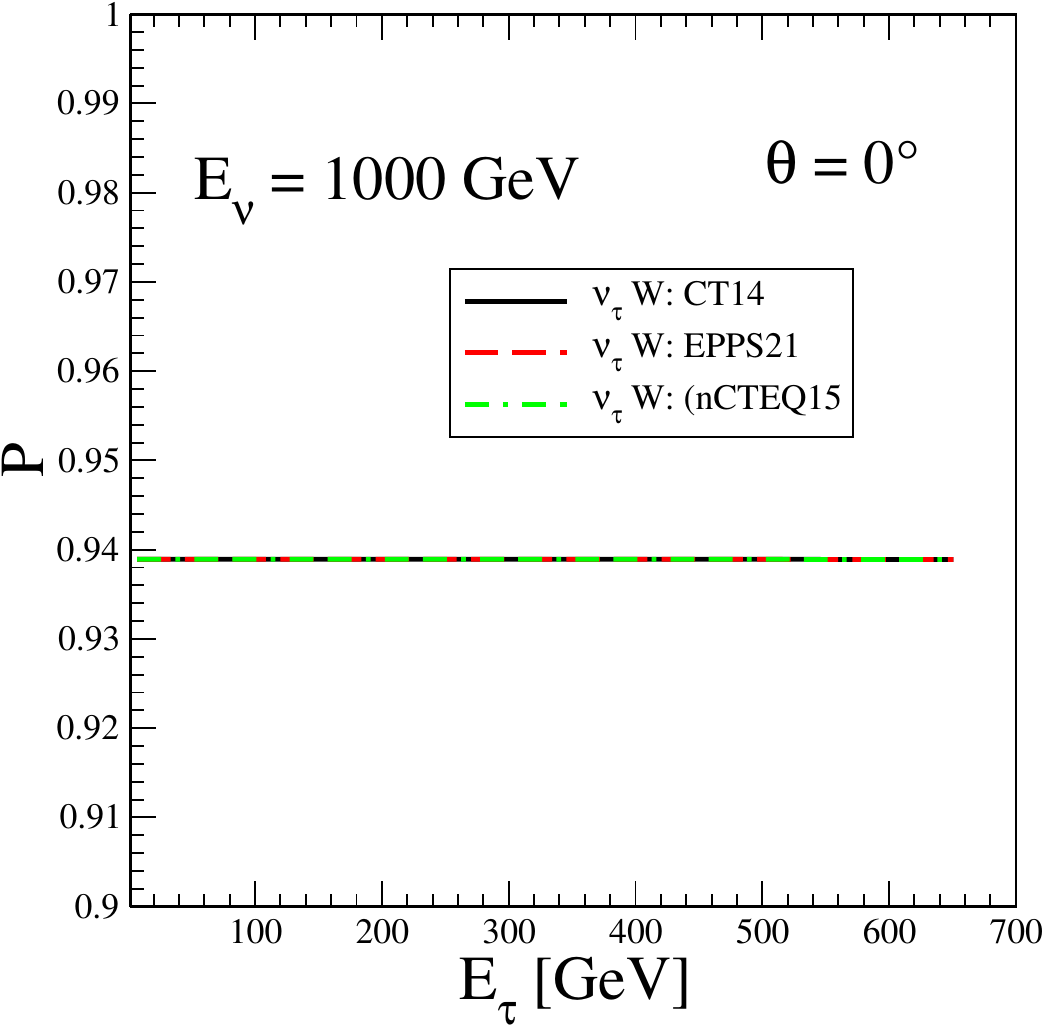} & \includegraphics[width=0.23\textwidth]{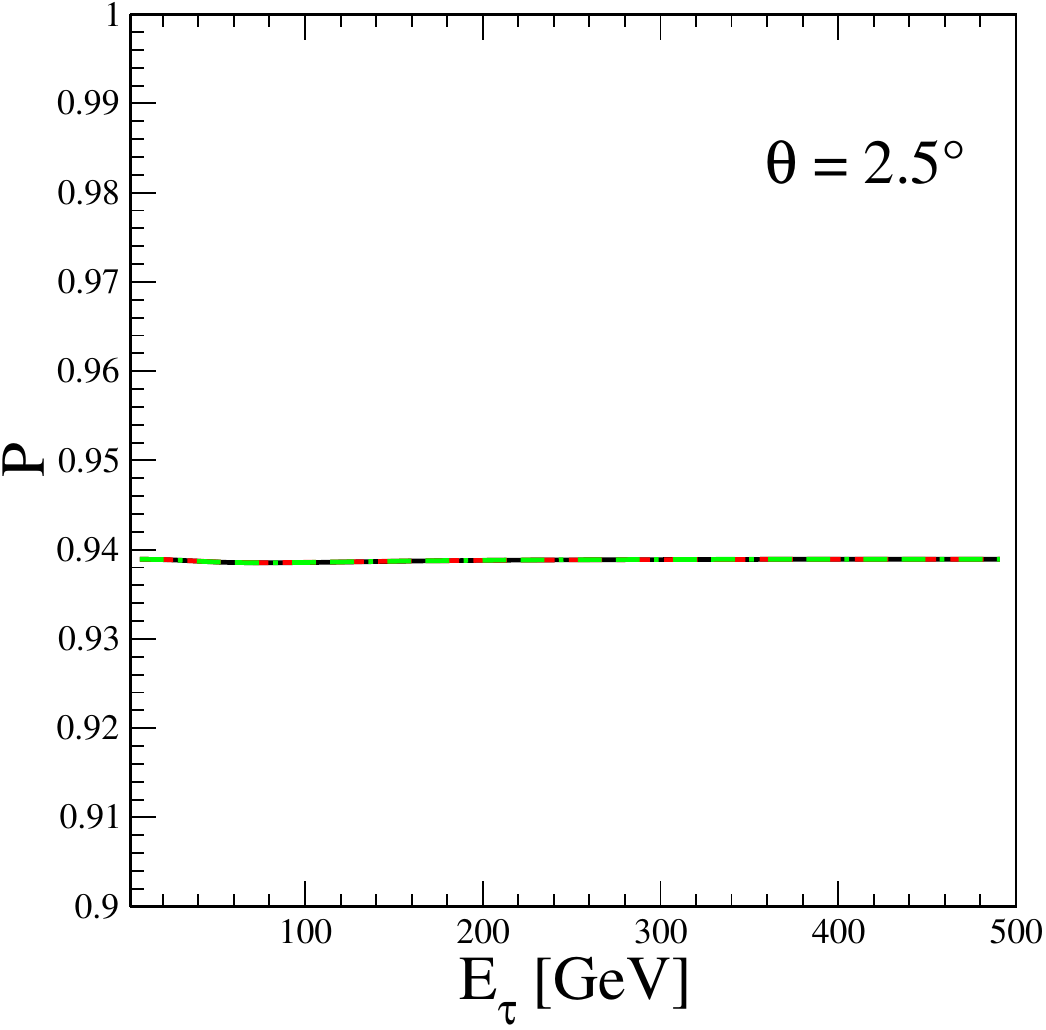} &  
    \includegraphics[width=0.23\textwidth]{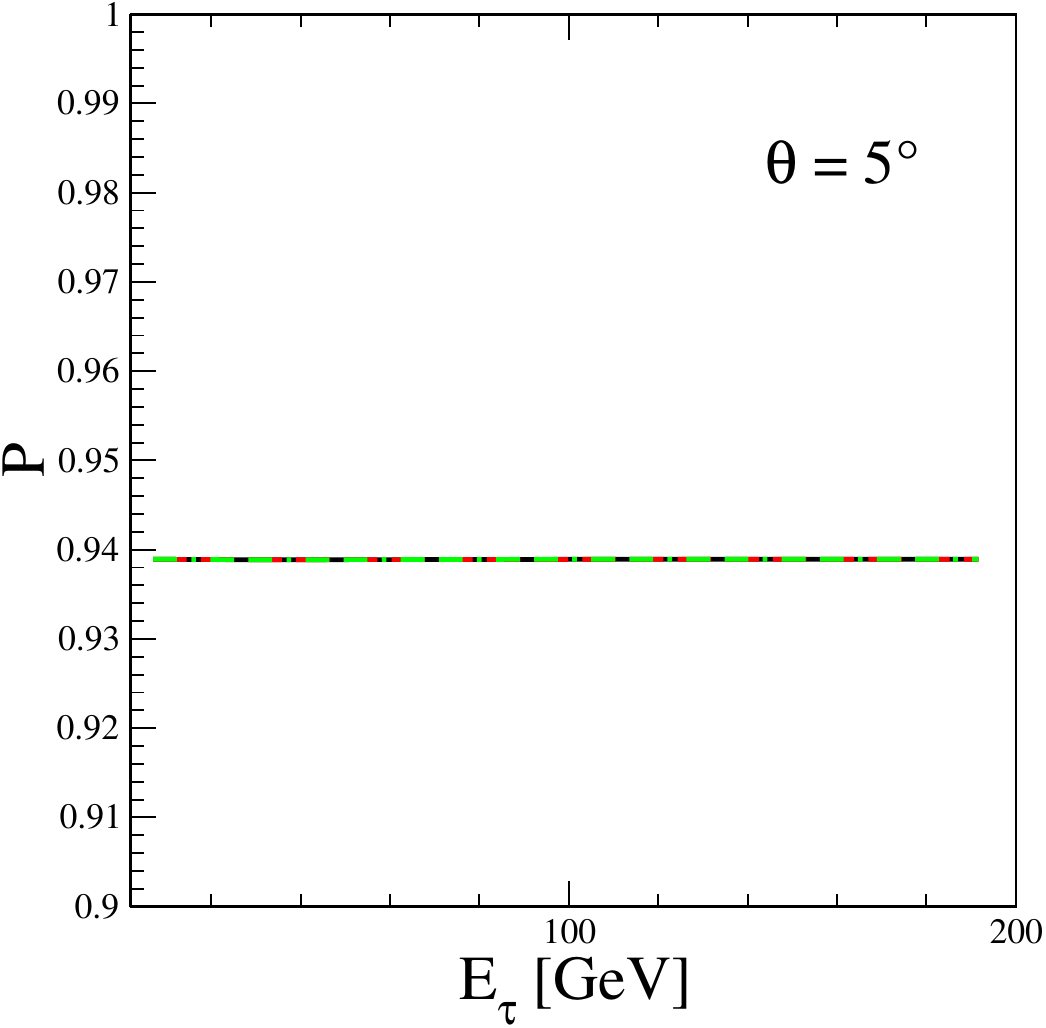} &  \includegraphics[width=0.23\textwidth]{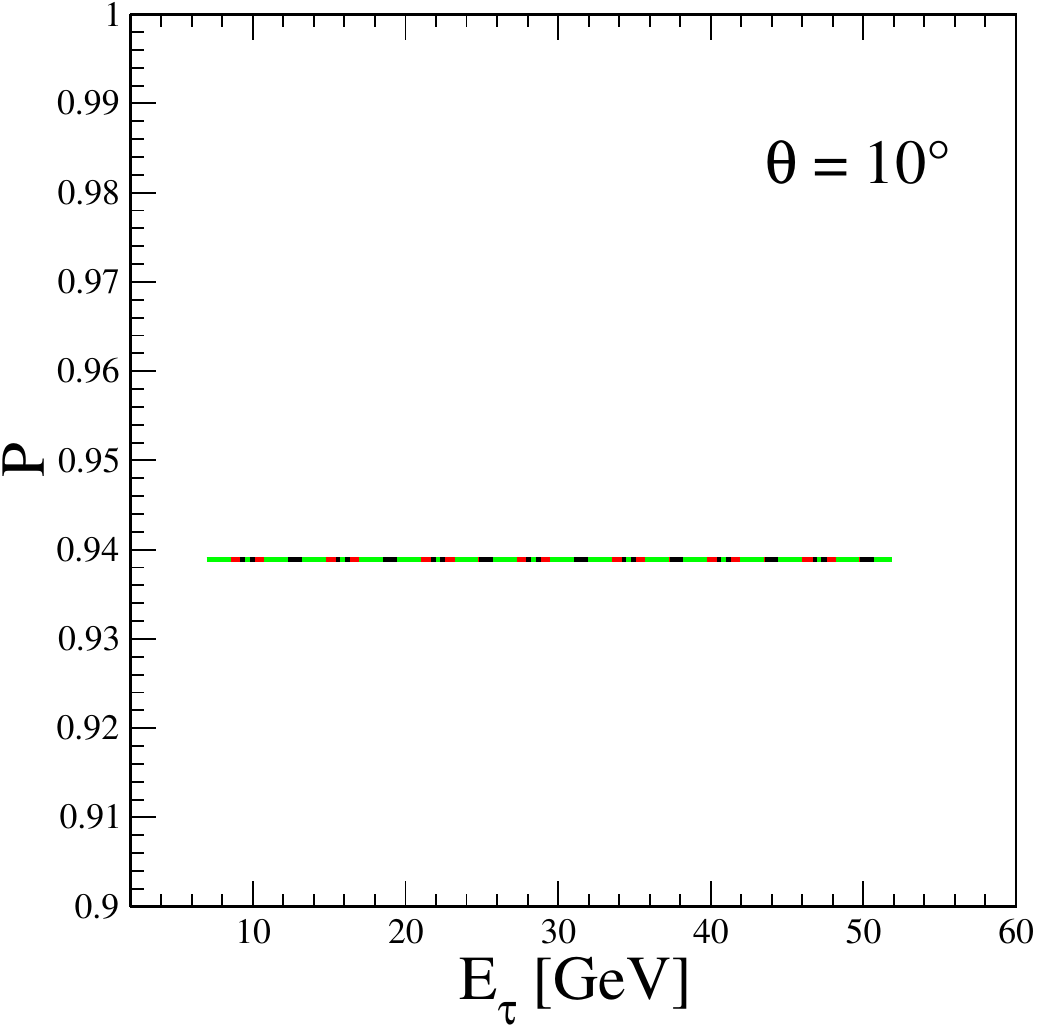}          \\
    \includegraphics[width=0.23\textwidth]{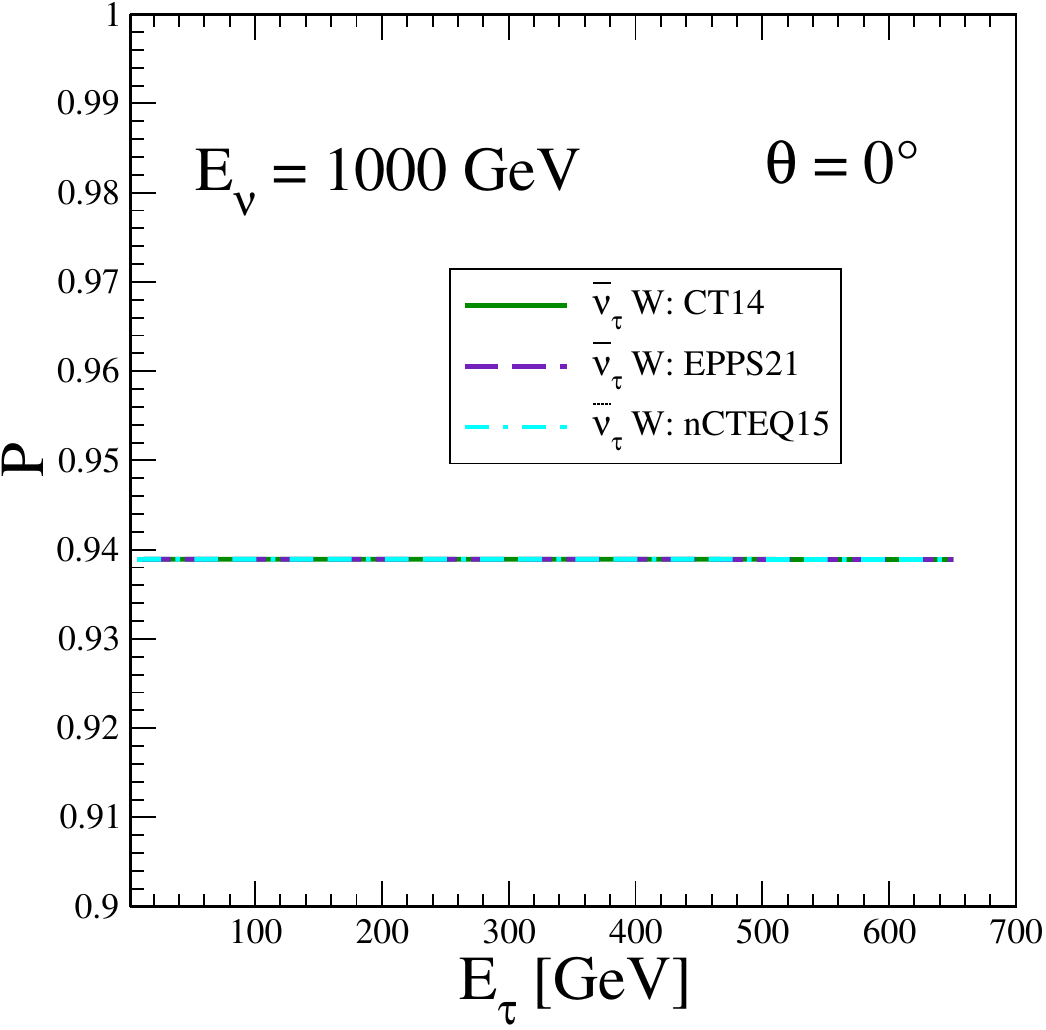} &  \includegraphics[width=0.23\textwidth]{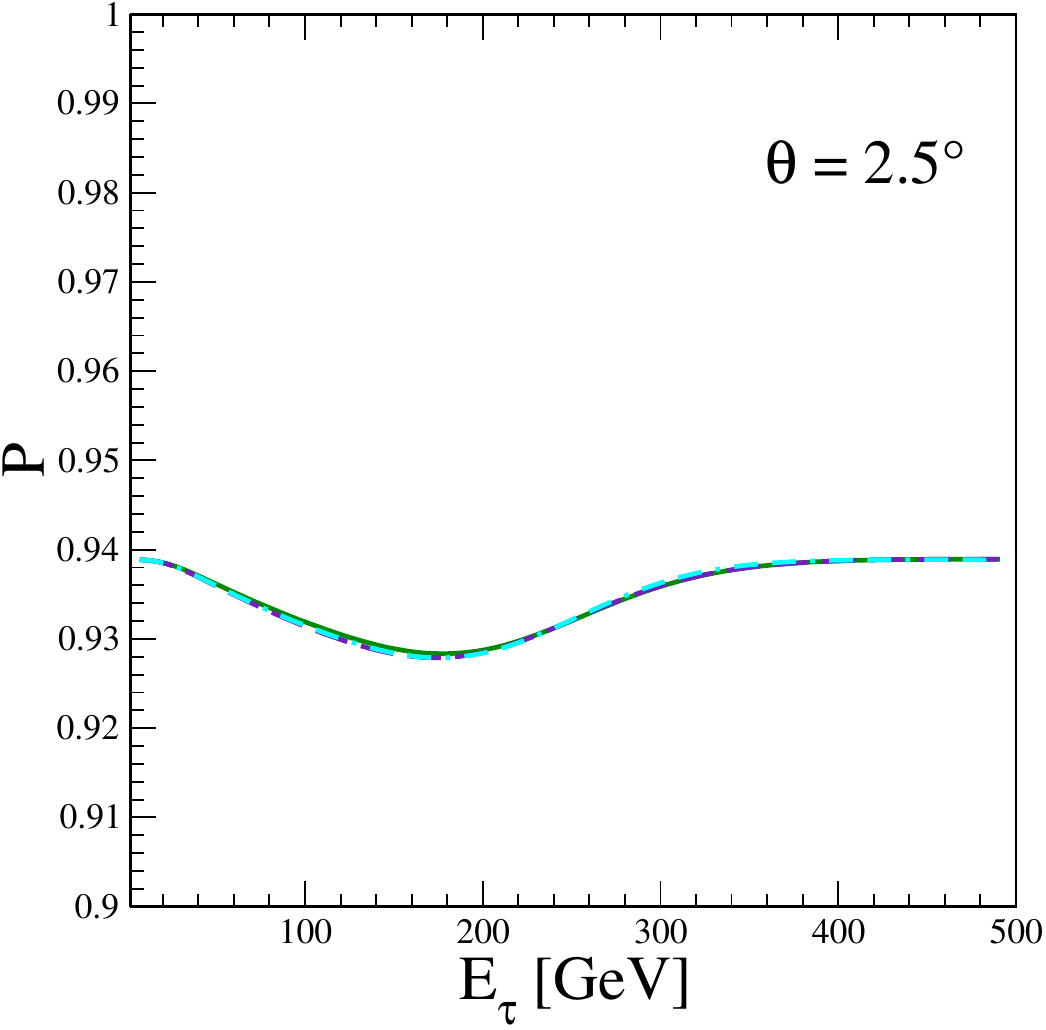} &  
    \includegraphics[width=0.23\textwidth]{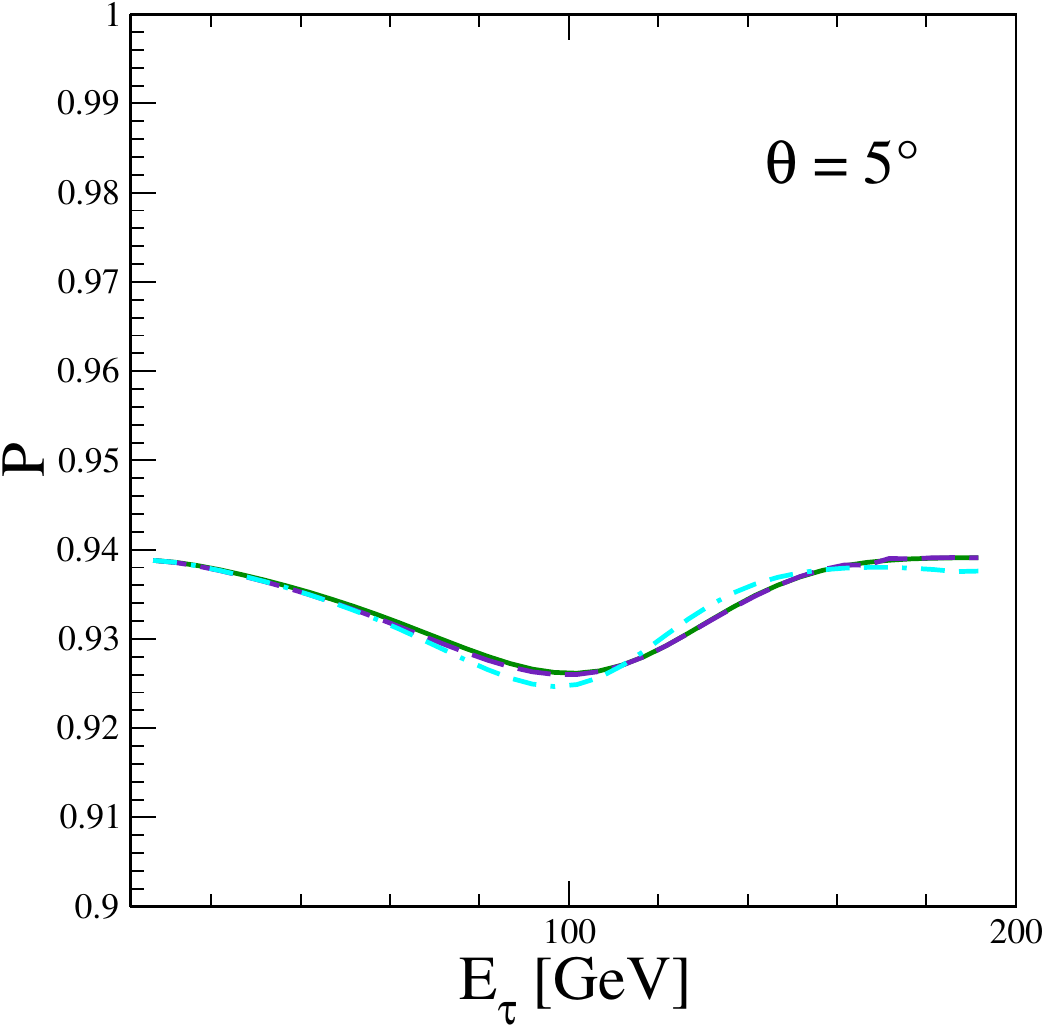} &  \includegraphics[width=0.23\textwidth]{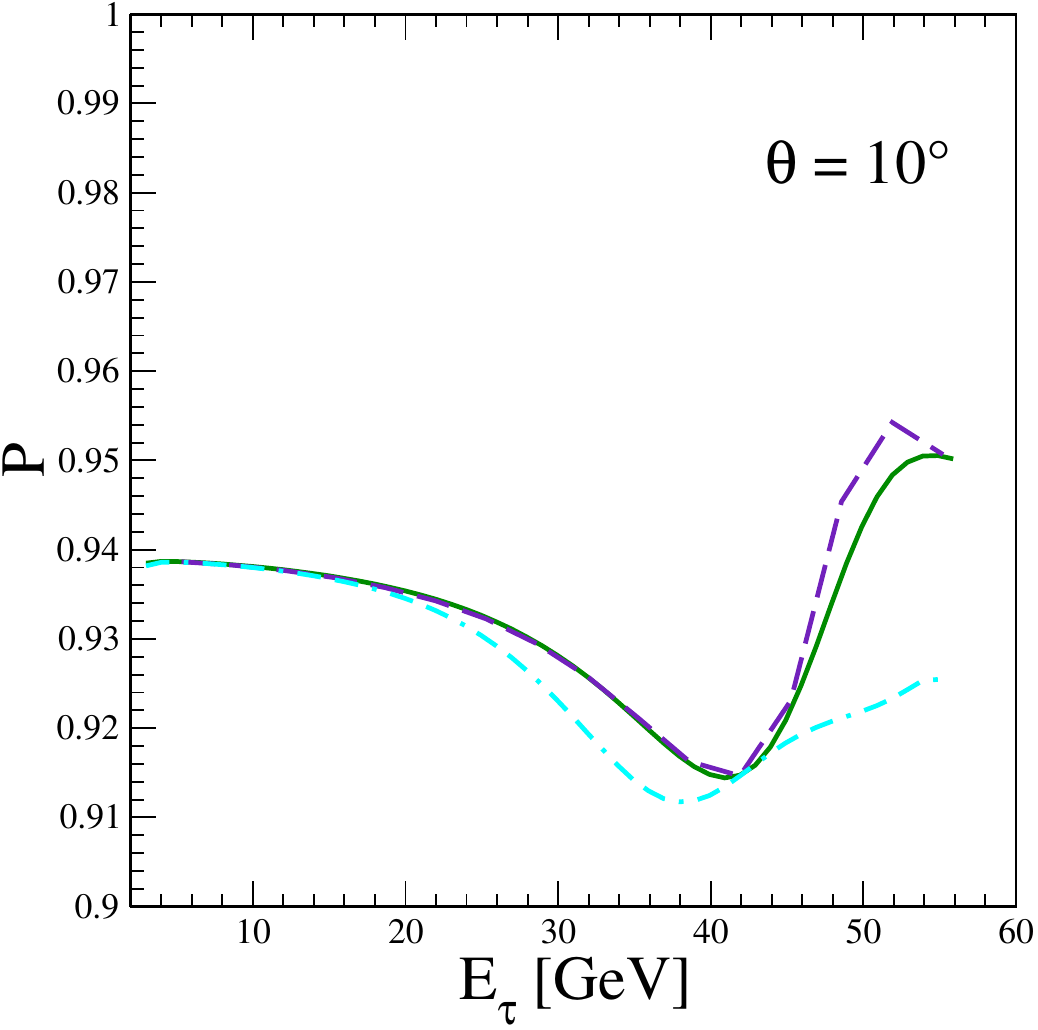} 
			\end{tabular}
\caption{ Degree of polarization of the tau (upper panels) and antitau (lower panels) produced in the ${\nu}_{\tau}W$ and $\bar{\nu}_{\tau}W$ interactions, respectively, as a function of the tau lepton energy for an incident (anti)neutrino with an energy of 1000 GeV. Results for different values of the angle $\theta$, derived assuming different nPDFs. }
\label{fig_4:P}
\end{figure}

With the tau polarization components presented above, we can estimate the degree of polarization of the tau produced, which is defined as
\begin{eqnarray}
    {\cal{P}} = \sqrt{P_L^2 + P_T^2} \, .
    \label{eq_4:P}
\end{eqnarray}
Solving the equation above, we obtain Figure \ref{fig_4:P}. In it, we calculate the degree of polarization of the tau as a function of its energy for different scattering angles, for incident neutrinos of 1000 GeV. Our results show that the taus produced in interactions of tauonic neutrinos at the LHC are not completely polarized, and that they have a degree of polarization of about 94\% for this neutrino energy analyzed. Furthermore, the degree of polarization for taus is weakly dependent on the tau energy, unlike antitaus, where this quantity has a local minimum at energies intermediate to the kinematic limits allowed for antitaus.

\section{Effects of tau polarization and $F_5$ on cross sections}

\begin{figure}
\centering
\includegraphics[scale=0.25]{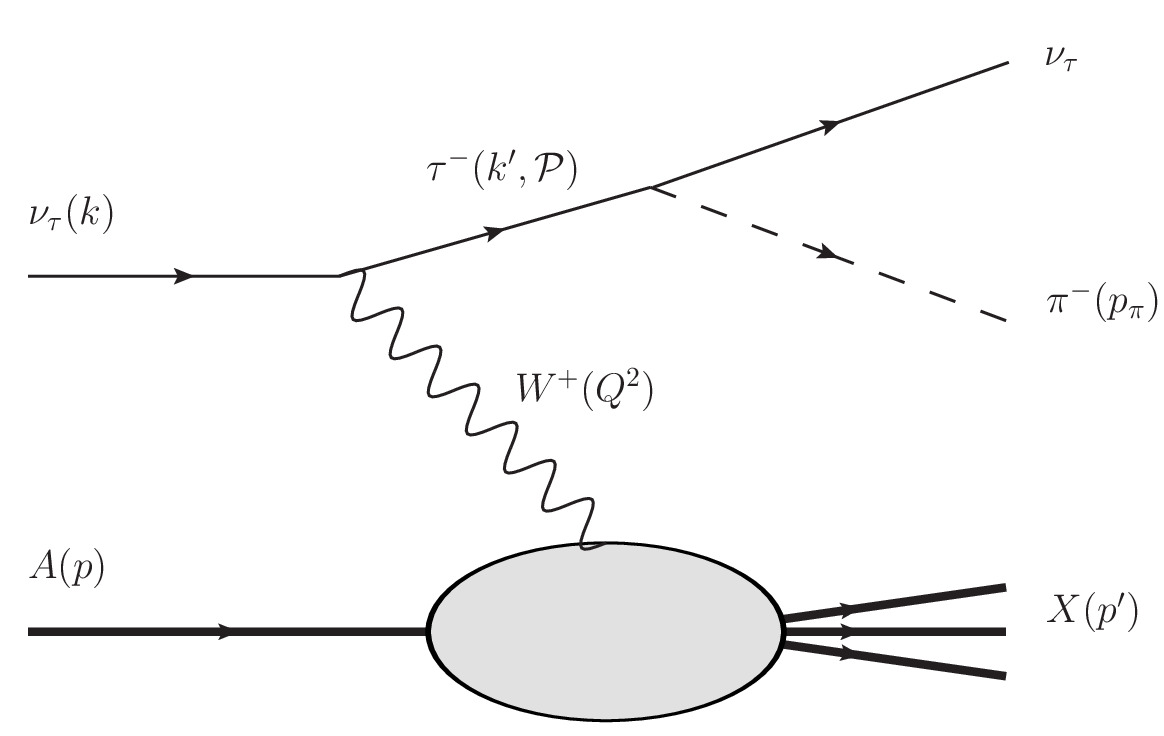} 
\caption{ Production of a tau lepton with momentum $k^{\prime}$ and polarization ${\cal{P}}$ in a  charged current deep inelastic scattering $\nu_{\tau} A$ and its subsequent decay into a pion with momentum $p_\pi$. }
\label{fig_4:diagramPion}
\end{figure}

In the previous section, we showed that taus produced by tau neutrinos at the LHC are not completely polarized. This motivates us to search for effects of realistic tau polarization in observables, such as cross sections. Observing taus in the LHC energy regime has an extra complication compared to other charged leptons: due to its high mass and short lifetime, tau must propagate only a few millimeters before decaying, preventing its energy and momentum from being precisely measured. After decaying, all channels have at least one neutrino in the final state, which prevents us from knowing the real energy and momentum of the tau through its decay products. In this study, we will focus on the cross sections of tau neutrinos in the decay channel into a tau neutrino plus a charged pion. This channel represents approximately 10\% of two tau decays, and is of interest to us because it has only one neutrino and one hadron in the final state, which facilitates detection and maximizes the preservation of the kinematic characteristics of the final state tau. This channel analyzed in this work is represented in the Feynman diagram in Figure \ref{fig_4:diagramPion}.

For this process, the differential cross section with respect to the kinematic variables of energy ($E_\pi$) and scattering angle ($\theta_\pi$) of the pion for the DIS in the nucleon's rest frame is given by \cite{Hernandez:2022nmp}
\begin{eqnarray}
    \begin{aligned}
    \frac{\mathrm{d}^2\sigma_A^{\pi}}{\mathrm{d}E_\pi\mathrm{d\,cos}\,\theta_\pi} = 
    {\cal{B}_{\pi}} 
    \frac{m_\tau^{2}}{m_\tau^{2}-m_\pi^{2}}
    \frac{G_F^{2} M_N}{2 \pi^{2}}
    \int_{E_{\tau}^{-}}^{E_{\tau}^{\mathrm{sup}}}\mathrm{d}E_{\tau}
    \int_{\mathrm{cos}\,(\theta_{\pi}+\theta_{\tau\pi})}^{\mathrm{cos}\,(\theta_{\pi}-\theta_{\tau\pi})} \\
    \frac{\mathrm{d}(\mathrm{cos}\,\theta_\tau)F(E_\tau, \mathrm{cos}\,\theta_\tau)}
    {\sqrt{[\mathrm{cos}\,(\theta_\pi-\theta_{\tau\pi})-\mathrm{cos}\,\theta_\tau]
    [\mathrm{cos}\,\theta_\tau-\mathrm{cos}\,(\theta_\pi-\theta_{\tau\pi})]}} \\
    \left\{ 
    1+\frac{2m_\tau}{m_\tau^{2}-m_\pi^{2}} 
    \frac{m_\tau^{2}-2 m_\pi^{2}}{m_\tau^{2}+2 m_\pi^{2}}
    \left[
    P_L(E_\tau, \mathrm{cos}\,\theta_\tau)
    \left(
    \frac{E_\pi|\vec{k}'|}{m_\tau} - \frac{E_\tau|\vec{p}_\pi|}{m_\tau}\mathrm{cos}\,\theta_\pi 
    \right) 
    \right. \right.
    \\
    \left.
    \left.
    + P_T(E_\tau, \mathrm{cos}\,\theta_\tau)
    \frac{|\vec{p}_\pi|(\mathrm{cos}_\pi-\mathrm{cos}\,\theta_\tau\mathrm{cos}\,\theta_{\tau\pi}))}{\mathrm{sin}\,\theta_\tau}
    \right] 
    \right\}\, ,
    \label{eq_4:sigma3}
    \end{aligned}
\end{eqnarray}
where ${\cal{B}}_\pi$ is the fraction of tau decays in this specific channel, $\vec{p}_\pi$ and $m_\pi$ are the momentum and mass of the pion, respectively, and $\theta_{\tau\pi}$ is the angle between $\vec{p}_\pi$ and $\vec{k}'$, defined by
\begin{eqnarray}
    \mathrm{cos}\, \theta_{\tau\pi} = 
    \frac{2E_\tau E_\pi -m_\tau^{2}-m_\pi^{2}}{2|\vec{k}'||\vec{p}_\pi|}
    \, .
    \label{eq:cos_theta_tau_pion}
\end{eqnarray}
For incident neutrino energies greater than $(m_{\tau}^{2}+m_{\pi}^{2})/2m_{\pi}$, the allowed limits for pion and tau energies are given by
\begin{eqnarray}
    \begin{aligned}
    m_{\pi} \leq E_{\pi} \leq E_{\pi}^{\mathrm{max}}
     , \,\,\,\,\,
    E_{\tau}^{\mathrm{sup}} = 
    \Theta(E_{\pi}^{\mathrm{int}} - E_{\pi})E_{\tau}^{+} 
    + \Theta(E_{\pi} - E_{\pi}^{\mathrm{int}})E_{\nu} \, ,
    \label{eq_4:limites2}
    \end{aligned}
\end{eqnarray}
where $\Theta$ is the step function, and we introduce the new variables
\begin{eqnarray}
    \begin{aligned}
    E_{\tau}^{\pm} = 
    \frac{(m_{\tau}^{2}+m_{\pi}^{2})E_{\pi}\pm (m_{\tau}^{2}-m_{\pi}^{2})|\vec{p}_{\pi}|}{2m_{\pi}^{2}} \, ,     
    \label{eq_4:limites1}
    \end{aligned}
\end{eqnarray} 
and
\begin{eqnarray}
    \begin{aligned}
    E_{\pi}^{\mathrm{max, int}} = 
    \frac{(m_{\tau}^{2}+m_{\pi}^{2})E_{\nu}\pm (m_{\tau}^{2}-m_{\pi}^{2})\sqrt{E_{\nu}^{2}-m_{\tau}^{2}}}{2m_{\tau}^{2}} \, .
    \label{eq_4:limites2}
    \end{aligned}
\end{eqnarray} 
Finally, to fully define Equation (\ref{eq_4:sigma3}), we have the function $F(E_\tau, \mathrm{cos}\,\theta_\tau)$, which gives us the dependence of the cross section on the kinematic variables of the DIS and on the structure functions, and is given by
\begin{eqnarray}
    F(E_\tau, \mathrm{cos}\,\theta_\tau)  =   
    \left\{ 2F^A_{1}(x,Q^2)(E_\tau -|\vec{k}'|\mathrm{cos}\,\theta_\tau) + 
    F^A_{2}(x,Q^2)\frac{M_A}{\nu}(E_\tau+|\vec{k}'|\mathrm{cos}\,\theta_\tau) \right. \nonumber \\
    \pm F^A_{3}(x,Q^2)\frac{1}{\nu}[| \vec{k}' |^2 + E_\nu E_\tau - (E_\nu + E_\tau )|\vec{k}'|\mathrm{cos}\,\theta_\tau] +   \\
    \left. 
    +F^A_{4}(x,Q^2)\frac{m_\tau^{2}}{\nu M_A x} (E_\tau - |\vec{k}'|\mathrm{cos}\,\theta_\tau) 
    - F^A_{5}(x,Q^2)\frac{2m_\tau^2}{\nu} \right\} \, . \nonumber
     \label{eq_4:F}
\end{eqnarray}
By the end of this chapter, our results will be obtained using the nCTEQ15 parameterization for the structure functions.

\begin{figure}
	\centering
	\begin{tabular}{c}
	\includegraphics[width=0.7\textwidth]{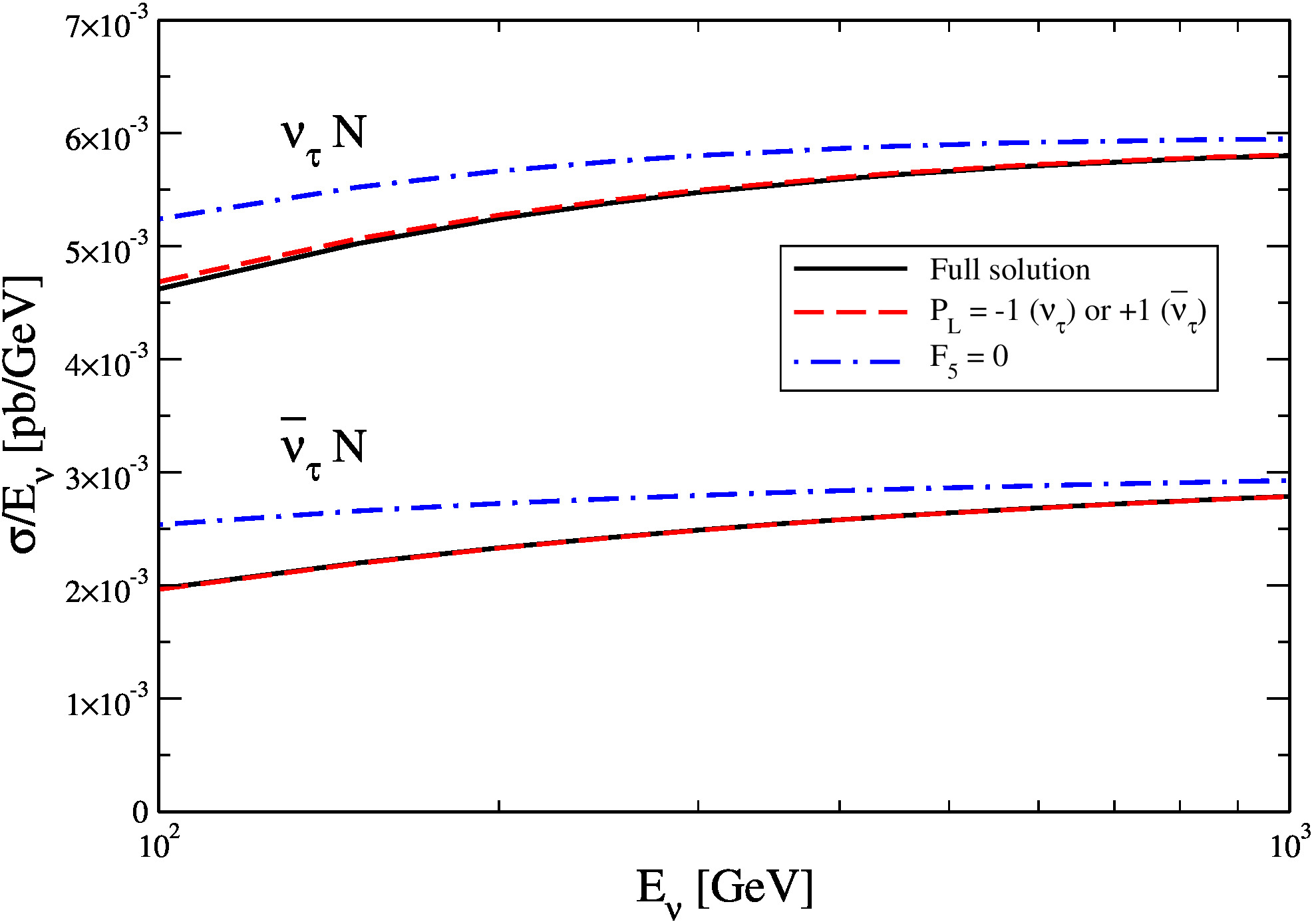} 
	\end{tabular}
\caption{ Predictions for the energy dependence of the $\nu_\tau / \bar{\nu}_\tau$ - tungsten cross section per nucleon, derived considering the complete solution of Equation (\ref{eq_4:sigma3}) (solid black line). For comparison, approximate results obtained assuming that the produced (anti)tau is fully polarized (dashed red line) or assuming that $F_5 = 0$ (dashed-dotted blue line) are also presented. }
\label{fig_4:sigmaPol}
\end{figure}

Our first analysis presents the total cross section as a function of the incident neutrino energy obtained from the solution of Equation (\ref{eq_4:sigma3}). We present this result in Figure \ref{fig_4:sigmaPol} in the energy regime from 100 GeV to 1000 GeV, which corresponds to the interval with the highest number of tau neutrino events expected in FASER$\nu$2 \cite{Anchordoqui:2021ghd,Feng:2022inv,FPFWorkingGroups:2025rsc,FPF:2025bor}. Our results are for three distinct configurations: (a) the continuous black line is for the complete solution of Equation (\ref{eq_4:sigma3}), considering the realistic polarization of the intermediate tau state; (b) to obtain the dashed black curve, we set $P_L = 1 (-1)$ for antineutrinos (neutrinos), which makes the tau completely longitudinally polarized; and (c) we solved Equation (\ref{eq_4:sigma3}) by setting $F_5 = 0$, to verify the impact of this structure function on the FASER$\nu$2 regime. The structure function $F_5$ has never been measured before, and only plays an important role for tau neutrino scattering, since it multiplies $m_\tau^{2}$ in Equation (\ref{eq_4:F}). Our results indicate that this structure function significantly impacts the total cross section, increasing it by 13.4\% (28.4\%) and 2.4\% (5.0\%) for (anti) neutrino-induced reactions with energies of 100 GeV and 1000 GeV, respectively. However, assuming the tau produced is fully polarized does not significantly alter the total cross section compared to the complete solution of Equation (\ref{eq_4:sigma3}).

\begin{figure}
	\centering
	\begin{tabular}{ccc}
    \includegraphics[width=0.48\textwidth]{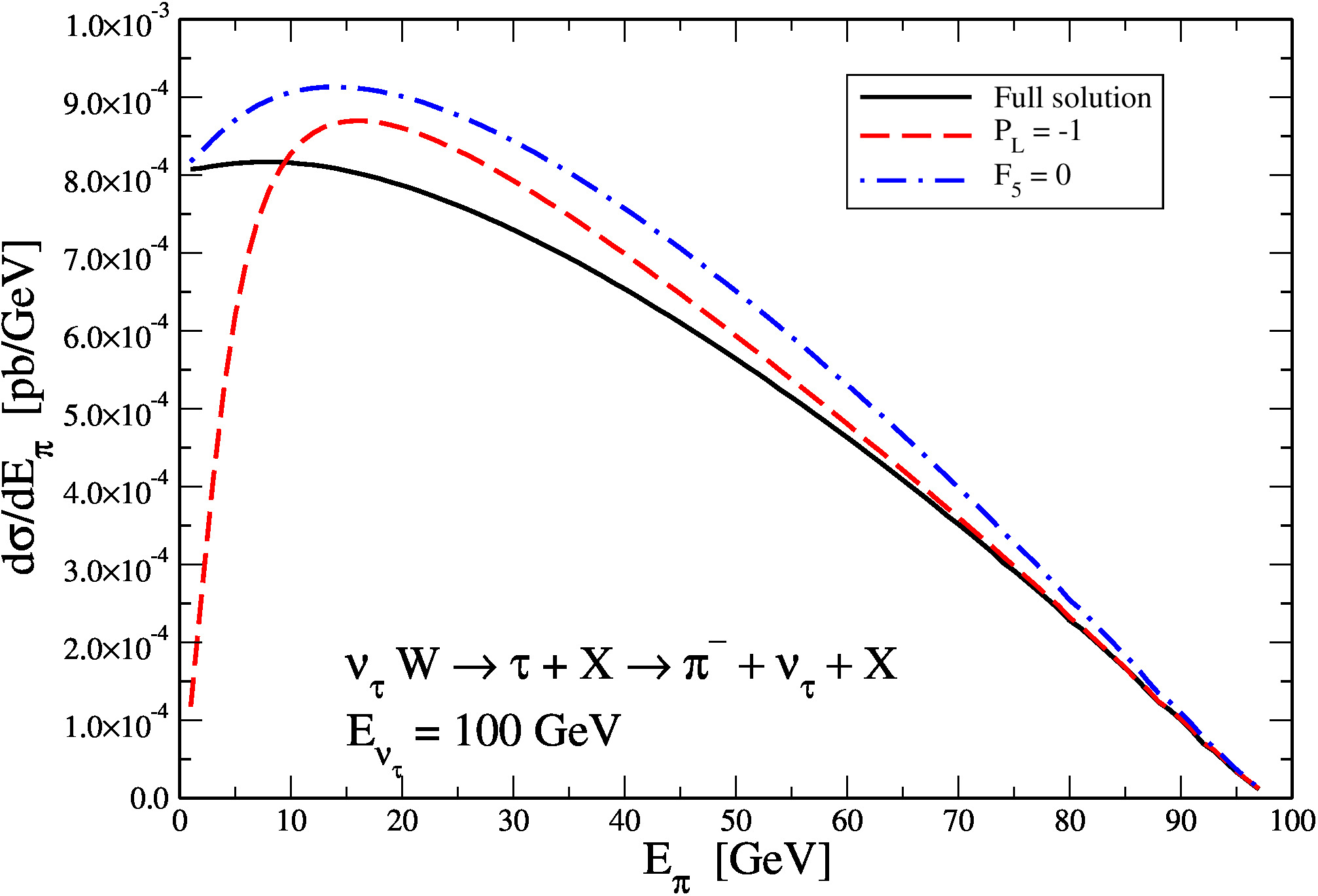} &
    \includegraphics[width=0.48\textwidth]{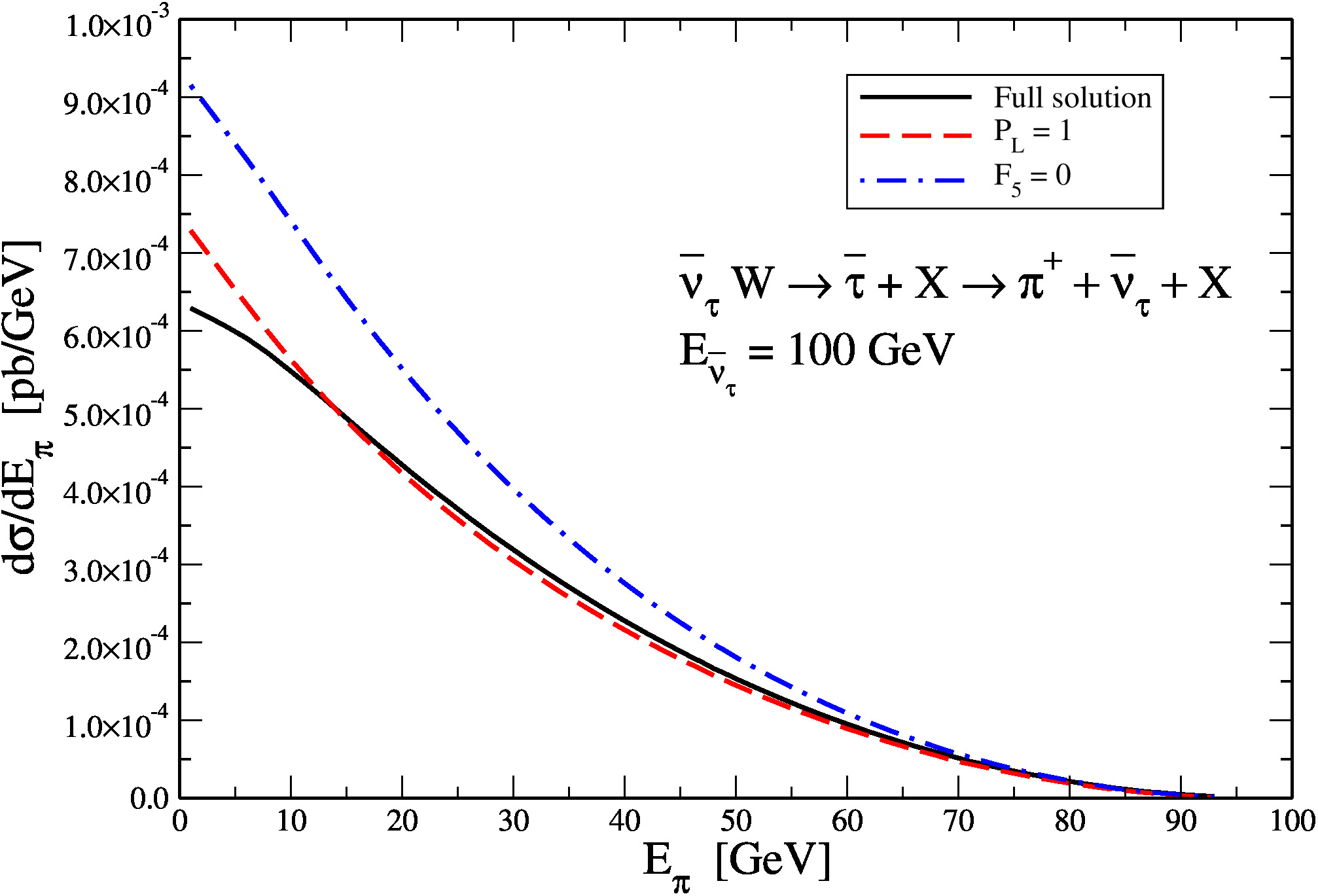} \\
    \includegraphics[width=0.48\textwidth]{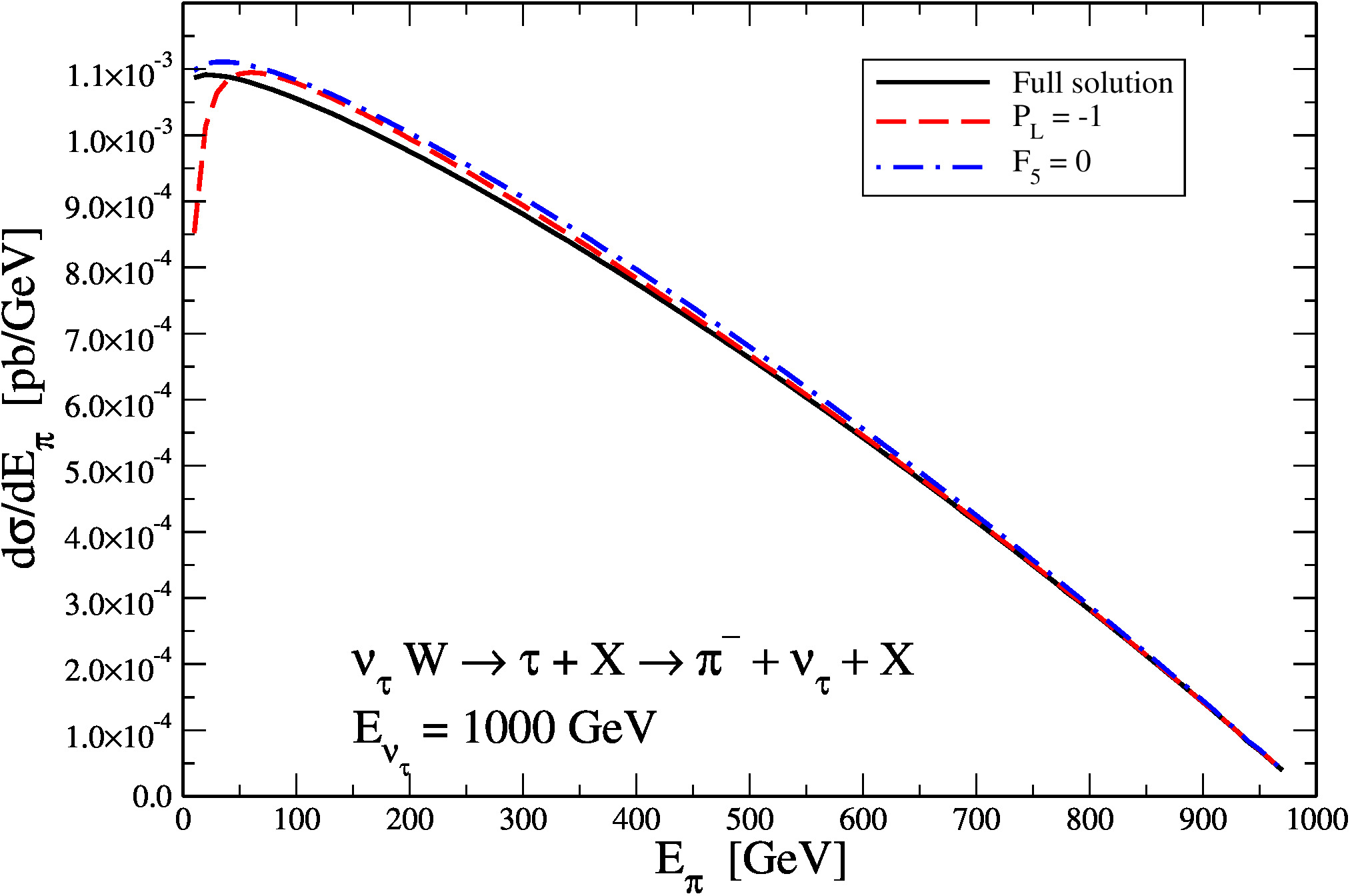} &
    \includegraphics[width=0.48\textwidth]{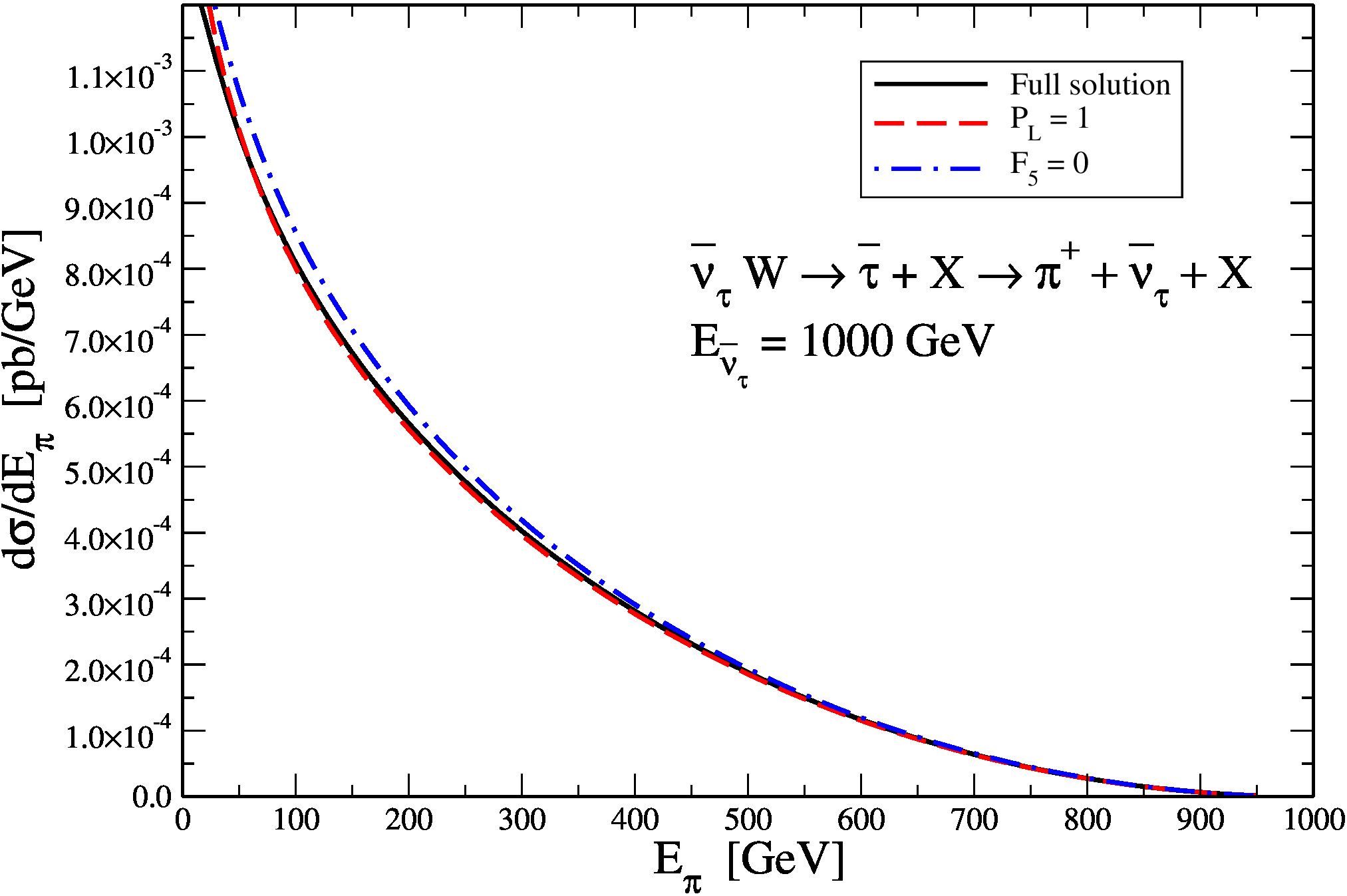} 
			\end{tabular}
\caption{ Differential cross section with respect to pion energy ($E_\pi$) for the processes $ \nu_\tau + W \rightarrow \tau^{-} + X \rightarrow \pi^{-} + \nu_\tau + X$ (left panels) and $ \bar{\nu}_\tau + W \rightarrow \tau^{+} + X \rightarrow \pi^{+} + \bar{\nu}_\tau + X$ (right panels). Predictions for an incident (anti)neutrino energy of 100 GeV (upper panels) and 1000 GeV (lower panels). }
\label{fig_4:dsdE}
\end{figure}

The absence of impact of tau polarization on the total neutrino-nucleon cross section motivates us to analyze the differential cross sections with respect to the energy and scattering angle of the pion resulting from tau decay. Initially, in Figure \ref{fig_4:dsdE}, we show the differential cross section of the pion energy as a function of its energy (integrating over the scattering angle across the entire interval). We present the results for incident neutrinos with 100 GeV (1000 GeV) on the left (right). We also show the results for neutrinos on the top, and antineutrinos on the bottom. Again, we compare the results of the complete solution of Equation (\ref{eq_4:sigma3}), making the tau longitudinally polarized and setting $F_5 = 0$. Our results show that when we assume $F_5 = 0$, the differential cross section increases across the entire pion energy regime. When we assume the tau is longitudinally polarized, the differential cross section decreases (increases) for low pion energies coming from the (anti)tau, and reverses this behavior for higher pion energies. Furthermore, this effect of tau polarization is more important for lower incident neutrino energies.

\begin{figure}
	\centering
	\begin{tabular}{ccc}
    \includegraphics[width=0.48\textwidth]{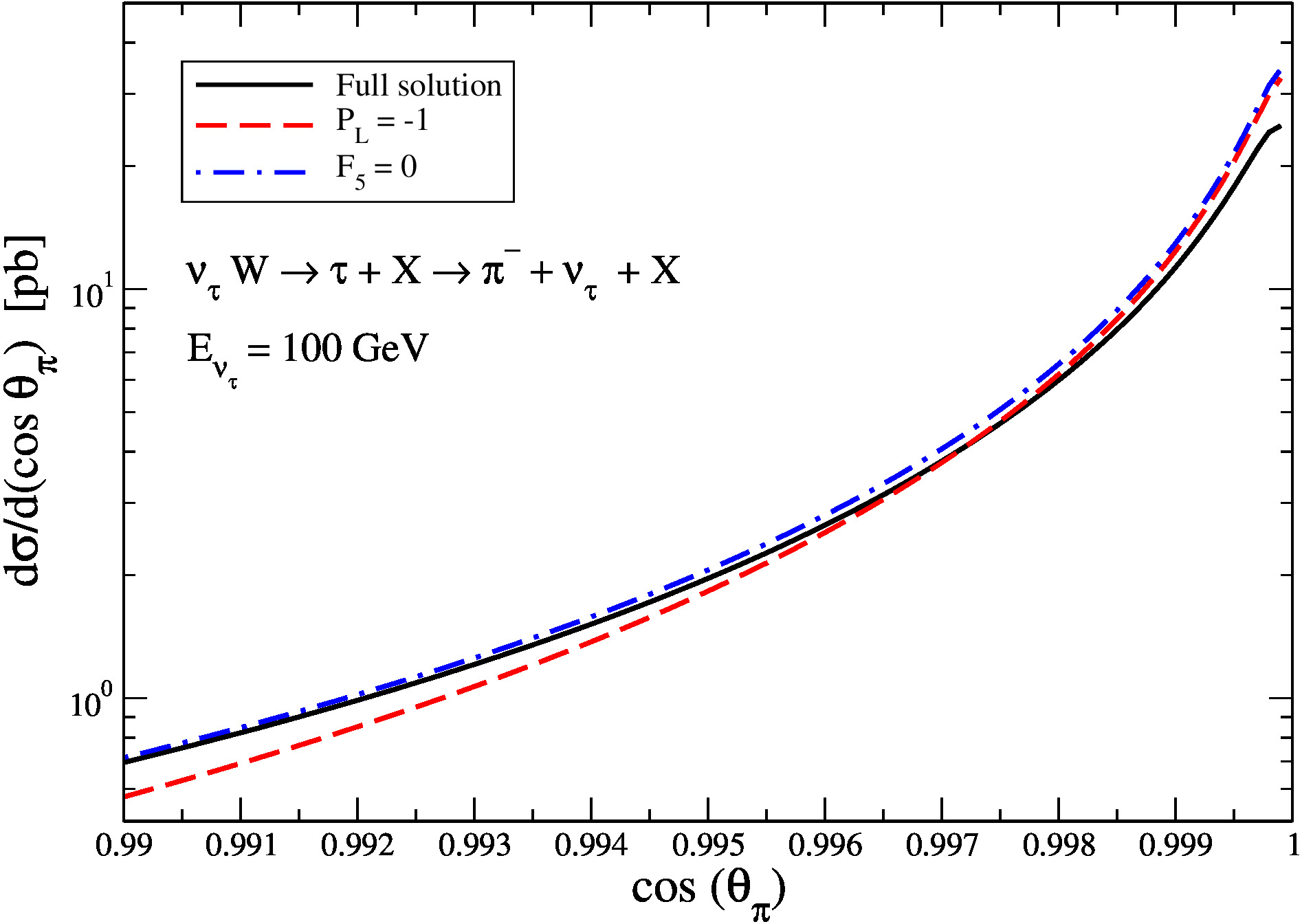} &
    \includegraphics[width=0.48\textwidth]{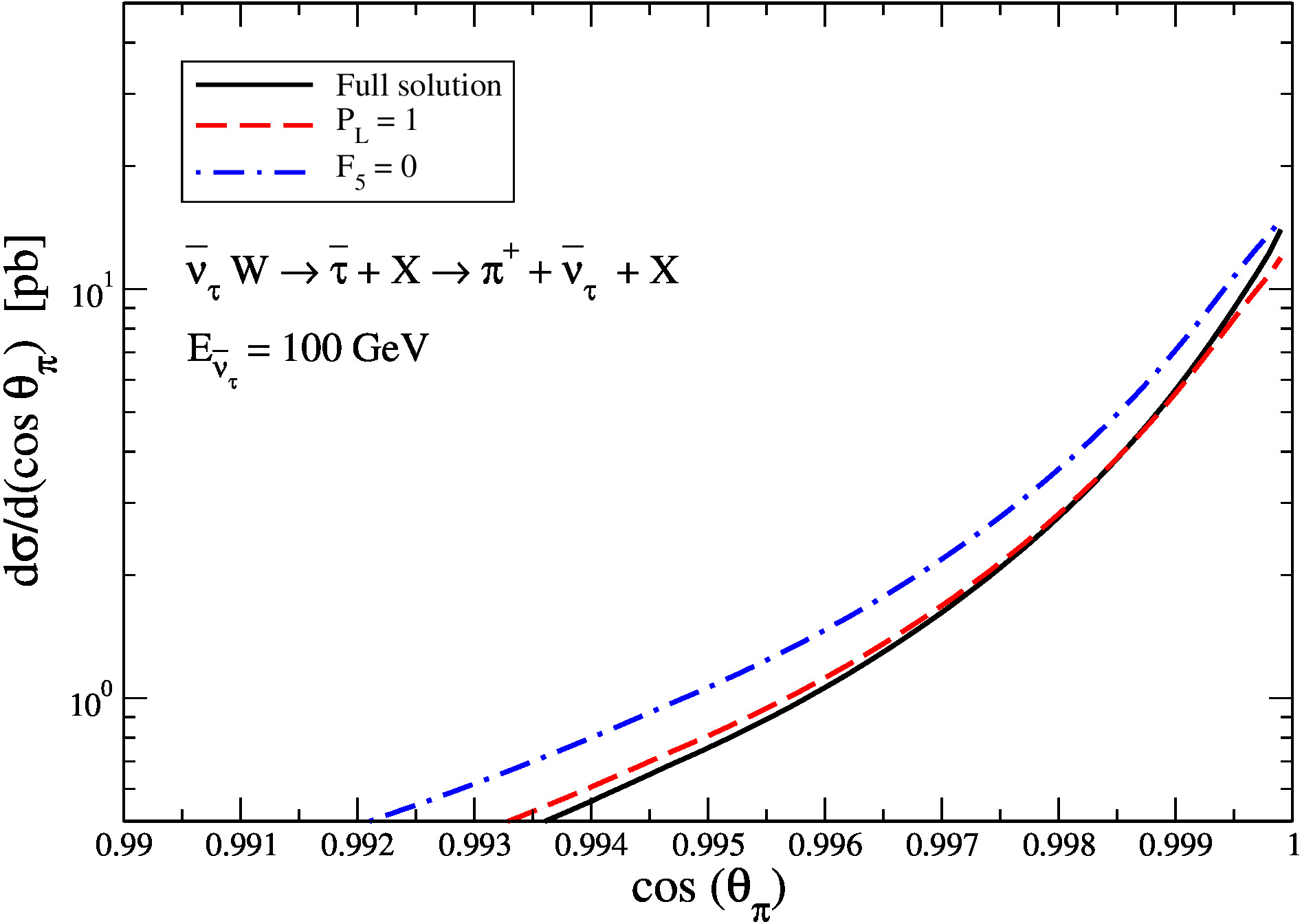} \\
    \includegraphics[width=0.48\textwidth]{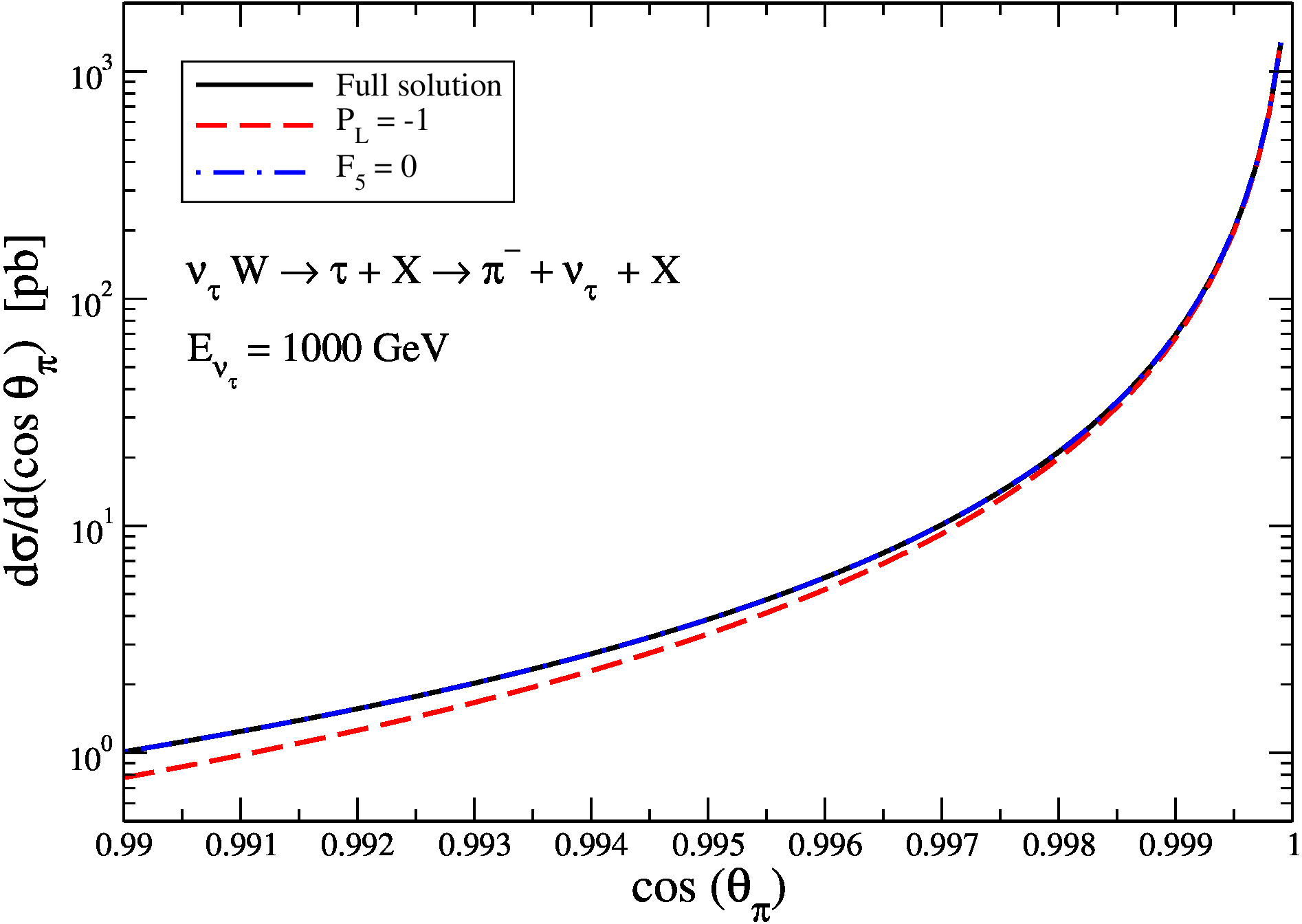} &
    \includegraphics[width=0.48\textwidth]{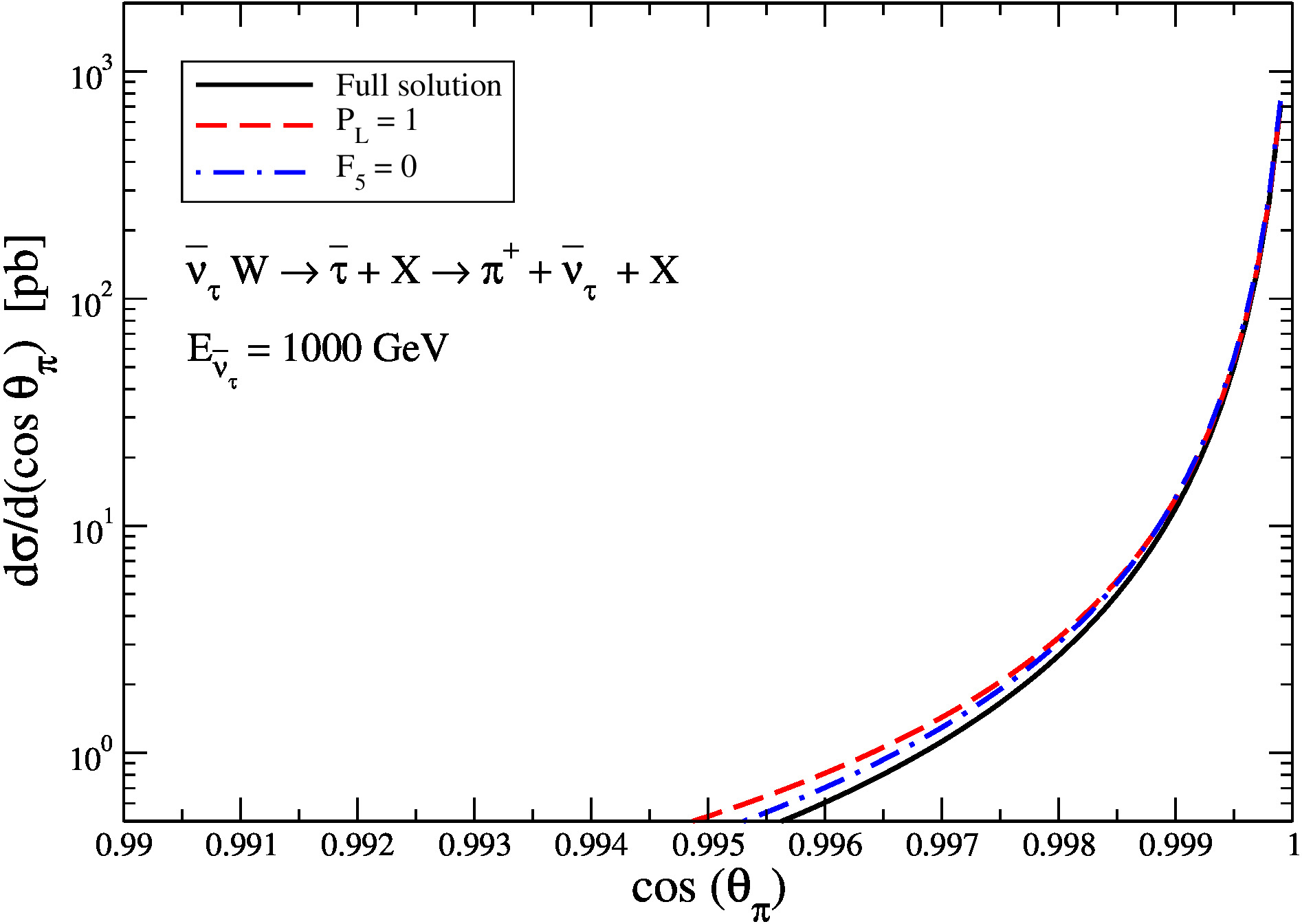} 
			\end{tabular}
\caption{ Differential cross section with respect to the cosine of the pion scattering angle ($\theta_\pi$) for the processes $ \nu_\tau + W \rightarrow \tau^{-} + X \rightarrow \pi^{-} + \nu_\tau + X$ (left panels) and $ \bar{\nu}_\tau + W \rightarrow \tau^{+} + X \rightarrow \pi^{+} + \bar{\nu}_\tau + X$ (right panels). Predictions for the incident (anti)neutrino energy of 100 GeV (upper panels) and 1000 GeV (lower panels). }
\label{fig_4:dsdcostd}
\end{figure}

In Figure \ref{fig_4:dsdcostd} we present results similar to Figure \ref{fig_4:dsdE}, but for the differential cross section in the cosine of the pion scattering angle relative to the neutrino's incidence axis. We present the results for $\mathrm{cos}\,\theta_\pi > 0.99$, since the total cross section is almost entirely included in this kinematic regime. Furthermore, FASER$\nu$2 will be sensitive to variations in mrad, which allows the investigation of processes with scattering angles as small as those presented in Figure \ref{fig_4:dsdcostd}.

Differential cross sections indicate that the effects of tau polarization, as well as the structure function $F_5$, are more feasible for experimental investigation than total cross sections, since when integrating the energy and scattering angle variables, some differences in the cross sections cancel out. Another approach to mapping the effects of tau polarization and the structure function $F_5$ is through the longitudinal and transverse moments of the pion produced. In this context, the double differential cross section of Equation (\ref{eq_4:sigma3}) becomes
\begin{eqnarray}
    \begin{aligned}
    \frac{\mathrm{d}^2\sigma_A^{\pi}}{\mathrm{d}p_{L\pi}\mathrm{d}p_{T\pi}} = 
    \frac{p_{T\pi}}{(p_{L\pi}^{2}+p_{T\pi}^{2})^{1/2}}
    \frac{1}{(p_{L\pi}^{2}+p_{T\pi}^{2} + m_{\pi}^{2})^{1/2}}
    \left[
    \frac{\mathrm{d}^2\sigma_A^{\pi}}{\mathrm{d}E_\pi\mathrm{d\,cos}\,\theta_\pi}
    \right]
    \, .
    \label{eq_4:sigma4}
    \end{aligned}
\end{eqnarray}

\begin{figure}
	\centering
	\begin{tabular}{cc}
    \includegraphics[width=0.48\textwidth]{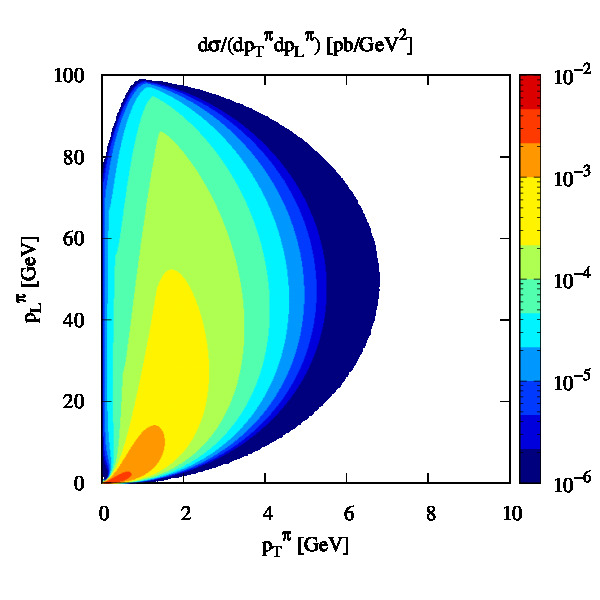} &
    \includegraphics[width=0.48\textwidth]{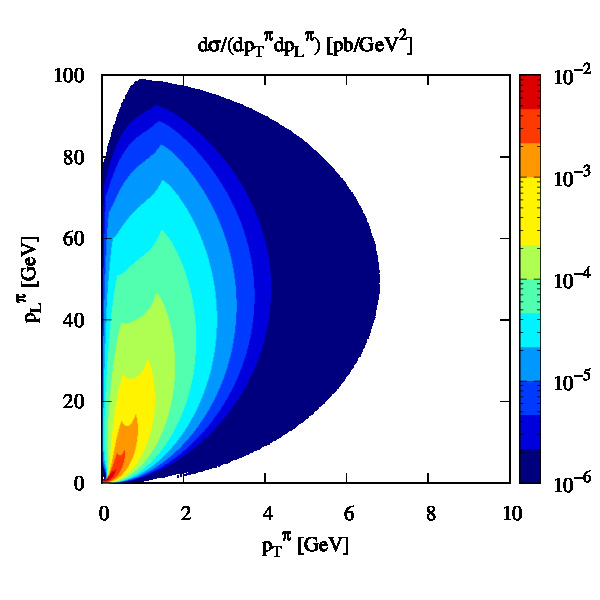} \\
    \includegraphics[width=0.48\textwidth]{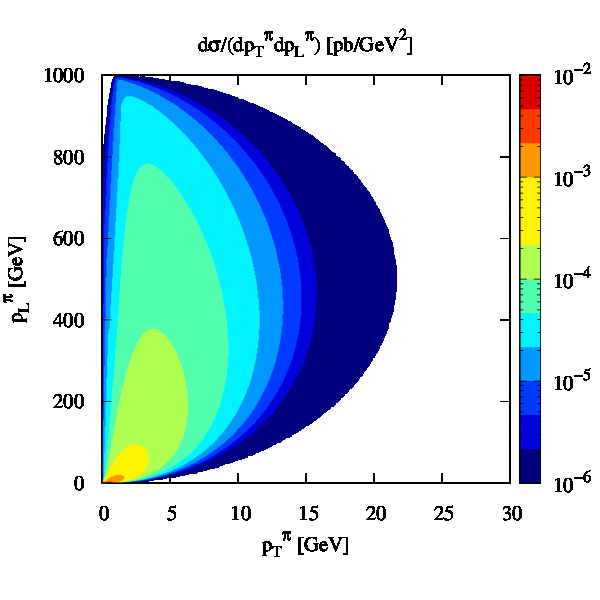} &
    \includegraphics[width=0.48\textwidth]{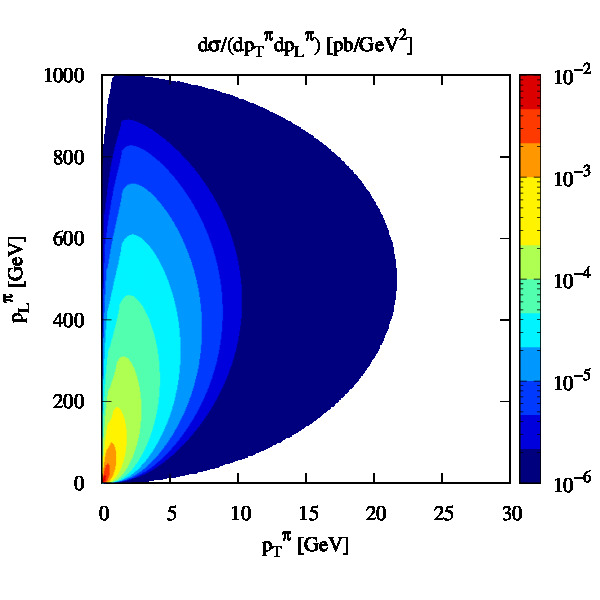}
			\end{tabular}
\caption{ Double differential cross section for the processes $\nu_\tau + W \rightarrow \tau^{-} + X \rightarrow \pi^{-} + \nu_\tau + X$ (left panels) and $\bar{\nu}_\tau + W \rightarrow \tau^+ + X \rightarrow \pi^{+} + \bar{\nu}_\tau + X$ (right panels) with respect to the longitudinal and transverse momentum of the pion. The tauonic neutrino energy is fixed at 100 GeV (upper panels) and 1000 GeV (lower panels). }
\label{fig_4:dsdptdpl}
\end{figure}

In Figure \ref{fig_4:dsdptdpl}, we present the results for the cross section of Equation (\ref{eq_4:sigma4}) for neutrinos (left) and antineutrinos (right) with energies of 100 GeV (top) and 1000 GeV (bottom). The results show that the cross section is mostly contained within regions of small transverse moments. In this figure, we show the results for the complete solution of Equation (\ref{eq_4:sigma4}), using the Albright-Jarlskog approximation \cite{Albright:1974ts} for $F_5$ and considering the realistic polarization of the tau of the intermediate state of the process.

\begin{figure}
	\centering
	\begin{tabular}{cc}
    \includegraphics[width=0.48\textwidth]{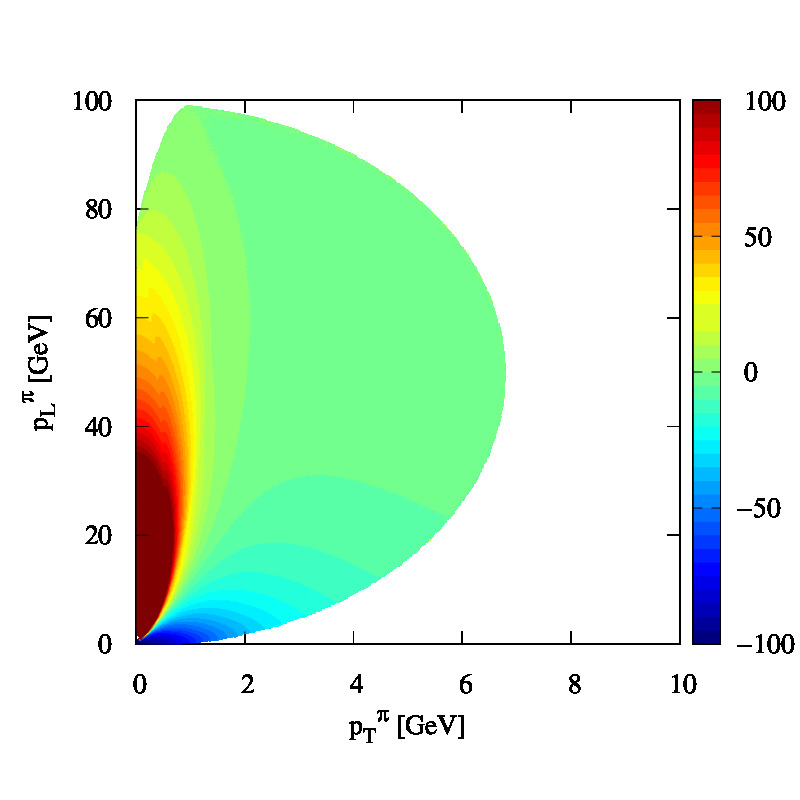} &
    \includegraphics[width=0.48\textwidth]{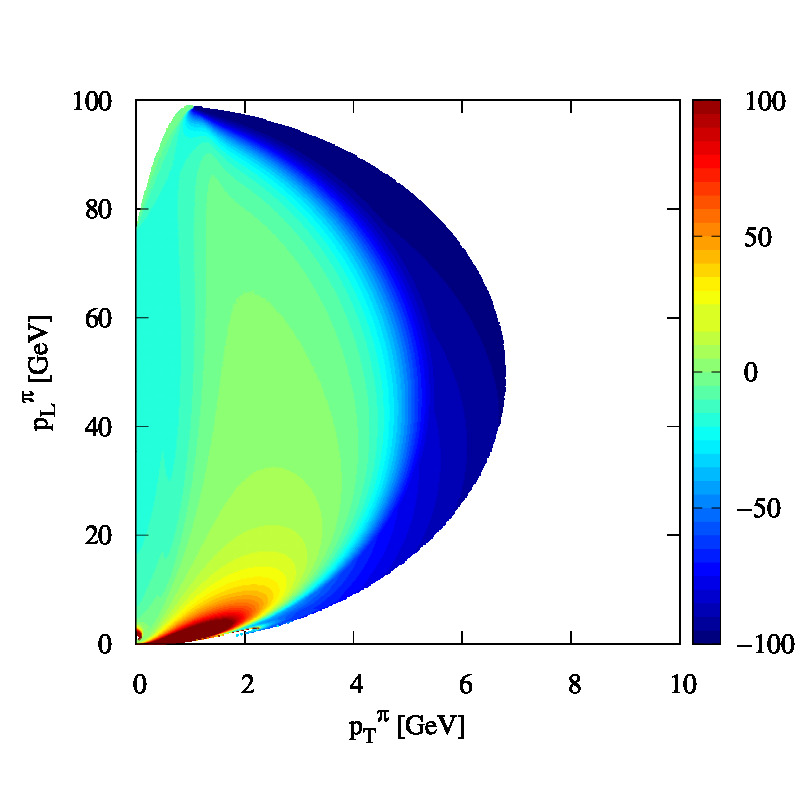} \\
    \includegraphics[width=0.48\textwidth]{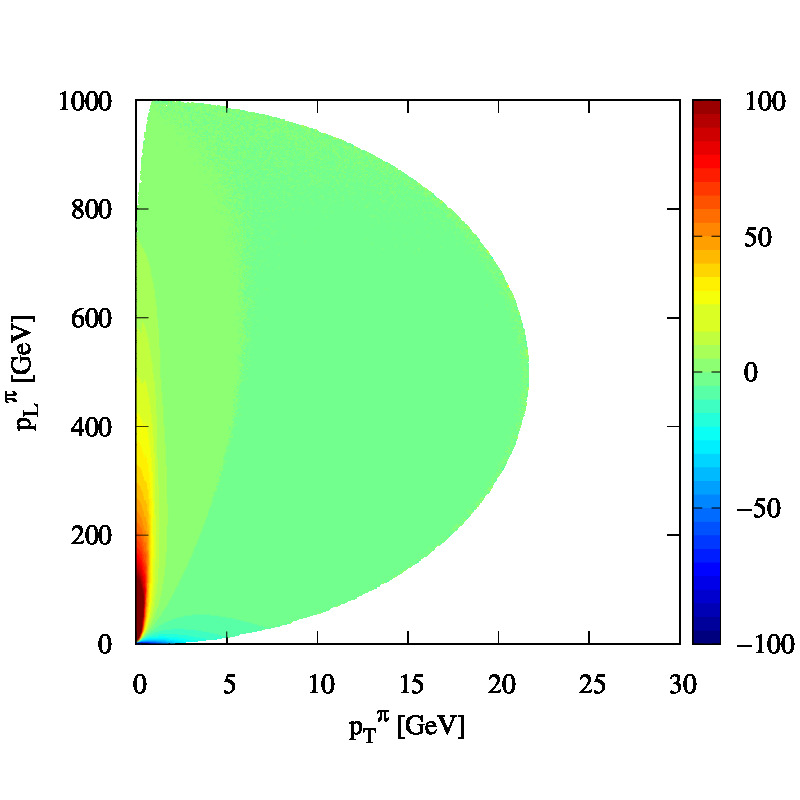} &
    \includegraphics[width=0.48\textwidth]{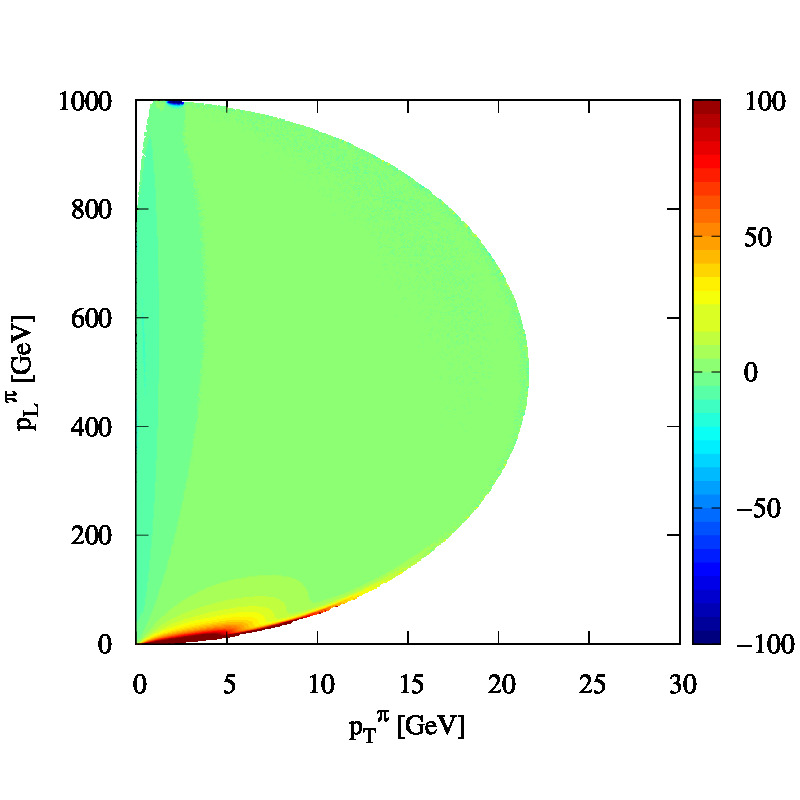}
			\end{tabular}
\caption{ Difference in \% between the predictions for $\mathrm{d}\sigma/(\mathrm{d}p_T^\pi \mathrm{d}p_L^\pi)$ derived assuming complete solution and obtained considering that the (anti)tau produced is fully longitudinally polarized on the left (right). Results for incident (anti)neutrino energy of 100 GeV (upper panels) and 1000 GeV (lower panels). }
\label{fig_4:diff_PL1}
\end{figure}

Now it is in our interest to map the regions where the double differential cross section of Equation (\ref{eq_4:sigma4}) is altered to a greater extent compared to the results in Figure \ref{fig_4:dsdptdpl} and when we assume $F_5 = 0$ and longitudinally polarized tau. To do this, we will calculate the percentage difference between the cross sections, given by
\begin{eqnarray}
    \begin{aligned}
\frac{\mathrm{d}\sigma/(\mathrm{d}p_T^\pi \mathrm{d}p_L^\pi)|_{F_5=0 (P_L=\pm 1)}-\mathrm{d}\sigma/(\mathrm{d}p_T^\pi \mathrm{d}p_L^\pi)}
{\mathrm{d}\sigma/(\mathrm{d}p_T^\pi \mathrm{d}p_L^\pi)} \times 100\% \, .
    \label{eq_4:diff_cs}
    \end{aligned}
\end{eqnarray}
In Figure \ref{fig_4:diff_PL1} we present this percentage difference assuming the tau is longitudinally polarized. Our results show that the cross section for neutrinos is sensitive to small longitudinal moments of the pion, while for antineutrinos the cross sections change more in regions of maximum allowed transverse moment.

\begin{figure}
	\centering
	\begin{tabular}{cc}
    \includegraphics[width=0.48\textwidth]{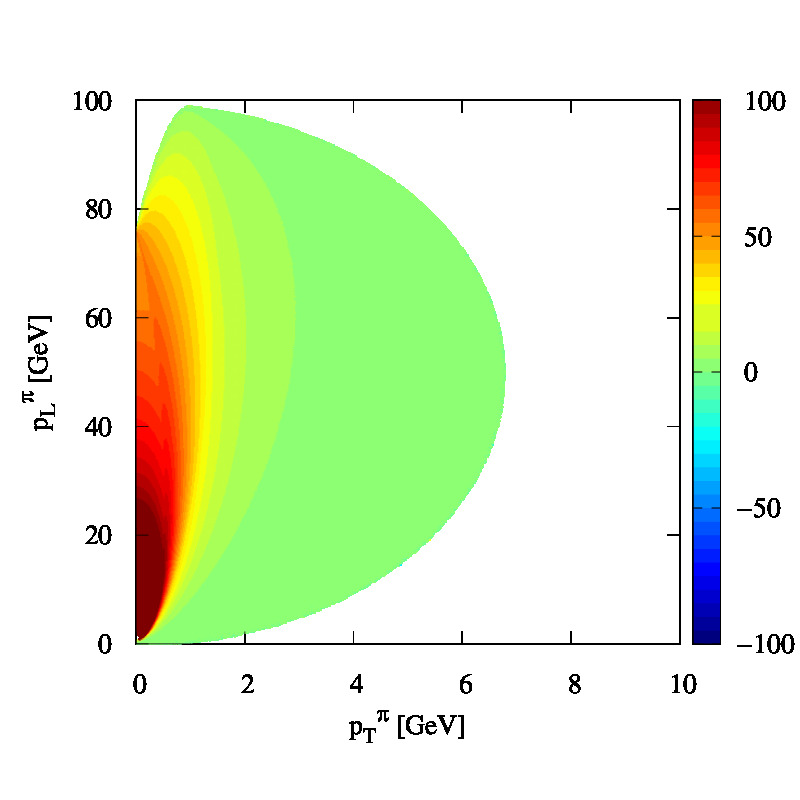} &
    \includegraphics[width=0.48\textwidth]{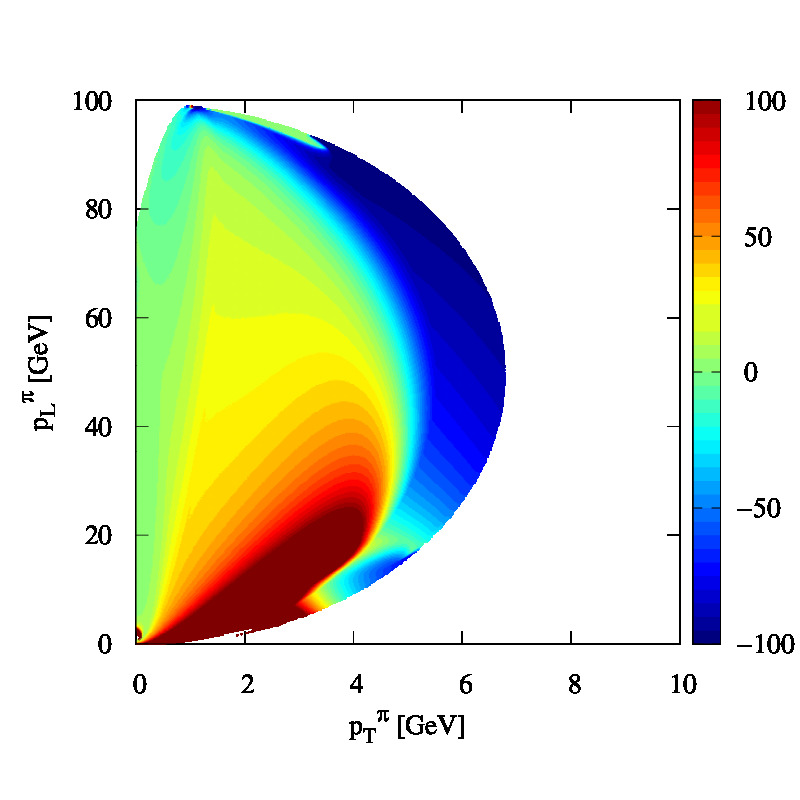} \\
    \includegraphics[width=0.48\textwidth]{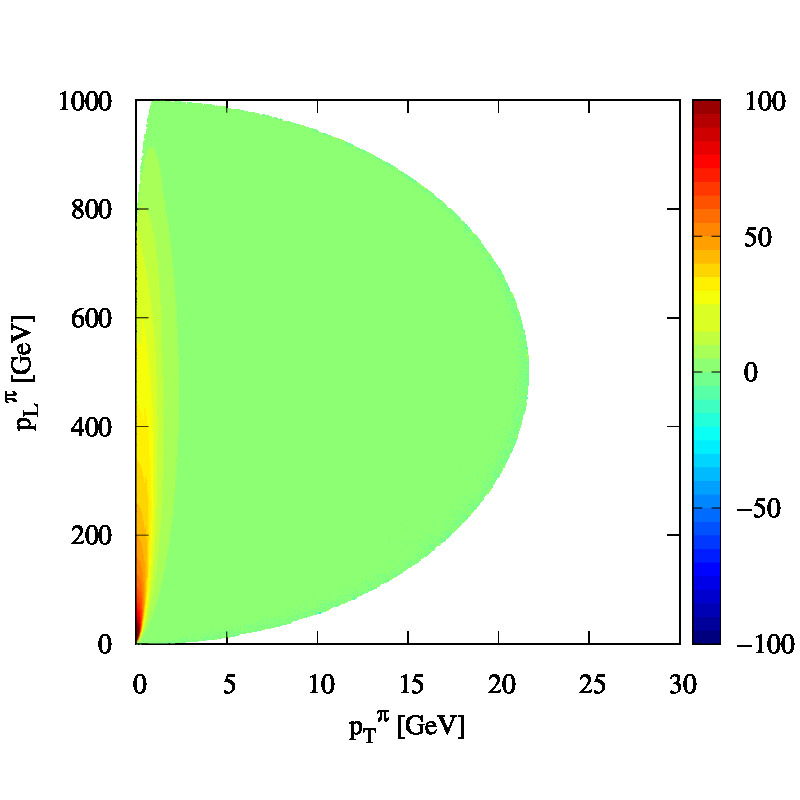} &
    \includegraphics[width=0.48\textwidth]{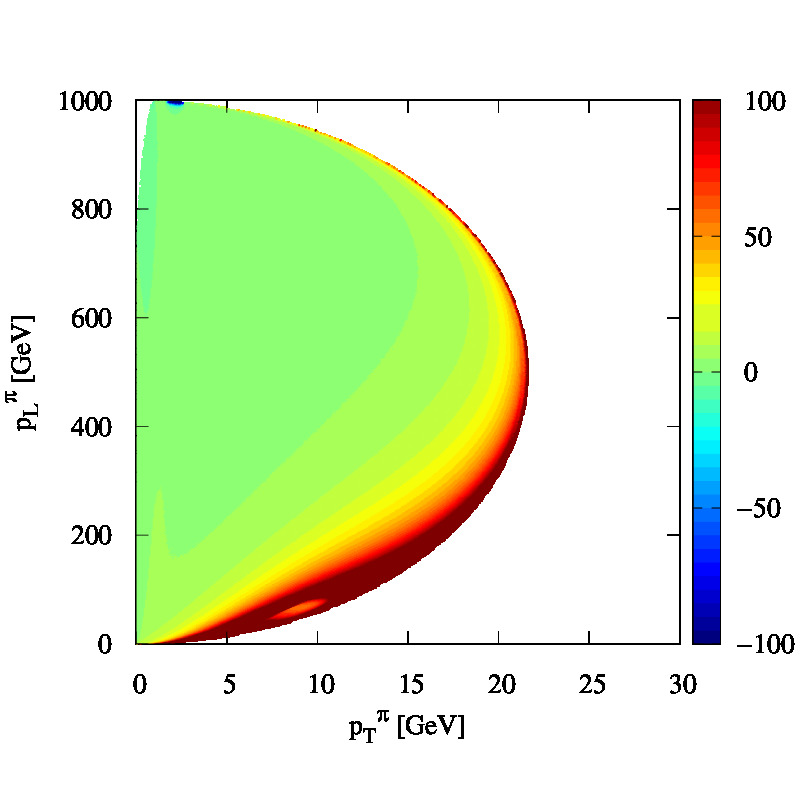}
			\end{tabular}
\caption{ Difference in \% between the predictions for $\mathrm{d}\sigma/(\mathrm{d}p_T^\pi \mathrm{d}p_L^\pi)$ derived assuming the complete solution and obtained considering that $F_5 = 0$. Results for the incident (anti)neutrino energy of 100 GeV (upper panels) and 1000 GeV (lower panels). }
\label{fig_4:diff_F50}
\end{figure}

Figure \ref{fig_4:diff_F50} shows the results of the percentage difference between the complete solution of the cross section of Equation (\ref{eq_4:sigma4}) and considering $F_5 = 0$. Our results show that the cross section has greater modifications for low (high) transverse moments of the pions produced by (anti)neutrinos.

\section{Conclusions}

In this chapter, we investigate how certain effects present only in tau neutrino interactions impact LHC neutrino observables, which can be measured in the proposed FASER$\nu$2 experiment during the high-luminosity LHC era. In particular, we show that taus produced in charged current neutrino interactions in the 100-1000 GeV range will not be completely polarized, but will have approximately 94\% polarization. Monte Carlo generators that simulate tau neutrino interactions usually consider taus to be completely polarized in the final state; therefore, our results encourage the inclusion of realistic polarizations for taus in these generators. In addition, our results showed that the differential cross sections with respect to the kinematic variables of the tau and also of the pion resulting from its decay are sensitive to the polarization state of the tau produced, as well as to the structure function $F_5$, which manifests itself more strongly in interactions of tauonic neutrinos. In the next chapter we will show that the frontal detectors of the LHC can measure muons in addition to neutrinos. We will discuss how experiments such as FASER$\nu$ can contribute to our understanding of the nucleon structure, especially its intrinsic charm and nuclear effects.

\chapter{DIS at the LHC with muons and neutrinos}	
\label{cap:MuonDIS}

In hadronic collisions that occur at the LHC, it is expected that all particles of the Standard Model will be produced, but among them, only neutrinos and muons are able to reach the far-forward detectors, such as FASER and SND@LHC. Neutrinos arrive with extreme ease, given that they have no electric charge and propagate over these distances (approximately 480 m) practically without interacting. On the other hand, muons are electrically charged, and their propagation along the LHC structure and the approximately 100 m of rock between FASER and ATLAS needs to be handled carefully. Typical LHC muons lose about 60 GeV of their energy as they pass through the rocks until they reach the far-forward detectors \cite{FPF:2025bor}.

Several recent studies have demonstrated the feasibility of using FASER to measure cross sections of deep inelastic scattering of all neutrino flavors in charged current \cite{FASER:2019dxq,Cruz-Martinez:2023sdv,vanBeekveld:2024ziz} and neutral current interactions \cite{Ismail:2020yqc}. In particular, the neutrino DIS data expected during Run 4 of the LHC may be important even in global analyses to reduce the current uncertainties present in the PDFs \cite{Cruz-Martinez:2023sdv}. In this work, we explore the possibilities of measuring the muon DIS at the LHC, and our results show that the kinematic region in $x$ and $Q^{2}$ of the muon DIS at the FASER$\nu$ will be similar to that covered by the Electron Ion Collider (EIC) \cite{AbdulKhalek:2021gbh}, which is being built in Brookhaven to operate in the next decade. We will show that the DIS of muons at FASER will extend the kinematic regimes tested by the experiments BCDMS \cite{BCDMS:1989ggw}, EMC \cite{Kullander:1990se}, NMC \cite{NewMuon:1996fwh} and COMPASS \cite{COMPASS:2007rjf} at CERN, HERA \cite{Klein:2008di} at DESY, and CLAS \cite{CLAS:2003umf} at Jefferson Lab, which also observed the DIS of charged leptons. As an application of the physical opportunities surrounding this process, we will study the possibility of discovering an intrinsic charm in the nucleon\footnote{The intrinsic charm is a component of charm that arises from non-perturbative mechanisms in the proton wavefunction. Conversely, the extrinsic component of charm (the perturbative component arising from gluon splitting) is expected to contribute mainly to a large $x$, and not disappear when $Q^2 \rightarrow 0$. We will provide more details about intrinsic charm in Subsection \ref{subsec_MuonDIS:charmContent}}, which has been debated since the results of the EMC experiment published in the 1980s \cite{EuropeanMuon:1982fow}.

Recent results for neutrino DIS at the LHC usually describe the structure of the tungsten target without considering nuclear effects \cite{Cruz-Martinez:2023sdv,vanBeekveld:2024ziz}. In this chapter, we will verify the nuclear effects in terms of total and differential numbers of neutrino and muon DIS events at the LHC at FASER$\nu$ and FASER$\nu$2. This measurement may be of particular interest in the study of PDF universality, given the current incompatibility between some neutrino experiment data compared to charged lepton scattering data \cite{Muzakka:2022wey}. Our main results presented in this chapter have been published in the references \cite{Francener:2025pnr,Francener:2025tyh}.

\section{Muon flux at the FASER$\nu$}
\label{sec_MuonDIS:flux}

The muon flux that reaches the LHC forward detectors is usually treated as a background that needs to be removed to search for neutrino interactions and physics beyond the Standard Model. Part of this flux is removed by the magnets that are installed to deflect protons \cite{FASER:2018bac}. The muon flux has already been measured by both collaborations of forward experiments at the LHC, FASER \cite{FASER:2018bac,FASER:2020gpr} and SND@LHC \cite{SNDLHC:2023mib}. In addition to being measured, the muon flux is simulated with the FLUKA Monte Carlo generator \cite{Sabate-Gilarte:2023aeg,Battistoni:2015epi}. Part of this flux comes directly from collisions in ATLAS and the decay of particles produced in these interactions, but a considerable portion comes from interactions of particles produced in ATLAS with the LHC infrastructure, as well as with the 100 meters of rock between the interaction point in ATLAS and the FASER experiment. Unlike neutrino flux, muon flux can be measured directly at the detectors, so we will consider the uncertainty in its modeling to be negligible.

We are considering two distinct cases in our analyses for the simulation of muon flux: (1) a rectangular area of  25 cm $\times$ 30 cm, which corresponds to the complete transverse area of the FASER$\nu$ detector and (2) a circular area of radius r = 9 cm, which corresponds to the overlap between the FASER$\nu$ and the FASER spectrometer. In both cases, the center of the both areas are located at $(x,y)$ = (1 cm, -3.3 cm). We will assume the expected muon flux during Run 3 of the LHC, in which the FASER is expected to operate with an integrated luminosity of $\mathcal{L}_{pp} = 250$~fb$^{-1}$ \cite{FASER:2024hoe}.

FASER$\nu$ is a detector with emulsion films interleaved with 730 tungsten plates of approximately 1.1 mm thickness, resulting in an approximately $L_T = 80$~ cm target with a 1.1 metric tons. In our results, we will assume only $L_T = 50$~cm of longitudinal length of the tungsten target, as we will keep the first and last 15~cm to be used in measuring the energy of the incident and final muons through multiple Coulomb scattering. Therefore, the associated event numbers presented here will be conservative, and the total number associated with FASER$\nu$ can be obtained by multiplying by 1.6.

The muon flux we are using can be represented by $N$, a Probability Density Function (PDF), normalized with the number of muons present in the flux. Then, to insert the detector characteristics, we rewrite the muon flux PDF as
\begin{eqnarray}
\label{eq_DIS:muon_fluxes}
f_\mu(x_\mu) \equiv  n_T L_T \frac{dN_{\mu}}{dx_\mu}(x_\mu) 
\, ,\qquad  x_\mu \equiv \frac{2E_\mu}{\sqrt{s_{\mathrm{pp}}}}\, ,
\end{eqnarray}
where $\sqrt{s_{\mathrm{pp}}}$ is the center-of-mass energy of proton-proton collisions at the LHC, which we are considering to be 14~TeV. When $x_\mu = 1$, the muon has the highest kinematically allowed energy, 7~TeV. $N_\mu$ is the number of muons and $n_T$ is the density of target nucleons, defined as the ratio between the density of tungsten (19.3~g/cm$^{3}$) and the mass of the nucleon. The Equation~(\ref{eq_DIS:muon_fluxes}) written in differential form for the muon flux as a function of the energy fraction varying between 0 and 1 allows us to use the LHAPDF \cite{Buckley:2014ana} to estimate its flux. Reference \cite{vanBeekveld:2024ziz} constructed the neutrino flux at the LHC in a similar way to that described for the muon flux, using the neutrino fluxes from light mesons described in \cite{Kling:2021gos} and from heavy mesons described in \cite{Kling:2021gos}.

\begin{figure}[t]
    \centering
\includegraphics[width=0.49\linewidth]{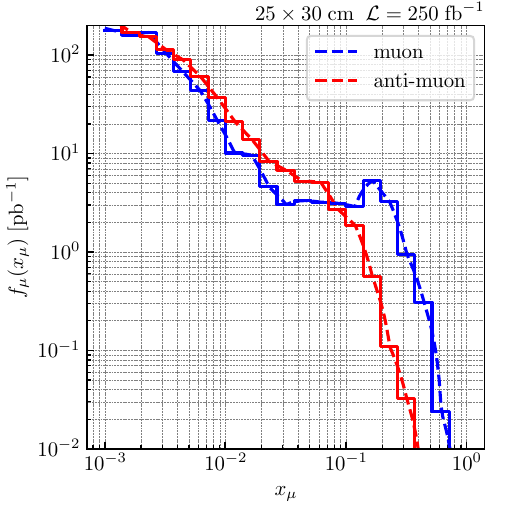}
\includegraphics[width=0.49\linewidth]{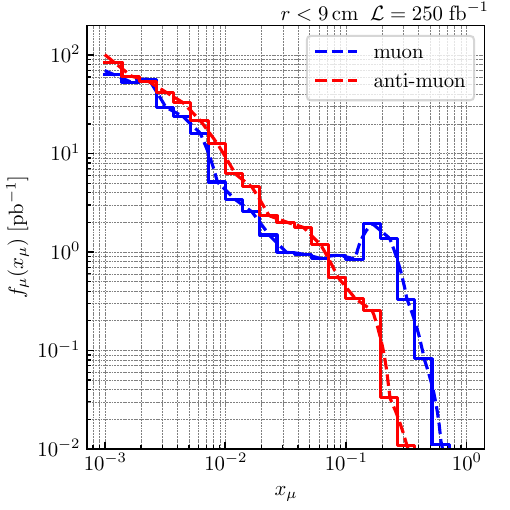}
    \caption{ Flux of (anti)muons through the rectangular (left) and circular (right) area of the FASER detector in blue (red). We are assuming an integrated luminosity of $\mathcal{L}_{\rm pp}=250$ fb$^{-1}$ for the FASER during Run 3 of the LHC. The dashed lines are obtained with LHAPDF interpolation, while the solid line histograms are the original results from the FLUKA simulations. The fluxes are normalized according to the Equation~(\ref{eq_DIS:muon_fluxes}).
    }
    \label{fig_DIS:muonflux}
\end{figure}

Figure \ref{fig_DIS:muonflux} shows the PDF of the muon flux at the FASER as described above, considering the expected luminosity for the FASER during Run 3 with $\mathcal{L}_{pp} = 250$~fb$^{-1}$ and using FLUKA simulations. We compare the FLUKA simulations in the histogram with our interpolation using LHAPDF, using dashed lines. Note the large asymmetry between the PDFs of the muon and antimuon fluxes, resulting from the passage of muons through the LHC magnets.

\section{DIS with muons at FASER$\nu$}
\label{sec_MuonDIS:config}

The muon flux described above that reaches the FASER detector is coupled with the lepton deep inelastic scattering event generator, POWHEG-BOX-RES \cite{Banfi:2023mhz,FerrarioRavasio:2024kem}, which provides the total and differential cross sections at the NLO of QCD coupling constant \cite{Frixione:2007vw,Alioli:2010xd}. The parton-level cross sections of POWHEG are interfaced with Pythia8 to perform hadronization of the hadronic final state \cite{Sjostrand:2014zea,FerrarioRavasio:2024kem,Bierlich:2022pfr}, with the Monash 2-13 tune \cite{Skands:2014pea}.

Our study in this section focuses on simulating the process
\begin{eqnarray}
\mu^\pm + W \to \mu^\pm + X_h \, ,
\end{eqnarray}
which includes semi-inclusive scattering with the production of charming hadrons
\begin{eqnarray}
\mu^\pm + W \to \mu^\pm + c\,(\bar{c}) +\widetilde{X}_h \, ,
\end{eqnarray}
with $W$ indicating the tungsten target, the material from which the FASER$\nu$ is composed, and $X_h,\widetilde{X}_h$ are the inclusive hadronic final states. In our studies, we will assume that the FASER$\nu$ is capable of measuring the initial ($E_\mu$) and final ($E_\mu '$) muon energies through multiple scatterings on the tungsten plates, as well as the muon scattering angle at the vertex of the DIS ($\theta_\mu$). With these three variables, energies and angle, it is possible to reconstruct the DIS variables of our interest, Bjorken-$x$, inelasticity $y$, and virtuality $Q^{2}$, through the definition of these variables, presented in more detail in Chapter \ref{cap:cs}.

The number of events expected at the detector can be expressed by
\begin{eqnarray}
    N_{\mathrm{events}} = 
    \int_{0}^{1} \int_{0}^{1} \int_{Q^{2}_{\mathrm{min}}}^{Q^{2}_{\mathrm{max}}}
    \mathrm{d}x\, \mathrm{d}x_{l_i}\,\mathrm{d}Q^{2}\,  \frac{\mathrm{d}^{2}\sigma_{l_i\, W}}{\mathrm{d}x\, \mathrm{d}Q^{2}}f(x_{l_i})\, \mathcal{A}(E_{l_f}, n_{\mathrm{tr}}) \, ,
    \label{eq:events}
\end{eqnarray}
where $\mathrm{d}^{2}\sigma_{l_i\, W}/\mathrm{d}x\,\mathrm{d}Q^{2}$ is the lepton-tungsten differential cross section, $x$ and $Q^{2}$ are the DIS variables, $f(x_{l_i})$ is the PDF of the lepton flux with energy fraction $x_{l_i}$, and $\mathcal{A}$ describes the detector efficiency as a function of the final lepton energy ($E_{l_f}$) and the number of charged hadrons with minimum momentum produced at the interaction vertex ($n_{\mathrm{tr}}$). The detector geometry and exposure time for data collection are contained in $f(x_{l_i})$.

For DIS events, in addition to the cuts of $Q > 1.65$ GeV and $W > 2$ GeV, which specify the DIS of our interest and ensure that we are considering the perturbative region of QCD, we will also apply some vertex selection cuts that describe the $\mathcal{A}(E_{l_f}, n_{\mathrm{tr}})$:

\begin{itemize}
\item The energy of the final muon must satisfy $E'_\mu > 100 \;\mathrm{GeV}$;

\item We are adopting a cut in the minimum number of charged particle tracks in the final state, $n_{\mathrm{tr}}$. Specifically, we are considering that each track has a minimum momentum of $|\vec{p}_{\rm tr}|=1\;\mathrm{GeV}$. We are considering three cases of minimum track numbers, denoted by $n_{\rm tr}\ge 3,4,$ or $5$. This number of tracks includes one muon track in the final state, in addition to 2, 3, and 4 hadronic tracks. The cut of at least 5 tracks in the final state is currently used by the FASER collaboration in the neutrino-initiated event selection algorithm. Given that muon-initiated DIS events are usually of lower $Q^2$ magnitude than neutrino-initiated events, hard cuts in the number of tracks exclude a significant portion of low and moderate $Q^2$ events, as well as the high $x$ region, which could potentially exclude some events with intrinsic charm in the nucleon;

\item In addition to the event selection kinematics, we are considering an efficiency of $\epsilon_c = 0.7$ for the detection of a charmed hadron that meets the required kinematic conditions, as suggested by members of the FASER collaboration \cite{FASERprivate}.

\end{itemize}

With the event selection settings listed above, we can estimate the number of events at the FASER$\nu$ in deep inelastic scattering of muons in neutral current interactions. The number of events is presented in Table~\ref{table:Nevents}, summing muon and antimuon contributions. In this table, as well as throughout the rest of the chapter, we are considering a time-integrated luminosity of $\mathcal{L}_{\rm pp}=250$ fb$^{-1}$ for the FASER$\nu$ during Run 3 of the LHC. These predictions for the events were obtained with the muon flux shown in Figure~\ref{fig_DIS:muonflux}, and with the cross section  modeled at NLO with POWHEG and hadronization done with Pythia8. Our results also take into account the non-isoscalarity corrections of the target, that is, the asymmetry between protons and neutrons in the tungsten nucleus. In this way, the tungsten nucleus is modeled as
\begin{eqnarray}
f_{i/N}(x,Q^2) = (Z\,f_{i/p}(x,Q^2) + (A - Z)\,f_{i/n}(x,Q^2))/A  \, , \qquad \mathrm{com}\,\, i = q,\,\bar{q}, \,g \, ,
\end{eqnarray}
where $f_{i/N}$ describes the PDF of the average tungsten nucleon, which has atomic number $Z$ and mass number $A$. $f_{i/p(n)}$ is the PDF of the free proton (neutron), and at this initial stage we will not consider its modifications due to nuclear effects. In this initial investigation of the number of DIS events with muons and the kinematic region covered by FASER$\nu$, we will be considering four distinct sets of PDFs provided at NNLO: NNPDF4.0~\cite{NNPDF:2021njg}, which contains the fitted (intrinsic) charm as standard, NNPDF4.0 with only the perturbative charm component, CT18 FC with fitted charm~\cite{Guzzi:2022rca} with the BHPS3 model~\cite{Brodsky:1980pb} for $\Delta\chi^2=30$, and the standard set from the collaboration CT18~\cite{Hou:2019efy} which includes only the perturbative charm. For each set of PDFs, we will present the predictions of total and differential numbers of events with and without charm identified in the final state.

\begin{table}[t]
\centering
\renewcommand{\arraystretch}{1.5}
\begin{tabularx}{\textwidth}{Xclcc}
\toprule
\multicolumn{5}{c}{Muon DIS events at FASER$\nu$ with $\mathcal{L}_{\rm pp}=250$ fb$^{-1}$}\\
\midrule
PDF     & Charm PDF  & Process		                 & DIS              & DIS + efi.\\	
\toprule
\multirow{2}{*}{NNPDF4.0} & \multirow{2}{*}{Fit. charm} &$\mu^\pm + W \rightarrow \mu^\pm + X$      & 2.7$\times10^{5}$         & 1.7$\times10^{5}$       \\
&& $\mu^\pm + W \rightarrow \mu^\pm + X + c(\bar{c})$  & 7.2$\times10^{3}$         & 5.8$\times10^{3}$ \\
\midrule
\multirow{2}{*}{NNPDF4.0~PC} & \multirow{2}{*}{Pert. charm} & $\mu^\pm + W \rightarrow \mu^\pm + X$      & 2.6$\times10^{5}$         & 1.8$\times10^{5}$      \\
& &$\mu^\pm + W \rightarrow \mu^\pm + X + c(\bar{c})$  & 7.5$\times10^{3}$         & 6.3$\times10^{3}$\\      
\midrule
\multirow{2}{*}{CT18 FC}     & \multirow{2}{*}{Fit. charm (BHPS3)} & $\mu^\pm + W \rightarrow \mu^\pm + X$      & 2.6$\times10^{5}$         & 1.8$\times10^{5}$        \\
& &$\mu^\pm + W \rightarrow \mu^\pm + X + c(\bar{c})$  & 1.6$\times10^{4}$         & 1.2$\times10^{4}$\\
\midrule
\multirow{2}{*}{CT18}     & \multirow{2}{*}{Pert. charm}& $\mu^\pm + W \rightarrow \mu^\pm + X$      & 2.6$\times10^{5}$         & 1.7$\times10^{5}$        \\
&& $\mu^\pm + W \rightarrow \mu^\pm + X + c(\bar{c})$  & 1.2$\times10^{4}$         & 9.5$\times10^{3}$\\
\bottomrule
\end{tabularx}
\vspace{0.3cm}
\caption{ Predicted number of neutral current muon DIS events at the FASER$\nu$, summing muon and antimuon scattering, and assuming an integrated luminosity of $\mathcal{L}_{\rm pp}=250$ fb$^{-1}$. The muon fluxes used in the calculation are shown in Figure~\ref{fig_DIS:muonflux} and the cross section modeled with POWHEG+Pythia8 NLO simulations. We consider four sets of NNLO PDFs: from top to bottom, NNPDF4.0 (which by default includes an intrinsic charm PDF), NNPDF4.0 with perturbative charm, CT18 FC with an intrinsic charm with the BHPS3 model (for $\Delta\chi^2=30$) and the reference CT18 set with perturbative charm. For each set of PDFs, we provide predictions for both inclusive DIS and charm production, in the latter case assuming a charm detection efficiency of $\epsilon_c$ = 70\%. We provide the events after applying only the DIS cuts and then after applying the kinematic cuts for detection efficiency as well.
}
\label{table:Nevents}
\end{table}

The results in Table \ref{table:Nevents} indicate a large number of muon scattering events at FASER$\nu$ during Run 3 of the LHC, both for inclusive events and for events with charm hadrons identified in the final state. Our results also indicate that the number of inclusive events is practically independent of the choice of PDF used in the calculations, and is on the order of $2 \times 10^{5}$ after applying the experimental cuts described in the previous section. The effect of the kinematic cuts decreases the events by about 30\% compared to events where only the DIS cuts are applied. In the semi-inclusive case, with at least one charm hadron identified in the final state, the number of events expected after applying cuts is on the order of $10^{4}$, about 20 times smaller than the inclusive events. However, the effects of cuts are less restrictive for events with detected charm, which is easily understood since these are usually more energetic events that more easily satisfy the minimum number of tracks.

Our predictions for the number of muon scattering events at FASER can be compared with the scattering observed in the emulsion detector pilot project. This pilot project exposed approximately 10 kg of tungsten plates interleaved with emulsion films to an integrated LHC luminosity of 12.2 fb$^{-1}$, and observed 78 muon scattering events \cite{FASER:2021mtu}, a number comparable to our prediction of 83 expected events. This indicates that our results obtained with POWHEG+Pythia8 are consistent with what was measured by the FASER$\nu$ pilot project.

\section{Inclusive DIS with muons}
\label{sec_MuonDIS:DISinclusivo}

In the previous section, we presented the total number of events from the muon DIS in neutral current interactions with and without detection cuts in the FASER$\nu$. The results indicate large numbers of expected events even for events with charm detected in the final state. In this section, we will take a closer look at the process
\begin{eqnarray}
\mu^\pm + W \rightarrow \mu^\pm + X_h \, ,
\end{eqnarray}
at the FASER$\nu$, considering the selection cuts for events described previously, but now focusing on the distributions of events in different kinematic variables. In the muon kinematic regime of the LHC, where the detected muons are generally not much more energetic than 1 TeV, muon scattering usually satisfies $Q^2\ll m_Z^2$ and the contribution of scattering via neutral current through the exchange of the $Z^{0}$ boson is not important. Although we are considering its contribution, we will not focus on its effects, nor on differences between muon and antimuon production rates due to the contribution of the antisymmetric part of the hadronic tensor.

\begin{figure}[t]
    \centering
\includegraphics[width=0.49\linewidth]{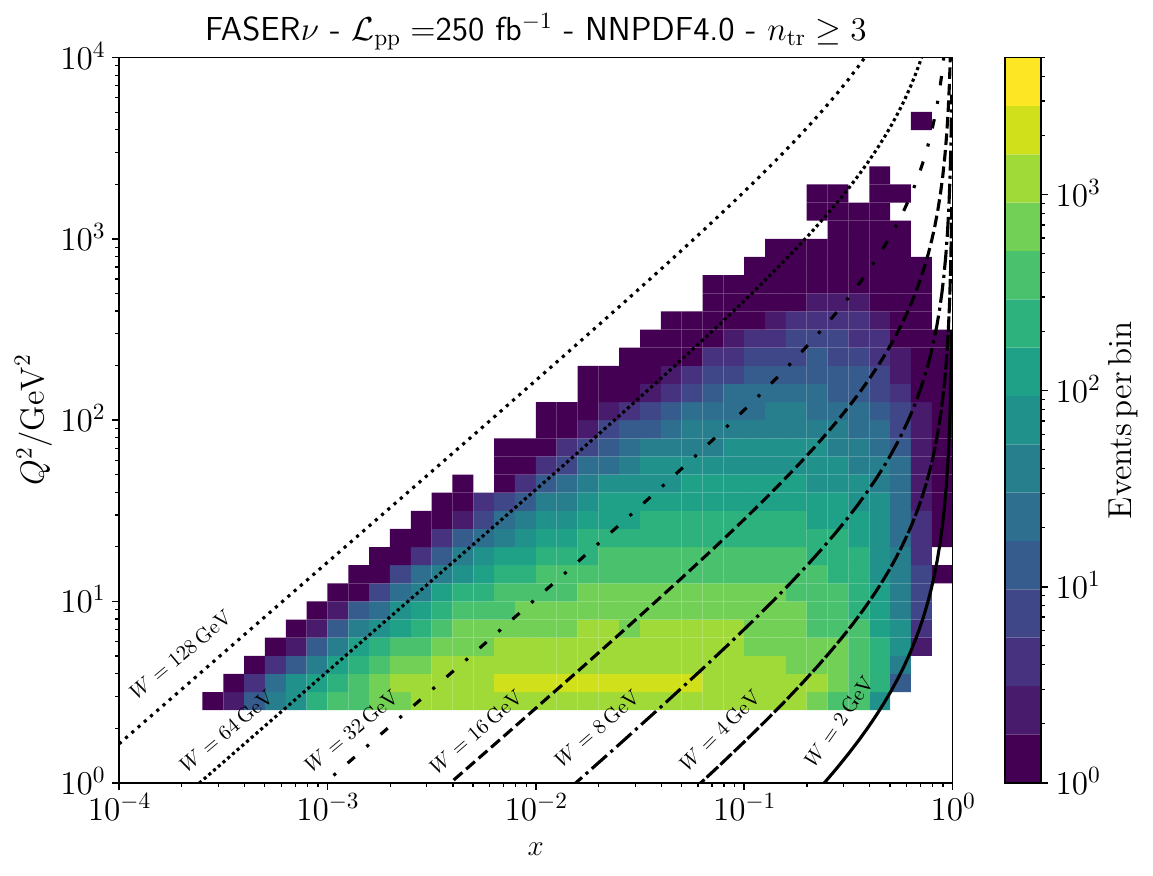}
\includegraphics[width=0.49\linewidth]{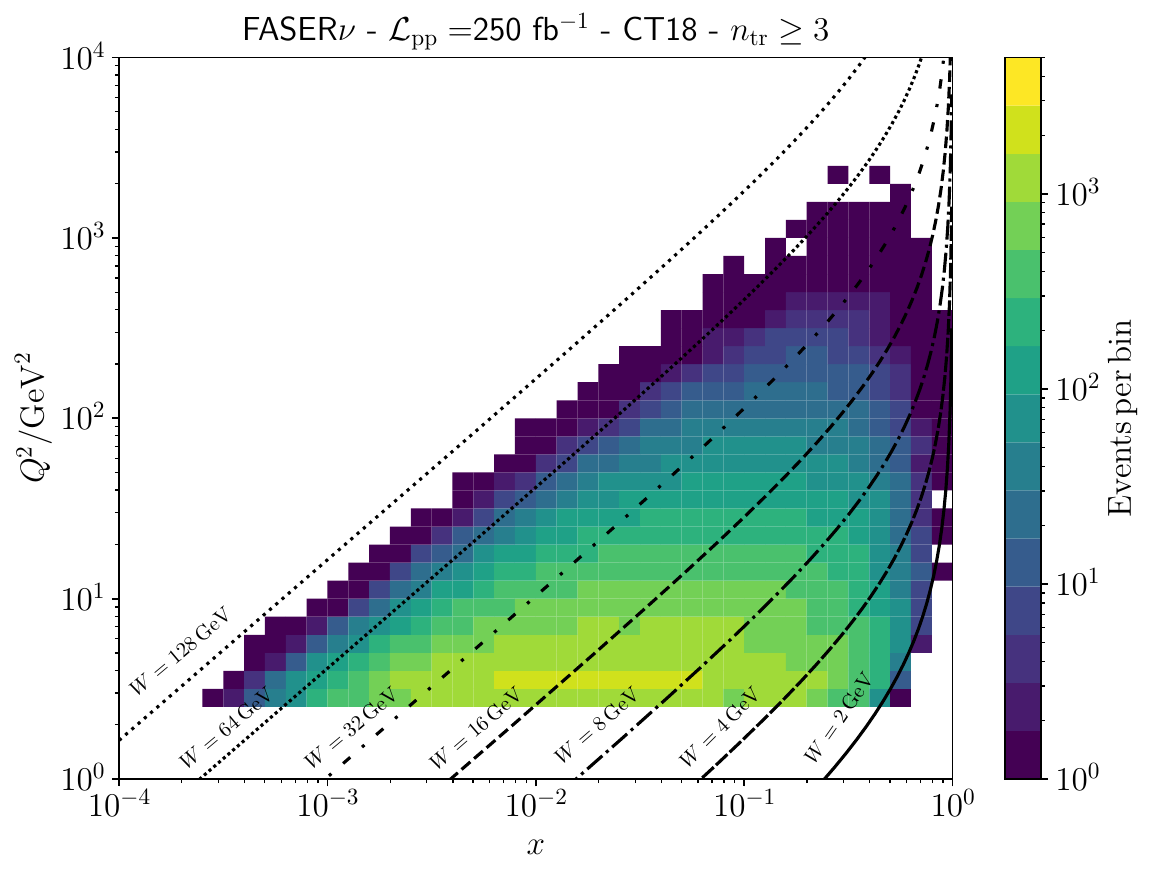}
\caption{ Number of inclusive muon neutral current DIS events predicted for FASER$\nu$ as a function of $Q^{2}$ and $x$ that satisfy the selection cuts described in Section \ref{sec_MuonDIS:config} for an integrated luminosity of $\mathcal{L}_{\rm pp}=250$ fb$^{-1}$. We present the results for the PDF sets NNPDF4.0 (left) and CT18 (right); see also Table~\ref{table:Nevents} for the corresponding integrated event numbers. Similar results are obtained for other PDF sets. The black curves indicate fixed values of the invariant mass $W$.
}
\label{fig:DIStotal_Q_x}
\end{figure}

Figure \ref{fig:DIStotal_Q_x} shows the number of neutral current muon deep inelastic scattering events that satisfy the DIS and selection cuts described in Section \ref{sec_MuonDIS:config} in bins of $Q^{2}$ and $x$ for the integrated luminosity $\mathcal{L}_{\rm pp}=250$ fb$^{-1}$, which is expected for FASER$\nu$ data collection during Run 3 of the LHC. We are presenting the results using two distinct parameterizations for the PDFs: NNPDF4.0 and CT18, which have fitted and perturbative charm, respectively. However, in this section we will not focus on verifying the number of events with and without intrinsic charm; we will restrict ourselves here to inclusive scattering events. The black curves in the figure are level curves in the invariant mass $W$ of the hadronic final state. We see in Figure \ref{fig:DIStotal_Q_x} that the muon DIS of the LHC at the FASER$\nu$ operating during Run 3 covers a region of $x \gsim 3\times 10^{-4}$, and extends to about $x\sim 0.8$ for $Q = 1.65$~GeV. While the range covered by $Q^{2}$ extends to about $Q^2_{\rm max}\sim 10^{3}$ GeV$^2$. This kinematic region covered by the FASER$\nu$ closely resembles the one that will be probed by the EIC.

The distribution of expected events in the ($x,Q^{2}$) plane also shows us that events are more concentrated at small values of $Q^{2}$, given that the interaction between charged leptons and hadronic targets usually occurs through the exchange of a virtual photon, and the cross section has a term of $1/Q^{4}$ due to the propagator of this photon, which suppresses contributions of high virtualities. This behavior is the opposite of that observed in the neutrino DIS, where contributions of small $Q^{2}$ are suppressed by a massive propagator \cite{Cruz-Martinez:2023sdv}. This difference between neutrino and muon DIS leads to a greater suppression of muon events than neutrino events when we apply cuts in $Q^{2}$, as well as in the number of charged particle tracks in the final state.

Figure~\ref{fig:DIStotal_Q_x} also shows the invariant mass ($W$) level curve of the hadronic final state in the plane ($x, Q^{2}$), which is given in terms of the DIS variables by
\begin{eqnarray}
\label{eq:W_def}
W^2 = m_p^2 + \frac{Q^2(1-x)}{x} \, .
\end{eqnarray}
Our results indicate that events are concentrated in $W$ values usually less than 32 GeV, and cuts of the $W \ge 8$ type would remove a significant portion of the events. These results indicate that a careful analysis of the low $Q^{2}$ and large $x$ region, i.e., low $W$, needs to be performed both theoretically and experimentally to explore important contributions of DIS with muons at FASER$\nu$. This region is sensitive to mass effects of the involved quark \cite{Buonocore:2024pdv}, as well as high twist corrections \cite{Virchaux:1991jc,Accardi:2011fa}, therefore these effects need to be understood and quantified prior to including the data in fits of global PDF analyses.

\begin{figure}[t]
\centering    \includegraphics[width=0.49\linewidth]{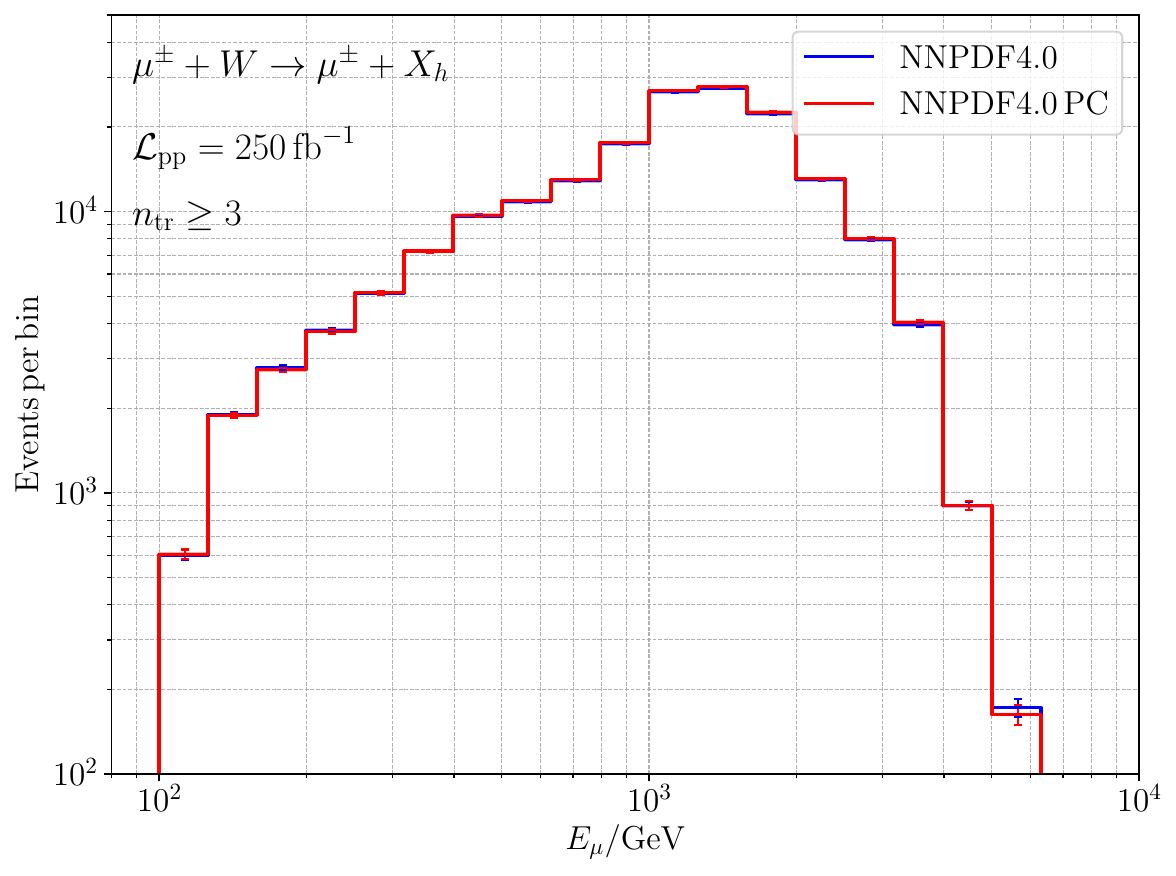}
\includegraphics[width=0.49\linewidth]{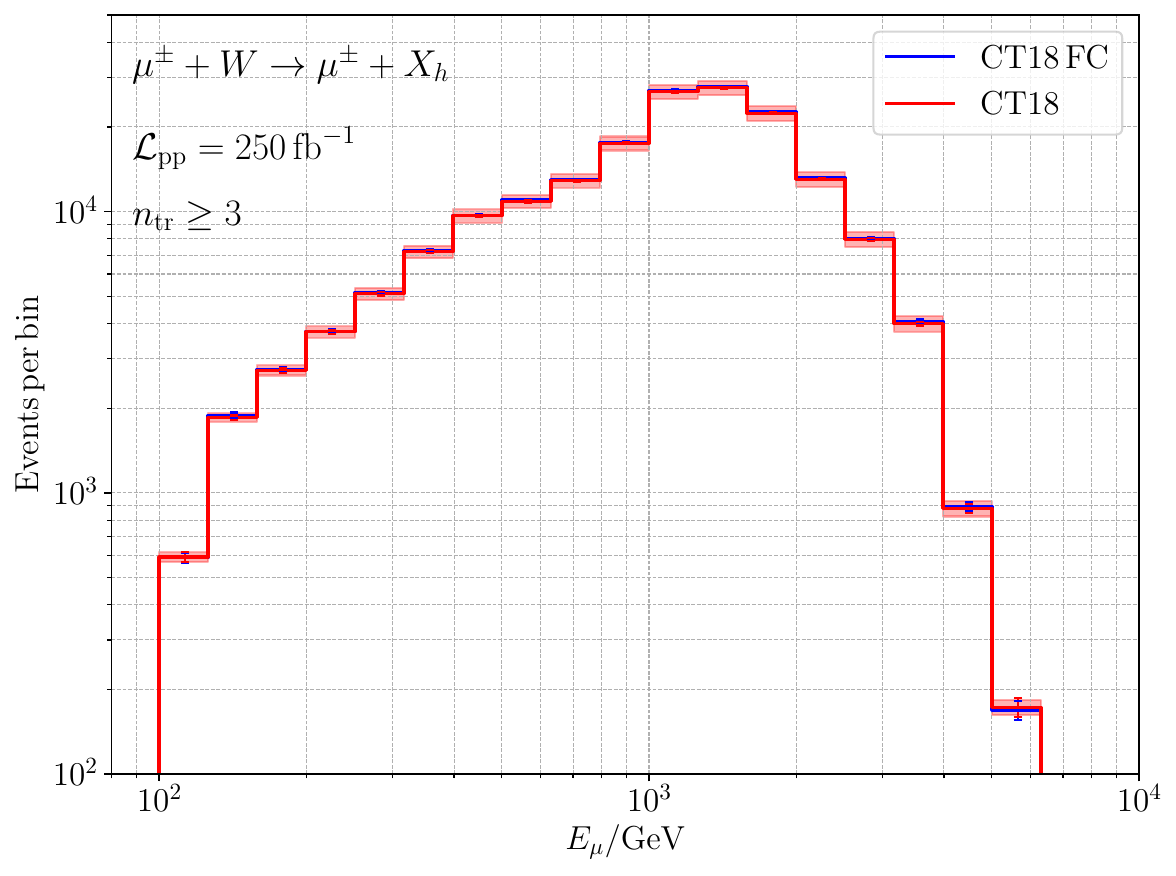} \\
    \caption{Number of events per bin of inclusive muon DIS at FASER$\nu$ for $\mathcal{L}_{\rm pp}=250$ fb$^{-1}$ as a function of the incident muon energy, $E_\mu$. The left (right) panel shows the predictions based on the NNPDF4.0 (CT18) parameterization. The uncertainty bars (bands) in each bin indicate the associated statistical (PDF) uncertainty.
}    
\label{fig:DIStotal_cuts_E_distrib}
\end{figure}

In Figure \ref{fig:DIStotal_cuts_E_distrib} we show the muon DIS events after fiducial cuts, binned in the incident muon energy, $E_\mu$. The left (right) panel shows the predictions obtained using the NNPDF4.0 (CT18) parameterization for the PDFs, and we can see similar results. The uncertainty bars represent the statistical uncertainty, where we are considering Gaussian statistics to obtain this quantity, and the uncertainty bands come from the current uncertainties we have in the PDFs, propagated to the number of events expected per bin. For the CT18 FC parameterization, there are no replicas provided by the CT18 collaboration to construct the associated uncertainties. Both uncertainties are set at a 68\% confidence level. The results indicate a maximum of events in the incident muon region between 1 and 2 TeV, dropping abruptly outside this region. These muons are the most energetic ever produced by humans, and have the potential to significantly expand the kinematic regions tested by previous muon DIS experiments. The event rate goes to zero for muons with less than 100 GeV, given that this is the threshold we are considering for the final muon energy in event selection.

\begin{figure}[t]
    \centering
\includegraphics[width=0.49\linewidth]{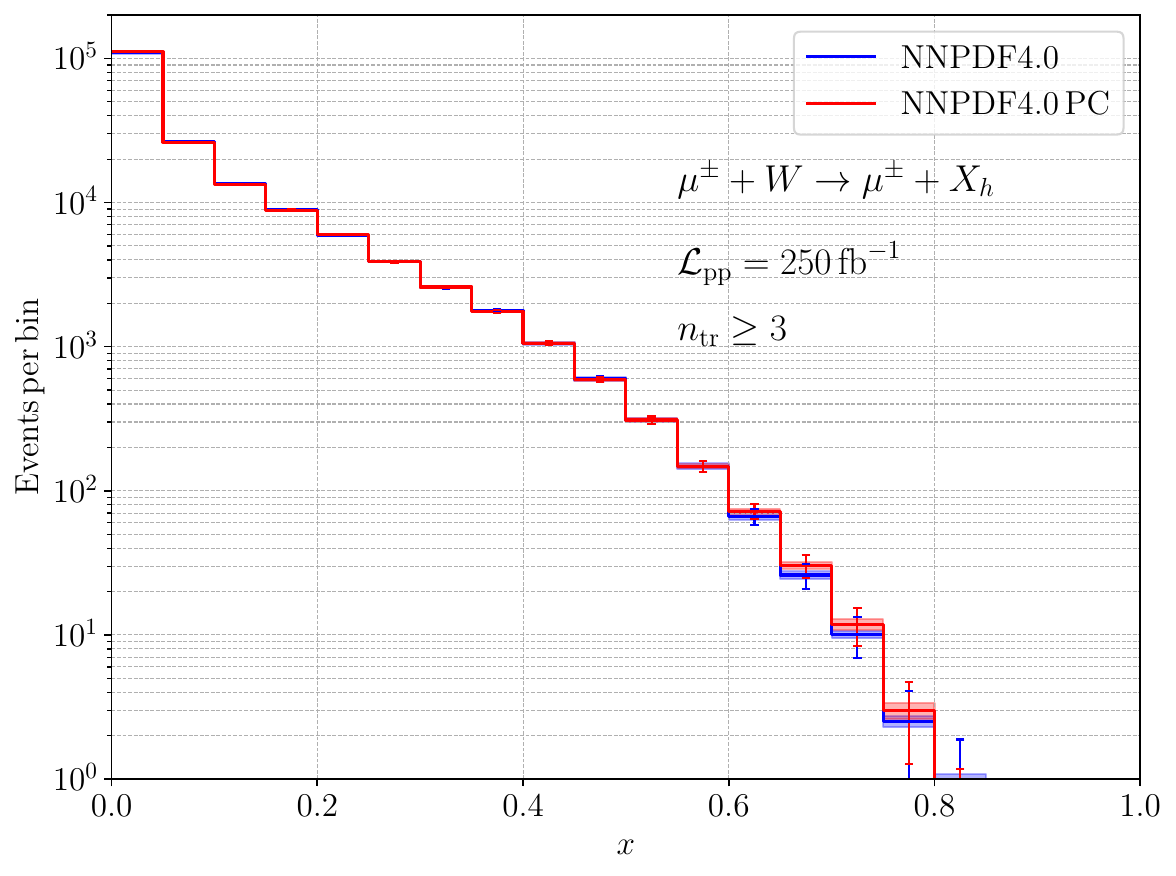}
\includegraphics[width=0.49\linewidth]{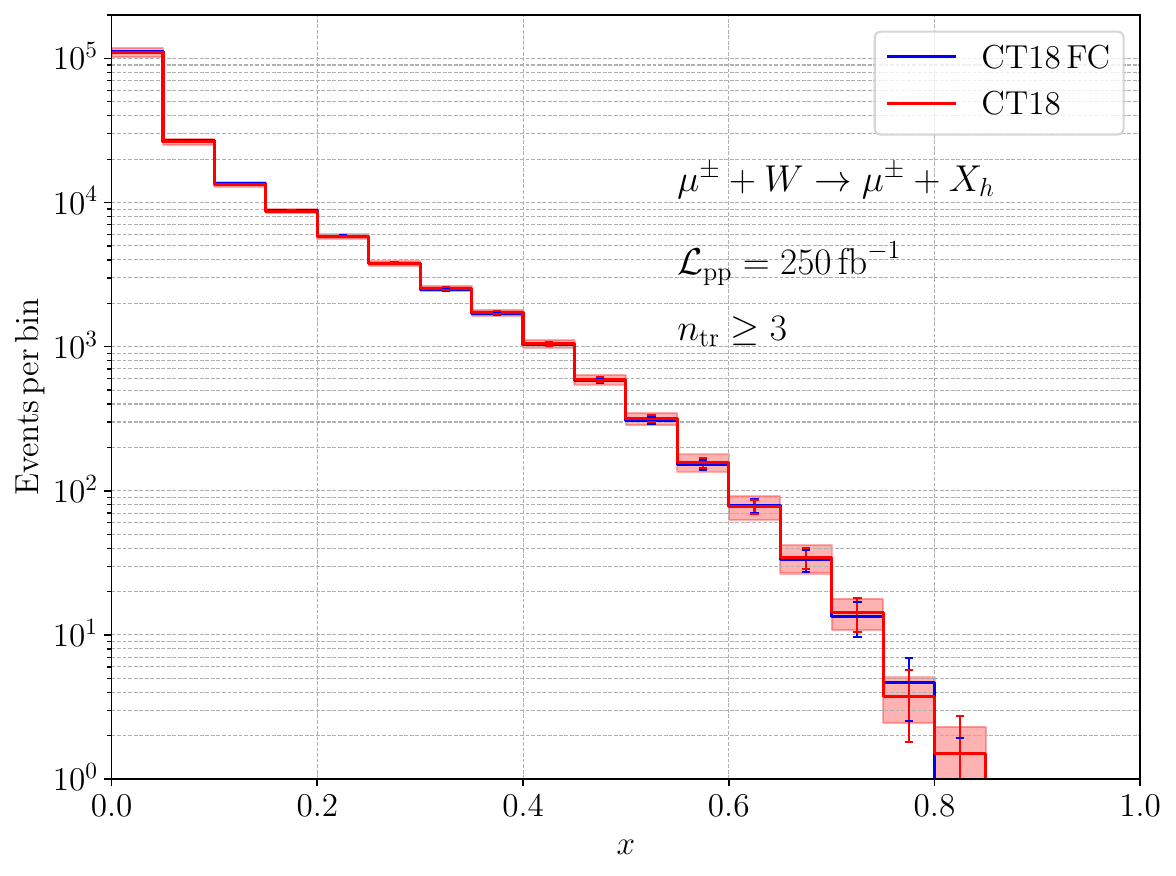} \\
    \caption{Number of events per bin of inclusive muon DIS at FASER$\nu$ for $\mathcal{L}_{\rm pp}=250$ fb$^{-1}$ as a function of Bjorken-$x$. The left (right) panel shows the predictions based on the NNPDF4.0 (CT18) parameterization. The uncertainty bars (bands) in each bin indicate the associated statistical (PDF) uncertainty. }
\label{fig:DIStotal_cuts_x}
\end{figure}

In Figure~\ref{fig:DIStotal_cuts_x} we show the same as presented in Figure~\ref{fig:DIStotal_cuts_E_distrib}, but now for events binned in Bjorken-$x$. Although most events are concentrated in small $x$, there are a significant number of them in the region up to 0.8, showing the potential of FASER$\nu$ to explore this kinematic region. The large $x$ region is where we currently have the greatest uncertainties in the PDFs, and our results show that the statistical uncertainty of the events will have a magnitude similar to the uncertainty of the current PDFs in this region. Furthermore, in the next section we will discuss the particular case of the intrinsic charm in the nucleon, which affects the large $x$ region and can be accessed experimentally.

\begin{figure}[t]
    \centering
\includegraphics[width=0.49\linewidth]{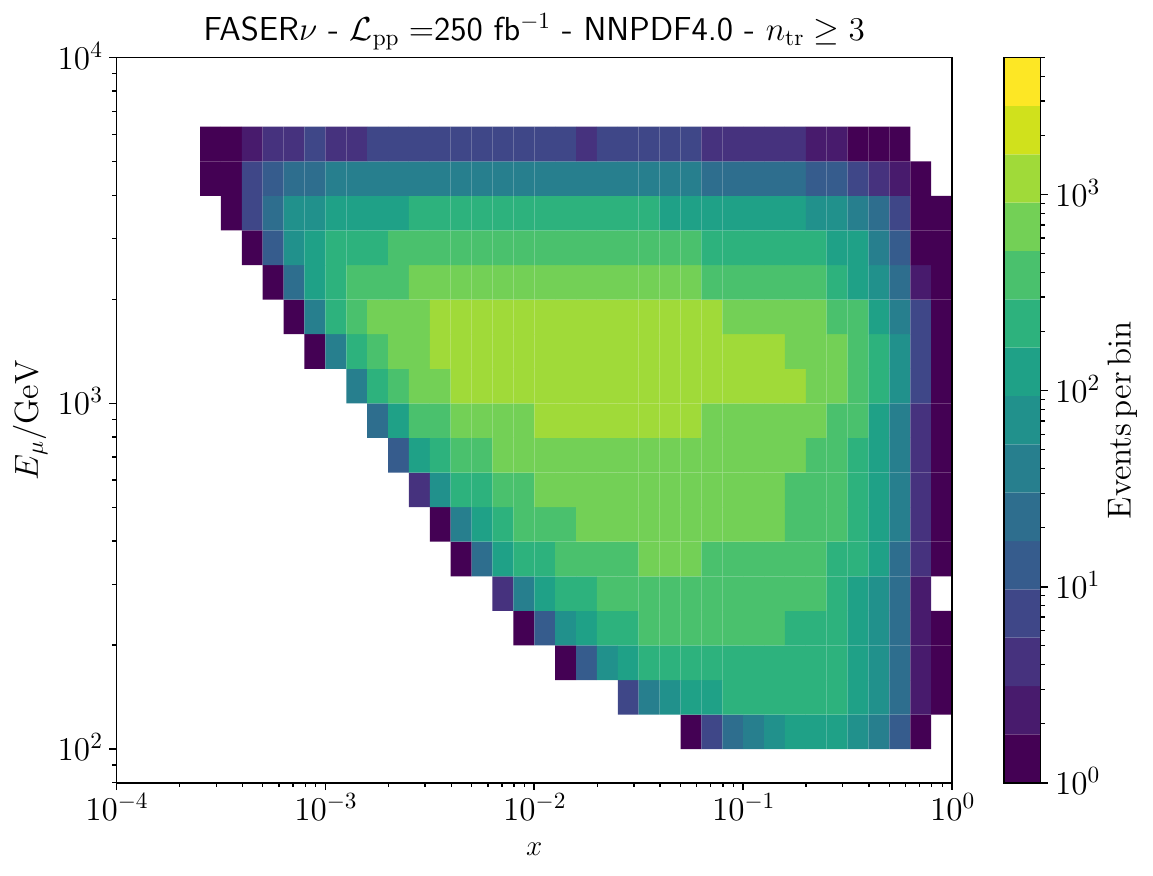}
\includegraphics[width=0.49\linewidth]{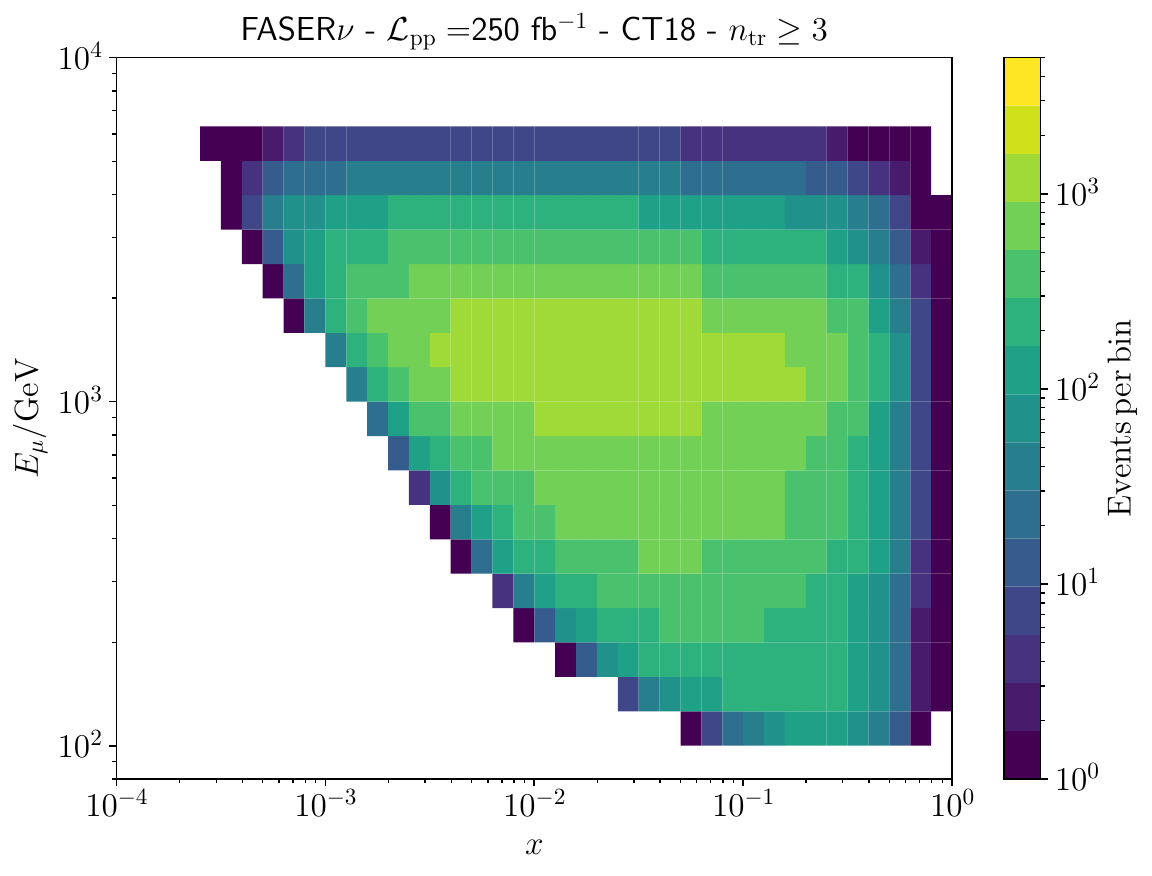} 
    \caption{ Number of inclusive neutral current muon DIS events predicted for FASER$\nu$ as a function of $E_\mu$ and $x$ that satisfy the selection cuts described in the previous section for an integrated luminosity of $\mathcal{L}_{\rm pp}=250$ fb$^{-1}$. We present the results for the PDF sets NNPDF4.0 (left) and CT18 (right); see also Table~\ref{table:Nevents} for the corresponding integrated event numbers. Similar results are obtained for other PDF sets. }
    \label{fig:DIStotal_correlation_E_x}
\end{figure}

Figures ~\ref{fig:DIStotal_cuts_E_distrib} and ~\ref{fig:DIStotal_cuts_x} show the events per bin in $E_\mu$ and $x$, respectively. Now we will explore the relations of these events in these two variables in the ($x,E_\mu$) plane, which is presented in Figure~\ref{fig:DIStotal_correlation_E_x}, similarly to what was presented in Figure~\ref{fig:DIStotal_Q_x} for the events in the ($x,Q^{2}$) plane. Again, we show comparisons between predictions constructed from the NNPDF4.0 (left) and CT18 (right) parameterizations. The figure shows a weak dependence between the $x$ and $E_\mu$ variables, and, as expected, more energetic muons are able to probe regions of lower $x$ of the target nucleon. In particular, 100 GeV muons have a typical $x$ of $\sim$ 0.2, while 1 TeV muons have a typical $x$ of $\sim$ 0.05, and can access regions up to $10^{-3}$.

\paragraph{Predictions for the HL-LHC:} Recently, FASER$\nu$ was approved to continue operating during Run 4 of the LHC. In recent years, as described in previous chapters, an upgrade of FASER$\nu$ to FASER$\nu$2, which would operate in the FPF, has also been proposed. Here we will present our projections for these experiments operating during the high-luminosity regime of the LHC, a period in which data is expected to be collected with an integrated luminosity of 3~ab$^{-1}$ \cite{Azzi:2019yne}. The pseudorapidity region probed by FASER$\nu$2 will be similar to that of FASER$\nu$, so we will consider the same muon flux for an initial approximation. The increase in the number of events will be due to the higher luminosity and the effective detection length by a factor of approximately 12.4 (we are considering the same 30 cm of tungsten plates assumed to measure the initial and final muon energy that we used for the FASER$\nu$ predictions during Run 3). In Table~\ref{table:NeventsFASERnuHL-LHC} we show the same predictions presented in Table~\ref{table:Nevents} for FASER$\nu$ and FASER$\nu$2 during the HL-LHC. The results correspond to the number of events expected applying the same DIS cuts and event selection described in the previous section. The number of events increases linearly with luminosity and detector length. For FASER$\nu$2 at the FPF during the high-luminosity era, up to $3 \times 10^7$ inclusive events and $2 \times 10^6$ events with charm detected are expected. This high number of events enables not only the investigation of the number of events, but a series of multi-differential measurements of the process.

\begin{table}[t]
\centering
\renewcommand{\arraystretch}{1.5}
\begin{tabularx}{\textwidth}{Xclcc}
\toprule
\multicolumn{5}{c}{Muon DIS at FASER$\nu$ (FASER$\nu$2) for $\mathcal{L}_{\rm pp}=250$ fb$^{-1}$ ($\mathcal{L}_{\rm pp}=3$ ab$^{-1}$)}\\
\midrule
PDF                    & charm PDF  & Process			                 & FASER$\nu$             & FASER$\nu$2 \\	
\toprule
\multirow{2}{*}{NNPDF4.0}   & \multirow{2}{*}{Fit. charm}        &$\mu^\pm + W \rightarrow \mu^\pm + X$             & 2.0$\times10^{6}$  
&
2.5$\times10^{7}$ 
\\
& & $\mu^\pm + W \rightarrow \mu^\pm + X + c(\bar{c})$              & 7.0$\times10^{4}$ 
&
8.6$\times10^{5}$
\\
\midrule
\multirow{2}{*}{NNPDF4.0PC} & \multirow{2}{*}{Pert. charm}         & $\mu^\pm + W \rightarrow \mu^\pm + X$             & 2.1$\times10^{6}$  
&
2.6$\times10^{7}$
\\
& &$\mu^\pm + W \rightarrow \mu^\pm + X + c(\bar{c})$              & 7.6$\times10^{4}$ 
&
9.4$\times10^{5}$
\\      
\midrule
\multirow{2}{*}{CT18 FC}    & \multirow{2}{*}{Fit. charm (BHPS3)} & $\mu^\pm + W \rightarrow \mu^\pm + X$      & 2.1$\times10^{6}$ 
&
2.6$\times10^{7}$ 
\\
& &$\mu^\pm + W \rightarrow \mu^\pm + X + c(\bar{c})$              &  1.4$\times10^{5}$
&
1.7$\times10^{6}$
\\
\midrule
\multirow{2}{*}{CT18}       & \multirow{2}{*}{Pert. charm}& $\mu^\pm + W \rightarrow \mu^\pm + X$              & 2.0$\times10^{6}$ & 
2.5$\times10^{7}$ 
\\
&& $\mu^\pm + W \rightarrow \mu^\pm + X + c(\bar{c})$              & 1.1$\times10^{5}$
&
1.4$\times10^{6}$
\\
\bottomrule
\end{tabularx}
\vspace{0.3cm}
\caption {Same result as Table \ref{table:Nevents} for the FASER$\nu$ and FASER$\nu$2 detectors operating during the high-luminosity regime of the LHC. Results for events that satisfy all the selection criteria for the DIS and detector that were listed in Section \ref{sec_MuonDIS:config}. We are considering $L_T=6.20$ m for the length of the tungsten target of the FASER$\nu$2 detector.
}
\label{table:NeventsFASERnuHL-LHC}
\end{table}

%
%
%
%
%
%
%
%

\paragraph{Integration into global (n)PDF fits:} The results presented in this section indicate that a large sample of muon DIS events with broad coverage in $x$ and $Q^2$ will be available at FASER$\nu$, as well as in its future upgrades. Compared with previous measurements of muon DIS in fixed-target experiments, FASER$\nu$ will broaden coverage in both small $x$ and large $x$, the latter region being especially relevant for searches beyond the Standard Model at the LHC \cite{Greljo:2021kvv,Hammou:2024xuj}. Quantifying the impact of muon DIS at FASER$\nu$ in global fits of proton and nucleus PDFs is possible by following the procedure of ~\cite{Cruz-Martinez:2023sdv} to estimate the predicted accuracy of the structure function determinations and then using the EKO~\cite{Candido:2022tld}, YADISM~\cite{Candido:2024rkr} and PineAPPL~\cite{Carrazza:2020gss} software to generate fast interpolation grids, allowing the inclusion of these projections in (n)PDF fits, as done for the EIC in ~\cite{AbdulKhalek:2021gbh,Khalek:2021ulf}. A more detailed study like this is left for future work, and here we restrict ourselves to demonstrating the impact of the muon DIS on the FASER$\nu$ to restrict the charm content of the proton in the large $x$ region.

\section{Muon DIS with charm production}
\label{sec_MuonDIS:DIScharm}

In the previous section, we focused on inclusive hadronic final states, that is, summing over all possible states without making distinctions between them. In this section, we will deal with a semi-inclusive case of DIS, with at least one charm hadron identified in the final state. Since we are interested in neutral current scattering, charm hadrons are sensitive to the initial amount of charm in the nucleon, unlike charged current interactions, where charmed hadrons mainly arise from the transition of a strange quark. The Lund string model \cite{Andersson:1977qx,Andersson:1977eb,Andersson:1978vj,Andersson:1979wj,Andersson:1980nj} allows charm quarks to be pulled from the sea during hadronization, but this process is suppressed by the charm mass. Therefore, the number of charm quarks resulting from hadronization is much smaller than the number resulting from interacting charm quarks.

A unique advantage of emulsion detectors, such as FASER$\nu$, is their ability to observe charm hadrons through their decay vertex, which at energies of a few tens of GeV occurs a few millimeters after the production vertex of these hadrons. Measurements by emulsion detectors differ, for example, from charm production measurements at HERA, where the charm is reconstructed through some unique decay modes. With this method, the H1 and ZEUS detectors made several measurements of the structure function $F_2^c$~\cite{H1:2018flt} covering a large regime in small $x$, but it was not possible to access the large $x$ region. The only existing measurement to date of $F_2^c$ in large $x$ is from the EMC collaboration~\cite{EuropeanMuon:1982fow}, which initiated discussions about the existence of an intrinsic charm in the nucleon, but these measurements have been contested since then (for a recent review see~\cite{Brodsky:2015fna}).

\subsection{Charm quark content in the nucleon}
\label{subsec_MuonDIS:charmContent}

Before calculating the expected events for the FASER$\nu$ detector with hadrons containing charm quarks in the final state, we first describe the sets of PDFs used and review some of their main aspects. Our results, as shown in the previous section, are based on the use of PDFs from two distinct collaborations: NNPDF4.0 \cite{NNPDF:2021njg} and CT18 \cite{Hou:2019efy}. For results derived using the fits obtained by NNPDF4.0, we will use three distinct variations of fits at NNLO: with only perturbative charm, denoted NNPDF4.0 PC; the standard version with symmetric fitted charm \cite{Ball:2022qks}, denoted NNPDF4.0 ($c=\bar{c}$); and its variation that allows for an asymmetry between the charm and anticharm components fitted in $x$ \cite{NNPDF:2023tyk}, denoted NNPDF4.0 ($c\neq \bar{c}$). In all cases, the uncertainties relating to the PDFs are calculated with the variance over the set of 100 replicas of the PDF and presented in 1$\sigma$ of statistical significance.

For the results derived using the CT18 fits at NNLO, we are considering its standard parameterization, which includes only perturbative charm, denoted CT18 \cite{Hou:2019efy}; in addition to its variant which has an intrinsic charm component, denoted CT18 FC \cite{Guzzi:2022rca}. There are three distinct models used by the CT18 collaboration for the inclusion of intrinsic charm: BHPS3, based on a partonic model that includes a $|{uudc\bar{c}}\rangle$ component for the proton wavefunction \cite{Brodsky:1980pb}; and for the meson-baryon model (MBM) based on confining (MBMC) and effective-mass (MCME) quark models \cite{Navarra:1995rq,Hobbs:2013bia,Hobbs:2017fom}. These meson-baryon models or meson cloud models are hadronic models for the intrinsic charm and include components for the proton wave function of the type $|{\bar{D} \Lambda_c}\rangle = |{(u\bar{c})(udc)}\rangle$, which allow for an asymmetry between charm and anticharm in $x$, arising from the charm being in the baryon state and the anticharm in the meson state. In addition to these three PDF models that include an intrinsic charm component, there are CT18 FC variants for different values of $\Delta\chi^2$ provided by the collaboration, where we will use as a standard the prediction with $\Delta\chi^2=30$, which is the maximum intrinsic charm tolerance allowed with 68\% confidence. It is important to emphasize again that there are no CT18 FC PDF replicas to build predictions with uncertainties associated with PDFs.

\begin{figure}[t]
    \centering
\includegraphics[width=0.99\linewidth]{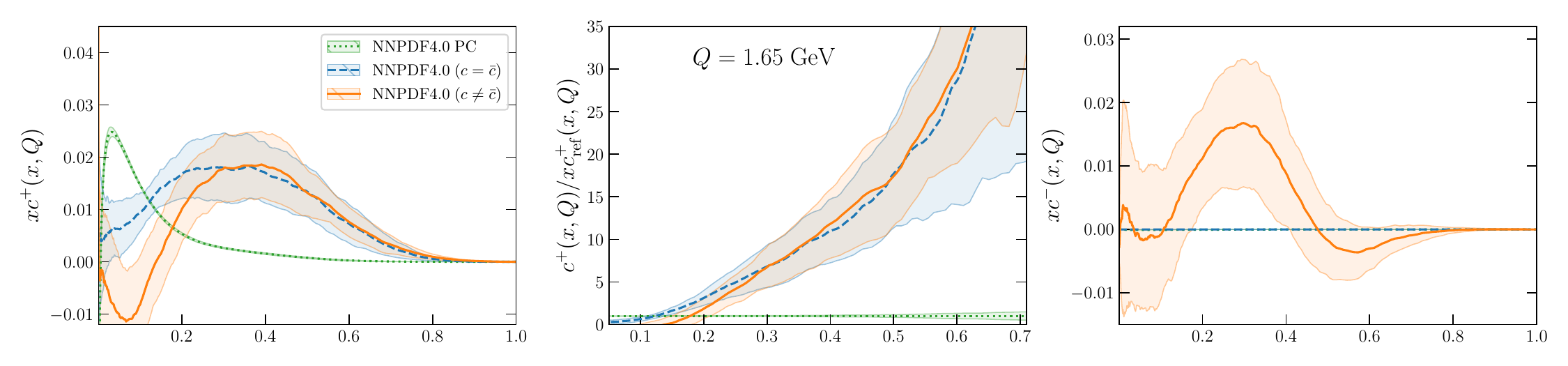}
\includegraphics[width=0.99\linewidth]{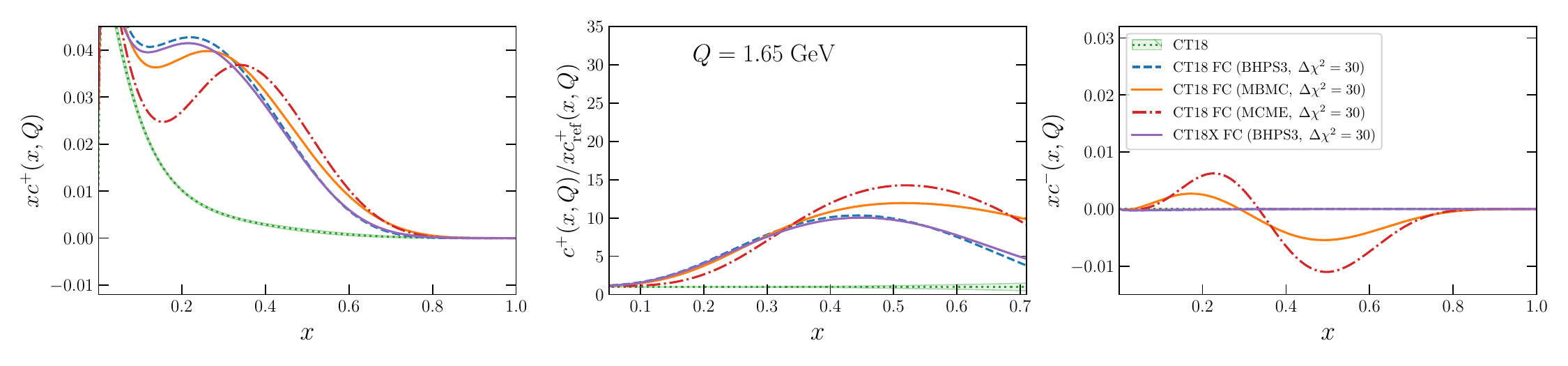}
    \caption{ Top: Comparison of three variants of NNLO fit from NNPDF4.0: with perturbative charm, the baseline result with fitted (symmetric) charm, and the variant that allows an asymmetry between the fitted charm and anticharm PDFs. From left to right, we show the total charm PDF $xc^+$ in absolute value and normalized to the perturbative charm fit, and then the valence charm PDF $xc^-$. Bottom: The same for the standard NNLO set from CT18 (with perturbative charm PDF) and the four FC variants from CT18 presented in~\cite{Guzzi:2022rca} for $\Delta\chi^2=30$ (corresponding to the standard tolerance of CT18). The uncertainties of the PDFs are displayed in 68\% confidence level intervals, and for the FC variants of CT18, no set of PDFs is provided for constructing the uncertainties.
    }
    \label{fig:charm-PDFs}
\end{figure}

In the upper panel of Figure~\ref{fig:charm-PDFs}, we present the comparison between the three sets mentioned above from NNPDF4.0 for the total and valence PDF of the charm quark, $xc^\pm(x,Q^2)= x[ c\pm \bar{c}(x,Q^2)]$. The lower panel shows the same comparisons, but now for the PDFs of the CT18 collaboration. The behaviors of the NNPDF4.0 and CT18 sets with fitted charm are similar for $xc^+$, but in general, larger amounts of intrinsic charm from CT18 FC can be seen. For both groups, we see that the peak contribution of intrinsic charm is positioned between 0.3 and 0.4, and then decreases for small values of $x$. Regarding the asymmetry in the valence charm $xc^-$, the CT18 FC set with the MBMC and MCME variants exhibits behavior similar to that verified in the NNPDF4.0 result, with asymmetry between charm and anticharm, being positive (negative) for regions close to 0.2 (0.5), but with reasonably different magnitudes for the central value of this quantity. Although we call the quantity $xc^-$ the valence charm, this is only a nomenclature adopted following the standard for light quarks, and does not constitute a real valence charm quark within the nucleon. It is also important to mention that the asymmetry between charm and anticharm is local in $x$, but the sum rules of the type $\int_{0}^{1}\mathrm{d}x\, xc^- = 0$ are respected.

The central panels of Figure~\ref{fig:charm-PDFs} show the ratio between the total charm PDF, $xc^+$, in the several sets considered with the perturbative charm result. This ratio highlights the increase in the charm PDF when a non-perturbative component is allowed and restricted by the data, compared to the scenario where the charm PDF is entirely generated by radiative QCD corrections. As is well known, although the differences between the perturbative and fitted charm PDFs are small for $x\lsim 0.1$, they grow rapidly with $x$, with an increase of a factor of 10 for $x\lsim 0.4$ for both NNPDF4.0 and CT18, and a factor of 30 for $x\lsim 0.6$ in NNPDF4.0, although the PDF uncertainties are significant. For CT18, the increase in intrinsic charm peaks between $x = 0.4$ and $0.5$, depending on the model, and then decreases for larger values of $x$. Given that the charm production cross sections in muon DIS are, to a first order, directly proportional to the charm PDF, it is expected that for FASER$\nu$ there will be a qualitatively similar increase to that found in Figure~\ref{fig:charm-PDFs} when comparing predictions binned in $x$ based on PDFs with intrinsic charm and with perturbative charm.

\subsection{Events with charm production}
\label{subsec_MuonDIS:charmEvents}

In this subsection, we are revisiting the analysis developed in the previous section, but now for events with charm production. As described in Section \ref{sec_MuonDIS:config}, we are interested in events that contain at least one charm hadron in the final state after the event passes through all DIS selection cuts, as well as detector acceptance. We are also considering the hadron identification efficiency with charm $\epsilon_c = 0.7$, as suggested by FASER collaboration members \cite{FASERprivate}.

\begin{figure}[t]
    \centering
\includegraphics[width=0.49\linewidth]{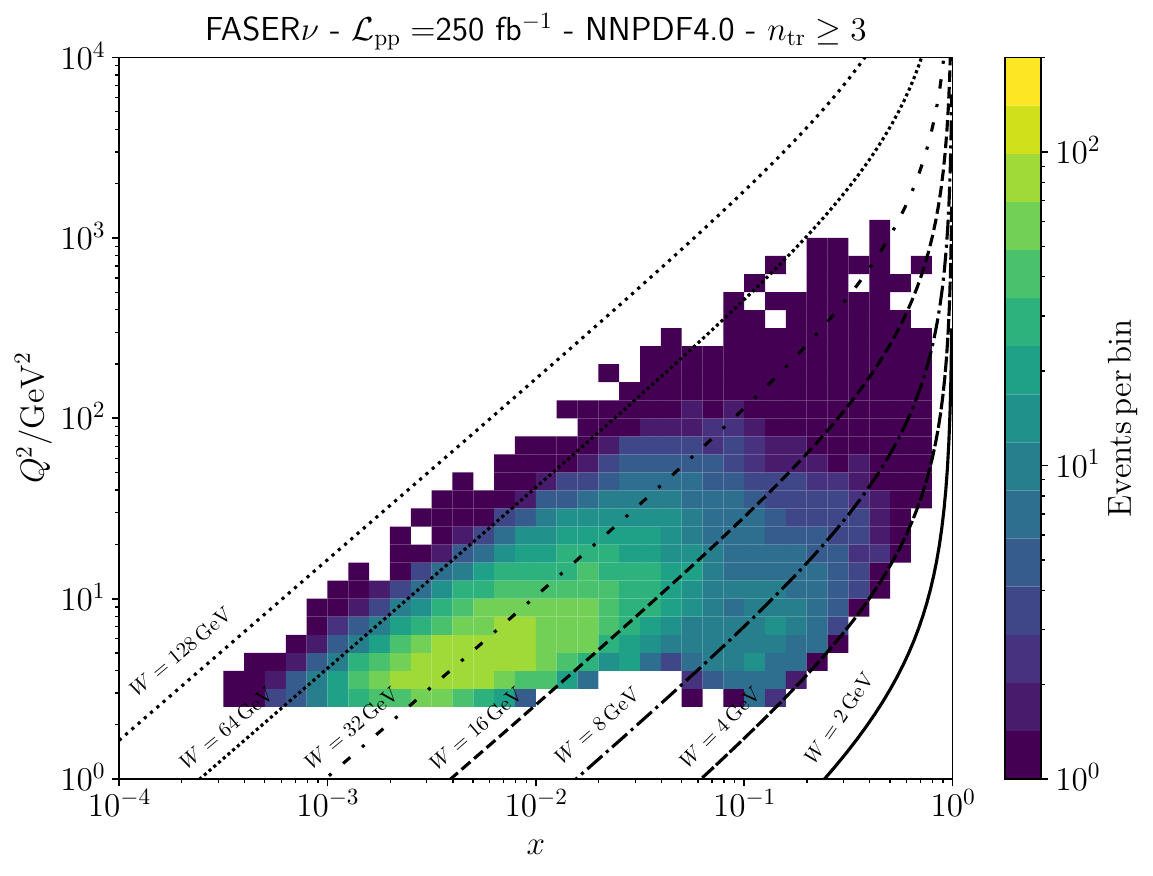}
\includegraphics[width=0.49\linewidth]{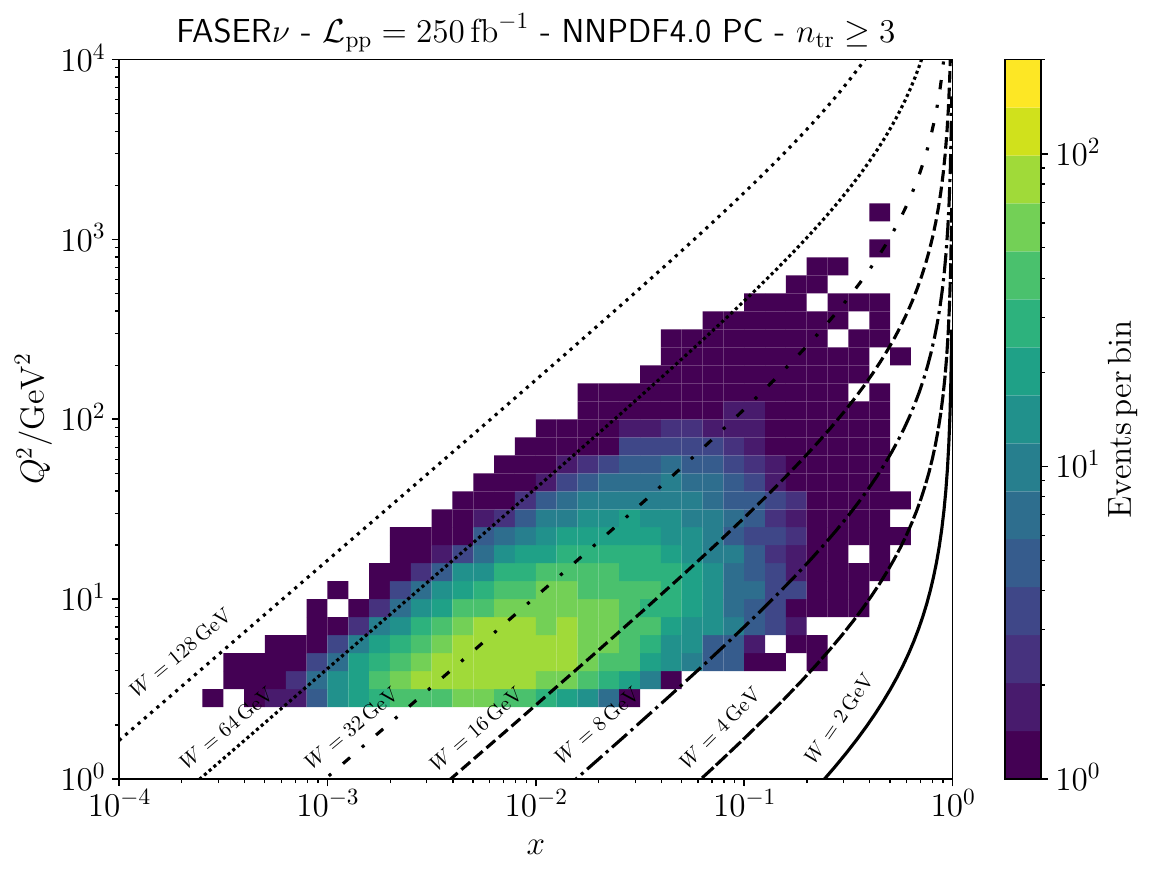}
    \caption{ Number of neutral current muon DIS interaction predicted for the FASER$\nu$ as a function of $Q^{2}$ and $x$ with at least one charm quark identified in the final state that satisfy the selection cuts described in Section \ref{sec_MuonDIS:config} for an integrated luminosity of $\mathcal{L}_{\rm pp}=250$ fb$^{-1}$. We present the results for the NNPDF4.0 PDF sets with intrinsic charm fitted (left) and only with perturbative charm (right). The black curves indicate fixed values of the invariant mass $W$. }
    \label{fig:DIStotal_charm_Q_x}
\end{figure}

Figure~\ref{fig:DIStotal_charm_Q_x} shows the same as presented in Figure~\ref{fig:DIStotal_Q_x}, now for events with at least one charm identified in the final state, in the process $\mu^\pm + W \rightarrow \mu^\pm + X + c(\bar{c})$, comparing the predictions of NNPDF4.0 (which has an intrinsic charm fitted as its standard) on the left, and its corresponding PDF which has only a perturbative component of the charm quark content on the right. We are considering at least one charm quark identified in the final state, and summing over contributions from muons and antimuons, as well as charm and anticharm in the final state. Compared to inclusive events, events with produced charm are completely suppressed in the region of $W\lsim 4$ GeV, given that this region is kinematically forbidden by the need for two charm quarks in the final state. Comparing the left and right panels, we clearly see the contribution of the intrinsic charm provided by NNPDF4.0: a clear increase in the event rate in the large $x$ region, especially in small and moderate $Q^{2}$.

\begin{figure}[t]
    \centering
\includegraphics[width=0.49\linewidth]{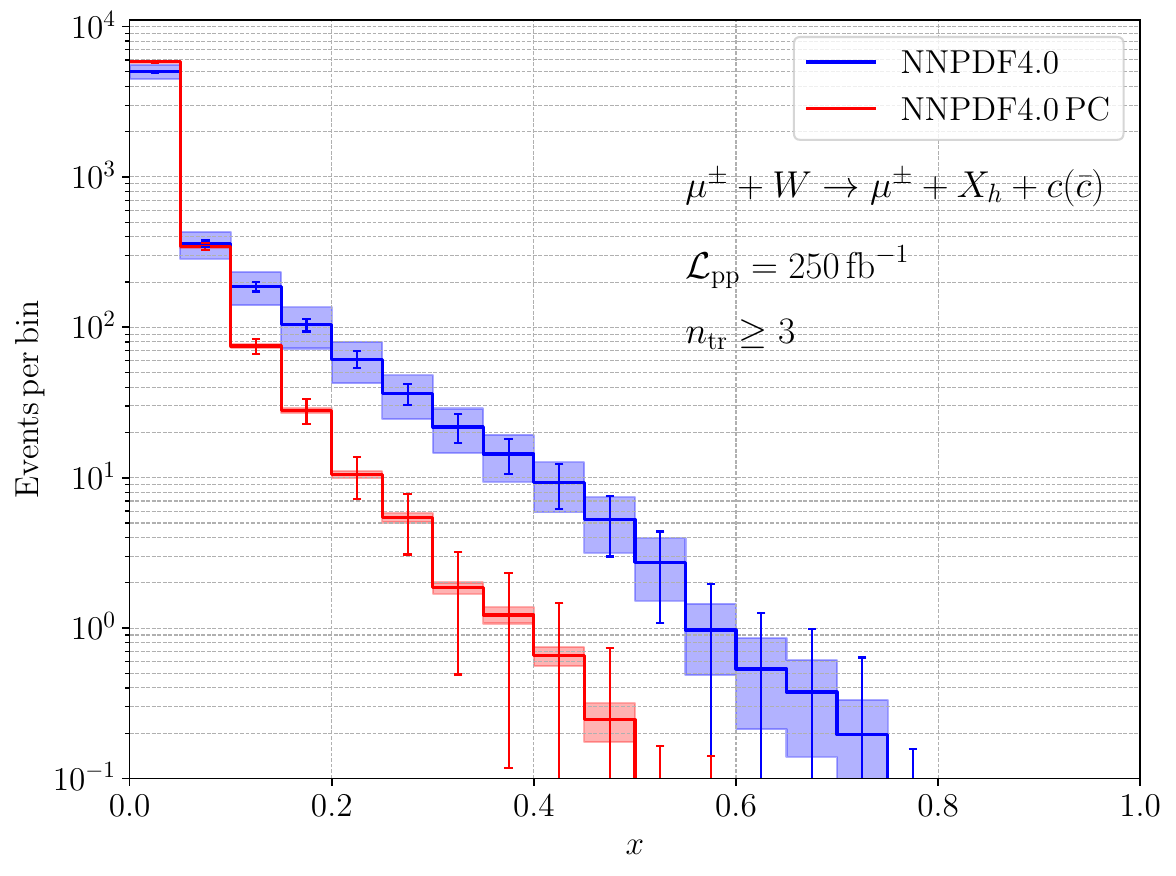}
\includegraphics[width=0.49\linewidth]{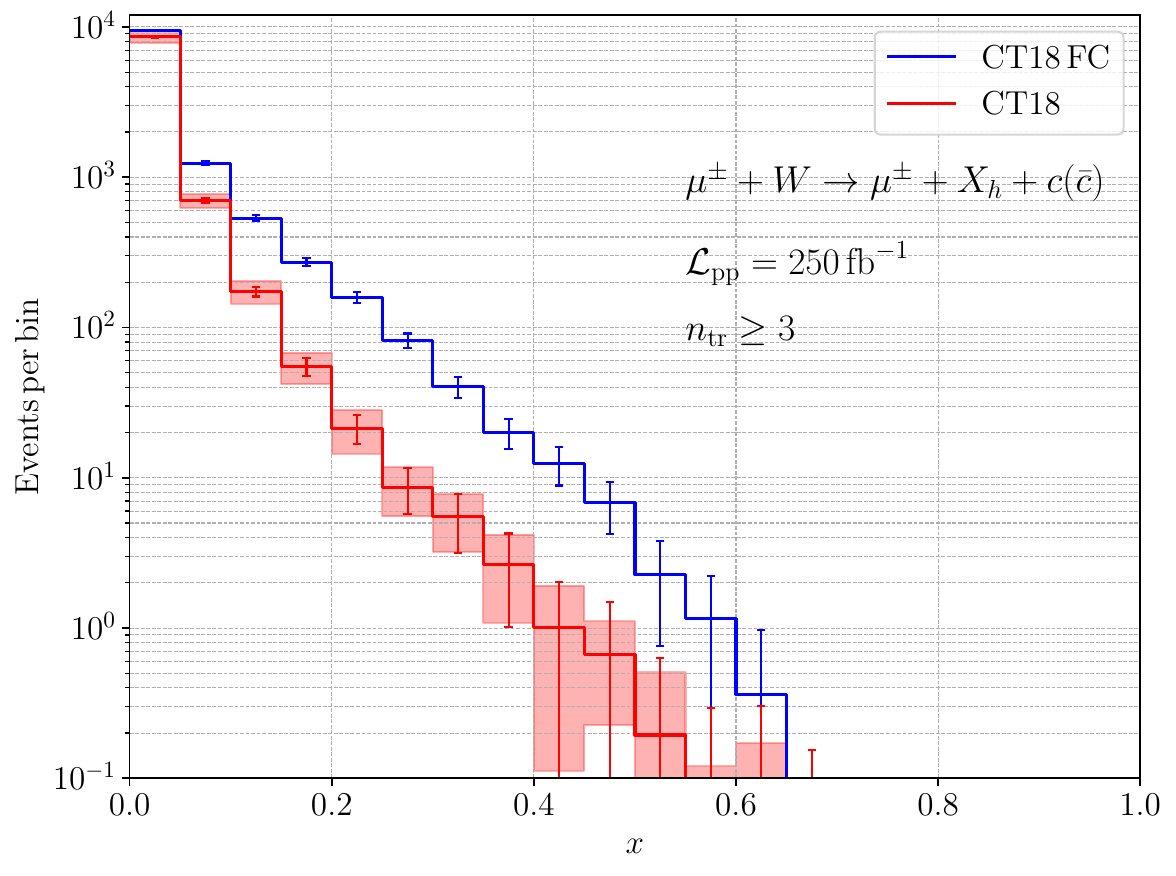}
    \caption{ Number of events per bin of muon DIS with charm identified in the final state at FASER$\nu$ for $\mathcal{L}_{\rm pp}=250$ fb$^{-1}$ as a function of Bjorken-$x$. The left (right) panel shows the predictions based on the NNPDF4.0 (CT18) parameterization. The uncertainty bars (bands) in each bin indicate the associated statistical (PDF) uncertainty.
    }
\label{fig:DIStotal_cuts_E_statBands}
\end{figure}

In Figure~\ref{fig:DIStotal_cuts_E_statBands} we show the same result obtained in Figure~\ref{fig:DIStotal_cuts_x}, but now for the case of events with at least one identified charm quark in the final state, obtained by summing the events presented in Figure~\ref{fig:DIStotal_charm_Q_x} across the entire kinematically allowed interval for $Q^{2}$. The uncertainty bands arise from the uncertainties in the PDFs propagated to the number of events, while the uncertainty bars are our predictions for the statistical uncertainty. As mentioned earlier, for the CT18 FC parameterization, no replicas of the PDFs are provided for constructing the uncertainty bands. In Table \ref{table:Nevents_charm} we show the expected event values at FASER$\nu$ considering regions of $x$ greater than 0.2 and 0.4, which are regions expected to have greater sensitivity to the existence of an intrinsic charm in the nucleon. The number in parentheses is the ratio of expected events using the PDFs with intrinsic charm and with only its perturbative contribution.

From Figure~\ref{fig:DIStotal_cuts_E_statBands} and Table~\ref{table:Nevents_charm} we observe that both PDFs with intrinsic charm, NNPDF4.0 and CT18 FC, show a large excess of events in the large $x$ region, where the contribution of intrinsic charm dominates over its perturbative component. Even in the region with $x\gsim 0.2$ there is already a large difference between the predictions with and without intrinsic charm, with approximately 150~(320) events predicted by the NNPDF4.0~(CT18 FC) parameterization, while its perturbative components predict only 20~(40) events. This increase in the event rate of approximately 8 is consistent with the difference present in the PDFs shown in Figure~\ref{fig:charm-PDFs}. For $x\gsim 0.4$, predictions using NNPDF4.0 (CT18 FC) show values with factors of 20~(12) greater than those predicted considering only the perturbative component of the charm content in the nucleon.

\begin{table}[t]
\centering
\renewcommand{\arraystretch}{1.5}
\begin{tabularx}{\textwidth}{XXlll}
\toprule
\multicolumn{5}{c}{Muon DIS events with identified charm at FASER$\nu$ for $\mathcal{L}_{\rm pp}=250$ fb$^{-1}$}\\
\midrule
PDF    & Charm PDF           & $x>0$                    & $x\ge 0.2$  & $x\ge 0.4$  \\	
\toprule
NNPDF4.0   & Fit. charm        & 5.8$\times10^{3}$ (0.9) & 153.2 (7.7) & 19.3 (19) \\
\midrule
NNPDF4.0~PC & Pert. charm         & 6.3$\times10^{3}$ (1.0)    & 20.0 (1.0)    & 1.0 (1.0)     \\ 
\midrule
CT18 FC    & Fit. charm  (BHPS3) & 1.2$\times10^{4}$ (1.2)  & 324.3 (8.1) & 23.0 (12) \\
\midrule
CT18       & Pert. charm         & 9.5$\times10^{3}$ (1.0)    & 40.2 (1.0)    & 2.0 (1.0)     \\
\bottomrule
\end{tabularx}
\vspace{0.3cm}
\caption{Similar to Table~\ref{table:Nevents} for charm production, with additional kinematic cuts at $x$ to highlight the differences between theoretical predictions based on intrinsic charm PDFs and only perturbative contribution. We consider $x>0$ (last column of Table~\ref{table:Nevents}), $x\ge 0.2$ and $x\ge 0.4$. The numbers in parentheses indicate the ratio between the number of events in the intrinsic charm PDF and perturbative PDF calculations.
}
\label{table:Nevents_charm}
\end{table}

\begin{figure}[t]
    \centering
\includegraphics[width=0.49\linewidth]{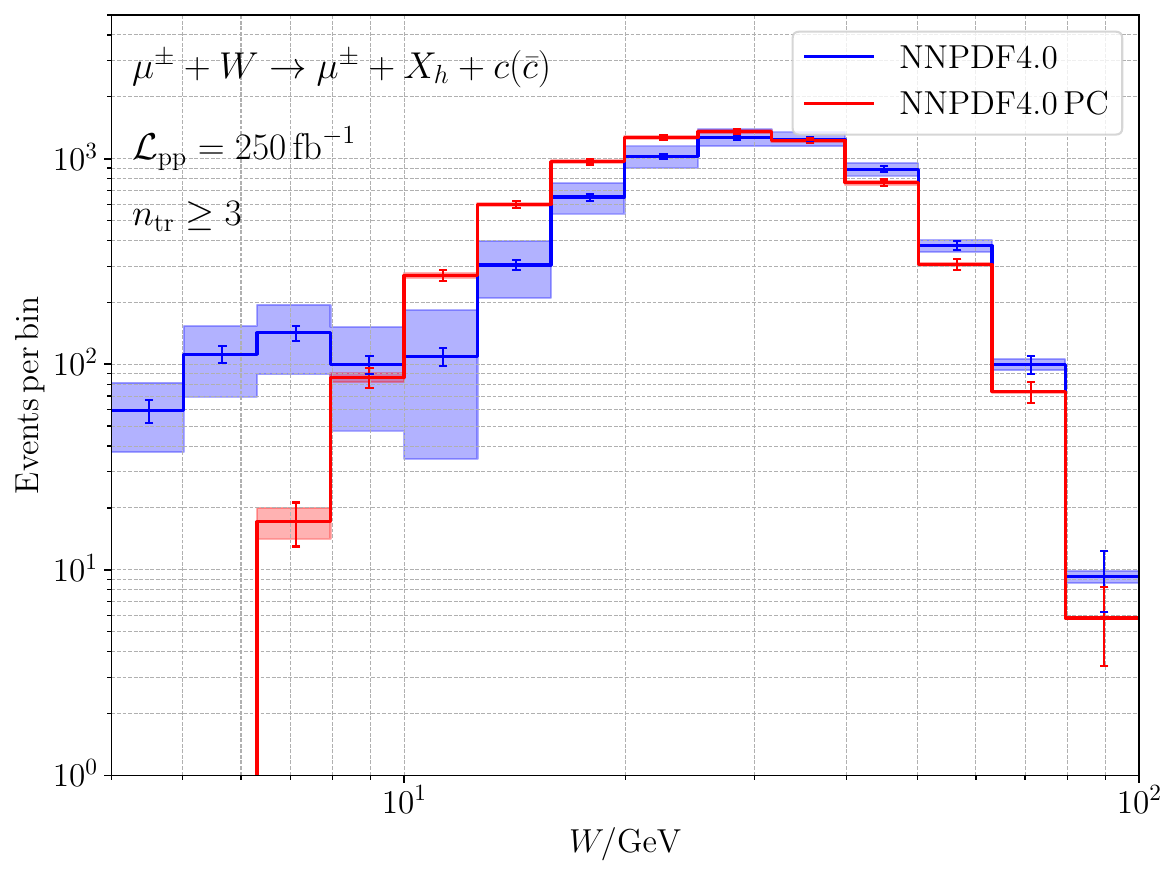}
\includegraphics[width=0.49\linewidth]{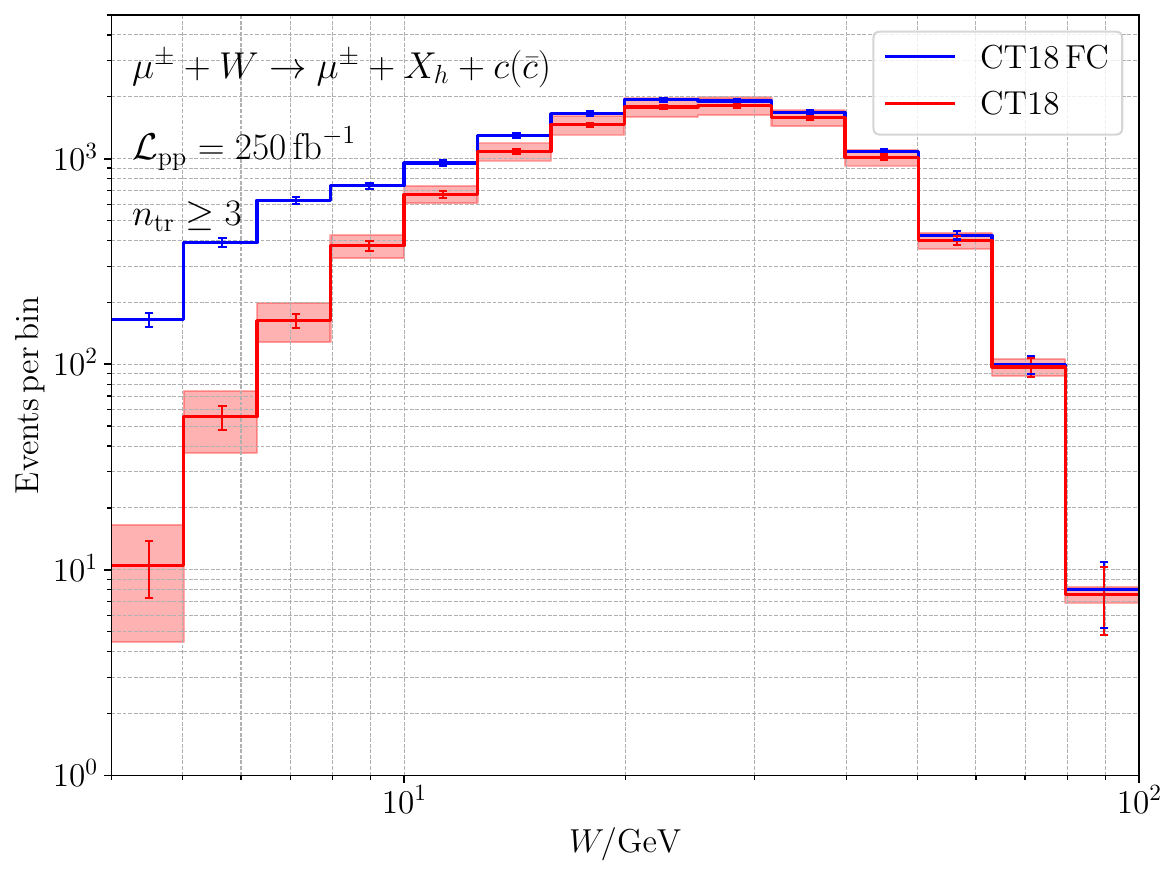}
    \caption{Number of events per bin of the muon DIS with charm identified in the final state at FASER$\nu$ for $\mathcal{L}_{\rm pp}=250$ fb$^{-1}$ as a function of the invariant mass $W$. The left (right) panel shows the predictions based on the NNPDF4.0 (CT18) parameterization. The uncertainty bars (bands) in each interval indicate the associated statistical (PDF) uncertainty.
    }
\label{fig:DIScharm_cuts_W}
\end{figure}

In addition to the distribution of events with charm in the final state at $x$, there is a large difference between theoretical predictions based on PDFs with and without intrinsic charm in events separated into bins of the invariant mass $W$, as we are showing in Figure \ref{fig:DIScharm_cuts_W}. Both predictions with the presence of intrinsic charm, NNPDF4.0 (left) and CT18 FC (right), have an increase in events at lower values of invariant mass, especially for $W \lsim 10$ GeV, where the contribution of perturbative charm is suppressed. Consequently, the invariant mass distributions of the hadronic final state provide a complementary signature of intrinsic charm. For the results derived with NNPDF4.0, there is a difference between the predictions for larger invariant mass values, given that the predictions of the charm quark content with and without intrinsic charm only coincide in the small $x$ and large $Q^{2}$ regions (see again in Figure~\ref{fig:charm-PDFs} and Reference~\cite{Ball:2022qks}). Conversely, the predictions based on the CT18 parameterization coincide for all small $x$, given that the intrinsic component of the charm is added on top of its perturbative part.

\begin{figure}[t]
    \centering
\includegraphics[width=0.49\linewidth]{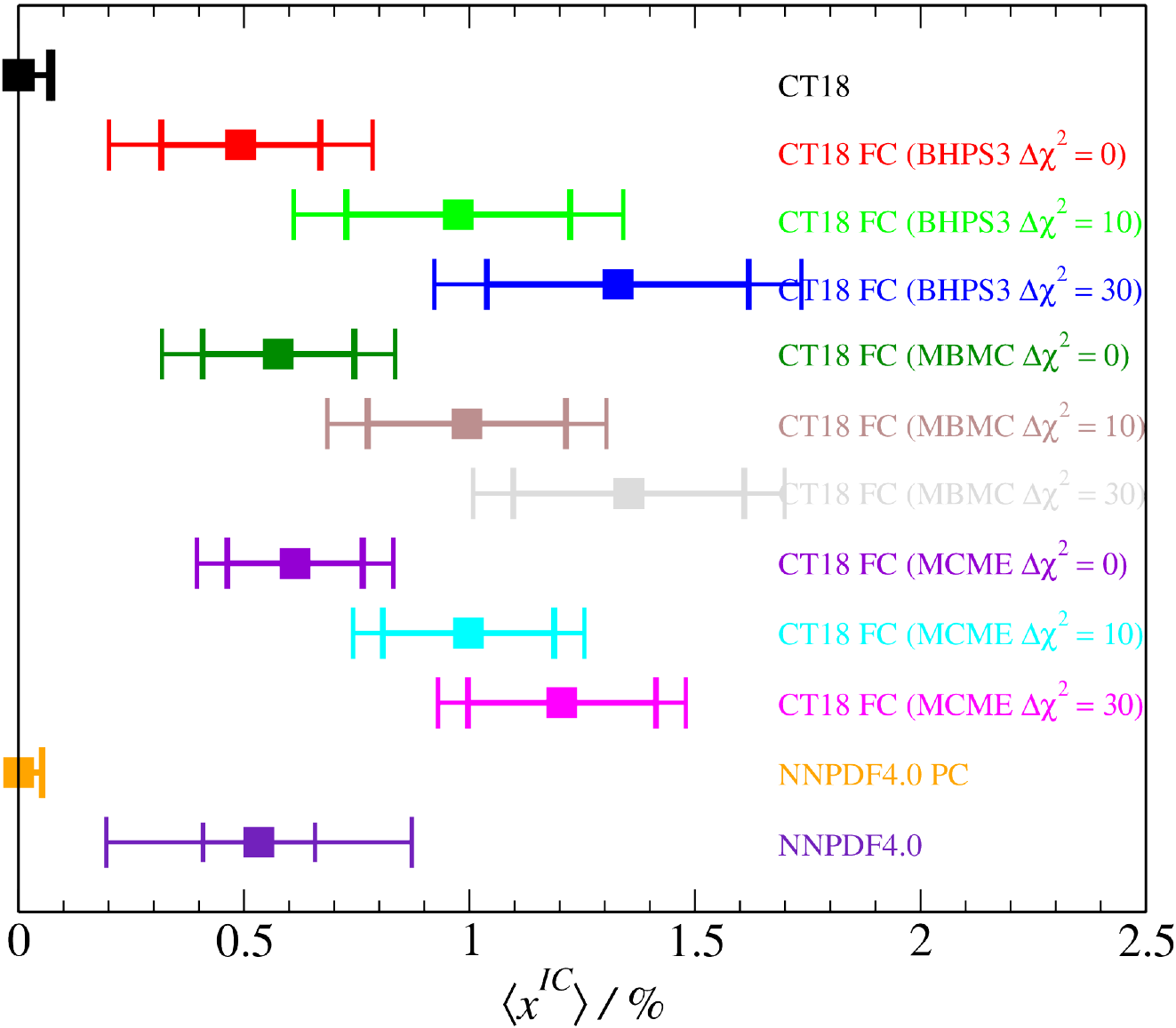}
\includegraphics[width=0.49\linewidth]{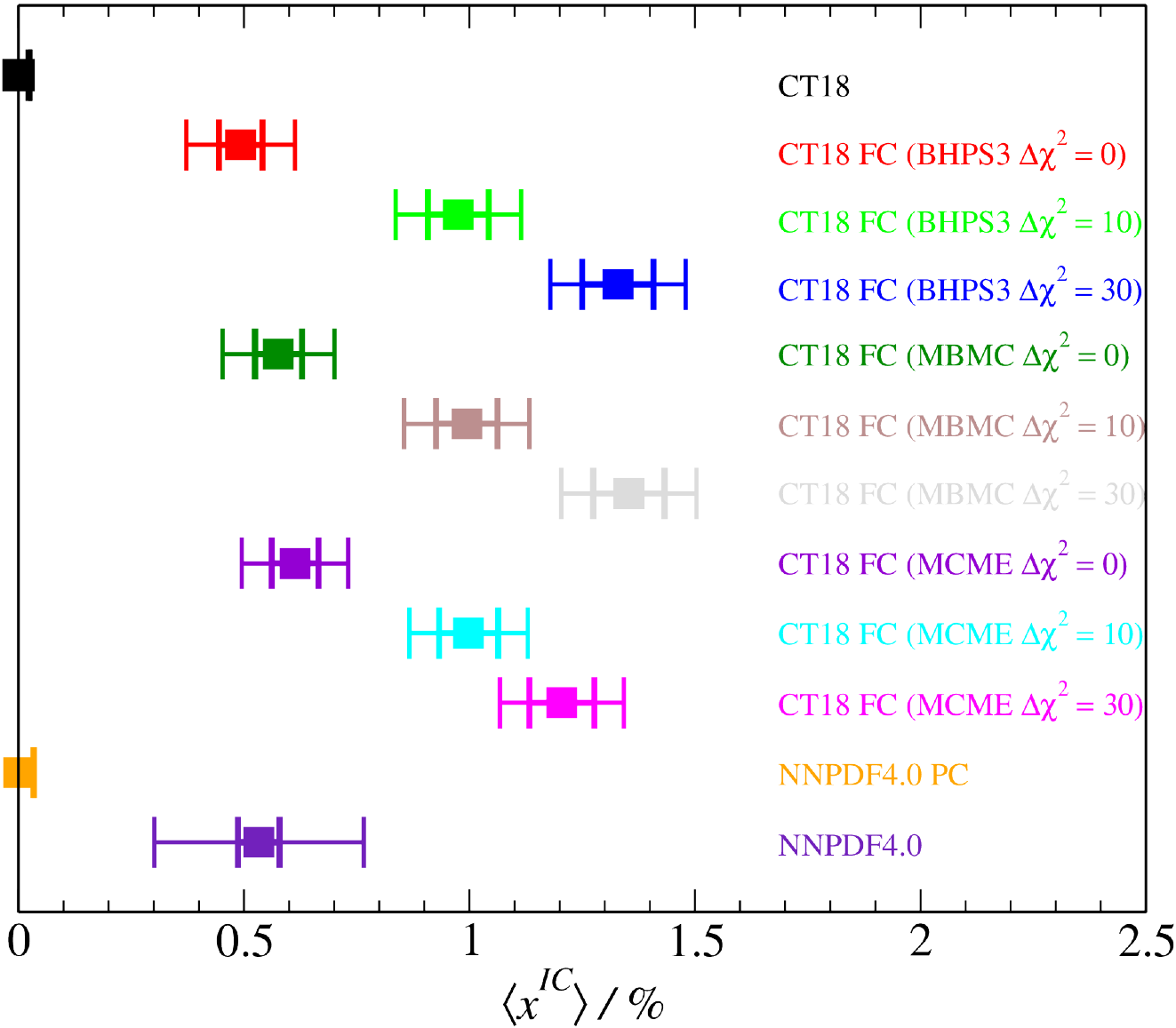}
    \caption{ Sensitivity to the intrinsic charm moment fraction (in percentage), defined in Equation (\ref{eq:ICmomfrac}), at FASER$\nu$ considering the selection cuts described in Section \ref{sec_MuonDIS:config} for an integrated luminosity of $\mathcal{L_\mathrm{pp}} = 250\,\mathrm{fb}^{-1}$ for events with $x\geq 0.2$ (left) and $x \geq 0.4$ (right). The inner uncertainty bars represent the statistical uncertainty, while the outer uncertainty bars include the statistical and PDF uncertainties at the 1$\sigma$ level of statistical significance.
    }
\label{fig:xcIC_FASERnu}
\end{figure}

Traditionally, the sensitivity of a given analysis to intrinsic charm in the nucleon is estimated in terms of the fraction of the nucleon momentum carried by its PDF. Given that PDFs that include intrinsic charm provide the sum of the perturbative and intrinsic components, we can define the fraction of momentum carried by the intrinsic component as
\begin{eqnarray}
\label{eq:ICmomfrac}
\langle x^{\mathrm{IC}}\rangle \equiv \int_{0}^{1}\mathrm{d}x\, x  \left( c+\bar{c} - c_\mathrm{PC}-\bar{c}_\mathrm{PC}\right)  \, ,
\end{eqnarray}
With the PDFs evaluated at $Q_0=1.65$ GeV, that is, the scale slightly above the limit for charm production $\mu_c=m_c$, and IC and PC are relative to the intrinsic and perturbative components of the charm quark amount. Subtracting the perturbative part, we obtain the momentum fraction relative only to the non-perturbative part. In Figure~\ref{fig:xcIC_FASERnu} we present the expected sensitivity of the FASER$\nu$ to the momentum fraction of the nucleon carried by the intrinsic charm defined in Equation~(\ref{eq:ICmomfrac}), considering events with $x\geq 0.2$ (left) and $x\geq 0.4$ (right) for NNPDF4.0 and CT18 FC. The internal uncertainty bar corresponds to the statistical uncertainty, while the external uncertainty was obtained by summing the statistical and propagated uncertainties of the PDFs. These results provide a complementary view to the previous figures of the potential of FASER$\nu$ to search for the existence of an intrinsic charm in the nucleon, even with data collected during Run 3 of the LHC.

\paragraph{Dependence on the $n_{\rm tr}$ cut:} As discussed in Section~\ref{sec_MuonDIS:config}, in the current experimental cuts strategy for detecting neutrino scattering in FASER$\nu$, events must have a minimum number of charged particle tracks ($n_{\rm tr}$) originating from the interaction vertex of that event. The values of $n_{\rm tr}$ are closely related to the values of the final hadronic state energy $E_h$, and the invariant mass $W$, since more energy transferred to the target, greater is the expected number of particles produced. In the case of the intrinsic charm contribution to the events, we saw above that its effect is found at low values of $Q^{2}$ and large $x$, a region that corresponds to low values of invariant mass. Therefore, preserving the sign of an intrinsic charm in the nucleon in the data strongly depends on our choices of cuts in $n_{\rm tr}$, since a very restrictive cut may end up disregarding a significant part of its contribution.

\begin{figure}[t]
\centering
\hspace{-12mm}
\includegraphics[width=0.49\linewidth]{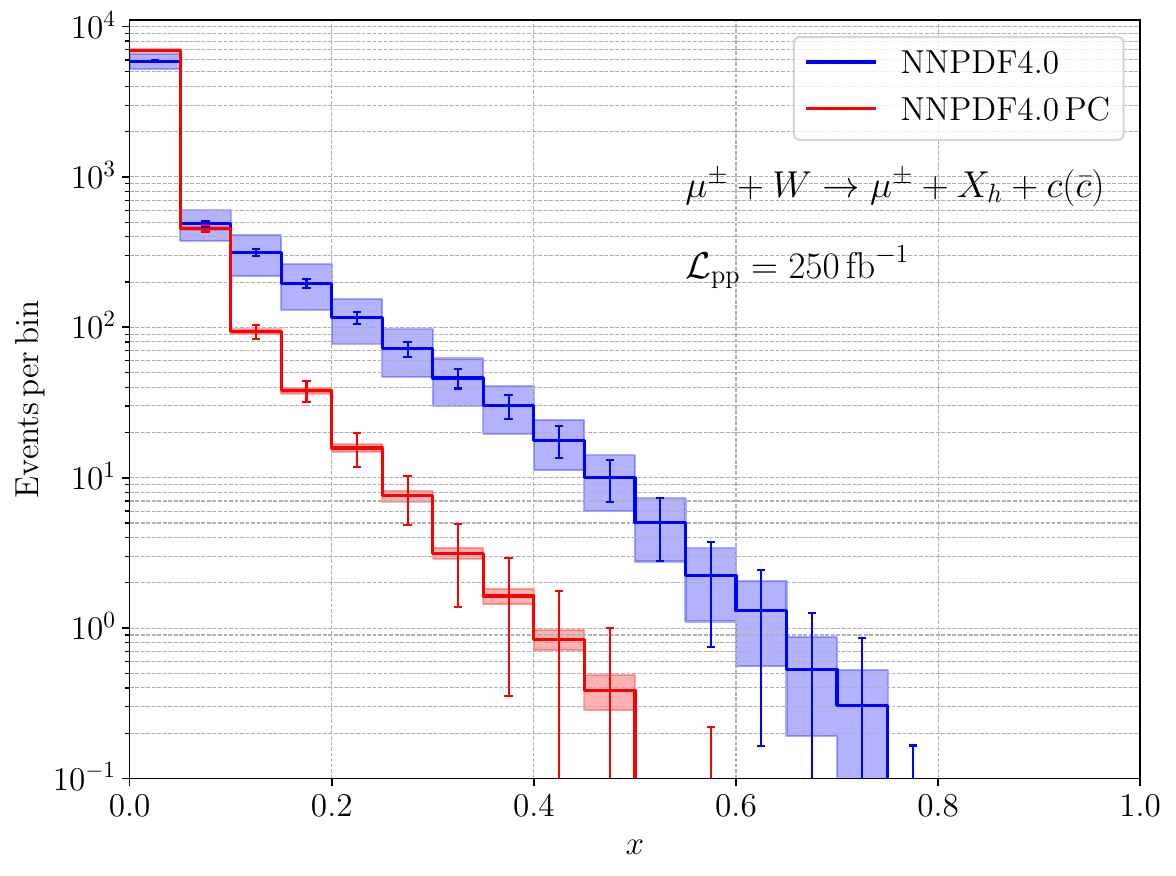}
\includegraphics[width=0.49\linewidth]{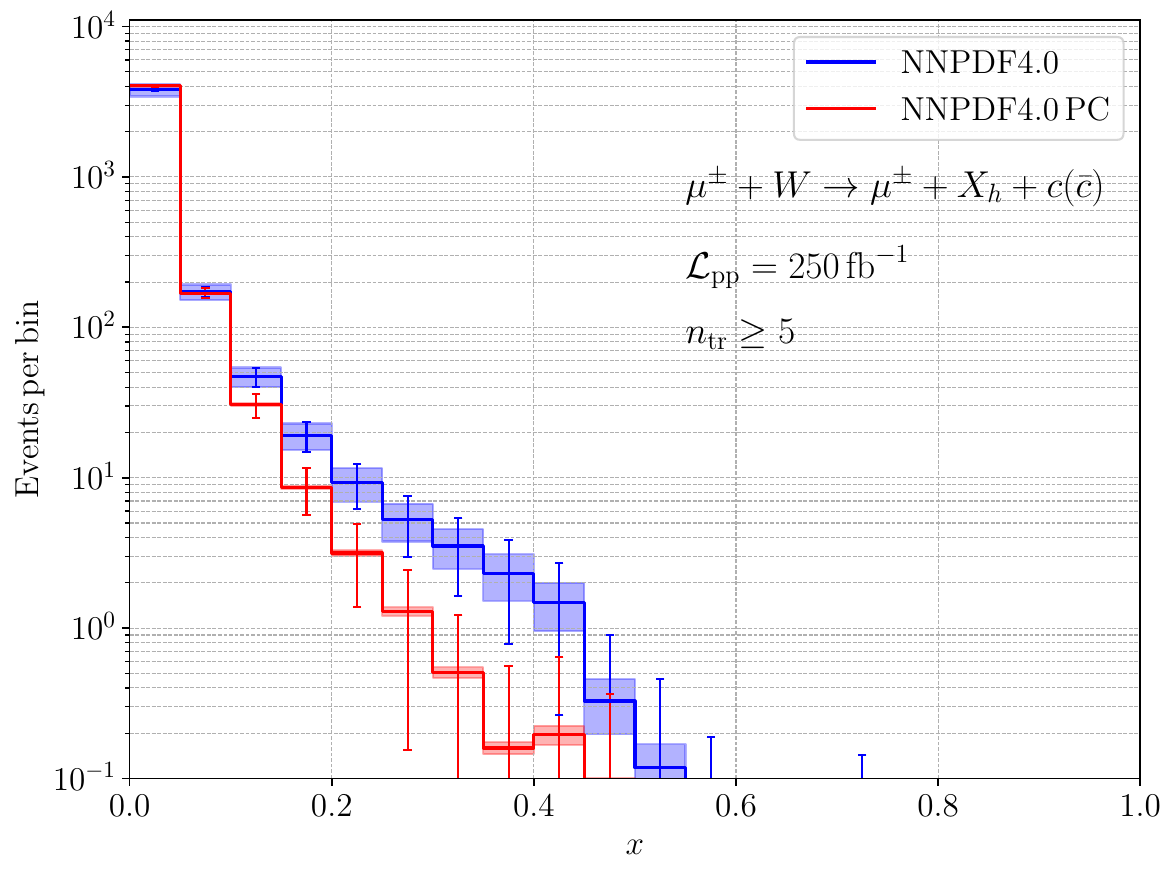}
\includegraphics[width=0.49\linewidth]{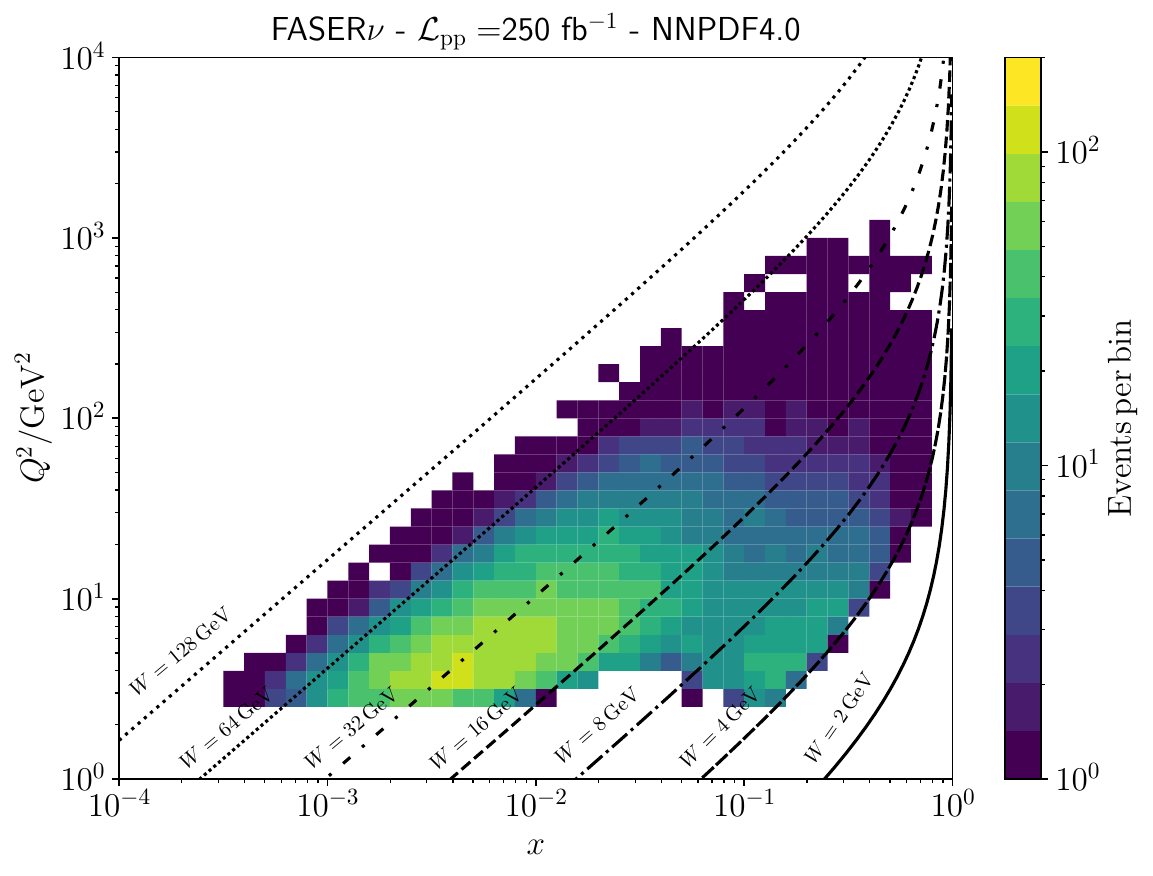}
\includegraphics[width=0.49\linewidth]{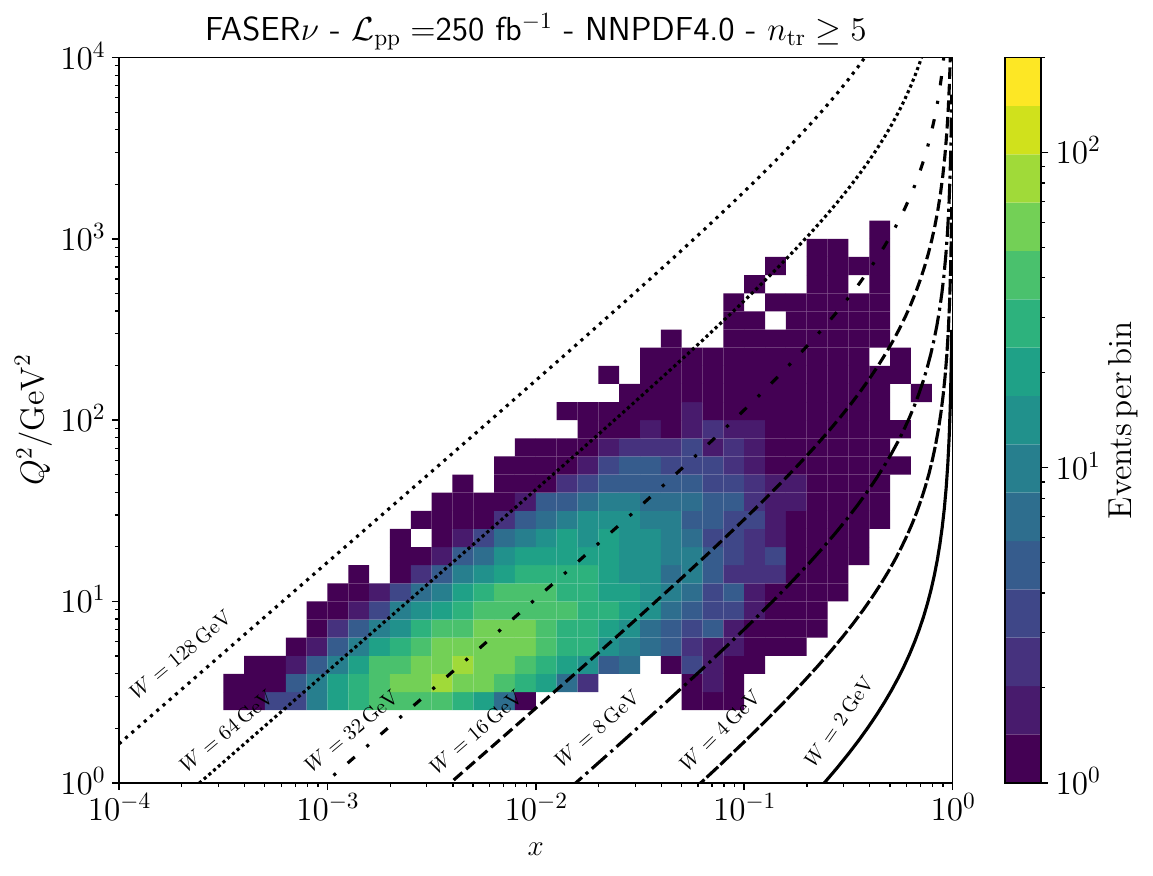}     
\caption{Top: similar to Figure~\ref{fig:DIStotal_cuts_E_statBands} in the case of NNPDF4.0 predictions comparing results where we require $n_{\rm tr}\ge 1$ (left) and $n_{\rm tr}\ge 5$ (right) for the number of charged particle tracks with at least 1 GeV momentum in the final state. Bottom: same analysis for events in the ($x,Q^{2}$ plane).
}
\label{fig:DIStotal_cuts_x_nTracks}
\end{figure}

To illustrate this dependence of the intrinsic charm event signal on $n_{\rm tr}$, we show Figure~\ref{fig:DIStotal_cuts_x_nTracks}, where its upper panel shows the same comparison that we presented in Figure~\ref{fig:DIStotal_cuts_E_statBands} (events binned in $x$) in the case of predictions using NNPDF4.0 and restricting $n_{\rm tr}\ge 1$ and $n_{\rm tr}\ge 5$. The lower panels show the same comparison adopting different values for the minimum number of tracks in the final state for events in the $(x,Q^2)$ plane. When we adopt $n_{\rm tr}\ge 5$, the difference between predictions with and without intrinsic charm in the PDFs becomes on the order of the expected statistical uncertainty for the number of events. The results presented in Figure~\ref{fig:DIStotal_cuts_x_nTracks} reinforce the importance of experimental optimization in the search for events with few particles in the final state, given that the large $x$ region is the most affected by more restrictive cuts in this observable.

\paragraph{Impact of systematic uncertainties:} While a complete simulation of event reconstruction at a more realistic detection level is beyond the scope of our study, we can estimate the impact of systematic uncertainties on events with charm observed in the final state, as described below. In the FASER$\nu$ experiment, it is possible to measure the energies associated with the initial and final muons, as well as their scattering angle, in addition to the energy of the final hadronic state. The precision in measuring the scattering angle is on the order of 0.05 mrad, given that the emulsion film has a smaller spatial resolution of a few hundred nm. Therefore, uncertainties related to this quantity can be disregarded. In the case of measuring the energy of muons present in the scattering, initial studies by the FASER collaboration point to uncertainties on the order of 30\% at 200 GeV and 50\% for more energetic muons. Muon energy is measured in the emulsion detector through multiple Coulomb scattering: more energetic muons tend to have straighter trajectories through the tungsten plates. Given that in this section we are interested in studying a QCD property in a specific regime of the plane ($x, Q^{2}$), and these variables are strongly dependent on muon energy, accurately reconstructing the energy means precisely accessing the regions of the DIS variables.

To model the finite resolution of the detector, we will apply Gaussian smearing to the initial and final muon energies, as well as to the hadronic state energy, with a fixed standard deviation of $\sigma_p$, in the events generated with POWHEG+Pythia8. For simplicity, we will adopt the same value of $\sigma_p$ for both leptonic and hadronic energies. We also impose that the relationship between the reconstructed variables must respect $E_\mu = E'_{\mu}+E_h$, minimizing a $\chi^2$. Having access to the aforementioned reconstructed energies, we can access the reconstructed DIS variables $x_{\rm reco}$ and $Q^2_{\rm reco}$. For illustration, in our analysis we will assume two distinct values of this energy smearing, given by $\sigma_p=10\%$ and $\sigma_p=30\%$. This simplified approach adopted here represents the worst-case scenario compared to a real-world analysis, and consequently overestimates the impact of the detector's finite resolution.

\begin{figure}[t]
    \centering
\includegraphics[width=0.49\linewidth]{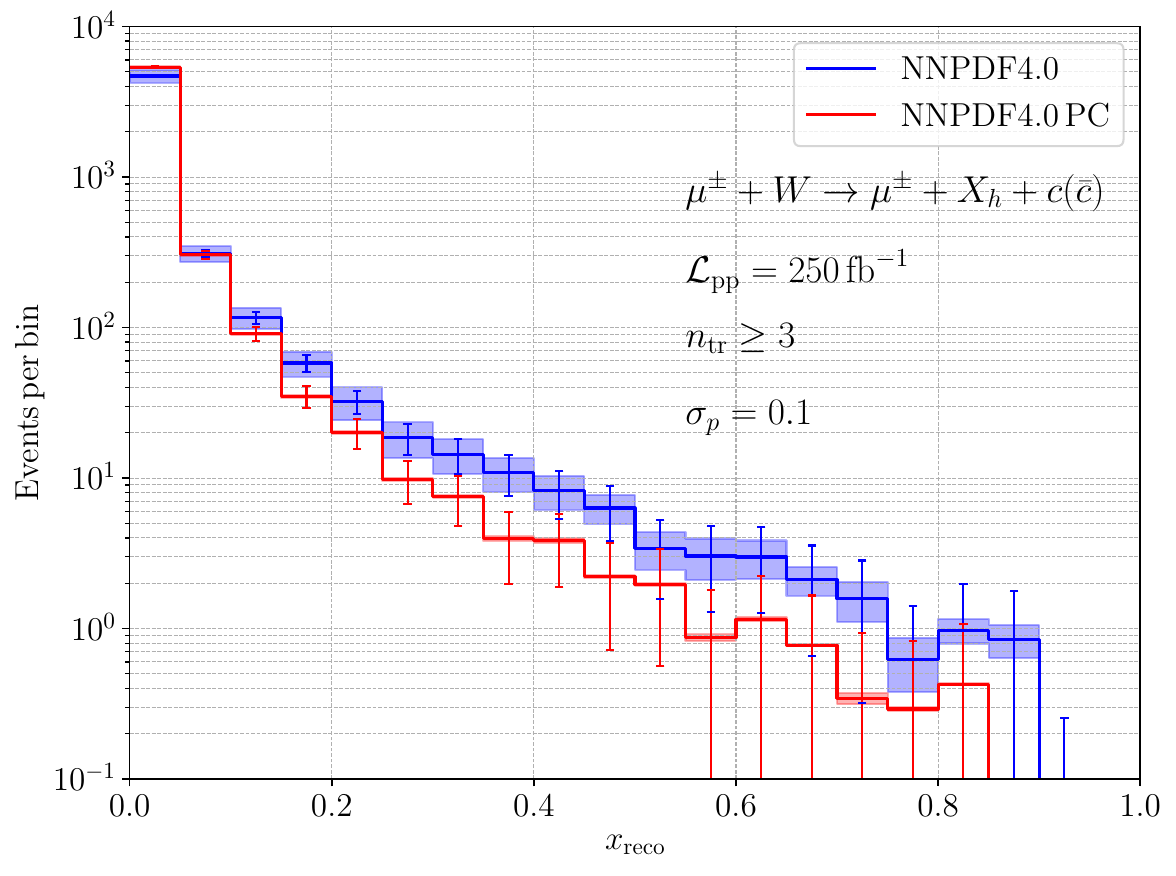}
\includegraphics[width=0.49\linewidth]{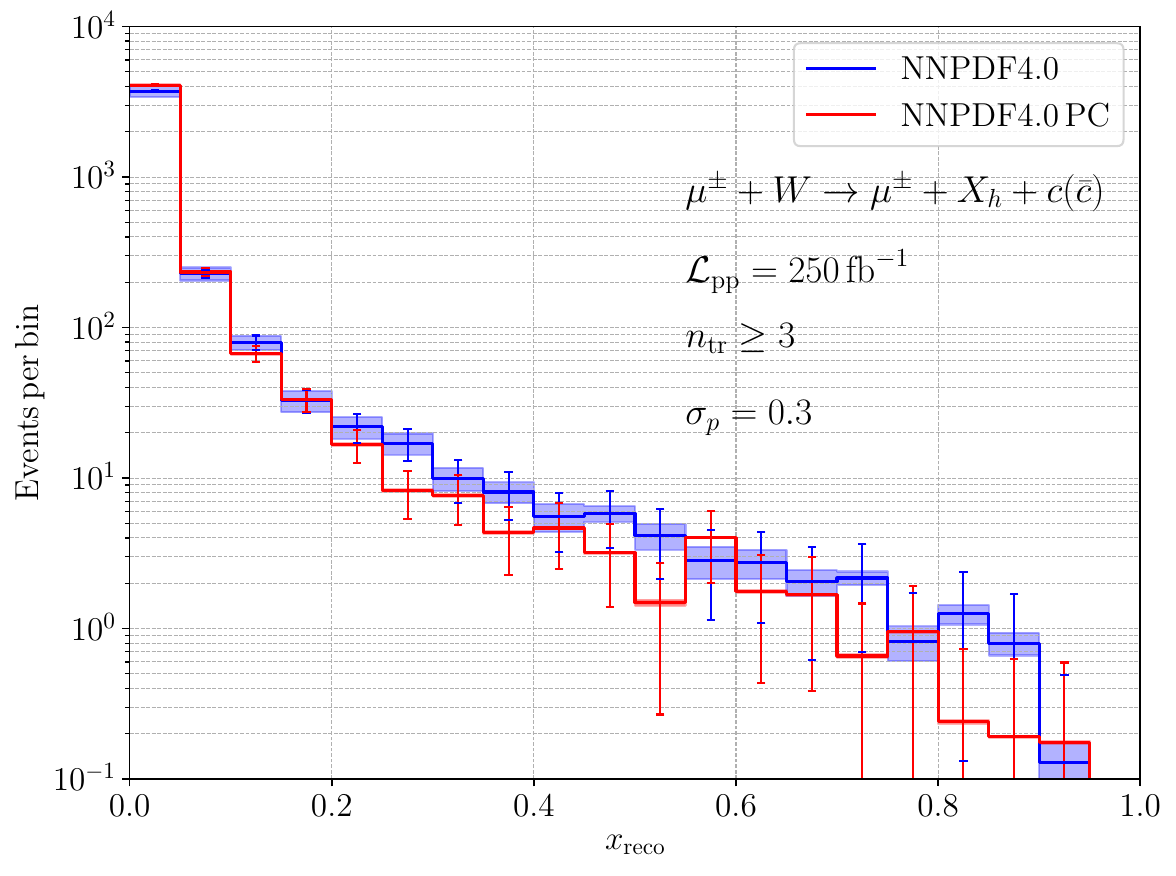}
  \caption{Similar analysis to Figure~\ref{fig:DIStotal_cuts_E_statBands} for the case of predictions assuming NNPDF4.0 comparing results with energy smearing with $\sigma_p=10\%$ (left) and $\sigma_p=30\%$ (right).
    }
    \label{fig:DIStotal_cuts_x_sigma}
\end{figure}

\begin{figure}[t]
    \centering
\includegraphics[width=0.49\linewidth]{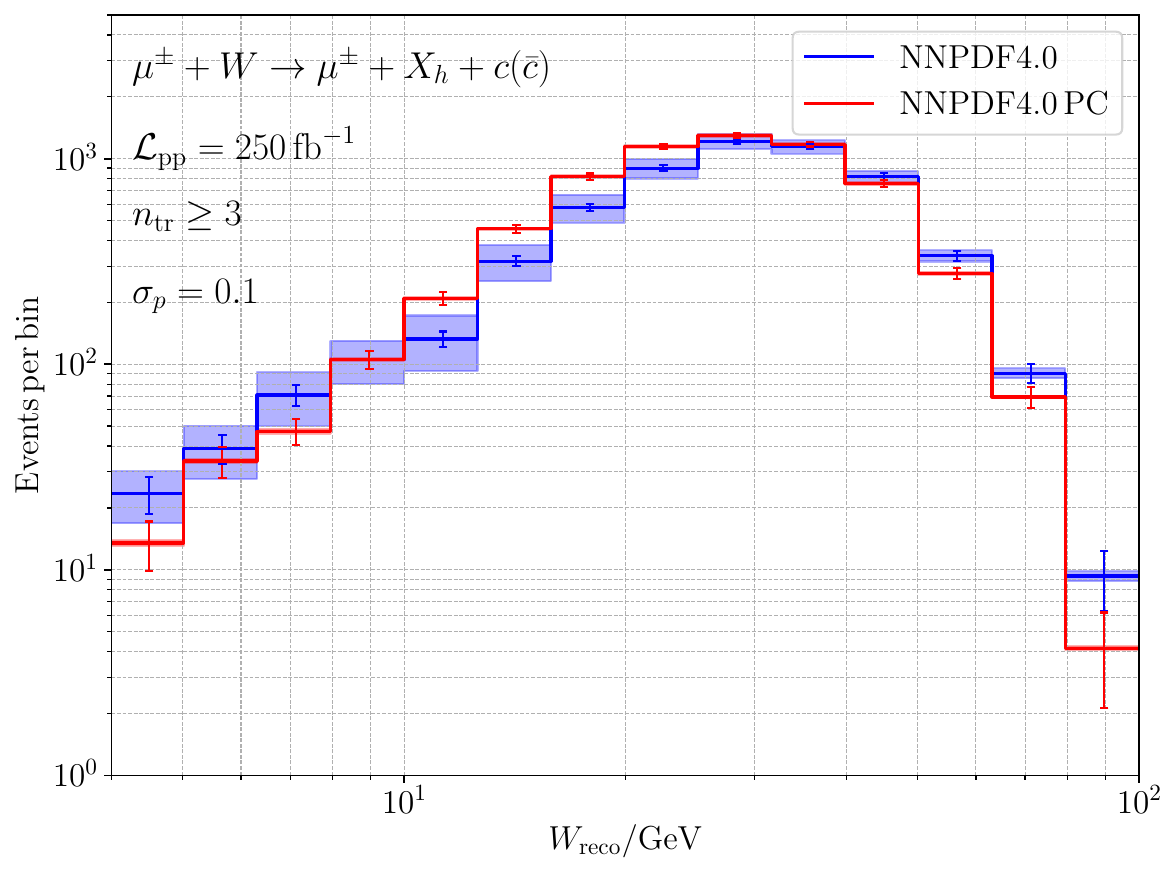}
  \caption{Same result shown on the left side of the Figure~\ref{fig:DIStotal_cuts_x_sigma} for events binned in $W_{\rm reco}$.
    }
    \label{fig:DIStotal_cuts_W_sigma}
\end{figure}

\begin{table}[t]
\centering
\renewcommand{\arraystretch}{1.5}
\begin{tabularx}{\textwidth}{XXXlll}
\toprule
\multicolumn{6}{c}{Muon DIS at FASER$\nu$ for $\mathcal{L}_{\rm pp}=250$ fb$^{-1}$}\\
\midrule
PDF    & Charm PDF    & $\sigma_p$ & $x_{\rm reco}>0$             & $x_{\rm reco}\ge 0.2$ & $x_{\rm reco}\ge 0.4$  \\	
\toprule
NNPDF4.0   & Fit. charm  & 0.1      & 5.3$\times10^{3}$ (0.9) & 104.6 (2.1)       & 27.6 (1.9) \\
\midrule
NNPDF4.0~PC & Pert. charm  & 0.1      & 5.8$\times10^{3}$ (1.0)    & 50.6 (1.0)          & 14.5 (1.0)     \\ 
\toprule
NNPDF4.0   & Fit. charm  & 0.3      & 4.2$\times10^{3}$ (0.9) & 81.4 (1.3)        & 22.7 (1.2) \\
\midrule
NNPDF4.0~PC & Pert. charm & 0.3      & 4.5$\times10^{3}$ (1.0)    & 60.6 (1.0)          & 18.4 (1.0)     \\ 
\bottomrule
\end{tabularx}
\vspace{0.3cm}
\caption{ Similar to Table~\ref{table:Nevents_charm}, now applying a smearing in the energy with $\sigma_p=10\%$ and $\sigma_p=30\%$ in the reconstruction of the variable $x$, assuming NNPDF4.0.
}
\label{table:Nevents_charm_smeared}
\end{table}

In Figure~\ref{fig:DIStotal_cuts_x_sigma} we again show our predictions for events distributed in bins of $x$ using the NNPDF4.0 PDF, but now with Gaussian smearing with standard deviation $\sigma_p=10\%$ and $30\%$ in the energy of the particles present in the scattering. The values for the number of events with cuts in $x_{\rm reco}$ are presented in Table~\ref{table:Nevents_charm_smeared}, and can be compared to those obtained and presented in Table~\ref{table:Nevents_charm}, where we did not consider any type of systematic uncertainty. Our predictions show that applying smearing to the particle energy decreases or even eliminates the effects of an intrinsic charm in large $x$ due to the mixing of events in different regions of the Bjorken variable. For $\sigma_p=10\%$, the increase in events due to intrinsic charm reduces by a factor of 2 for both regions with $x_{\rm reco}\ge 0.2$ and $\ge 0.4$, while without this systematic effect we had factors of 8 and 19, respectively. Our predictions still indicate that assuming $\sigma_p=30\%$ it is no longer possible to distinguish between the predictions with and without intrinsic charm. Similar conclusions are obtained by analyzing Figure~\ref{fig:DIStotal_cuts_W_sigma}, where we use $\sigma_p=10\%$ to obtain the binned events in the invariant mass $W$. The differences in low invariant mass between the predictions with and without the fitted intrinsic charm are notably reduced. Here we use the muon energies, as well as their scattering angle, to reconstruct the invariant mass of the hadronic state. An independent measure for $W_{\rm reco}$ can be obtained experimentally by directly accessing the hadronic final state, which can offer different systematic uncertainties and help improve the accuracy of a measurement.

\begin{figure}[t]
    \centering
\includegraphics[width=0.49\linewidth]{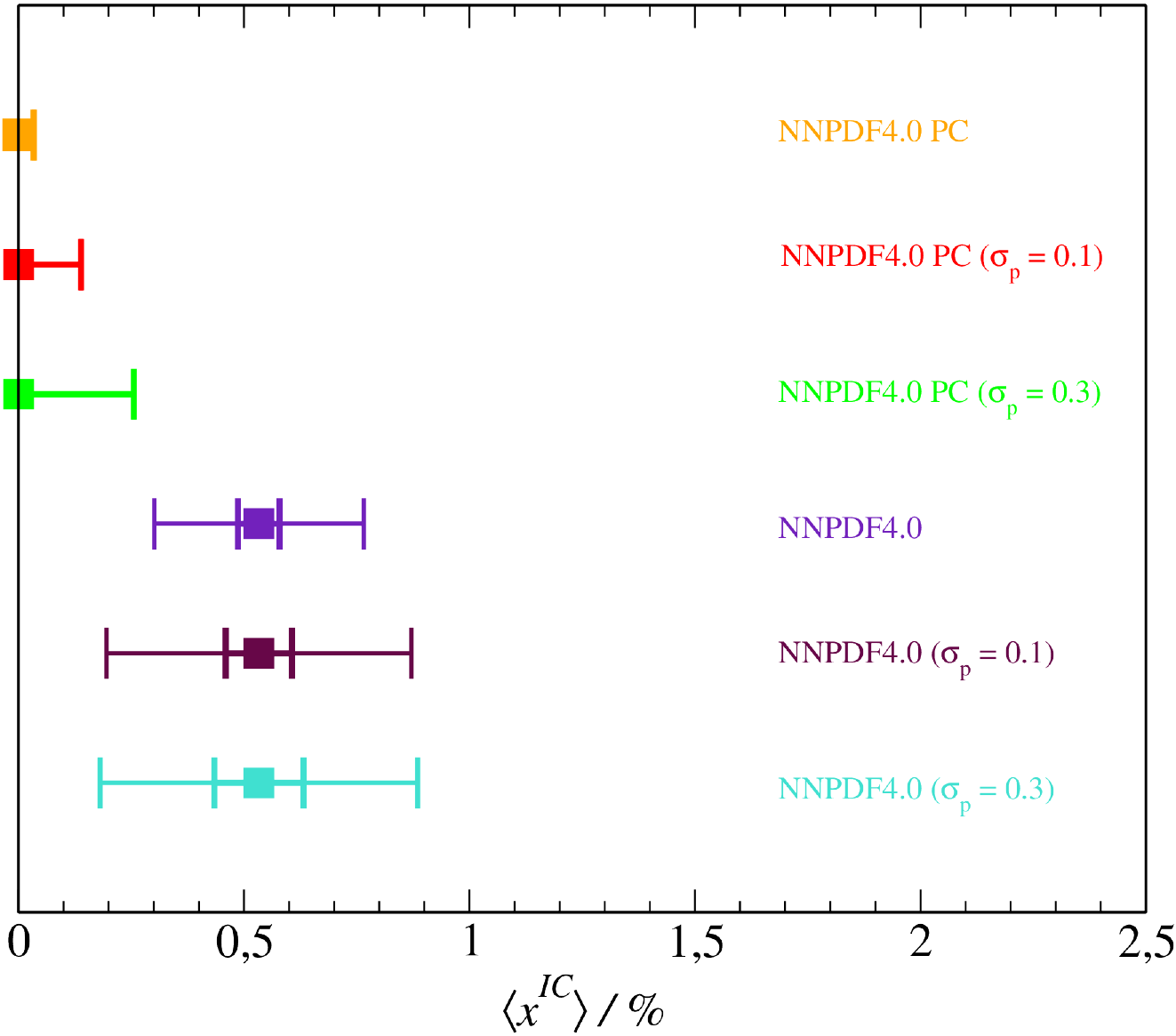}
\includegraphics[width=0.49\linewidth]{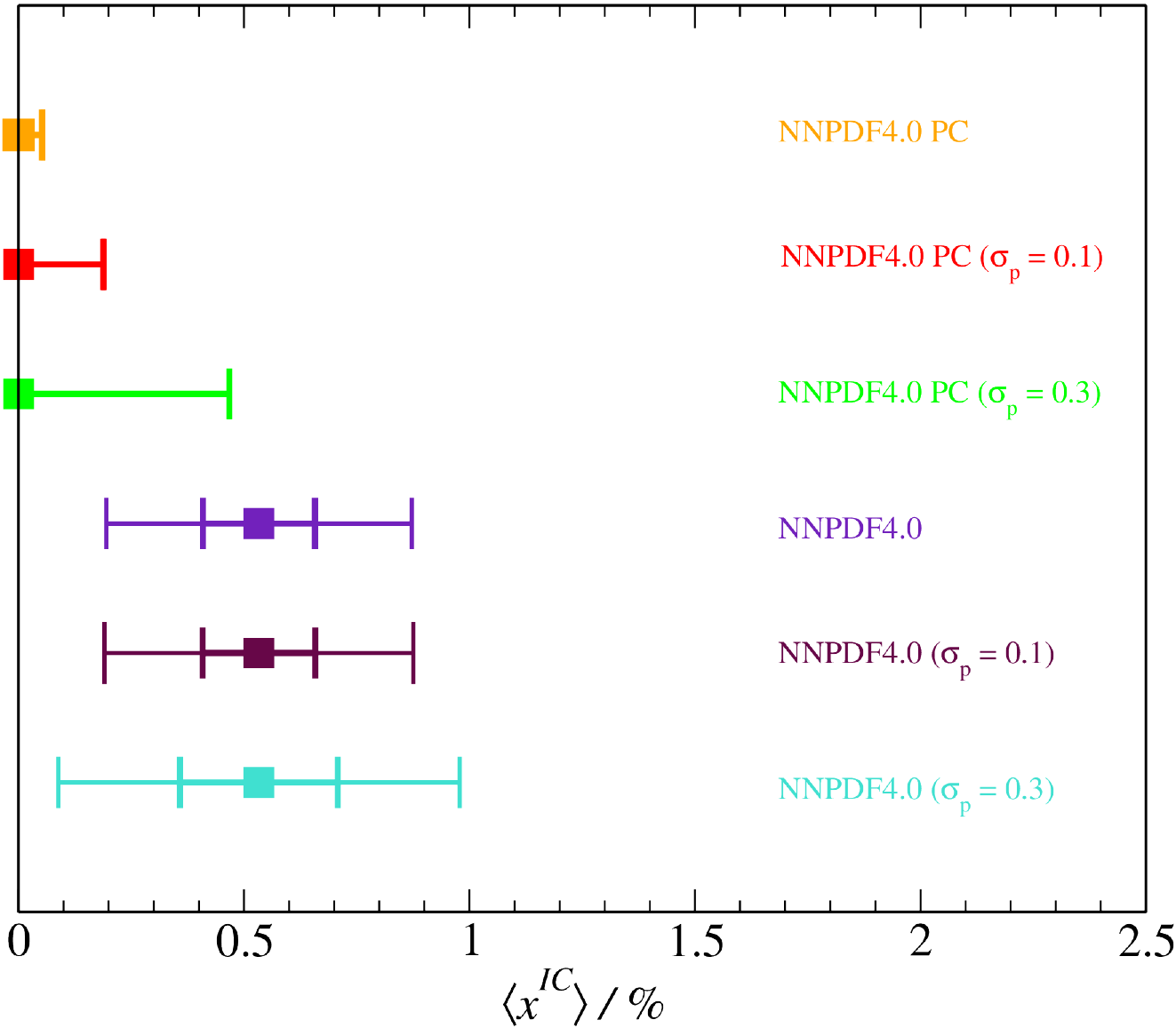}
    \caption{ Similar to Figure~\ref{fig:xcIC_FASERnu} for NNPDF4.0, comparing results without treatment of systematic uncertainties with results assuming $\sigma_p=10\%$ and $\sigma_p=30\%$ in the uncertainty of the determination of muon energies and also of the hadronic final state.
    }
\label{fig:xcIC_smearing_FASERnu}
\end{figure}

In Figure~\ref{fig:xcIC_smearing_FASERnu} we estimate the impact of systematic uncertainty on the precision with which FASER$\nu$ can measure the fraction of the nucleon momentum carried by the intrinsic charm, again assuming $\sigma_p = 10\%$ and $\sigma_p = 30\%$ for the predictions constructed with the NNPDF4.0 parameterization and events with $x \gsim 0.2$ (left) and $x \gsim 0.4$ (right). The systematic uncertainty we are modeling makes it difficult to distinguish between predictions with and without intrinsic charm even for the most optimistic case with $\sigma_p = 10\%$, and particularly challenging for the most pessimistic case. Our results indicate that for the FASER$\nu$ data during Run 3, the momentum fraction carried by this quark component that was defined in Equation~(\ref{eq:ICmomfrac}) will no longer be sensitive to different intrinsic charm models, and studies to reduce systematic uncertainties are needed to prove it at the nucleon.

\begin{figure}[t]
    \centering
\includegraphics[width=0.49\linewidth]{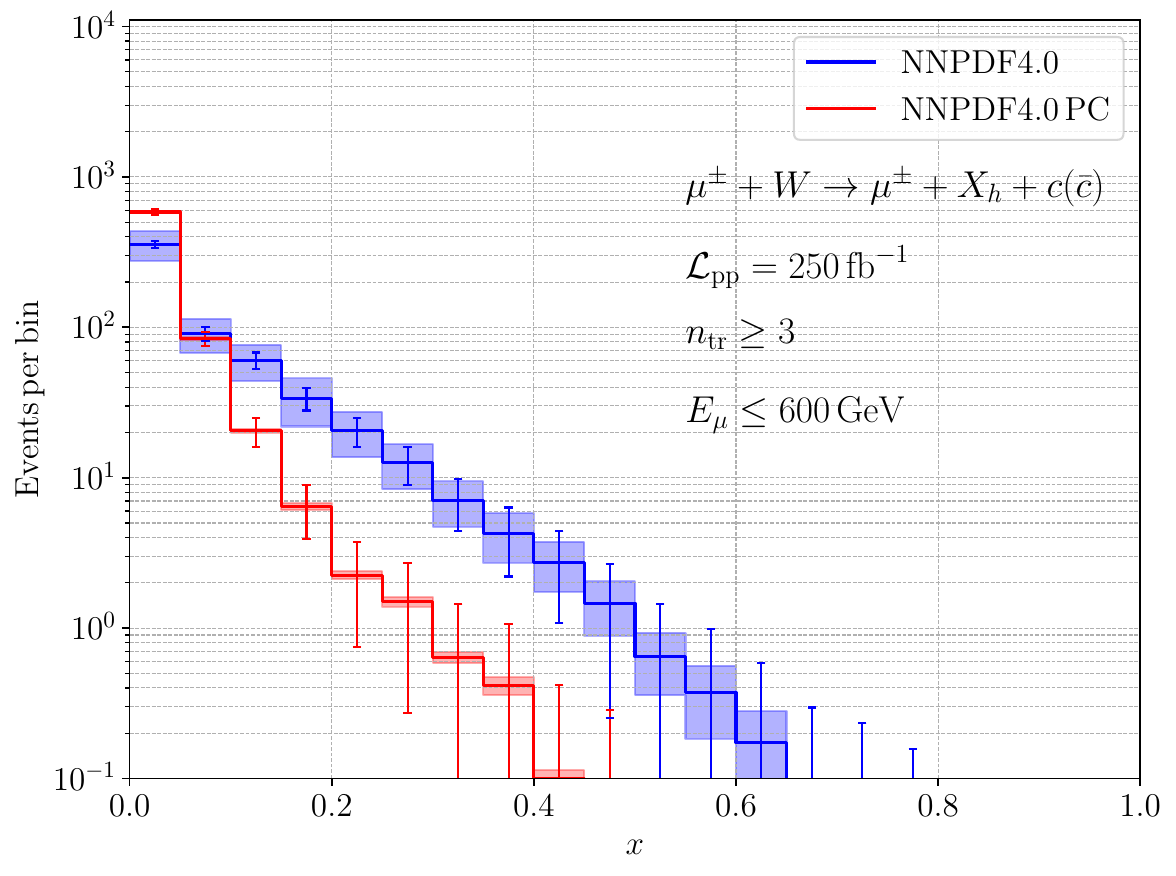}
\includegraphics[width=0.49\linewidth]{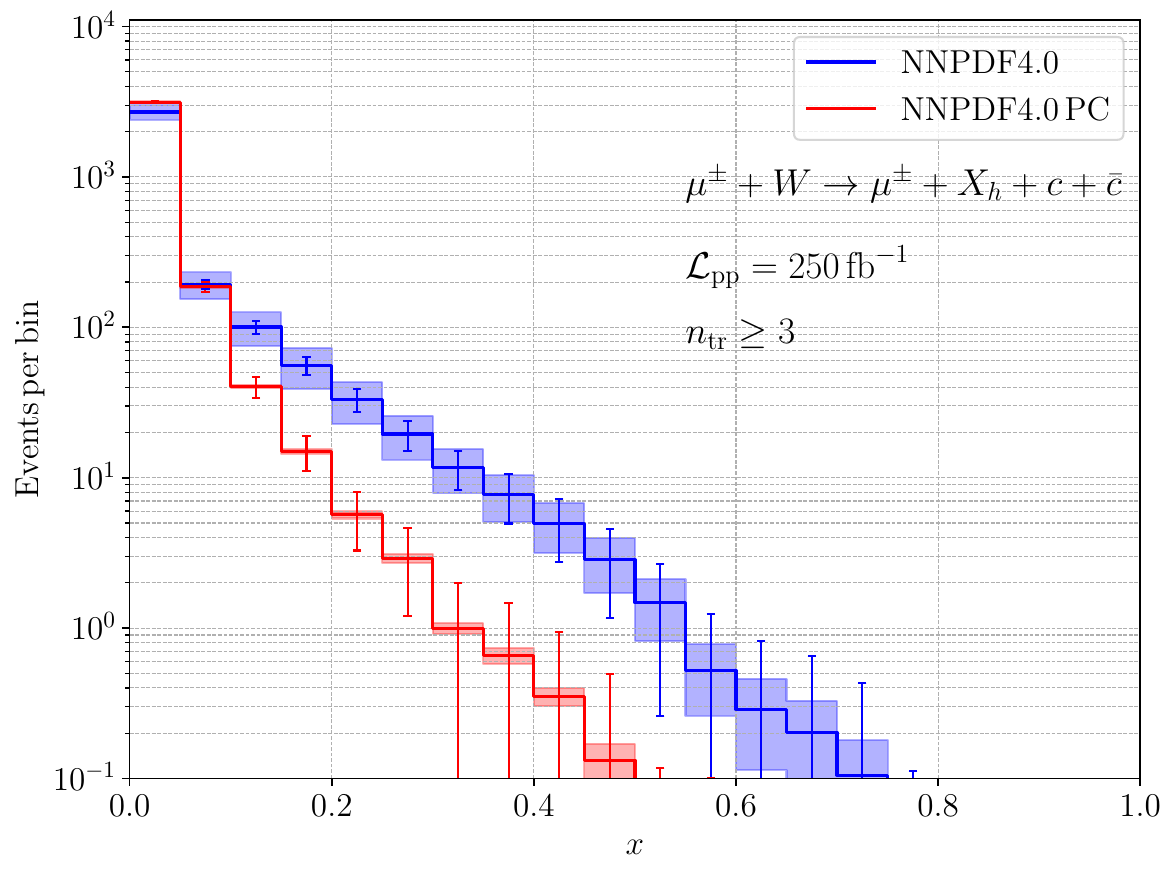}
    \caption{ Similar to Figure~\ref{fig:DIStotal_cuts_E_statBands} with additional cuts: initial muon with energy less than 600 GeV (left) and requiring observation of the two charm quarks produced in the final state (right).
    }
\label{fig:DIScharm_cuts_x_Ei_600GeV}
\end{figure}

Another strategy to reduce systematic uncertainties in intrinsic charm predictions for events in the FASER$\nu$ is to select only events with $E_\mu \le 600$ GeV, which have more precisely measured momenta compared to events of more energetic muons. In Figure~\ref{fig:DIScharm_cuts_x_Ei_600GeV} (left) we show the events in the FASER$\nu$ binned at $x$ applying this additional cut to the maximum energy of the muons. Our predictions indicate that in the large $x$ region the events decrease by a factor of approximately three when compared to the results without this additional cut, however the percentage difference between predictions with and without intrinsic charm in the nucleon remains stable. On the right panel of Figure \ref{fig:DIScharm_cuts_x_Ei_600GeV}, we consider only events with observation of the two charm quarks present in the final state. Considering the identification of the two charm quarks can significantly reduce the systematic uncertainties arising from the misidentification of a charm by the rescattering of light hadrons from the final state, forming a dataset genuinely dominated by charm production events.

Beyond traditional data bin analysis, significant improvements in event reconstruction and interpretation can be achieved using machine learning methods with complete detector-level event information, such as neural simulation based inference~\cite{ATLAS:2024jry} or transformer architectures~\cite{ATLAS:2025dkv} deployed in ATLAS. The benefits of data bin-free measurements in DIS for proton structure and QCD studies have been demonstrated, for example, in~\cite{Aggarwal:2022cki,H1:2021wkz}. These works, along with the study of the $n_{tr}$ dependence in Figure \ref{fig:DIStotal_cuts_x_nTracks}, indicate that specific improvements in event selection and kinematic reconstruction in muon DIS would significantly increase the prospects for impactful measurements for QCD analyses.

\subsection{ Intrinsic charm asymmetry at FASER$\nu$}
\label{subsec_MuonDIS:charmAsymmetry}

The analysis presented in Subsection \ref{subsec_MuonDIS:charmEvents} showed that the production of charm hadrons at FASER$\nu$ for events in the large $x$ region is highly sensitive to the amount of charm quark in the nucleon, being able to separate between predictions with and without intrinsic charm in the PDFs. In this analysis, we summed the charm to anticharm contributions, without worrying about the electric charge of the particles produced. However, an ultimate smoking gun for searching for effects of a non-perturbative charm would be the asymmetry between the charm and anticharm PDFs, which cannot be generated by any known perturbative QCD mechanism\footnote{Except for a very small asymmetry induced by DGLAP evolution at NNLO~\cite{Catani:2004nc}.}.

While the FASER$\nu$ emulsion detector is unable to identify the charge of charged particles produced by the absence of a magnetic field, the combination of FASER$\nu$ with information obtained by the FASER electronic detector, particularly the spectrometer, can provide information about the charge of charm quarks produced in the interactions when they decay semileptonically into muons, and these reach the electronic detector. The decay fraction of charm quarks in this channel, according to the most recent version of the Particle Data Group \cite{ParticleDataGroup:2022pth}, is
\begin{eqnarray}
\label{eq:charm_BR}
{\rm BR}( c \to \ell^+ +X)= 
{\rm BR}( \bar{c} \to \ell^- +X) = 
0.096 \pm 0.004 \, ,
\end{eqnarray}
a suppression factor of approximately 10 in relation to the total amount of charm detected in the emulsion films. In this electronic detector scenario, the signal would be the reconstruction of charm hadrons visible in the emulsion, and two muons passing through the spectrometer.

The correlation between muons observed in the FASER$\nu$ and the FASER spectrometer is performed through the track interface, a scintillator dedicated to this purpose between the two experiments. Assuming that the muon track from charm decay can be reconstructed in the spectrometer, the charm charge can be obtained. In this scenario, it becomes possible to identify charm and anticharm initiated events at the interaction vertex with muons from ATLAS. Obviously, this only applies to the fraction of charm hadrons that decay semileptonically according to Equation~(\ref{eq:charm_BR}).

Specifically, we can estimate the sensitivity of FASER$\nu$ and its successors by analogy to the analysis developed for EIC proposed in~\cite{NNPDF:2023tyk}. The observable is the asymmetry between final states with charm and with anticharm in certain regions of $x$, according to
\begin{equation}
\label{eq:charm_asymmetry}
    \mathcal{A}_c \equiv \frac{ N(\mu^\pm + W \to \mu^\pm + \widetilde{X}_h + c)-N(\mu^\pm + W \to \mu^\pm + \widetilde{X}_h + \bar{c})}{N(\mu^\pm + W \to \mu^\pm + \widetilde{X}_h + c) + N(\mu^\pm + W \to \mu^\pm + \widetilde{X}_h + \bar{c})} \, ,
\end{equation}
where $N$ is the number of muons scattering that meet all the selection criteria described previously, including the charm detection efficiency described with $\epsilon_c$ and with semileptonic decay given by Equation~(\ref{eq:charm_BR}). In summary, Equation~(\ref{eq:charm_asymmetry}) measures the difference between events initiated by charm and by anticharm. At leading order, this is directly proportional to the charm valence PDF, $xc^-=x(c-\bar{c})$, which is dominated by the non-perturbative quark component. In the calculation of Equation~(\ref{eq:charm_asymmetry}) we are using the most energetic (anti)charm quark of the final state, which generally corresponds to the interacting quark, given that it receives most of the energy transferred by the muon to the nuclear target.

\begin{figure}[t]
    \centering
\includegraphics[width=0.45\linewidth]{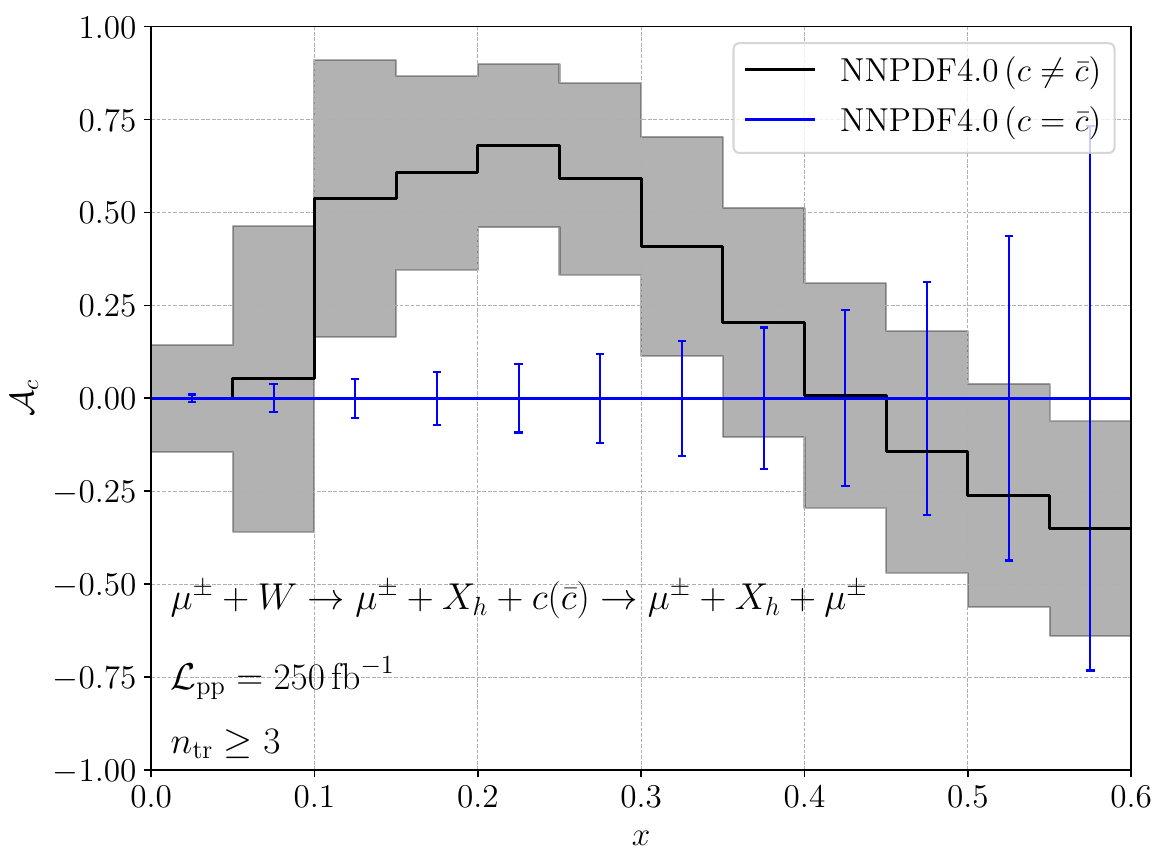} 
\includegraphics[width=0.45\linewidth]{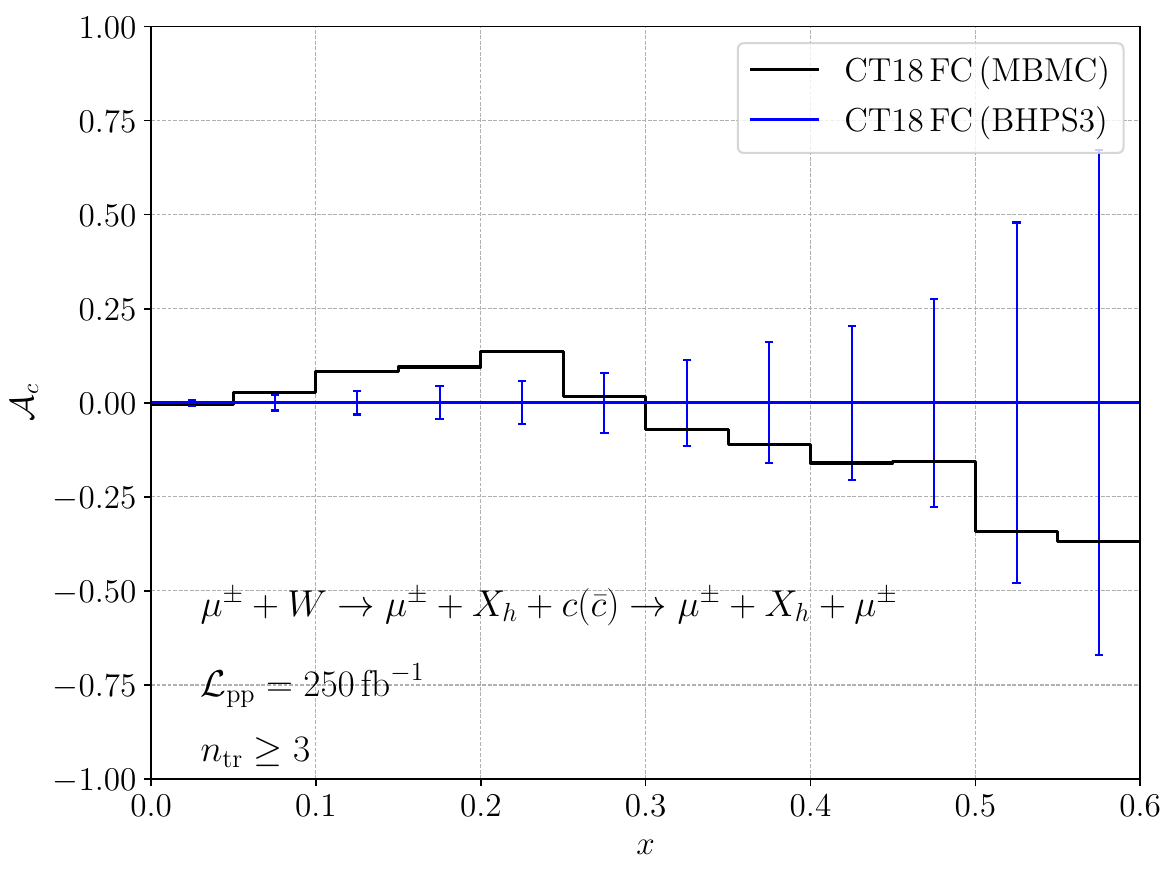}
    \caption{ 
    Left: charm-anticharm asymmetry $\mathcal{A}_c$ defined in Eq.~(\ref{eq:charm_asymmetry}) for the NNPDF4.0 parameterization set that has intrinsic charm fitted with asymmetry~\cite{NNPDF:2023tyk}, along with the predictions of the standard NNPDF4.0 set (with symmetric intrinsic charm fitted), see also Figure~\ref{fig:charm-PDFs}. The uncertainty bands correspond to the uncertainties in the PDFs, while the uncertainty bars represent the expected statistical uncertainty for this dataset at FASER$\nu$. Right: same analysis using the CT18 FC PDF set with the MBMC model that allows asymmetry between charm and anticharm, compared with predictions based on the BHPS3 model that has symmetric charm.
     }
    \label{fig:asy_x_charm}
\end{figure}

In Figure~\ref{fig:asy_x_charm} we show the asymmetry described by Equation~(\ref{eq:charm_asymmetry}) for the NNPDF4.0 (left) and CT18 FC (right) parameterizations that include an intrinsic charm component, for both sets of PDFs: $c=\bar{c}$ and $c\ne \bar{c}$. The uncertainty bands in the prediction using NNPDF4.0 correspond to the uncertainty we currently have in the PDFs, while the bars correspond to the statistical uncertainty projected for the FASER$\nu$ using the PDFs without asymmetry in the intrinsic charm. For simplicity, the results presented here do not take into account any systematic uncertainty, but a more realistic analysis can be done following the procedure presented in the previous subsection with uncertainties in the energy measurements that propagate to the $x_{\mathrm{reco}}$.

Both parameterizations, NNPDF4.0 ($c\ne\bar{c}$) and CT18 FC (MBMC), predict a non-zero asymmetry, with the peak of the positive part near $x\sim 0.2$, and with a peak in the negative part at $x\gsim 0.4$ (for NNPDF4.0) and $x\gsim 0.3$ for CT18 FC. In the case of NNPDF4.0 (CT18 FC), the expected value of this positive asymmetry is 75\% (15\%) at $x\sim 0.2$ and $-30\%$ ($-35\%$) at $x\sim 0.6$, although in NNPDF4.0 large variations are still possible within the estimated uncertainties of the PDF. For sets of PDFs with symmetric intrinsic charm, the asymmetry satisfies $\mathcal{A}_c\sim 0$, providing a basis for comparison. By comparing the predictions of the POWHEG+Pythia8 simulation with the expected statistical uncertainties, we see that the accuracy of FASER$\nu$ should be enough to identify positive values of $\mathcal{A}_c$ in the region around $x\sim 0.2$, especially in the case of NNPDF4.0. The results in Figure~\ref{fig:asy_x_charm} establish the measurement of $\mathcal{A}_c$ from the Run~3 dataset as an important and accessible target for the FASER$\nu$ neutrino program, enabling a prominent QCD analysis that would otherwise be delayed until the construction and data collection of the EIC.

\subsection{Predictions for HL-LHC}
\label{subsec_MuonDIS:charmHL-LHC}

\begin{figure}[t]
    \centering
\includegraphics[width=0.45\linewidth]{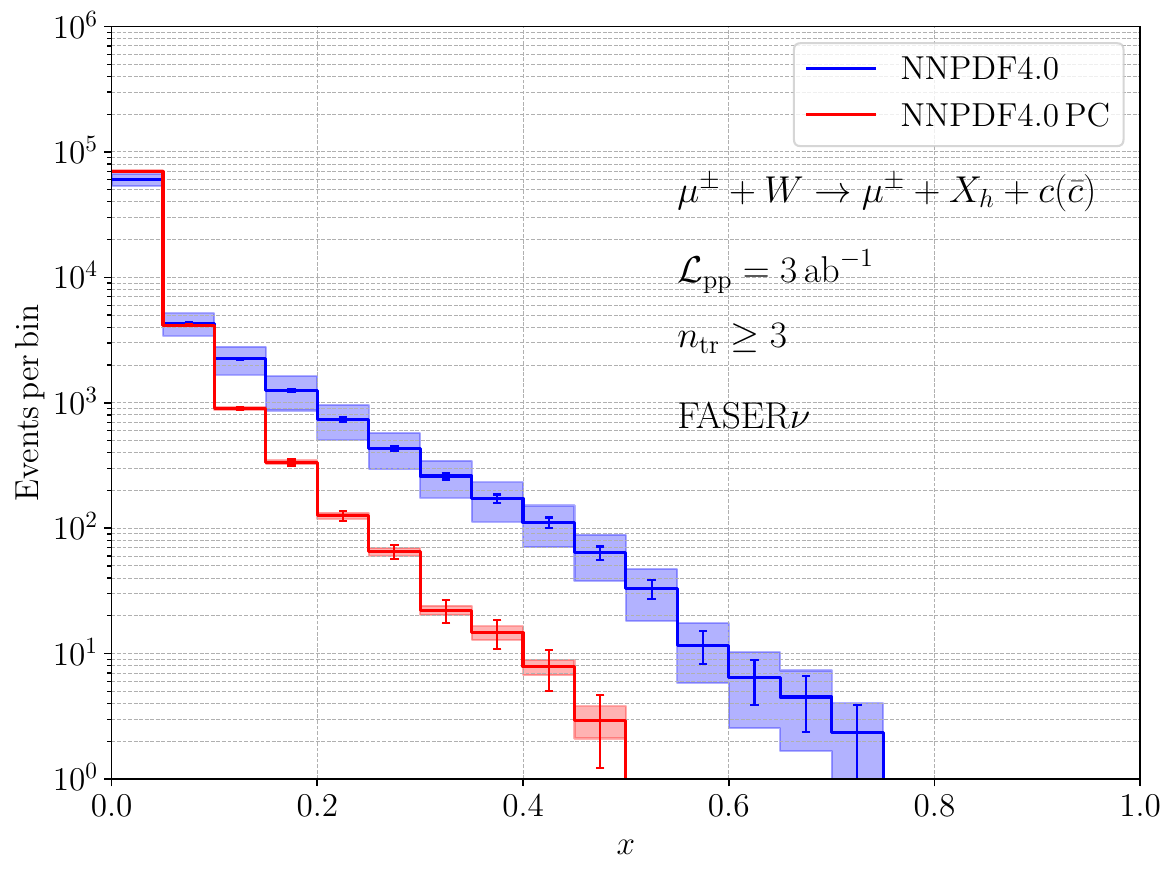} 
\includegraphics[width=0.45\linewidth]{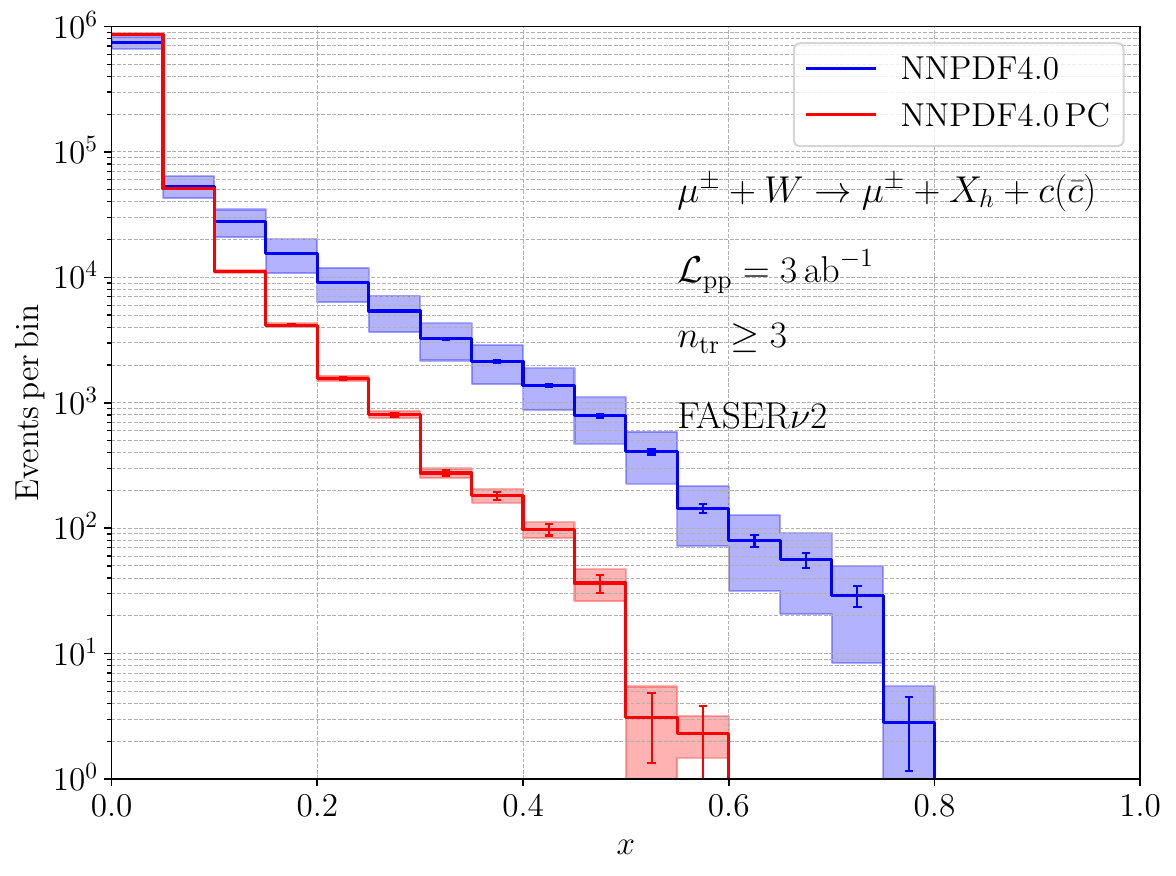}
\includegraphics[width=0.45\linewidth]{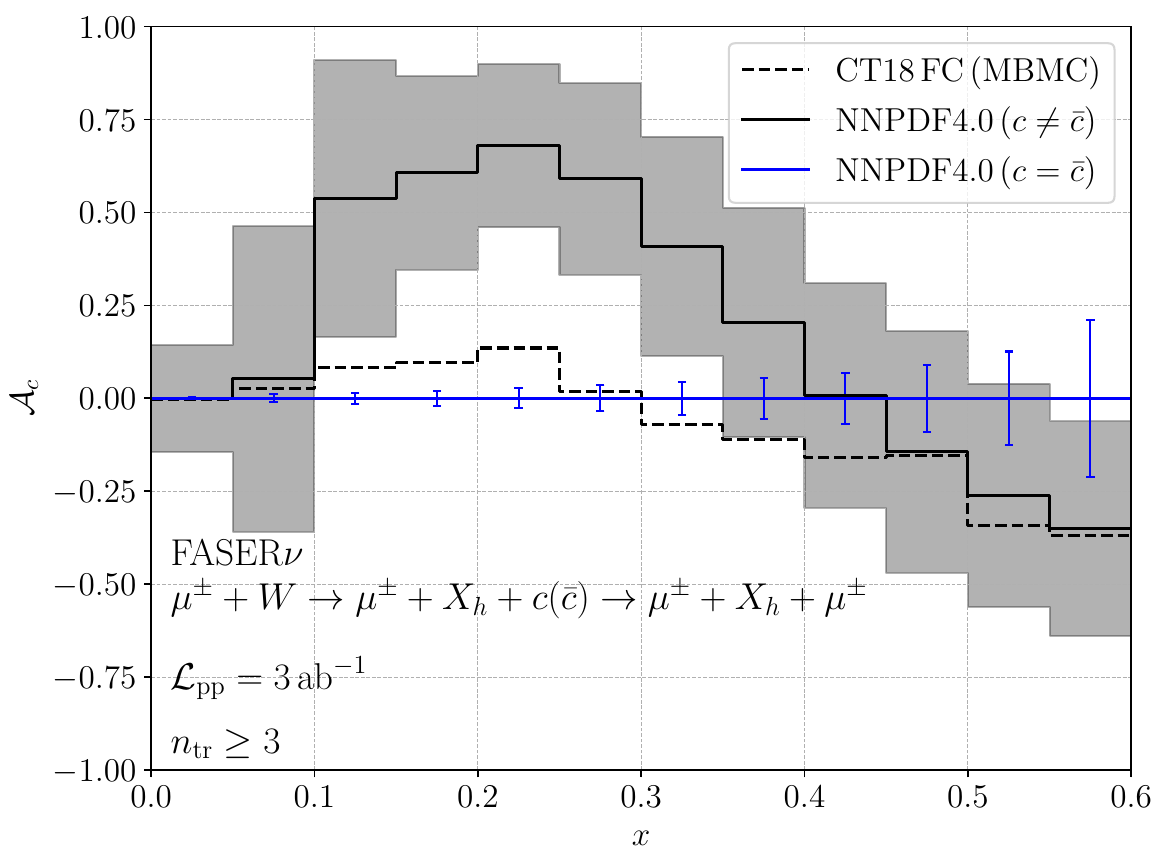} 
\includegraphics[width=0.45\linewidth]{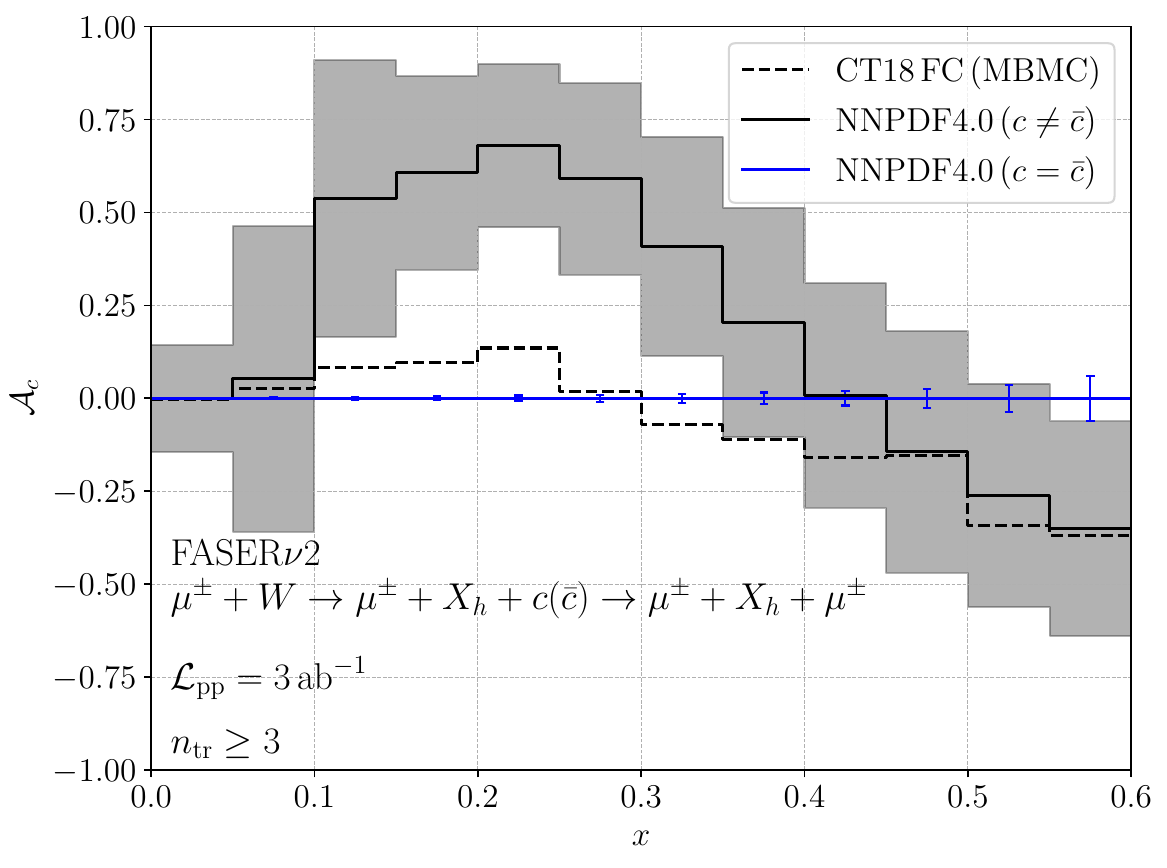}
    \caption{ 
    Top: similar to Figure~\ref{fig:DIStotal_cuts_E_statBands} for the FASER$\nu$ (left) and FASER$\nu$2 (right) detectors during the HL-LHC assuming $\mathcal{L}_{\rm pp}=3$ ab$^{-1}$. Bottom: similar to Figure~\ref{fig:asy_x_charm} for charm production with asymmetry $\mathcal{A}_c$. In both cases, the predictions are obtained using the NNPDF4.0 parameterization set, and in the bottom panel we also have the predictions obtained using the CT18 FC (MBMC) parameterization.
     }
\label{fig:Events_x_charm_HL-LHC}
\end{figure}

Finally, following Section ~\ref{sec_MuonDIS:DISinclusivo}, we provide predictions for charm production in the muon DIS and for the asymmetry in charm production $\mathcal{A}_c$ for the HL-LHC data collection era. We consider both the current FASER$\nu$ detector and the larger FASER$\nu$2 detector hosted at the FPF, in both cases for an integrated luminosity of $\mathcal{L}_{\rm pp}=3$ ab$^{-1}$. Figure~\ref{fig:Events_x_charm_HL-LHC} displays these projections, in the same format as Figure~\ref{fig:DIStotal_cuts_E_statBands}, for the expected events for total charm production binned in $x$ (top) and for the charm production asymmetry $\mathcal{A}_c$ (bottom).

Figure~\ref{fig:Events_x_charm_HL-LHC} illustrates the benefits of the higher expected statistics for the HL-LHC data collection period. For example, while for NNPDF4.0 at FASER$\nu$ (Run~3) only 20 charm production events are expected for $x\gsim 0.4$, where intrinsic charm has a more pronounced contribution, more than 2000 events would be recorded by FASER$\nu$2 in the same kinematic region, offering unparalleled sensitivity to charm PDF. This higher statistic also offers direct access to the large ``background-free'' $x$ region; for example, at FASER$\nu$2 no events are expected in $x\gsim 0.6$ in the perturbative charm scenario, while more than 100 events would be recorded for the charm case fitted by NNPDF4.0, proving an unequivocal signal.

\begin{figure}[t]
    \centering
\includegraphics[width=0.49\linewidth]{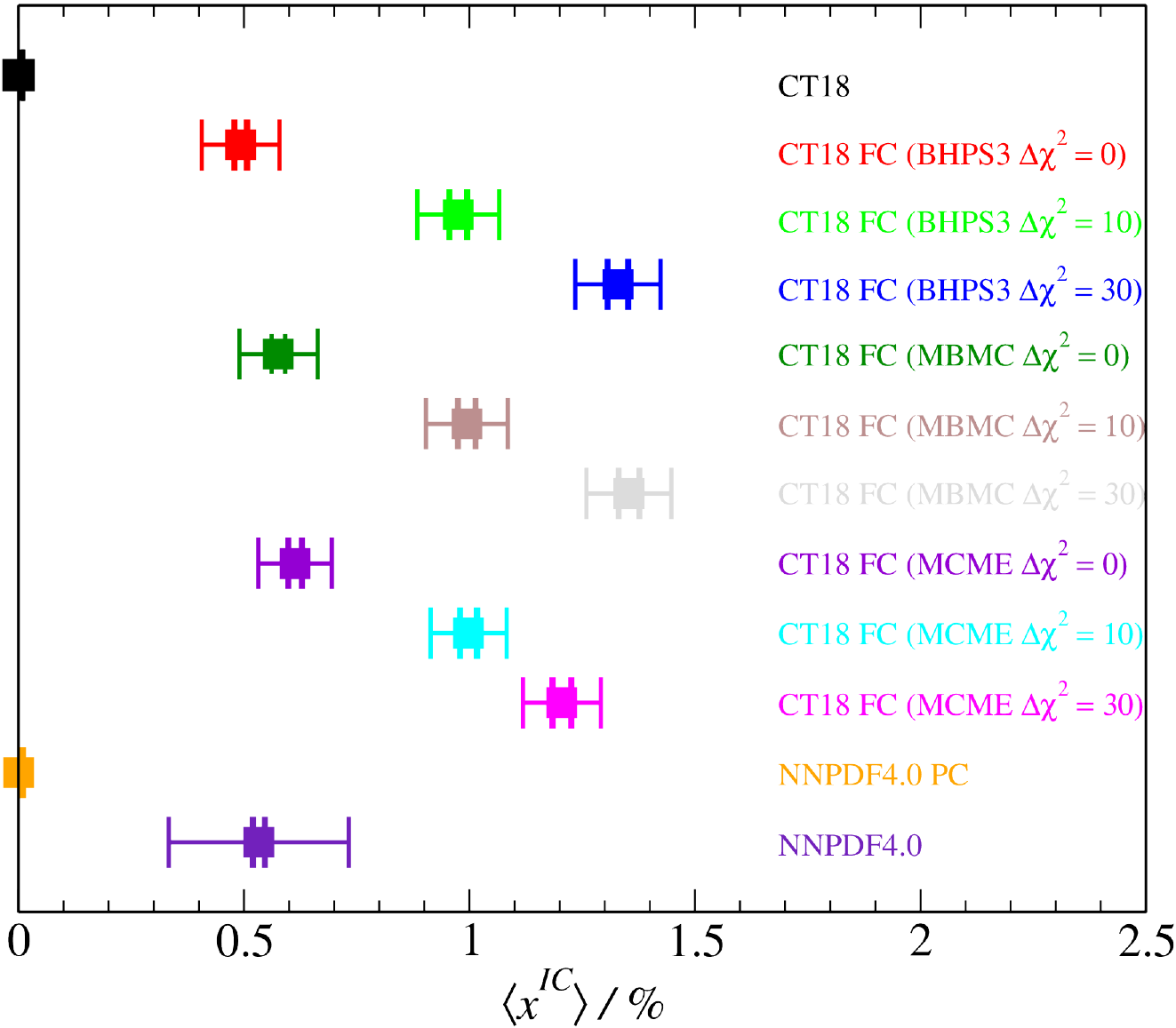}
\includegraphics[width=0.49\linewidth]{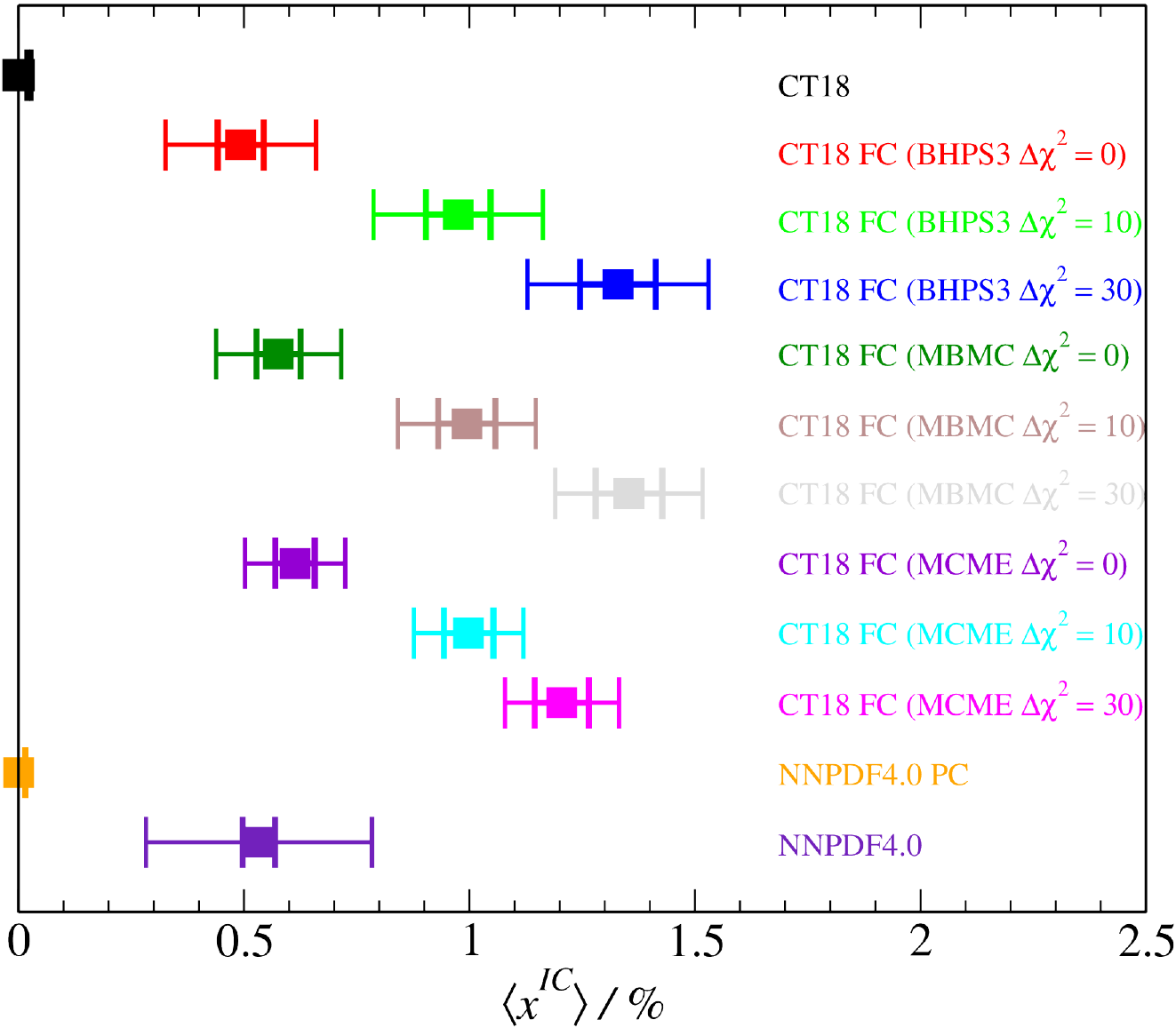} \\
\includegraphics[width=0.49\linewidth]{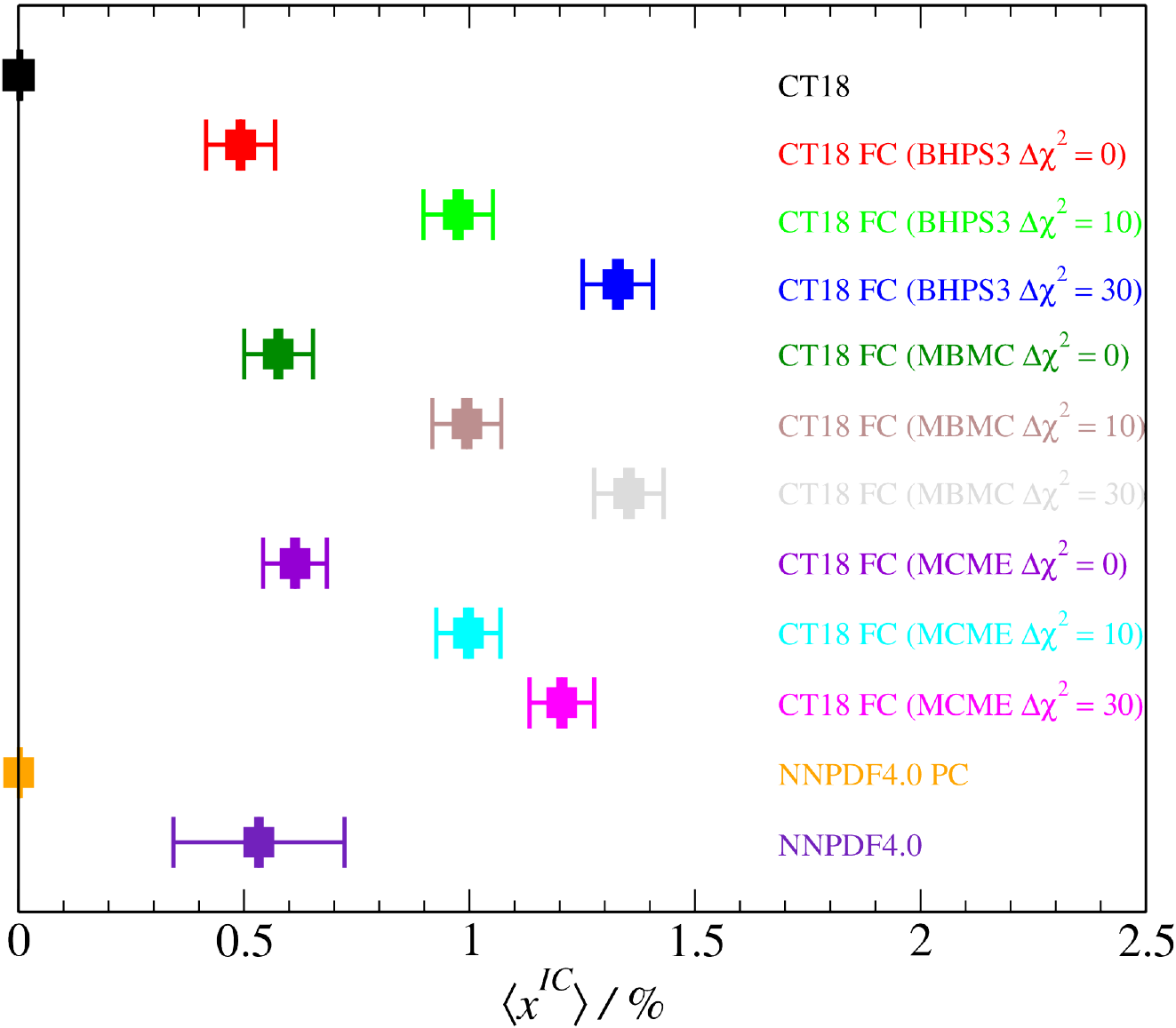}
\includegraphics[width=0.49\linewidth]{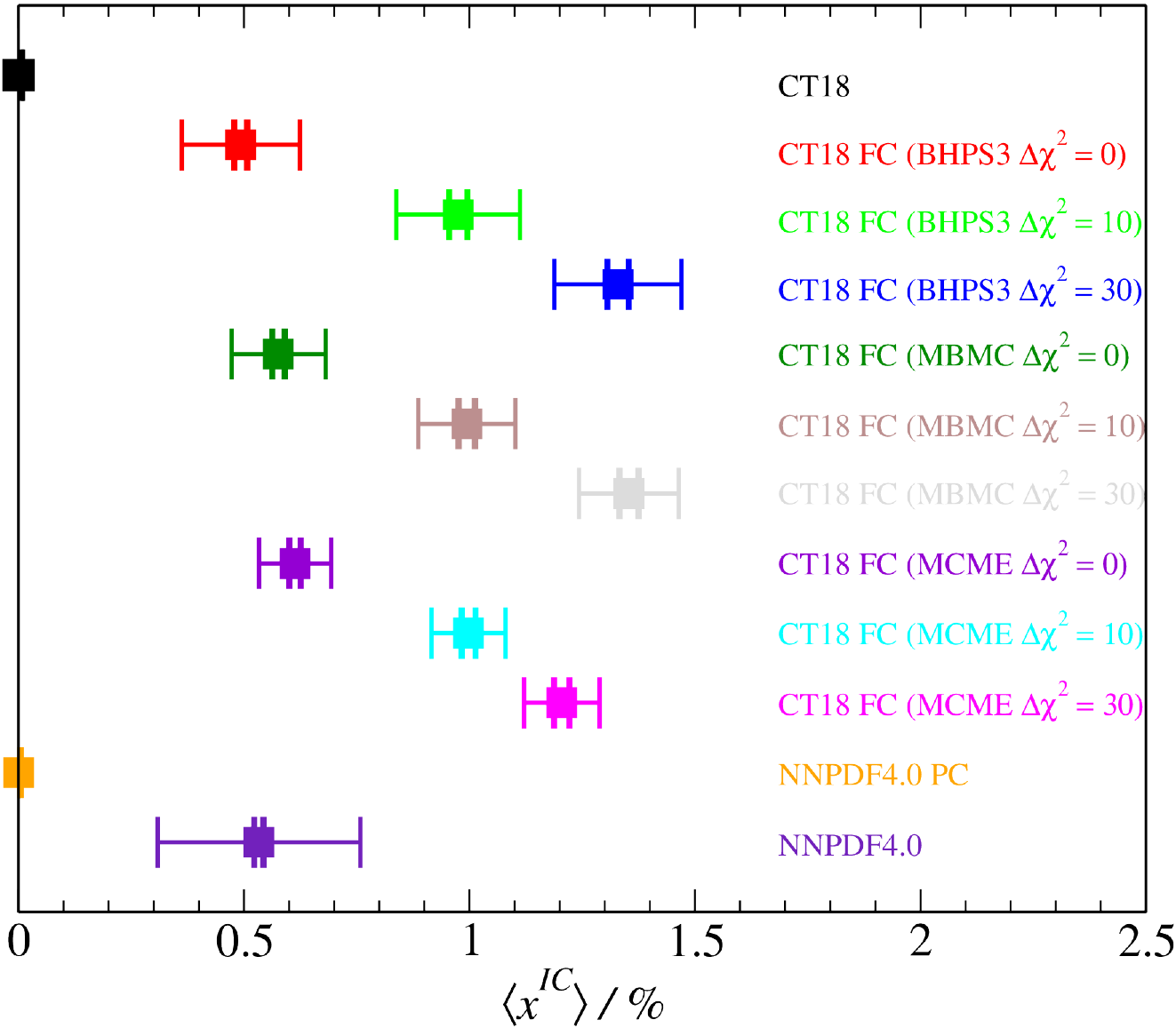}
    \caption{ Same analysis presented in Figure~\ref{fig:xcIC_FASERnu} for FASER$\nu$ (upper) and FASER$\nu2$ (lower) during HL-LHC assuming $\mathcal{L_{\mathrm{pp}}} = 3\,\mathrm{ab}^{-1}$.
    }
\label{fig:xcIC_FASERnu_HL}
\end{figure}

Similar considerations apply to projections of asymmetry in charm production. Although theoretical predictions based on NNPDF4.0 remain essentially unchanged, given that the magnitude of the predicted asymmetry does not depend on the magnitude of the total events, the estimated statistical uncertainties for measuring $\mathcal{A}_c$ become much smaller in future LHC runs. For the specific case of FASER$\nu$2, statistical errors on the order of a few percent in the relevant region of $x$ are expected, and even for $x\sim 0.6$ it is possible to measure $\mathcal{A}_c$ with uncertainties of 5\%. These results imply that FASER$\nu$(2) at the HL-LHC will accurately identify a possible charm asymmetry in the proton, with excellent sensitivity to the signals predicted by the vast majority of principal PDF models that allow $c\ne \bar{c}$, including the MBMC model adopted in the CT18 FC study. Figure~\ref{fig:xcIC_FASERnu_HL} presents the same as Figure~\ref{fig:xcIC_FASERnu} discussed previously, but now for FASER$\nu$ (upper) and FASER$\nu 2$ (lower) in the HL-LHC. In this case, the total uncertainties are dominated by the uncertainties of the PDFs, especially when considering events with $x\geq 0.2$. The high statistics available in the HL-LHC, therefore, will result in strong constraints on the intrinsic charm moment fraction, even considering only events with $x \geq 0.4$.

\section{Nuclear effects in lepton DIS at the FASER$\nu$(2)}
\label{sec_MuonDIS:nuclearEffects}

All the results derived so far in this chapter consider that the target nucleus in the scattering can be modeled as the sum of nucleons parameterized with free proton PDFs, that is, without taking into account effects of nuclear PDFs (nPDFs).

One of the main goals of current hadronic colliders and future electron-ion colliders is the determination of nPDFs with greater precision, given that these quantities are fundamental to improving our understanding of the structure of nuclei and to determining the initial conditions and properties of the Quark-Gluon Plasma (QGP) formed in heavy-ion collisions \cite{Rabelo-Soares:2025dfu}. In recent decades, different measurements in fixed-target and collider experiments have demonstrated that nPDFs are not a simple superposition of the distributions of partons of free nucleons, with the difference depending on the value of the Bjorken-$x$ variable (for a recent review, see, for example, reference~\cite{Klasen:2023uqj}). The first DIS data for the nuclear structure function $F_2^A$, measured in DIS of charged leptons in different nuclear targets, indicated that $F_2^A \lesssim A F_N$ for moment fractions $x \lesssim 0.1$ (shadowing region) and $0.3 \lesssim x \lesssim 0.7$ (EMC effect region), and that $F_2^A \gtrsim A F_N$ for $0.1 \lesssim x \lesssim 0.3$ (anti-shadowing region) and $x \gtrsim 0.7$ (Fermi motion).

\begin{figure}[H]
	\centering
	\begin{tabular}{ccc}
	\includegraphics[width=0.49\textwidth]{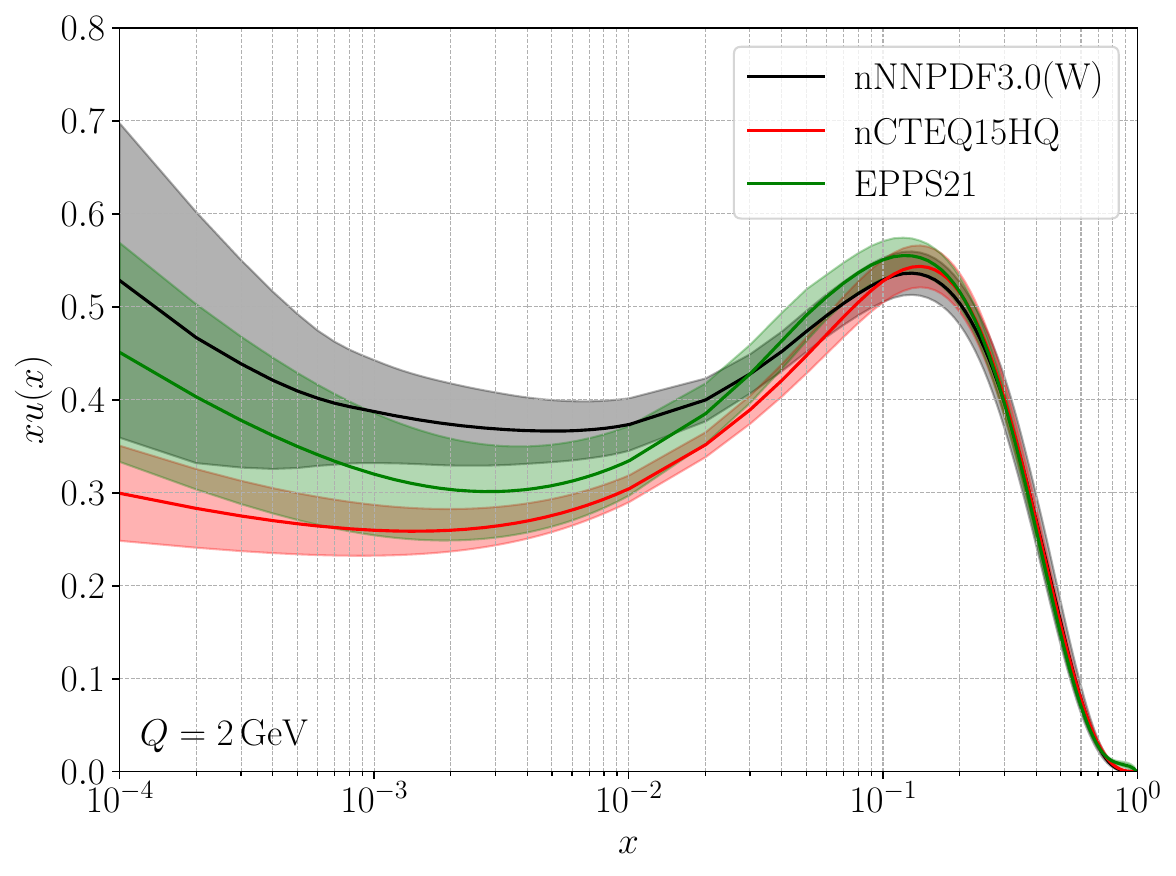}
	\includegraphics[width=0.49\textwidth]{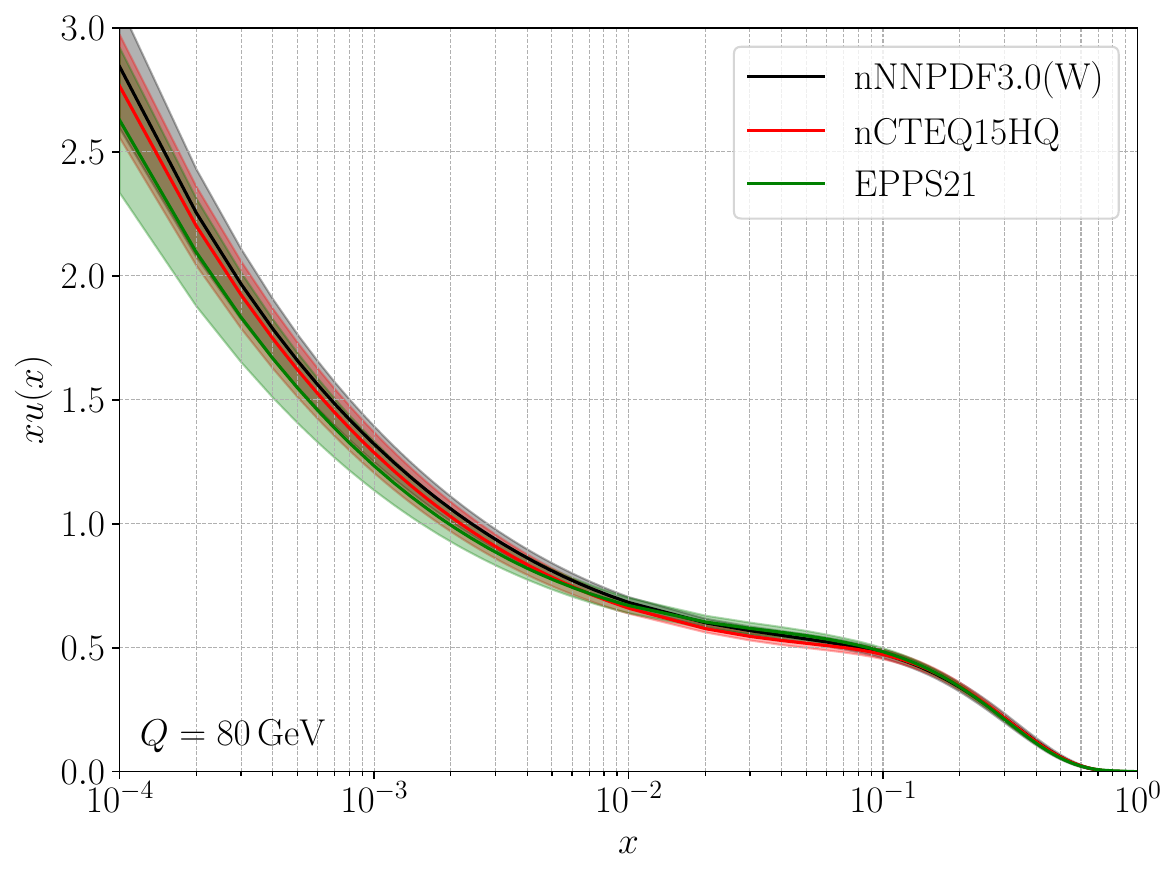}  \\
	\includegraphics[width=0.49\textwidth]{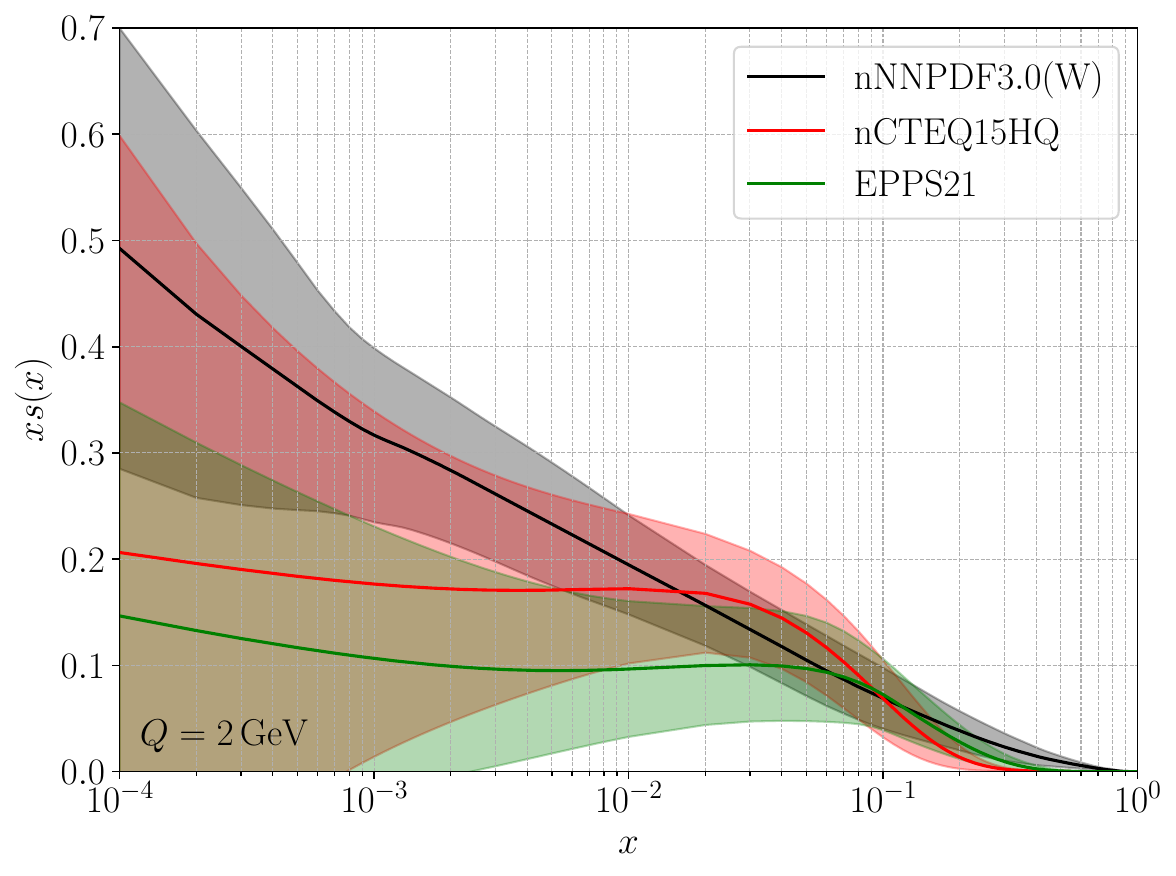} 
	\includegraphics[width=0.49\textwidth]{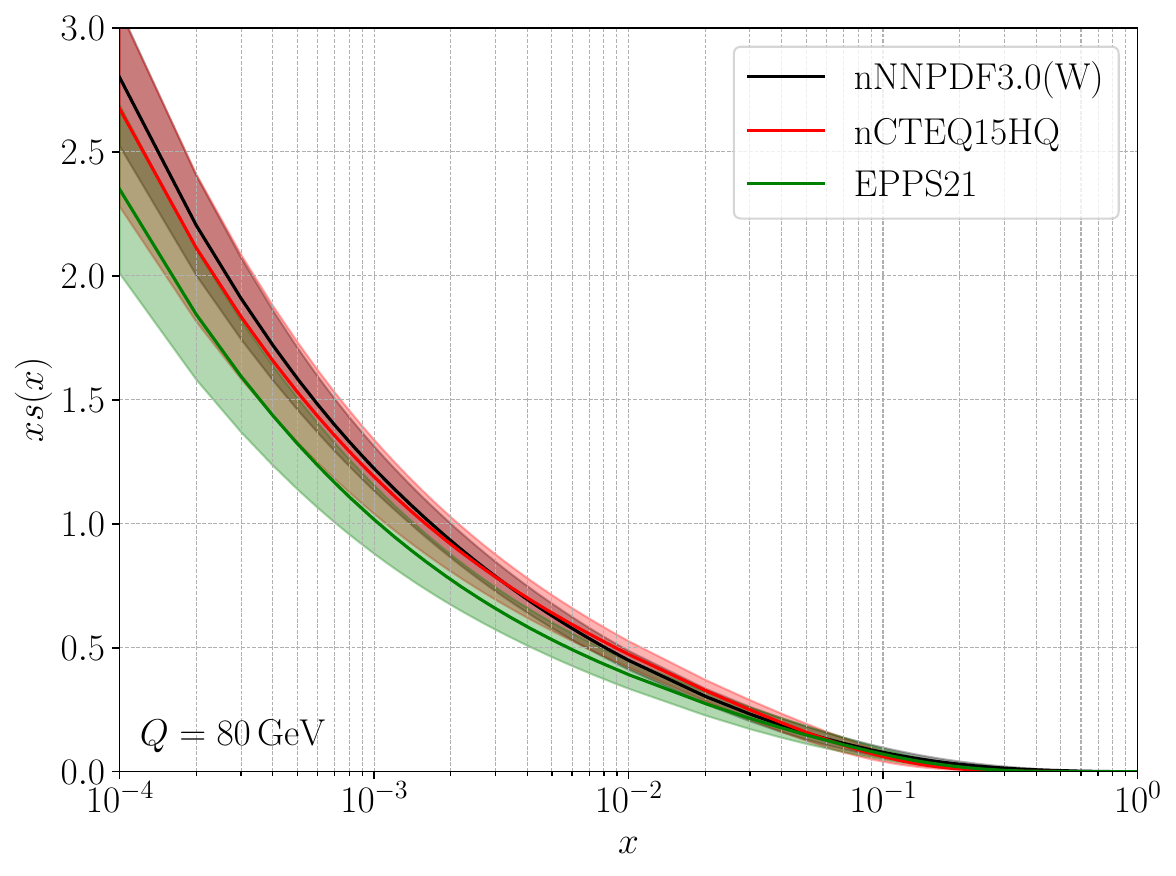} \\
	\includegraphics[width=0.49\textwidth]{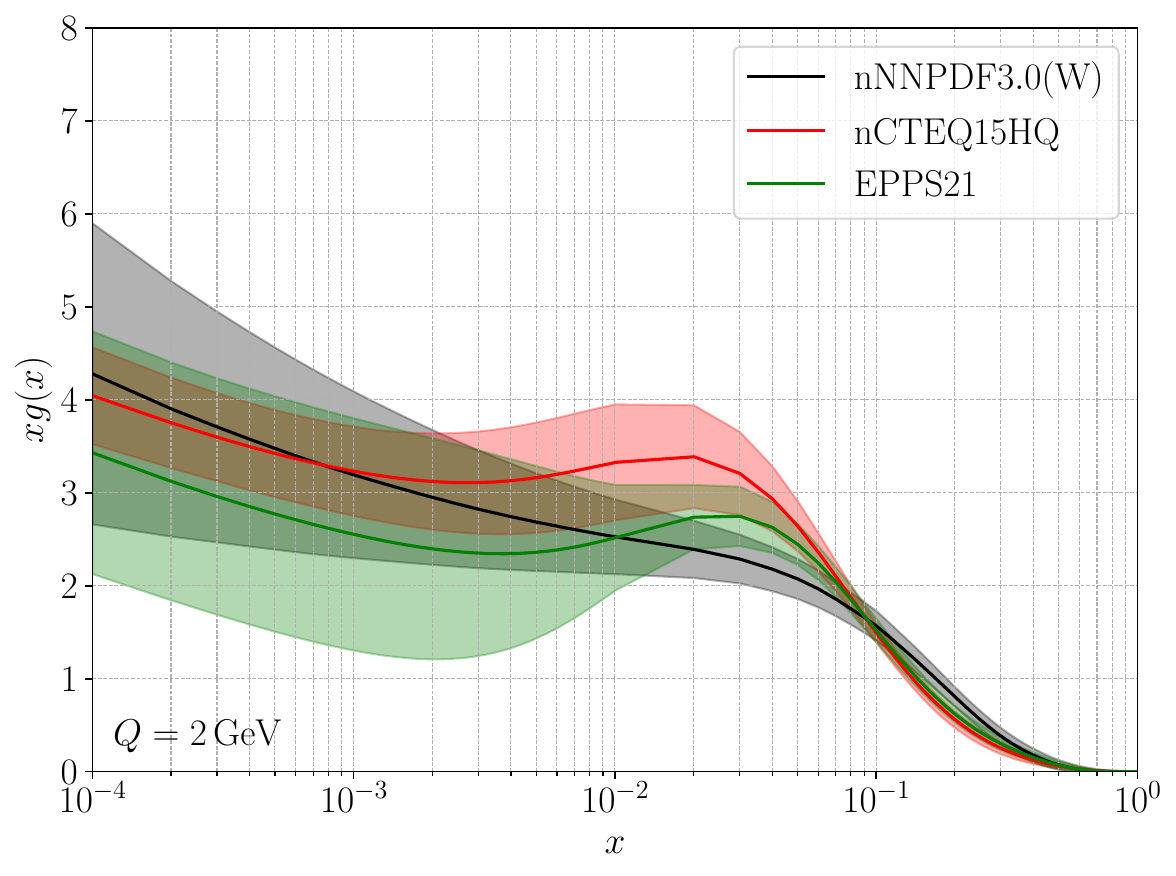}  
	\includegraphics[width=0.49\textwidth]{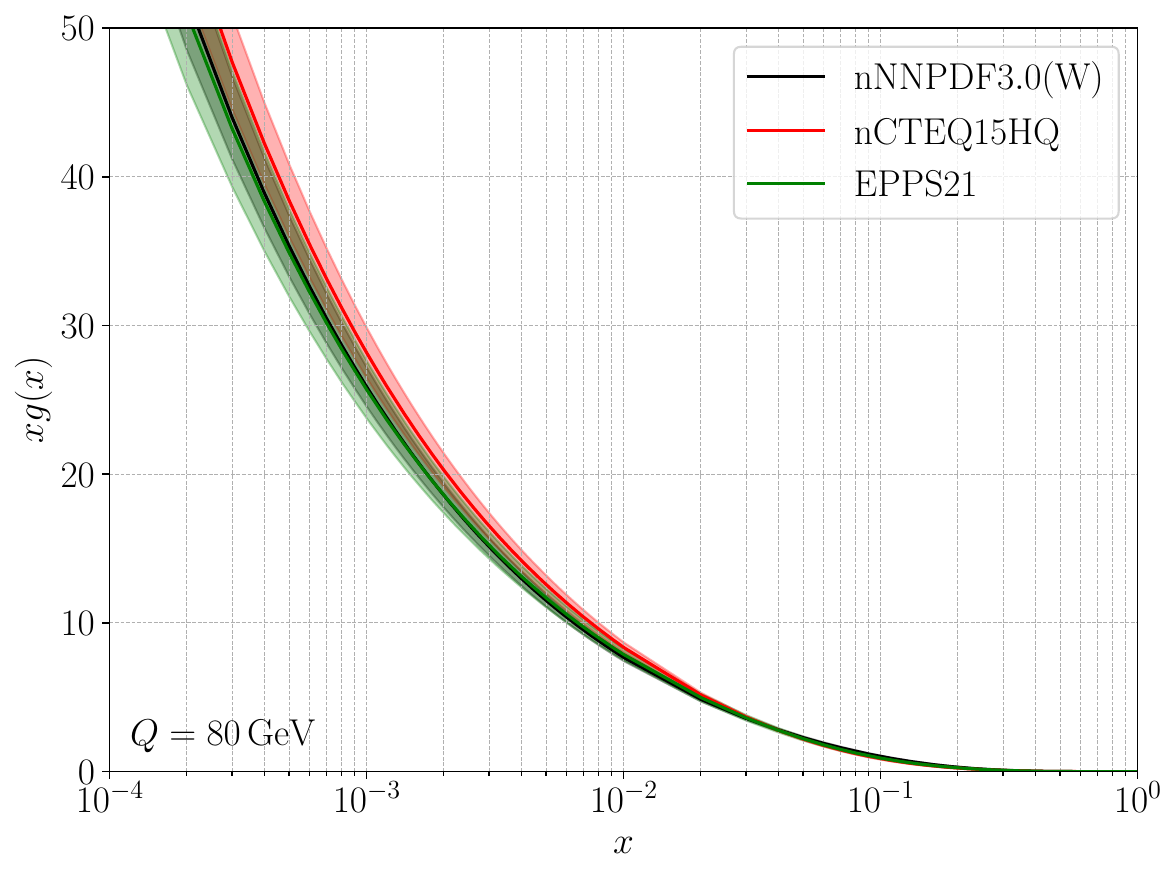}
    
			\end{tabular}
\caption{ Comparison between predictions for the PDFs of up (top), strange (center), and gluon (bottom) of tungsten, derived considering the parameterizations EPPS21 \cite{Eskola:2021nhw}, nNNPDF 3.0(W) \cite{AbdulKhalek:2022fyi}, and nCTEQ15HQ \cite{Duwentaster:2022kpv}. Results for two values of the virtuality of the boson: $Q=2$~GeV (left) and $Q=80$~GeV (right). The uncertainty bands correspond to confidence levels of 68\%. }
\label{fig:pdfs_nuclear}
\end{figure}

Nuclear effects have also been observed in neutrino scattering in charged current processes in heavy nuclear targets and in the production of dileptons via the Drell-Yan process in fixed-target proton-nucleus collisions\footnote{Annihilation process of a quark pair into a lepton pair \cite{Drell:1970wh}.}. More recently, the study of nuclear collisions at RHIC and LHC has provided a large amount of data for various observables sensitive to environmental effects, which has allowed us to reduce uncertainties in nuclear PDFs. However, the difference between the predictions derived by different groups \cite{Eskola:2021nhw,AbdulKhalek:2022fyi,Duwentaster:2022kpv} performing a global analysis of the current data\footnote{for a detailed discussion of the different methodologies used by these groups, see reference~\cite{Klasen:2023uqj}.} is still quite significant, as demonstrated in Figure \ref{fig:pdfs_nuclear}, especially for small values of the factorization scale $Q$, low $x$, and in the strange particle and gluon distributions.

Another important result obtained in these global analyses is that the inclusion of certain neutrino DIS data decreases the quality of the fit, indicating a tension between different datasets. In particular, Reference~\cite{Muzakka:2022wey} confirmed that existing charged lepton-ion DIS data are incompatible with most $\nu A$ DIS data, which motivated an intense debate about the breakdown of the universality of nuclear PDFs some years ago ~\cite{Schienbein:2007fs,Paukkunen:2010hb,Kovarik:2010uv,Paukkunen:2013grz}. However, the interpretation that this tension is associated with problems in the acquisition of neutrino-ion interaction data cannot be disregarded ~\cite{Muzakka:2022wey}. Furthermore, it is important to emphasize that any tension was observed in the analysis performed by the NNPDF group. These results demonstrate that new data on deep inelastic scattering of neutrinos in ions and charged leptons in ions in the GeV to TeV range are needed to improve our understanding of nuclear effects and to decide on the universality (or not) of nPDFs.

In this section, we will investigate the impact of nuclear effects on muon and neutrino scattering in nuclei at FASER$\nu$ and its proposed upgrade, FASER$\nu$2, which is expected to operate during the high-luminosity era of the LHC \cite{Feng:2022inv,FPFWorkingGroups:2025rsc,FPF:2025bor}. Our analysis is motivated by studies presented in previous sections, which showed large numbers of muon scattering events at FASER$\nu$ covering a wide range of $x$ ($3 \times 10^{-4} - 0.9$), that is, allowing us to investigate the impact of different nuclear effects (shadowing, anti-shadowing, EMC, and Fermi motion) on observables. Furthermore, since FASER$\nu$ will be able to measure neutrino-induced scattering of all flavors and muons using the same detector and nuclear target, this will allow us to perform a direct test of the universality of nPDFs. Several recent works have shown the possibility of studying nuclear structure, such as PDF uncertainty reduction, with neutrinos at FASER$\nu$(2) \cite{Cruz-Martinez:2023sdv}. However, these previous works have been based on free proton PDFs for the description of nuclear matter.

In our analysis, we will consider two classes of events generated in neutrino-tungsten and muon-tungsten interactions: (a) inclusive events and (b) events with charm detected in the final state. Our motivation for selecting these events is associated with the fact that they probe different kinematic ranges of $x$ and are sensitive to distinct parton distributions. In particular, events with charm detected are strongly dependent on the charm (strange) PDF in muon (neutrino)-ion interactions, which are the distributions with the greatest uncertainties, as shown in Figure \ref{fig:pdfs_nuclear}. Furthermore, we will present results for the ratio between events with charm detected and inclusive events, which may be useful for discriminating between the different parameterizations of nuclear effects.

\subsection{Simulation settings at the FASER$\nu$}
\label{sec_MuonDIS:nucSimulation}

In this section we will describe the procedures for calculating the number of events associated with the deep inelastic scattering of leptons in tungsten targets at the FASER$\nu$(2). In general, we will extend what we did at the beginning of the chapter for muon scattering also to neutrino-induced scattering. The reaction we are interested in is
\begin{eqnarray}
 l_i + W \rightarrow l_f + X_h \,,   
\end{eqnarray}
where $l_i$ is the incident lepton scattering in the tungsten nucleus $W$. The final state of the interaction is an inclusive state of hadrons $X_h$ in addition to a charged lepton $l_f$. Given that we are interested in electrically charged leptons in the final state, we are only interested in muon-neutral current interactions and neutrino-charged current interactions.

The number of events expected at the detector can be described again by the Equation (\ref{eq:events}). Among its main ingredients is the lepton flux $f(x_{l_i})$. These fluxes for muons and antimuons in the forward direction of ATLAS were simulated with FLUKA and are shown in Figure~\ref{fig_DIS:muonflux}. For the case of the (anti)muon flux, $f(x_{\mu})$, we have $x_{\mu}$ being the fraction of the muon energy compared to the incident proton energy at the LHC. A similar analysis for the neutrino flux was performed in the reference \cite{vanBeekveld:2024ziz}, which also provided a PDF for the neutrino flux in LHAPDF format, but in this case $x_{\nu} = E_\nu / \sqrt{s_{\mathrm{pp}}} = E_\nu / 2E_p$. This neutrino PDF was constructed using the neutrino flux of light mesons obtained from \cite{Kling:2021gos}, while the neutrino flux from the decay of heavy mesons was obtained from \cite{Buonocore:2023kna}.

Lepton-tungsten cross sections will be estimated using the POWHEG-BOX-RES event generator \cite{Banfi:2023mhz,FerrarioRavasio:2024kem}, which simulates fixed-target lepton DIS with $\alpha_s^{2}$ (NLO) QCD coupling corrections. Partonic cross sections are then integrated into Pythia8 for hadronization \cite{vanBeekveld:2024ziz,Bierlich:2022pfr}. Furthermore, we will assume the three distinct parameterizations for the nuclear PDFs presented in Fig. \ref{fig:pdfs_nuclear}: EPPS21 \cite{Eskola:2021nhw}, nCTEQ15HQ \cite{Duwentaster:2022kpv} and nNNPDF 3.0 \cite{AbdulKhalek:2022fyi}, which will be denoted as nNNPDF 3.0(W) from now on. In particular, we will use the derived nNNPDF 3.0 set including the LHCb data for the production of the $D$ meson at forward rapidity. For comparison, we will also present the derived predictions disregarding nuclear effects, calculated using the CT18ANLO proton parameterization \cite{Hou:2019qau} and the baseline for the free nucleon of the nNNPDF 3.0 parameterization \cite{AbdulKhalek:2022fyi}, which will be referred to as NNPDF 3.0(p) from now on. Both parameterizations are rescaled for a tungsten target, i.e., inserting the necessary non-isoscalarity corrections.

The last ingredient in Equation~(\ref{eq:events}) needed to estimate the number of events is the detector efficiency, $\mathcal{A}$. In our analysis, we will consider the same efficiency used in previous sections: final leptonic energy greater than 100 GeV and at least two charged tracks in the hadronic final state with momentum greater than 1 GeV each. For semi-inclusive final states, where a charm is produced, we will assume a charm identification efficiency of $\epsilon = 0.7$. In addition to detector acceptance, we will select DIS events with $Q \geq 1.65\, \mathrm{GeV}$ and invariant mass of hadronic final state greater than $2\, \mathrm{GeV}$. We will use these same cuts for both neutrino and muon scattering. Although the FASER collaboration is considering in its current algorithm the reconstruction of neutrino events with at least 5 tracks in the final state, we are considering only three in our analysis to remain consistent with the muon scattering case. Finally, in our analysis, we defined the detector target length at 50 cm for the muon DIS, since the identification of muons, as well as the measurements of the momentum of incident and emitted muons, require propagation over a few centimeters. For the neutrino case, we are considering the full size of the detector, as done in the references \cite{Cruz-Martinez:2023sdv,vanBeekveld:2024ziz}.

\subsection{Results for the FASER$\nu$}
\label{sec_MuonDIS:nucResultsFASERnu}

One of the main goals of this study is to investigate the potential of leptons produced in $pp$ collisions at the LHC and measured by detectors positioned in forward regions to improve our understanding of nuclear effects in DIS. Next, we will present our results for event rates considering muon-tungsten and neutrino-tungsten interactions in the FASER$\nu$ and FASER$\nu 2$ detectors. Motivated by recent analyses from FASER \cite{FASER:2024hoe,FASER:2024ref}, which measured differential cross sections in energy and neutrino rapidity, we will also present predictions for the number of events binned in the Bjorken-$x$ variable. This distribution allows us to estimate the impact of nuclear effects on cross sections in different kinematic regimes.

\begin{figure}[H]
	\centering
	\begin{tabular}{ccc}
	\includegraphics[width=0.49\textwidth]{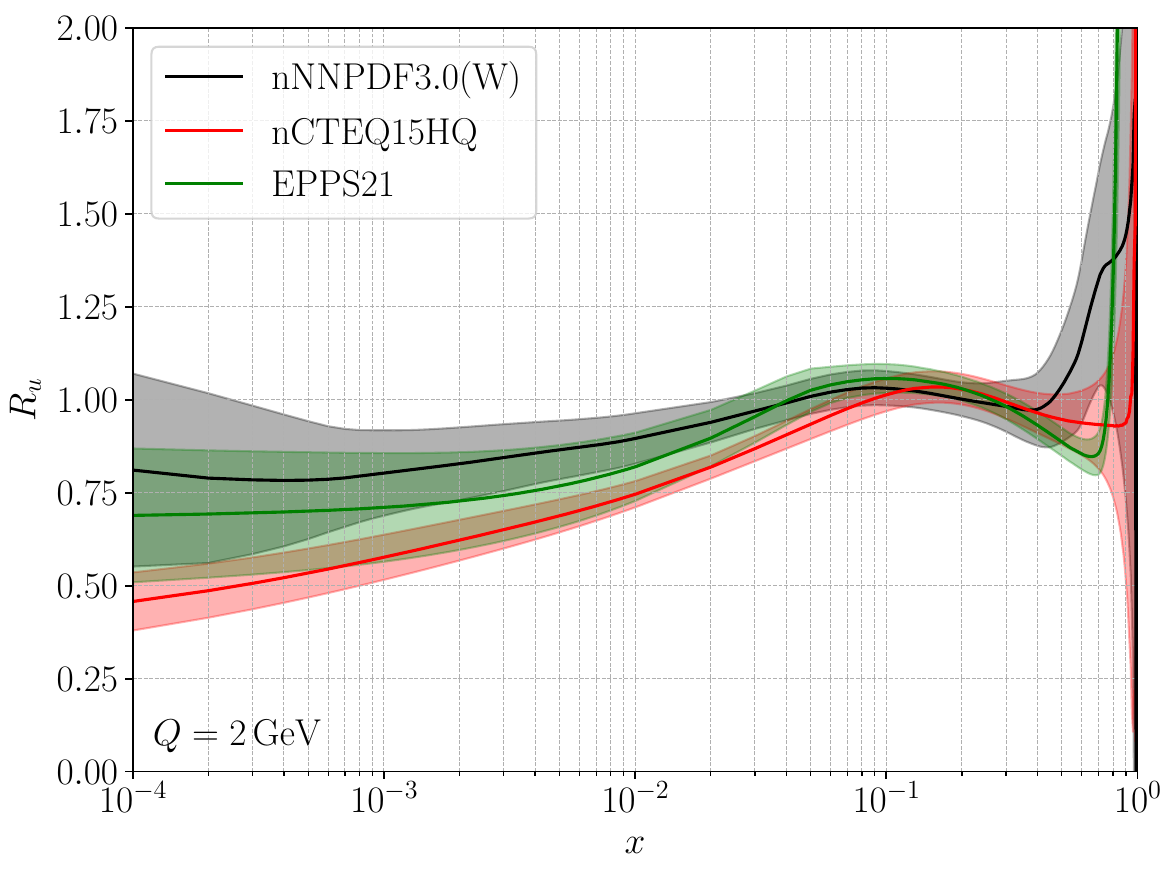}
	\includegraphics[width=0.49\textwidth]{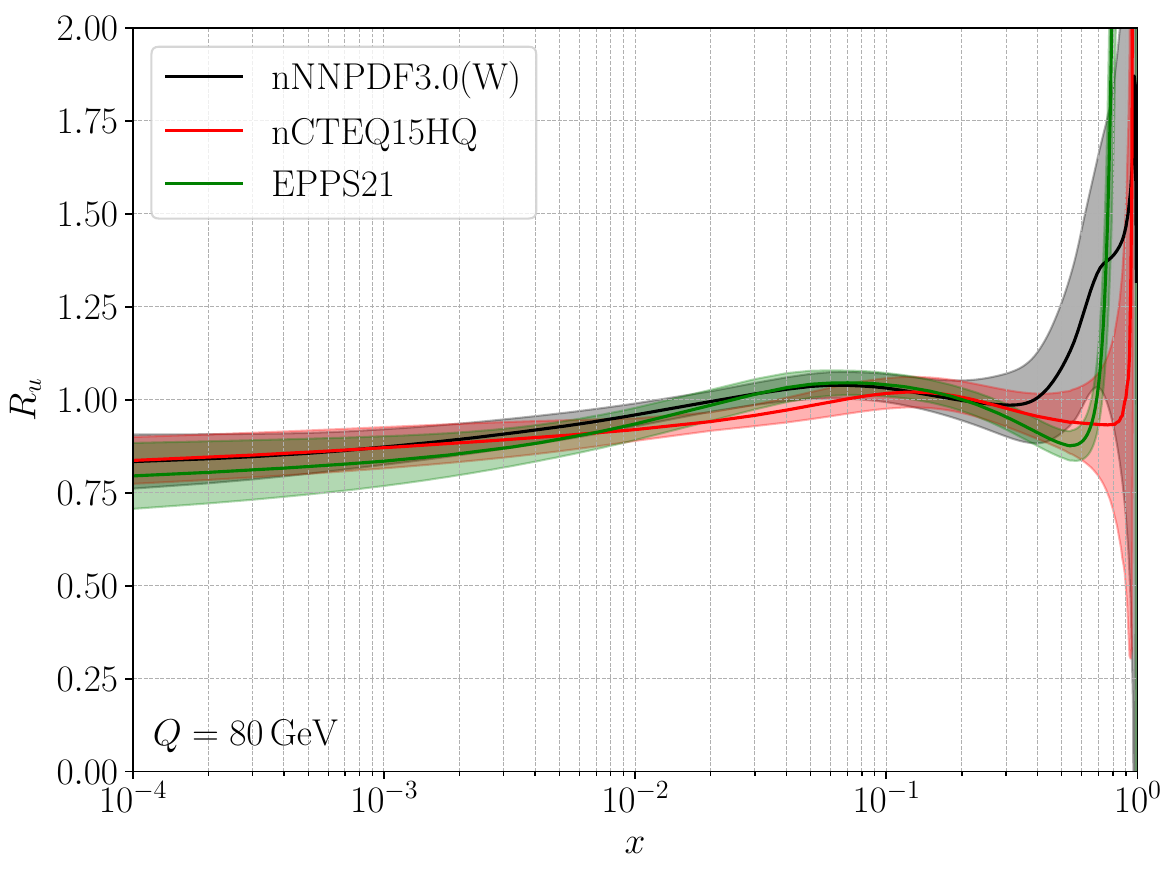} \\
	\includegraphics[width=0.49\textwidth]{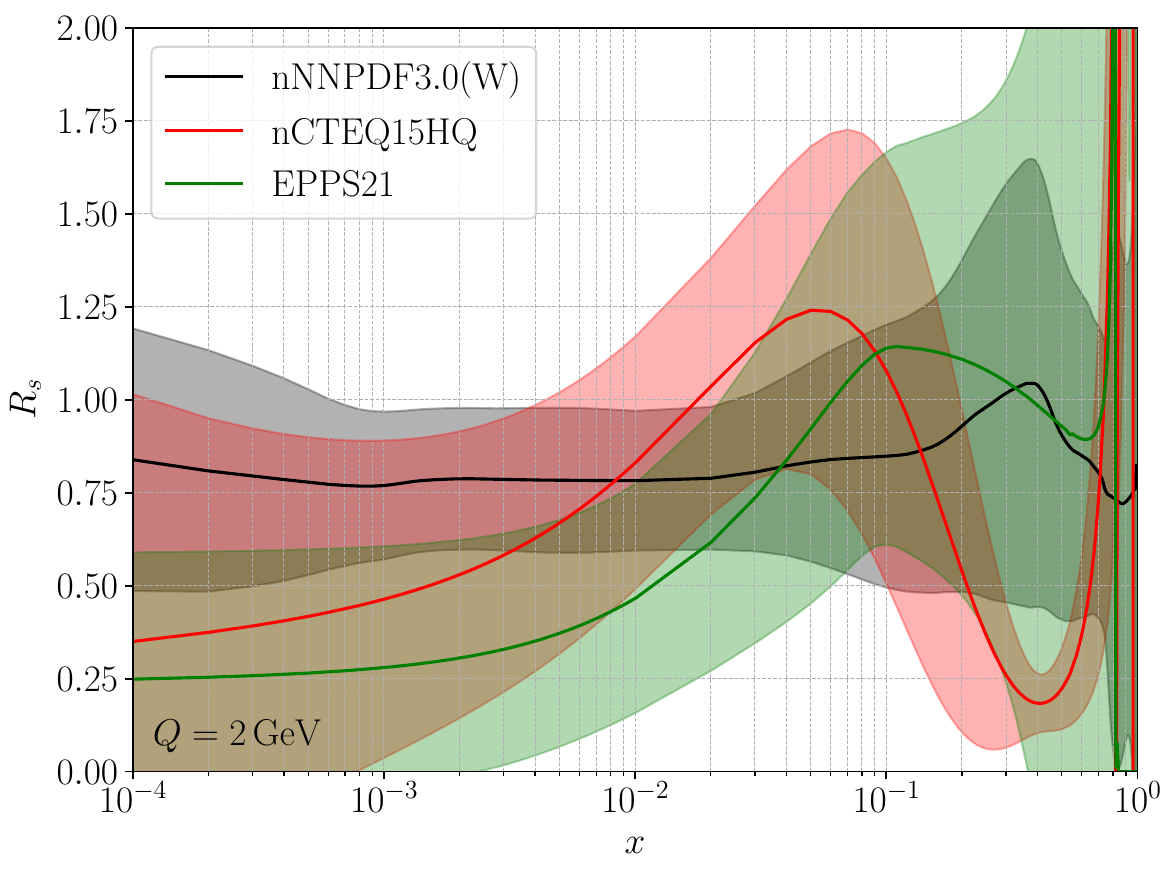}
	\includegraphics[width=0.49\textwidth]{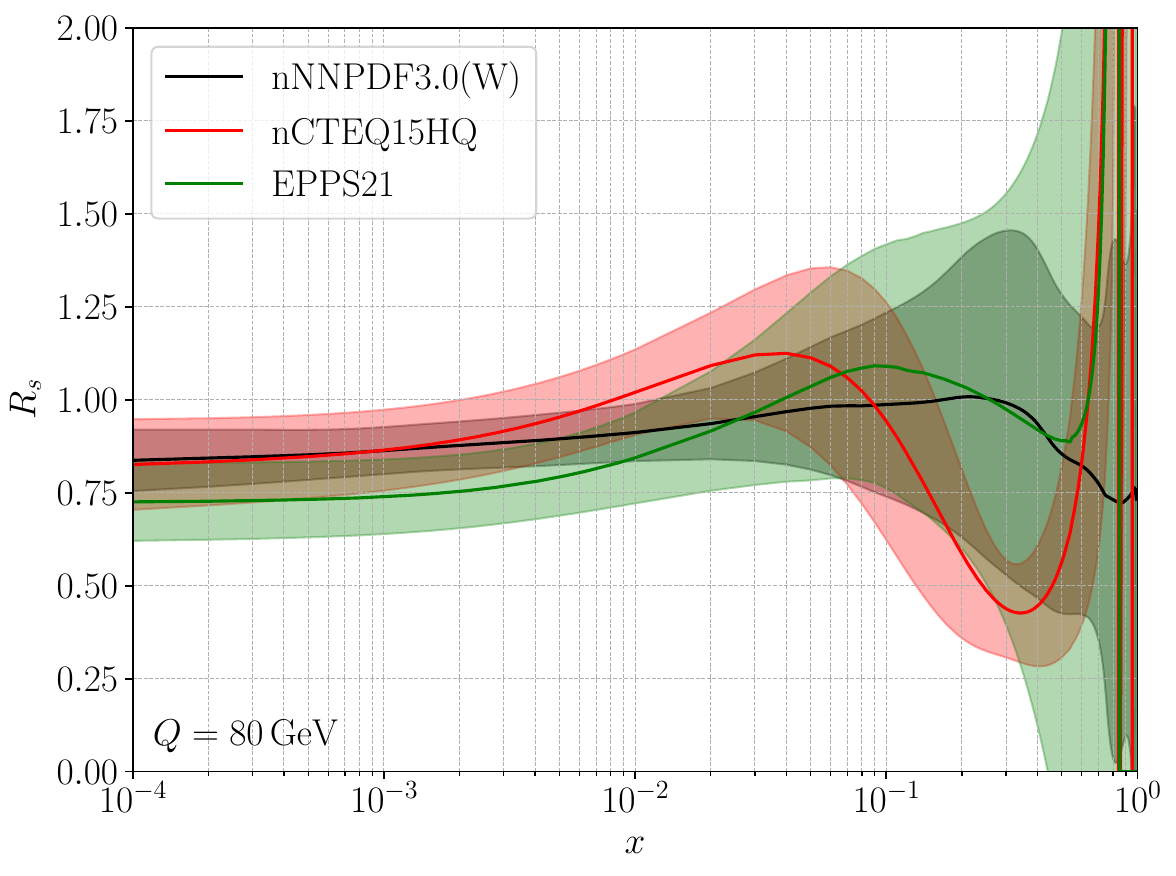} \\
	\includegraphics[width=0.49\textwidth]{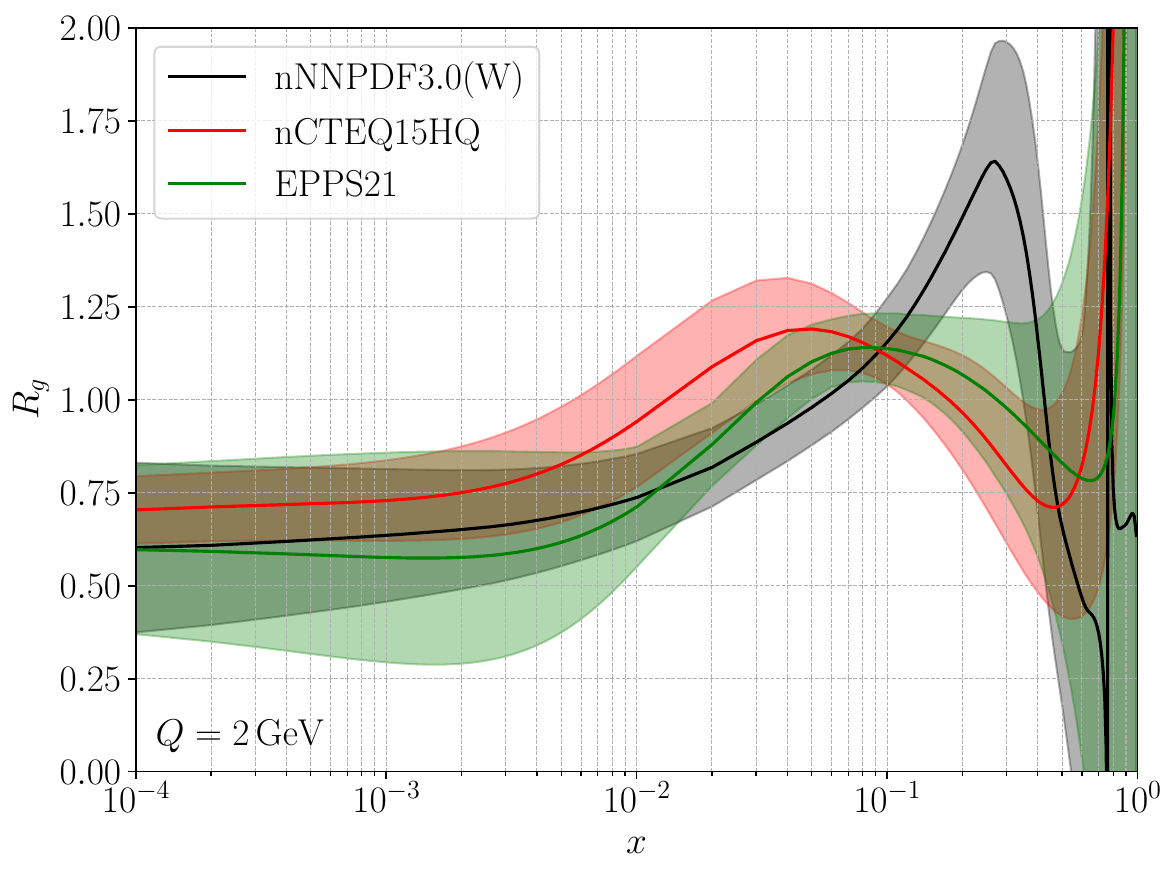}  
	\includegraphics[width=0.49\textwidth]{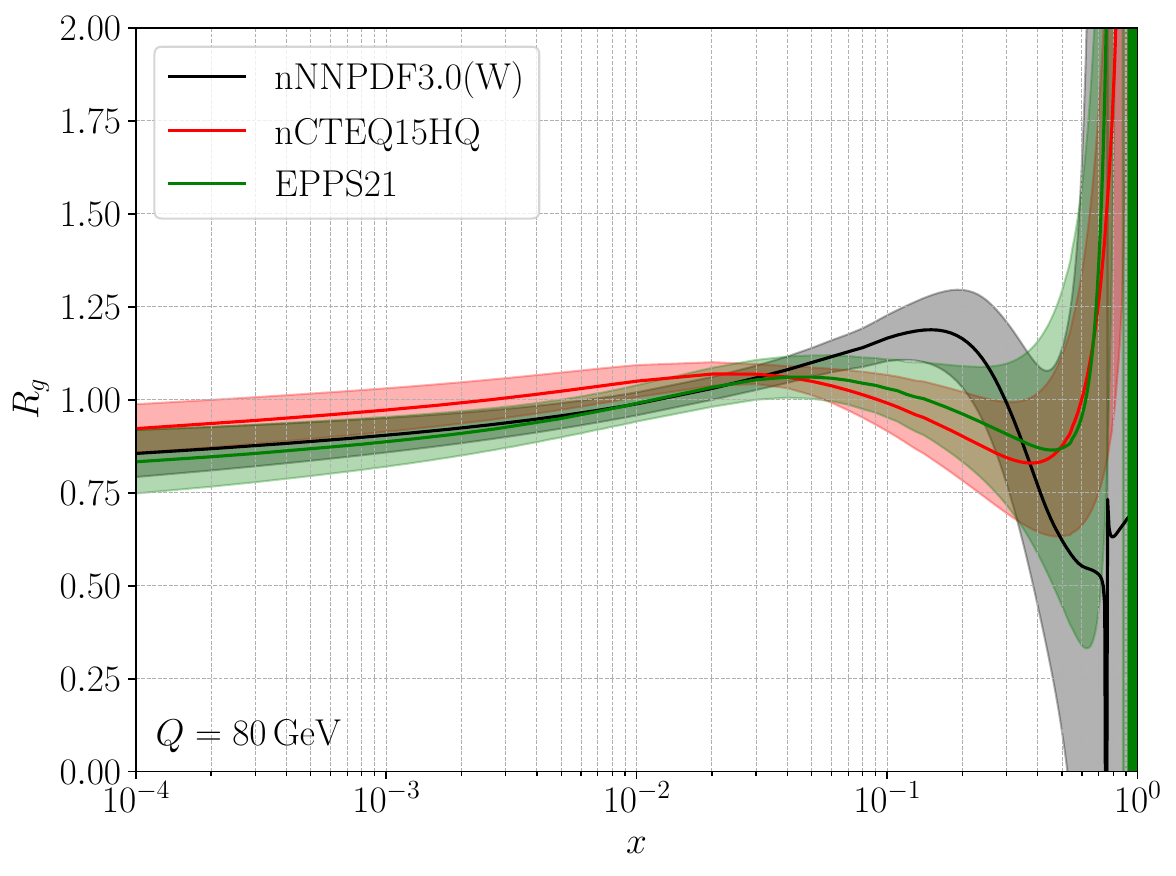}
    
			\end{tabular}
\caption{ Ratio between predictions with and without nuclear effects for the PDFs of the tungsten up (top), strange (center), and gluon (bottom) partons, derived considering the parameterizations EPPS21 \cite{Eskola:2021nhw}, nNNPDF 3.0(W) \cite{AbdulKhalek:2022fyi}, and nCTEQ15HQ \cite{Duwentaster:2022kpv}. Results for two values of the boson virtuality: $Q=2$~GeV (left) and $Q=80$~GeV (right). The uncertainty bands correspond to confidence levels of 68\%. }
\label{fig:ratios_pdfs_nuclear}
\end{figure}

Initially, to illustrate the magnitude of nuclear effects on the distributions of up quarks, strange quarks, and gluons, predicted by the different parameterizations used in our calculations, we present in Figure \ref{fig:ratios_pdfs_nuclear} the results for the nuclear ratio $R_i = f_i^A / (A \times f_i^p)$ ($i = u, \,s,\,g$) considering two values for the virtuality $Q$. We observe that the impact of nuclear effects decreases for larger values of $Q$. Furthermore, for the up quark (left panels), the predictions are similar for $x \approx 0.1$, but become distinct for smaller values of $x$, with the nCTEQ15HQ (nNNPDF 3.0(W)) parameterization predicting a greater (smaller) amount of shadowing. In contrast, for the gluon case (right panels), the central predictions are similar for small values of $x$, but differ considerably in the magnitude and position of the maximum anti-shadowing. Finally, for the strange distribution case (central panels), we observe that the predictions of the different parameterizations are quite distinct, especially for low values of $Q$.

\begin{table}[t]
\centering
\renewcommand{\arraystretch}{1.5}
\begin{tabularx}{\textwidth}{XXcc}
\toprule
\hline
\multicolumn{3}{c}{\bf Lepton DIS events -- Inclusive case} \\
\hline
\midrule
{\bf PDF }                         & {\bf Process}		                                & {\bf Event number}\\
\hline	
\toprule   
\multirow{2}{*}{CT18ANLO}       &  $\mu^\pm + W \rightarrow \mu^\pm + X$        & 1.75$\times10^{5}$ (2.60$\times10^{7}$)         \\
                                &  $\nu_l + W \rightarrow l^\pm + X$            & 5.66$\times10^{3}$ (7.67$\times10^{5}$)         \\
\hline
\midrule
\multirow{2}{*}{nCTEQ15HQ}      &  $\mu^\pm + W \rightarrow \mu^\pm + X$        & 1.54$\times10^{5}$ (2.29$\times10^{7}$)         \\
                                &  $\nu_l + W \rightarrow l^\pm + X$            & 5.66$\times10^{3}$ (7.68$\times10^{5}$)         \\
\hline
\midrule
\multirow{2}{*}{EPPS21}         &  $\mu^\pm + W \rightarrow \mu^\pm + X$        & 1.60$\times10^{5}$ (2.38$\times10^{7}$)         \\
                                &  $\nu_l + W \rightarrow l^\pm + X$            & 5.55$\times10^{3}$ (7.52$\times10^{5}$)         \\
\hline   
\midrule
\multirow{2}{*}{nNNPDF3.0(p)}   &  $\mu^\pm + W \rightarrow \mu^\pm + X$        & 1.77$\times10^{5}$ (2.64$\times10^{7}$)         \\
                                &  $\nu_l + W \rightarrow l^\pm + X$            & 5.81$\times10^{3}$ (7.88$\times10^{5}$)         \\
\hline
\midrule
\multirow{2}{*}{nNNPDF3.0(W)}   &  $\mu^\pm + W \rightarrow \mu^\pm + X$        & 1.71$\times10^{5}$ (2.55$\times10^{7}$)         \\
                                &  $\nu_l + W \rightarrow l^\pm + X$            & 6.12$\times10^{3}$ (8.29$\times10^{5}$)         \\
\hline
\bottomrule
\end{tabularx}
\vspace{0.3cm}
\caption{Predictions for the number of inclusive events in $\mu W$ and $\nu W$ interactions at the FASER$\nu$ (FASER$\nu$2), derived considering different parameterizations for the nPDFs and assuming an integrated luminosity of $\mathcal{L}_{\rm pp}=250$ fb$^{-1}$ (3 ab$^{-1}$).
}
\label{table:Nevents2}
\end{table}

\begin{table}[t]
\centering
\renewcommand{\arraystretch}{1.5}
\begin{tabularx}{\textwidth}{XXcc}
\toprule
\hline
\multicolumn{3}{c}{\bf Lepton DIS events  -- Charm detected case}\\
\hline
\midrule
{\bf PDF }                         & {\bf Process}		                                & {\bf Event number}\\
\hline	
\toprule   
\multirow{2}{*}{CT18ANLO}       &  $\mu^\pm + W \rightarrow \mu^\pm + X + c(\bar{c})$        & 9.15$\times10^{3}$ (1.36$\times10^{6}$)         \\
                                &  $\nu_l + W \rightarrow l^\pm + X + c(\bar{c})$            & 281 (3.82$\times10^{4}$)                        \\
\hline
\midrule
\multirow{2}{*}{nCTEQ15HQ}      &  $\mu^\pm + W \rightarrow \mu^\pm + X + c(\bar{c})$        & 9.91$\times10^{3}$ (1.47$\times10^{6}$)         \\
                                &  $\nu_l + W \rightarrow l^\pm + X + c(\bar{c})$            & 263 (3.56$\times10^{4}$)                        \\
\hline
\midrule
\multirow{2}{*}{EPPS21}         &  $\mu^\pm + W \rightarrow \mu^\pm + X + c(\bar{c})$        & 9.37$\times10^{3}$ (1.39$\times10^{6}$)         \\
                                &  $\nu_l + W \rightarrow l^\pm + X + c(\bar{c})$            & 263 (3.57$\times10^{4}$)                        \\
\hline   
\midrule
\multirow{2}{*}{nNNPDF3.0(p)}   &  $\mu^\pm + W \rightarrow \mu^\pm + X + c(\bar{c})$        & 6.41$\times10^{3}$ (9.53$\times10^{5}$)         \\
                                &  $\nu_l + W \rightarrow l^\pm + X + c(\bar{c})$            & 411 (5.58$\times10^{4}$)                        \\
\hline
\midrule
\multirow{2}{*}{nNNPDF3.0(W)}   &  $\mu^\pm + W \rightarrow \mu^\pm + X + c(\bar{c})$        & 7.14$\times10^{3}$ (1.06$\times10^{6}$)         \\
                                &  $\nu_l + W \rightarrow l^\pm + X + c(\bar{c})$            & 381 (5.13$\times10^{4}$)                        \\
\hline
\bottomrule
\end{tabularx}
\vspace{0.3cm}
\caption{Predictions for the number of events with charm in the final state in $\mu W$ and $\nu W$ interactions at the FASER$\nu$ (FASER$\nu$2), derived considering different parameterizations for the nPDFs and assuming an integrated luminosity of $\mathcal{L}_{\rm pp}=250$ fb$^{-1}$ (3 ab$^{-1}$).
}
\label{table:Nevents_charm2}
\end{table}

In Tables \ref{table:Nevents2} and \ref{table:Nevents_charm2} we present our predictions for the number of events in the inclusive and charm-detected cases, respectively, derived considering muon-tungsten and neutrino-tungsten interactions at the FASER$\nu$ and FASER$\nu2$ detectors. The results for the FASER$\nu2$ detector are in parentheses. We observe that the number of events associated with $\nu W$ interactions is almost two orders of magnitude lower compared to the $\mu W$ case. Furthermore, as expected from the analyses performed in the References~\cite{Francener:2025pnr,Cruz-Martinez:2023sdv}, the number of events at FASER$\nu2$ will be a factor $\approx 100$ greater than in FASER$\nu$. Another important aspect is the large number of events with charm detection in $\mu W$ interactions, which are expected to be sensitive to gluon and charm distributions in the target. For the inclusive case, we observed that the CT18ANLO (nCTEQ15HQ) parameterization provides the highest value for the number of events in $\mu W$ ($\nu W$) interactions. In contrast, for events with charm detection, the highest values for the number of events in $\mu W$ and $\nu W$ interactions are generated by the nNNPDF3.0(W) and nNNPDF3.0(p) parameterizations, respectively. These results demonstrate the dependence of the predictions on the nPDF parameterization considered in the calculations and motivate a more detailed analysis of less inclusive observables.

\begin{figure}[H]
	\centering
	\begin{tabular}{ccc}
	\includegraphics[width=0.49\textwidth]{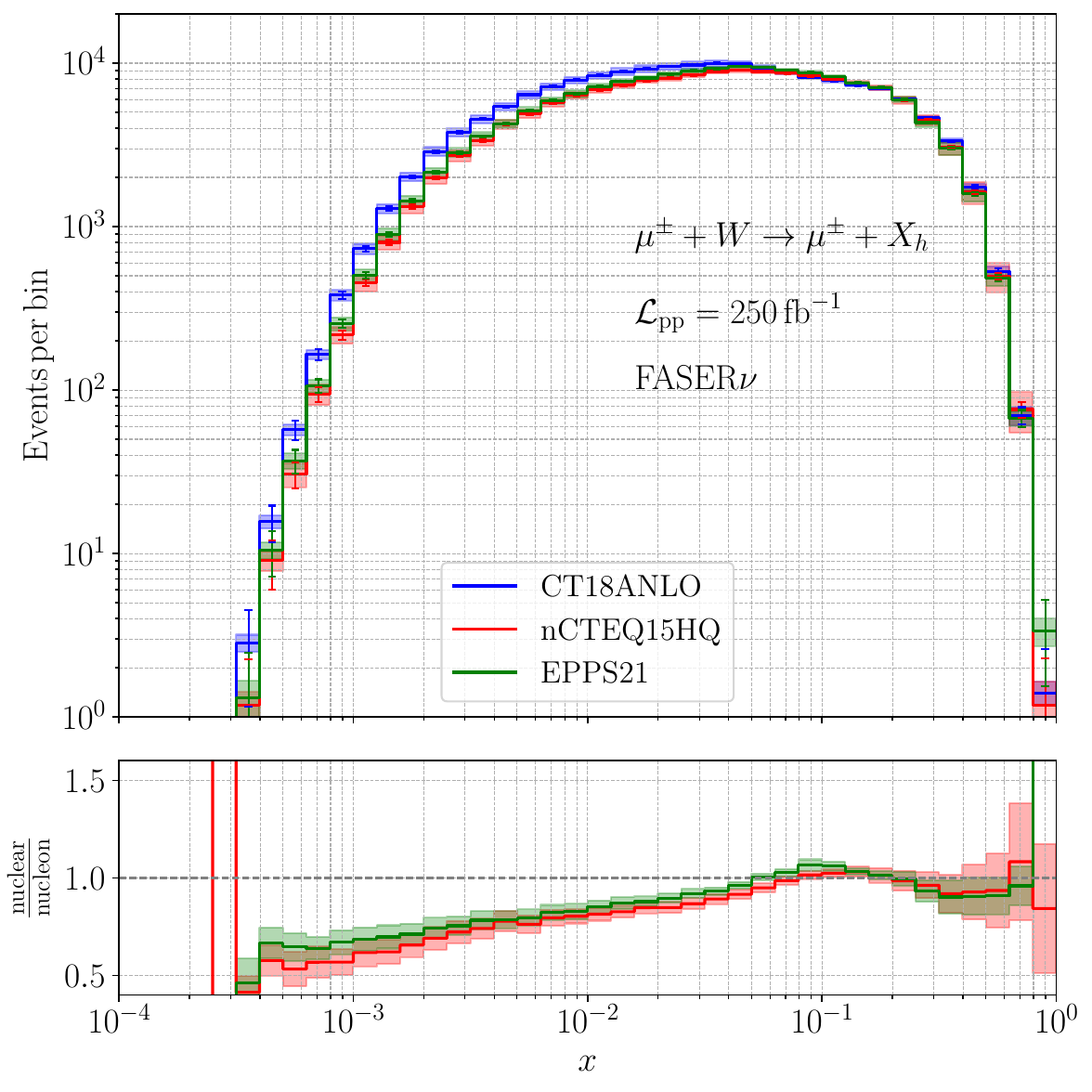}
	\includegraphics[width=0.49\textwidth]{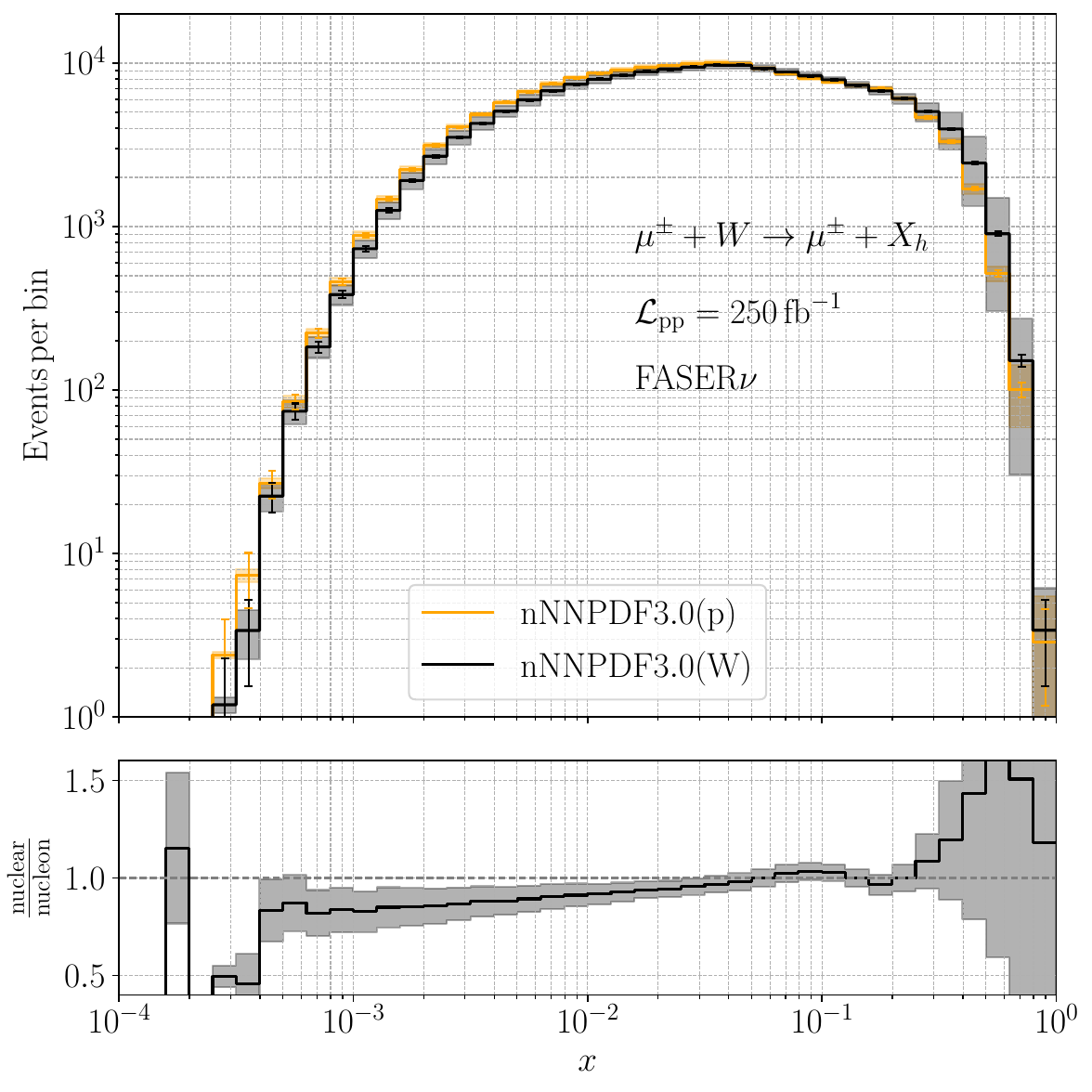} \\
	\includegraphics[width=0.49\textwidth]{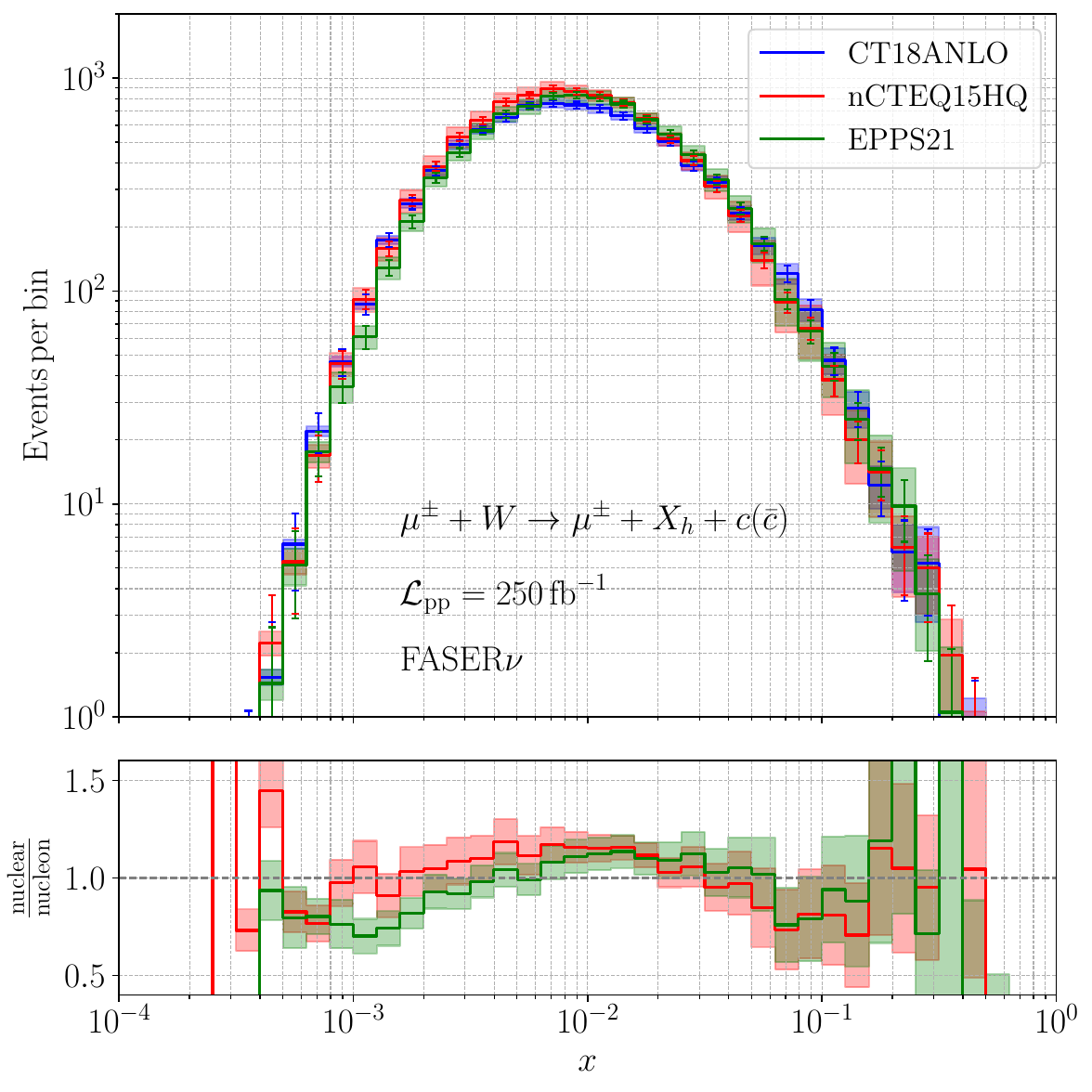}
	\includegraphics[width=0.49\textwidth]{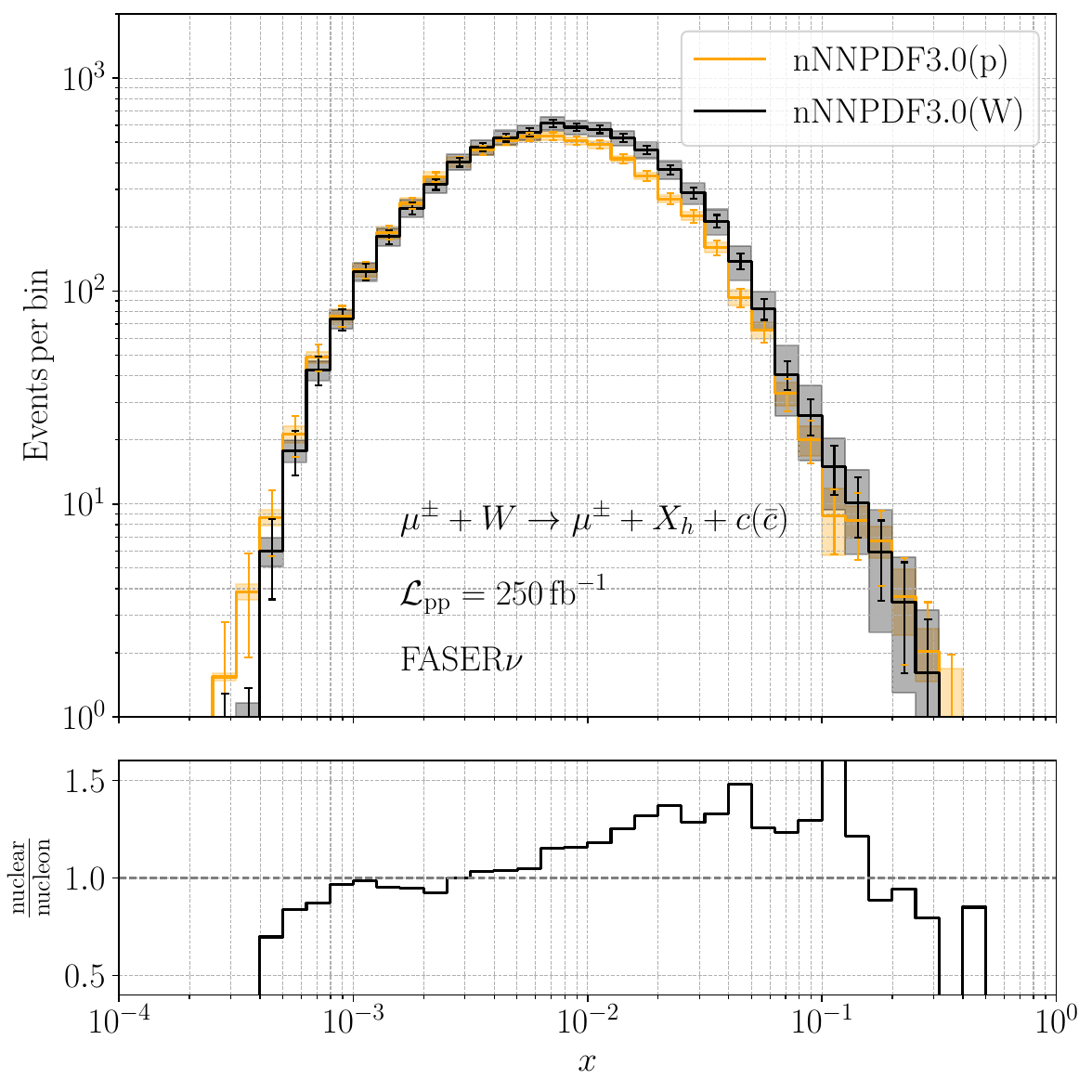}
	\end{tabular}
\caption{ Predictions for the number of muon DIS events at the FASER$\nu$ binned in $x$ for the inclusive case (upper panels) and with an observed charm (lower panels) in the final state. Results derived assuming different parameterizations for the nPDFs. Predictions for the ratio between nuclear and nucleonic results are presented in the lower panels of the plots.}
\label{fig:muonDIS_x_FASERnu}
\end{figure}

In Figure~\ref{fig:muonDIS_x_FASERnu} we present our predictions for muon plus antimuon DIS events binned in $x$ for the FASER$\nu$ detector. In the upper panel, we present our predictions for inclusive events, while in the lower panel we have the results for events with at least one charm hadron observed in the final state. The ratio between the predictions derived with and without nuclear effects is also presented at the bottom of each panel. As in our analysis we are considering PDFs based on different groups of parameterizations, we will present two plots for each event class. In the left panels, we present the predictions based on the CTEQ parameterization, that is, the results associated with the CT18ANLO parameterization, which disregards nuclear effects, and the nCTEQ15HQ and EPPS21 predictions, obtained by modifying the CT18ANLO parameterization to include nuclear effects. On the other hand, in the right panels, we present the results for the nNNPDF parameterization, where we compare the predictions of nNNPDF3.0 with those derived using its baseline for a free nucleon (NNPDF3.0). The uncertainty bands represent the uncertainties of the PDF with a 68\% confidence level, while the bars represent the expected statistical uncertainty, also with a 68\% confidence level, constructed considering Gaussian statistics. The events presented here are subsequent to the acceptance cuts discussed earlier.

\begin{figure}[H]
	\centering
	\begin{tabular}{ccc}
	\includegraphics[width=0.49\textwidth]{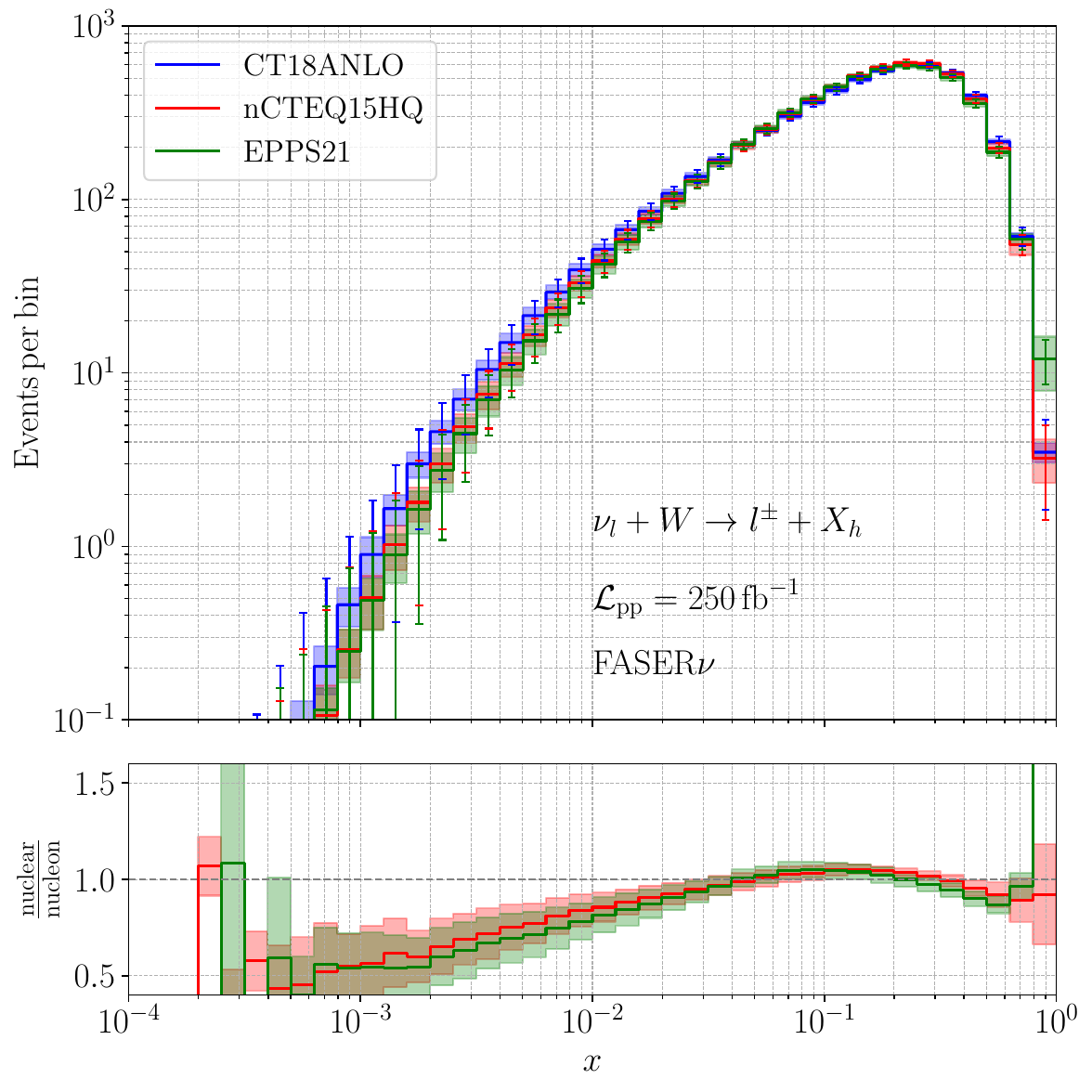}
	\includegraphics[width=0.49\textwidth]{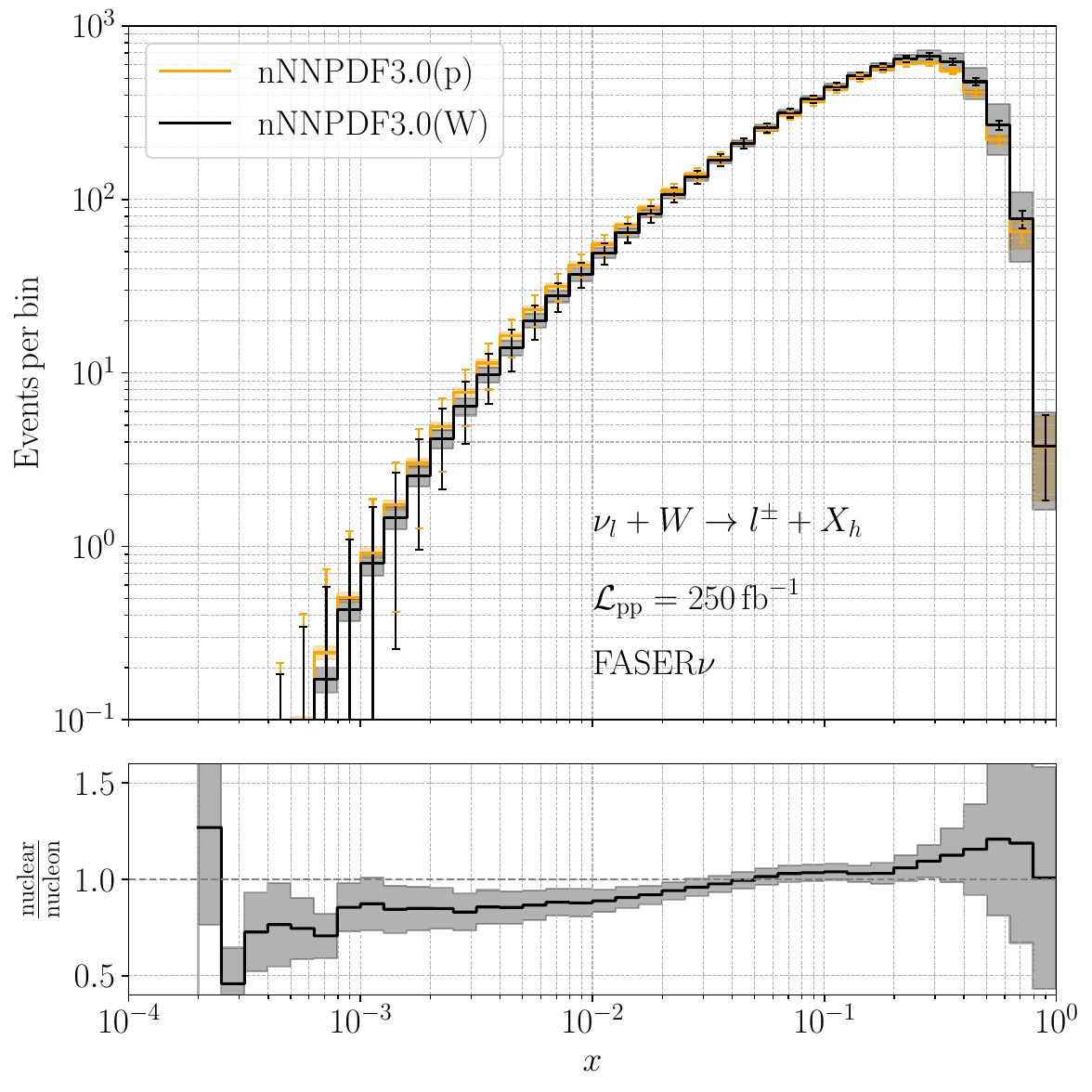} \\
	\includegraphics[width=0.49\textwidth]{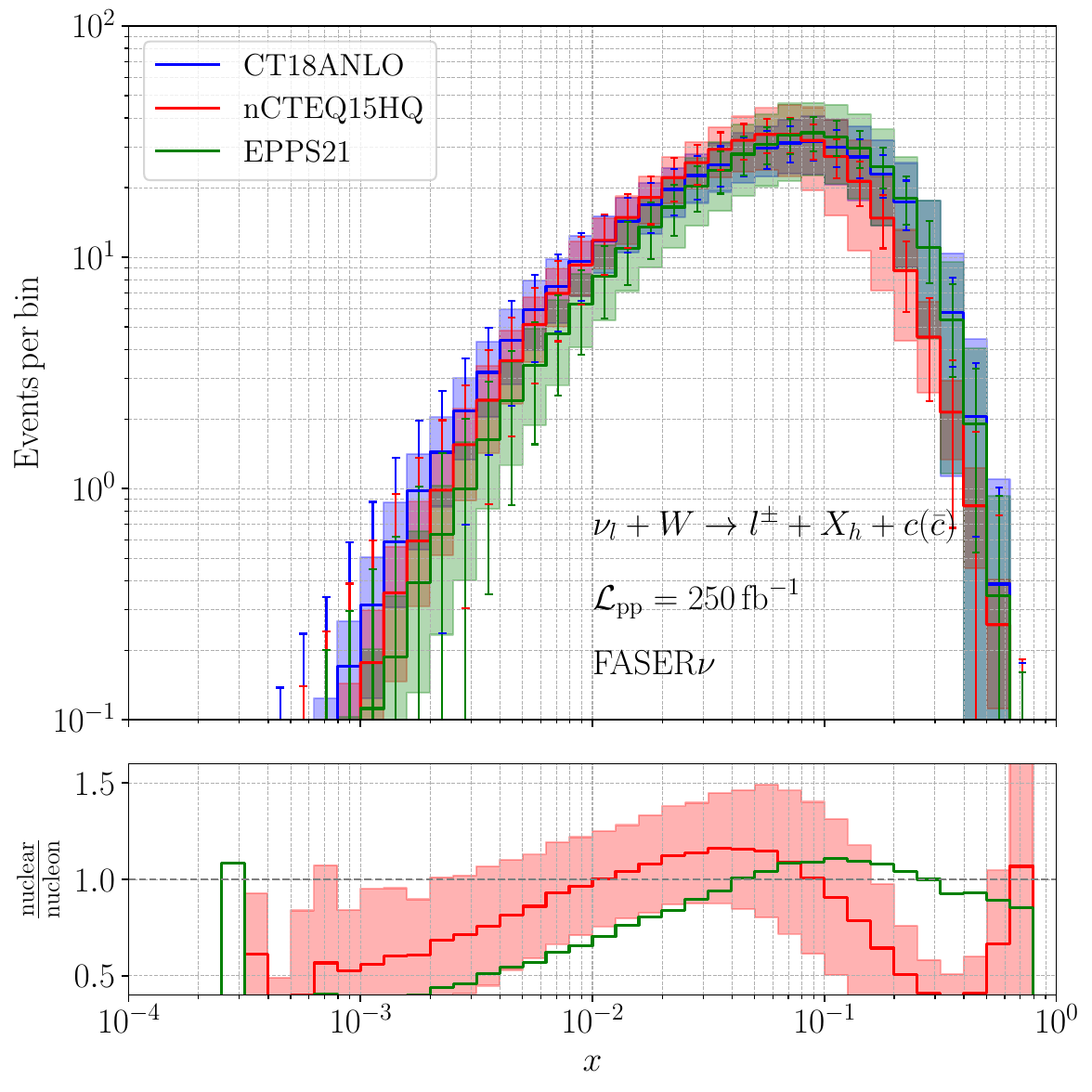}
	\includegraphics[width=0.49\textwidth]{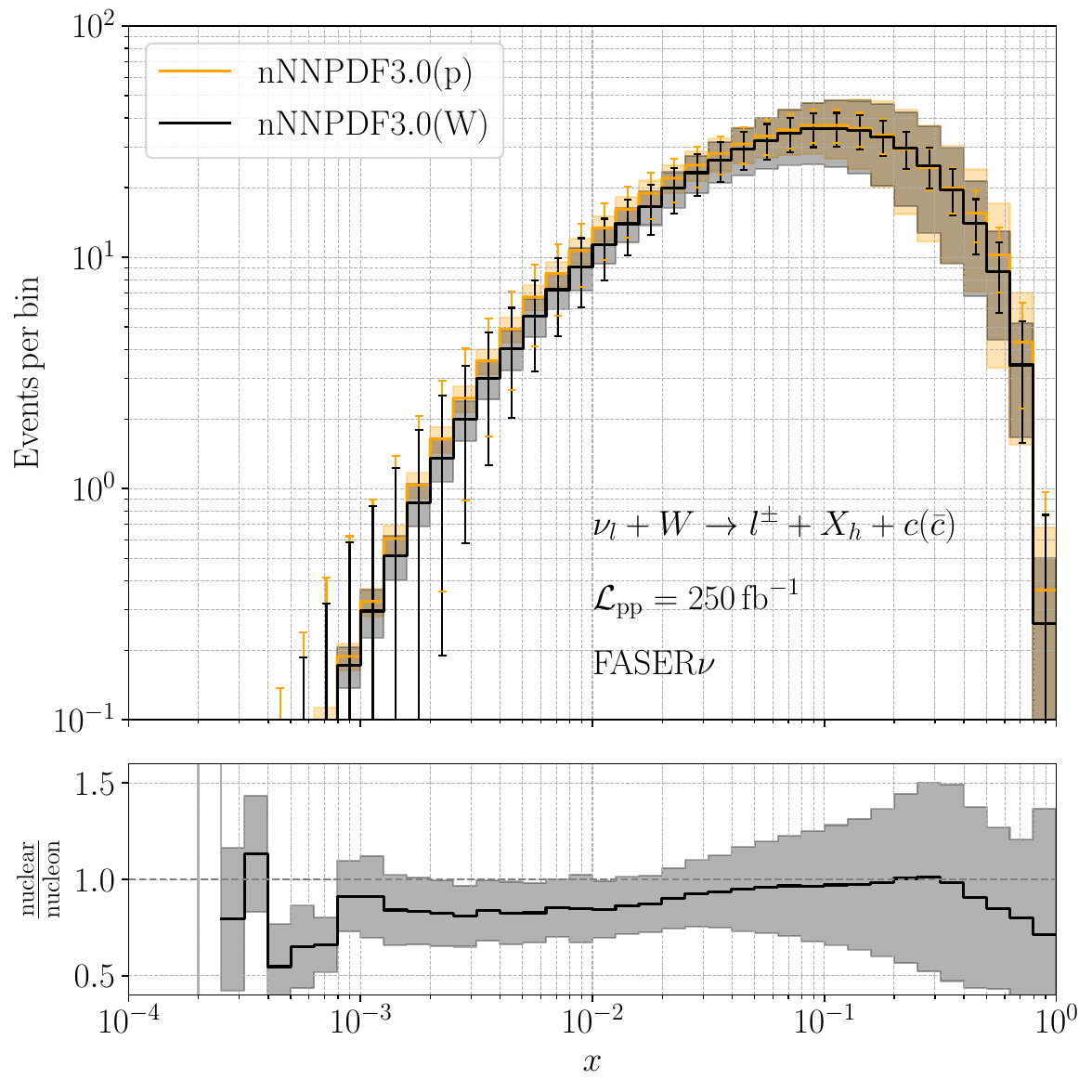}
    
			\end{tabular}
\caption{Predictions for the number of neutrino DIS events at the FASER$\nu$ binned in $x$ in the inclusive case (upper panels) and with a charm detected in the final state (lower panels). Results derived assuming different parameterizations for the nPDFs. Predictions for the ratio between nuclear and nucleonic results are presented in the lower panels of the plots. }
\label{fig:neutrinoDIS_x_FASERnu}
\end{figure}

The results presented in the upper panels of Figure \ref{fig:muonDIS_x_FASERnu} show that predictions with and without nuclear effects can be distinguished depending on the region in $x$ and the models compared. For inclusive events and for small values of $x$, where we expect the shadowing effect, we observe a large difference between the predictions of CT18ANLO and nCTEQ15HQ/EPPS21. In contrast, this difference does not occur for nNNPDF3.0, where the impact of shadowing is smaller. For large values of $x$, we observe a difference between the predictions for the expected number of events with and without nuclear effects caused by the EMC effect, with similar results obtained using nCTEQ15HQ and EPPS21. The magnitude of anti-shadowing for $x \approx 0.1$ depends on the nPDF considered in the calculation, as well as the position in $x$ of its maximum.

In the lower panels of Figure~\ref{fig:muonDIS_x_FASERnu}, we present our results for events with a charm detected in the final state. As discussed previously, these events are generated, at leading order, by photon-gluon interactions and are strongly sensitive to the nuclear gluon distribution. We observe that, compared to the inclusive case, the number of events is reduced by an order of magnitude. Furthermore, the maximum number of events occurs for a smaller value of $x$. We find that the EPPS21 parameterization predicts a smaller number of events for small values of $x$ compared to the nCTEQ15HQ result. In contrast, these two parameterizations imply a similar number of events for $x \gtrsim 10^{-2}$. On the other hand, for the NNPDF case (lower right panel), the impact of nuclear effects on small $x$ is small, but the number of events is increased by these effects for $x$ in the range of $10^{-2}$ to $10^{-1}$.

We now present our predictions for the number of events associated with charged current neutrino-tungsten interactions, considering the sum of neutrino and antineutrino fluxes, as well as electron and muon flavors. In the upper panels of Figure \ref{fig:neutrinoDIS_x_FASERnu}, we present the results for the inclusive case. We observe that the expected number of events per bin is smaller when compared to muon-induced events. Furthermore, the maximum number of events occurs at higher values of $x$, which is associated with the fact that neutrino interactions (DIS) are dominated by higher values of $Q^2$. This dominance also implies a smaller impact of nuclear effects for $x \gtrsim 10^{-2}$. Consequently, the analysis of these events can be useful for constraining the description of nucleon parton distributions. The results for events with charm detected (lower panels of Figure \ref{fig:neutrinoDIS_x_FASERnu}), indicate that the predictions have a large uncertainty, which is mainly associated with the current uncertainty in the strange quark PDF (see Figures \ref{fig:pdfs_nuclear} and \ref{fig:ratios_pdfs_nuclear}). Furthermore, we predict few events per bin ($\mathcal{O}(30)$ for $x \approx 0.1$).

\begin{figure}[H]
	\centering
	\begin{tabular}{ccc}
	\includegraphics[width=0.49\textwidth]{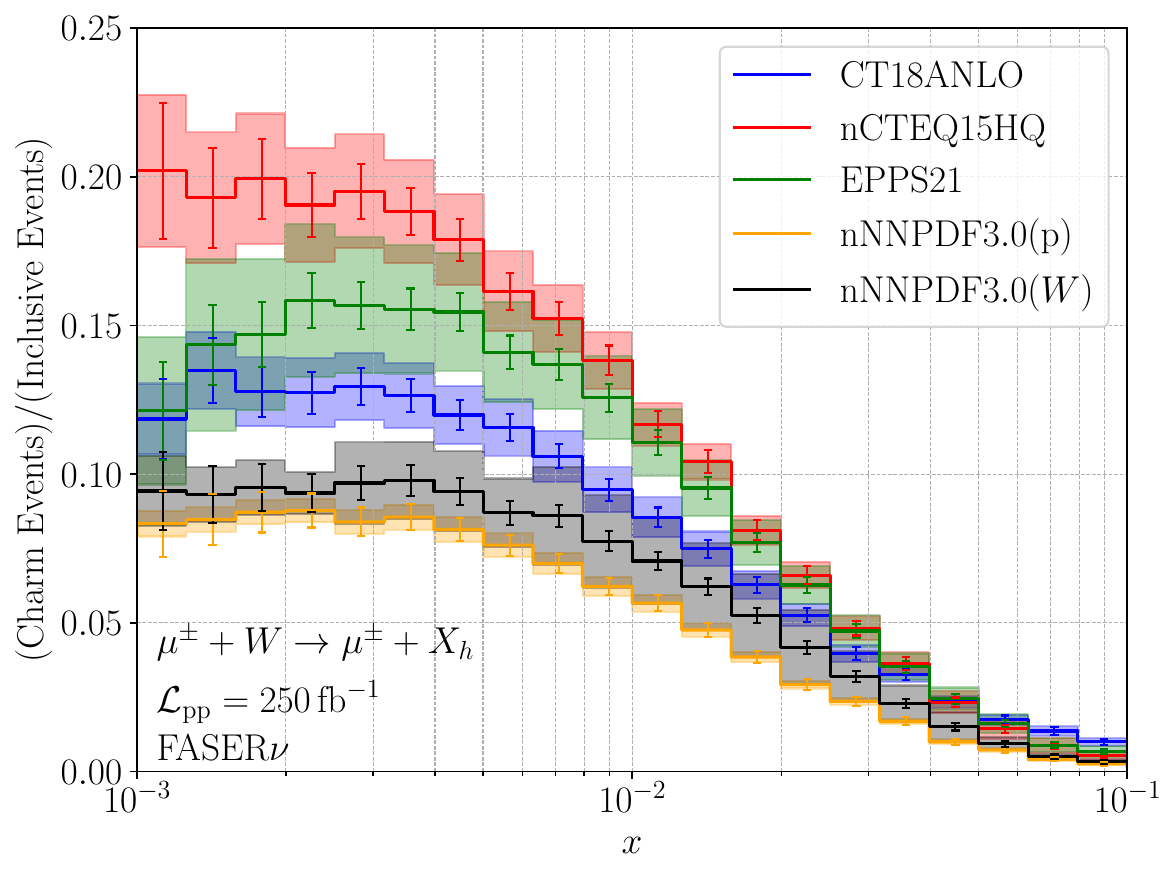}
	\includegraphics[width=0.49\textwidth]{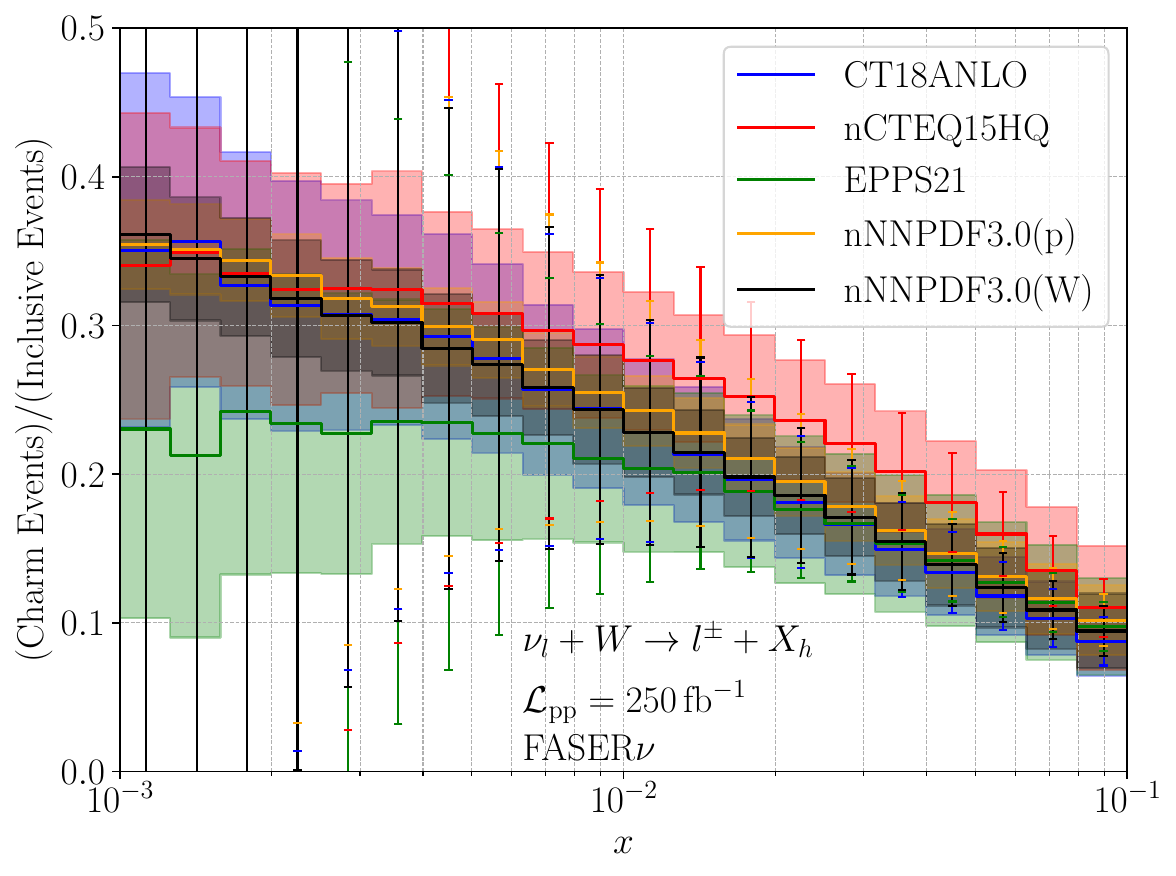}

			\end{tabular}
\caption{ Ratio between events with charming hadrons and inclusive events for muon (left) and neutrino (right) DIS at the FASER$\nu$. Results derived assuming different parameterizations for the nPDFs.}
\label{fig:Ratio_charm_total_FASERnu}
\end{figure}

Considering that the parameterizations predict distinct amounts of nuclear effects in the different parton distributions (see Figure \ref{fig:ratios_pdfs_nuclear}), an alternative to discriminate between these models is to consider the ratio between the cross sections whose behaviors are determined by distinct PDFs. Here, we propose the analysis of the ratio between the rates for events with charm detected and inclusive events. We observe that inclusive events are mainly determined by valence quarks and sea quarks, while events with charm detection are sensitive to the charm (strange) distribution in the case of muon-induced (neutrino) interactions. Our results for the ratio are presented in Figure \ref{fig:Ratio_charm_total_FASERnu}. We observed that, for $\mu W$ events (left panel), the magnitude of the ratio for $x \lesssim 10^{-2}$ depends on the nPDF used as input in the calculation, with nCTEQ15HQ predicting the largest value. In contrast, the results obtained using the NNPDF framework are smaller by a factor of approximately 2. Another important aspect is that we expect a small statistical uncertainty in the predictions, represented by the vertical lines in the distinct curves. These results indicate that a future experimental analysis of this ratio may be useful to improve our understanding of nuclear effects at small values of $x$. On the other hand, in the results for $\nu W$ DIS (right panel), we observed that the value of the ratio increases by a factor of $\approx 2$, with the EPPS21 parameterization (nCTEQ15HQ) predicting the smallest (largest) value for $x \lesssim 10^{-2}$ ($x \gtrsim 10^{-2}$). However, in this case, we expect a large statistical uncertainty at FASER$\nu$, which implies that it is not possible to reach a more robust conclusion about the most appropriate description of nuclear effects.

\subsection{Results for the FASER$\nu 2$}
\label{sec_MuonDIS:nucResultsFASERnu2}

The results presented in the previous subsection strongly motivate the analysis of how an improved experimental scenario can enhance the constraint on nuclear effects in lepton-ion interaction at the LHC. As discussed in Subsection \ref{subsec_MuonDIS:charmHL-LHC}, the FASER$\nu 2$ detector was proposed to be installed in the Forward Physics Facility and operate during the HL-LHC era. It is expected that the larger size of this future detector and the enormous increase in luminosity will increase event rates and decrease statistical uncertainties. This expectation is confirmed by the results we will show below.

\begin{figure}[H]
	\centering
	\begin{tabular}{ccc}
	\includegraphics[width=0.49\textwidth]{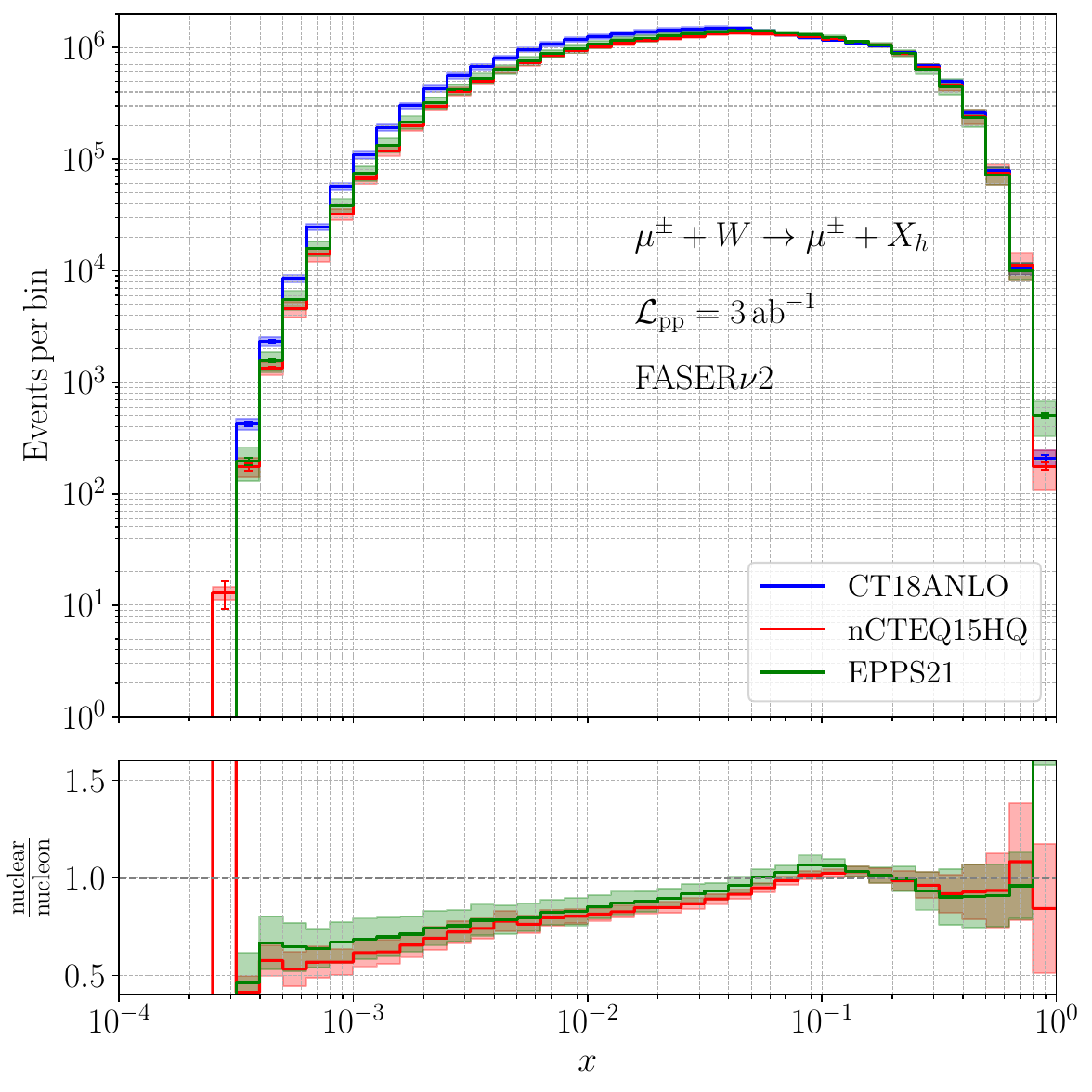}
	\includegraphics[width=0.49\textwidth]{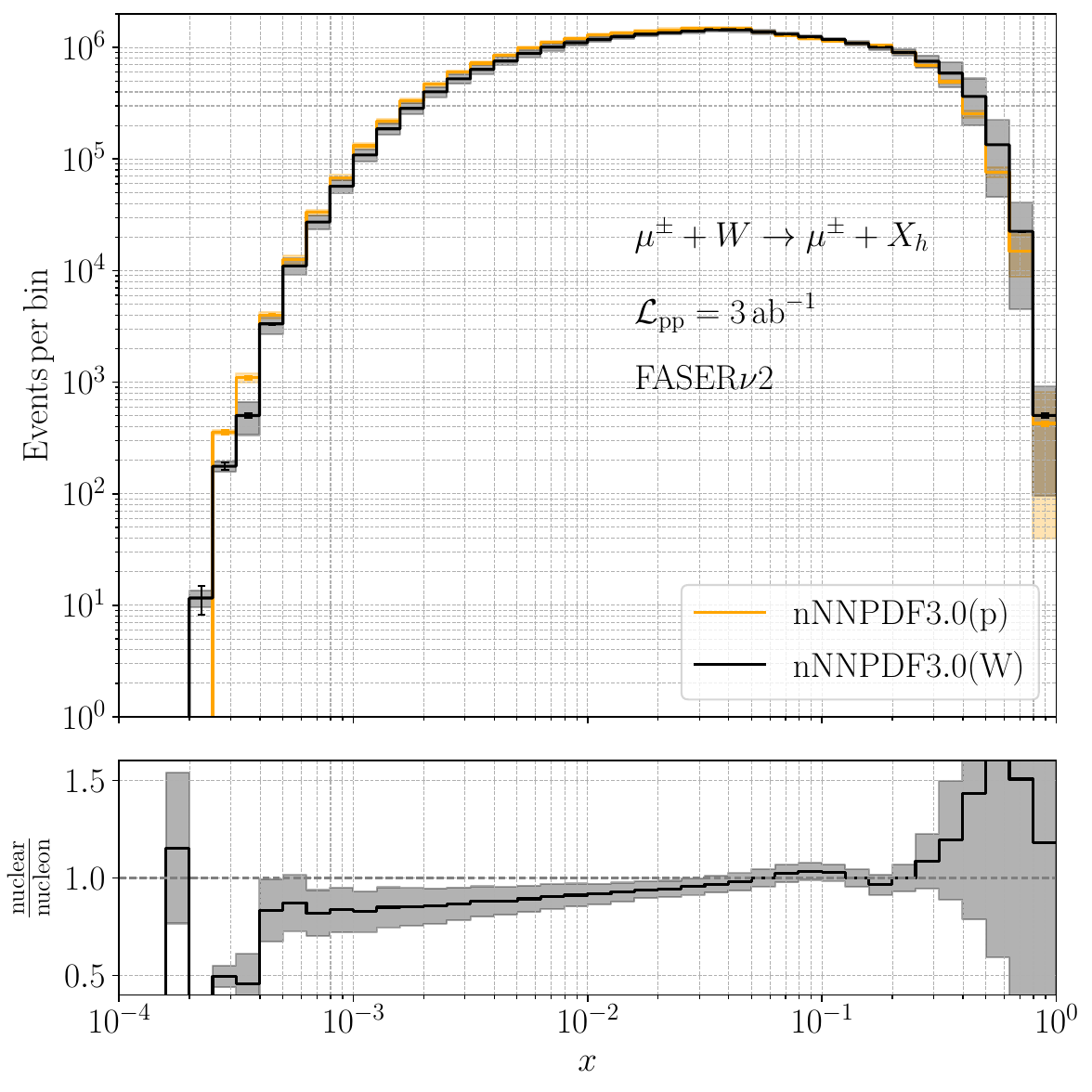} \\
	\includegraphics[width=0.49\textwidth]{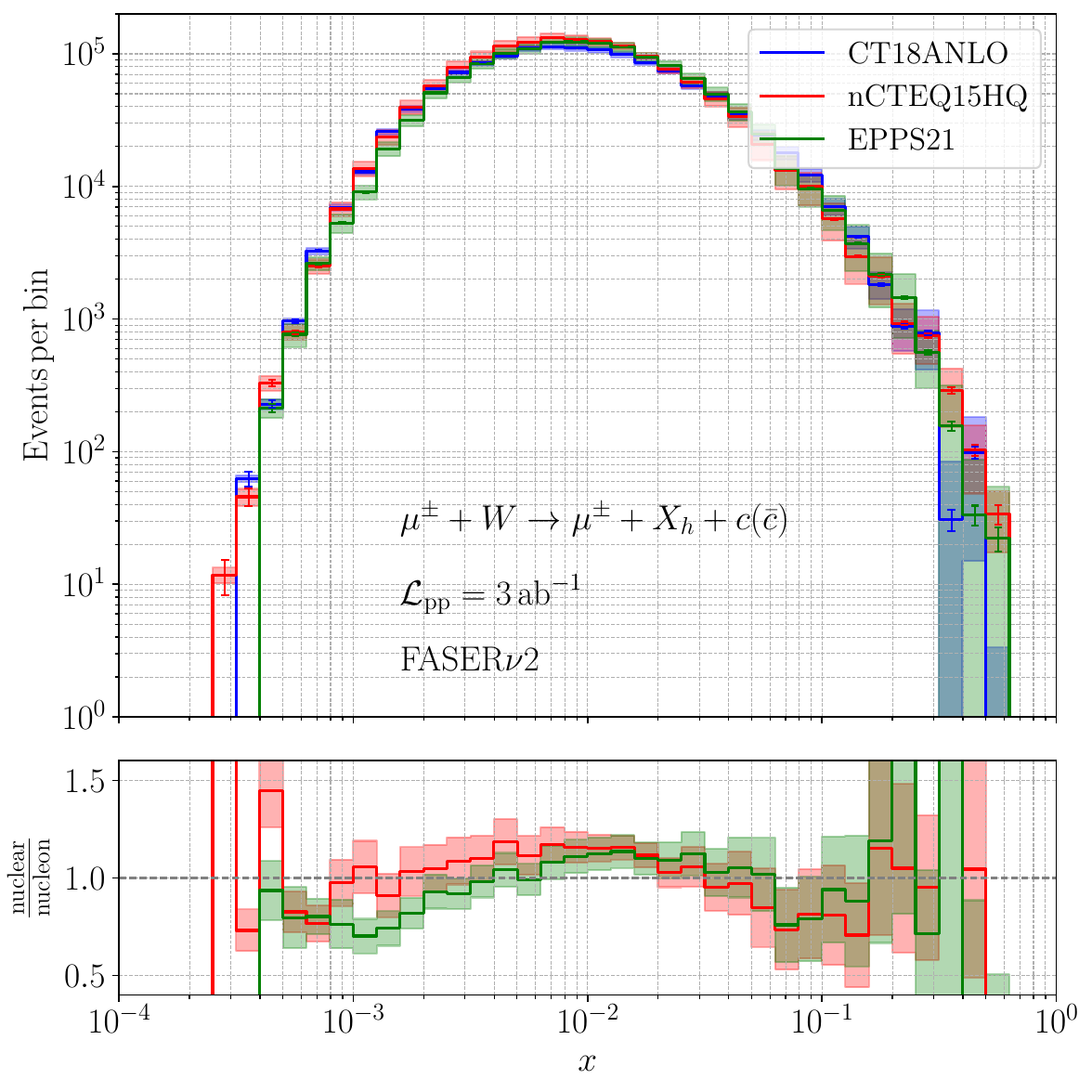}
	\includegraphics[width=0.49\textwidth]{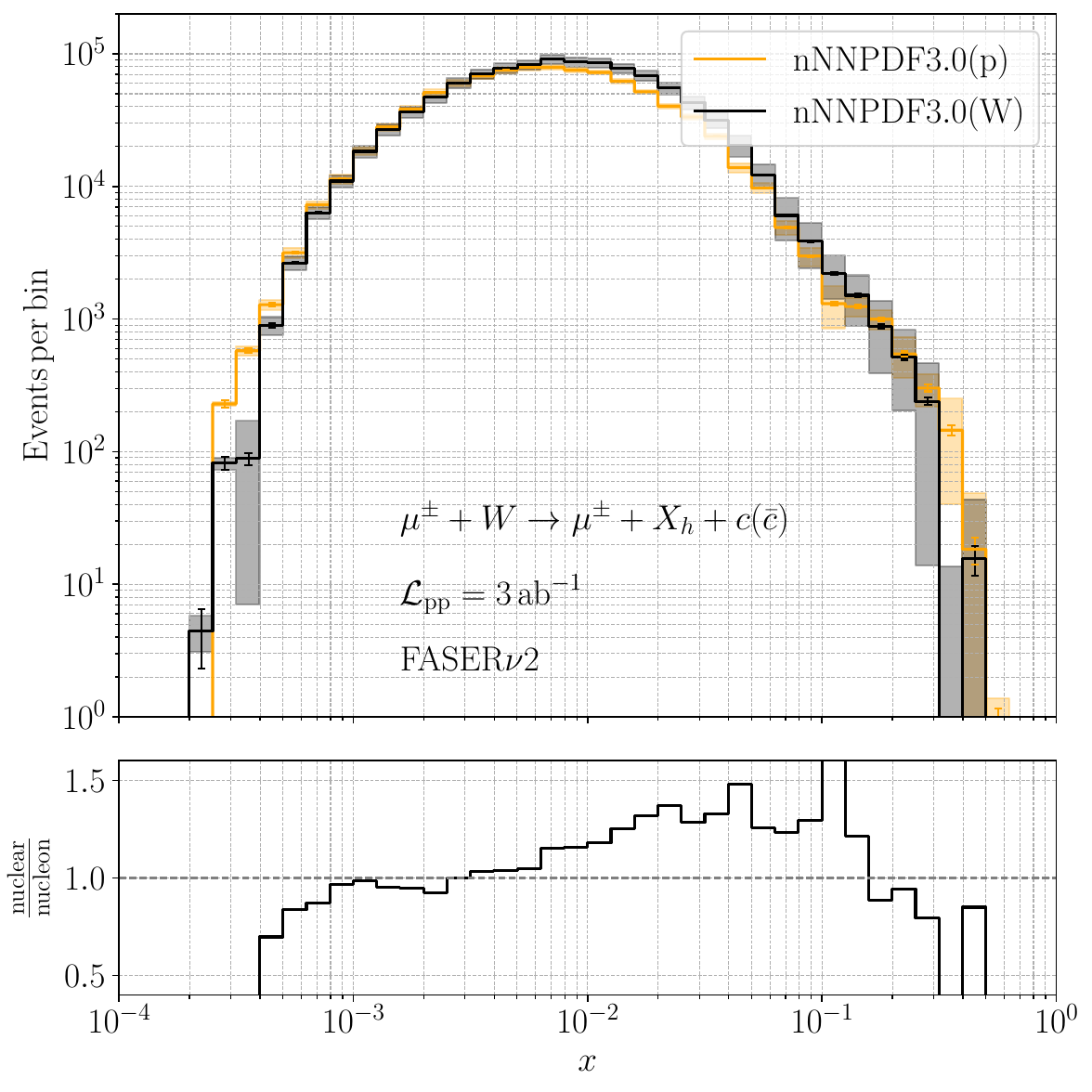}
    
			\end{tabular}
\caption{Predictions for the number of muon DIS events at the FASER$\nu2$ binned in $x$ in the inclusive case (upper panels) and with a charm detected in the final state (lower panels). Results derived assuming different parameterizations for the nPDFs. Predictions for the ratio between nuclear and nucleonic results are presented in the lower panels of the plots.}
\label{fig:muonDIS_x_FASERnu2HL}
\end{figure}

\begin{figure}[H]
	\centering
	\begin{tabular}{ccc}
	\includegraphics[width=0.49\textwidth]{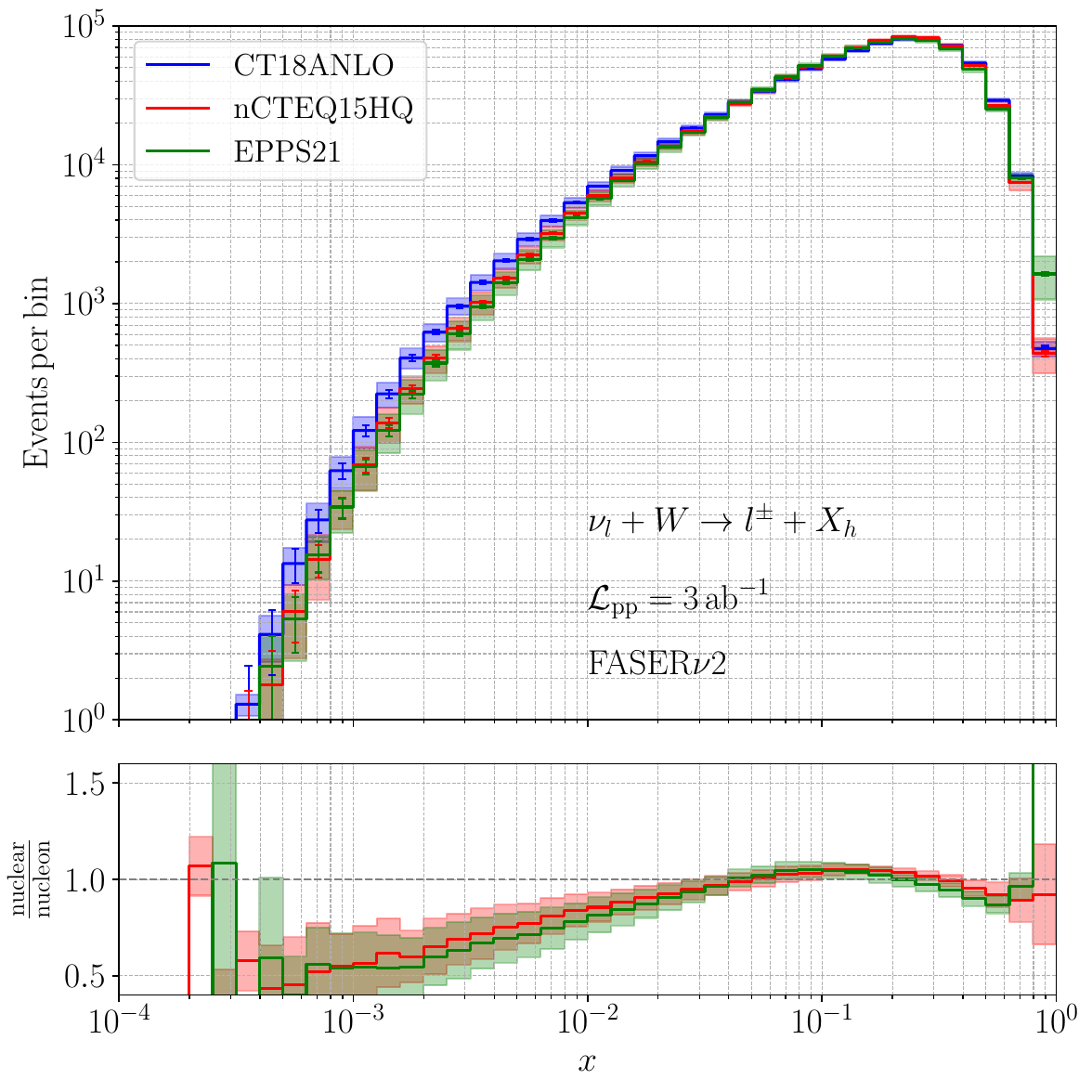}
	\includegraphics[width=0.49\textwidth]{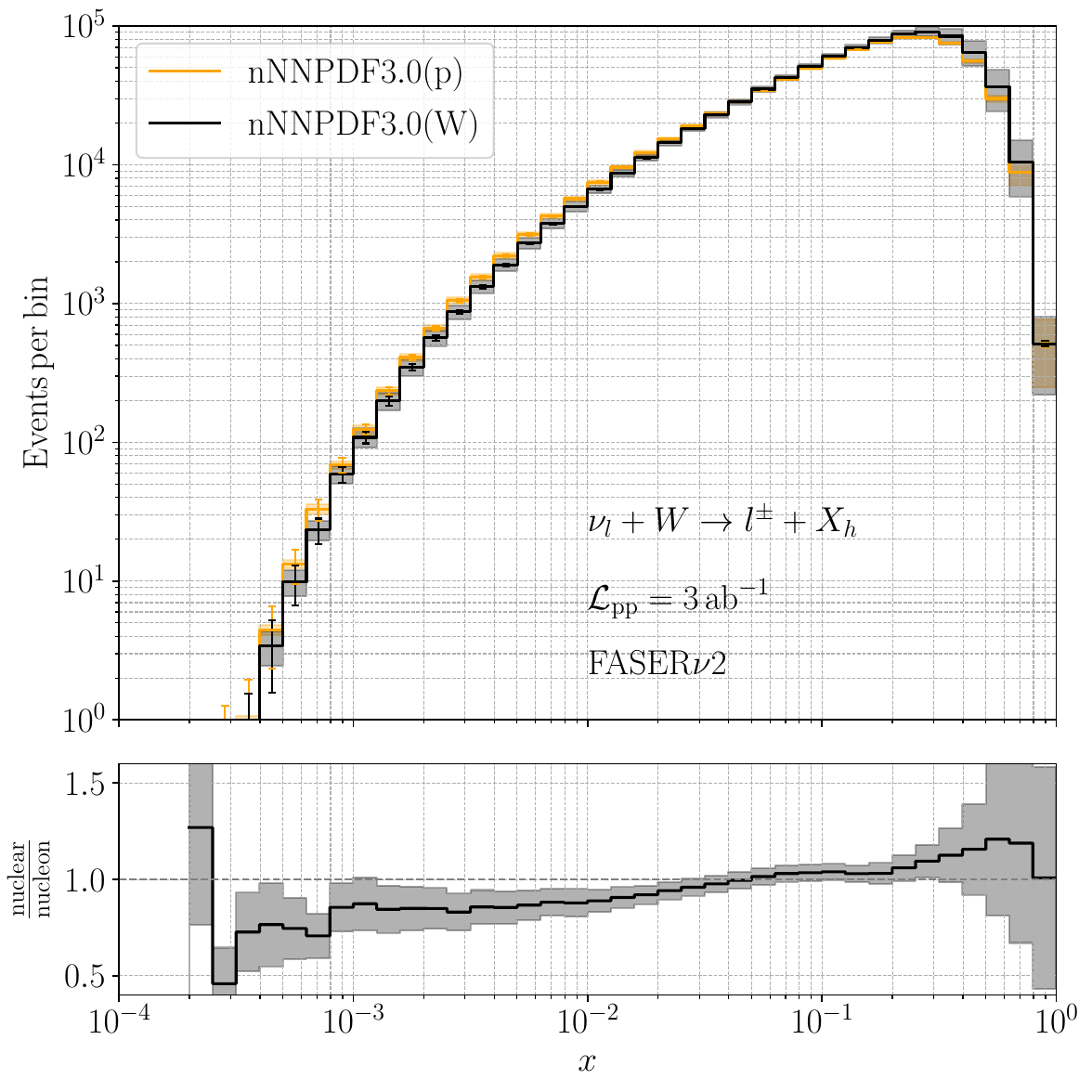} \\
	\includegraphics[width=0.49\textwidth]{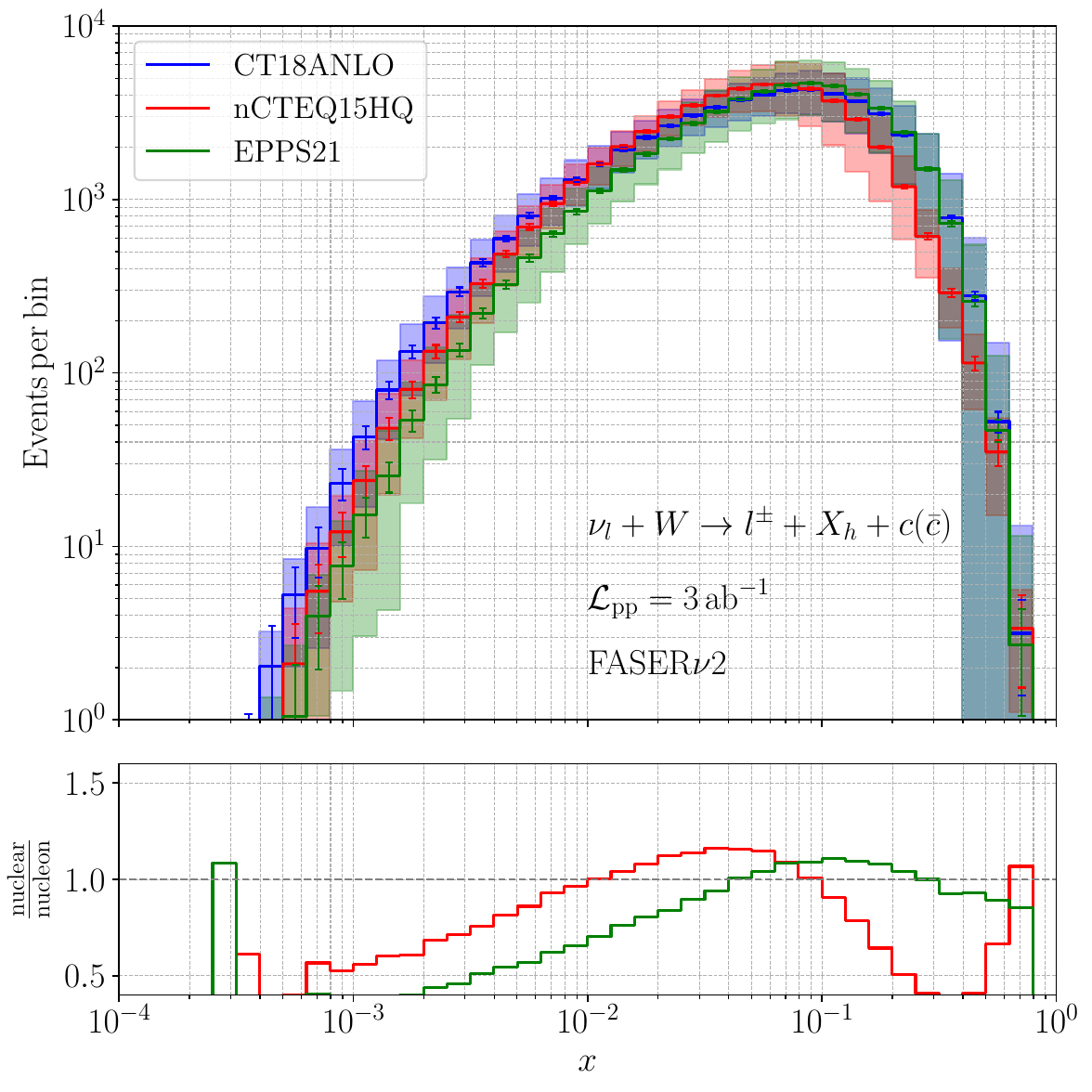}
	\includegraphics[width=0.49\textwidth]{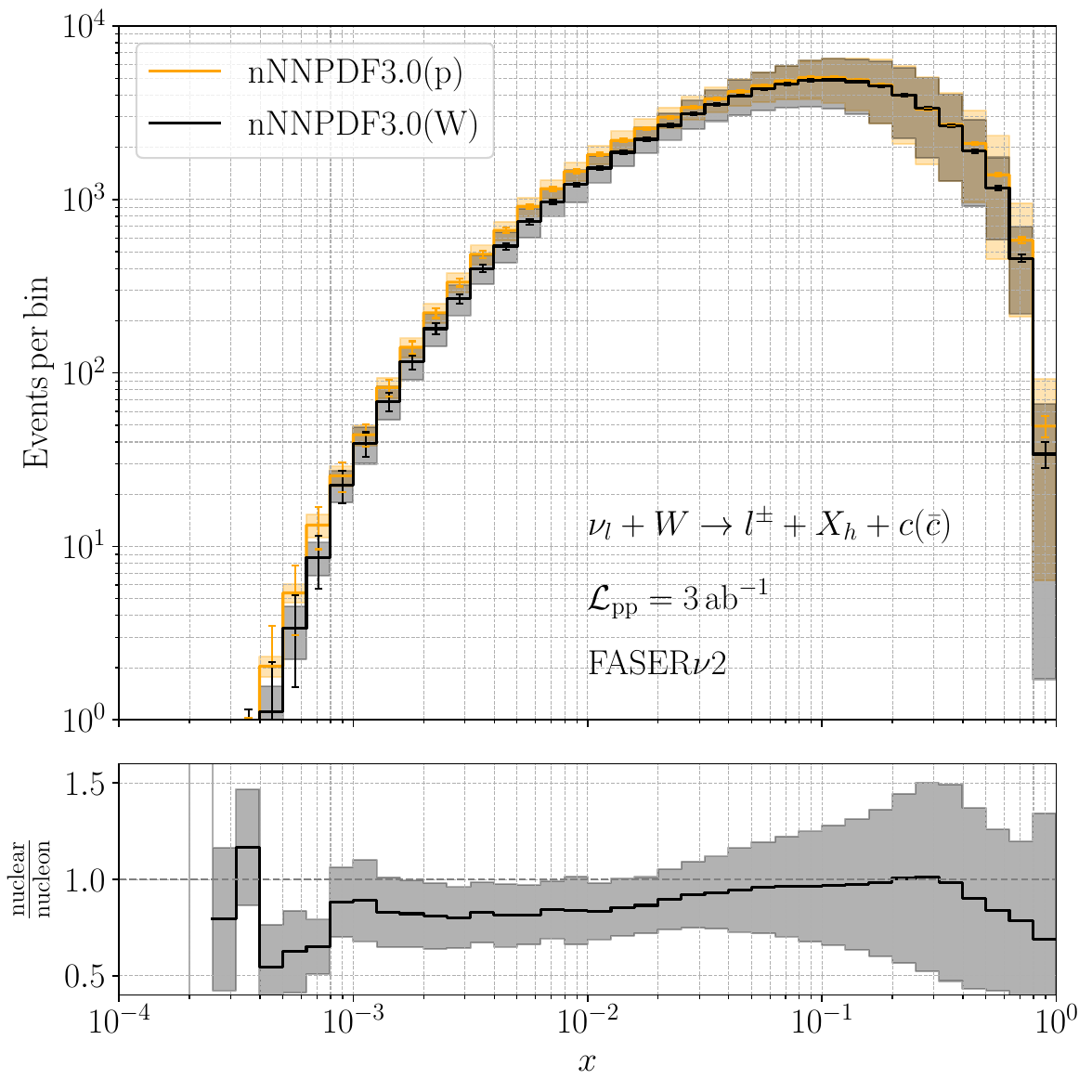}
    
			\end{tabular}
\caption{Predictions for the number of neutrino DIS events at the FASER$\nu2$ binned in $x$ in the inclusive case (upper panels) and with a charm detected in the final state (lower panels). Results derived assuming different parameterizations for the nPDFs. Predictions for the ratio between nuclear and nucleonic results are presented in the lower panels of the plots. }
\label{fig:neutrinoDIS_x_FASERnu2HL}
\end{figure}

In Figure \ref{fig:muonDIS_x_FASERnu2HL}, we present our results for muon DIS events binned in $x$ at the FASER$\nu 2$, considering an integrated luminosity of 3 ab$^{-1}$. We observed an increase in event rates of approximately two orders of magnitude compared to the results for FASER$\nu$ during Run 3. Furthermore, the relative expected statistical uncertainty bars decrease by a factor of $\approx 10$, making the statistical uncertainties smaller than the current PDF uncertainty across the entire range of $x$ considered. The differences between the predictions for the impact of nuclear effects on inclusive charm events and charm-detected events in the final state, derived using the different nPDFs, are also expected at FASER$\nu 2$. The same conclusion also holds true for neutrino DIS events, as demonstrated by the results presented in Figure \ref{fig:neutrinoDIS_x_FASERnu2HL}. The results with different PDF parameterizations provide similar predictions, but the expected statistical uncertainties are smaller than the current PDF uncertainty, indicating that FASER$\nu 2$ can be a powerful tool to reduce the current uncertainties of nuclear PDFs.

\begin{figure}[H]
	\centering
	\begin{tabular}{ccc}
	\includegraphics[width=0.49\textwidth]{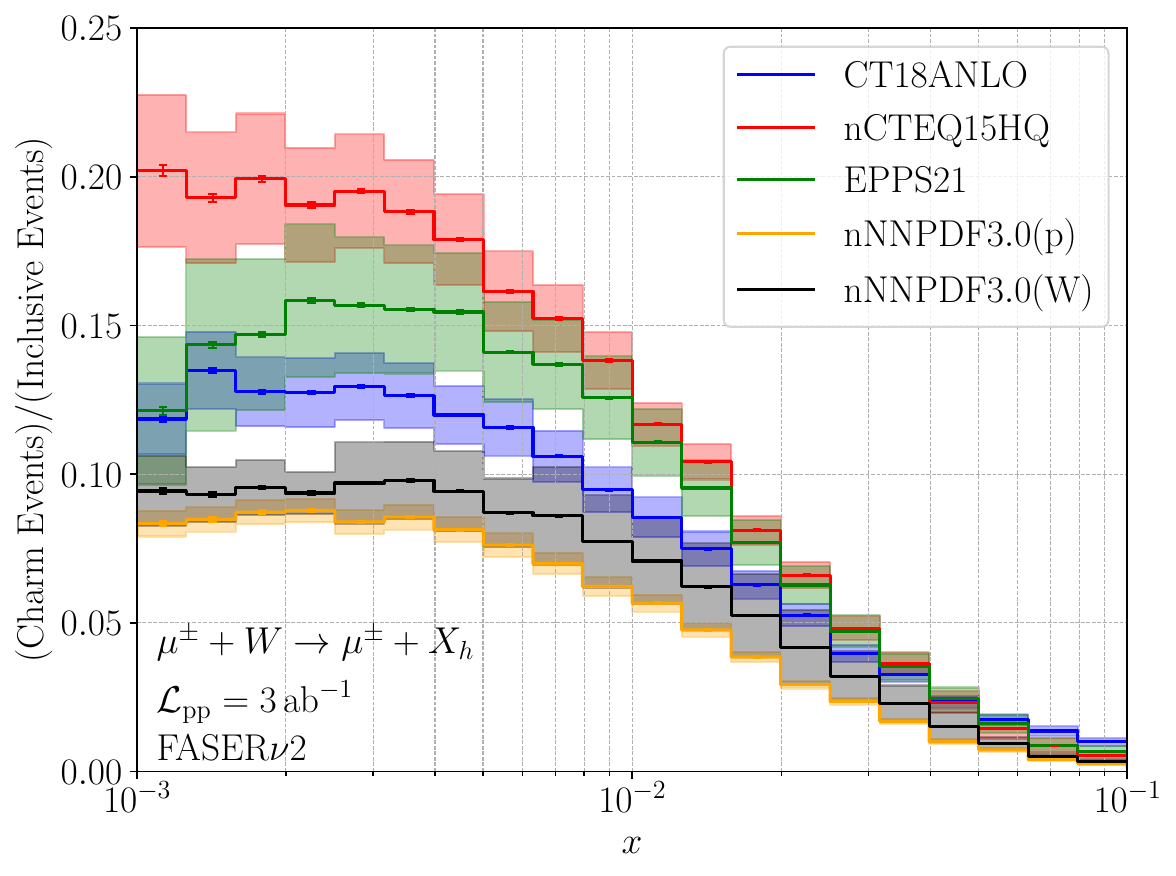}
	\includegraphics[width=0.49\textwidth]{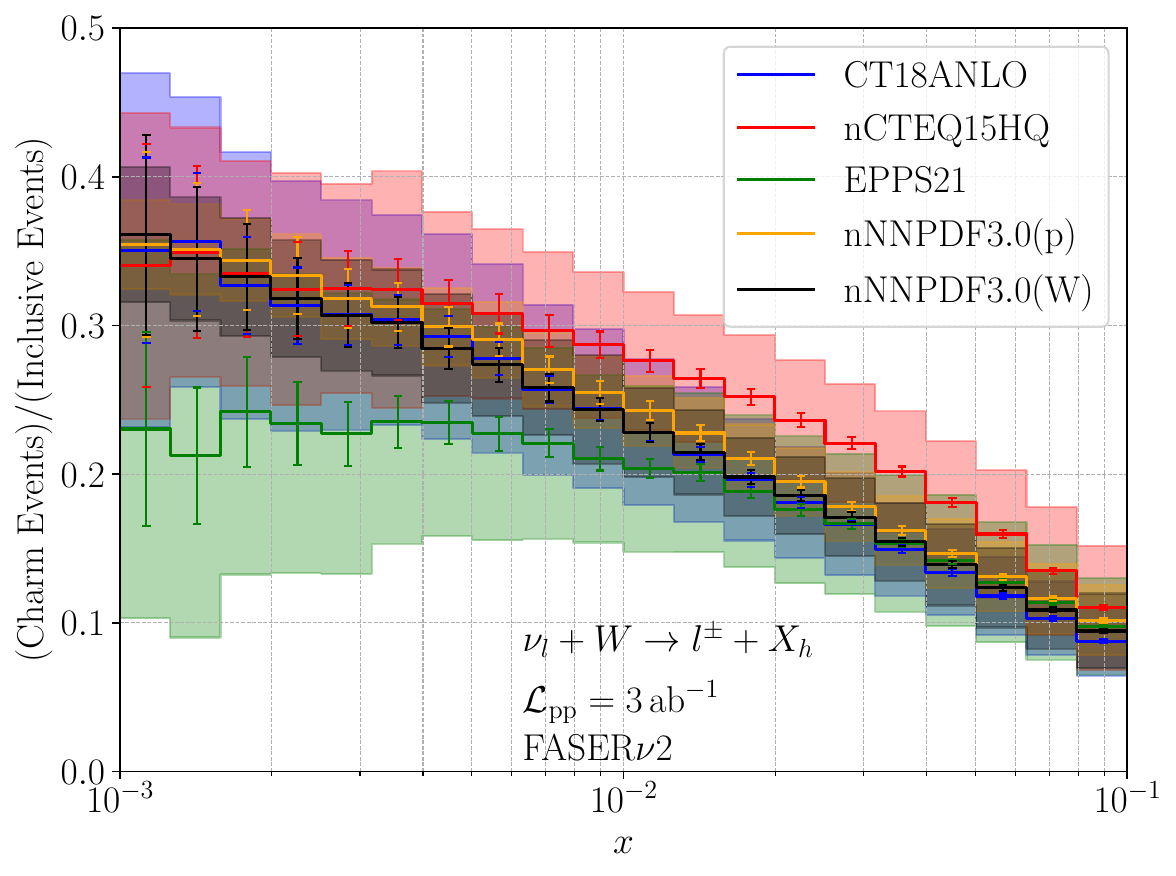}

			\end{tabular}
\caption{ Ratio between events with charming hadrons and inclusive events for muon (left) and neutrino (right) DIS at the FASER$\nu 2$ during the high-luminosity era of the LHC. Results derived assuming different parameterizations for the nPDFs. }
\label{fig:Ratio_charm_total_FASERnu2_HL}
\end{figure}

Finally, in Figure \ref{fig:Ratio_charm_total_FASERnu2_HL} we present our predictions for the ratio between events with charm detected and inclusive events for muon interactions (left panel) and neutrino interactions (right panel). Compared to the results shown in Figure \ref{fig:Ratio_charm_total_FASERnu}, we now find that the difference between the EPPS21 prediction for neutrino DIS and those derived using the other PDFs will be greater than the expected statistical uncertainties. Consequently, a future experimental analysis of the proposed ratio in $\mu W$ and $\nu W$ events, measured by the same detector, could be used to constrain the description of nuclear effects, as well as to test the universality of the nPDFs.

\section{Conclusions}
\label{sec_MuonDIS:conclusion}

In this section, we explore the possibility of studying muon and neutrino DIS at the FASER$\nu$ detector, as well as its proposed upgrade for the high-luminosity era of the LHC, FASER$\nu$2. We use POWHEG+Pythia8 to simulate muon-tungsten interactions and show that DIS measurements at FASER$\nu$ will cover a range in $x$ and $Q^{2}$ that considerably overlaps with the expected observations at the EIC. As possible applications of observing these processes, we show that FASER$\nu$ can observe the existence of an intrinsic charm quark in the nucleon, even with data collected during Run 3. We also show that FASER$\nu$ will be sensitive to different models of intrinsic charm in the nucleon and constrain the fraction of nucleon momentum carried by this quark component. By coupling the observations from FASER$\nu$ with the FASER electronic detector, it will be possible to extract information regarding a possible asymmetry between intrinsic charm and anticharm in the nucleon. In addition, the observation of both muon and neutrino DIS can make important contributions to the understanding of nuclear effects, particularly in understanding the universality of PDFs. In the next chapter we will study a rare process that can be explored in FASER-like detectors, the neutrino trident scattering.

\chapter{Neutrino trident scattering at the LHC energy regime}
\label{cap:tridente}

The production of neutrino beams in hadronic colliders was predicted as early as the 1980s \cite{DeRujula:1984ns}, but remained unexplored until recently. The first neutrino detections from colliders were made at CERN in 2022 with the FASER and SND@LHC experiments \cite{SNDLHC:2023pun,FASER:2021mtu,FASER:2023zcr,FASER:2024hoe,FASER:2024ref} and initiated a new era in neutrino physics. With a controlled environment of high-energy hadronic collisions, new hadrons are produced which subsequently decay, producing a highly collimated neutrino beam, with the highest neutrino energies ever produced by humans. These highly dense neutrino fluxes allow the investigation of hadronic interaction effects, given that neutrinos originate from hadrons produced in the frontal region of the LHC interaction points, and also, of course, allow us to investigate aspects of neutrino physics not yet explored, such as neutrino-nucleon cross sections in the TeV region \cite{FASER:2024ref}, as well as models beyond the Standard Model involving neutrinos.

\begin{figure}[H]
	\centering
	\begin{tabular}{ccccc}
\includegraphics[width=0.5\textwidth]{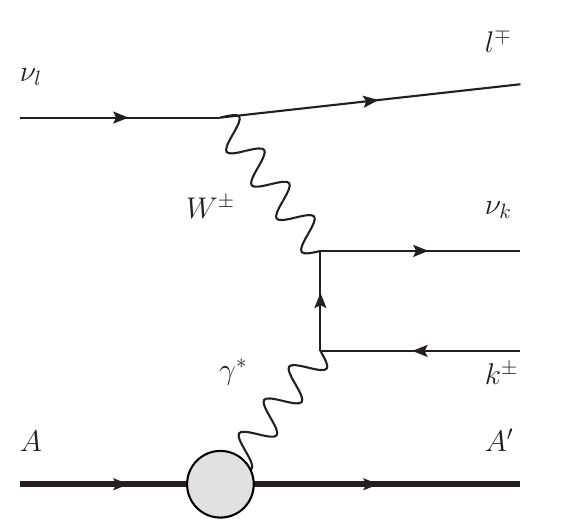} &
\includegraphics[width=0.5\textwidth]{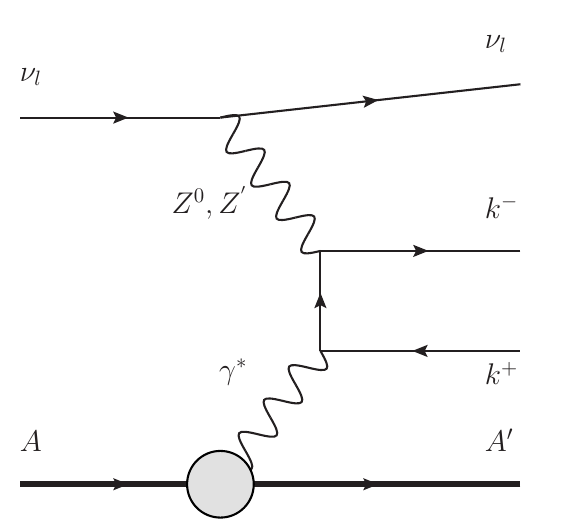} 
	\end{tabular}
\caption{ Feynman diagrams for the neutrino trident scattering in the Coulomb field of the target nucleus associated with the exchange of $W^\pm$ (left), $Z^0$ and $Z'$ (right) bosons. The flavor of the $l$ and $k$ leptons can be the same or different, and $A^{\prime} = A$ ($A^{\prime} \neq A$) characterizes coherent (incoherent) interactions. }
\label{fig_3:diagrams}
\end{figure}

A rare process predicted by the Standard Model, and not yet observed with sufficient statistical evidence to characterize a discovery, is the trident scattering of neutrinos. This process, shown in the Feynman diagrams in Figure \ref{fig_3:diagrams}, is characterized by three leptons in the final state, two of which are electrically charged. One of the leptons comes from the vertex of the initial neutrino, and the two additional ones are produced in the fusion of two gauge bosons: the $W^{\pm}$ or $Z^{0}$ from the neutrino, and the photon from the target nucleus. In the $Z^{0}\, \gamma^{*}$ fusion, charged leptons of the same family are always produced, while in the process where the neutrino interacts by exchanging a $W^{\pm}$ boson, the charged leptons may or may not be of the same generation. Trident scattering is said to be coherent when the nuclear target interacts as a whole in the photon exchange. On the other hand, it is said to be incoherent if the photon interacts with individual nucleons, whether protons or neutrons, which is common for more energetic photons. There is also DIS for trident scattering, when the photon interacts with individual partons within the nuclear target, but this scattering has an important contribution at higher energies than those studied here, and therefore will not be considered.

This scattering has already been investigated by three distinct experiments, CHARM-II \cite{CHARM-II:1990dvf}, CCFR \cite{CCFR:1991lpl} and NuTeV \cite{NuTeV:1999wlw}, all in the last century and with muon pairs produced in the final state. The measurements of the number of events associated with trident scattering in the cited experiments and the ratio of the measured and predicted neutrino-target cross sections by the Standard Model are presented in Table \ref{table_3:medidas}. The measurements of the cross sections were not very precise, not enough for these experiments to claim the discovery of the process, but they are within the expected in the Standard Model.

	\begin{table}
	\centering
		\begin{tabular}{|c|c|c|c|c|}
			\hline 
Experiment	    & Observed events    & $\sigma_{exp} / \sigma_{MP}$ \\
			\hline
CHARM-II 	    & 55$\pm$16          & 1.58$\pm$0.64
\tabularnewline
CCFR 		    & 37$\pm$12          & 0.82$\pm$0.28
\tabularnewline
NuTeV	            & 7$^{+17}_{-7}$      & 0.72$^{+1.73}_{-0.72}$
\tabularnewline
			\hline		
		\end{tabular}
		\caption{ Number of events in the neutrino trident scattering. We have a comparison of the ratio of the measurement to the associated theoretical prediction of the Standard Model. }
		\label{table_3:medidas}
	\end{table}

Currently, several phenomenological studies point to the possibility of observing trident scattering in new experiments \cite{Altmannshofer:2019zhy,Zhou:2019frk,Ballett:2018uuc,Bigaran:2025vea}. Given that this is a process calculated precisely in the Standard Model, its observation is of great interest for searching for physics beyond the Standard Model. One model of interest in recent works is the $L_\mu - L_\tau$ theory, which we investigated in Chapter \ref{cap:Ice} in the context of astrophysical neutrinos measured by IceCube. This process contributes to trident scattering according to the diagram on the right of Figure~\ref{fig_3:diagrams}, that is, to processes with leptons of the same generation in the final state, as well as initial muonic or tauonic neutrinos.

In this chapter, we present our phenomenological studies of the neutrino trident scattering process at FASER$\nu$2, which is one of the experiments proposed for the Forward Physics Facility (FPF) \cite{Feng:2022inv,FPFWorkingGroups:2025rsc,FPF:2025bor}. Our interest in this detector stems from its expected volume compared to other FPF experiments, and because it is planned to be situated in the direction of the ATLAS collision axis, making it sensitive to the most intense part of the neutrino flux produced in hadronic collisions.

\section{Cross section for the neutrino trident scattering}
\label{sec_tridente:cs}

The neutrino-target cross section for trident scattering initiated by a four-momentum (energy) neutrino $p_1$ ($E_\nu$) with a four-momentum (mass) hadronic target $P$ ($M_P$) can be calculated from
\begin{eqnarray}
\begin{aligned}
\sigma_{\mathrm{tridente}} = & \int \frac{1}{M_P E_\nu} |\mathcal{M}|^{2} \frac{\mathrm{d^3}\vec{p}_2}{(2\pi)^{3}2 E_2} 
\frac{\mathrm{d^3}\vec{p}_3}{(2\pi)^{3}2 E_3} \frac{\mathrm{d^3}\vec{p}_4}{(2\pi)^{3}2 E_4} \frac{\mathrm{d^3}\vec{P}'}{(2\pi)^{3}2 E'}  \\
& (2\pi)^{4} \delta (p_1 + P - p_2-p_3-p_4-P') \, ,
 \label{eq_3:cs}
\end{aligned}
\end{eqnarray}
where $p_2$ ($E_2$), $p_3$ ($E_3$) and $p_4$ ($E_4$) are the momenta (energies) of the leptons present in the final state, $P'$ ($E'$) is the momenta (energy) of the final nucleus, and $\mathcal{M}$ is the invariant matrix element of the process.

The total and differential cross sections for this scattering have already been estimated by different authors, and in this work we will use the Monte Carlo generator from the reference \cite{Altmannshofer:2019zhy}, which uses the variable and integration limits derived in \cite{Czyz:1964zz,Lovseth:1971vv}. In our work, we implement the tungsten target in the cited generator, given that our interest lies in trident processes at FASER$\nu$2, which will have this material as its target. Our calculations use the tungsten nucleus described by the Woods-Saxon nuclear charge distribution \cite{Woods:1954zz}, parameterized in \cite{DeVries:1987atn}. In addition to the coherent interactions with the tungsten nucleus, we will calculate the cross sections and events for incoherent scattering with protons and neutrons within the target nucleus. This process has already been implemented in the generator, and uses dipole form factors for nucleons within the nucleus \cite{Budnev:1975poe,Drees:1988pp} modeled as a Fermi gas \cite{Bell:1963ogq}.

The nuclear form factor needed to describe the target nucleus is the Fourier transform of the spatial charge distribution, usually presented normalized by the total charge of the nucleus. We can interpret it as the charge distribution in the momentum space transferred by the nucleus:

\begin{equation}
F(q)=\dfrac{1}{N}\int \rho (\vec{r})e^{-i\vec{q}\cdot \vec{r}}d^{3}\vec{r} \, ,
\label{eq_3.82}
\end{equation}

\noindent with

\begin{equation}
N=\int \rho (\vec{r})d^{3}\vec{r} \, ,
\end{equation}
where $\rho (\vec{r})$ is the spatial charge density, and $\vec{q}$ is the momentum transferred by the nucleus in the collision. There are several form factors present in the literature, and as mentioned in the previous paragraph, we will use the Woods-Saxon distribution in our analyses, parameterized with elastic electron scattering data in nuclei. This distribution is given by
\begin{equation}
\rho (r)=\rho_{0}\left[1+\mathrm{exp}\left(\dfrac{r-R}{a} \right) \right]^{-1},
\label{eq_3.87}
\end{equation}
with parameters $a=0.535\;$fm and $R=6.51\;$fm for the tungsten nucleus. For a detailed review of the form factors and approximations usually adopted, see reference \cite{Francener:2022sfw}.

\begin{figure}
	\centering
	\begin{tabular}{ccccc}
\includegraphics[width=0.6\textwidth]{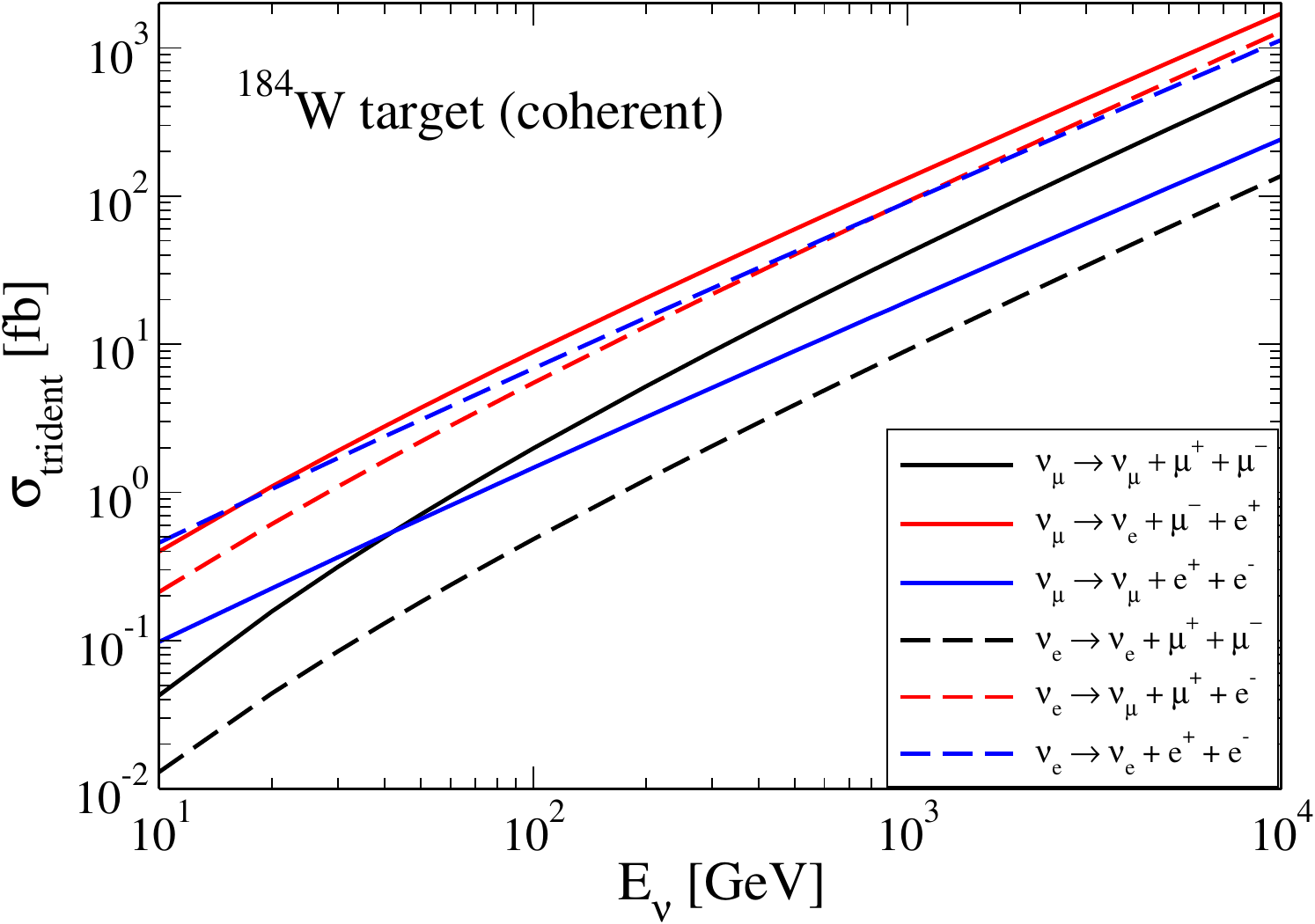} \\
\includegraphics[width=0.6\textwidth]{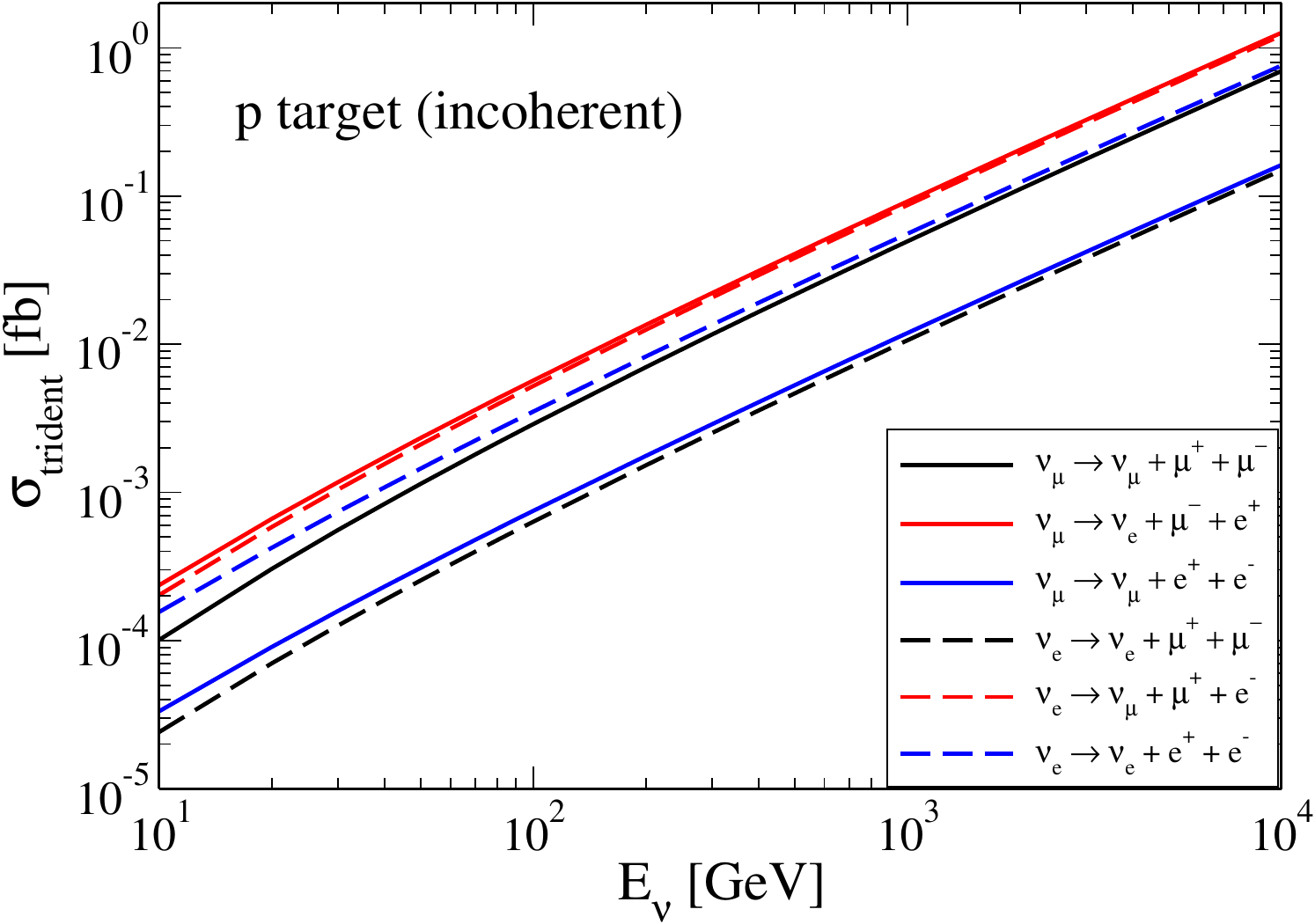} \\
\includegraphics[width=0.6\textwidth]{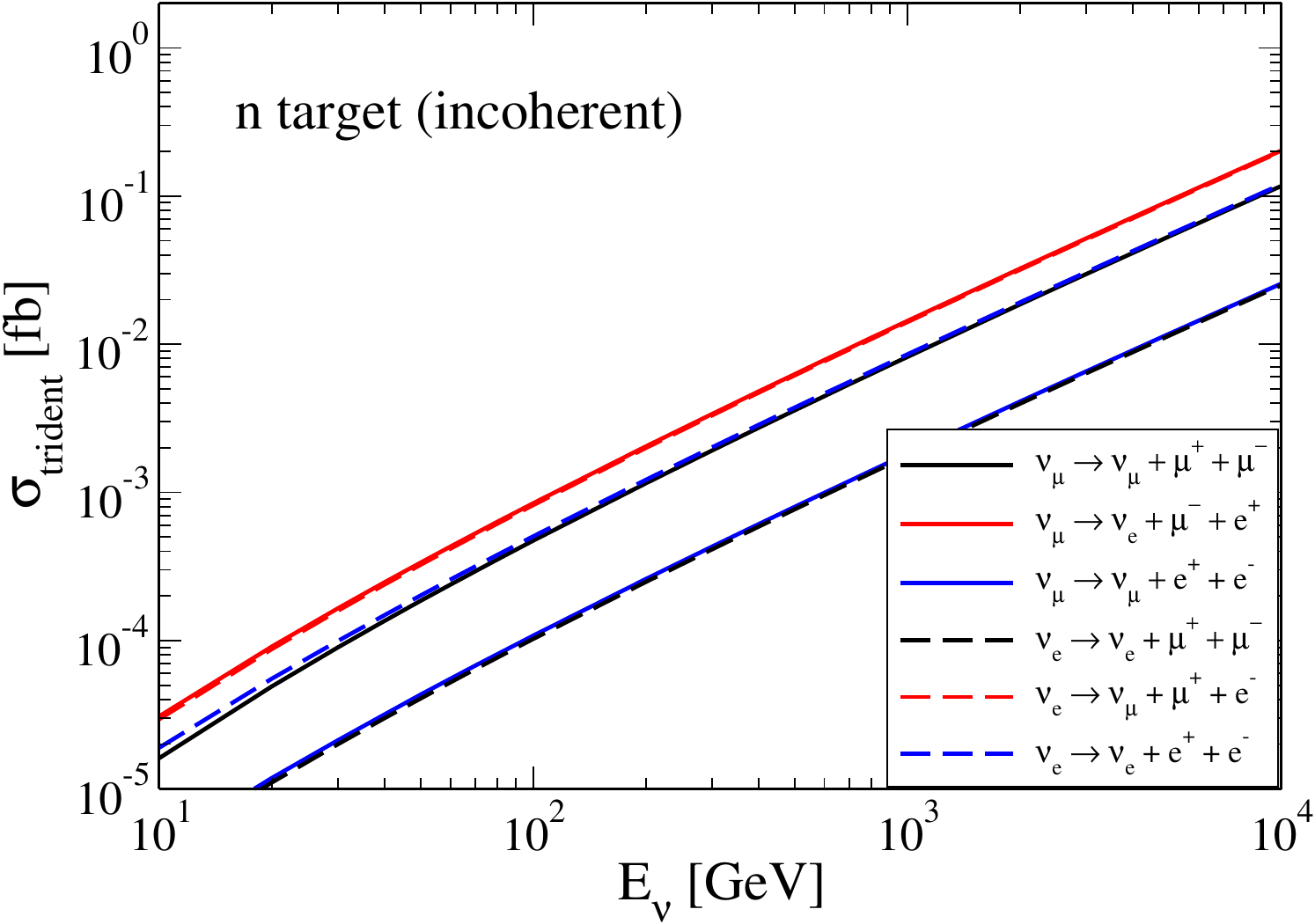} 
	\end{tabular}
\caption{ Cross sections for neutrino trident scattering as a function of incident neutrino energy. We consider muonic neutrino interactions (solid lines) and electronic neutrino interactions (dashed lines). Our results are for muon pairs (black), muon plus electron pairs (red), and electron pairs (blue) in the final state. The results are for coherent scattering in the top panel and incoherent scattering in the middle panel (proton target) and bottom panel (neutron target). }
\label{fig_3:cs}
\end{figure}

Firstly, in Figure \ref{fig_3:cs}, we present the cross section of the trident process as a function of the incident neutrino energy, in the range 10 - 10000 GeV, expected for LHC neutrinos. Our results were obtained using the Monte Carlo generator described in reference \cite{Altmannshofer:2019zhy}, where the authors considered Argon as the target and the DUNE experiment for the neutrino flux. We changed the neutrino and target energy configurations to FASER$\nu$2 as described above. In the upper panel, we present the cross sections for the tungsten target (coherent scattering), in the middle panel for the proton target, and in the lower panel for the neutron target (incoherent scatterings). Our results are for events induced by muonic neutrinos (solid lines) and electronic neutrinos (dashed lines). We do not consider scattering by tauonic neutrinos, given that the flux of this neutrino flavor is expected to be subdominant at the LHC frontal experiments. We also have in Figure \ref{fig_3:cs} different final states for charged leptons: muon pairs in the black lines, electron pairs in the blue lines, and electron plus muon in the red lines. We do not consider final states with taus here, given that these are kinematically suppressed by the need for sufficient energy to produce the tau mass. Coherent scattering dominates in this energy regime, being 3 (4) orders of magnitude above incoherent proton (neutron) scattering.

Our results indicate that processes involving leptons with distinct flavors in the final state have larger cross sections. These processes can only occur through the exchange of the $W^{\pm}$ boson, while processes involving leptons from the same family also have a contribution from the $Z^{0}$ boson exchange. These processes, which can occur through both weak boson exchanges, have slightly smaller cross sections due to destructive interference between the matrix elements of the scattering mediated by the two weak interaction bosons. It is also possible to see that the cross sections with muons in the final state have slightly smaller values than with electrons in the final state, especially for lower incident neutrino energies. This suppression arises from the muon mass being approximately 200 times greater than the electron mass, resulting in kinematic suppression in the process.

\begin{figure}
	\centering
	\begin{tabular}{ccccc}
\includegraphics[width=0.48\textwidth]{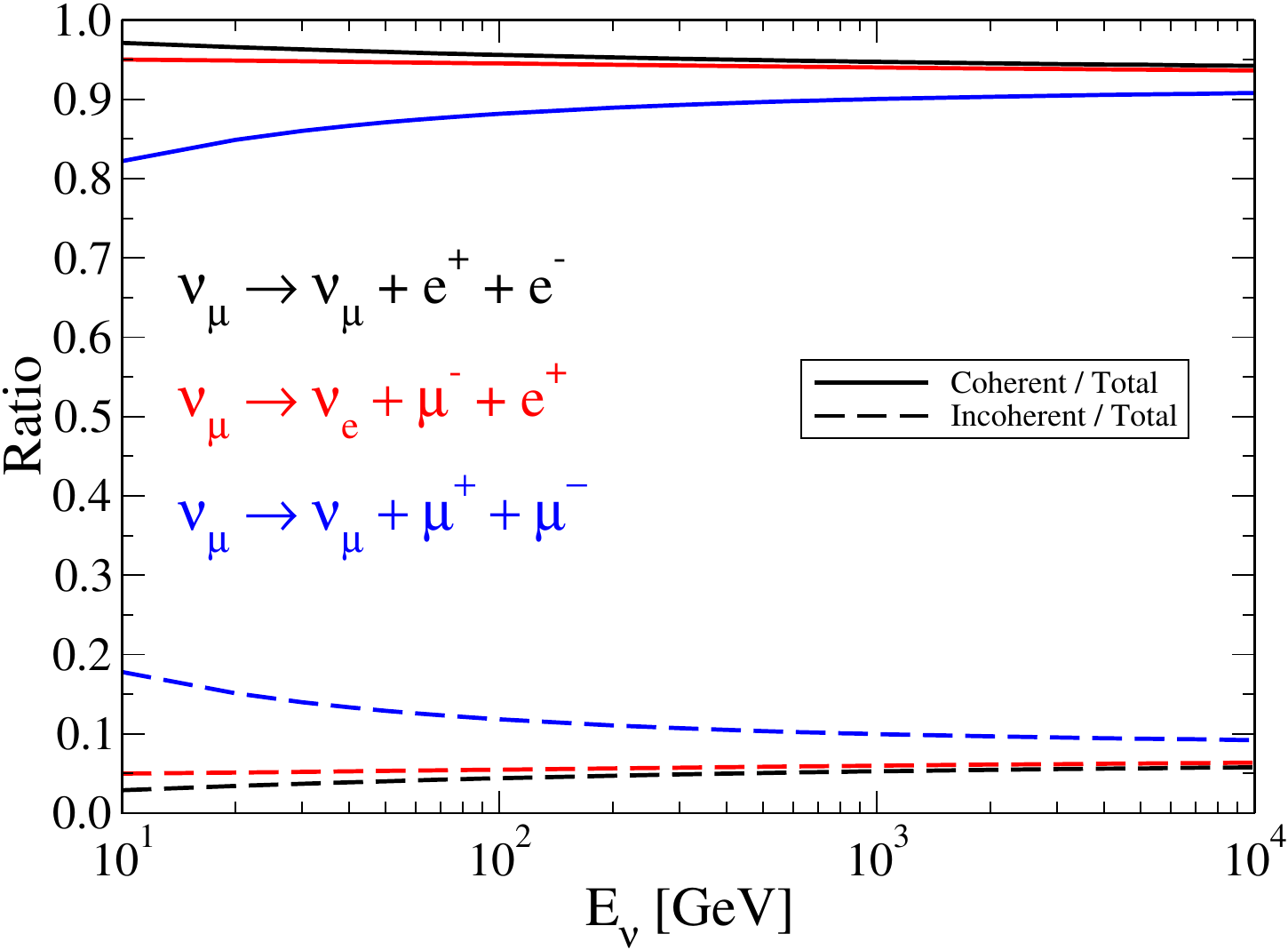} &
\includegraphics[width=0.48\textwidth]{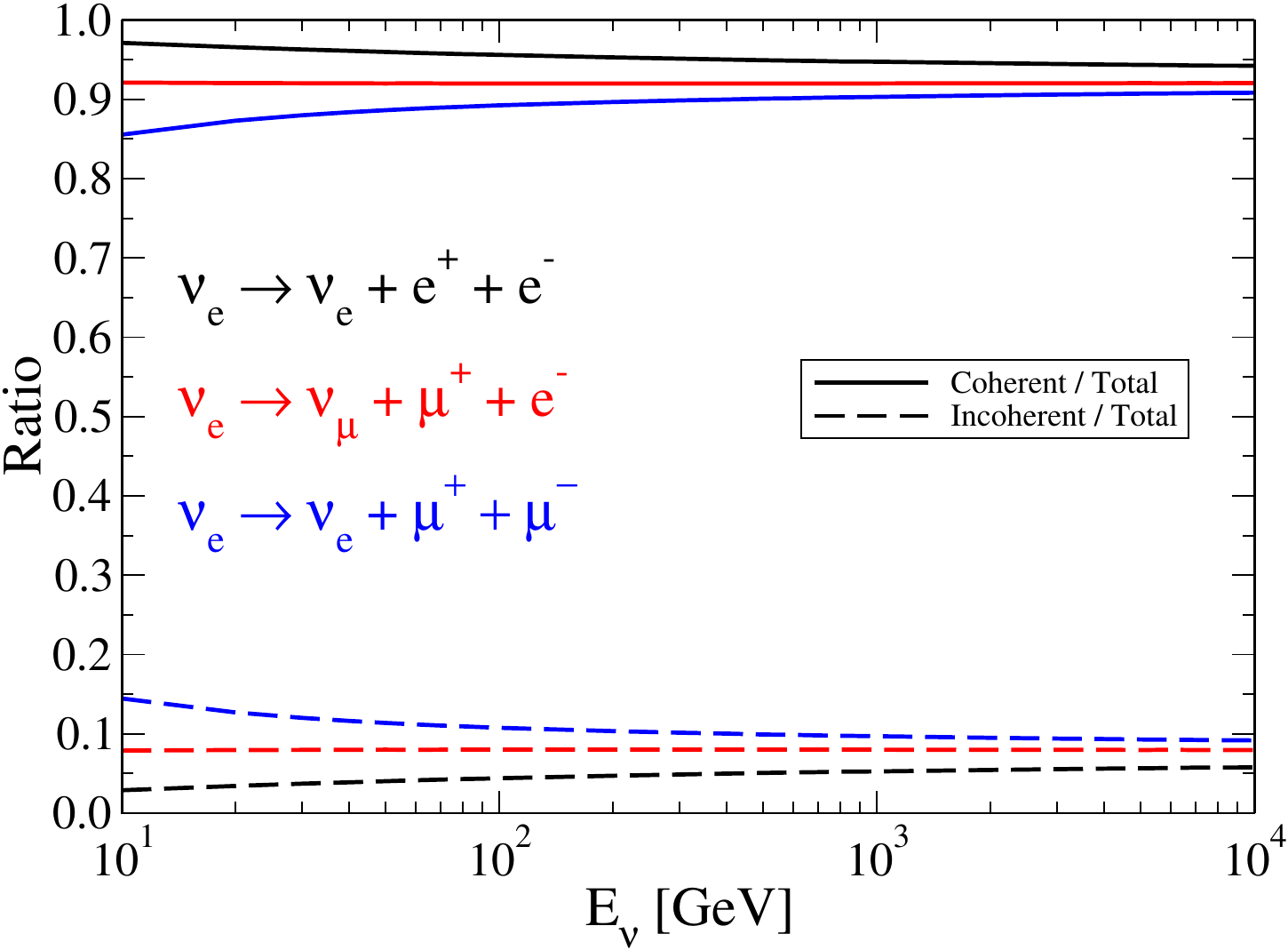} 
	\end{tabular}
\caption{ Ratio between coherent (solid lines) and incoherent (dashed lines) contributions to the total cross section (sum of coherent and incoherent contributions) of trident scattering for events initiated by muonic neutrinos (left) and electronic neutrinos (right) on tungsten targets. }
\label{fig_3:ratio}
\end{figure}

To better analyze the importance of cross sections in incoherent and coherent channels for the total cross section, we present Figure \ref{fig_3:ratio}. In it, we show the ratio of contributions of coherent processes (solid lines) and incoherent processes (dashed lines) to the total cross section of the trident processes described above. On the left, we present the results for processes induced by muonic neutrinos, while on the right, for reactions induced by electron neutrinos. Our results indicate that coherent processes dominate the total cross section in the energy regime studied here, accounting for more than 90\% of the total cross section. Processes with more massive particles in the final state, i.e., with muons, have a greater contribution from incoherent scattering, given that the production of heavier particles requires more energetic photons, which are more abundant in incoherent processes. We emphasize that the ratio presented in Figure \ref{fig_3:ratio} is for the tungsten nucleus as the target, that is, we sum the incoherent contributions of all the nucleons of the target.

\section{Trident scattering events at FASER$\nu$2}
\label{sec_tridente:events}

Knowing the cross sections for the neutrino trident scattering process presented in the previous section, we are able to estimate the main observable in an experiment: the number of events. For detectors like the FASER$\nu$2, this number is obtained from the expression
\begin{eqnarray}
N_{\mathrm{events}} =  \sigma_{\nu A}(E_\nu) \times  \phi(E_\nu) \times \rho L/m_{A},
\label{eq:eventosFASER}
\end{eqnarray}
where $\sigma_{\nu A}$ is the neutrino-target cross section, $\rho$ is the density of the target material in the detector ($19.3\,\mathrm{g/cm}^{3}$ for tungsten), $L$ is the length of the detector (6.6 m), $m_A$ is the mass of the particle target and $\phi(E_\nu)$ is the time-integrated neutrino flux that will cross the front surface of the detector.

As described in Equation~(\ref{eq:eventosFASER}), in addition to the detector characteristics and the cross section presented earlier for scattering, an indispensable ingredient is the neutrino flux that reaches the detector. The neutrino flux that will reach FASER$\nu$2 will be produced in proton-proton collisions at ATLAS with a center-of-mass energy of 14 TeV, and detected approximately 620 meters from the interaction point at ATLAS. The FASER$\nu$2 detector will be aligned with the collision axis, capturing the most collimated part of the neutrinos and the most intense flux. Other detectors, with other purposes, are being proposed for installation at the FPF, some outside the collision axis, with the intention of searching for rare events and other processes. Proton-proton interactions at the LHC are known to produce a large number of hadrons, which subsequently decay giving rise to the neutrinos that will be detected by FASER$\nu$2. These hadrons can be light, formed by up, down, and strange quarks, or heavy, containing charm and bottom quarks. The production of light hadrons in the frontal regions is described by non-perturbative QCD, while heavy hadrons in this kinematic region can be modeled with perturbative QCD. Neutrino fluxes have been obtained by different research groups, which use different methods and phenomenological models in their acquisition. In our work, we are using neutrino fluxes obtained from the reference \cite{Kling:2023tgr}, which uses several Monte Carlo generators in the simulation of hadronic collisions, and subsequently decays giving rise to the neutrino flux. For the simulation of neutrinos originating from light hadrons, we are using the results derived with the Monte Carlo generators SIBYLL \cite{Ahn:2009wx,Riehn:2015oba}, EPOS-LHC \cite{Pierog:2013ria}, DPMJET \cite{Roesler:2000he}, Pythia8 \cite{Sjostrand:2014zea,Fieg:2023kld} and QGSJET \cite{Ostapchenko:2010vb}. On the other hand, for the neutrino flux from heavy hadrons, we are using the results derived from the references \cite{Bai:2020ukz,Bai:2021ira,Bai:2022xad} (denoted BDGJKR), \cite{Buonocore:2023kna} (denoted BKRS), \cite{Bhattacharya:2023zei} (denoted BKSS $k_T$), \cite{Maciula:2022lzk} (denoted MS $k_T$) and \cite{Ahn:2011wt,Fedynitch:2018cbl} (denoted SIBYLL 2.3d). In particular, we will use the same combinations of fluxes from light and heavy hadrons used in the reference \cite{Kling:2023tgr}.

\begin{figure}
	\centering
	\begin{tabular}{ccccc}
\includegraphics[width=0.48\textwidth]{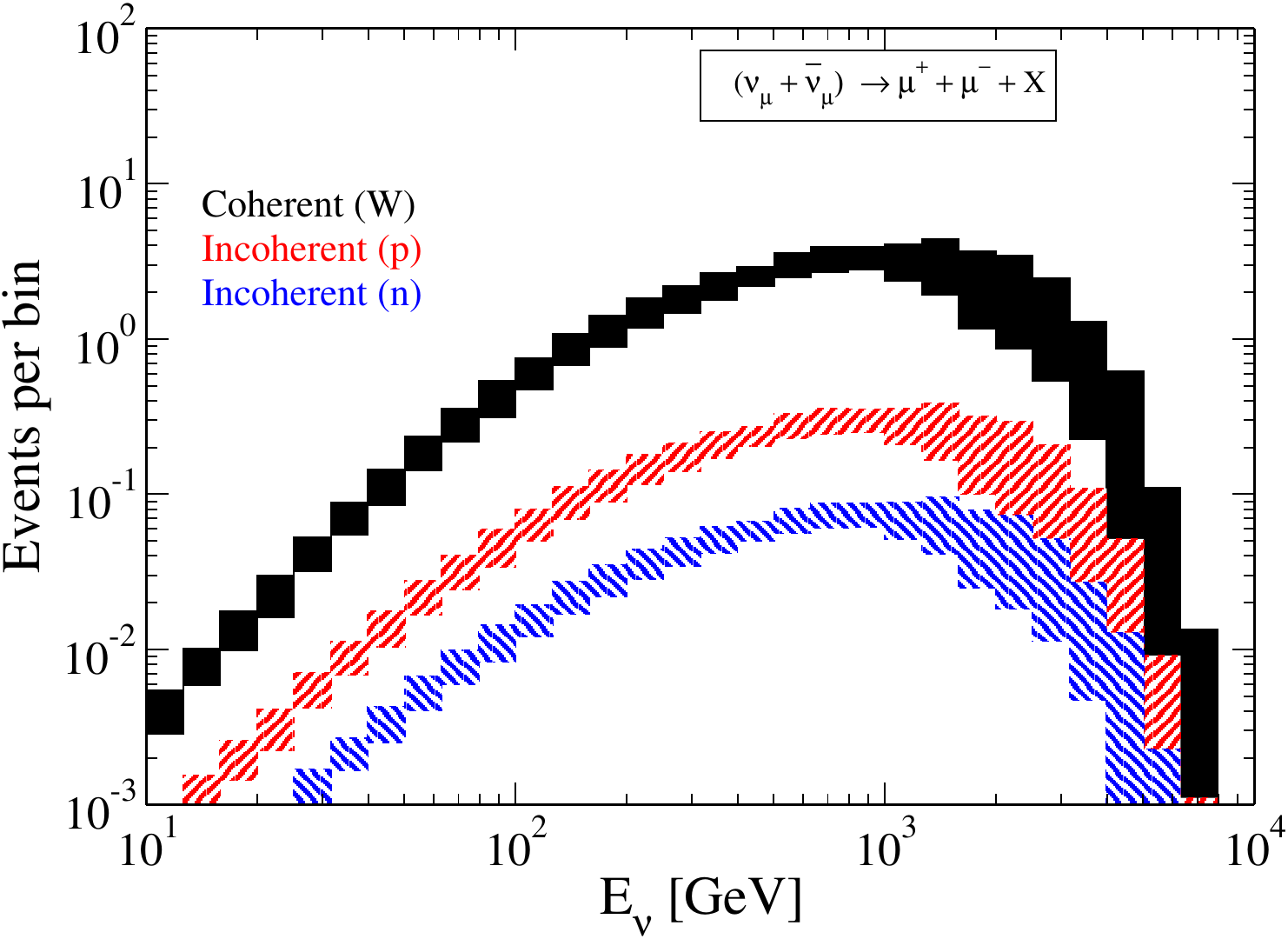} &
\includegraphics[width=0.48\textwidth]{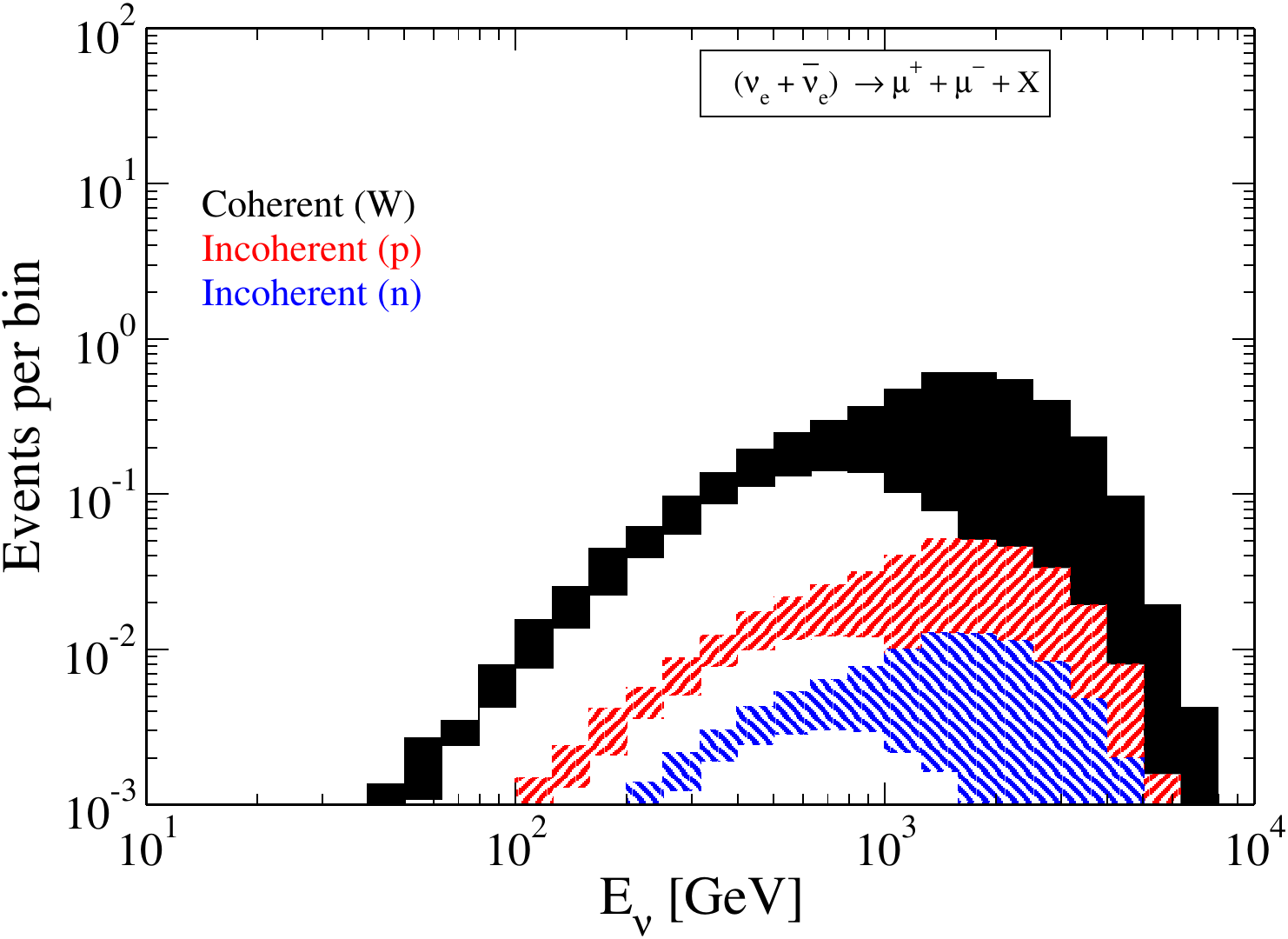} \\
\includegraphics[width=0.48\textwidth]{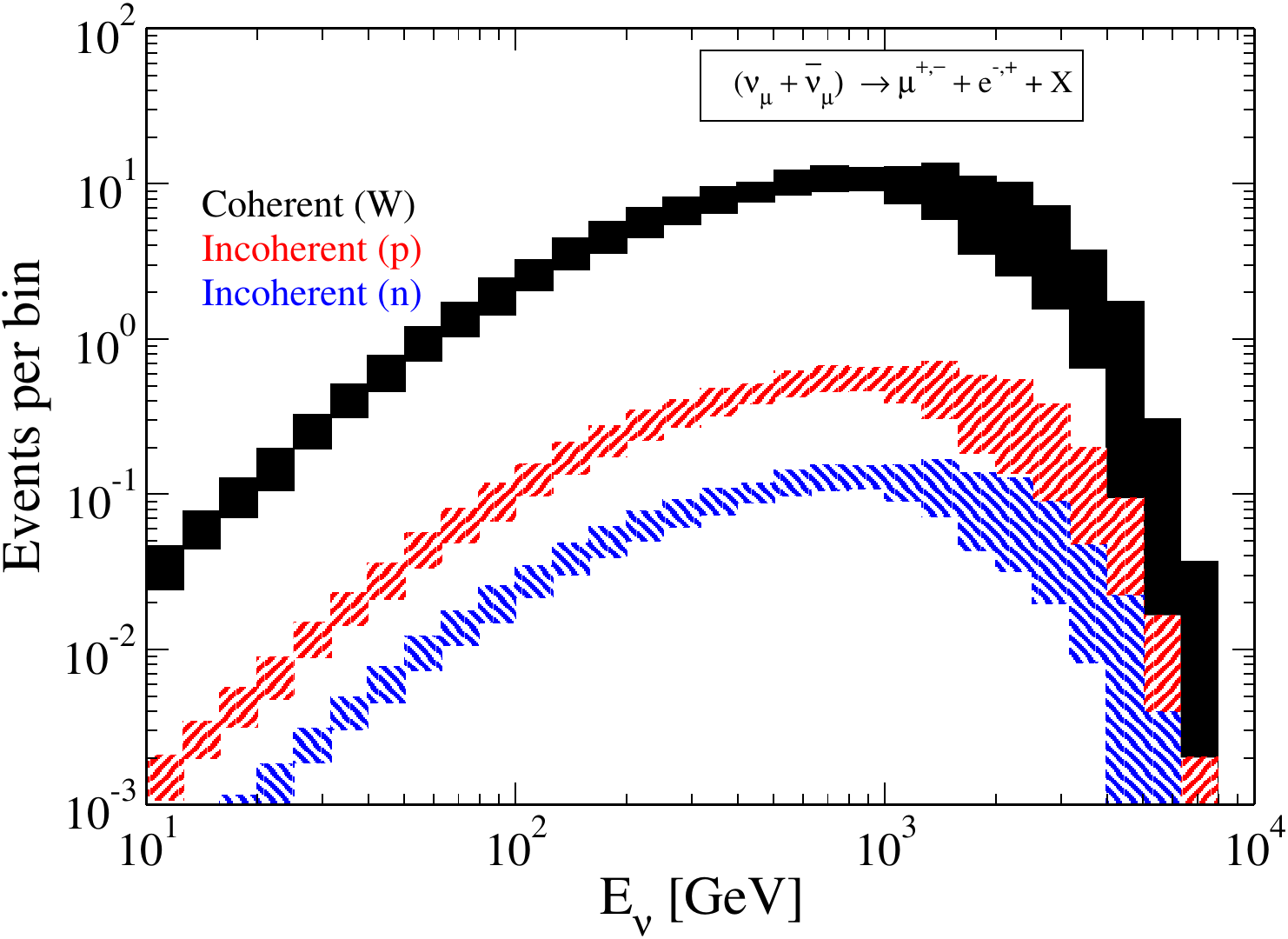} &
\includegraphics[width=0.48\textwidth]{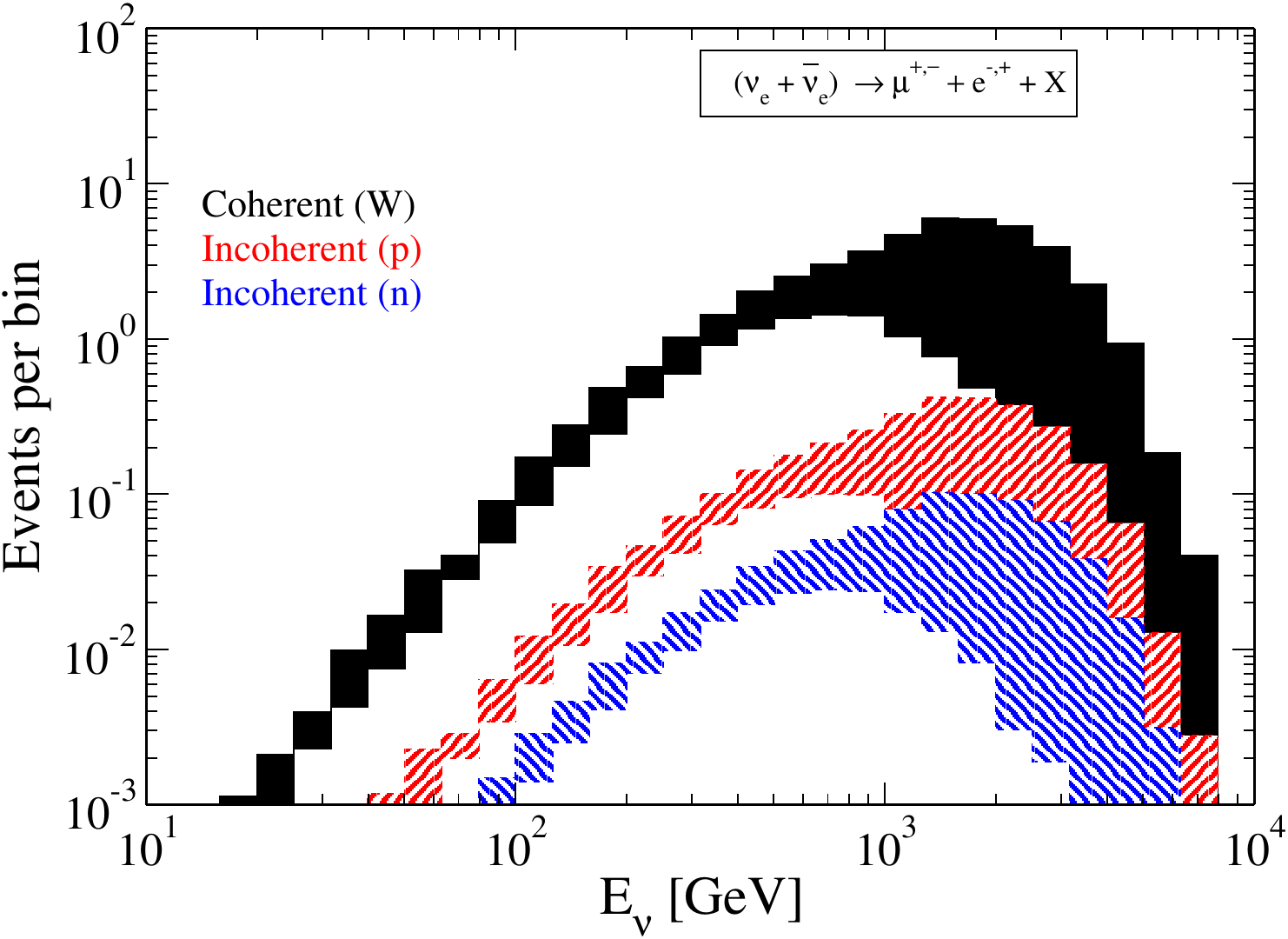} \\
\includegraphics[width=0.48\textwidth]{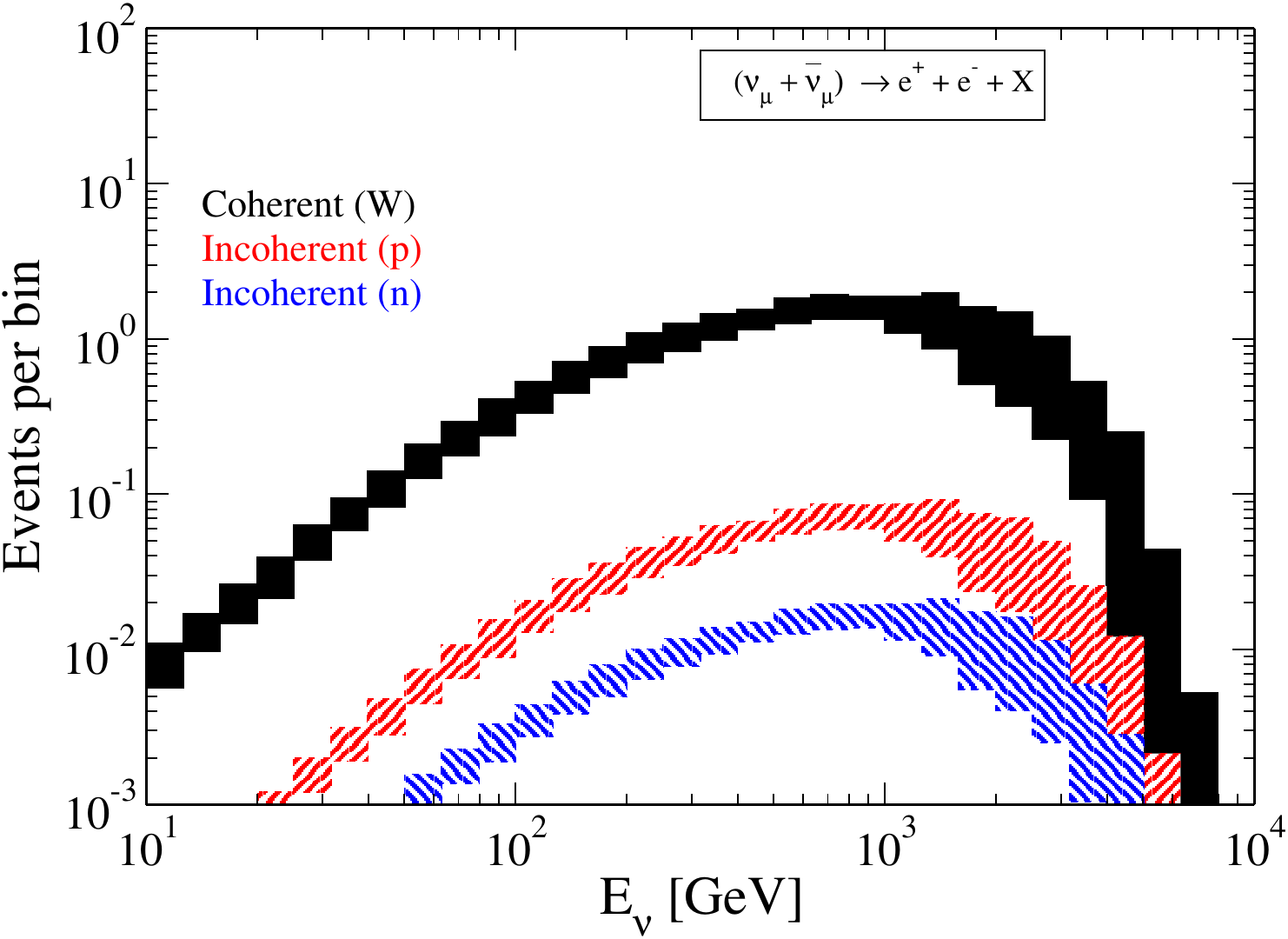} &
\includegraphics[width=0.48\textwidth]{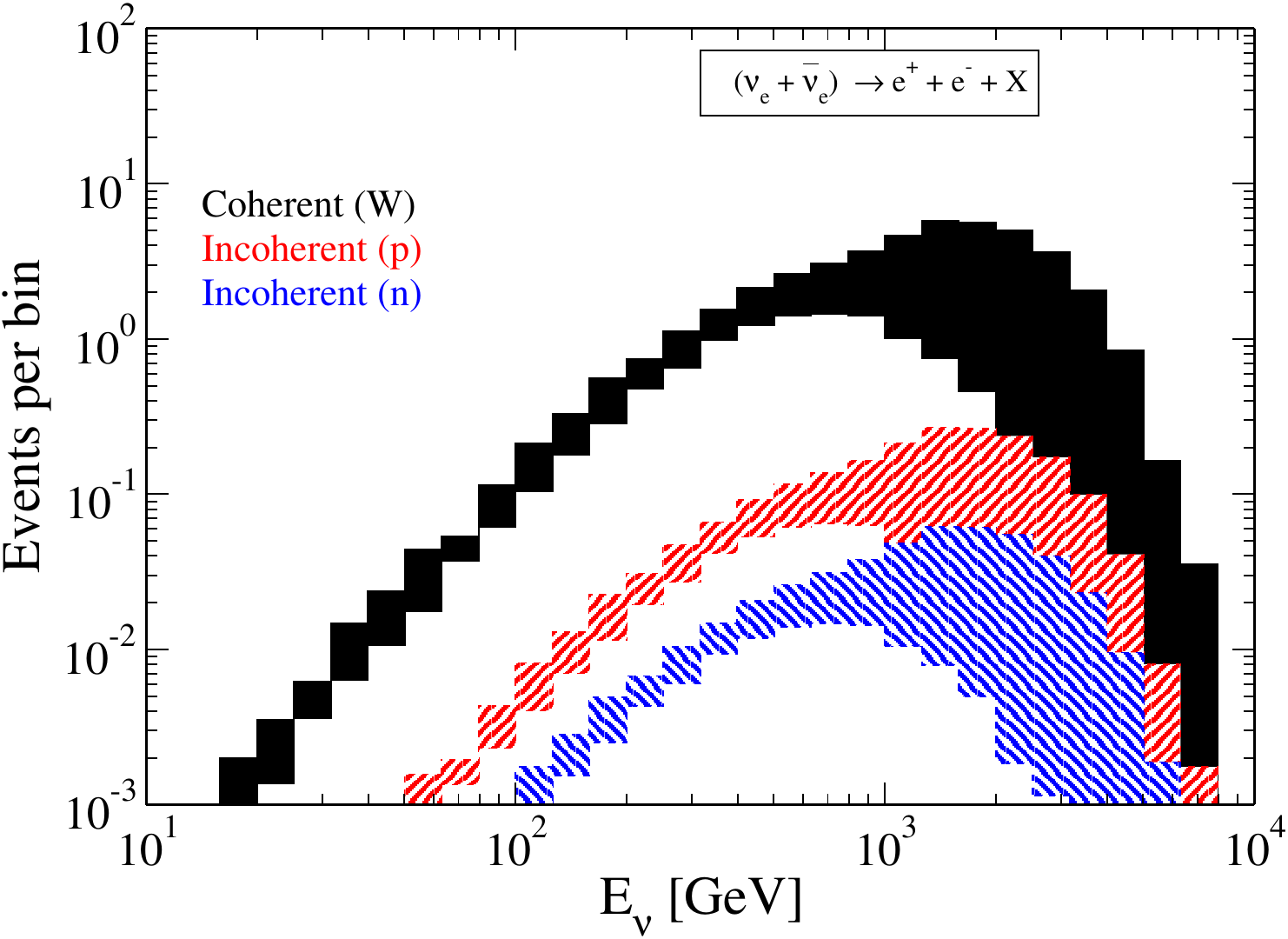} 
	\end{tabular}
\caption{ Number of trident events expected per bin at the FASER$\nu$2 detector, initiated by muonic (top) and electronic (bottom) neutrino interactions, derived assuming the expected luminosity for the LHC high-luminosity period of $3\, \mathrm{ab}^{-1}$. Our results are for different leptons in the final state: muon pairs (top), muon plus electron (middle), and electron pairs (bottom). The uncertainty band was constructed assuming different Monte Carlo generators for the incident neutrino flux. }
\label{fig_3:events1}
\end{figure}

Figure \ref{fig_3:events1} shows the results for the number of events expected per bin for trident scattering at FASER$\nu$2 during the high-luminosity regime of the LHC, assuming a time-integrated luminosity of 3~ab$^{-1}$. The top panel shows muon pairs, the middle panel muon-electron pairs, and the bottom panel electron pairs in the final state. On the left, the results are for events induced by muonic neutrinos, while on the right, they are from electronic neutrinos. Additionally, we sum the contributions of neutrinos and antineutrinos. Our results for the number of events indicate the dominance of events induced in coherent scattering, these being at least an order of magnitude greater than incoherent scattering. The uncertainty bands in the histograms were constructed with different Monte Carlo generators for the neutrino flux. The bands become larger for events induced by more energetic neutrinos (> 1 TeV), as the differences in flux models become greater for more frontal regions. The uncertainty bands also become larger for electron neutrino events, as these have a greater contribution from the flux resulting from the decay of heavy hadrons, which also differ more between the models adopted here.

\begin{figure}
	\centering
	\begin{tabular}{ccccc}
\includegraphics[width=0.6\textwidth]{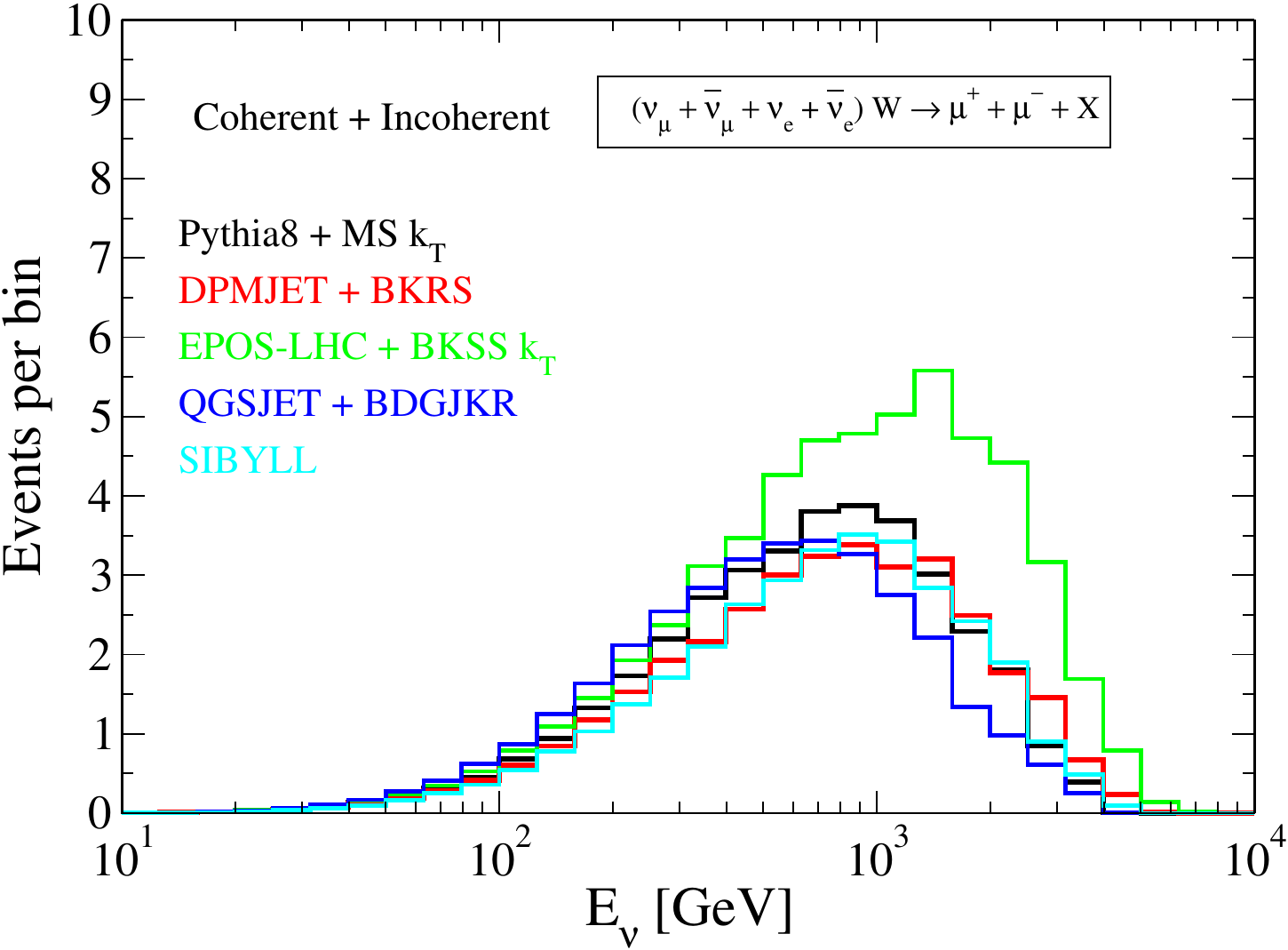} \\
\includegraphics[width=0.6\textwidth]{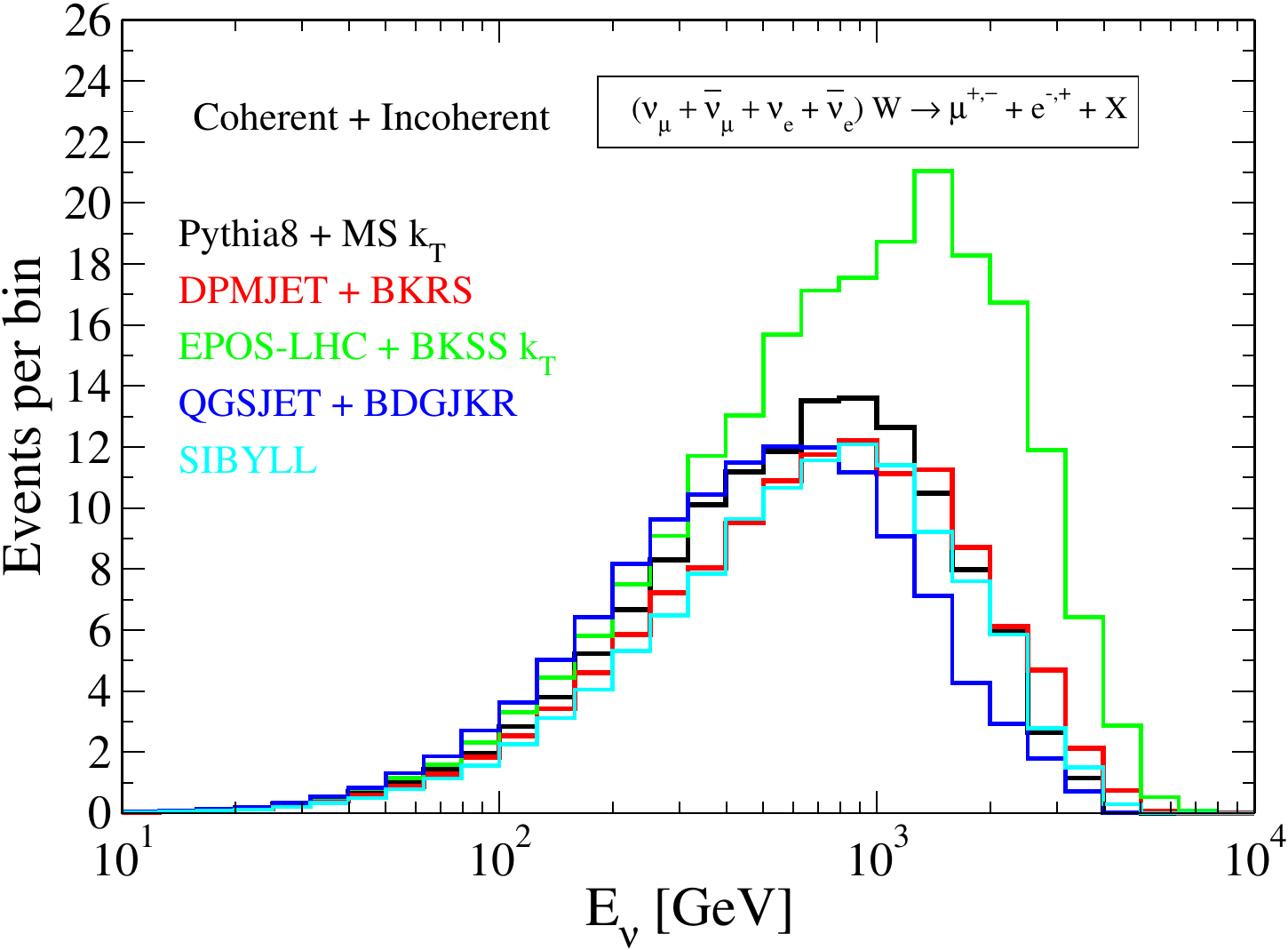} \\
\includegraphics[width=0.6\textwidth]{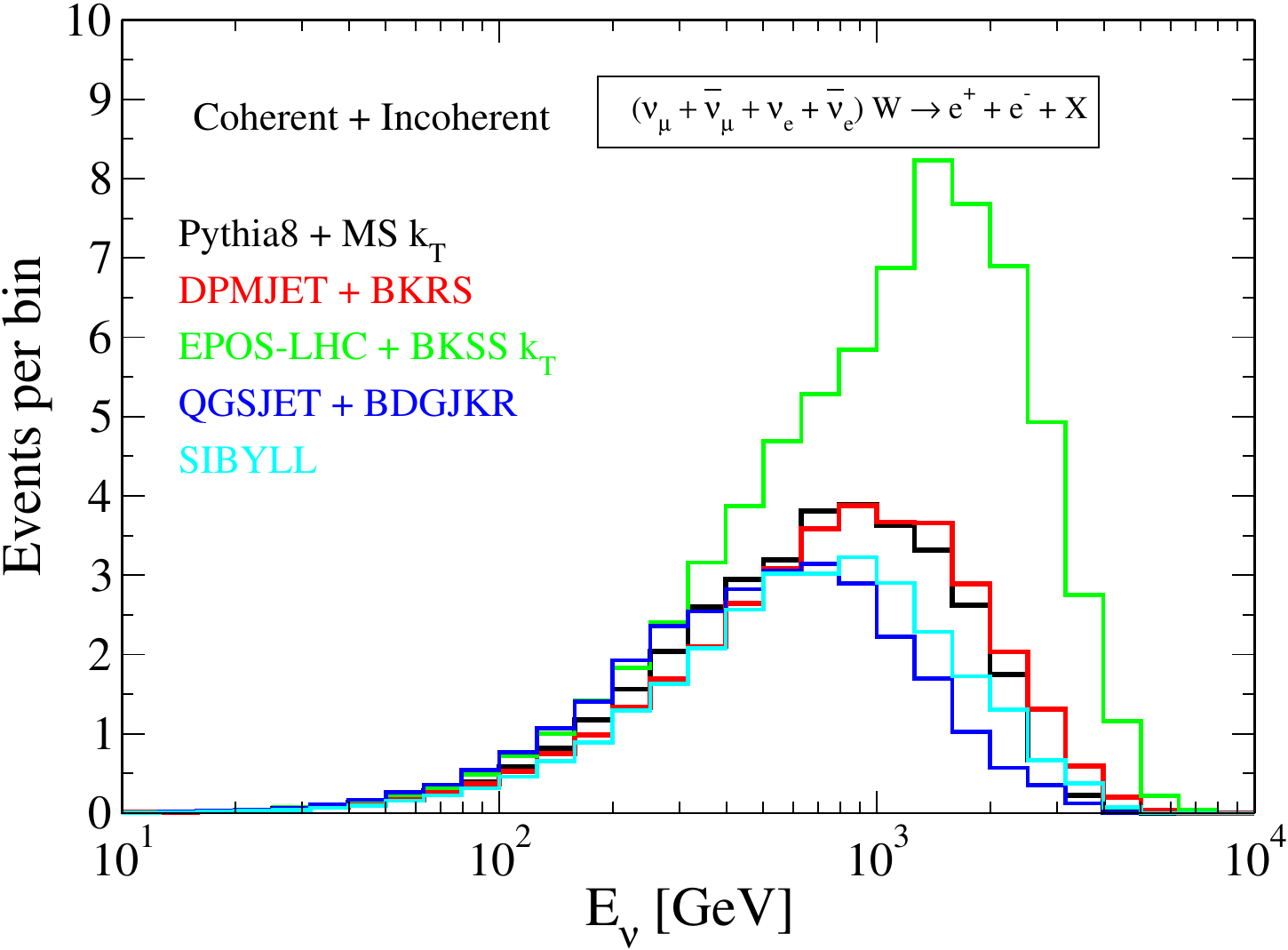} 
	\end{tabular}
\caption{ Total number of neutrino trident events at FASER$\nu$2 for different final states, estimated considering different Monte Carlo generators for the neutrino flux. Results for muon pairs (top), muon plus electron (center), and electron pairs (bottom).  }
\label{fig_3:events2}
\end{figure}

Figure \ref{fig_3:events2} shows the total contribution to the channels of the different final states of neutrino trident scattering at FASER$\nu$2 during the high-luminosity regime of the LHC, with different combinations of Monte Carlo generators for the neutrino flux. Our results indicate that the number of events per bin is highly modified by the different neutrino flux models, with the result using EPOS-LHC combined with BKSS $k_T$ differing most from the others for the neutrino flux components arising from light and heavy mesons, respectively. There are about 4 (14) events per bin for the regions with more events in the case of electron pairs and muon pairs (muon plus electron) produced, which indicates that this process can be explored in the future at the FASER$\nu$2.

Summing the events per bin in Figure \ref{fig_3:events2}, we obtain the total expected number of events at FASER$\nu$2 during the high-luminosity regime of the LHC for each of the final state topologies using the different combinations of Monte Carlo generators, presented in Table \ref{table_3:Nevents}. The number of events varies between 33 and 55, 116 and 209, and 29 and 70 for muon pairs, muon plus electron, and electron pairs in the final state, respectively. The results for muon plus electron show a significantly higher number of events, even though for this channel there is only a contribution from the process with the exchange of the $W^{\pm}$ boson for the incident neutrino. For processes involving charged leptons of the same family, there is a contribution from both weak gauge bosons channels, but also destructive interference between these channels, resulting in a lower number of expected events.

	\begin{table}
	\centering
		\begin{tabular}{|c|c|c|c|c|}
			\hline 
			& $\mu^{+} + \mu^{-}$ & $\mu^{\pm} + e^{\mp}$ & $e^{+} + e^{-}$ \\
			\hline
Pythia + MS $k_T$ 	    & 37.0 & 134.2 & 36.1
\tabularnewline
DPMJET + BKRS 		    & 34.6 & 126.5 & 36.1
\tabularnewline
EPOS-LHC + BKSS $k_T$	    & 55.0 & 209.0 & 70.5
\tabularnewline
QGSJET + BDGJKR 	    & 34.4 & 124.0 & 29.6
\tabularnewline
SIBYLL 		       	    & 33.0 & 116.5 & 29.1
\tabularnewline
			\hline		
		\end{tabular}
		\caption{ Total number of trident events at FASER$\nu$2 by final state topology during the high-luminosity era of the LHC, estimated considering different Monte Carlo generators for the incident neutrino flux. }
		\label{table_3:Nevents}
	\end{table}

Our results indicate that the number of events at FASER$\nu$2 is reasonably large, greater than 30 for each topology, at the detector level. This motivates the investigation of the detector's efficiency for this process, to estimate how many of the expected events can be effectively detected. Such an analysis was developed in the reference \cite{Altmannshofer:2024hqd} for the detection of muon pairs in the final state. The authors showed that the trident process can be detected at FASER$\nu$2, and that the measurement may, for the first time, show separation of the process from the backgrounds with statistical significance greater than 5 $\sigma$.

\section{Beyond the Standard Model effects in the trident process }
\label{sec_tridente:Zprime}

The possibility of measuring trident scattering in future neutrino detectors at the FPF also motivates the search for physics effects beyond the Standard Model in this process. Reference \cite{Altmannshofer:2024hqd} showed that this process will be sensitive to variations in the coupling constant of leptons with weak bosons in regimes not yet explored in other detectors. The lepton coupling only with weak bosons present in the interaction at the two upper vertices of Figure \ref{fig_3:diagrams} also motivates the search for new gauge bosons from leptophilic theories, that is, gauge bosons beyond the Standard Model that can only interact with leptons. There are several works in the literature that expand the Standard Model with new symmetries; our interest here will be restricted to the $L_\mu - L_\tau$ theory \cite{He:1990pn,He:1991qd}, in which the charge of the theory is the difference between the leptonic numbers of the second and third generations of leptons, and which we studied in the context of IceCube in Chapter \ref{cap:Ice}. This new symmetry, if it exists, would imply a new massive and electrically neutral gauge boson, which we will call $Z'$. The $Z'$ boson would interact only with leptons of the second and third leptonic generations. There are also models involving the electron family: $L_e - L_\mu$ and $L_e - L_\tau$. What motivates us to investigate the effects of the $L_\mu - L_\tau$ theory, and not other similar theories, is the fact that this theory continues to have the mass and coupling parameter space of $Z'$ with fewer experimentally excluded regions, given the difficulty in producing and observing heavier leptons. Another motivation is that the flux of muonic neutrinos is dominant in FPF experiments, therefore neutrino events of this type will be the most common.

To investigate the possibility of searching for effects of the $L_\mu - L_\tau$ theory at FASER$\nu$2, we will initially discuss experimental cuts and methods for excluding backgrounds for observing the trident at FASER$\nu$2. In this work, we will only analyze the detection of muon pairs in the final state, since the muon is a particle that travels long distances in matter and is easily identified in detectors like FASER$\nu$.

The main channel that produces two muons in the final state is the charged current interaction of muonic neutrinos with nuclear targets, with the subsequent production of heavy mesons that promptly decay into muons. These channels are part of the DIS, and generally produce a large number of charged particles in the final state. The trident process with two muons in the final state produces only two charged particles in the final state (disregarding the target nucleus, which generally exchanges very little energy and does not excite the detector). Therefore, the first cut applied for background elimination of the DIS is to exclude processes with three or more particles detected in the final state. The second cut applied for separating DIS processes is to exclude processes where the two muons do not originate from the same interaction vertex. This excludes a large portion of DIS events, as the heavy mesons that decay into muons propagate a few millimeters before decaying, leaving a gap between the vertices of the two detected muons. We also applied cuts to the angular separation between the two muons in the final state, requiring it to be smaller than 0.1 rad, so that the two muons propagate greater distances in the detector and can be identified. These three experimental cuts discussed in this paragraph virtually eliminate all background events from the DIS, and practically do not exclude trident scattering events, as demonstrated in the reference \cite{Altmannshofer:2024hqd}.

Another process that can leave signals similar to trident scattering in the detector is the exclusive production of charged pions by muonic neutrino interactions. These processes produce only one muon in the final state, originating from the neutrino's vertex. However, given that the mass of pions is similar to that of muons, they can be mistakenly identified as muons. To exclude these processes and separate events from trident scattering, it is necessary to use the FASER2 spectrometer, which will be installed at the end of the 6.6~m tungsten plates of FASER$\nu$2. This spectrometer will be able to separate muons from pions, but for this to be possible, they must pass through all the tungsten and reach FASER2. Therefore, it is necessary to apply cuts to the minimum energy of both muons produced in the trident process, so that we only consider the processes in which both muons have enough energy to reach the spectrometer and be identified as muons. The efficiency of muon identification as a function of the energy we are using is shown in Figure 4 of the reference \cite{Altmannshofer:2024hqd}.

Assuming the experimental energy and angular separation cuts of the two muons, the reference \cite{Altmannshofer:2024hqd} shows that the trident events are orders of magnitude above the background of other processes. We implemented the same experimental cuts in our simulations to study their impact on the expected trident events at FASER$\nu$2. In Table \ref{table_3:Nevents2}, we show the events with two muons in the final state without experimental cuts in the left column, applying the cut at the maximum angular separation of the two muons in the final state in the middle, and applying the cuts of angular separation and minimum energy for the muon on the right. Our results indicate that the cut that most impacts the number of events is the one that imposes an efficiency in the detection of muons as a function of energy. However, we also found that, even after the experimental cuts, between 17 and 31 events remain in FASER$\nu$2 during the high-luminosity regime of the LHC. Given that the background is orders of magnitude lower, this number becomes sufficient for observing the process with more than 5$\sigma$ of statistical significance.

	\begin{table}
	\centering
		\begin{tabular}{|c|c|c|c|c|}
			\hline 
			& No cuts & Angular separation  &  $\theta \leq 0.1$ rad + \\
            		&  &  $\theta \leq 0.1$ rad & Minimum muon energy cut \\
			\hline
Pythia8 + MS $k_T$ 	    & 36.3 & 32.4 & 19.6
\tabularnewline
DPMJET + BKRS 		    & 34.0 & 30.4 & 18.7
\tabularnewline
EPOS-LHC + BKSS $k_T$ 	& 54.2 & 48.7 & 30.8
\tabularnewline
QGSJET + BDGJKR 	    & 33.8 & 29.9 & 17.4
\tabularnewline
SIBYLL 		       	    & 32.4 & 29.0 & 17.8
\tabularnewline
			\hline		
		\end{tabular}
		\caption{ Total number of trident events at FASER$\nu$2 for muon pairs in the final state during the high-luminosity regime of the LHC, estimated considering different Monte Carlo generators for the neutrino flux. We present the total number of events in the detector and the number expected to be observed after experimental selection and process identification cuts.
         }
		\label{table_3:Nevents2}
	\end{table}

With the number of events at FASER$\nu$2 after the detector efficiency analysis, we are able to estimate the sensitivity of this detector to the $L_\mu - L_\tau$ theory. The first step is to obtain the cross section and number of events for different combinations of mass and coupling of the $Z'$ boson. We did this using the Monte Carlo generator built and presented in the reference \cite{Altmannshofer:2019zhy}, where the authors study the possibility of detecting this process at DUNE. In the reference \cite{Shimomura:2020tmg}, the authors show that we can estimate the cross section of the neutrino trident with the Standard Model plus $Z'$ by modifying the cross section of the Standard Model with the substitution
\begin{eqnarray}
g_{L(R)} \rightarrow g_{L(R)} \mp \frac{\sqrt{2}}{4G_F} \frac{g'}{q_{Z'}^{2}-m_{Z'}^{2}} \, ,
\label{eq_cap1:subsZprime}
\end{eqnarray}
where $g_{L(R)}$ are the Standard Model couplings and $g'$ the coupling of the $Z'$ with mass $m_{Z'}$ and four-momentum $q_{Z'}$.

With the number of trident scattering events in the Standard Model and assuming the $L_\mu - L_\tau$ theory, we can estimate the detector sensitivity to exclude or validate this theory as a function of the mass and coupling of $Z'$. For the statistical analysis, we are considering that the number of events follows a Poisson distribution for each new parameter, which implies the likelihood function given by
\begin{eqnarray}
L(\vec{\theta}) = \prod_{i}^{N}\frac{\mu_{i}^{n_{i}}\mathrm{e}^{-\mu_{i}}}{n_{i}!} \, ,
    \label{eq_cap3:likelihood}
\end{eqnarray}
where $\vec{\theta}$ is the set of parameters that maximize the likelihood, $\mu_i$ is the number of events expected in the theory that adds physics beyond the Standard Model, and $n_i$ is the number of events expected in the Standard Model for bin $i$. Using the Neyman-Pearson lemma \cite{Neyman:1933wgr} for the statistical validation test of the model, $\lambda = -2\,\mathrm{ln}[L(\vec{\theta}_1)/L(\vec{\theta}_2)]$, the best curve for the pseudodata $\vec{\theta}$ will be those that minimize $\lambda$, which is given by
\begin{eqnarray}
\lambda = 2\sum_{i}^{N}\left[ \mu_{i} - n_{i} + n_{i}\,\mathrm{ln} \left( \frac{n_{i}}{\mu_{i}} \right) \right] \, .
    \label{eq_cap3:chi2}
\end{eqnarray}
Finally, we assume the validity of Wilks' theorem \cite{Wilks:1938dza}, which implies that $\lambda$ follows a $\chi^2$ distribution.

\begin{figure}
	\centering
	\begin{tabular}{ccccc}
\includegraphics[width=0.8\textwidth]{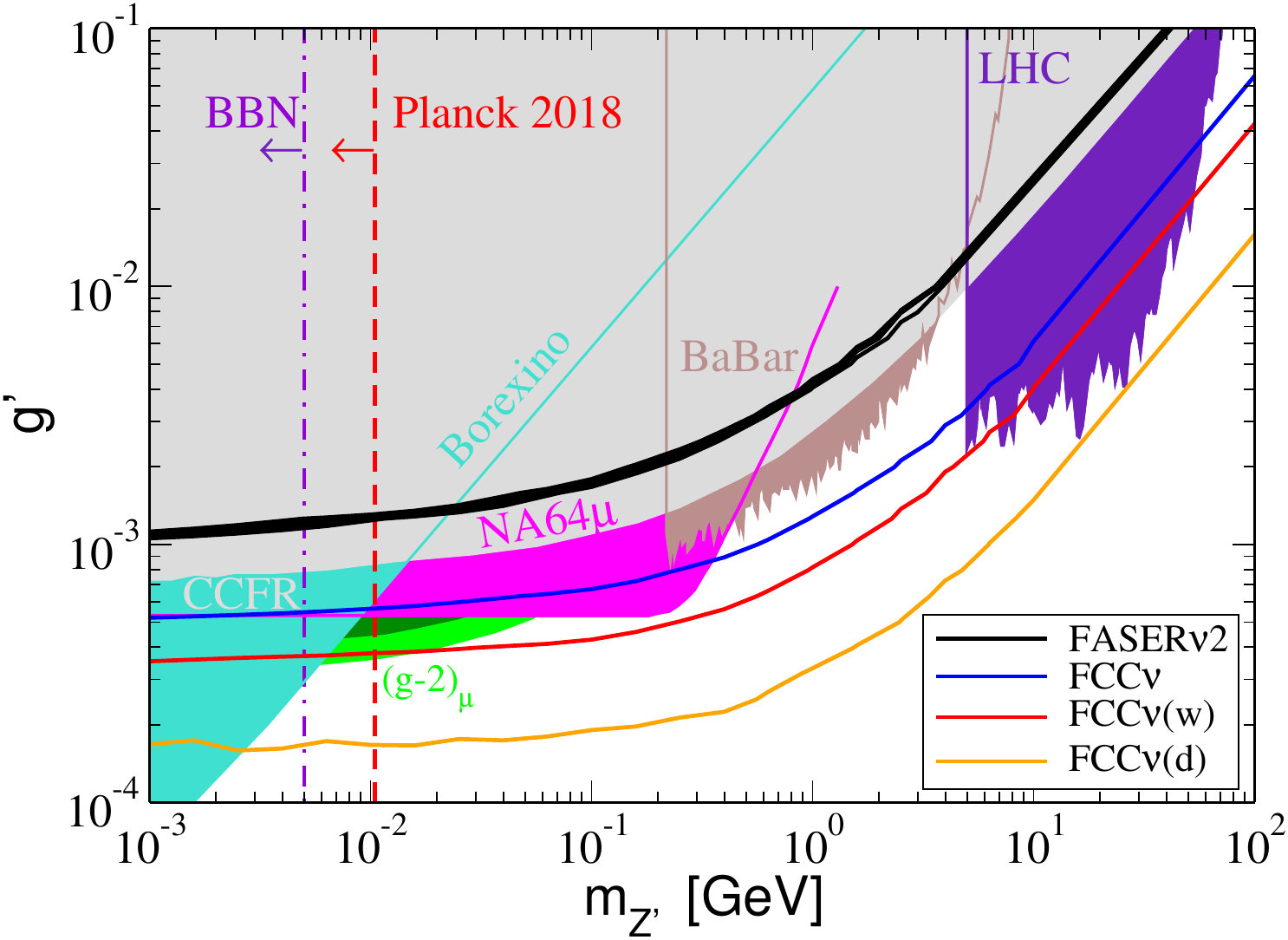} 
	\end{tabular}
\caption{ Sensitivity of the FASER$\nu$2 detector at the LHC and FCC for neutrino trident events associated with the $Z'$ gauge boson predicted by the $L_\mu - L_\tau$ model. Results, at $2\sigma$ level, are derived considering different Monte Carlo generators for the incident neutrino flux. For comparison, existing constraints from other processes and experiments are also presented. The green bands represent the parameter space in which the presence of a $Z'$ solves the anomalous magnetic moment anomaly of the muon at the $1\sigma$ and $2\sigma$ levels. The LHC lane represents the region excluded in $pp$ collisions at the LHC, considering the analysis carried out by the CMS \cite{CMS:2018yxg} and ATLAS \cite{ATLAS:2023vxg,ATLAS:2024uvu} collaborations on the production of $Z^{0}$ and $W^{\pm}$ bosons and their subsequent decay in the $Z^{0} \rightarrow Z' \mu^{\pm} \mu^{\mp} \rightarrow \mu^{\pm} \mu^{\mp} \mu^{\pm} \mu^{\mp}$ and $W^{\pm} \rightarrow Z' \mu^{\pm} \nu_{\mu} \rightarrow \mu^{\pm} \mu^{\mp} \mu^{\pm} \nu_{\mu}$ channels. The BaBar band represents the excluded region associated with the study of the decay process of the $Z^{0}$ boson \cite{BaBar:2016sci}. We also included in the figure the excluded regions from the Borexino experiment on neutrino-electron interactions \cite{Bellini:2011rx,Harnik:2012ni}, which were constructed considering that the interaction rate should not be more than $8\%$ greater than the Standard Model prediction. The magenta region presents the recent results of the NA64$\mu$ experiment with 90\% CL \cite{NA64:2024klw}. The Big Bang Nucleosynthesis (BBN) restrictions derived in \cite{Escudero:2019gzq} and Planck 2018 \cite{Planck:2018vyg,Ghosh:2024cxi}, which imply that $m_{Z'} \gtrsim 5 \,\mathrm{MeV}$ and $m_{Z'} \gtrsim 10 \,\mathrm{MeV}$, respectively, are also presented. Finally, the region excluded by the CCFR experiment in the neutrino trident process with two muons in the final state is also presented \cite{CCFR:1991lpl,Altmannshofer:2014pba}. }
\label{fig_3:mapa1}
\end{figure}

In Figure \ref{fig_3:mapa1} we present our results for the sensitivity of FASER$\nu$2 during the high-luminosity regime of the LHC for the $L_\mu - L_\tau$ theory via the neutrino trident process with muon pairs in the final state. Our results are presented together with the experimental constraints for the parameter space already excluded by other experiments with a 95\% confidence level. More details of these processes are in the figure caption. Our results form a narrow band, shown in the solid black line, for the limit of the region that will be excluded by FASER$\nu$2. This band was obtained using the different Monte Carlo generators mentioned earlier for the incident neutrino flux, and given that each generator predicts a slightly different number of events, the generators that predict more events also predict a slightly higher sensitivity to physics beyond the Standard Model. Our results indicate that FASER$\nu$2 will not be able to exclude any new regions of the parameter space that have not yet been excluded by other experiments. This low sensitivity is a result of the still low number of events, on the order of 20, for the trident in FASER$\nu$2.

The lack of sensitivity of FASER$\nu$2 during the high-luminosity regime of the LHC motivates us to study in which scenarios of future colliders proposed to operate at CERN there will be sensitivity to the $L_\mu - L_\tau$ theory via the trident process. Recent studies show that the proposed Future Circular Collider (FCC), a collider that could replace the LHC in the coming decades, will produce even higher neutrino fluxes than the LHC, both by increasing the center-of-mass energy in proton-proton collisions and by further increasing the expected luminosity for the high-luminosity regime of the LHC \cite{MammenAbraham:2024gun}.

In our analysis, we will consider neutrino detectors at the FCC using FASER$\nu$2 as a baseline: we will assume the same material and the same detection efficiency as at the LHC. The neutrino fluxes for these possible detectors were simulated in the reference \cite{MammenAbraham:2024gun}, where the authors considered \textit{pp} collisions in the FCC with a center-of-mass energy of 100 TeV and an integrated luminosity of 30 ab$^{-1}$. The neutrino detectors would be located 1500 meters from the proton collision point. We will consider three FASER$\nu$2 type detectors in our analysis: (a) identical to FASER$\nu$2, called FCC$\nu$, (b) with a ten times increase in cross-sectional area, called FCC$\nu$(w), and (c) with a ten times increase in detector length, called FCC$\nu$(d).

In Table \ref{table_3:Nevents_FCC} we show our results for the number of events expected at FASER$\nu$2 at the FCC. The values are for events with muon pairs in the final state, presented with and without the experimental cuts discussed above. Furthermore, we consider the configurations of FASER$\nu$2 proposed for operation at the LHC, as well as versions with larger area and length.

In Figure \ref{fig_3:mapa1} we present the sensitivity to the $L_\mu - L_\tau$ theory through the trident process in these proposed detectors. Our results indicate that, unlike the FASER$\nu$2 at the LHC, the FASER$\nu$2 at the FCC will be able to test new regions of the $Z'$ parameter space of this beyond the Standard Model theory. The new regions covered compared to current experiments are seen both in regions of lower $Z'$ masses (approximately 1 GeV) and in the region of mass greater than the masses of the $W^{\pm}$ and $Z^0$ bosons of the electroweak theory. We also see that the ideal scenario for searching for BSM effects in this process would be with a FASER$\nu$2 with an upgrade in its length, which would allow capturing the most intense part of the neutrino flux and, consequently, observing more trident scattering processes.

\begin{center}
	\begin{table}[t]
		\begin{tabular}{|c|c|c|c|c|}
			\hline 
			& No cuts & Angular separation  &  $\theta \leq 0.1$ rad + final muon energy cut \\
            		&  &  $\theta \leq 0.1$ rad &   \\
			\hline
FCC$\nu$                & 6305.5 & 5799.3 & 4078.8
\tabularnewline
FCC$\nu$(w)             & 19006.9 & 17363.8 & 11840.4
\tabularnewline
FCC$\nu$(d)             & 63220.6 & 58148.6 & 40907.4
\tabularnewline
            \hline
		\end{tabular}
		\caption{ Total number of trident events at FASER$\nu$2 for muon pairs in the final state during FCC operation. We present the total number of events in the detector and the number expected to be observed after experimental selection and process identification cuts. }
		\label{table_3:Nevents_FCC}
	\end{table}
\end{center}

\section{Conclusions}
\label{sec_tridente:conclusion}

In this chapter, we investigate the possibility of observing the neutrino trident scattering in detectors proposed for CERN. The trident process has about 100 candidate events, all from observations made in the last century, but none with sufficient statistics for the discovery of the process. Our results indicate that this process can be discovered at the FASER$\nu$2 experiment during the high-luminosity regime of the LHC. We also show that the detection of this process in experiments like FASER$\nu$2 operating at the FCC can be sensitive to physics beyond the Standard Model, such as the $L_\mu - L_\tau$ theory, which predicts the existence of the $Z'$ gauge boson. In the next chapter, we will explore the muon-initiated trident process at FASER$\nu$.

\chapter{Muon trident scattering at the LHC energy regime}	
\label{cap:MuonTridente}

The production of lepton pairs has been extensively studied experimentally in hadron-hadron collisions at the RHIC \cite{STAR:2004bzo} and at the LHC \cite{ALICE:2013wjo,Dyndal:2017wbv,ATLAS:2022yad,ALICE:2023mfc}. This process is of particular interest because it is useful for probing the charge distribution in the nucleus \cite{Azevedo:2019fyz}, validating some approaches such as the Equivalent Photon Approximation (EPA) \cite{Jentschura:2009mb}, and also for searching for scenarios beyond the Standard Model \cite{Duarte:2024zjh}. As a notable result, the ATLAS and CMS collaborations observed the production of tau pairs through the two-photon fusion channel in $PbPb$ \cite{ATLAS:2022ryk,CMS:2022arf} and $pp$ \cite{CMS:2024qjo} collisions, and significantly constrained the anomalous magnetic moment of the tau lepton.

In recent decades, several groups have conducted studies of the trident process from a theoretical \cite{Bjorken:1966kh,Tannenbaum:1968zz,Albright:1977zn,Albright:1977rk,Albright:1978mg,Ganapathi:1978qm,Ganapathi:1979mn,Ganapathi:1979dj,Barger:1979eg,Ganapathi:1980wh,Kelner:1998mh,Altmannshofer:2019zhy,Ballett:2018uuc,Zhou:2019vxt,Zhou:2019frk,Francener:2024wul,Altmannshofer:2024hqd,Abbiendi:2024swt,Lee:2023cwb,Bigaran:2025vea,Francener:2024jra} and experimental \cite{Roe:1959zz,Russell:1971mf,CHARM-II:1990dvf,CCFR:1991lpl,NuTeV:1999wlw,Maciuc:2006xb,SNDLHC:2024bzp} point of view. In other words, they investigated the production of a pair of charged leptons by the interaction of lepton beams on a hadronic target. As in the case of dilepton production in hadronic interactions, these studies are also motivated by the possibility of testing the Standard Model and searching for signs of new physics. The interest in this process has recently been renewed with the discovery of collider neutrinos by the FASER \cite{FASER:2023zcr,FASER:2024hoe,FASER:2024ref} and SND@LHC \cite{SNDLHC:2023pun} collaborations, which verified that an intense and collimated flux of neutrinos is produced in the forward direction of $pp$ collisions at the LHC, as predicted in the reference \cite{DeRujula:1984ns}. In particular, in Chapter \ref{cap:tridente} we estimate the production of neutrino tridents, a weak process characterized by the production of a pair of charged leptons through neutrino scattering in the Coulomb field of a heavy nucleus, considering neutrino scattering in tungsten in the detectors planned for the LHC in the future, and we demonstrate that a future observation of this rare process is feasible.

In addition to the neutrino flux, $pp$ collisions at the LHC also generate a muon flux that can reach the LHC's forward detectors, as we presented and discussed in Chapter \ref{cap:MuonDIS}. The measurement of this flux by the SND@LHC detector \cite{SNDLHC:2023mib} inaugurated a new era in the study of deep inelastic muon scattering in nucleus \cite{Francener:2025pnr,Francener:2025tyh}, which will allow us to improve the description of the hadronic structure and constrain the magnitude of nuclear effects in a new kinematic range. Furthermore, several recent works have explored the potential of this muon flux for the search for signals beyond the Standard Model \cite{Ariga:2023fjg,Batell:2024cdl,MammenAbraham:2025gai}. Motivated by these results, in this chapter, we will investigate the muon trident process, which is the electromagnetic production of a dilepton pair in muon-nucleus scattering, the results of which are presented in reference \cite{Francener:2025wzh}. In particular, we will consider the production of $e^+ e^-$, $\mu^+ \mu^-$ and $\tau^+ \tau^-$ in muon-tungsten scattering in the existing FASER$\nu$ detector and in its proposed upgrade to be installed at the Forward Physics Facility \cite{Anchordoqui:2021ghd,Feng:2022inv,FPFWorkingGroups:2025rsc,FPF:2025bor}, called the FASER$\nu$2 detector.

Electron-positron pairs have been observed in muon-nucleus interactions since the 1950s \cite{Roe:1959zz}, while muon pairs have only been observed at Brookhaven National Laboratory \cite{Russell:1971mf}, by the ALEPH detector \cite{Maciuc:2006xb} and, more recently, by the SND@LHC collaboration \cite{SNDLHC:2024bzp}. On the other hand, tau pairs have never been observed in muon-nucleus interactions to date.

As we will demonstrate below, the predicted event rates for the production of $e^+ e^-$ and $\mu^+ \mu^-$ are very high, especially for the FASER$\nu$2 detector, which will allow us to constrain the magnitude of the associated cross sections, which are important, for example, to estimate the electromagnetic energy loss of a muon in transit through materials \cite{Bulmahn:2008fa}. Furthermore, our results indicate that a future measurement of the electromagnetic production of $\tau^+ \tau^-$ will, in principle, be feasible at FASER$\nu$ during Run 3 of the LHC.

\section{Cross sections for muon trident scattering}

In this section, we will present a brief review of the formalism necessary to describe the electromagnetic production of a dilepton pair in a muon-nucleus scattering reaction at LHC energies. This process is represented by the reaction:
\begin{eqnarray}
\mu^{\pm}_{i} + A \rightarrow \mu^{\pm}_{f} + A + l^{+} + l^{-},
\label{eq_MuonTrident:reaction}
\end{eqnarray}
where $\mu^{\pm}_i$ ($\mu^{\pm}_f$) is the muon in the initial (final) state and $l= e, \mu, \tau$ is the lepton produced in the process. We will restrict our analysis to coherent scattering, where the leptonic system scatters in the entire nucleus and the nucleus remains intact in the final state. In this case, the cross section is proportional to the square of the nuclear charge $Z^{2}$. Our results will not take into account the contribution associated with incoherent scattering with individual nucleons within the nucleus, since its contribution is proportional to $Z$ and is therefore expected to be subdominant. In principle, these two contributions can be experimentally separated, since coherent scattering should be characterized by a clean environment in the final state, i.e., without hadronic activity, in contrast to the incoherent case, where the nucleus fragments. For a more detailed discussion of the incoherent contribution to the neutrino trident case, see, for example, the reference ~\cite{Francener:2024wul}.

\begin{figure}[t]
	\centering
	\begin{tabular}{ccc}
    \includegraphics[scale=0.6]{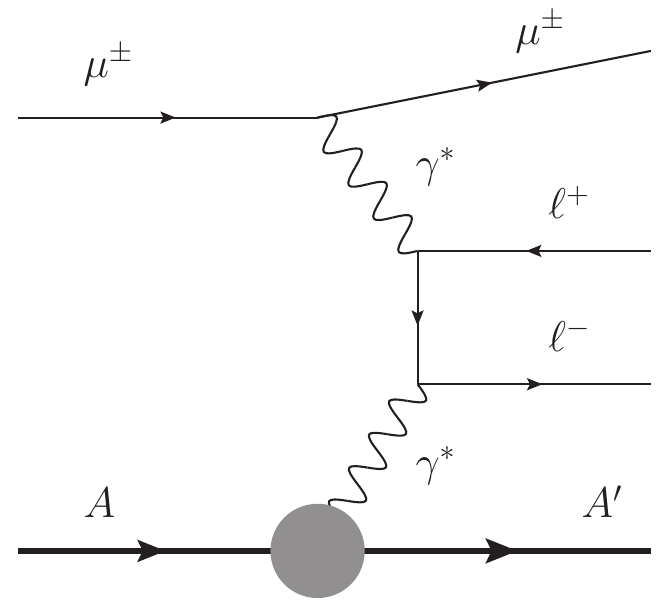} \;\;\;\;\;\;\;\;\;
    \includegraphics[scale=0.6]{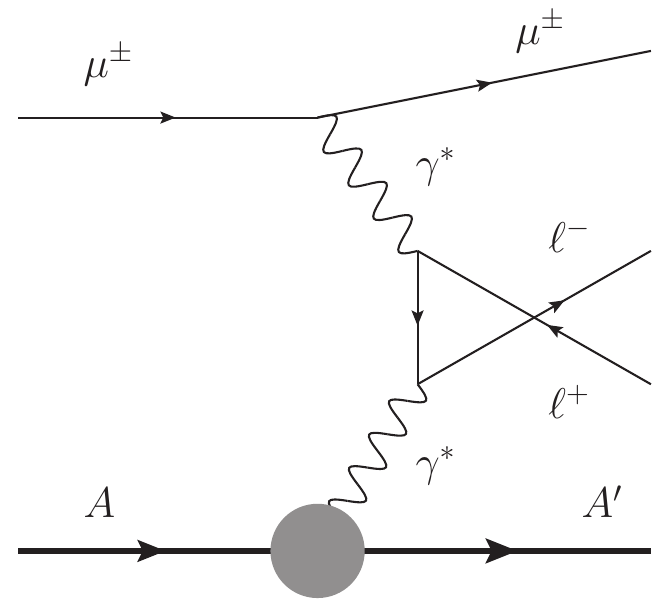} \\
    \includegraphics[scale=0.6]{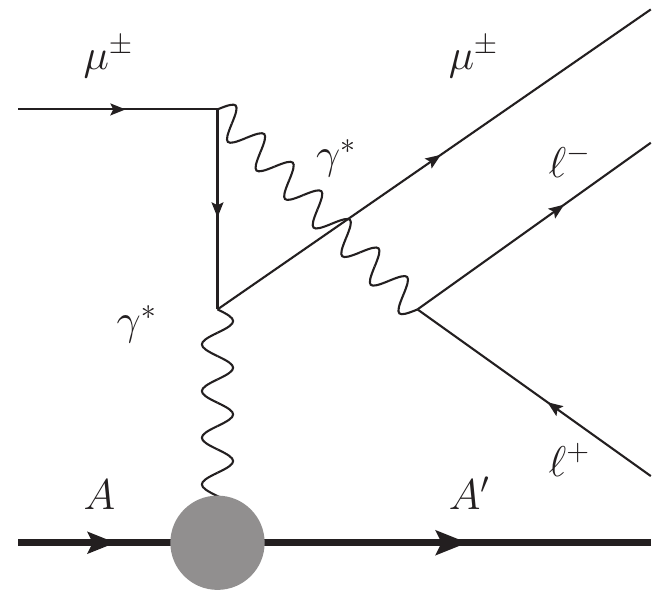} \;\;\;\;\;\;\;\;\;
    \includegraphics[scale=0.6]{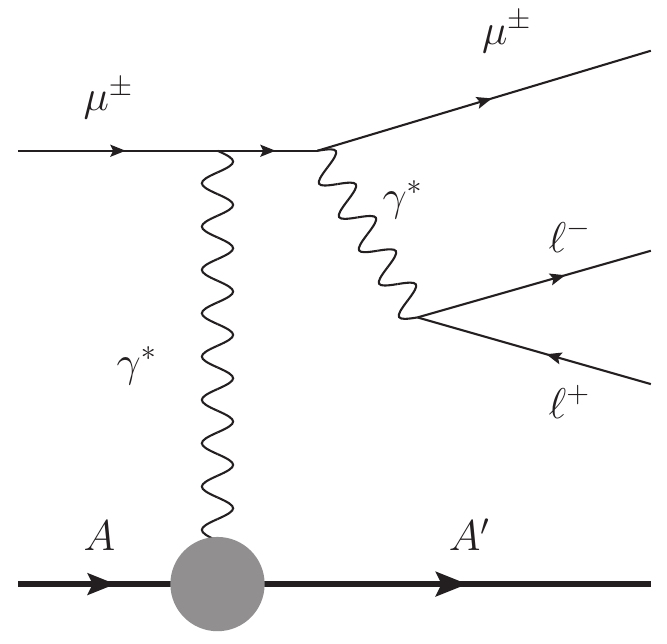}
    \end{tabular}
    \caption{ Leading order diagrams for the electromagnetic production of a lepton pair in muon-nucleus scattering. The lepton pair can be produced by the Bethe-Heitler process (upper diagrams) and by the muon bremsstrahlung channel (lower diagrams). }
    \label{fig_MuonTrident:diagramas}
\end{figure}

At leading order, lepton pairs can be created electromagnetically by the Bethe-Heitler process  \cite{Bethe:1934za} (photon-photon fusion) and by bremsstrahlung of the interacting particles (muon and nucleus). Previous studies have shown that the contribution of nuclear bremsstrahlung is negligible. Consequently, in our calculations, we will only consider the contributions of the diagrams represented in Figure \ref{fig_MuonTrident:diagramas}. Due to charge conjugation invariance, there is no interference between the Bethe-Heitler and bremsstrahlung diagrams in the total cross section, but it is present in the differential distributions.

Experimentally, the muon trident process is characterized by a primary track associated with the incident muon and three nearly collinear tracks for the emergent leptons ($\mu_f^{\pm} l^+l^-$). This signature arises from the small scattering deflection, along with the low momentum transfer to the nucleus. It is important to emphasize that a similar topology can also arise from the combination of the photon emission process ($\mu^{\pm}_{i} + A \rightarrow \mu^{\pm}_{f} + A + \gamma$) followed by conversion ($\gamma \rightarrow l^+l^-$). In this process, when the photon has an interaction length of a few millimeters in matter, it would be possible to separate its contribution if we could adequately reconstruct the interaction vertex. However, such an analysis is beyond the scope of this study.


Let us now focus on the muon trident process at the LHC far-forward detectors. In particular, we will consider FASER$\nu$ and its proposed upgrade, FASER$\nu$2. FASER$\nu$ operates with 1.1 metric tons of tungsten distributed across 730 plates measuring 25 cm $\times$ 30 cm $\times$ 1.1 mm \cite{FASER:2018bac,FASER:2019dxq,FASER:2020gpr,FASER:2022hcn}. In our analysis, we will define the detector target length as 50 cm, since muon identification, as well as measurements of the momentum of incident and emergent muons, require propagation over a few centimeters, as we also did in Chapter \ref{cap:MuonDIS}. We will assume an integrated luminosity of 250 fb$^{-1}$, which is expected for data collection by FASER$\nu$ during Run 3 of LHC operation. It is important to note that the FASER$\nu$ detector will also operate during Run 4 with a total integrated luminosity of 1000 fb$^{-1}$. In contrast, FASER$\nu$2 should operate during the high-luminosity era of the LHC at the Forward Physics Facility \cite{Anchordoqui:2021ghd,Feng:2022inv,FPFWorkingGroups:2025rsc,FPF:2025bor}, and will be characterized by approximately 20 metric tons of tungsten and a time-integrated luminosity of 3 ab$^{-1}$. Next, we will discuss the production of muon tridents at FASER$\nu$, but the formalism can be directly extended to other LHC frontal detectors.

The associated number of muon trident events at the FASER experiment can be expressed as \cite{Francener:2025pnr}
\begin{eqnarray}
N_{\mathrm{events}} = \int^{1}_{0} \mathrm{d} x_{\mu_i^{\pm}} f(x_{\mu_i^{\pm}}) \sigma_{\mu^{\pm} + W} (E_{\mu^{\pm}_{i}}) \, ,
\label{eq_MuonTrident:events}
\end{eqnarray}
where $\sigma_{\mu^{\pm} + W} (E_{\mu^{\pm}})$ is the muon-tungsten cross section for the process of interest, $f(x_{\mu^{\pm}})$ is the PDF of the muon flux, $E_{\mu^{\pm}_{i}}$ is the energy of the incident muon, and $x_{\mu^{\pm}_{i}} = E_{\mu^{\pm}_{i}} / E_{p}$, where $E_p$ is the energy of the proton that collides at the ATLAS interaction point. The muon PDF $f(x_{\mu_i^{\pm}})$ is the muon flux that reaches the detector and contains information associated with the detector geometry and its exposure time for data collection. For this quantity, we will use the muon and antimuon fluxes in the forward direction of ATLAS simulated with FLUKA in the references \cite{Sabate-Gilarte:2023aeg,Battistoni:2015epi}. Recently, the reference ~\cite{Francener:2025pnr} provided the PDF of the muon flux in LHAPDF format \cite{Buckley:2014ana}.

The muon-tungsten cross section for the trident process, Equation~(\ref{eq_MuonTrident:reaction}), will be estimated using a modified version of the event generator presented in \cite{Altmannshofer:2019zhy} for the neutrino trident process. This generator has already been adapted for a tungsten target in the reference ~\cite{Francener:2024wul}, where we also implemented a more precise description of the nuclear form factor, which is expressed in terms of the Fourier transform of the Woods-Saxon nuclear charge distribution \cite{Woods:1954zz}, with the parameterization provided in \cite{DeVries:1987atn}. For the analysis performed in this paper, we generalized this improved version of the Monte Carlo generator by including electromagnetic interactions initiated by a charged lepton.

The muon-target cross section for trident scattering initiated by a four-momentum (energy) muon $p_1$ ($E_\mu$) with a four-momentum (mass) hadronic target $P$ ($M_P$) can be calculated with
\begin{eqnarray}
\begin{aligned}
\sigma_{\mathrm{trident}} = & \int \frac{1}{M_P E_\mu} |\mathcal{M}|^{2} \frac{\mathrm{d^3}\vec{p}_2}{(2\pi)^{3}2 E_2} 
\frac{\mathrm{d^3}\vec{p}_3}{(2\pi)^{3}2 E_3} \frac{\mathrm{d^3}\vec{p}_4}{(2\pi)^{3}2 E_4} \frac{\mathrm{d^3}\vec{P}'}{(2\pi)^{3}2 E'}  \\
& (2\pi)^{4} \delta (p_1 + P - p_2-p_3-p_4-P') \, ,
 \label{eq_3:cs}
\end{aligned}
\end{eqnarray}
where $p_2$ ($E_2$), $p_3$ ($E_3$) and $p_4$ ($E_4$) are the momenta (energies) of the leptons present in the final state, $P'$ ($E'$) is the momenta (energy) of the final nucleus, and $\mathcal{M}$ is the invariant matrix element of the process. We are obtaining the cross section taking into account all the masses of the leptons, and the matrix element was calculated with the FeynCalc package \cite{Mertig:1990an,Shtabovenko:2016sxi,Shtabovenko:2020gxv} without any approximation. In the case of a pair of muons produced, there would be, in addition to the four Feynman diagrams in Figure~\ref{fig_MuonTrident:diagramas}, four more diagrams that take into account the effects of two identical particles in the final state. Given that we are interested in a high-energy incident muon ($\gg 10$~GeV), the effects of Fermi-Dirac statistics in this kinematic range are negligible, as shown in the references ~\cite{Kelner:1998mh,Russell:1971mf}.

\begin{figure}[t]
	\centering
	\begin{tabular}{ccc}
    \includegraphics[scale=1]{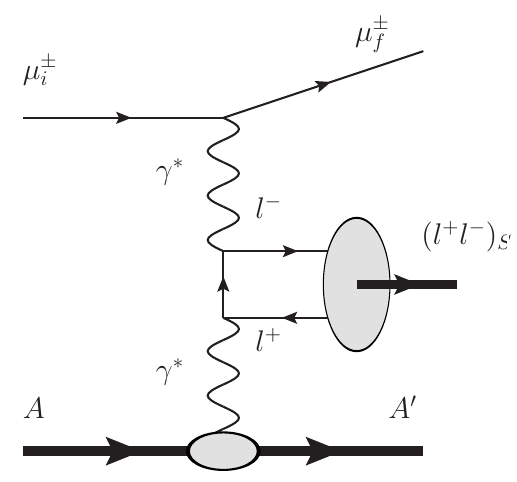} 
    \end{tabular}
    \caption{ Diagram for the electromagnetic production of a singlet bound state of leptons in muon-nucleus scattering. }
    \label{fig_MuonTrident:QEDbound}
\end{figure}

The formalism above can be directly extended to the calculation of the number of events associated with the production of QED bound states. In our analysis, we will focus on the production of singlet lepton bound states, $(l^{+} l^{-})_S$, which can be produced by the fusion of two photons, whose Feynman diagram is presented in Figure \ref{fig_MuonTrident:QEDbound}. Among the QED bound states with leptons of the same family, the only one observed experimentally is the positronium \cite{Deutsch:1951zza}. Recently, several works have discussed the possibility of discovering the true muonium, denoted by $(\mu^{+} \mu^{-})_S$, both through direct production in colliders \cite{Ginzburg:1998df,Brodsky:2009gx,Azevedo:2019hqp,Francener:2021wzx,Francener:2022egx,Dai:2024imb,Bertulani:2023nch,Gninenko:2025hsv,Liang:2025sol,Francener:2024eep,Gargiulo:2023tci,Gargiulo:2024zyc}, and through rare decays of heavier mesons into true muonium \cite{CidVidal:2019qub,Dai:2024yhy,Zhao:2025gia,Gargiulo:2025pmu}. The heaviest bound state of QED, tauonium, has gained attention in recent years \cite{Francener:2021wzx,Francener:2022egx,dEnterria:2022ysg,dEnterria:2022alo,dEnterria:2023yao,Fu:2023uzr,Francener:2024eep}, but due to its high mass, its discovery is not trivial. Using the Equivalent Photon Approximation, also known as the Weizsäcker-Williams formalism \cite{vonWeizsacker:1934nji,Williams:1934ad,Budnev:1975poe}, the cross section for the production of bounded leptons is factored with \cite{Dobrich:2015jyk}
\begin{equation}
\begin{aligned}
    \sigma_{\mu^{\pm}_{i} + W \rightarrow \mu^{\pm}_{f} + W + (l^{+} l^{-})_S} =
    \frac{1}{2\pi E_{\mu^{\pm}_{i}}} \int   \mathrm{d}p_{t}^{2} \, \mathrm{d}\phi \, \mathrm{d}E_{(l^{+} l^{-})_S} \, \mathrm{d(cos}\, \theta )
     f_{\gamma/\mu} \left(\frac{E_{(l^{+} l^{-})_S}}{ E_{\mu^{\pm}_i}}, q_{t}^{2} \right) \frac{\mathrm{d}\sigma_{\gamma W}}{\mathrm{d(cos}\,\theta )}, 
\end{aligned}
\label{eq_MuonTrident:muonium}
\end{equation}
with $p_{t}$ being the transverse moment of the bound state, $\theta$ the angle between the produced state and the direction of the incident muon, and $\phi$ the angle between the transverse momenta of the bound state and the photon. The lower integration in $p_{t}$ was performed from zero, since there are no infrared divergences. We emphasize here that the use of the equivalent photon approximation was restricted to the description of incident muons, while its use has been shown to overestimate the cross sections to describe the target nucleus in this energy regime \cite{Ballett:2018uuc}. The photon flux $f_{\gamma/\mu}(x, q_{t}^{2})$ associated with the incident muon can be written as \cite{vonWeizsacker:1934nji,Williams:1934ad,Budnev:1975poe,Bertulani:1987tz,Krauss:1997vr,Bertulani:2005ru,Goncalves:2005sn,Contreras:2015dqa,LHCForwardPhysicsWorkingGroup:2016ote,Klein:2019qfb,Klein:2020fmr}
\begin{eqnarray}
     f_{\gamma/\mu}(x, q_{t}^{2}) = 
     \frac{\alpha}{2 \pi} 
     \frac{1+(1-x)^{2}}{x}
     \frac{q_{t}^{2}}{(q_{t}^{2}+ x^{2}q_{t}^{2})^{2}},
     \label{eq_MuonTrident:EPA}
\end{eqnarray}
where $q_{t}$ is the transverse momentum transferred by the incident muon. We will not focus on discussing or deriving the EPA in detail here, but we recommend the reference \cite{Francener:2022sfw} to the interested reader, where this approximation was derived by both classical and quantum approaches. The last ingredient in Equation~({\ref{eq_MuonTrident:muonium}}) is the photonuclear cross section $\sigma_{\gamma W}$, which is evaluated with
\begin{eqnarray}
    \frac{\mathrm{d}\sigma_{\gamma W}}{\mathrm{d(cos}\,\theta )} = 
    \frac{Z^{2}\alpha^{6} (-4 E_{(l^{+} l^{-})_S}^{2} t - 16m_l^{4}) }{4 m_l^{2} t^{2}} F(|t|)^{2} \, ,
    \label{eq_MuonTrident:photonuclear}
\end{eqnarray}
where $F(|t|)^{2}$ is the nuclear form factor of the target nucleus with atomic number $Z$, which is a function of the square of the four-momentum transferred by the nucleus, $t$, which can be approximated by
\begin{eqnarray}
    t = 
    \frac{4m^{2}_l}{E_{(l^{+} l^{-})_S}^{2}} - p_{t}^{2}
    + 2 E_{(l^{+}  l^{-})_S} \, p_{t} \, \theta \, \mathrm{cos}\,\phi - E_{(l^{+} l^{-})_S}^{2}\, \theta^{2} \, .  
    \label{eq_MuonTrident:t}
\end{eqnarray}
Next, we will estimate the cross sections for the production of positronium, muonium, and tauonium states in $\mu W$ interactions at the FASER$\nu$ and FASER$\nu$2 detectors. Our main motivation is to verify whether these detectors would be able to measure the $(\mu^+ \mu^-)_S$ state, which has not yet been observed.

\section{Results for muon trident events at the LHC}

In this section, we will present our predictions for the trident scattering of muons with a tungsten target at the FASER$\nu$ and FASER$\nu$2 experiments. Initially, to verify our implementation of the trident process, we estimated the cross section for the production of muon pairs in the muon-carbon scattering for an incident muon of 160 GeV, which was previously calculated in the references ~\cite{Abbiendi:2024swt,Lee:2023cwb}. Neglecting the nuclear form factor, as in the references \cite{Abbiendi:2024swt,Lee:2023cwb}, we obtained 195(1) nb for the corresponding cross section, in agreement with the results of 196.3(9) nb and 196.74 nb derived using the MESMER monte carlo generator \cite{Abbiendi:2024swt} and an analytical expression presented in \cite{Lee:2023cwb}, respectively.

\begin{figure}[H]
	\centering
	\begin{tabular}{ccc}
    \includegraphics[scale=0.5]{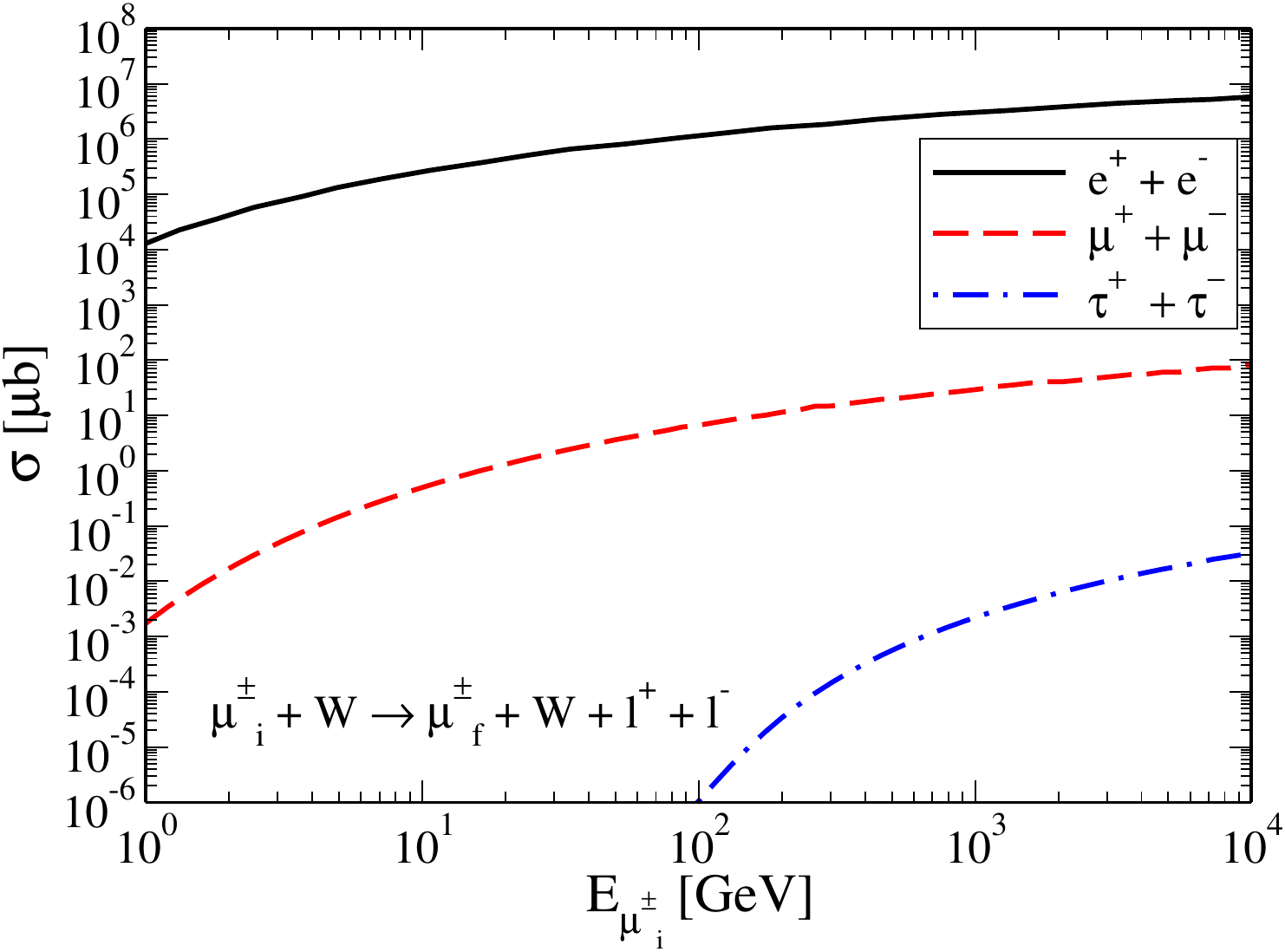}
    \end{tabular}
    \caption{ Total cross section as a function of the incident muon energy for different final states: electron pairs (solid black line), muon pairs (dashed red line), and tau pairs (dashed-dotted blue line). }
    \label{fig_MuonTrident:sigmaTotal}
\end{figure}

\begin{figure}[H]
	\centering
	\begin{tabular}{ccc}
    \includegraphics[scale=0.4]{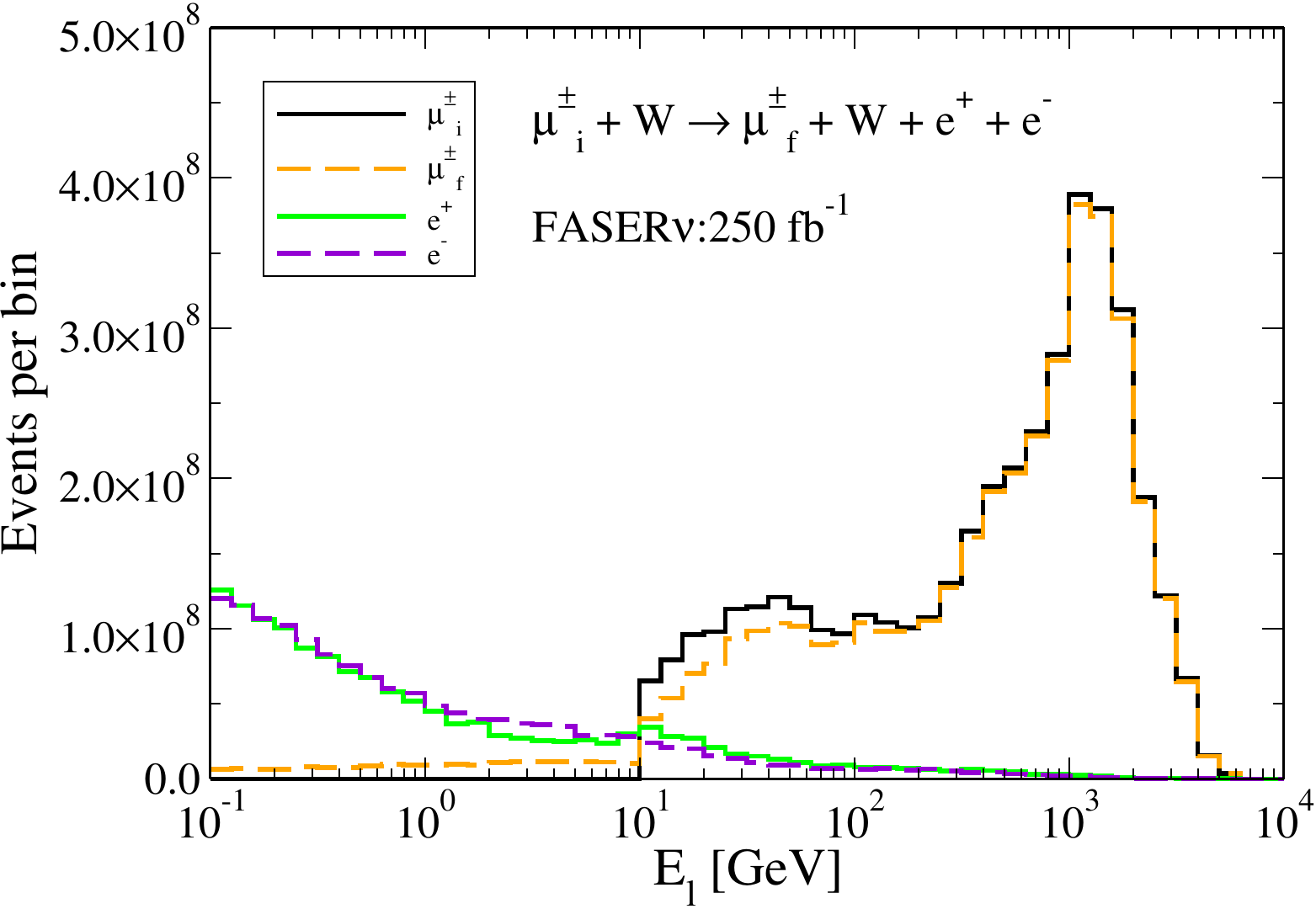} \\
    \includegraphics[scale=0.4]{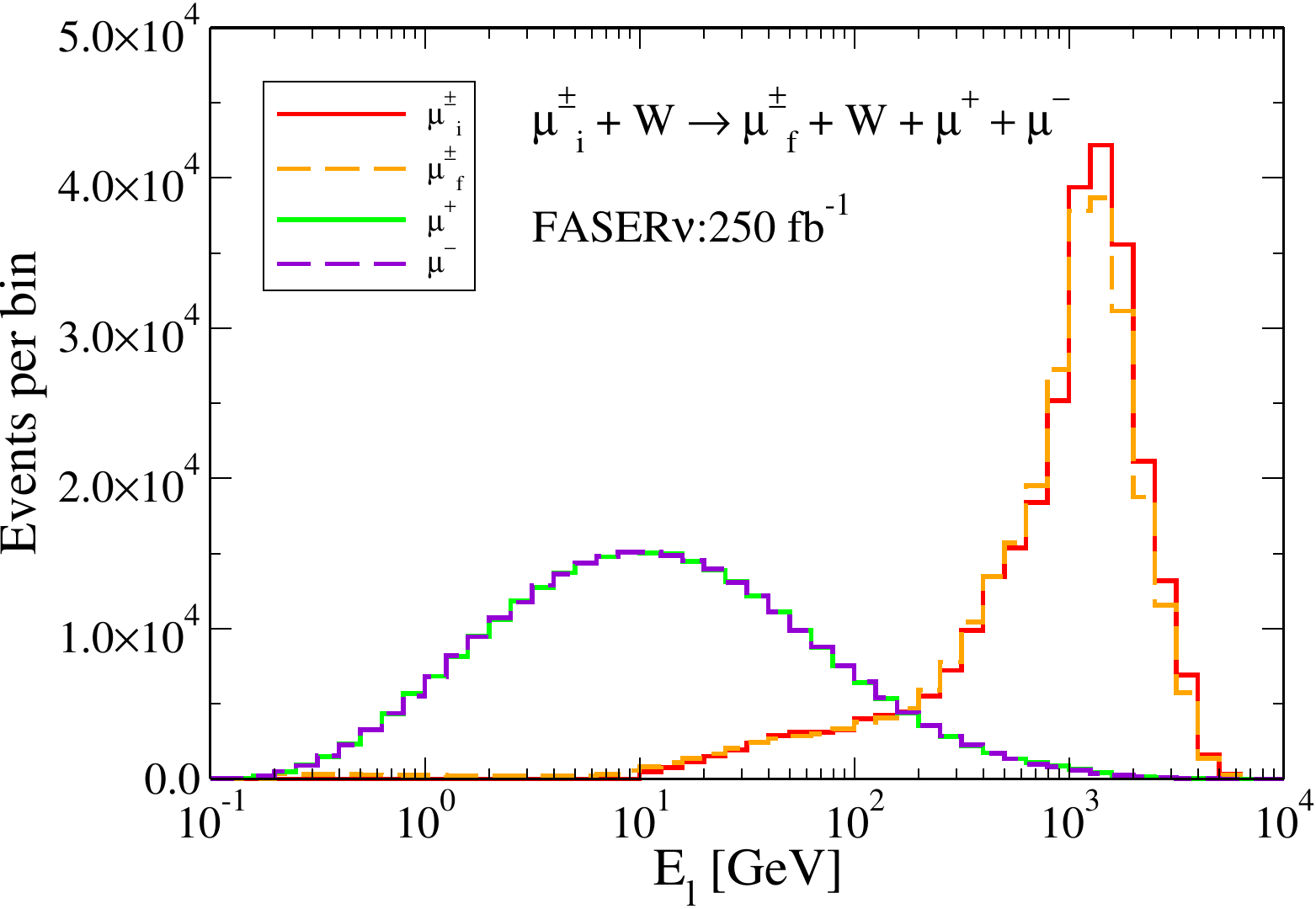} \\
    \includegraphics[scale=0.4]{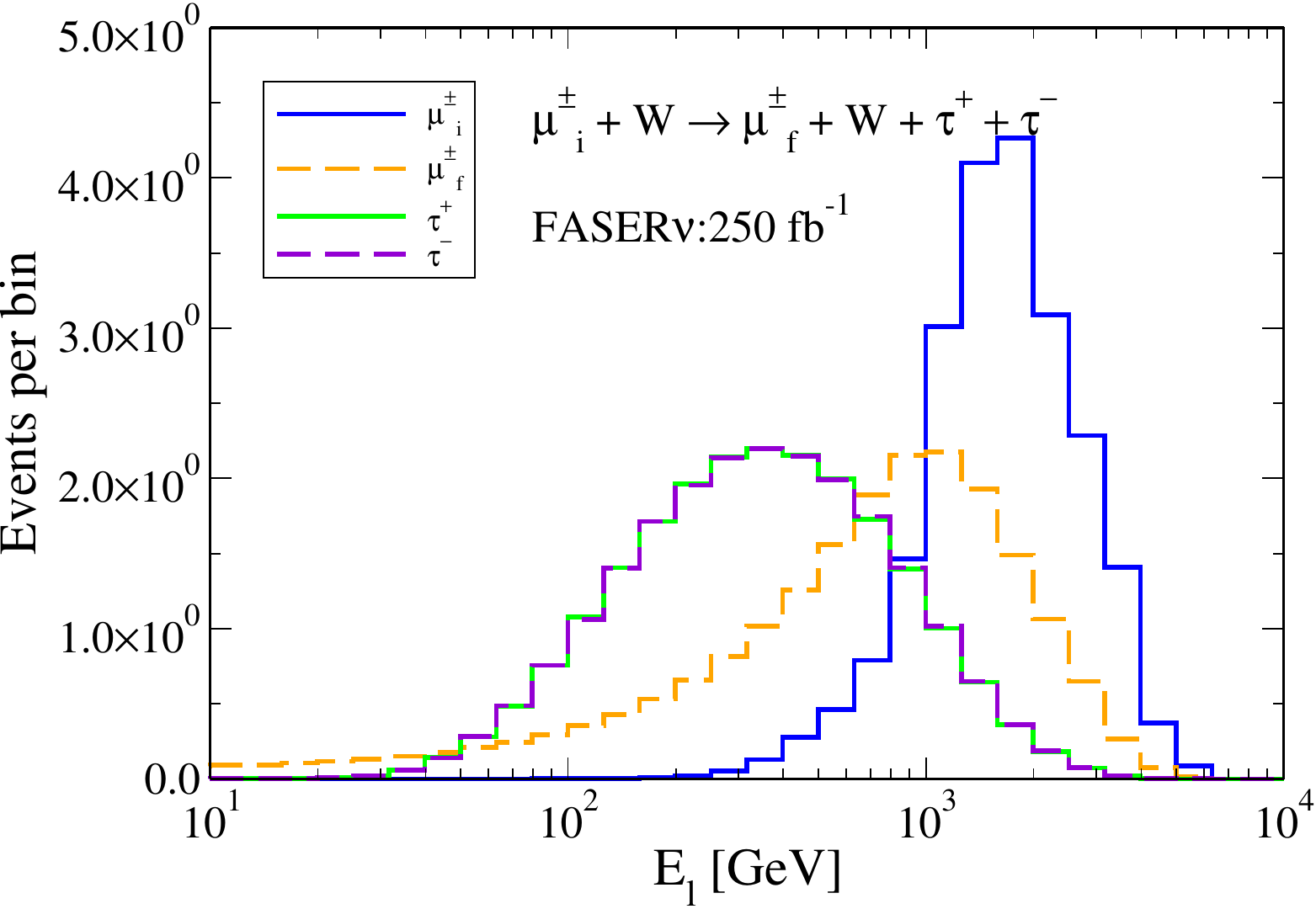}
    \end{tabular}
    \caption{ Predictions for the number of events associated with the production of an electron pair (top panel), muons (center panel), and taus (bottom panel) by an incident muon, binned at the energy of each lepton. The results are for the FASER$\nu$ during the Run 3 of the LHC, considering an integrated luminosity of 250 $^{-1}$. }
    \label{fig_MuonTrident:eventosBinEnergia}
\end{figure}

In Figure \ref{fig_MuonTrident:sigmaTotal} we present our results for the energy dependence of the total muon-tungsten cross section in the regime covered by the LHC forward detectors. We present the cross section for an electron pair (solid black line), muon pair (dashed red line), and tau pair (dashed-dotted blue line) produced in a coherent scattering, considering all the diagrams represented in Figure \ref{fig_MuonTrident:diagramas}. The cross section for an electron-positron pair is approximately $10^{5}$ times larger than that of the process with a muon pair produced, which is about $10^{3}$ times larger than that of the process with a tau pair produced by TeV energy muons. For lower incident muon energies, the cross sections for taus in the final state decrease rapidly, suppressed by the kinematic condition of $W_{\gamma^{*}\gamma^{*}} \geq 2 m_{\tau}$.

Next, using the total cross section shown in Figure \ref{fig_MuonTrident:sigmaTotal} and Equation \ref{eq:events}, we will calculate the expected number of events at the FASER$\nu$ detector during Run 3 of the LHC, considering a time-integrated luminosity of 250 fb$^{-1}$. It is important to note that the muon flux of \cite{Francener:2025pnr} does not take into account the entire length of the FASER$\nu$, but only 50 cm of the total 80 cm. The first and last 15 cm of the target are supposedly used to measure the momentum of incident and emergent muons through multiple scattering in tungsten. Therefore, the results for the number of events are underestimated by a factor of 1.6 if we consider the full length of the FASER$\nu$. Our results are obtained by summing the contributions of muons and antimuons and considering only incident (anti)muons with magnitudes greater than 10 GeV.

In Figure \ref{fig_MuonTrident:eventosBinEnergia} we present our predictions for the number of muon trident events at the FASER$\nu$ binned in the energy of the leptons involved in the scattering, considering the production of electron pairs (top panel), muon pairs (center panel), and tau pairs (bottom panel). We observe that the expected number of events per bin is approximately $10^{8}$ for electron pairs, decreasing by a factor of $10^{4}$ ($10^{8}$) for muon (tau) pairs. Our results indicate that the energies of the electron and positron produced in the interaction are small in the tungsten rest frame, typically less than 1 GeV, with the incident muon losing a small fraction of its energy. In contrast, the $\mu^+$ and $\mu^-$ leptons, produced by muons at the LHC, have energies on the order of $\mathcal{O}$(10 GeV).

Finally, for the production of the $\tau^+ \tau^-$ pair, we have that the taus in the final state will have an energy of $\mathcal{O}$(300~GeV) and are produced by muons with an energy of TeV. As a consequence, in this case, the incident muon loses a large fraction of its energy. At lower energies, the number of events is strongly suppressed due to the behavior of the cross section in this kinematic range (see Figure \ref{fig_MuonTrident:sigmaTotal}). It is important to emphasize that a tau with energy in the range of 100~GeV$-$1~TeV has a decay length on the order of 5~mm$-$5~cm, and it is possible to reconstruct its decay with the emulsion detector between the tungsten plates present in the FASER$\nu$.

\begin{figure}[t]
	\centering
	\begin{tabular}{ccc}
    \includegraphics[scale=0.32]{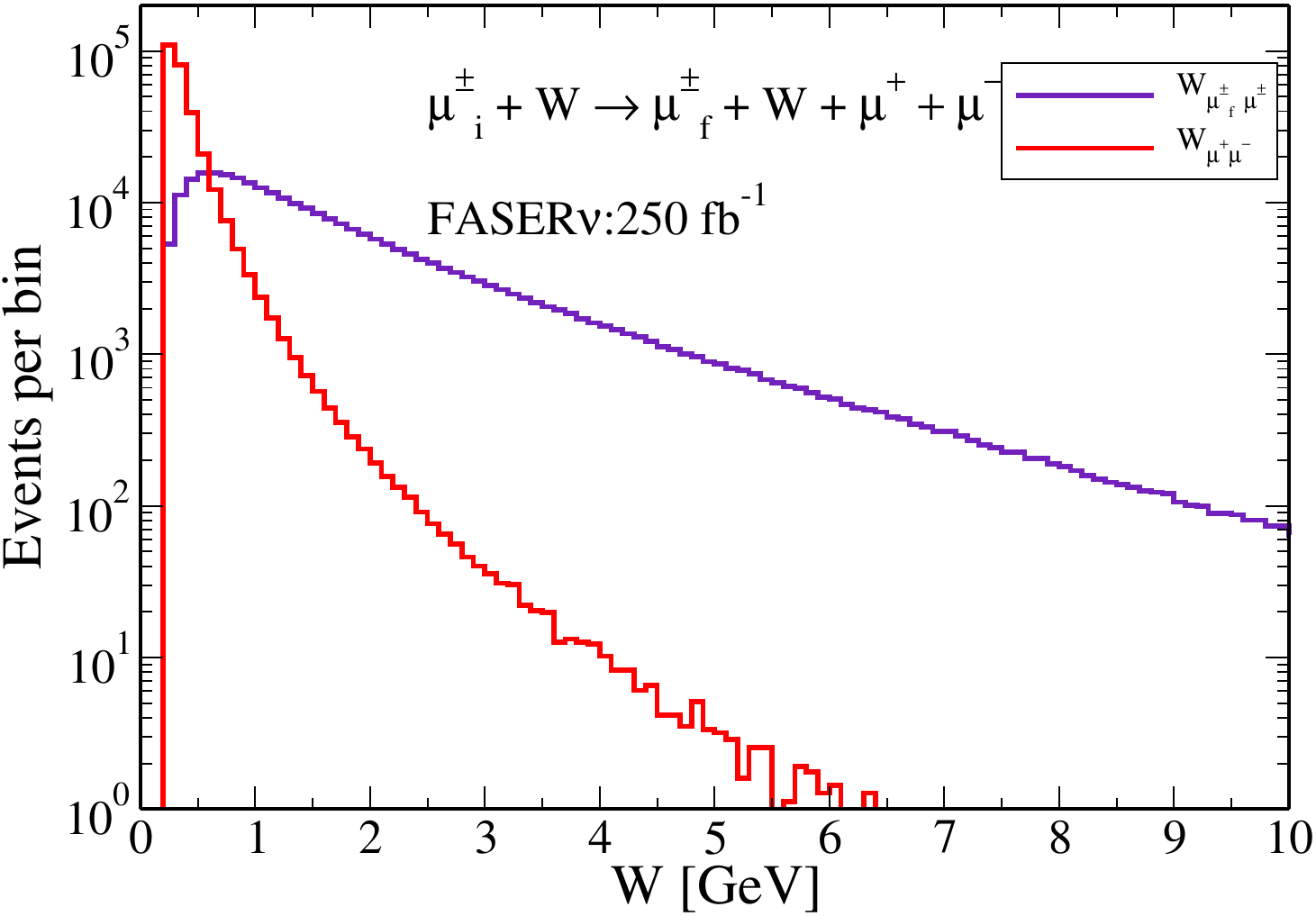} 
    \includegraphics[scale=0.32]{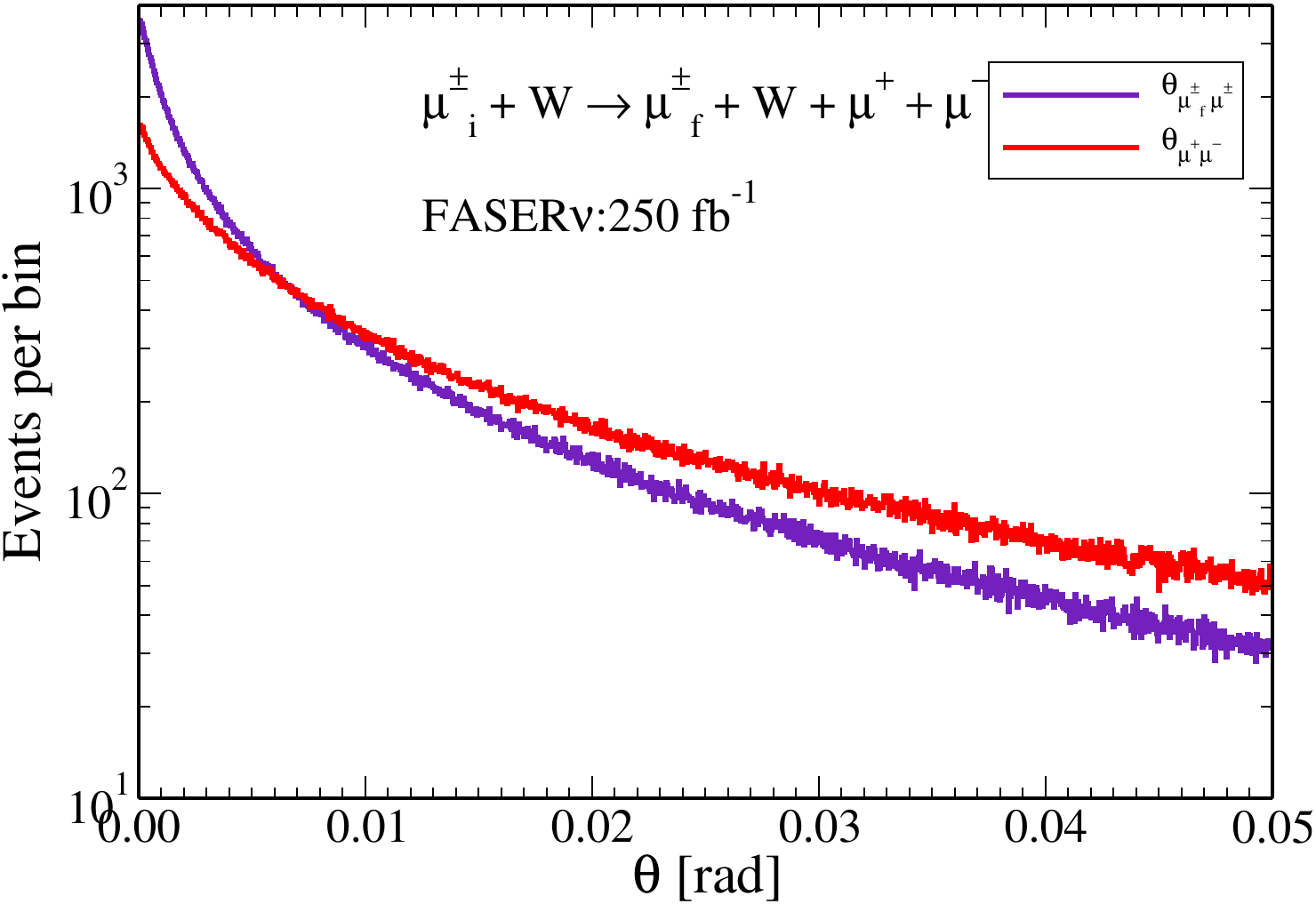}
    \end{tabular}
    \caption{ Predictions for the number of events associated with the electromagnetic production of $\mu^+\mu^-$ by an incident muon, binned at the invariant mass (left panel) and opening angle (right panel) of two muons in the final state. The results are for the FASER$\nu$ during Run 3 of the LHC, considering an integrated luminosity of 250~fb$^{-1}$. }
    \label{fig_MuonTrident:events_W_theta}
\end{figure}

The electromagnetic production of $\mu^+\mu^-$ by an incident muon is characterized by two identical particles in the final state. Since detectors cannot know which muon is produced at the vertex of the incident muon, it is interesting to analyze whether it is possible to distinguish the produced pair ($\mu^+\mu^-$) from a pair associated with a produced muon plus a muon originating from the vertex of the initial muon ($\mu^{\pm}\mu_f^\pm$). In Figure~\ref{fig_MuonTrident:events_W_theta} we show our results for the muon trident binned in the invariant mass (left panel) and opening angle (right panel) for these muon pair systems. We consider a range of 0.1 GeV for the invariant mass and 0.05 mrad for the opening angle, which is the angular sensitivity of the FASER$\nu$ detector \cite{FASER:2025qaf}. We are comparing the predictions for the produced muon pair (red) with the identical muon pair (violet).

The produced pair has a smaller invariant mass than the $\mu^{\pm}\mu_f^\pm$ pair, which is associated with the higher energy of the muon in the final state that was produced at the vertex of the incident muon. The produced pair also generally has a larger opening angle than the  $\mu^{\pm}\mu_f^\pm$ pair, given the small deflection of the incident muon after scattering.

	\begin{table}[t]
		\begin{tabular}{|c|c|c|c|c|}
			\hline
			\hline 
Final state		      & FASER$\nu$           & FASER$\nu$ ($E_l \ge 10$~GeV)& FASER$\nu$2 & FASER$\nu$2 ($E_l \ge 10$~GeV) \\
			\hline 	
			\hline
$e^{+} + e^{-}$ 	  & 4.10$\times 10^{10}$ & 7.23$\times 10^{8}$ & 6.10$\times 10^{12}$ & 1.08$\times 10^{10}$
\tabularnewline
$\mu^{+} + \mu^{-}$   & 2.61$\times 10^{5}$  & 1.03$\times 10^{5}$  & 3.87$\times 10^{7}$ & 1.53$\times 10^{7}$ 
\tabularnewline
$\tau^{+} + \tau^{-}$ & 21.83                & 19.97                & 3.25$\times 10^{3}$ & 2.97$\times 10^{3}$ 
\tabularnewline
			\hline 
            \hline
		\end{tabular}
		\caption{ Number of events associated with the electromagnetic production of dileptons in $\mu W$ interactions at the FASER$\nu$ and FASER$\nu$2 detectors, derived assuming integrated luminosities of $\mathcal{L}_{\rm pp}=250$ fb$^{-1}$ and 3 ab$^{-1}$, respectively. Results obtained with and without the inclusion of cuts in the minimum energy of each lepton in the final state. }
		\label{table_MuonTrident:Nevents_open}
	\end{table}

In Table \ref{table_MuonTrident:Nevents_open} we present the total number of events expected at the FASER$\nu$ and FASER$\nu$2 during Run 3 and high-luminosity era of the LHC, considering an integrated luminosity of 250 fb$^{-1}$ and 3 ab$^{-1}$, respectively. The impact of a cut in the energy of the final state leptons is also estimated. For the energy cut, we are assuming final state leptons with at least 10 GeV each. The total number of events expected at the FASER$\nu$ detector for electron, muon, and tau pairs is greater than $10^{10}$, $10^{5}$, and $10$, respectively. Considering the energy cut in the final state leptons, the number associated with electron, muon, and tau pairs decreases by factors of $56.7$, $2.5$, and $1.1$, respectively. The events at the FASER$\nu$2 at the HL-LHC increase by a factor of approximately 150. These results indicate that FASER$\nu$ will be able to perform a detailed study of the electromagnetic production of $e^+e^-$ and $\mu^+\mu^-$ pairs. Furthermore, the first observation of the production of the $\tau^+ \tau^-$ pair in the trident muon process is, in principle, feasible using this detector.

\begin{figure}[t]
	\centering
	\begin{tabular}{ccc}
    \includegraphics[scale=0.31]{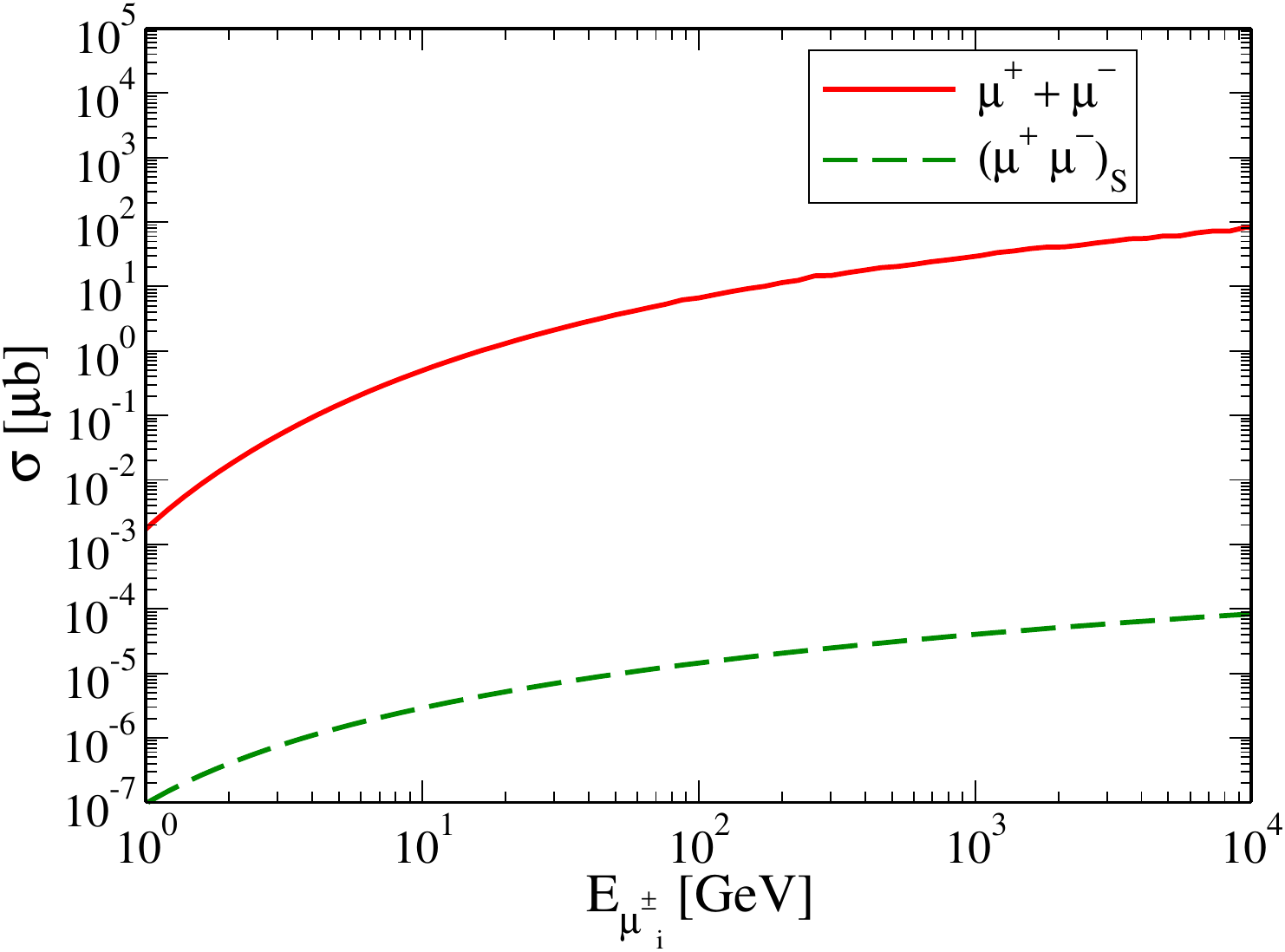} 
    \includegraphics[scale=0.32]{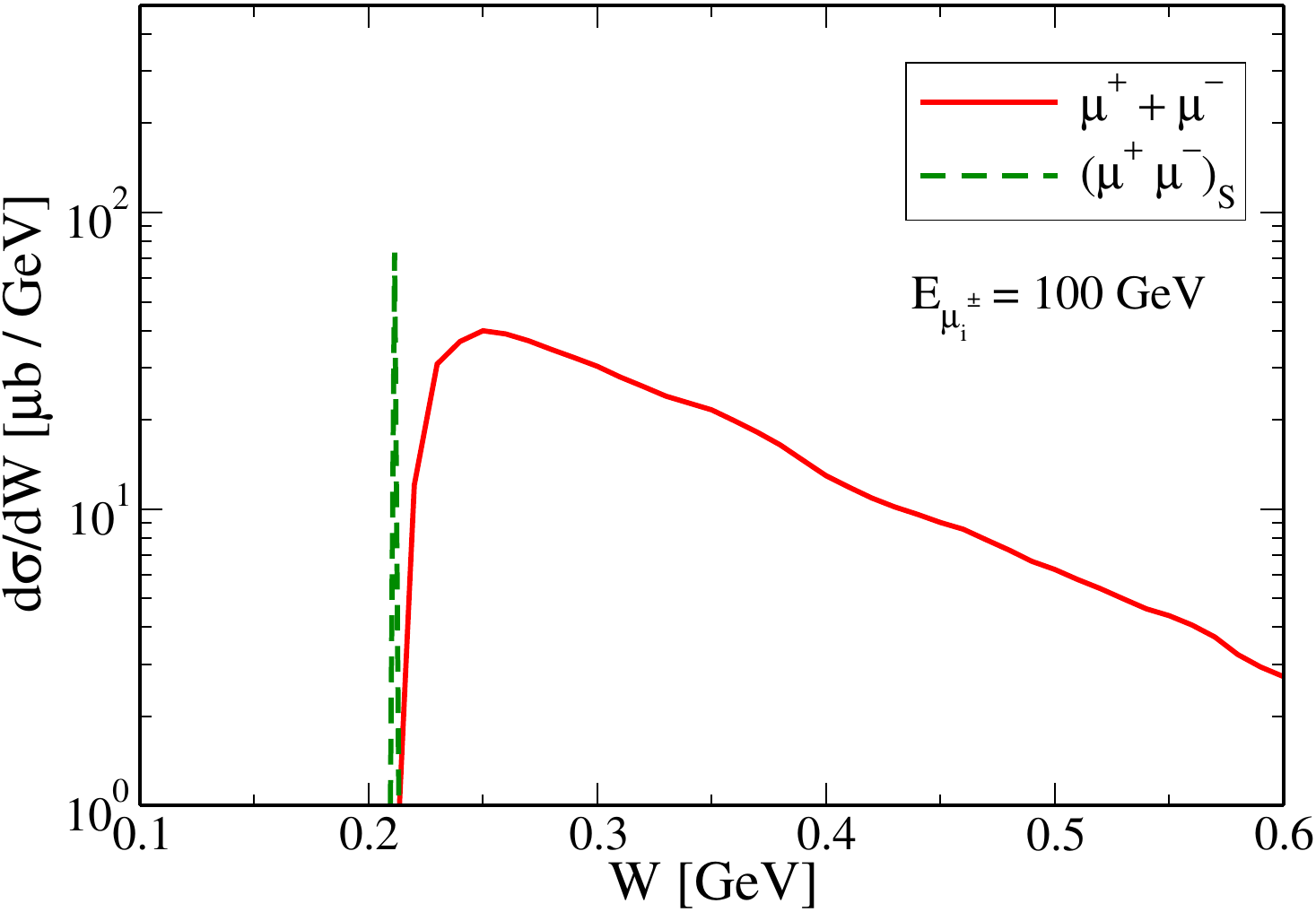}
    \end{tabular}
    \caption{ On the left, we present the total cross sections for the production of true muonium and open muon pairs in $\mu W$ interactions as a function of the incident muon energy. On the right, we present the differential cross section as a function of the invariant mass of the open and bound muon-antimuon pair in a singlet state, produced in $\mu W$ interactions for an incident muon of 100 GeV.
    }
    \label{fig_MuonTrident:sigmaMuonium}
\end{figure}

Finally, let us consider the electromagnetic production of a true muonium $(\mu^+ \mu^-)_S$ by an incident muon and estimate the associated cross section and the expected number of events. The signal of the true muonium at FASER-type detectors would be a pair of muons with approximately the same energy and highly collimated, since the true muonium is expected to open in the tungsten plates after propagating less than 0.1 mm, which before its decay \cite{CidVidal:2019qub}.

In Figure~\ref{fig_MuonTrident:sigmaMuonium} (left) we present the total cross section as a function of the incident muon energy. For comparison, the results for the production of the open pair $\mu^+\mu^-$ are also presented. Our results indicate that the cross section for the true muonium is smaller than that of the open-pair production by a factor of $10^{4}$ and $10^{6}$ for an incident muon of 1 GeV and 10 TeV, respectively. The invariant mass distribution of the cross section for a pair of muons produced by an incident muon of 100 GeV is presented in the right panel of Figure~\ref{fig_MuonTrident:sigmaMuonium}. While the open muon pair has a continuous spectrum starting at the production threshold, the bound state is characterized by a resonance peak at $W \approx 2m_\mu$ with a width on the order of $\alpha^5 m_\mu / 2$.

\begin{figure}[t]
	\centering
	\begin{tabular}{ccc}
    \includegraphics[scale=0.5]{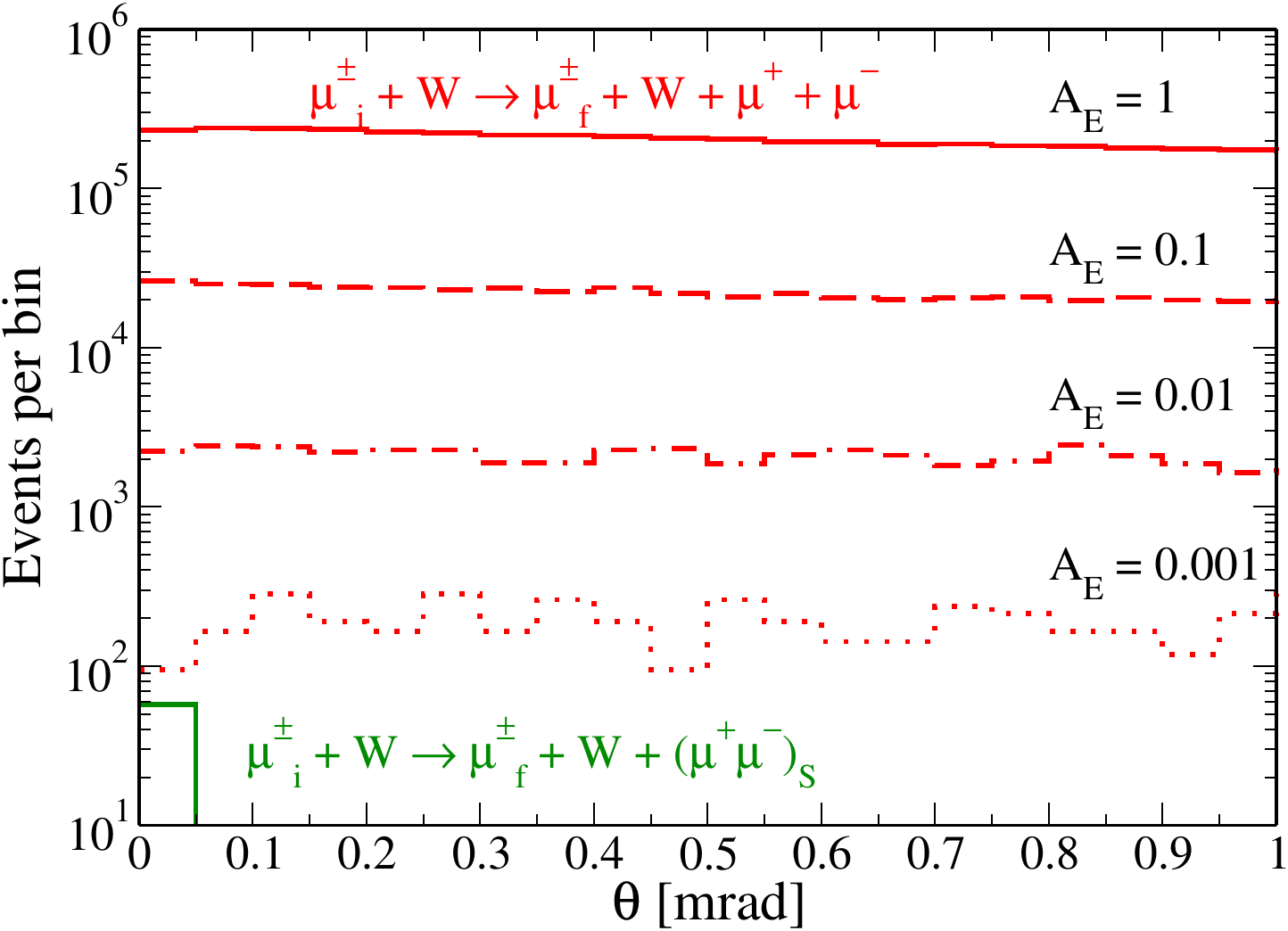}
    \end{tabular}
    \caption{ Predictions for true muonium production events and open-pair binned in the opening angle between the two muons produced at FASER$\nu$2. We compare the true muonium production with open-pair background, considering different cuts in the energy asymmetry of the produced pair.
    }
    \label{fig_MuonTrident:sigmaMuonium_theta}
\end{figure}

In Figure~\ref{fig_MuonTrident:sigmaMuonium_theta} we show the events of the true muonium in green, binned in the opening angle of the two muons produced. In red we show the background of muons resulting from open pairs produced, with different asymmetries in the energy of the two muons, which is defined as
\begin{eqnarray}
    A_E = \frac{E_{\mu^{+}}-E_{\mu^{-}}}{E_{\mu^{+}}+E_{\mu^{-}}} \, .  
    \label{eq_MuonTrident:assimetria}
\end{eqnarray}
The results in Figure \ref{fig_MuonTrident:sigmaMuonium_theta} indicate that to reduce the background to the order of magnitude of the muonium signal, measurements of the muon energies with an accuracy of approximately 0.1\% are necessary.

The number of expected QED bound state events at the FASER$\nu$ and FASER$\nu$2 during Run 3 and the high-luminosity era of the LHC, considering an integrated luminosity of 250 fb$^{-1}$ and 3 ab$^{-1}$, respectively, is presented in Table \ref{table_MuonTrident:Nevents}. Our estimate for true muonium events at FASER$\nu$ is less than one, but around 60 events at FASER$\nu$2. In contrast, for positronium production, we predict that the number of events will be greater than $10^5$ ($10^7$) at FASER$\nu$ (FASER$\nu$2). For completeness, we estimate the number of tauonium produced. This number is much less than 1 even at the FASER$\nu$2 during the high-luminosity regime of the LHC, making its observation unfeasible. When considering a 10 GeV cut in the energies of the final-state leptons, the number of events decreases by a factor of approximately $1.1$, $1.9$, and approximately $16.2$ for the production of tauonium, muonium, and positronium, respectively.

	\begin{table}[t]
		\begin{tabular}{|c|c|c|c|c|}
			\hline
			\hline 
Final state		        & FASER$\nu$          & FASER$\nu$ ($E_l \ge 10$~GeV) & FASER$\nu$2         & FASER$\nu$2($E_l \ge 10$~GeV) \\
			\hline 	
			\hline
$(e^{+} e^{-})_{S}$     & 1.51$\times 10^{5}$ & 9.30$\times 10^{3}$ & 2.24$\times 10^{7}$ & 1.38$\times 10^{6}$
\tabularnewline
$(\mu^{+} \mu^{-})_{S}$ & 0.38                & 0.20                & 57.04               & 29.76
\tabularnewline
$(\tau^{+} \tau^{-})_{S}$ & 2.58$\times 10^{-5}$ & 2.40$\times 10^{-5}$ & 3.83$\times 10^{-3}$ & 3.52$\times 10^{-3}$
\tabularnewline
			\hline 
            \hline
		\end{tabular}
		\caption{ Number of events associated with the electromagnetic production of QED bound states in $\mu W$ interactions at the FASER$\nu$ and FASER$\nu$2 detectors, derived assuming integrated luminosities of $\mathcal{L}_{\rm pp}=250$ fb$^{-1}$ and 3 ab$^{-1}$, respectively. Results obtained with and without the inclusion of cuts in the minimum energy of each lepton in the final state. }
		\label{table_MuonTrident:Nevents}
	\end{table}

\section{Conclusions}

In this chapter, we continued our studies of lepton tridents interactions with tungsten at the LHC far-forward detectors. Here, we extended our studies to muon-induced tridents in purely electromagnetic interactions. Our results indicate a large number of electron-positron and muon pairs during Run 3 of the LHC, suggesting that this process can be observed. In the case of produced muon pairs, this process has only been observed in two experiments to date, and its experimental cross section has large uncertainties. We also show that there will be dozens of tau pair events produced at the FASER$\nu$, indicating that this process may be observed for the first time. Using the Equivalent Photon Approximation formalism, we extended our studies to the production of QED bound states (positronium, muonium, and tauonium). The true muonium has not yet been discovered, and our results indicate approximately 60 events at the FASER$\nu$2 during the high-luminosity era of the LHC.

\chapter{Conclusions and perspectives}
\label{cap:conclusions}

In this work, we presented our phenomenological studies regarding the interaction of charged and neutral leptons, exploring different theoretical contexts focused on LHC experiments, as well as neutrino observatories. In particular, we studied effects arising from the polarization state of the tau produced in charged current interactions of tauonic neutrinos in an energy regime that has been observed at the LHC; the rare neutrino scattering process called the neutrino trident; in addition to effects on the deep inelastic scattering of muons and neutrinos at the LHC arising from QCD effects, such as intrinsic charm and nuclear effects in the modeling of parton distribution functions. Focusing on astrophysical neutrino physics in neutrino observatories such as IceCube, we studied the impacts of different descriptions of the Earth's structure on the transmission of high-energy neutrinos; subdominant channels in events characterized as tracks and beyond the Standard Model neutrino scattering in the universe until they reach the IceCube observatory.

In Chapter \ref{cap:Ice} we focused on studies related to neutrino interaction at the IceCube observatory. The results can be directly extended to other observatories that are beginning operations, such as Baikal-GVD, KM3NeT, as well as others proposed for the coming decades, such as P-ONE. In a first analysis, we showed that simplified models of the Earth, with three, two, and even one layer, can be reasonable alternatives for estimating neutrino transmission through the Earth in certain energy regimes and directions of incident neutrinos, compared to PREM. However, our results show that with longer data collection times and increasingly larger observatories, such as IceCube-Gen2, these simplified approximations of the Earth's structure may fail to provide satisfactory results.

Recent works on processes known as tracks at IceCube, characterized by highly energetic muons in the final state, has gained prominence in the literature in recently for several reasons. Among other things, these processes are interesting because high-energy muons travel several kilometers before decaying, so muons produced in charged current neutrino interactions outside the instrumented volume of IceCube can be measured. There is also interest in measuring the inelasticity of each event characterized as a track for those that start within IceCube, separating the energies deposited by the final muon and the other particles produced in the scattering. In this sense, in Chapter \ref{cap:Ice} we presented our results for subdominant processes in IceCube that can be identified as tracks, but which do not originate from muonic and tauonic neutrino interactions, as usually treated in previous works. In particular, we included in our analyses processes that produce heavy quarks that readily decay into semileptonic channels with muons in the final state, as well as channels producing real $W^{\pm}$ bosons. Our results showed that these channels, which are usually disregarded, make significant contributions, greater than 10\% of the events identified as tracks. We showed that one way to extract the contribution of these channels in the future is through reconstructed average inelasticity, given that the muon in these subdominant channels usually carries a smaller fraction of the event energy compared to the dominant muon channel resulting from muonic neutrino interactions.

In Chapter \ref{cap:Ice} of this thesis, we presented our results for the current sensitivity of IceCube, and also the projected sensitivity for IceCube-Gen2 with respect to the massive gauge boson $Z'$, which is predicted by the $L_\mu - L_\tau$ theory. This boson, beyond the Standard Model, can only be exchanged between second and third lepton generations, therefore it does not directly affect the interaction of neutrinos with targets in IceCube. However, this new boson can affect the propagation of high-energy neutrinos in interactions propagating through the cosmic neutrino background. We estimate the impacts of this boson on the astrophysical neutrino flux and on IceCube events classified in the HESE. Our results were obtained considering different scenarios for the redshift distribution of astrophysical neutrino sources, for the ordering of neutrino masses, as well as for the sum of the masses of neutrino mass eigenstates. Our analysis showed that, depending on the scenario adopted, IceCube can exclude with 95\% confidence level the existence of $Z'$ in mass and coupling regions that had not yet been probed by other experiments (except cosmological observations), and that IceCube-Gen2 has the potential to significantly increase these parameter space regions.

Starting in Chapter \ref{cap:pol}, we turned our attention to the neutrinos and muons produced at the LHC and which can be measured by the current FASER$\nu$ and SND@LHC experiments, as well as future detectors that are being proposed. In Chapter \ref{cap:pol}, we investigated some properties of the electroweak theory that do not manifest in electron and muon neutrino interactions, but may be important in tau neutrino interactions. Our studies focused on neutrinos on the order of 100 GeV -- 1 TeV, which is the expected region for most neutrino measurements at the LHC. First, we investigated the degree of polarization of taus produced in charged current neutrino interactions. Our results showed that these neutrinos are not completely polarized, and that the degree of polarization is around 94\%. We further showed that this degree of polarization of the taus affects the kinematic distributions of the pions resulting from the tau decay, with less energetic pions being more sensitive to the degree of polarization of the taus produced. In our analysis, we also investigated the impacts of the structure function $F_5$, which has never been measured before, on the total and differential cross sections of tau neutrinos with tungsten. We showed that the cross sections for neutrinos increase by 28\% (5\%) for initial neutrinos with energies of 100 GeV (1 TeV) when we insert the structure function $F_5$, therefore, it should play an important role in describing tau neutrino events at the LHC.

We know that in addition to neutrinos, muons produced at ATLAS are also capable of reaching frontal detectors like FASER at the LHC. In Chapter \ref{cap:MuonDIS} we showed that TeV-energy muons detected at FASER$\nu$ cover a kinematic region of $x$ and $Q^2$ similar to that expected for the EIC, and that more than $10^5$ DIS events are expected to be collected during Run 3 after applying realistic selection cuts with the detector. As applications of the advantages offered by this fixed-target collider, we showed that these data may be able to prove the existence of an intrinsic charm in the nucleon, which was proposed after EMC data in the 1980s and remains contested to this day. In addition to studying the large $x$ charm in the PDFs, we showed that muon-induced events at the FASER$\nu$ are sensitive to nuclear effects such as shadowing, anti-shadowing, and EMC effects. Conversely, neutrino-initiated events are not very sensitive to nuclear effects, given that they are usually events with higher $Q^2$. Our results indicate that the FASER and its proposed updates can be important tools for studying the universality of PDFs. In particular, measuring neutrinos and muons with the same experiment, subject to the same systematic uncertainties, can be a good way to verify the (in)compatibility of existing neutrino and charged lepton DIS data shown in \cite{Muzakka:2022wey}.

In Chapter \ref{cap:tridente} we explored the possibility of detecting the trident neutrino process with the FASER$\nu$2 detector, a process characterized by the production of a lepton pair in the fusion of a weak boson with a photon. Our results showed a total number of events on the order of 35 in this detector, and after some kinematic cuts aimed at excluding the main backgrounds, about 20 events remain, sufficient for the discovery of this process during the high luminosity regime of the LHC. In addition to studying the trident process in the Standard Model, we estimated the impact of new physics on events in the FASER$\nu$2 detector operating at the LHC and the FCC. We restricted ourselves to the impact of the $Z'$ of the $L_\mu - L_\tau$ theory. Our results showed that during the HL-LHC, FASER$\nu$2 via the trident process will not be able to probe any new regions of the $Z'$ parameter space, however its operation in the FCC will be able to test new regions, including for masses beyond the mass of the $Z^0$ boson, which are not accessible in the current ATLAS and CMS analyses.

Finally, in Chapter \ref{cap:MuonTridente} we studied the muon-induced trident at FASER$\nu$ and FASER$\nu$2. In the case of the muon-induced trident, at leading order the process is completely described with QED, with one photon exchanged by the projectile muon and one by the target nucleus. Our results show that for final states with electron-positron and muon-antimuon pairs, FASER$\nu$, even during Run 3 of the LHC, can improve existing measurements and, in addition, observe for the first time the production of taus pairs in muon-nucleus interactions. Besides the open lepton pairs produced, we studied the production of lepton bound states in the interaction of muons with tungsten. We show that positronium production will have high rates already in Run 3 of the LHC. However, the production of true muonium will only be significant in the FASER$\nu$2 operating during the high-luminosity regime of the LHC, and its separation from the background of open pairs produced tends to be a hard task.

Our results, discussed throughout this thesis, motivate several analyses that we can conduct in the future. As a particular case, we intend to study new processes involving LHC muons, such as bottom production, and testing the existence of intrinsic bottom in the nucleon. In another line of research, we are pursuing studies on the production of lepton pairs by these energetic muons traversing the FASER$\nu$, aiming at the application of realistic experimental cuts for the process. In another topic, we are studying the possibility of observing final-state interactions in exclusive processes of proton ejection in tungsten through interaction with the muon. Still focused on the potential of FASER$\nu$, now for neutrino detection, we are considering extensions of previous work to the effects of intrinsic charm in charged current interactions, where the interacting charm changes its flavor, but there is a low-energy charm in the fragmented nucleon that can be detected. There is also the possibility of exploring the strange-antistrange asymmetry in the nucleon, a topic closely related to intrinsic charm and which has evidence from both experimental measurements and lattice-QCD calculations. We aim to study this asymmetry in neutrino interaction events in the strange-charm transition, with charm detected in the final state.

Beyond the advances we aim to achieve regarding neutrino and muon topics at the LHC, there are topics we intend to explore related to IceCube. A natural extension of our analysis of the sensitivity of different Earth structure models is to explore more subtle variations in Earth models, such as PREM, similar to what was done in recent work on neutrino oscillation, but for the transmission of high-energy neutrinos. Another topic we can advance is the impacts of the $L_\mu - L_\tau$ theory, but now looking at the fraction of neutrino flavors that reach Earth, given that this quantity will be affected by the regeneration of the neutrino flux after the decay of $Z'$.

\renewcommand{\bibname}{ 
        \MakeUppercase{
        Bibliography
        } 
}

\bibliographystyle{naturemag}
\bibliography{Chapters/Bibliografia}

\appendix

\end{document}